# The Evolution and Development of the Universe

Edited by
Clément Vidal

Special Issue of the
First International Conference on
the Evolution and Development
of the Universe, 8-9 October 2008,
Ecole Normale Supérieure, Paris.

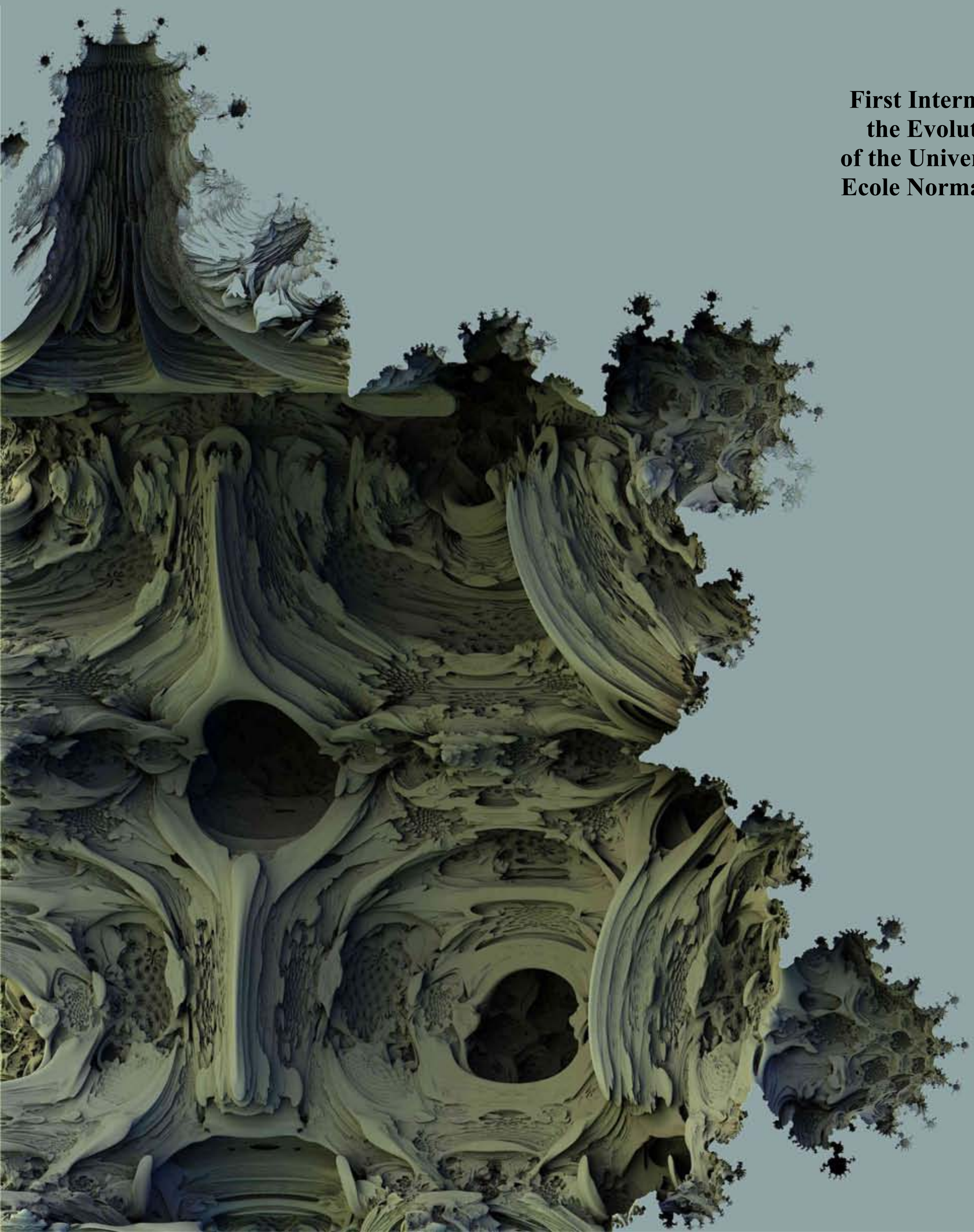

# Contents

## Part II – Biology

## Part III - Philosophy and Big Questions

Evo Devo Universe

# Preface

*The Evo Devo Universe Community*

In 2007, John M. Smart and I entered in contact discussing issues about universal change and broad cosmological and futuristic views. We noticed that scholars studying the cosmos where mainly into theoretical physics. It is of course an important approach, but it does not connect with life, intelligence and technology. Yet, we were also aware of dispersed insights in cosmology, theoretical and evolutionary developmental (evo-devo) biology and the complexity sciences, which are providing ways to understand our universe within a broader framework.

We thought that these results and hypotheses deserved to be explored, criticized, and analyzed by an international interdisciplinary research community which we set up in 2008: '**Evo Devo Universe (EDU)**'. Such a framework promises to advance our understanding of both unpredictable "evolutionary" processes and predictable "developmental" processes at all scales, including the human scale.

I welcome any researcher interested in these topics to join the research community at http://evodevouniverse.com

*The Conference on the Evolution and Development of the Universe*

John and I first started to work actively on building the EDU website in 2008 (http://www.evodevouniverse.com). However, we understood that having a website and a virtual community was not enough. To effectively collaborate, human beings still need to meet in flesh. We thus focused our energy into setting up "The First International Conference on the Evolution and Development of the Universe" project[1].

With the early support and interest of Alain Prochiantz (neurobiologist), Jean-Pierre Luminet (cosmologist) and Francis Heylighen (systems theorist) our conference project was funded by *The Complex Systems Institute, Paris (ISC-PIF)*.

Speakers at the conference were selected by our scientific committee based on their abstract. Yet, after the conference, I organized a more in depth peer-review on all papers, except one. Exceptionally, we decided not to peer-review Crane's paper, because it was since 1994 on the popular arXiv repository as a non-published pre-print, and researchers already referred to it "as it was". However, in this Special Issue, Crane's position was updated and clarified in his response to my commentary on his paper.

The special issue you have here is thus a selection of papers presented at the conference, plus a few other papers by Crane, Heylighen and Salthe. This volume is also the result of time and efforts from many referees (about 40 referee reports in total).

I admire and endorse Richard Gordon's courage and position to refuse anonymous refereeing. I find the arguments for non-anonymous refereeing very

1 EDU 2008, http://evodevouniverse.com/wiki/Conference_2008

compelling. Thomas Durt and Nicolás Lori also engaged in a transparent peer-review, which was long and sometimes difficult. Their disagreement about the definition of information in physics gave rise to a discussion which is reflected in this proceedings by Lori's commentary: *On definitions of Information in Physics*. I think that referees should be more recognized in this indispensable task for the scientific community. It is not normal that they do it on a voluntary basis and almost without recognition. A system to academically and publicly recognize this effort is yet to be invented.

After the peer-review, I organized a public invitation to comment on the selected papers[2]. I used the "Open Peer Commentary" (OPC) model for this purpose, which was initially used by journals *Current Anthropology* and *Behavioral and Brain Sciences*.

The purpose of OPC is to provide a concentrated constructive interaction between author and commentators. The target article, commentaries, and authors' responses are published together. I am delighted with the outcome of this call for commentaries, which generated 12 commentaries. Counting author's replies, this makes a total of 20 additional manuscripts to the main articles, giving additional discussion, critique, debate to this special issue. These exchanges show research as it is done, with explicit disagreements and debates. If authors agree to provide further responses to new commentaries, we would be glad to also continue this OPC process for later issues of *Foundations of Science*.

Finally, I apologize for the varying layouts of this proceedings preprint. I invite the demanding reader to wait a few months for the publication of all articles in *Foundations of Science*.

December 2009,
Clément Vidal.

2 http://evodevouniverse.com/wiki/EDU_2008_Call_for_Commentaries

# List of Contributors

Charles Auffray
Functional Genomics and Systems Biology for Health, CNRS, Institute of Biological Sciences, Villejuif, France.
charles.auffray@vjf.cnrs.fr

Alex Blin
Physics Department, Coimbra University, Coimbra, Portugal.
alex@fis.uc.pt

Jean Chaline
Biogeosciences Laboratory, Bourgogne University, Dijon, France.
jean.chaline@wanadoo.fr

Louis Crane
Mathematics Department, Kansas State University, Manhattan KS, USA.
crane@math.ksu.edu

Thomas Durt
Department of Physics, Vrije Universiteit Brussel, Brussels, Belgium.
thomdurt@vub.ac.be

Börje Ekstig
Department of Curriculum Studies, Uppsala University, Uppsala, Sweden.
borje.ekstig@did.uu.se

Horace Fairlamb
Departement of Arts & Science, University of Houston-Victoria, Victoria, TX, USA.
horus471@gmail.com

Jan Greben
Council for Scientific and Industrial Research, Pretoria, South Africa.
jGreben@csir.co.za

Rob Hengeveld
Institute of Ecological Science, Vrije Universiteit, Amsterdam, The Netherlands.
rob.hengeveld@gmail.com

Francis Heylighen,
Department of Philosophy, Vrije Universiteit Brussel, Brussels, Belgium.
fheyligh@vub.ac.be

Gerard Jagers op Akkerhuis
Centre for Ecosystem Studies, Wageningen University and Research Centre, Wageningen, The Netherlands.
Gerard.Jagers@wur.nl

Giuseppe Longo
Computer Science Department, CNRS, Ecole Normale Supérieure, Paris, France.
Giuseppe.Longo@ens.fr

Nicolás Lori
Institute of Biomedical Research in Light and Image, Coimbra University, Coimbra, Portugal.
nicolas.lori@gmail.com

Denis Noble
Department of Physiology, Anatomy and Genetics, Oxford University, Oxford, United Kingdom.
denis.noble@physiol.ox.ac.uk

Laurent Nottale
Universe and Theories Laboratory (LUTH), CNRS, Paris Observatory and Paris-Diderot University, Meudon, France.
laurent.nottale@obspm.fr

Stanley Salthe
Biological Sciences, Binghamton University Emeritus, Biology, City University of New York, New York, USA.
ssalthe@binghamton.edu

John Stewart
Member of Evolution, Complexity and Cognition Research Group, The Free University of Brussels (VUB), Melbourne, Australia.
future.evolution@gmail.com

Rüdiger Vaas
Center for Philosophy and Foundations of Science, University of Giessen, Germany.
Ruediger.Vaas@t-online.de

Gertrudis Van de Vijver
Centre for Critical Philosophy, Ghent University, Ghent, Belgium.
Gertrudis.VandeVijver@ugent.be

Nico M. van Straalen
Department of Ecological Science, VU University, Amsterdam, The Netherlands.
nico.van.straalen@falw.vu.nl

Clément Vidal
Department of Philosophy, Vrije Universiteit Brussel, Brussels, Belgium.
clement.vidal@philosophons.com

# Introduction

This introduction provides a quick overview of what you will find in this volume.

*Scale Relativity - Nottale, Chaline, Auffray and Noble.*

**Laurent Nottale** presents a review of Scale Relativity (ScR), a natural extension of Einstein's general relativity. The principle of relativity has been successfully applied to position, orientation and motion and is at the core of all physical theories. However, problems related to scales are ubiquitous in science. In particular, physics struggles for decades to connect in a meaningful way quantum theory (microphysics) with classical physics (macrophysics). In a similar manner, relating our knowledge of macrophysics to the cosmological scale leads to arduous problems in cosmology, for example about dark matter or vacuum energy density.

Scale Relativity proposes "to extend theories of relativity by including the scale in the very definition of the coordinate system, then to account for these scale transformations in a relativistic way." (p57) How is it possible? And why did Einstein not found this extension before? As often in the history of physics, part of the answer lies in the mathematical tools.

Einstein struggled years to develop the general relativity theory of gravitation, because it involved non-euclidian geometries. These geometries are (or were) counter-intuitive to manipulate and understand, and they were not used in physics before. Similarly, ScR uses a fundamental mathematical tool to deal with scales: fractals. This leads to an extension of general relativity (i.e. it includes its previous results) by constructing a theory of fractal space-time. Fractals were only studied in depth by Mandelbrot in the 1950's, although they were known by mathematicians much before (e.g. Georg Cantor's triadic set).

This simple yet fundamental approach generates a proliferation of results, which are both theoretical and with concrete applications and validated predictions. Let us mention of few of them. A new light on quantum mechanics can be thrown, since it is possible to derivate postulates of quantum mechanics with ScR[3]. A macroscopic Schrödinger equation is derived, which brings statistical predictability characteristic of QM into other scales in nature. For example, the position of exoplanets can be predicted in a statistical manner. The theory predicts that they have more chances to be found at such or such distance from their star. On cosmological scales, ScR also predicted with great precision the value of the cosmological constant (see section 3.1.2). There are many other fascinating implications of the theory, not only in physics, but also in earth sciences, history, geography and biology. All these are reviewed in Nottale's paper.

The theory can also be applied in systems biology, as testified in more details in **Charles Auffray**'s and **Denis Noble**'s commentary. In his reply, Nottale describes quite in general how ScR can be applied to various systems,

3 Nottale, L., and M. N. Celerier. 2007. Derivation of the postulates of quantum mechanics from the first principles of scale relativity. *Journal of Physics A-Mathematical and Theoretical* 40, no. 48: 14471-14498. http://arxiv.org/abs/0711.2418.

and gives an indication on how this can be applied in biology more specifically. Using updated data, he also corrects the prediction of the full arctic ice melting, which is now predicted to be in 2014-2015 instead of 2011...

Scale Relativity is a fundamental approach to science and has consequences for nearly all sciences. ScR suggests that cosmology, fundamental particle physics, structure formation, biology, geology, economy and other fields might be approached with tools derived from the same few principles exposed in this paper. Although, as Nottale explains, a lot of work still has to be done, the exposed vision is extraordinarily far reaching and inspiring. For these reasons, I am delighted to deliver Laurent Nottale the **EDU 2008 Best Paper Award**.

**Jean Chaline**'s paper applies ScR principles to biological and paleontological data. He shows in his paper that log-periodic behaviors of acceleration or deceleration can be applied to branching macroevolution and to the time sequences of major evolutionary leaps. This includes the global tree of life, sauropod and theropod dinosaurs postural structures, North American fossil equids, rodents, primates, echinoderms clades and human ontogeny.

*Causality and symmetries - Heylighen and Longo*

Back to fundamental physics, **Francis Heylighen** conducts a reflection on our most fundamental scientific concepts: time and causality. He explains how the concept of self-organization can be applied in this context. Starting from a random graph of events, he shows how a transitive closure can transform it into a partial order relation of precedence and thus generate a causal structure.

In his commentary, the mathematician **Giuseppe Longo** explores the relationship between the properties of symmetry captured by mathematical group structures and logical structures used by Heylighen. In his response, Heylighen formulates another fundamental problem, which is that causal laws of classical mechanics are reversible in their formulations, whereas "the laws of thermodynamics, on the other hand, are intrinsically irreversible as they imply a maximization of entropy".

*Greben's Cosmological Model*

**Jan Greben** presents a cosmological model where the vacuum energy dominates the universe. He claims that this model avoids the horizon and cosmological constant problems, and also provides a possible explanation for dark matter. Greben also includes a model of the evolution of the early universe, which describes the formation of elementary particles from a supposed classical state with perfect symmetry and zero entropy.

*Quantum Darwinism – Durt, Lori and Blin*

**Thomas Durt**'s paper *Anthropomorphic Quantum Darwinism as an explanation for Classicality* tackles a very basic question about our nature as observers. Why is our representation of the world classical although at smallest scales, it is quantum? Durt reminds us that "millions of years of evolution modelled our vision of the world, leading us to become blind to aspects of it that are not advantageous from the point of view of the acquisition of useful information." He proposes along the lines of Zurek's Quantum Darwinism approach that the basis in which we measure the external world obeys an optimality principle that results from a biological selection mechanism: "the fittest is the best-informed." More precisely, he aims at establishing the identity between "classical islands" and our cognitive representation of what elementary particles are.

**Nicolás Lori**'s commentary *On definitions of information in physics* and Durt's reply *Competing definitions of Information versus Entropy in Physics* is a discussion about the concept of **information** in physics. It is important to note that this concept, although fundamental, is difficult to use, and not always defined in the same manner.

**Nicolás Lori** and **Alex Blin**'s paper *Application of Quantum Darwinism to Cosmic Inflation: an example of the limits imposed in Aristotelian logic by information-based approach to Gödel's incompleteness* constitutes another application of Quantum Darwinism, here in cosmology. To formalize quantum darwinism and cope with the random extinction of information, they define and distinguish between Formal Axiomatic Systems (FAS) and Darwinian Axiomatic Systems (DAS). After reminding us that "Gödel's incompleteness theorems showed that a non-trivial axiomatic system cannot be both complete and consistent" they add that "the FAS is the choice for consistency and the DAS is the choice for completeness". This approach is then applied to cosmic inflation. This constitute a very original reflection yet involving speculative hypotheses like cosmic inflation or baby universes.

*Life and Complexity: Jagers op Akkerhuis, Hengeveld, Van Straalen and Ekstig*

In "Life, the organism and death", **Gerard Jagers op Akkerhuis** conducts a reflection on the definition of life. He first constructs a broader context, a "theory of life", from which he derives a definition of life. He uses hierarchy theory to this end, and defines a ranking called the "operator hierarchy". From this hierarchical perspective, he argues that to define life, construction is more important than metabolism, growth or reproduction. In his commentary *Definitions of life are not only unnecessary, but they can do harm to understanding*, Rob Hengeveld radically criticises this endeavour, which he considers of no scientific interest. Nico van Straalen insists on the importance of the transition from non-life to life in his commentary *The issue of "closure" in Jagers op Akkerhuis's operator theory.* In his response, Jagers op Akkerhuis addresses this issues and argues in particular that, as his title expresses, *Explaining the origin of life is not enough for a definition of life.*

**Börje Ekstig** argues that there is an acceleration of complexity in evolution. His paper, *Complexity and Evolution - a study of the growth of complexity in organic and cultural evolution*, presents a model integrating evolution on large time scales, and the development of an individual. He suggests some methods to estimate the growth of complexity in evolution, and also connects biological and cultural evolution.

*Natural Philosophy with Salthe, Fairlamb and Van de Vijver.*

**Stanley Salthe**'s reflection on the *Development (and evolution) of the Universe* presents a broad and challenging outlook in natural philosophy. He argues that Aristotelian causal analysis, and especially final causes, can be helpful for dealing with complex systems. More precisely, he takes a developmental perspective to model our world, and uses a "specification hierarchy" to describe the emergence of systems of higher levels in the universe. **Horace Fairlamb** questions the metaphysics behind Salthe's paper, and asks whether implications of a supposed developmental trajectory are necessarily materialistic. In her commentary, **Gertrudis Van de Vijver** explores the concept of "critique" in relation to objectivism and dogmatism. Instead of Aristotle, she also suggests that Kant and Husserl might be more appropriate philosophers to be inspired by to tackle the issue of final causes, or teleology.

*Big Questions with Crane, Vidal, Greben, Vaas, Stewart and Rottiers.*

**Louis Crane**'s paper explores philosophical implications of a supposed quantum theory of gravity. He builds on Lee Smolin's Cosmological Natural Selection (CNS) and conjectures that future civilizations will want to create black holes. In my commentary to this seminal paper, I distinguished two purposes of such a black hole engineering endeavor: either for (i) energy production or for (ii) universe production. In his reply, Crane makes clear that the purpose of energy production now interests him most, as "a practical suggestion for the middle-term future".

In my paper, I explore computational and biological analogies to address the fine-tuning issue in cosmology. I show that current solutions are not actually satisfying, to motivate other kinds of approaches. First, I analyze what are physical constants from a physical perspective. Then, I explore computational and biological analogies to tackle this issue, and propose and extension of CNS, stimulated by ideas from Crane and other authors. Inspired by a biological analogy, I named this extension of CNS "Cosmological Artificial Selection" (CAS).

In his commentary *On the nature of initial conditions and fundamental parameters in physics and cosmology*, **Jan Greben** criticizes the idea of fine-tuning of initial conditions, suggesting that his cosmological model does not need such fine-tuning. In my reply, I argue that this reasoning only holds if we take seriously the idea that “nature is quantum mechanical”. I also clarify some epistemological issues related to fine-tuning.

**Rüdiger Vaas** in his commentary *Cosmological Artificial Selection: Creation out of something?* discusses some far-reaching problems of CAS. I address them and clarify their scientific or philosophical nature in my response.

In *The meaning of life in a developing Universe*, **John Stewart** takes a broad evolutionary view on the cosmos. After summarizing the arguments supporting the proposition that biological evolution has a trajectory, he attempts to extend it to the universe. He also builds on the work of Crane and other authors to explore the idea that "our universe and the evolution of life within it is a developmental process that has itself been shaped by evolutionary processes of even wider scale". He further argues that at a particular point, evolution will continue to advance only if we decide to advance the evolutionary process intentionally.

**Franc Rottiers** criticizes Stewart's proposition that humanity has discovered the trajectory of past evolution, as just one possible perspective. Stewart suggests to replace postmodern scepticism and relativism "with an evolutionary grand narrative that can guide humanity to participate successfully in the future evolution of life in the universe".

My commentary to Stewart's paper is, as the title suggests, an *Analysis of Some Speculations Concerning the Far-Future of Intelligent Civilizations*. It concerns some (relatively) minor speculative issues where we disagree. Those disagreements are clarified and qualified in Stewart's response.

I hope the reader will be inspired by the insights gathered in this volume to further build an even more comprehensive view on the cosmos.

December 2009,
Clément Vidal.

# Acknowledgements

John Smart and I would like to thank the following people. First, we would like to dedicate this volume to the memory of Peter Winiwarter, who tragically ended his life this summer 2009. We have always been very grateful of his active help and enthusiasm, especially for the first steps of EDU community building. He will always remain in our memories.

We are grateful to Daniel White for his authorization to use his wonderful 3D fractal rendering of the Mandelbrot set, which is on the cover of this proceedings. I find those 3D fractals very fascinating, because they stem from simple mathematical formulas, and generate incredibly complex structures, which are quasi-biological. More beautiful images can be found on his website: http://www.skytopia.com/project/fractal/mandelbulb.html

I warmly thank Alain Prochiantz for his trust and support for organizing this conference; Jean-Pierre Luminet for his early interest and *The Complex Systems Institute, Paris (ISC-PIF)* for setting up this interdisciplinary "call for ideas" and for funding EDU2008.

We also very much appreciated Georges Ellis' early endorsement of the EDU project, and 45 other researchers from all disciplines[4]. I am grateful to Diederik Aerts for his willingness to set up with me open peer commentary in his journal, *Foundations of Science*. The advices and support of Stevan Harnad were also very helpful in this process.

A special thank is due to the EDU scientific board[5], James A. Coffman, James N. Gardner, Carlos Gershenson, Richard Gordon, Stevan Harnad, Francis Heylighen, David Holcman, Nicolás Lori, Laurent Nottale for their numerous advices. We also thank Mehdi Khamassi, Arnaud Blanchard, Mariam Chammat, Elizabeth Molière, Lina Boitier and Dominique Borg for their enthousiastic help during the conference.

I am very grateful for the remarkable work anonymous referees did with their critical assessment and in their help to select and improve the papers.

I am immensely grateful to my colleague and friend John Stewart for his efforts of coming from Australia to attend the conference and for sharing such inspiring insights and visions on evolution.

I am most grateful to my colleague and friend John Smart for his energy, perseverance, insights, vision, and for being an amazing colleague to work with. It has been an empowering adventure to work with him, and I would like to express my deepest gratitude for everything I learnt thanks to him.

December 2009,
Clément Vidal and John Smart.

4 http://evodevouniverse.com/wiki/People
5 http://evodevouniverse.com/wiki/Scientific_Board

# Scale relativity and fractal space-time: theory and applications

Laurent Nottale
CNRS, LUTH, Paris Observatory and Paris-Diderot University
92190 Meudon, France
laurent.nottale@obspm.fr

December 7, 2009

**Abstract**

In the first part of this contribution, we review the development of the theory of scale relativity and its geometric framework constructed in terms of a fractal and nondifferentiable continuous space-time. This theory leads (i) to a generalization of possible physically relevant fractal laws, written as partial differential equation acting in the space of scales, and (ii) to a new geometric foundation of quantum mechanics and gauge field theories and their possible generalisations.

In the second part, we discuss some examples of application of the theory to various sciences, in particular in cases when the theoretical predictions have been validated by new or updated observational and experimental data. This includes predictions in physics and cosmology (value of the QCD coupling and of the cosmological constant), to astrophysics and gravitational structure formation (distances of extrasolar planets to their stars, of Kuiper belt objects, value of solar and solar-like star cycles), to sciences of life (log-periodic law for species punctuated evolution, human development and society evolution), to Earth sciences (log-periodic deceleration of the rate of California earthquakes and of Sichuan earthquake replicas, critical law for the arctic sea ice extent) and tentative applications to systems biology.

## 1 Introduction

One of the main concern of the theory of scale relativity is about the foundation of quantum mechanics. As it is now well known, the principle of relativity (of motion) underlies the foundation of most of classical physics. Now, quantum mechanics, though it is harmoniously combined with special relativity in the framework of relativistic quantum mechanics and quantum field theories, seems, up to now, to be founded on different grounds. Actually, its present foundation is mainly axiomatic, i.e., it is based on postulates and rules which are not derived from any underlying more fundamental principle.

The theory of scale relativity [67, 68, 69, 72, 79, 95] suggests an original solution to this fundamental problem. Namely, in its framework, quantum mechanics may indeed be founded on the principle of relativity itself, provided this principle (applied up to now to position, orientation and motion) be extended to scales. One generalizes the definition of reference systems by including variables characterizing their scale, then one generalizes the possible transformations of these reference systems by adding, to the relative transformations already accounted for (translation, velocity and acceleration of the origin, rotation of the axes), the transformations of these scale variables, namely, their relative dilations and contractions. In the framework of such a newly generalized relativity theory, the laws of physics may be given a general form that transcends and includes both the classical and the quantum laws, allowing in particular to study in a renewed way the poorly understood nature of the classical to quantum transition.

A related important concern of the theory is the question of the geometry of space-time at all scales. In analogy with Einstein's construction of general relativity of motion, which is based on the generalization of flat space-times to curved Riemannian geometry, it is suggested, in the framework of scale relativity, that a new generalization of the description of space-time is now needed, toward a still continuous but now nondifferentiable and fractal geometry (i.e., explicitly dependent on the scale of observation or measurement). New mathematical and physical tools are therefore developed in order to implement such a generalized description, which goes far beyond the standard view of differentiable manifolds. One writes the equations of motion in such a space-time as geodesics equations, under the constraint of the principle of relativity of all scales in nature. To this purpose, covariant derivatives are constructed that implement the various effects of the nondifferentiable and fractal geometry.

As a first theoretical step, the laws of scale transformation that describe the new dependence on resolutions of physical quantities are obtained as solutions of differential equations acting in the space of scales. This leads to several possible levels of description for these laws, from the simplest scale invariant laws to generalized laws with variable fractal dimensions, including log-periodic laws and log-Lorentz laws of "special scale-relativity", in which the Planck scale is identified with a minimal, unreachable scale, invariant under scale transformations (in analogy with the special relativity of motion in which the velocity $c$ is invariant under motion transformations).

The second theoretical step amounts to describe the effects induced by the internal fractal structures of geodesics on motion in standard space (of positions and instants). Their main consequence is the transformation of classical dynamics into a generalized, quantum-like self-organized dynamics. The theory allows one to define and derive from relativistic first principles both the mathematical and physical quantum tools (complex, spinor, bispinor, then multiplet wave functions) and the equations of which these wave functions are solutions: a Schrodinger-type equation (more generally a Pauli equation for spinors) is derived as an integral of the geodesic equation in a fractal space, then Klein-Gordon and Dirac equations in the case of a full fractal space-time. We then briefly recall that gauge fields and gauge charges can also be constructed from a geometric re-interpretation of gauge transformations as scale transformations in fractal space-time.

In a second part of this review, we consider some applications of the theory to various sciences, particularly relevant to the questions of evolution and development. In the realm of physics and cosmology, we compare the various theoretical predictions obtained at the beginning of the 90's for the QCD coupling constant and for the cosmological constant to their present experimental and observational measurements. In astrophysics, we discuss applications to the formation of gravitational structures over many scales, with a special emphasis on the formation of planetary systems and on the validations, on the new extrasolar planetary systems and on Solar System Kuiper belt bodies discovered since 15 years, of the theoretical predictions of scale relativity (made before their discovery). This is completed by a validation of the theoretical prediction obtained some years ago for the solar cycle of 11 yrs on other solar-like stars whose cycles are now measured. In the realm of life sciences, we discuss possible applications of this extended framework to the processes of morphogenesis and the emergence of prokaryotic and eukaryotic cellular structures, then to the study of species evolution, society evolution, embryogenesis and cell confinement. This is completed by applications in Earth sciences, in particular to a prediction of the Arctic ice rate of melting and to possible predictivity in earthquake statistical studies.

# 2 Theory

## 2.1 Foundations of scale relativity theory

The theory of scale relativity is based on the giving up of the hypothesis of manifold differentiability. In this framework, the coordinate transformations are continuous but can be nondifferentiable. This implies several consequences [69], leading to the following steps of construction of the theory:

(1) One can prove the following theorem [69, 72, 7, 22, 23]: a continuous and nondifferentiable curve is fractal in a general meaning, namely, its length is explicitly dependent on a scale variable $\varepsilon$, i.e., $\mathcal{L} = \mathcal{L}(\varepsilon)$, and it diverges, $\mathcal{L} \to \infty$, when $\varepsilon \to 0$. This theorem can be readily extended to a continuous and nondifferentiable manifold, which is therefore fractal, not as an hypothesis, but as a consequence of the giving up of an hypothesis (that of differentiability).

(2) The fractality of space-time [69, 104, 66, 67] involves the scale dependence of the reference frames. One therefore adds to the usual variables defining the coordinate system, new variables $\varepsilon$ characterizing its 'state of scale'. In particular, the coordinates themselves become functions of these scale variables, i.e., $X = X(\varepsilon)$.

(3) The scale variables $\varepsilon$ can never be defined in an absolute way, but only in a relative way. Namely, only their ratio $\rho = \varepsilon'/\varepsilon$ does have a physical meaning. In experimental situations, these scales variables amount to the resolution of the measurement apparatus (it may be defined as standard errors, intervals, pixel size, etc...). In a theoretical analysis, they are the space and time differential elements themselves. This universal behavior leads to extend the principle of relativity in such a way that it applies also to the transformations (dilations and contractions) of these resolution variables [67, 68, 69].

## 2.2 Laws of scale transformation

### 2.2.1 Fractal coordinate and differential dilation operator

Consider a variable length measured on a fractal curve, and, more generally, a non-differentiable (fractal) curvilinear coordinate $\mathcal{L}(s,\varepsilon)$, that depends on some parameter $s$ which characterizes the position on the curve (it may be, e.g., a time coordinate), and on the resolution $\varepsilon$. Such a coordinate generalizes to nondifferentiable and fractal space-times the concept of curvilinear coordinates introduced for curved Riemannian space-times in Einstein's general relativity [69].

Such a scale-dependent fractal length $\mathcal{L}(s,\varepsilon)$, remains finite and differentiable when $\varepsilon \neq 0$, namely, one can define a slope for any resolution $\varepsilon$, being aware that this slope is itself a scale-dependent fractal function. It is only at the limit $\varepsilon \to 0$ that the length is infinite and the slope undefined, i.e., that nondifferentiability manifests itself.

Therefore the laws of dependence of this length upon position and scale may be written in terms of a double differential calculus, i.e., it can be the solution of differential equations involving the derivatives of $\mathcal{L}$ with respect to both $s$ and $\varepsilon$.

As a preliminary step, one needs to establish the relevant form of the scale variables and the way they intervene in scale differential equations. For this purpose, let us apply an infinitesimal dilation $d\rho$ to the resolution, which is therefore transformed as $\varepsilon \to \varepsilon' = \varepsilon(1+d\rho)$. The dependence on position is omitted at this stage in order to simplify the notation. By applying this transformation to a fractal coordinate $\mathcal{L}$, one obtains, to first order in the differential element,

$$\mathcal{L}(\varepsilon') = \mathcal{L}(\varepsilon + \varepsilon\, d\rho) = \mathcal{L}(\varepsilon) + \frac{\partial \mathcal{L}(\varepsilon)}{\partial \varepsilon}\, \varepsilon\, d\rho = (1 + \tilde{D}\; d\rho)\, \mathcal{L}(\varepsilon), \tag{1}$$

where $\tilde{D}$ is, by definition, the dilation operator.

Since $d\varepsilon/\varepsilon = d\ln\varepsilon$, the identification of the two last members of equation (1) yields

$$\tilde{D} = \varepsilon\, \frac{\partial}{\partial \varepsilon} = \frac{\partial}{\partial \ln \varepsilon}\ . \tag{2}$$

This form of the infinitesimal dilation operator shows that the natural variable for the resolution is $\ln\varepsilon$, and that the expected new differential equations will indeed involve quantities such as $\partial\mathcal{L}(s,\varepsilon)/\partial\ln\varepsilon$. This theoretical result agrees and explains the current knowledge according to which most measurement devices (of light, sound, etc..), including their physiological counterparts (eye, ear, etc..) respond according to the logarithm of the intensity (e.g., magnitudes, decibels, etc..).

#### 2.2.2 Self-similar fractals as solutions of a first order scale differential equation

Let us start by writing the simplest possible differential equation of scale, then by solving it. We shall subsequently verify that the solutions obtained comply with the principle of relativity. As we shall see, this very simple approach already yields a fundamental result: it gives a foundation and an understanding from first principles for self-similar fractal laws, which have been shown by Mandelbrot and many others to be a general description of a large number of natural phenomena, in particular biological ones (see, e.g., [60, 103, 59], other volumes of these series and references therein). In addition, the obtained laws, which combine fractal and scale-independent behaviours, are the equivalent for scales of what inertial laws are for motion [60]. Since they serve as a fundamental basis of description for all the subsequent theoretical constructions, we shall now describe their derivation in detail.

The simplest differential equation of explicit scale dependence which one can write is of first order and states that the variation of $\mathcal{L}$ under an infinitesimal scale transformation $d\ln\varepsilon$ depends only on $\mathcal{L}$ itself. Basing ourselves on the previous derivation of the form of the dilation operator, we thus write

$$\frac{\partial\mathcal{L}(s,\varepsilon)}{\partial\ln\varepsilon} = \beta(\mathcal{L}). \tag{3}$$

The function $\beta$ is *a priori* unknown. However, still looking for the simplest form of such an equation, we expand $\beta(\mathcal{L})$ in powers of $\mathcal{L}$, namely we write $\beta(\mathcal{L}) = a + b\mathcal{L} + \ldots$. Disregarding for the moment the $s$ dependence, we obtain, to the first order, the following linear equation, in which $a$ and $b$ are constants:

$$\frac{d\mathcal{L}}{d\ln\varepsilon} = a + b\mathcal{L}. \tag{4}$$

In order to find the solution of this equation, let us change the names of the constants as $\tau_F = -b$ and $\mathcal{L}_0 = a/\tau_F$, so that $a + b\mathcal{L} = -\tau_F(\mathcal{L} - \mathcal{L}_0)$. We obtain the equation

$$\frac{d\mathcal{L}}{\mathcal{L} - \mathcal{L}_0} = -\tau_F \, d\ln\varepsilon. \tag{5}$$

Its solution reads

$$\mathcal{L}(\varepsilon) = \mathcal{L}_0 \left\{1 + \left(\frac{\lambda}{\varepsilon}\right)^{\tau_F}\right\}, \tag{6}$$

where $\lambda$ is an integration constant. This solution corresponds to a length measured on a fractal curve up to a given point. One can now generalize it to a variable length that also depends on the position characterized by the parameter $s$. One obtains

$$\mathcal{L}(s,\varepsilon) = \mathcal{L}_0(s) \left\{1 + \zeta(s) \left(\frac{\lambda}{\varepsilon}\right)^{\tau_F}\right\}, \tag{7}$$

in which, in the most general case, the exponent $\tau_F$ may itself be a variable depending on the position.

The same kind of result is obtained for the projections on a given axis of such a fractal length [69]. Let $X(s,\varepsilon)$ be one of these projections, it reads

$$X(s,\varepsilon) = x(s) \left\{1 + \zeta_x(s) \left(\frac{\lambda}{\varepsilon}\right)^{\tau_F}\right\}. \tag{8}$$

In this case $\zeta_x(s)$ becomes a highly fluctuating function which may be described by a stochastic variable.

The important point here and for what follows is that the solution obtained is the sum of two terms, a classical-like, "differentiable part" and a nondifferentiable "fractal part", which is explicitly scale-dependent and tends to infinity when $\varepsilon \to 0$ [69, 17]. By differentiating these two parts in the above projection, we obtain the differential formulation of this essential result,

$$dX = dx + d\xi, \tag{9}$$

where $dx$ is a classical differential element, while $d\xi$ is a differential element of fractional order. This relation plays a fundamental role in the subsequent developments of the theory.

Consider the case when $\tau_F$ is constant. In the asymptotic small scale regime, $\varepsilon \ll \lambda$, one obtains a power-law dependence on resolution that reads

$$\mathcal{L}(s,\varepsilon) = \mathcal{L}_0(s) \left(\frac{\lambda}{\varepsilon}\right)^{\tau_F}. \tag{10}$$

We recognize in this expression the standard form of a self-similar fractal behaviour with constant fractal dimension $D_F = 1 + \tau_F$, which have already been found to yield a fair description of many physical and biological systems [60]. Here the topological dimension is $D_T = 1$, since we deal with a length, but this can be easily generalized to surfaces ($D_T = 2$), volumes ($D_T = 3$), etc.., according to the general relation $D_F = D_T + \tau_F$. The new feature here is that this result has been derived from a theoretical analysis based on first principles, instead of being postulated or deduced from a fit of observational data.

It should be noted that in the above expressions, the resolution is a length interval, $\varepsilon = \delta X$ defined along the fractal curve (or one of its projected coordinate). But one may also travel on the curve and measure its length on constant time intervals, then change the time scale. In this case the resolution $\varepsilon$ is a time interval, $\varepsilon = \delta t$. Since they are related by the fundamental relation

$$\delta X^{D_F} \sim \delta t, \tag{11}$$

the fractal length depends on the time resolution as

$$X(s, \delta t) = X_0(s) \times \left(\frac{T}{\delta t}\right)^{1-1/D_F}. \tag{12}$$

An example of the use of such a relation is Feynman's result according to which the mean square value of the velocity of a quantum mechanical particle is proportional to $\delta t^{-1}$ [33, p. 176], which corresponds to a fractal dimension $D_F = 2$, as later recovered by Abbott and Wise [1] by using a space resolution.

More generally, (in the usual case when $\varepsilon = \delta X$), following Mandelbrot, the scale exponent $\tau_F = D_F - D_T$ can be defined as the slope of the $(\ln \varepsilon, \ln \mathcal{L})$ curve, namely

$$\tau_F = \frac{d \ln \mathcal{L}}{d \ln(\lambda/\varepsilon)}. \tag{13}$$

For a self-similar fractal such as that described by the fractal part of the above solution, this definition yields a constant value which is the exponent in Eq. (10). However, one can anticipate on the following, and use this definition to compute an "effective" or "local" fractal dimension, now variable, from the complete solution that includes the differentiable and the nondifferentiable parts, and therefore a transition to effective scale independence. Differentiating the logarithm of Eq. (6) yields an effective exponent given by

$$\tau_{\text{eff}} = \frac{\tau_F}{1 + (\varepsilon/\lambda)^{\tau_F}}. \tag{14}$$

The effective fractal dimension $D_F = 1 + \tau_F$ therefore jumps from the nonfractal value $D_F = D_T = 1$ to its constant asymptotic value at the transition scale $\lambda$.

### 2.2.3 Galilean relativity of scales

The above scale laws have been obtained as solutions of the simplest possible differential equation acting in scale space. Now the main method of the scale relativity theory consists of constraining the various laws which are obtained by mathematical and/or physical tools by the principle of relativity applied to scale transformations.

In order to check whether the obtained laws come indeed under the principle of scale relativity, one should verify that these laws are covariant under a transformation of scale. We have found that these simple scale laws are the sum of a scaling (fractal) part and of a scale-independent part. The question of the compatibility with the principle of scale relativity concerns the scale-dependent part.

It reads $\mathcal{L} = \mathcal{L}_0(\lambda/\varepsilon)^{\tau_F}$ (Eq. 10), and it is therefore a law involving two variables ($\ln \mathcal{L}$ and $\tau_F$) in function of one parameter ($\varepsilon$) which, according to the relativistic view, characterizes the state of scale of the system (its relativity is apparent in the fact that we need another scale $\lambda$ to define it by their ratio). Note that, to be complete, we anticipate on what follows and consider *a priori* $\tau_F$ to be a variable, even if, in the simple law first considered here, it takes a constant value.

Let us take the logarithm of Eq. (10). It yields $\ln(\mathcal{L}/\mathcal{L}_0) = \tau_F \ln(\lambda/\varepsilon)$. The two quantities $\ln \mathcal{L}$ and $\tau_F$ then transform, under a finite scale transformation $\varepsilon \to \varepsilon' = \rho\, \varepsilon$, as

$$\ln \frac{\mathcal{L}(\varepsilon')}{\mathcal{L}_0} = \ln \frac{\mathcal{L}(\varepsilon)}{\mathcal{L}_0} - \tau_F \ln \rho \,, \tag{15}$$

and, to be complete,

$$\tau_F' = \tau_F. \tag{16}$$

These transformations have exactly the same mathematical structure as the Galilean group of motion transformation (applied here to scale rather than motion), which reads

$$x' = x - t\, v, \quad t' = t. \tag{17}$$

This is confirmed by the dilation composition law, $\varepsilon \to \varepsilon' \to \varepsilon''$, which writes

$$\ln \frac{\varepsilon''}{\varepsilon} = \ln \frac{\varepsilon'}{\varepsilon} + \ln \frac{\varepsilon''}{\varepsilon'} \,, \tag{18}$$

and is therefore similar to the law of composition of velocities between three reference systems $K$, $K'$ and $K$",

$$V''(K''/K) = V(K'/K) + V'(K''/K'). \tag{19}$$

Since the Galileo group of motion transformations is known to be the simplest group that implements the principle of relativity, the same is true for scale transformations.

It is important to realize that this is more than a simple analogy: the same physical problem is set in both cases, and is therefore solved under similar mathematical structures (since the logarithm transforms what would have been a multiplicative group into an additive group). Indeed, in both cases, it amounts to find the law of transformation of a position variable ($X$ for motion in a Cartesian system of coordinates, $\ln \mathcal{L}$ for scales in a fractal system of coordinates) under a change of the state of the coordinate system (change of velocity $V$ for motion and of resolution $\ln \rho$ for scale), knowing that these state variables are defined only in a relative way. Namely, $V$ is the relative velocity between the reference systems $K$ and $K'$, and $\rho$ is the relative scale: note that $\varepsilon$ and $\varepsilon'$ have indeed disappeared in the transformation law, only their ratio remains. This remark founds the status of resolutions as (relative) "scale velocities" and of the scale exponent $\tau_F$ as a "scale time".

Recall finally that, since the Galilean group of motion is only a limiting case of the more general Lorentz group, a similar generalization is expected in the case of scale transformations, which we shall briefly consider in Sec. 2.2.6.

### 2.2.4 Breaking of scale invariance

The standard self-similar fractal laws can be derived from the scale relativity approach. However, it is important to note that Eq. (6) provides us with another fundamental result. Namely, it also contains a spontaneous breaking of the scale symmetry. Indeed, it is characterized by the existence of a transition from a fractal to a non-fractal behaviour at scales larger than some transition scale $\lambda$. The existence of such a breaking of scale invariance is also a fundamental feature of many natural systems, which remains, in most cases, misunderstood.

The advantage of the way it is derived here is that it appears as a natural, spontaneous, but only effective symmetry breaking, since it does not affect the underlying scale symmetry. Indeed, the obtained solution is the sum of two terms, the scale-independent contribution (differentiable part), and the explicitly scale-dependent and divergent contribution (fractal part). At large scales the scaling part becomes dominated by the classical part, but it is still underlying even though it is hidden. There is therefore an apparent symmetry breaking, though the underlying scale symmetry actually remains unbroken.

The origin of this transition is, once again, to be found in relativity (namely, in the relativity of position and motion). Indeed, if one starts from a strictly scale-invariant law without any transition, $\mathcal{L} = \mathcal{L}_0(\lambda/\varepsilon)^{\tau_F}$, then adds a translation in standard position space ($\mathcal{L} \to \mathcal{L} + \mathcal{L}_1$), one obtains

$$\mathcal{L}' = \mathcal{L}_1 + \mathcal{L}_0 \left(\frac{\lambda}{\varepsilon}\right)^{\tau_F} = \mathcal{L}_1 \left\{ 1 + \left(\frac{\lambda_1}{\varepsilon}\right)^{\tau_F} \right\}. \tag{20}$$

Therefore one recovers the broken solution (that corresponds to the constant $a \neq 0$ in the initial scale differential equation). This solution is now asymptotically scale-dependent (in a scale-invariant way) only at small scales, and becomes independent of scale at large scales, beyond some relative transition $\lambda_1$ which is partly determined by the translation itself.

### 2.2.5 Generalized scale laws

**Discrete scale invariance, complex dimension and log-periodic laws** Fluctuations with respect to pure scale invariance are potentially important, namely the log-periodic correction to power laws that is provided, e.g., by complex exponents or complex fractal dimensions. It has been shown that such a behaviour provides a very satisfactory and possibly predictive model of the time evolution of many critical systems, including earthquakes and market crashes ([121] and references therein). More recently, it has been applied to the analysis of major event chronology of the evolutionary tree of life [19, 86, 87], of human development [14] and of the main economic crisis of western and precolumbian civilizations [44, 86, 50, 45].

One can recover log-periodic corrections to self-similar power laws through the requirement of covariance (i.e., of form invariance of equations) applied to scale differential equations [75]. Consider a scale-dependent function $\mathcal{L}(\varepsilon)$. In the applications to temporal evolution quoted above, the scale variable is identified with the time interval $|t-t_c|$, where $t_c$ is the date of a crisis. Assume that $\mathcal{L}$ satisfies a first order differential equation,

$$\frac{d\mathcal{L}}{d\ln\varepsilon} - \nu\mathcal{L} = 0, \tag{21}$$

whose solution is a pure power law $\mathcal{L}(\varepsilon) \propto \varepsilon^\nu$ (cf Sect. 2.2.2). Now looking for corrections to this law, one remarks that simply incorporating a complex value of the exponent $\nu$ would lead to large log-periodic fluctuations rather than to a controllable correction to the power law. So let us assume that the right-hand side of Eq. (21) actually differs from zero

$$\frac{d\mathcal{L}}{d\ln\varepsilon} - \nu\mathcal{L} = \chi. \tag{22}$$

We can now apply the scale covariance principle and require that the new function $\chi$ be solution of an equation which keeps the same form as the initial equation

$$\frac{d\chi}{d\ln\varepsilon} - \nu'\chi = 0. \tag{23}$$

Setting $\nu' = \nu + \eta$, we find that $\mathcal{L}$ must be solution of a second-order equation

$$\frac{d^2\mathcal{L}}{(d\ln\varepsilon)^2} - (2\nu+\eta)\frac{d\mathcal{L}}{d\ln\varepsilon} + \nu(\nu+\eta)\mathcal{L} = 0. \tag{24}$$

The solution reads $\mathcal{L}(\varepsilon) = a\varepsilon^\nu(1 + b\varepsilon^\eta)$, and finally, the choice of an imaginary exponent $\eta = i\omega$ yields a solution whose real part includes a log-periodic correction:

$$\mathcal{L}(\varepsilon) = a\,\varepsilon^\nu\,[1 + b\cos(\omega\ln\varepsilon)]. \tag{25}$$

As previously recalled in Sect. 2.2.4, adding a constant term (a translation) provides a transition to scale independence at large scales.

**Lagrangian approach to scale laws** In order to obtain physically relevant generalizations of the above simplest (scale-invariant) laws, a Lagrangian approach can be used in scale space, in analogy with its use to derive the laws of motion, leading to reverse the definition and meaning of the variables [75].

This reversal is an analog to that achieved by Galileo concerning motion laws. Indeed, from the Aristotle viewpoint, "time is the measure of motion". In the same way, the fractal dimension, in its standard (Mandelbrot's) acception, is defined from the topological measure of the fractal object (length of a curve, area of a surface, etc..) and resolution, namely (see Eq. 13)

$$t = \frac{x}{v} \quad \leftrightarrow \quad \tau_F = D_F - D_T = \frac{d \ln \mathcal{L}}{d \ln(\lambda/\varepsilon)}. \tag{26}$$

In the case, mainly considered here, when $\mathcal{L}$ represents a length (i.e., more generally, a fractal coordinate), the topological dimension is $D_T = 1$ so that $\tau_F = D_F - 1$. With Galileo, time becomes a primary variable, and the velocity is deduced from space and time, which are therefore treated on the same footing, in terms of a space-time (even though the Galilean space-time remains degenerate because of the implicitly assumed infinite velocity of light).

In analogy, the scale exponent $\tau_F = D_F - 1$ becomes, in this new representation, a primary variable that plays, for scale laws, the same role as played by time in motion laws (it is called "djinn" in some publications which therefore introduce a five-dimensional 'space-time-djinn' combining the four fractal fluctuations and the scale time).

Carrying on the analogy, in the same way as the velocity is the derivative of position with respect to time, $v = dx/dt$, we expect the derivative of $\ln \mathcal{L}$ with respect to scale time $\tau_F$ to be a "scale velocity". Consider as reference the self-similar case, that reads $\ln \mathcal{L} = \tau_F \ln(\lambda/\varepsilon)$. Derivating with respect to $\tau_F$, now considered as a variable, yields $d \ln \mathcal{L}/d\tau_F = \ln(\lambda/\varepsilon)$, i.e., the logarithm of resolution. By extension, one assumes that this scale velocity provides a new general definition of resolution even in more general situations, namely,

$$\mathbb{V} = \ln\left(\frac{\lambda}{\varepsilon}\right) = \frac{d \ln \mathcal{L}}{d\tau_F} \, . \tag{27}$$

One can now introduce a scale Lagrange function $\widetilde{L}(\ln \mathcal{L}, \mathbb{V}, \tau_F)$, from which a scale action is constructed

$$\widetilde{S} = \int_{\tau_1}^{\tau_2} \widetilde{L}(\ln \mathcal{L}, \mathbb{V}, \tau_F) \, d\tau_F. \tag{28}$$

The application of the action principle yields a scale Euler-Lagrange equation that writes

$$\frac{d}{d\tau_F} \frac{\partial \widetilde{L}}{\partial \mathbb{V}} = \frac{\partial \widetilde{L}}{\partial \ln \mathcal{L}} \, . \tag{29}$$

One can now verify that, in the free case, i.e., in the absence of any "scale force" (i.e., $\partial \widetilde{L}/\partial \ln \mathcal{L} = 0$), one recovers the standard fractal laws derived hereabove. Indeed, in this case the Euler-Lagrange equation becomes

$$\partial \widetilde{L}/\partial \mathbb{V} = \text{const} \Rightarrow \mathbb{V} = \text{const}. \tag{30}$$

which is the equivalent for scale of what inertia is for motion. Still in analogy with motion laws, the simplest possible form for the Lagrange function is a quadratic dependence on the scale velocity, (i.e., $\widetilde{L} \propto \mathbb{V}^2$). The constancy of $\mathbb{V} = \ln(\lambda/\varepsilon)$ means that it is independent of the scale time $\tau_F$. Equation (27) can therefore be integrated to give the usual power law behaviour, $\mathcal{L} = \mathcal{L}_0(\lambda/\varepsilon)^{\tau_F}$, as expected.

But this reversed viewpoint has also several advantages which allow a full implementation of the principle of scale relativity:

(i) The scale time $\tau_F$ is given the status of a fifth dimension and the logarithm of the resolution, $\mathbb{V} = \ln(\lambda/\varepsilon)$, its status of scale velocity (see Eq. 27). This is in accordance with its scale-relativistic definition, in which it characterizes the state of scale of the reference system, in the same way as the velocity $v = dx/dt$ characterizes its state of motion.

(ii) This allows one to generalize the formalism to the case of four independent space-time resolutions, $\mathbb{V}^\mu = \ln(\lambda^\mu/\varepsilon^\mu) = d\ln\mathcal{L}^\mu/d\tau_F$.

(iii) Scale laws more general than the simplest self-similar ones can be derived from more general scale Lagrangians [74, 75] involving "scale accelerations" $\mathbb{\Gamma} = d^2\ln\mathcal{L}/d\tau_F^2 = d\ln(\lambda/\varepsilon)/d\tau_F$, as we shall see in what follows.

Note however that there is also a shortcoming in this approach. Contrarily to the case of motion laws, in which time is always flowing toward the future (except possibly in elementary particle physics at very small time scales), the variation of the scale time may be non-monotonic, as exemplified by the previous case of log-periodicity. Therefore this Lagrangian approach is restricted to monotonous variations of the fractal dimension, or, more generally, to scale intervals on which it varies in a monotonous way.

**Scale dynamics** The previous discussion indicates that the scale invariant behaviour corresponds to freedom (i.e. scale force-free behaviour) in the framework of a scale physics. However, in the same way as there are forces in nature that imply departure from inertial, rectilinear uniform motion, we expect most natural fractal systems to also present distorsions in their scale behaviour with respect to pure scale invariance. This implies taking non-linearity in the scale space into account. Such distorsions may be, as a first step, attributed to the effect of a dynamics of scale ("scale dynamics"), i.e., of a "scale field", but it must be clear from the very beginning of the description that they are of geometric nature (in analogy with the Newtonian interpretation of gravitation as the result of a force, which has later been understood from Einstein's general relativity theory as a manifestation of the curved geometry of space-time).

In this case the Lagrange scale-equation takes the form of Newton's equation of dynamics,

$$F = \mu\,\frac{d^2\ln\mathcal{L}}{d\tau_F^2}, \tag{31}$$

where $\mu$ is a "scale mass", which measures how the system resists to the scale force, and where $\mathbb{\Gamma} = d^2\ln\mathcal{L}/d\tau_F^2 = d\ln(\lambda/\varepsilon)/d\tau_F$ is the scale acceleration.

In this framework one can therefore attempt to define generic, scale-dynamical behaviours which could be common to very different systems, as corresponding to a given form of the scale force.

**Constant scale force** A typical example is the case of a constant scale force. Setting $G = F/\mu$, the potential reads $\varphi = G \ln \mathcal{L}$, in analogy with the potential of a constant force $f$ in space, which is $\varphi = -fx$, since the force is $-\partial\varphi/\partial x = f$. The scale differential equation writes

$$\frac{d^2 \ln \mathcal{L}}{d\tau_F^2} = G. \tag{32}$$

It can be easily integrated. A first integration yields $d \ln \mathcal{L}/d\tau_F = G\tau_F + \mathbb{V}_0$, where $\mathbb{V}_0$ is a constant. Then a second integration yields a parabolic solution (which is the equivalent for scale laws of parabolic motion in a constant field),

$$\mathbb{V} = \mathbb{V}_0 + G\tau_F \quad ; \quad \ln \mathcal{L} = \ln \mathcal{L}_0 + \mathbb{V}_0\tau_F + \frac{1}{2} G\, \tau_F^2, \tag{33}$$

where $\mathbb{V} = d \ln \mathcal{L}/d\tau_F = \ln(\lambda/\varepsilon)$.

However the physical meaning of this result is not clear under this form. This is due to the fact that, while in the case of motion laws we search for the evolution of the system with time, in the case of scale laws we search for the dependence of the system on resolution, which is the directly measured observable. Since the reference scale $\lambda$ is arbitrary, the variables can be re-defined in such a way that $\mathbb{V}_0 = 0$, i.e., $\lambda = \lambda_0$. Indeed, from Eq. (33) one gets $\tau_F = (\mathbb{V} - \mathbb{V}_0)/G = [\ln(\lambda/\varepsilon) - \ln(\lambda/\lambda_0)]/G = \ln(\lambda_0/\varepsilon)/G$. Then one obtains

$$\tau_F = \frac{1}{G} \ln \left( \frac{\lambda_0}{\varepsilon} \right), \qquad \ln \left( \frac{\mathcal{L}}{\mathcal{L}_0} \right) = \frac{1}{2G} \ln^2 \left( \frac{\lambda_0}{\varepsilon} \right). \tag{34}$$

The scale time $\tau_F$ becomes a linear function of resolution (the same being true, as a consequence, of the fractal dimension $D_F = 1 + \tau_F$), and the $(\ln \mathcal{L}, \ln \varepsilon)$ relation is now parabolic instead of linear. Note that, as in previous cases, we have considered here only the small scale asymptotic behaviour, and that we can once again easily generalize this result by including a transition to scale-independence at large scale. This is simply achieved by replacing $\mathcal{L}$ by $(\mathcal{L} - \mathcal{L}_0)$ in every equations.

There are several physical situations where, after careful examination of the data, the power-law models were clearly rejected since no constant slope could be defined in the $(\log \mathcal{L}, \log \varepsilon)$ plane. In the several cases where a clear curvature appears in this plane, e.g., turbulence [29], sandpiles [11], fractured surfaces in solid mechanics [13], the physics could come under such a scale-dynamical description. In these cases it might be of interest to identify and study the scale force responsible for the scale distorsion (i.e., for the deviation from standard scaling).

### 2.2.6 Special scale-relativity

Let us close this section about the derivation of scale laws of increasing complexity by coming back to the question of finding the general laws of scale transformations that meet the principle of scale relativity [68]. It has been shown in Sec. 2.2.3 that the standard self-similar fractal laws come under a Galilean group of scale transformations. However, the Galilean relativity group is known, for motion laws, to be only a degenerate form of the Lorentz group. It has been proven that a similar result holds for scale laws [68, 69].

The problem of finding the laws of linear transformation of fields in a scale transformation $\mathbb{V} = \ln \rho$ $(\varepsilon \to \varepsilon')$ amounts to finding four quantities, $a(\mathbb{V}), b(\mathbb{V}), c(\mathbb{V})$, and $d(\mathbb{V})$, such that

$$\ln \frac{\mathcal{L}'}{\mathcal{L}_0} = a(\mathbb{V}) \ln \frac{\mathcal{L}}{\mathcal{L}_0} + b(\mathbb{V})\, \tau_F, \tag{35}$$

$$\tau_F' = c(\mathbb{V}) \ln \frac{\mathcal{L}}{\mathcal{L}_0} + d(\mathbb{V})\, \tau_F.$$

Set in this way, it immediately appears that the current 'scale-invariant' scale transformation law of the standard form of constant fractal dimension (Eq. 15), given by $a = 1, b = \mathbb{V}, c = 0$ and $d = 1$, corresponds to a Galilean group.

This is also clear from the law of composition of dilatations, $\varepsilon \to \varepsilon' \to \varepsilon''$, which has a simple additive form,

$$\mathbb{V}'' = \mathbb{V} + \mathbb{V}'. \tag{36}$$

However the general solution to the 'special relativity problem' (namely, find $a, b, c$ and $d$ from the principle of relativity) is the Lorentz group [58, 68]. This result has led to the suggestion of replacing the standard law of dilatation, $\varepsilon \to \varepsilon' = \varrho \times \varepsilon$ by a new Lorentzian relation, namely, for $\varepsilon < \lambda_0$ and $\varepsilon' < \lambda_0$

$$\ln \frac{\varepsilon'}{\lambda_0} = \frac{\ln(\varepsilon/\lambda_0) + \ln \varrho}{1 + \ln \varrho \, \ln(\varepsilon/\lambda_0)/\ln^2(\lambda_H/\lambda_0)}. \tag{37}$$

This relation introduces a fundamental length scale $\lambda_H$, which is naturally identified, toward the small scales, with the Planck length (currently $1.6160(11) \times 10^{-35}$ m) [68],

$$\lambda_H = l_{\mathbb{P}} = (\hbar G/c^3)^{1/2}, \tag{38}$$

and toward the large scales (for $\varepsilon > \lambda_0$ and $\varepsilon' > \lambda_0$) with the scale of the cosmological constant, $\lambda_H = \mathbb{L} = \Lambda^{-1/2}$ [69, Chap. 7.1].

As one can see from Eq. (37), if one starts from the scale $\varepsilon = \lambda_H$ and applies any dilatation or contraction $\varrho$, one obtains again the scale $\varepsilon' = \lambda_H$, whatever the initial value of $\lambda_0$. In other words, $\lambda_H$ can be interpreted as a limiting lower (or upper) length-scale, impassable, invariant under dilatations and contractions, which has the nature of a horizon.

As concerns the length measured along a fractal coordinate which was previously scale-dependent as $\ln(\mathcal{L}/\mathcal{L}_0) = \tau_0 \ln(\lambda_0/\varepsilon)$ for $\varepsilon < \lambda_0$, it becomes in the new framework, in the simplified case when one starts from the reference scale $\mathcal{L}_0$

$$\ln \frac{\mathcal{L}}{\mathcal{L}_0} = \frac{\tau_0 \ln(\lambda_0/\varepsilon)}{\sqrt{1 - \ln^2(\lambda_0/\varepsilon)/\ln^2(\lambda_0/\lambda_H)}}. \tag{39}$$

The main new feature of special scale relativity respectively to the previous fractal or scale-invariant approaches is that the scale exponent $\tau_F$ and the fractal dimension $D_F = 1+\tau_F$, which were previously constant ($D_F = 2, \tau_F = 1$ ), are now explicitly varying with scale, following the law (given once again in the simplified case when we start from the reference scale $\mathcal{L}_0$):

$$\tau_F(\varepsilon) = \frac{\tau_0}{\sqrt{1 - \ln^2(\lambda_0/\varepsilon)/\ln^2(\lambda_0/\lambda_H)}}. \tag{40}$$

Under this form, the scale covariance is explicit, since one keeps a power law form for the length variation, $\mathcal{L} = \mathcal{L}_0(\lambda/\varepsilon)^{\tau_F(\varepsilon)}$, but now in terms of a variable fractal dimension.

For a more complete development of special relativity, including its implications as regards new conservative quantities and applications in elementary particle physics and cosmology, see [68, 69, 72, 102].

The question of the nature of space-time geometry at Planck scale is a subject of intense work (see e.g. [5, 57] and references therein). This is a central question for practically all theoretical attempts, including noncommutative geometry [20, 21], supersymmetry and superstrings theories [43, 111], for which the compactification scale is close to the Planck scale, and particularly for the theory of quantum gravity. Indeed, the development of loop quantum gravity by Rovelli and Smolin [113] led to the conclusion that the Planck scale could be a quantized minimal scale in Nature, involving also a quantization of surfaces and volumes [114].

Over the last years, there has also been significant research effort aimed at the development of a 'Doubly-Special-Relativity' [4] (see a review in [5]), according to which the laws of physics involve a fundamental velocity scale $c$ and a fundamental minimum length scale $L_p$, identified with the Planck length.

The concept of a new relativity in which the Planck length-scale would become a minimum invariant length is exactly the founding idea of the special scale relativity theory [68], which has been incorporated in other attempts of extended relativity theories [15, 16]. But, despite the similarity of aim and analysis, the main difference between the 'Doubly-Special-Relativity' approach and the scale relativity one is that the question of defining an invariant length-scale is considered in the scale relativity/fractal space-time theory as coming under a relativity of scales. Therefore the new group to be constructed is a multiplicative group, that becomes additive only when working with the logarithms of scale ratios, which are definitely the physically relevant scale variables, as one can show by applying the Gell-Mann-Levy method to the construction of the dilation operator (see Sec. 2.2.1).

## 2.3 Fractal space and quantum mechanics

The first step in the construction of a theory of the quantum space-time from fractal and nondifferentiable geometry, which has been described in the previous sections, has consisted of finding the laws of explicit scale dependence at a given "point" or "instant" (under their new fractal definition).

The next step, which will now be considered, amounts to write the equation of motion in such a fractal space(-time) in terms of a geodesic equation. As we shall see, this equation takes, after integration, the form of a Schrödinger equation (and of the Klein-Gordon and Dirac equations in the relativistic case). This result, first obtained in Ref. [69], has later been confirmed by many subsequent physical [72, 74, 28, 17] and mathematical works, in particular by Cresson and Ben Adda [22, 24, 8, 9] and Jumarie [51, 52, 53, 54], including attempts of generalizations using the tool of the fractional integro-differential calculus [9, 26, 54].

In what follows, we consider only the simplest case of fractal laws, namely, those characterized by a constant fractal dimension. The various generalized scale laws considered in the previous section lead to new possible generalizations of quantum mechanics [72, 102].

### 2.3.1 Critical fractal dimension 2

Moreover, we simplify again the description by considering only the case $D_F = 2$. Indeed, the nondifferentiability and fractality of space implies that the paths are random walks of the Markovian type, which corresponds to such a fractal dimension. This choice is also justified by Feynman's result [33], according to which the typical paths of quantum particles (those which contribute mainly to the path integral) are nondifferentiable and of fractal dimension $D_F = 2$ [1]. The case $D_F \neq 2$, which yields generalizations to standard quantum mechanics has also been studied in detail (see [72, 102] and references therein). This study shows that $D_F = 2$ plays a critical role in the theory, since it suppresses the explicit scale dependence in the motion (Schrödinger) equation – but this dependence remains hidden and reappears through, e.g., the Heisenberg relations and the explicit dependence of measurement results on the reolution of the measurement apparatus.

Let us start from the result of the previous section, according to which the solution of a first order scale differential equation reads for $D_F = 2$, after differentiation and reintroduction of the indices,

$$dX^\mu = dx^\mu + d\xi^\mu = v^\mu ds + \zeta^\mu \sqrt{\lambda_c \, ds}, \tag{41}$$

where $\lambda_c$ is a length scale which must be introduced for dimensional reasons and which, as we shall see, generalizes the Compton length. The $\zeta^\mu$ are dimensionless highly fluctuating functions. Due to their highly erratic character, we can replace them by stochastic variables such that $<\zeta^\mu>= 0$, $<(\zeta^0)^2>= -1$ and $<(\zeta^k)^2>= 1$ ($k$ =1 to 3). The mean is taken here on a purely mathematic probability law which can be fully general, since the final result does not depend on its choice.

### 2.3.2 Metric of a fractal space-time

Now one can also write the fractal fluctuations in terms of the coordinate differentials, $d\xi^\mu = \zeta^\mu \sqrt{\lambda^\mu \, dx^\mu}$. The identification of this expression with that of Eq. (41) leads to recover the Einstein-de Broglie length and time scales,

$$\lambda_x = \frac{\lambda_c}{dx/ds} = \frac{\hbar}{p_x}, \quad \tau = \frac{\lambda_c}{dt/ds} = \frac{\hbar}{E}. \tag{42}$$

Let us now assume that the large scale (classical) behavior is given by Riemannian metric potentials $g_{\mu\nu}(x, y, z, t)$. The invariant proper time $dS$ along a geodesic writes, in terms of the complete differential elements $dX^\mu = dx^\mu + d\xi^\mu$,

$$dS^2 = g_{\mu\nu} dX^\mu dX^\nu = g_{\mu\nu}(dx^\mu + d\xi^\mu)(dx^\nu + d\xi^\nu). \tag{43}$$

Now replacing the $d\xi$'s by their expression, one obtains a fractal metric [69, 85] (which is valid only on the geodesics). Its two-dimensional and diagonal expression, neglecting the terms of zero mean (in order to simplify its writing) reads

$$dS^2 = g_{00}(x,t) \left(1 + \zeta_0^2 \, \frac{\tau_F}{dt}\right) c^2 dt^2 - g_{11}(x,t) \left(1 + \zeta_1^2 \, \frac{\lambda_x}{dx}\right) dx^2. \tag{44}$$

We therefore obtain generalized fractal metric potentials which are divergent and explicitly dependent on the coordinate differential elements [67, 69]. Another equivalent way to understand this metric consists in remarking that it is no longer only quadratic in the space-time differental elements, but that it also contains them in a linear way. Now this metric being valid only on the fractal geodesics, the question of finding its general expression for the whole fractal space-time remains an open question.

As a consequence, the curvature is also explicitly scale-dependent and divergent when the scale intervals tend to zero. This property ensures the fundamentally non-Riemannian character of a fractal space-time, as well as the possibility to characterize it in an intrinsic way. Indeed, such a characterization, which is a necessary condition for defining a space in a genuine way, can be easily made by measuring the curvature at smaller and smaller scales. While the curvature vanishes by definition toward the small scales in Gauss-Riemann geometry, a fractal space can be characterized from the interior by the verification of the divergence toward small scales of curvature, and therefore of physical quantities like energy and momentum.

Now the expression of this divergence is nothing but the Heisenberg relations themselves, which therefore acquire in this framework the status of a fundamental geometric test of the fractality of space-time [66, 67, 69].

### 2.3.3 Geodesics of a fractal space-time

The next step in such a geometric approach consists in the identification of wave-particles with fractal space-time geodesics. Any measurement is interpreted as a selection of the geodesics bundle linked to the interaction with the measurement apparatus (that depends on its resolution) and/or to the information known about it (for example, the which-way-information in a two-slit experiment [72].

The three main consequences of nondifferentiability are:

(i) The number of fractal geodesics is infinite. This leads to adopt a generalized statistical fluid-like description where the velocity $V^\mu(s)$ is replaced by a scale-dependent velocity field $V^\mu[X^\mu(s,ds),s,ds]$.

(ii) There is a breaking of the reflexion invariance of the differential element $ds$. Indeed, in terms of fractal functions $f(s,ds)$, two derivatives are defined,

$$X'_+(s,ds) = \frac{X(s+ds,ds) - X(s,ds)}{ds}, \quad X'_-(s,ds) = \frac{X(s,ds) - X(s-ds,ds)}{ds}, \qquad (45)$$

which transform one in the other under the reflection ($ds \leftrightarrow -ds$), and which have a priori no reason to be equal. This leads to a fundamental two-valuedness of the velocity field.

(iii) The geodesics are themselves fractal curves of fractal dimension $D_F = 2$ [33].

This means that one defines two divergent fractal velocity fields, $V_+[x(s,ds),s,ds] = v_+[x(s),s] + w_+[x(s,ds),s,ds]$ and $V_-[x(s,ds),s,ds] = v_-[x(s),s] + w_-[x(s,ds),s,ds]$, which can be decomposed in terms of differentiable parts $v_+$ and $v_-$, and of fractal parts $w_+$ and $w_-$. Note that, contrarily to other attempts such as Nelson's stochastic quantum mechanics which introduces forward and backward velocities [63] (and which has been later disproved [42, 125]), the two velocities are here both forward, since they do not correspond to a reversal of the time coordinate, but of the time differential element now considered as an independent variable.

More generally, we define two differentiable parts of derivatives $d_+/ds$ and $d_-/ds$, which, when they are applied to $x^\mu$, yield the differential parts of the velocity fields, $v_+^\mu = d_+x^\mu/ds$ and $v_-^\mu = d_-x^\mu/ds$.

### 2.3.4 Covariant total derivative

We now come to the definition of the main tool of the scale relativity theory, a covariant derivative which includes in its very construction the various effects of the space-time nondifferentiable geometry. Such a method is inspired from Einstein's general relativity, in which the effects of the curved geometry are included into the construction of a covariant derivative $DA^i = dA^i + \Gamma^i_{jk}A^j dx^k$ allowing to write the equation of motion as a geodesic equation, $Du^k/ds = 0$, which keeps the form of Galileo's equation of inertial motion (strong covariance). In the scale relativity theory, a similar mathematical tool is constructed, which is based on the same concept (namely, include the effect of geometry in the differentiation process allowing to write a geodesic equation under the strongly covariant form $\widehat{d}\mathcal{V}^k/ds = 0$), but which is also different since it accounts for the three above manifestations of nondifferentiability instead of that of curvature.

We mainly consider here the non-relativistic case. It corresponds to a three-dimensional fractal space, without fractal time, in which the invariant $ds$ is therefore identified with the time differential element $dt$. One describes the elementary displacements $dX^k$, $k = 1, 2, 3$, on the geodesics of a nondifferentiable fractal space in terms of the sum of two terms (omitting the indices for simplicity) $dX_\pm = d_\pm x + d\xi_\pm$, where $dx$ represents the differentiable part and $d\xi$ the fractal (nondifferentiable) part, defined as

$$d_\pm x = v_\pm\, dt, \quad d\xi_\pm = \zeta_\pm \sqrt{2\mathcal{D}}\, dt^{1/2}. \tag{46}$$

Here $\zeta_\pm$ are stochastic dimensionless variables such that $<\zeta_\pm>= 0$ and $<\zeta_\pm^2>= 1$, and $\mathcal{D}$ is a parameter that generalizes, up to the fundamental constant $c/2$, the Compton scale (namely, $\mathcal{D} = \hbar/2m$ in the case of standard quantum mechanics). The two time derivatives are then combined in terms of a complex total time derivative operator [69],

$$\frac{\widehat{d}}{dt} = \frac{1}{2}\left(\frac{d_+}{dt} + \frac{d_-}{dt}\right) - \frac{i}{2}\left(\frac{d_+}{dt} - \frac{d_-}{dt}\right). \tag{47}$$

Applying this operator to the differentiable part of the position vector yields a complex velocity

$$\mathcal{V} = \frac{\widehat{d}}{dt}x(t) = V - iU = \frac{v_+ + v_-}{2} - i\,\frac{v_+ - v_-}{2}\,. \tag{48}$$

In order to find the expression for the complex time derivative operator, let us first calculate the derivative of a scalar function $f$. Since the fractal dimension is 2, one needs to go to second order of expansion. For one variable it reads

$$\frac{df}{dt} = \frac{\partial f}{\partial t} + \frac{\partial f}{\partial X}\frac{dX}{dt} + \frac{1}{2}\frac{\partial^2 f}{\partial X^2}\frac{dX^2}{dt}. \tag{49}$$

The generalization of this writing to three dimensions is straighforward.

Let us now take the stochastic mean of this expression, i.e., we take the mean on the stochastic variables $\zeta_\pm$ which appear in the definition of the fractal fluctuation $d\xi_\pm$. By definition, since $dX = dx + d\xi$ and $<d\xi>= 0$, we have $<dX>= dx$, so that the second term is reduced (in 3 dimensions) to $v.\nabla f$. Now concerning the term $dX^2/dt$, it is infinitesimal and therefore it would not be taken into account in the standard differentiable case. But in the nondifferentiable case considered here, the mean square fluctuation is non-vanishing and of order $dt$, namely, $<d\xi^2>= 2\mathcal{D}dt$, so that the last term of Eq. (49) amounts in three dimensions to a Laplacian operator. One obtains, respectively for the (+) and (-) processes,

$$\frac{d_\pm f}{dt} = \left(\frac{\partial}{\partial t} + v_\pm.\nabla \pm \mathcal{D}\Delta\right) f\,. \tag{50}$$

Finally, by combining these two derivatives in terms of the complex derivative of Eq. (47), it reads [69]

$$\frac{\widehat{d}}{dt} = \frac{\partial}{\partial t} + \mathcal{V}.\nabla - i\mathcal{D}\Delta. \tag{51}$$

Under this form, this expression is not fully covariant [110], since it involves derivatives of the second order, so that its Leibniz rule is a linear combination of the first and second order Leibniz rules. By introducing the velocity operator [90]

$$\widehat{\mathcal{V}} = \mathcal{V} - i\mathcal{D}\nabla, \tag{52}$$

it may be given a fully covariant expression,

$$\frac{\widehat{d}}{dt} = \frac{\partial}{\partial t} + \widehat{\mathcal{V}}.\nabla, \tag{53}$$

namely, under this form it satisfies the first order Leibniz rule for partial derivatives.

We shall now see that $\widehat{d}/dt$ plays the role of a "covariant derivative operator" (in analogy with the covariant derivative of general relativity), namely, one may write in its terms the equation of physics in a nondifferentiable space under a strongly covariant form identical to the differentiable case.

### 2.3.5 Complex action and momentum

The steps of construction of classical mechanics can now be followed, but in terms of complex and scale dependent quantities. One defines a Lagrange function that keeps its usual form, $\mathcal{L}(x, \mathcal{V}, t)$, but which is now complex, then a generalized complex action

$$\mathcal{S} = \int_{t_1}^{t_2} \mathcal{L}(x, \mathcal{V}, t)dt. \tag{54}$$

Generalized Euler-Lagrange equations that keep their standard form in terms of the new complex variables can be derived from this action [69, 17], namely

$$\frac{\widehat{d}}{dt}\frac{\partial L}{\partial \mathcal{V}} - \frac{\partial L}{\partial x} = 0. \tag{55}$$

From the homogeneity of space and Noether's theorem, one defines a generalized complex momentum given by the same form as in classical mechanics, namely,

$$\mathcal{P} = \frac{\partial \mathcal{L}}{\partial \mathcal{V}}. \tag{56}$$

If the action is now considered as a function of the upper limit of integration in Eq. (54), the variation of the action from a trajectory to another nearby trajectory yields a generalization of another well-known relation of classical mechanics,

$$\mathcal{P} = \nabla \mathcal{S}. \tag{57}$$

### 2.3.6 Motion equation

Consider, as an example, the case of a single particle in an external scalar field of potential energy $\phi$ (but the method can be applied to any situation described by a Lagrange function). The Lagrange function , $L = \frac{1}{2}mv^2 - \phi$, is generalized as $\mathcal{L}(x, \mathcal{V}, t) = \frac{1}{2}m\mathcal{V}^2 - \phi$. The Euler-Lagrange equations then keep the form of Newton's fundamental equation of dynamics $F = m\, dv/dt$, namely,

$$m \frac{\widehat{d}}{dt} \mathcal{V} = -\nabla\phi, \tag{58}$$

which is now written in terms of complex variables and complex operators.

In the case when there is no external field ($\phi = 0$), the covariance is explicit, since Eq. (58) takes the free form of the equation of inertial motion, i.e., of a geodesic equation,

$$\frac{\widehat{d}}{dt} \mathcal{V} = 0. \tag{59}$$

This is analog to Einstein's general relativity, where the equivalence principle leads to write the covariant equation of motion of a free particle under the form of an inertial motion (geodesic) equation $Du_\mu/ds = 0$, in terms of the general-relativistic covariant derivative $D$, of the four-vector $u_\mu$ and of the proper time differential $ds$.

The covariance induced by the effects of the nondifferentiable geometry leads to an analogous transformation of the equation of motions, which, as we show below, become after integration the Schrödinger equation, which can therefore be considered as the integral of a geodesic equation in a fractal space.

In the one-particle case the complex momentum$\mathcal{P}$ reads

$$\mathcal{P} = m\mathcal{V}, \tag{60}$$

so that, from Eq. (57), the complex velocity $\mathcal{V}$ appears as a gradient, namely the gradient of the complex action

$$\mathcal{V} = \nabla\mathcal{S}/m. \tag{61}$$

### 2.3.7 Wave function

Up to now the various concepts and variables used were of a classical type (space, geodesics, velocity fields), even if they were generalized to the fractal and nondifferentiable, explicitly scale-dependent case whose essence is fundamentally not classical.

We shall now make essential changes of variable, that transform this apparently classical-like tool to quantum mechanical tools (without any hidden parameter or new degree of freedom). The complex wave function $\psi$ is introduced as simply another expression for the complex action $\mathcal{S}$, by making the transformation

$$\psi = e^{i\mathcal{S}/S_0}. \tag{62}$$

Note that, despite its apparent form, this expression involves a phase and a modulus since $\mathcal{S}$ is complex. The factor $S_0$ has the dimension of an action (i.e., an angular momentum) and must be introduced because $\mathcal{S}$ is dimensioned while the phase should be dimensionless. When this formalism is applied to standard quantum mechanics, $\mathcal{S}_0$ is nothing but the fundamental constant $\hbar$. As a consequence, since

$$\mathcal{S} = -i\mathcal{S}_0 \ln \psi, \tag{63}$$

one finds that the function $\psi$ is related to the complex velocity appearing in Eq. (61) as follows

$$\mathcal{V} = -i\,\frac{S_0}{m}\,\nabla \ln \psi. \tag{64}$$

This expression is the fondamental relation that connects the two description tools while giving the meaning of the wave function in the new framework. Namely, it is defined here as a velocity potential for the velocity field of the infinite family of geodesics of the fractal space. Because of nondifferentiability, the set of geodesics that defines a ‘particle’ in this framework is fundamentally non-local. It can easily be generalized to a multiple particle situation, in particular to entangled states, which are described by a single wave function $\psi$, from which the various velocity fields of the subsets of the geodesic bundle are derived as $\mathcal{V}_k = -i\,(\mathcal{S}_0/m_k)\,\nabla_k \ln \psi$, where $k$ is an index for each particle. The indistinguishability of identical particles naturally follows from the fact that the ‘particles’ are identified with the geodesics themselves, i.e., with an infinite ensemble of purely geometric curves. In this description there is no longer any point-mass with ‘internal‘ properties which would follow a ‘trajectory’, since the various properties of the particle – energy, momentum, mass, spin, charge (see next sections) – can be derived from the geometric properties of the geodesic fluid itself.

### 2.3.8 Correspondence principle

Since we have $\mathcal{P} = -i\mathcal{S}_0\nabla \ln \psi = -i\mathcal{S}_0(\nabla\psi)/\psi$, we obtain the equality [69]

$$\mathcal{P}\psi = -i\hbar\nabla\psi \tag{65}$$

in the standard quantum mechanical case $\mathcal{S}_0 = \hbar$, which establishes a correspondence between the classical momentum $p$, which is the real part of the complex momentum in the classical limit, and the operator $-i\hbar\nabla$.

This result is generalizable to other variables, in particular to the Hamiltonian. Indeed, a strongly covariant form of the Hamiltonian can be obtained by using the fully covariant form Eq. (53) of the covariant derivative operator. With this tool, the expression of the relation between the complex action and the complex Lagrange function reads

$$\mathcal{L} = \frac{\widehat{d}\mathcal{S}}{dt} = \frac{\partial\mathcal{S}}{\partial t} + \widehat{\mathcal{V}}.\nabla\mathcal{S}\ . \tag{66}$$

Since $\mathcal{P} = \nabla\mathcal{S}$ and $\mathcal{H} = -\partial\mathcal{S}/\partial t$, one obtains for the generalized complex Hamilton function the same form it has in classical mechanics, namely [95, 102],

$$\mathcal{H} = \widehat{\mathcal{V}}.\mathcal{P} - \mathcal{L}\,. \tag{67}$$

After expansion of the velocity operator, one obtains $\mathcal{H} = \mathcal{V}.\mathcal{P} - i\mathcal{D}\nabla.\mathcal{P} - \mathcal{L}$, which includes an additional term [110], whose origin is now understood as an expression of nondifferentiability and strong covariance.

### 2.3.9 Schrödinger equation and Compton relation

The next step of the construction amounts to write the fundamental equation of dynamics Eq. (58) in terms of the function $\psi$. It takes the form

$$iS_0\,\frac{\widehat{d}}{dt}(\nabla\ln\psi) = \nabla\phi. \tag{68}$$

As we shall now see, this equation can be integrated in a general way under the form of a Schrödinger equation. Replacing $\widehat{d}/dt$ and $\mathcal{V}$ by their expressions yields

$$\nabla\Phi = iS_0\left[\frac{\partial}{\partial t}\nabla\ln\psi - i\left\{\frac{S_0}{m}(\nabla\ln\psi.\nabla)(\nabla\ln\psi) + \mathcal{D}\Delta(\nabla\ln\psi)\right\}\right]. \tag{69}$$

This equation may be simplified thanks to the identity [69],

$$\nabla\left(\frac{\Delta\psi}{\psi}\right) = 2(\nabla\ln\psi.\nabla)(\nabla\ln\psi) + \Delta(\nabla\ln\psi). \tag{70}$$

We recognize, in the right-hand side of Eq. (70), the two terms of Eq. (69), which were respectively in factor of $S_0/m$ and $\mathcal{D}$. This leads to definitely define the wave function as

$$\psi = e^{i\mathcal{S}/2m\mathcal{D}}, \tag{71}$$

which means that the arbitrary parameter $S_0$ (which is identified with the constant $\hbar$ in standard QM) is now linked to the fractal fluctuation parameter by the relation

$$S_0 = 2m\mathcal{D}. \tag{72}$$

This relation (which can actually be proved instead of simply being set as a simplifying choice, see [99, 95]) is actually a generalization of the Compton relation, since the geometric parameter $\mathcal{D} = <d\xi^2> /2dt$ can be written in terms of a length scale as $\mathcal{D} = \lambda c/2$, so that, when $S_0 = \hbar$, it becomes $\lambda = \hbar/mc$. But a geometric meaning is now given to the Compton length (and therefore to the inertial mass of the particle) in the fractal space-time framework.

The fundamental equation of dynamics now reads

$$\nabla\phi = 2im\mathcal{D}\left[\frac{\partial}{\partial t}\nabla\ln\psi - i\left\{2\mathcal{D}(\nabla\ln\psi.\nabla)(\nabla\ln\psi) + \mathcal{D}\Delta(\nabla\ln\psi)\right\}\right]. \tag{73}$$

Using the above remarkable identity and the fact that $\partial/\partial t$ and $\nabla$ commute, it becomes

$$-\frac{\nabla\phi}{m} = -2\mathcal{D}\nabla\left\{i\frac{\partial}{\partial t}\ln\psi + \mathcal{D}\frac{\Delta\psi}{\psi}\right\}. \tag{74}$$

The full equation becomes a gradient,

$$\nabla\left\{\frac{\phi}{m} - 2\mathcal{D}\nabla\left(\frac{i\,\partial\psi/\partial t + \mathcal{D}\Delta\psi}{\psi}\right)\right\} = 0. \tag{75}$$

and it can be easily integrated, to finally obtain a generalized Schrödinger equation [69]

$$\mathcal{D}^2\Delta\psi + i\mathcal{D}\frac{\partial}{\partial t}\psi - \frac{\phi}{2m}\psi = 0, \tag{76}$$

up to an arbitrary phase factor which may be set to zero by a suitable choice of the $\psi$ phase. One recovers the standard Schrödinger equation of quantum mechanics for the particular case when $\mathcal{D} = \hbar/2m$.

### 2.3.10 Von Neumann's and Born's postulates

In the framework described here, "particles" are identified with the various geometric properties of fractal space(-time) geodesics. In such an interpretation, a measurement (and more generally any knowledge about the system) amounts to a selection of the sub-set of the geodesics family in which are kept only the geodesics having the geometric properties corresponding to the measurement result. Therefore, just after the measurement, the system is in the state given by the measurement result, which is precisely the von Neumann postulate of quantum mechanics.

The Born postulate can also be inferred from the scale-relativity construction [17, 99, 95]. Indeed, the probability for the particle to be found at a given position must be proportional to the density of the geodesics fluid at this point. The velocity and the density of the fluid are expected to be solutions of a Euler and continuity system of four equations, for four unknowns, $(\rho, V_x, V_y, V_z)$.

Now, by separating the real and imaginary parts of the Schrödinger equation, setting $\psi = \sqrt{P} \times e^{i\theta}$ and using a mixed representation $(P, V)$, where $V = \{V_x, V_y, V_z\}$, one obtains precisely such a standard system of fluid dynamics equations, namely,

$$\left(\frac{\partial}{\partial t} + V\cdot\nabla\right)V = -\nabla\left(\phi - 2\mathcal{D}^2\frac{\Delta\sqrt{P}}{\sqrt{P}}\right), \qquad \frac{\partial P}{\partial t} + \mathrm{div}(PV) = 0. \tag{77}$$

This allows one to univoquely identify $P = |\psi|^2$ with the probability density of the geodesics and therefore with the probability of presence of the 'particle'. Moreover,

$$Q = -2\mathcal{D}^2 \frac{\Delta\sqrt{P}}{\sqrt{P}} \tag{78}$$

can be interpreted as the new potential which is expected to emerge from the fractal geometry, in analogy with the identification of the gravitational field as a manifestation of the curved geometry in Einstein's general relativity. This result is supported by numerical simulations, in which the probability density is obtained directly from the distribution of geodesics without writing the Schrödinger equation [47, 102].

#### 2.3.11 Nondifferentiable wave function

In more recent works, instead of taking only the differentiable part of the velocity field into account, one constructs the covariant derivative and the wave function in terms of the full velocity field, including its divergent nondifferentiable part of zero mean [81, 99]. This still leads to the standard form of the Schrödinger equation. This means that, in the scale relativity framework, one expects the Schrödinger equation to have fractal and nondifferentiable solutions. This result agrees with a similar conclusion by Berry [10] and Hall [46], but it is considered here as a direct manifestation of the nondifferentiability of space itself. The research of such a behavior in laboratory experiments is an interesting new challenge for quantum physics.

### 2.4 Generalizations

#### 2.4.1 Fractal space time and relativistic quantum mechanics

All these results can be generalized to relativistic quantum mechanics, that corresponds in the scale relativity framework to a full fractal space-time. This yields, as a first step, the Klein-Gordon equation [70, 72, 17].

Then the account of a new two-valuedness of the velocity allows one to suggest a geometric origin for the spin and to obtain the Dirac equation [17]. Indeed, the total derivative of a physical quantity also involves partial derivatives with respect to the space variables, $\partial/\partial x^\mu$. From the very definition of derivatives, the discrete symmetry under the reflection $dx^\mu \leftrightarrow -dx^\mu$ is also broken. Since, at this level of description, one should also account for parity as in the standard quantum theory, this leads to introduce a bi-quaternionic velocity field [17], in terms of which Dirac bispinor wave function can be constructed.

We refer the interested reader to the detailed papers [72, 17, 18].

### 2.4.2 Gauge fields as manifestations of fractal geometry

The scale relativity principles has been also applied to the foundation of gauge theories, in the Abelian [70, 72] and non-Abelian [92, 102] cases.

This application is based on a general description of the internal fractal structures of the "particle" (identified with the geodesics of a nondifferentiable space-time) in terms of scale variables $\eta_{\alpha\beta}(x, y, z, t) = \varrho_{\alpha\beta}\, \varepsilon_\alpha\, \varepsilon_\beta$ whose true nature is tensorial, since it involves resolutions that may be different for the four space-time coordinates and may be correlated. This resolution tensor (similar to a covariance error matrix) generalizes the single resolution variable $\varepsilon$. Moreover, one considers here a more profound level of description in which the scale variables may now be function of the coordinates. Namely, the internal structures of the geodesics may vary from place to place and during the time evolution, in agreement with the non-absolute character of the scale space.

This generalization amounts to construct a 'general scale relativity' theory. The various ingredients of Yang-Mills theories (gauge covariant derivative, gauge invariance, charges, potentials, fields, etc...) can be recovered in such a framework, but they are now founded from first principles and are found to be of geometric origin, namely, gauge fields are understood as manifestations of the fractality of space-time [70, 72, 92, 102].

### 2.4.3 Quantum mechanics in scale space

One may go still one step further, and also give up the hypothesis of differentiability of the scale variables. Another generalization of the theory then amounts to use in scale space the method that has been built for dealing with nondifferentiability in space-time [90]. This results in scale laws that take quantum-like forms instead of classical ones, and which may have several applications, as well in particle physics [90] as in biology [100].

# 3 Applications

## 3.1 Applications to physics and cosmology

### 3.1.1 Application of special scale relativity: value of QCD coupling

In the special scale relativity framework, the new status of the Planck length-scale as a lowest unpassable scale must be universal. In particular, it applies also to the de Broglie and Compton relations themselves. They must therefore be generalized, since in their standard definition they may reach the zero length, which is forbidden in the new framework.

A fundamental consequence of these new relations for high energy physics is that the mass-energy scale and the length-time scale are no longer inverse as in standard quantum field theories, but they are now related by the special scale-relativistic generalized

Compton formula, that reads [68]

$$\ln \frac{m}{m_0} = \frac{\ln(\lambda_0/\lambda)}{\sqrt{1 - \ln^2(\lambda_0/\lambda)/\ln^2(\lambda_0/l_{\mathbb{P}})}}, \tag{79}$$

where $m_0\lambda_0 = \hbar/c$. This relation generalizes, for $m > m_0$ and $\lambda < \lambda_0$, the Compton relation $m\lambda = \hbar/c$ which connects any mass-energy scale $m$ to a length-scale $\lambda$.

As a consequence of this new relation, one finds that the grand unification scale becomes the Planck energy scale [68, 69]. We have made the conjecture [68, 69] that the SU(3) inverse coupling reaches the critical value $4\pi^2$ at this unification scale, i.e., at an energy $m_{\mathbb{P}}c^2/2\pi$ in the special scale-relativistic modified standard model.

By running the coupling from the Planck to the $Z$ scale, this conjecture allows one to get a theoretical estimate for the value of the QCD coupling at $Z$ scale. Indeed its renormalization group equation yields a variation of $\alpha_3 = \alpha_s$ with length scale given to second order (for six quarks and $N_H$ Higgs doublets) by [69]

$$\begin{aligned} \alpha_3^{-1}(r) &= \alpha_3^{-1}(\lambda_Z) + \frac{7}{2\pi} \ln \frac{\lambda_Z}{r} \\ &+ \frac{11}{4\pi(40 + N_H)} \ln \left\{ 1 - \frac{40 + N_H}{20\pi} \alpha_1(\lambda_Z) \ln \frac{\lambda_Z}{r} \right\} \\ &- \frac{27}{4\pi(20 - N_H)} \ln \left\{ 1 + \frac{20 - N_H}{12\pi} \alpha_2(\lambda_Z) \ln \frac{\lambda_Z}{r} \right\} \\ &+ \frac{13}{14\pi} \ln \left\{ 1 + \frac{7}{2\pi} \alpha_3(\lambda_Z) \ln \frac{\lambda_Z}{r} \right\}. \end{aligned} \tag{80}$$

The variation with energy scale is obtained by making the transformation given by Eq. (79) in which we take as reference scale the $Z$ boson scale, i.e., $\lambda_0 = \lambda_Z$ and $m_0 = m_Z$. This led in 1992 to the expectation [68] $\alpha_3(m_Z) = 0.1165 \pm 0.0005$, that compared well with the experimental value at that time, $\alpha_3(m_Z) = 0.112 \pm 0.010$, and was more precise by a factor 20.

This calculation has been more recently reconsidered [12, 102], by using improved experimental values of the $\alpha_1$ and $\alpha_2$ couplings at $Z$ scale (which intervene at second order), and by a better account of the top quark contribution. Indeed, its mass was unknown at the time of our first attempt in 1992, so that the running from $Z$ scale to Planck scale was performed by assuming the contribution of six quarks on the whole scale range.

However, the now known mass of the top quark, $m_t = 174.2 \pm 3.3$ GeV [107] is larger than the $Z$ mass, so that only five quarks contribute to the running of the QCD coupling between $Z$ scale and top quark scale, then six quarks between top and Planck scale. Moreover, the possibility of a threshold effect at top scale cannot be excluded. This led to an improved estimate :

$$\alpha_s(m_Z) = 0.1173 \pm 0.0004, \tag{81}$$

which agrees within uncertainties with our initial estimate 0.1165(5) [68]. This expectation is in very good agreement with the recent experimental average $\alpha_s(m_Z) = 0.1176 \pm 0.0009$ [107], where the quoted uncertainty is the error on the average. We give in Fig. 1 the evolution of the measurement results of the strong coupling at $Z$ scale, which compare very well with the theoretical expectation.

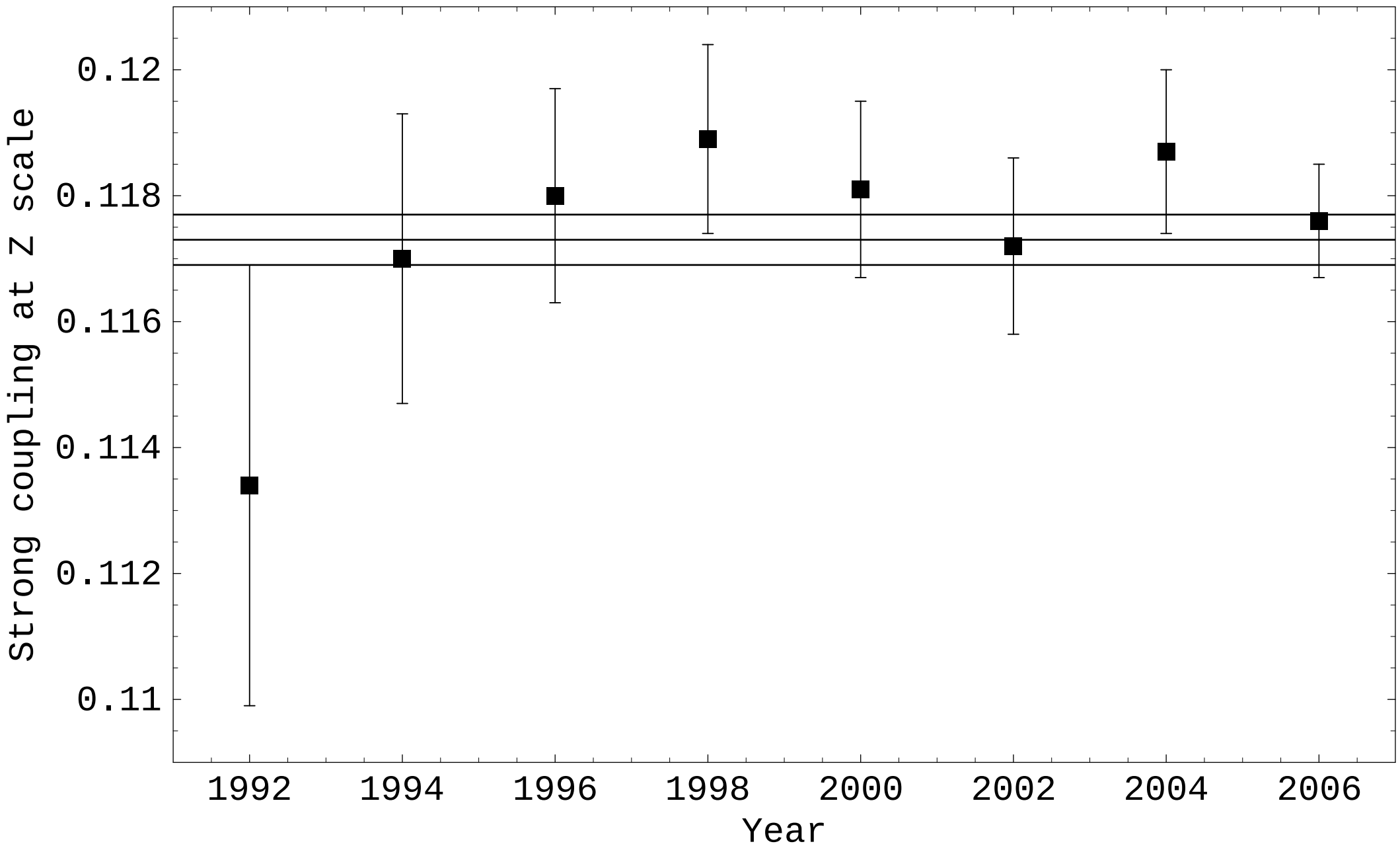

Figure 1: Measured values of $\alpha_s(M_Z)$ from 1992 (date of the theoretical prediction) to 2006 [107] compared with the expectation $\alpha_s(m_Z) = 0.1173 \pm 0.0004$ made from assuming that the inverse running coupling reaches the value $4\pi^2$ at Planck scale (see text). The 2008 experimental value is unchanged ($0.1176 \pm 0.0009$).

### 3.1.2 Value of the cosmological constant

One of the most difficult open questions in present cosmology is the problem of the vacuum energy density and its manifestation as an effective cosmological constant. In the framework of the theory of scale relativity a new solution can be suggested to this problem, which also allows one to connect it to Dirac's large number hypothesis [69, Chap. 7.1], [72].

The first step toward a solution has consisted in considering the vacuum as fractal, (i.e., explicitly scale dependent). As a consequence, the Planck value of the vacuum energy density is relevant only at the Planck scale, and becomes irrelevant at the cosmological scale. One expects such a scale-dependent vacuum energy density to be solution of a scale differential equation that reads

$$d\varrho/d\ln r = \Gamma(\varrho) = a + b\varrho + O(\varrho^2), \tag{82}$$

where $\varrho$ has been normalized to its Planck value, so that it is always $< 1$, allowing a Taylor expansion of $\Gamma(\varrho)$. This equation is solved as:

$$\varrho = \varrho_c \left[1 + \left(\frac{r_0}{r}\right)^{-b}\right]. \tag{83}$$

This solution is the sum of a fractal, power law behavior at small scales, that can be identified with the quantum scale-dependent contribution, and of a scale-independent term at large scale, that can be identified with the geometric cosmological constant observed at cosmological scales. The new ingredient here is a fractal/non-fractal transition about some scale $r_0$ that comes out as an integration constant, and which allows to connect the two contributions.

The second step toward a solution has been to realize that, when considering the various field contributions to the vacuum density, we may always chose $< E >= 0$ (i.e., renormalize the energy density of the vacuum). But consider now the gravitational self-energy of vacuum fluctuations. It writes:

$$E_g = \frac{G}{c^4}\frac{< E^2 >}{r}. \tag{84}$$

The Heisenberg relations prevent from making $< E^2 >= 0$, so that this gravitational self-energy *cannot* vanish. With $< E^2 >^{1/2} = \hbar c/r$, we obtain an asymptotic high energy behavior due to quantum effects

$$\varrho_g = \varrho_{\mathbb{P}} \left(\frac{l_{\mathbb{P}}}{r}\right)^6, \tag{85}$$

where $\varrho_{\mathbb{P}}$ is the Planck energy density and $l_{\mathbb{P}}$ the Planck length. From this equation one can make the identification $-b = 6$, so that one obtains $\varrho = \varrho_c \left[1 + (r_0/r)^6\right]$.

Therefore one of Dirac's large number relations is proved from this result [69]. Indeed, introducing the characteristic length scale $\mathbb{L} = \Lambda^{-1/2}$ of the cosmological constant $\Lambda$ (which is a curvature, i.e. the inverse of the square of a length), one obtains the relation:

$$\mathbb{K} = \mathbb{L}/l_{\mathbb{P}} = (r_0/l_{\mathbb{P}})^3 = (m_{\mathbb{P}}/m_0)^3, \tag{86}$$

where the transition scale $r_0$ can be identified with the Compton length of a particle of mass $m_0$. Then the power 3 in Dirac's large number relation is understood as coming from the power 6 of the gravitational self-energy of vacuum fluctuations and of the power 2 that relies the invariant scale $\mathbb{L}$ to the cosmological constant, following the relation $\Lambda = 1/\mathbb{L}^2$. The important point here is that in this new form of the Eddington-Dirac's relation, the cosmological length is no longer the time-varying $c/H_0$ (which led to theories of variation of constants), but the invariant cosmological length $\mathbb{L}$, which can therefore be connected to an invariant elementary particle scale without any longer a need for fundamental constant variation.

Now, a complete solution to the problem can be reached only provided the transition scale $r_0$ be identified. Our first suggestion [69, Chap. 7.1] has been that this scale is given by the classical radius of the electron.

Let us give an argument in favor of this conjecture coming from a description of the evolution of the primeval universe. Despite its name (which comes from historical reasons), the classical radius of the electron $r_e$ is of a quantum nature, since it actually defines the $e^+e^-$ annihilation cross section and the $e^-e^-$ cross section $\sigma = \pi r_\mathrm{e}^2$ at energy $m_\mathrm{e}c^2$. This length corresponds to an energy $E_\mathrm{e} = \hbar c/r_\mathrm{e} = 70.02$ MeV. This means that it yields the 'size' of an electron viewed by another electron. Therefore, when two electrons are separated by a distance smaller than $r_\mathrm{e}$, they can no longer be considered as different, independent objects.

The consequence of this property for the primeval universe is that $r_\mathrm{e}$ should be a fundamental transition scale. When the Universe scale factor was so small that the inter-distance between any couple of electrons was smaller than $r_\mathrm{e}$, there was no existing genuine separated electron. Then, when the cooling and expansion of the Universe separates the electron by distances larger than $r_\mathrm{e}$, the electrons that will later combine with the protons and form atoms appear for the first time as individual entities. Therefore the scale $r_\mathrm{e}$ and its corresponding energy 70 MeV defines a fundamental phase transition for the universe, which is the first appearance of electrons as we know them at large scales. Moreover, this is also the scale of maximal separation of quarks (in the pion), which means that the expansion, at the epoch this energy is reached, stops to apply to individual quarks and begins to apply to hadrons. This scale therefore becomes a reference static scale to which larger variable scales driven with the expansion can now be compared. Under this view, the cosmological constant would be a 'fossil' of this phase transition, in similarity with the 3K microwave radiation being a fossil of the combination of electrons and nucleons into atoms.

One obtains with the CODATA 2002 values of the fundamental constants a theoretical estimate

$$\mathbb{K}(\mathrm{pred}) = (5.3000 \pm 0.0012) \times 10^{60}, \tag{87}$$

i.e. $\mathbb{C}_U = \ln \mathbb{K} = 139.82281(22)$, which corresponds to a cosmological constant (see [69] p. 305)

$$\Lambda(\mathrm{pred}) = (1.3628 \pm 0.0004) \times 10^{-56}\ \mathrm{cm}^{-2} \tag{88}$$

i.e., a scaled cosmological constant

$$\Omega_\Lambda(\mathrm{pred}) = (0.38874 \pm 0.00012)\, h^{-2}. \tag{89}$$

Finally the corresponding invariant cosmic length scale is theoretically predicted to be

$$\mathbb{L}(\text{pred}) = (2.77608 \pm 0.00042)\ \text{Gpc}, \tag{90}$$

i.e., $\mathbb{L}(\text{pred}) = (8.5661 \pm 0.0013) \times 10^{25}$ m.

Let us compare these values with the most recent determinations of the cosmological constant, sometimes now termed, in a somewhat misleading way, 'dark energy' (see Fig. 2). The WMAP three year analysis of 2006 [122] has given $h = 0.73 \pm 0.03$ and $\Omega_\Lambda(\text{obs}) = 0.72 \pm 0.03$. These results, combined with the recent Sloan (SDSS) data [123], yield, assuming $\Omega_{\text{tot}} = 1$ (as supported by its WMAP determination, $\Omega_{\text{tot}} = 1.003 \pm 0.010$)

$$\Omega_\Lambda(\text{obs}) = \frac{\Lambda c^2}{3H_0^2} = 0.761 \pm 0.017, \quad h = 0.730 \pm 0.019. \tag{91}$$

Note that these recent results have also reinforced the cosmological constant interpretation of the 'dark energy' with a measurement of the coefficient of the equation of state $w = -0.941 \pm 0.094$ [123], which encloses the value $w = -1$ expected for a cosmological constant.

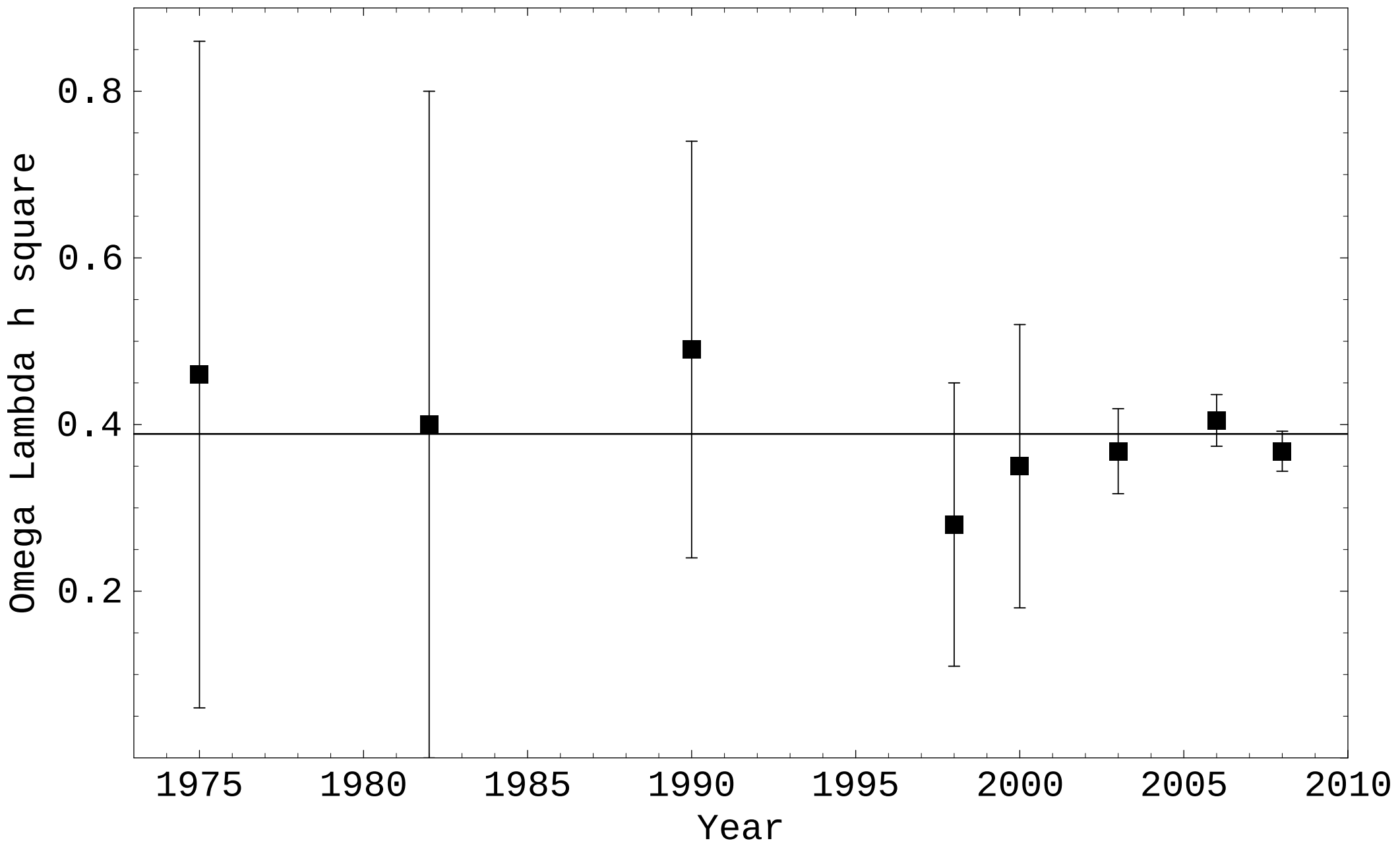

Figure 2: Evolution of the measured values of the dimensionless cosmological constant $\Omega_\Lambda h^2 = \Lambda c^2/3H_{100}^2$, from 1975 to 2008, compared to the theoretical expectation $\Lambda = (m_e/\alpha\, m_{\mathbb{P}})^6\ (1/l_{\mathbb{P}})^2$ [69] that gives numerically $\Omega_\Lambda h^2(\text{pred}) = 0.38874 \pm 0.00012$.

With these values one finds a still improved cosmological constant

$$\Omega_\Lambda h^2(\text{obs}) = 0.406 \pm 0.030, \tag{92}$$

which corresponds to a cosmic scale

$$\mathbb{L}(\text{obs}) = (2.72 \pm 0.10)\ \text{Gpc}, \quad \text{i.e.,}\ \ \mathbb{K}(\text{obs}) = (5.19 \pm 0.19) \times 10^{60}, \tag{93}$$

in excellent agreement with the predicted values $\mathbb{L}(\text{pred}) = 2.7761(4)$ Gpc, and $\mathbb{K}(\text{pred}) = 5.300(1) \times 10^{60}$.

The evolution of these experimental determinations [102] is shown in Fig. 2 where they are compared with the theoretical expectation

$$\Omega_\Lambda h^2(\text{pred}) = 0.38874 \pm 0.00012. \tag{94}$$

The convergence of the observational values toward the theoretical estimate, despite an improvement of the precision by a factor of more than 20, is striking, although this estimate is partly phenomenological, since it remains dependent on a conjecture about the transition scale, which clearly needs to be investigated and comprehended more profoundly. The 2008 value from the Five-Year WMAP results is $\Omega_\Lambda h^2(\text{obs}) = 0.384 \pm 0.043$ [49] and is once again in very good agreement with the theoretical expectation made 16 years ago [69], before the first genuine measurements in 1998.

## 3.2 Applications to astrophysics

### 3.2.1 Gravitational Schrödinger equation

Let us first briefly recall the basics of the scale-relativistic theoretical approach. It has been reviewed in Sec. 2.3 in the context of the foundation of microphysics quantum mechanics. We shall now see that some of its ingredients, leading in particular to obtain a generalized Schrödinger form for the equation of motion, also applies to gravitational structure formation.

Under three general conditions, namely, {(i) infinity of geodesics (which leads to introduce a non-deterministic velocity field), (ii) fractal dimension $D_F = 2$ of each geodesic, on which the elementary displacements are described in terms of the sum $dX = dx + d\xi$ of a classical, differentiable part $dx$ and of a fractal, non-differentiable fluctuation $d\xi$, (iii) two-valuedness of the velocity field, which is a consequence of time irreversibility at the infinitesimal level issued from non-differentiability, one can construct a complex covariant derivative that reads

$$\frac{\widehat{d}}{dt} = \frac{\partial}{\partial t} + \mathcal{V}.\nabla - i\mathcal{D}\Delta\ , \tag{95}$$

where $\mathcal{D}$ is a parameter that characterizes the fractal fluctuation, which is such that $< d\xi^2 >= 2\mathcal{D}dt$, and where the classical part of the velocity field, $\mathcal{V}$ is complex as a consequence of condition (iii) (see [17, 95] for more complete demonstrations).

Then this covariant derivative, that describes the non-differentiable and fractal geometry of space-time, can be combined with the covariant derivative of general relativity, that describes the curved geometry. We shall briefly consider in what follows only the Newtonian limit. In this case the equation of geodesics keeps the form of Newton's fundamental equation of dynamics in a gravitational field,

$$\frac{\widehat{D}\mathcal{V}}{dt} = \frac{\widehat{d}\mathcal{V}}{dt} + \nabla\left(\frac{\phi}{m}\right) = 0, \tag{96}$$

where $\phi$ is the Newtonian potential energy. Introducing the action $S$, which is now complex, and making the change of variable $\psi = e^{iS/2m\mathcal{D}}$, this equation can be integrated under the form of a generalized Schrödinger equation [69]:

$$\mathcal{D}^2\Delta\psi + i\mathcal{D}\frac{\partial}{\partial t}\psi - \frac{\phi}{2m}\,\psi = 0. \tag{97}$$

Since the imaginary part of this equation is the equation of continuity (Sec. 3), and basing ourselves on our description of the motion in terms of an infinite family of geodesics, $P = |\psi|^2$ naturally gives the probability density of the particle position [17, 95].

Even though it takes this Schrödinger-like form, equation (97) is still in essence an equation of gravitation, so that it must come under the equivalence principle [73, 2], i.e., it is independent of the mass of the test-particle. In the Kepler central potential case ($\phi = -GMm/r$), $GM$ provides the natural length-unit of the system under consideration. As a consequence, the parameter $\mathcal{D}$ reads:

$$\mathcal{D} = \frac{GM}{2w}, \tag{98}$$

where $w$ is a constant that has the dimension of a velocity. The ratio $\alpha_g = w/c$ actually plays the role of a macroscopic gravitational coupling constant [2, 82].

### 3.2.2 Formation and evolution of structures

Let us now compare our approach with the standard theory of gravitational structure formation and evolution. By separating the real and imaginary parts of the Schrödinger equation we obtain, after a new change of variables, respectively a generalized Euler-Newton equation and a continuity equation, namely,

$$m\,(\frac{\partial}{\partial t} + V\cdot\nabla)V = -\nabla(\phi + Q), \qquad \frac{\partial P}{\partial t} + \mathrm{div}(PV) = 0, \tag{99}$$

where $V$ is the real part of the complex velocity field $\mathcal{V}$ and where the gravitational potential $\phi$ is given by the Poisson equation. In the case when the density of probability is proportional to the density of matter, $P \propto \rho$, this system of equations is equivalent to the classical one used in the standard approach of gravitational structure formation, except for the appearance of an extra potential energy term $Q$ that writes:

$$Q = -2m\mathcal{D}^2\frac{\Delta\sqrt{P}}{\sqrt{P}}. \tag{100}$$

The existence of this potential energy, (which amount to the Bohm potential in standard quantum mechanics) is, in our approach, readily demonstrated and understood: namely, it is the very manifestation of the fractality of space, in similarity with Newton's potential being a manifestation of curvature. We have suggested [83, 91, 93] that it could be the origin of the various effects which are usually attributed to an unseen, 'dark' matter.

In the case when actual particles achieve the probability density distribution (structure formation), we have $\rho = m_0 P$. Then the Poisson equation (i.e., the field equation) becomes $\Delta\phi = 4\pi G m m_0 |\psi|^2$ and it is therefore strongly interconnected with the Schrödinger equation (which is here a new form for the equation of motion). Such a system of equations is similar to that encountered in the description of superconductivity (Hartree equation). We expect its solutions to provide us with general theoretical predictions for the structures (in position and velocity space) of self-gravitating systems at multiple scales [74, 27]. This expectation is already supported by the observed agreement of several of these solutions with astrophysical observational data [69, 73, 82, 76, 80, 77, 78, 48].

### 3.2.3 Planetary systems

Let us briefly consider the application of the theory to the formation of planetary systems. The standard model of formation of planetary systems can be reconsidered in terms of a fractal description of the motion of planetesimals in the protoplanetary nebula. On length-scales much larger than their mean free path, we have assumed [69] that their highly chaotic motion satisfy the three conditions upon which the derivation of a Schrödinger equation is based (large number of trajectories, fractality and time symmetry breaking). In modern terms, our proposal is but a 'migration' theory, since it amounts to take into account the coupling between planetesimals (or proto-planets) and the remaining disk. But, instead of considering a mean field coupling, we consider the effect of the closest bodies to be the main one, leading to Brownian motion and irreversibility.

This description applies to the distribution of planetesimals in the proto-planetary nebula at several embedded levels of hierarchy [73]. Each hierarchical level ($k$) is characterized by a length-scale defining the parameter $\mathcal{D}_k$ (and therefore the velocity $w_k$) that appears in the generalized Schrödinger equation describing this sub-system. Through matching of the wave functions for these different subsystems (for example, the inner solar system in its whole constitutes the fundamental 'orbital' $n = 1$ of the outer solar system), the ratios of their structure constants $w_k$ are expected to be themselves given by integer numbers [73]. This expectation is supported by the observed sub-structures of our solar system, which are organised according to constants $w_0 = 144.7 \pm 0.7$ km/s (inner system), $3 \times w_0$ (Sun and intramercurial system), $w_0/5$ (outer solar system), $w_0/(5 \times 7)$ (distant Kuiper belt). This hierarchical model has allowed us to recover the mass distribution of planets and small planets in the inner and outer solar systems [76].

**The Sun** One can apply this approach to the organization of the Sun surface itself. One expects the distribution of the various relevant physical quantities that characterize the solar activity at the Sun surface (Sun spot number, magnetic field, etc...) to be described by a wave function whose stationary solutions read $\psi = \psi_0 \, e^{iEt/2m\mathcal{D}}$. In this relation, the parameter $\mathcal{D} = GM/2w$ must now be directly related to the Sun itself, which naturally leads to take $M = M_\odot$ and $w = w_\odot = 437.1$ km/s, which is the Keplerian velocity at the Sun radius $R_\odot = 0.00465$ AU. It is also remarkable that this velocity is very close to $3 \times 144.7 = 434.1$, where $w = 144.7$ km/s is the structural constant of the inner solar system (in accordance with the expectation of integer ratios for the gravitational structure constants [73], and of most of the extrasolar planetary systems discovered up to now (see what follows and Fig. 6).

The energy $E$ results from the rotational velocity and, to be complete, should also include the turbulent velocity, so that $E = (v_{\rm rot}^2 + v_{\rm turb}^2)/2$. This means that we expect the solar surface activity to be subjected to a fundamental period:

$$\tau = \frac{2\pi m \mathcal{D}}{E} = \frac{4\pi \mathcal{D}}{v_{\rm rot}^2 + v_{\rm turb}^2}, \tag{101}$$

The parameter $\mathcal{D}$ at the Sun radius is $\mathcal{D} = GM_\odot/2w_\odot$, then we obtain:

$$\tau = \frac{2\pi G M_\odot}{w_\odot (v_{\rm rot}^2 + v_{\rm turb}^2)}. \tag{102}$$

The average sideral rotation period of the Sun is 25.38 days, yielding a velocity of 2.01 km/s at equator [108]. The turbulent velocity has been found to be $v_{\rm turb} = 1.4 \pm 0.2$ km/s [56]. Therefore we find numerically

$$\tau = (10.2 \pm 1.0) \text{ yrs}. \tag{103}$$

The observed value of the period of the Solar activity cycle, $\tau_{\rm obs} = 11.0$ yrs, nicely supports this theoretical prediction. This is an interesting result, owing to the fact that there is, up to now, no existing theoretical prediction of the value of the solar cycle period, except in terms of very rough order of magnitude [127].

Moreover, since we have now at our disposal a simple and precise formula for a stellar cycle which precisely accounts for the solar period, the advantage is that it can be tested with other stars. The observation of the magnetic activity cycle of distant solar-like stars remains a difficult task, but it has now been performed on several stars. A first attempt gives very encouraging results (see Fig. 3), since we obtain indeed a satisfactory agreement between the observed and predicted periods, in a statistically significant way, despite the small number of objects.

**The intramercurial system** organized on the constant $w_\odot = 3 \times 144 = 432$ km/s. The existence of an intramercurial subsystem is supported by various stable and transient structures observed in dust, asteroid and comet distributions (see [27]). We have in particular suggested the existence of a new ring of asteroids, the 'Vulcanoid belt', at a preferential distance of about 0.17 AU from the Sun.

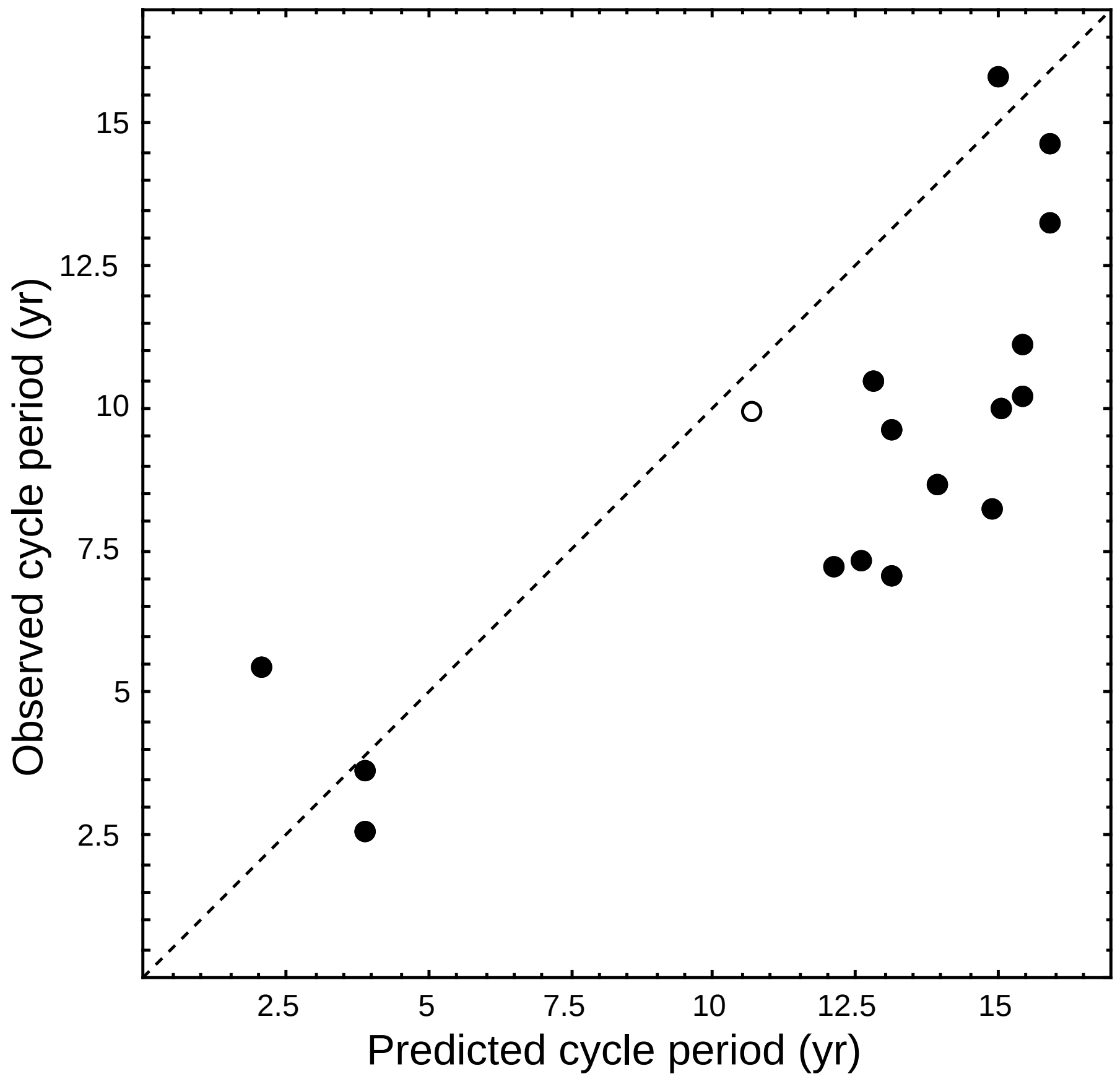

Figure 3: Comparison between the observed values of the period of solar-like star cycles (inactive stars with better determined behavior in Table 1 of Ref. [115]) and the predicted periods (see text). The open point is for the Sun. The correlation is significant at a probability level $P \approx 10^{-4}$ (Student variable $t \approx 5$).

**The inner solar system** (earth-like planets), organized with a constant $w_i = 144$ km/s (see Fig. 6).

**The outer solar system** organized with a constant $w_o = 144/5 = 29$ km/s (see Fig. 4), as deduced from the fact that the mass peak of the inner solar system lies at the Earth distance ($n = 5$). The Jovian giant planets lie from $n = 2$ to $n = 5$. Pluton lies on $n = 6$, but is now considered to be a dwarf planet part of the Kuiper belt.

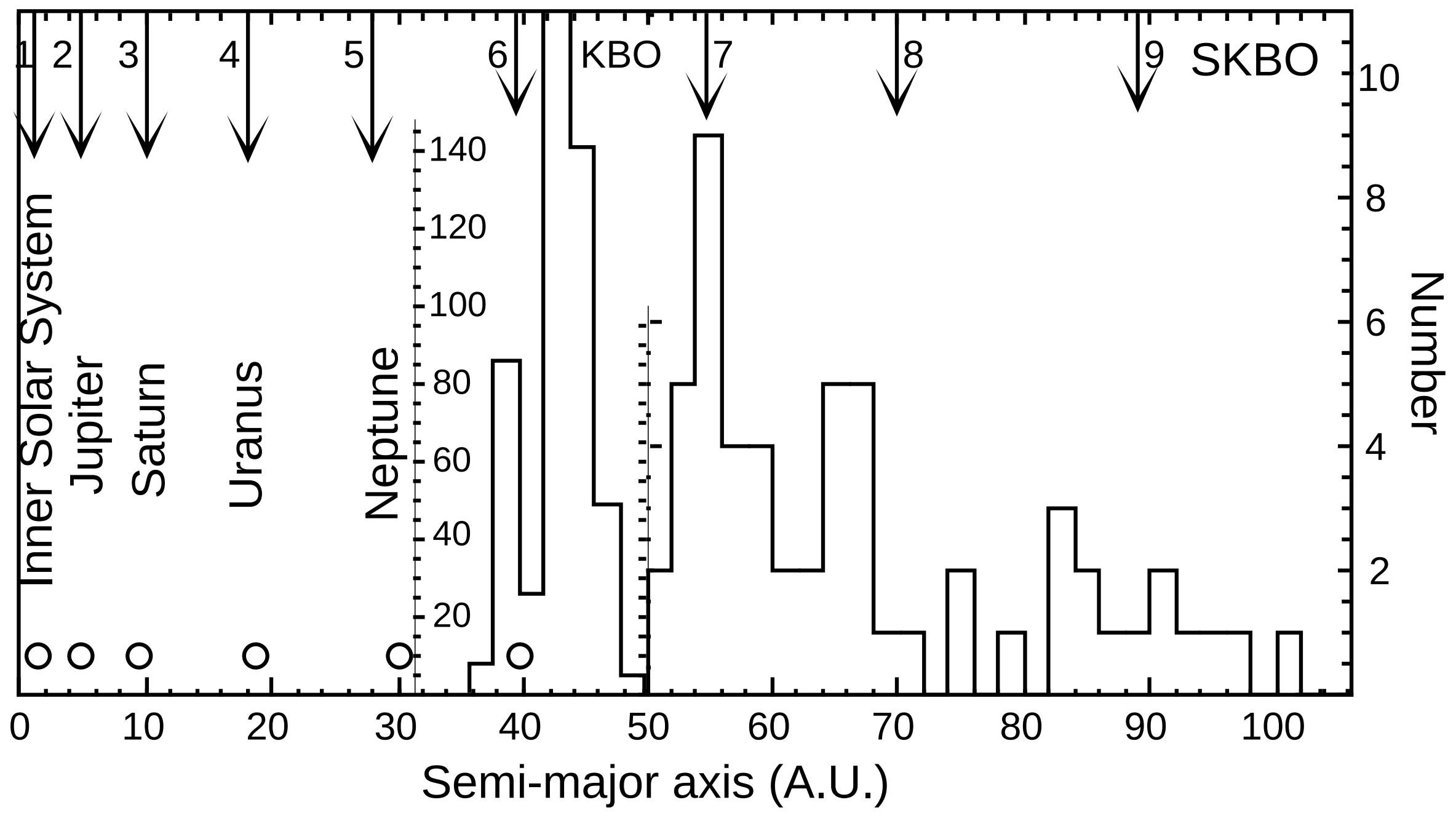

Figure 4: Distribution of the semi-major axis of Kuiper belt objects (KBO) and scattered Kuiper belt objects (SKBO), compared with the theoretical predictions (arrows) of probability density peaks for the outer solar system [27] (see text). The existence of probability density peaks for the Kuiper belt at ≈ 40, 55, 70, 90 AU, etc..., has been theoretically predicted in 1993 before the discovery of these objects [71], and it is now supported by the observational data, in particular by the new small planet Eris at 68 AU, whose mass is larger than Pluto, and which falls close to the expected probability peak $n = 8$ at 70 AU (see text).

**Kuiper belt** The recently discovered Kuiper and scattered Kuiper belt objects (Fig. 4) show peaks of probability at $n = 6$ to 9 [27], as predicted before their discovery [71]. In particular, the predicted peak around 57 AU ($n = 7$) is the main observed peak in the SKBO distribution of semi-major axes. The following peak ($n = 8$), predicted to be around 70 AU, has received a spectacular verification with the discovery of the dwarf planet Eris (2003 UB313) at 68 AU, whose mass larger than Pluton has recently led to a revision of planetary nomenclature.

**Distant Kuiper belt** Beyond these distances, we have been able to predict a new level of hierarchy in the Solar System whose main SKBO peak at 57 AU would be the fundamental level ($n = 1$) [37]. The following probability peaks are expected, according to the $n^2$ law, to lie for semi-major axes of 228, 513, 912, 1425, 2052 AU, etc.... Once again this prediction has been validated by the observational data in a remark-able way (see Fig. 5), since 4 bodies, including the very distant small planet Sedna, have now been discovered in the 513 AU peak ($n = 3$), 7 bodies in the 228 AU peak ($n = 2$) , and now one very distant object at about 1000 AU (data Minor Planet Center, http://www.cfa.harvard.edu/iau/mpc.html).

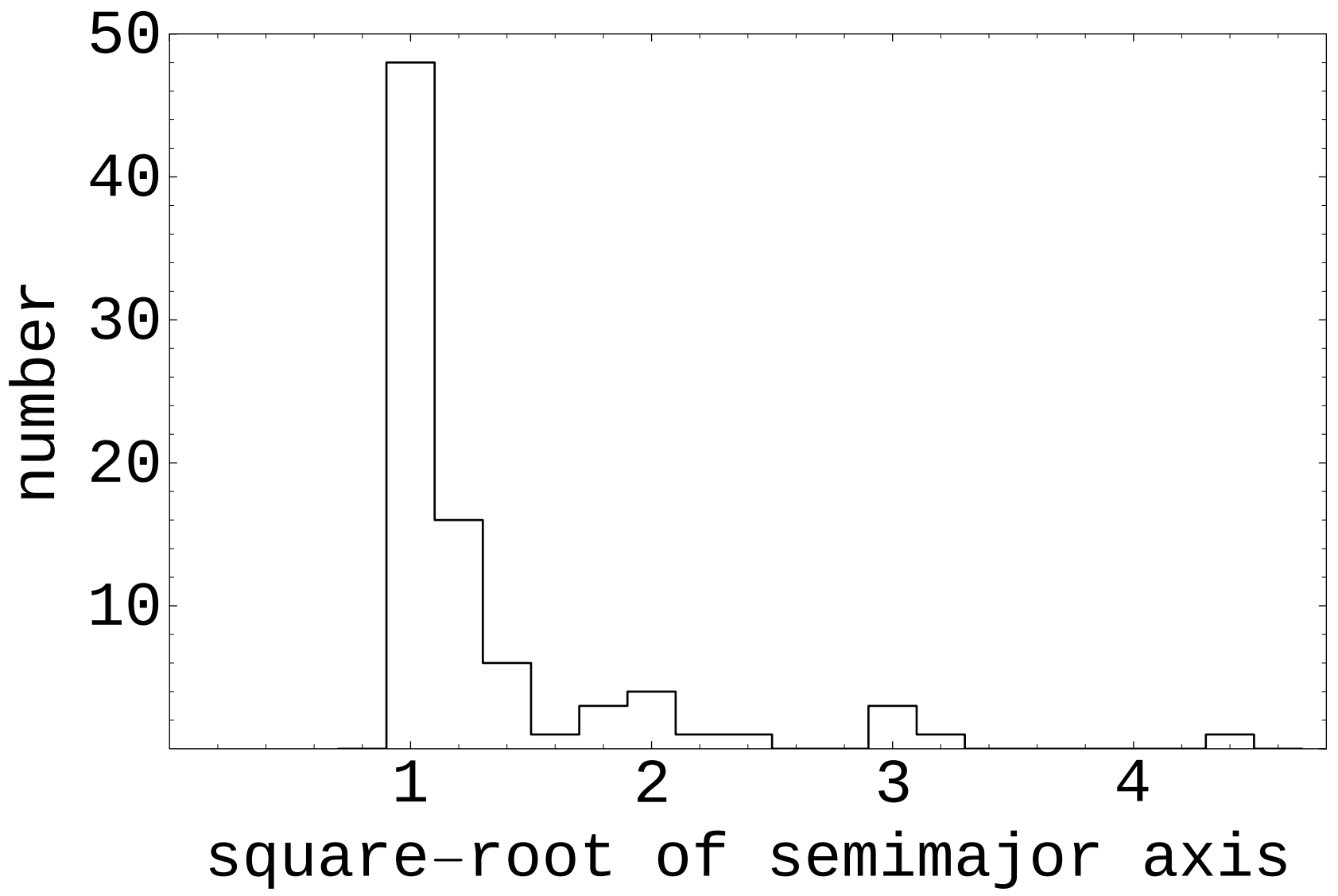

Figure 5: Distribution of the semi-major axis of very distant scattered Kuiper belt objects (SKBO) , compared with the theoretical predictions of probability density peaks (see text). We have taken the main SKBO peak at ≈57 AU (which is the predicted $n = 7$ peak of the outer solar system) as fundamental level ($n = 1$) for this new level of hierarchy of the solar system. The figure plots the histogram of the variable $(a/57)^{1/2}$, where $a$ is the semimajor axis of the object orbit in AU. The theoretical prediction, done before the discovery of the distant objects, is that the distribution of this variable should show peaks for integer values, as now verified by the observational data.

**Extrasolar planets** We have suggested more than 16 years ago [69, 71], before the discovery of exoplanets, that the theoretical predictions from this approach of planetary formation should apply to all planetary systems, not only our own solar system. Meanwhile more than 300 exoplanets have now been discovered, and the observational data support this prediction in a highly statistically significant way (see [73, 82, 27] and Fig. 6).

The presently known exoplanets mainly correspond to the intramercurial and inner solar systems. The theoretical prediction, made in 1993 [69, Chap. 7.2], according to which the distribution of semi-major axes $a$ is expected to show peaks of probability for integer values of the variable $4.83(a/M)^{1/2}$, where $M$ is the star mass, remains validated with a high statistical significance (see Fig. 6). In particular, in addition to the peaks of probability corresponding to the inner solar system planets ($n = 3$ Mercury, $n = 4$

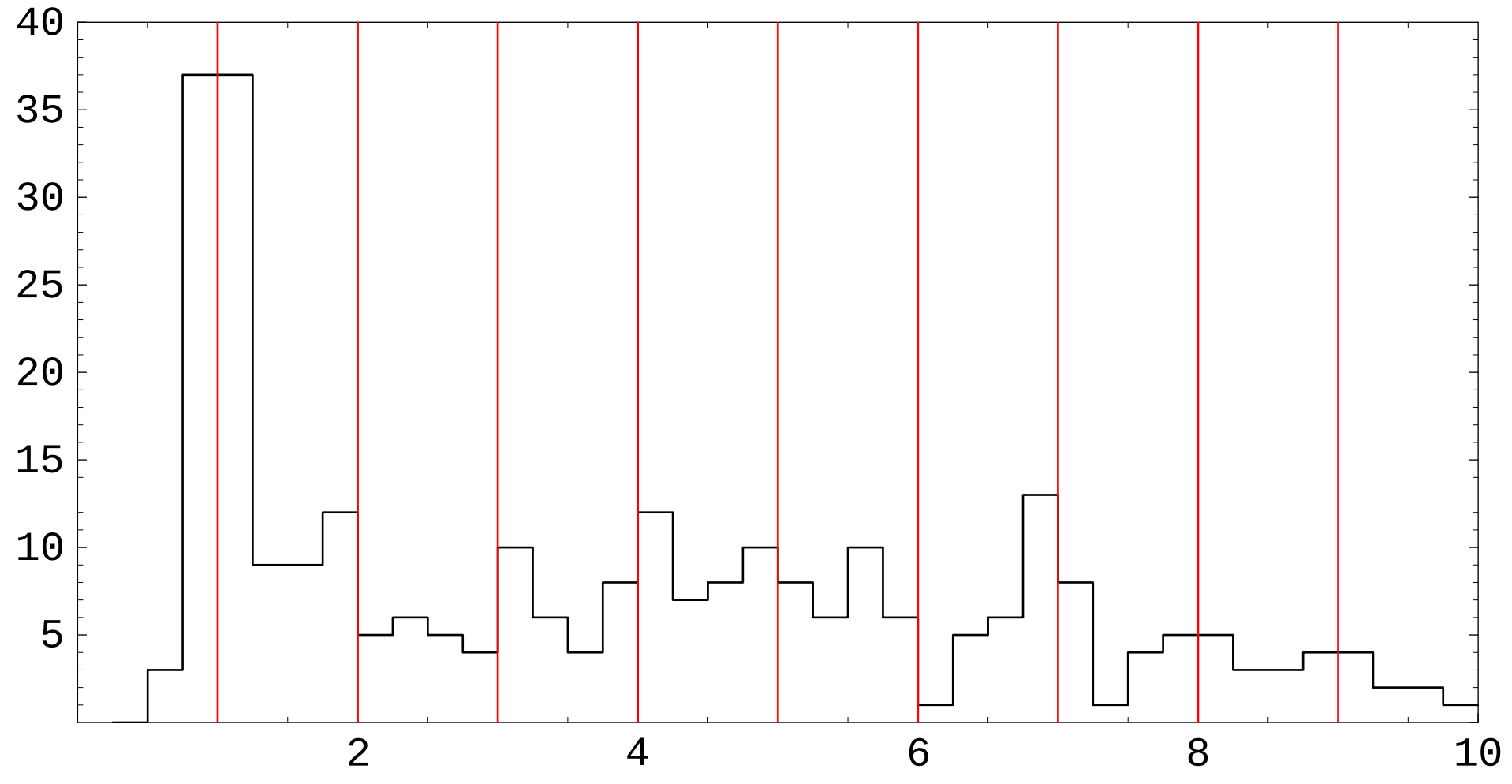

Figure 6: Observed distribution of the semi-major axes of 300 exoplanets (June 2008 data [117]) and inner solar system planets, compared with the theoretical prediction (vertical lines). The figure gives the histogram of the distribution of the variable $4.83(a/M)^{1/2}$, where $a$ is the semi-major axis of the planet orbit and $M$ is the star mass. One predicts the occurence of peaks of probability density for semimajor axes $a_n = GM(n/w_0)^2$, where $n$ is integer and $w_0 = 144.7 \pm 0.7$ km/s is a gravitational coupling constant (see text). The planets of the inner solar system (Mercury, Venus, the Earth and Mars) fall respectively in $n = 3$, 4, 5 and 6. The probability to obtain such an agreement by chance, measured by comparing the number of exoplanets falling around the predicted peaks (integer values $n$, red vertical lines) to those which fall between the predicted peaks $(n + 1/2)$ is now found to be $P = 5 \times 10^{-7}$.

Venus, $n = 5$ Earth, $n = 6$ Mars), two additional predicted peaks of probability, the 'fundamental' one at 0.043 AU/M$_\odot$ and the second one at 0.17 AU/M$_\odot$, have been made manifest in extrasolar planetary systems. In particular, the validation of the principal prediction of the SR approach, namely, the main peak at the fundamental level $n = 1$, is striking since it now contains more than 70 exoplanets. A power spectrum analysis of the distribution of exoplanets of Fig. 6 yields a definite peak with a power $p = 16$ for the predicted periodicity of $(a/M)^{1/2}$, which corresponds to the very low probability $P = 1.1 \times 10^{-7}$ that such a periodicity be obtained by chance. This value of the probability is supported by another method (see legend of Fig. 6). It is important to note that this observed distribution now combines exoplanets found from different methods which have their own limitations, and it is therefore strongly biased; however this bias is expected to change only their large scale distribution (for example the larger number of exoplanets at intramercurial distances and its decrease at large distance probably come from such an observational bias), so that it does not affect the SR prediction and its test, which concerns a small scale modulation in terms of $4.83(a/M)^{1/2}$ (i.e., the differences between the peaks at integer values and the holes between the peaks).

## 3.3 Applications to sciences of life

The scale relativity theory has also been recently applied to sciences other than physical sciences, including sciences of life, sciences of societies, historical [34] and geographical sciences [61, 35, 36] and human sciences [124, 88, 94, 101]. We refer the interested reader to the books [86, 98], to parts of review papers or books [84, 90, 96] and full review papers on this specific subject [6, 100] for more details.

### 3.3.1 Applications of log-periodic laws

**Species evolution** Let us first consider the application of log-periodic laws to the description of critical time evolution. Recall that a log-periodic generalization to scale invariance has been obtained as a solution to wave-like differential scale equations, which can themselves be constructed from the requirement of scale covariance (see Sec. 2.2.5). Interpreted as a distribution of probability, such solutions therefore lead to a geometric law of progression of probability peaks for the occurence of events.

Now several studies have shown that many biological, natural, sociological and economic phenomena obey a log-periodic law of time evolution such as can be found in some critical phenomena : earthquakes [119], stock market crashes [120], evolutionary leaps [19, 86, 87], long time scale evolution of western and other civilizations [86, 87, 45], world economy indices dynamics [50], embryogenesis [14], etc... Thus emerges the idea that this behaviour typical of temporal crisis could be extremely widespread, as much in the organic world as in the inorganic one [121].

In the case of species evolution, one observes the occurrence of major evolutionary leaps leading to bifurcations among species, which proves the existence of punctuated evolution [41] in addition to the gradual one. The global pattern is assimilated to a 'tree of life', whose bifurcations are identified to evolutionary leaps, and branch lengths to the time intervals between these major events [19]. As early recognized by Leonardo da Vinci, the branching of vegetal trees and rivers may be described as a first self-similar approximation by simply writing that the ratio of the lengths of two adjacent levels is constant in the mean. We have made a similar hypothesis for the time intervals between evolutionary leaps, namely, $(T_n - T_{n-1})/(T_{n+1} - T_n) = g$. Such a geometric progression yields a log-periodic acceleration for $g > 1$, a deceleration for $g < 1$, and a periodicity for $g = 1$. Except when $g = 1$, the events converge toward a critical time $T_c$ which can then be taken as reference, yielding the following law for the event $T_n$ in terms of the rank $n$:

$$T_n = T_c + (T_0 - T_c)\, g^{-n}, \tag{104}$$

where $T_0$ is any event in the lineage, $n$ the rank of occurrence of a given event and $g$ is the scale ratio between successive time intervals. Such a chronology is periodic in terms of logarithmic variables, i.e., $\log |T_n - T_c| = \log |T_0 - T_c| - n \log g$.

This law is dependent on two parameters only, $g$ and $T_c$, which of course have no reason a priori to be constant for the entire tree of life. Note that $g$ is not expected to be an absolute parameter, since it depends on the density of events chosen, i.e., on the adopted threshhold in the choice of their importance (namely, if the number of events is doubled, $g$ is replaced by $\sqrt{g}$). Only a maximal value of $g$, corresponding to the very major events, could possibly have a meaning. On the contrary, the value of $T_c$ is expected to be a characteristic of a given lineage, and therefore not to depend on such a choice. This expectation is supported by an analysis of the fossil record data under various choices of the threshold on the events, which have yielded the same values of $T_c$ within error bars [87].

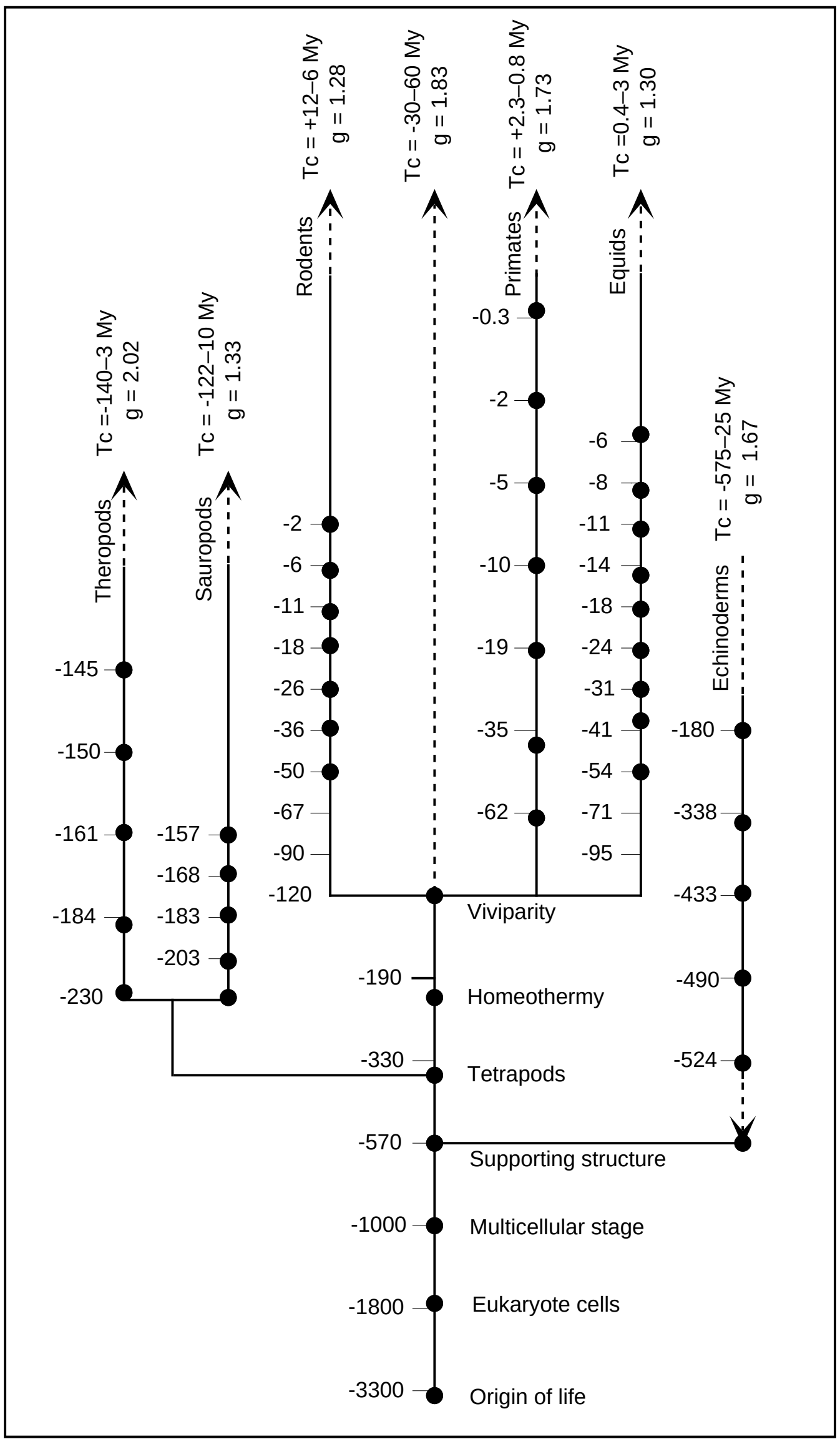

Figure 7: The dates of major evolutionary events of seven lineages (common evolution from life origin to viviparity, Theropod and Sauropod dinosaurs, Rodents, Equidae, Primates including Hominidae, and Echinoderms) are plotted as black points in terms of $\log(T_c - T)$, and compared with the numerical values from their corresponding log-periodic models (computed with their best-fit parameters). The adjusted critical time $T_c$ and scale ratio $g$ are indicated for each lineage (figure adapted from Refs. [19, 86, 87]).

A statistically significant log-periodic acceleration has been found at various scales for global life evolution, for primates, for sauropod and theropod dinosaurs, for rodents and North American equids. A deceleration law was conversely found in a statistically significant way for echinoderms and for the first steps of rodents evolution (see Fig. 7 and more detail in Refs. [19, 86, 87]). One finds either an acceleration toward a critical date $T_c$ or a deceleration from a critical date, depending on the considered lineage.

It must be remarked that the observed dates follow a log-periodic law only in the mean, and show a dispersion around this mean (see [86, p. 320]. In other words, this is a statistical acceleration or deceleration, so that the most plausible interpretation is that the discrete $T_n$ values are nothing but the dates of peaks in a continuous probability distribution of the events. Moreover, it must also be emphasized that this result does not put the average constancy of the mutation rate in question. This is demonstrated by a study of the cytochrome c tree of branching (in preparation), which is based on genetic distances instead of geological chronology, and which nevertheless yields the same result, namely, a log-periodic acceleration of most lineages, and a periodicity (which corresponds to a critical time tending to infinity) in some cases. The average mutation rate remains around 1/20 Myr since about 1 Gyr, so that one cannot escape the conclusion that the number of mutations needed to obtain a major evolutionary leap decreases with time among many lineages, and increases for some of them.

**Embryogenesis and human development** Considering the relationships between phylogeny and ontogeny, it appeared interesting to verify whether the log-periodic law describing the chronology of several lineages of species evolution may also be applied to the various stages in human embryological development. The result, (see Figure in Chaline's contribution), is that a statistically significant log-periodic deceleration with a scale ratio $g = 1.71 \pm 0.01$ is indeed observed, starting from a critical date that is consitent with the conception date [14].

**Evolution of societies** Many observers have commented on the way historical events accelerate. Grou [44] has shown that the economic evolution since the neolithic can be described in terms of various dominating poles which are submitted to an accelerating crisis-nocrisis pattern, which has subsequently been quantitatively analysed using log-periodic laws.

For the Western civilization since the Neolithic (i.e., on a time scale of about 8000 years), one finds that a log-periodic acceleration with scale factor $g = 1.32 \pm 0.018$ occurs toward $T_c = 2080 \pm 30$ (see Fig. 8), in a statistically highly significant way. This result has been later confirmed by Johansen and Sornette [50] by an independent study on various market, domestic, research and development, etc... indices on a time scale of about 200 years, completed by demography on a time scale of about 2000 years. They find critical dates for these various indices in the range 2050-2070, which support the longer time scale result.

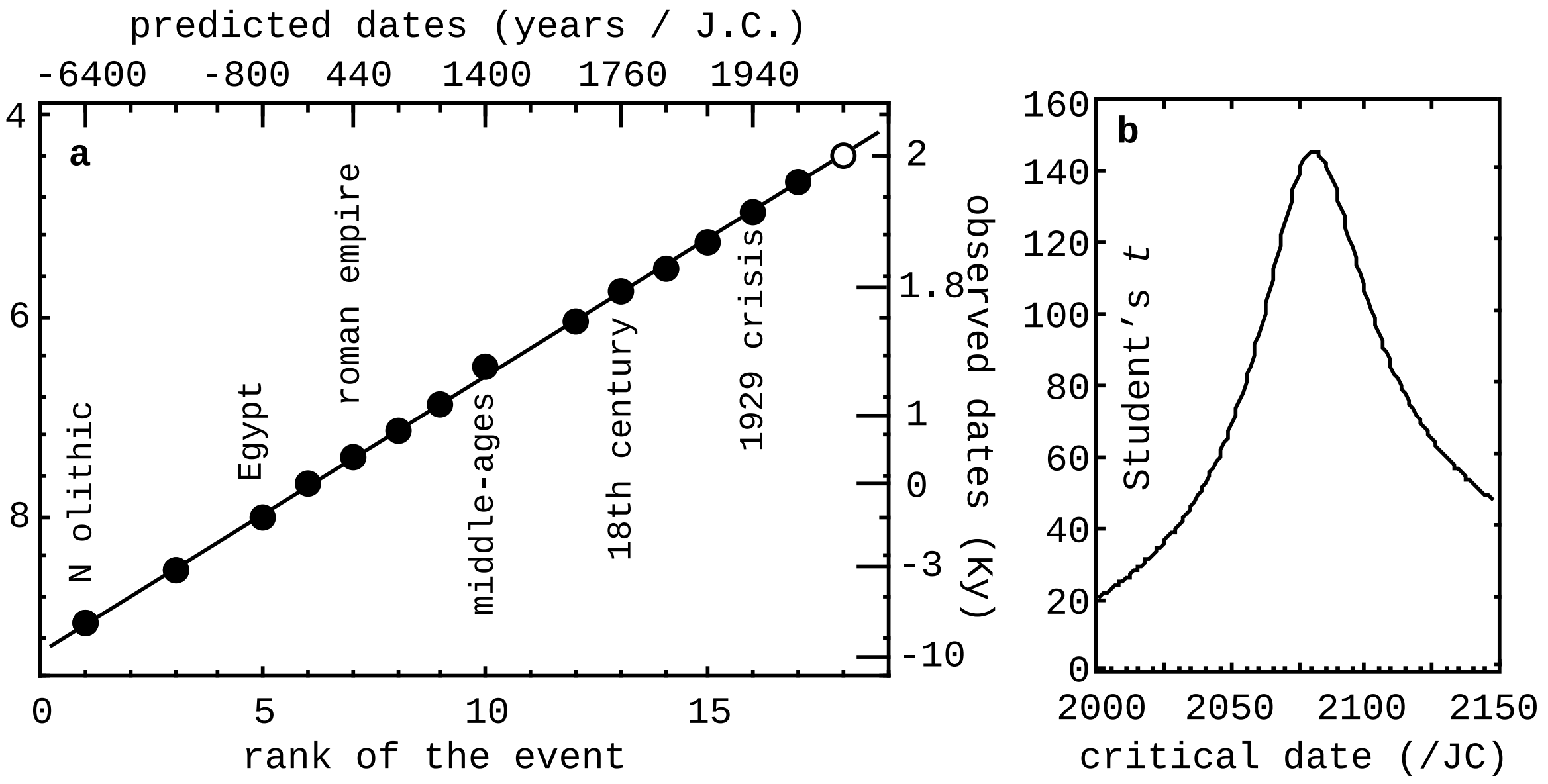

Figure 8: Comparison of the median dates of the main economic crises of western civilization with a log-periodic accelerating law of critical date $T_c = 2080$ and scale ratio $g = 1.32$ (figure a). The last white point corresponds to the predicted next crisis (1997-2000) at the date of the study (1996), as has been later supported in particular by the 1998 and 2000 market crashes, while the next crises are now predicted for (2015-2020), then (2030-2035). Figure b shows the estimation of the critical date through the optimisation of the Student's $t$ variable. This result is statistically significant, since the probability to obtain such a high peak by chance is $P < 10^{-4}$ (figure adapted from Fig. 47a of Ref. [86]).

One of the intriguing features of all these results is the frequent occurence of values of the scale ratio $g$ close to $g = 1.73$ and its square root 1.32 (recall that one passes from a value of $g$ to its square root by simply doubling the number of events). This suggests once again a discretization of the values of this scale ratio, that may be the result of a probability law (in scale space) showing quantized probability peaks . We have considered the possibility that $g = 1.73 \approx \sqrt{3}$ could be linked to a most probable branching ratio of 3 [19, 86], while Queiros-Condé [112] has proposed a 'fractal skin' model for understanding this value.

### 3.3.2 History and geography

The application of the various tools and methods of the scale relativity theory to history and geography has been proposed by Martin and Forriez [34, 61, 35, 36]. Forriez has shown that the chronology of some historical events (various steps of evolution of a given site) recovered from archeological and historical studies can be fitted by a log-periodic deceleration law with again $g \approx 1.7$ and a retroprediction of the foundation date of the site from the critical date [34, 35]. Moreover, the various differential equation tools developed in the scale relativity approach both in scale and position space, including the nonlinear cases of variable fractal dimensions, have been found to be particularly well adapted to the solution of geographical problems [61].

### 3.3.3 Predictivity

Although these studies remain, at that stage, of an empirical nature (it is only a purely chronological analysis which does not take into account the nature of the events), they nevertheless provide us with a beginning of predictivity. Indeed, the fitting law is a two parameter function ($T_c$ and $g$) that is applied to time intervals, so that only three events are needed to define these parameters. Therefore the subsequent dates are predicted after the third one, in a statistical way. Namely, as already remarked, the predicted dates should be interpreted as the dates of the peaks of probability for an event to happen.

Examples of such a predictivity (or retropredictivity) are:
(i) the retroprediction that the common Homo-Pan-Gorilla ancestor (expected, e.g., from genetic distances and phylogenetic studies), has a more probable date of appearance at $\approx -10$ millions years [19]; its fossil has not yet been discovered (this is one of the few remaining 'missing links');
(ii) the prediction of a critical date for the long term evolution of human societies around the years 2050-2080 [86, 50, 87, 45];
(iii) the finding that the critical dates of rodents may reach +60 Myrs in the future, showing their large capacity of evolution, in agreement with their known high biodiversity;
(iv) the finding that the critical dates of dinosaurs are about $-150$ Myrs in the past, indicating that they had reached the end of their capacity of evolution (at least for the specific morphological characters studied) well before their extinction at $-65$ Myrs;
(v) the finding that the critical dates of North american Equids is, within uncertainties, consistent with the date of their extinction, which may mean that, contrarily to the dinosaur case, the end of their capacity of evolution has occured during a phase of environmental change that they have not been able to deal with by the mutation-selection process;
(vi) the finding that the critical date of echinoderms (which decelerate instead of accelerating) is, within uncertainties, the same as that of their apparition during the PreCambrian-Cambrian radiation, this supporting the view of the subsequent events as a kind of "scale wave" expanding from this first shock.

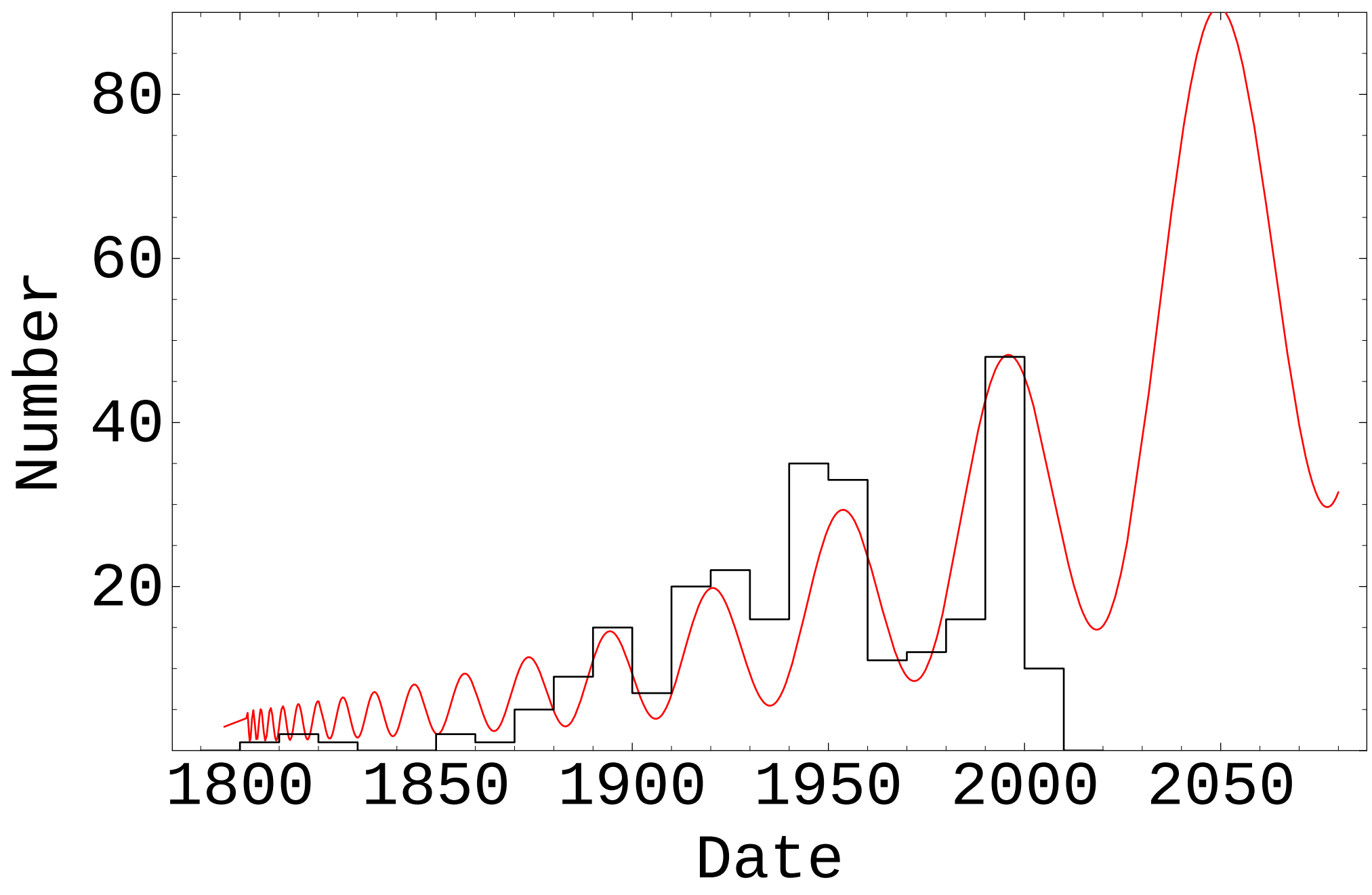

Figure 9: Observed rate of Southern California earhquakes of magnitude larger than 5 (histogram). The data are taken from the U.S. Geological Survey EarthQuake Data Center (years 1932-2006) and EarthQuake Data Base (Historical earthquakes, years 1500-1932). This rate is well fitted by a power law subjected to a log-periodic fluctuation decelerating since a critical date $T_c = 1796$ (red fluctuating line). The model predicts the next probability peak around the years 2050 [97].

#### 3.3.4 Applications in Earth sciences

As last examples of such a predictivity, let us give some examples of applications of critical laws (power laws in $|T - T_c|^\gamma$ and their log-perodic generalizations) to problems encountered in Earth sciences, namely, earthquakes (California and Sichuan) and decline of Arctic sea ice.

**California earthquakes** The study of earthquakes has been one of the first domain of application of critical and log-periodic laws [119, 3]. The rate of California earthquakes is found to show a very marked log-periodic deceleration [97, 98]. We show indeed in Fig. 9 the observed rate of Southern California earhquakes of magnitude larger than 5, compared with a log-periodic deceleration law. This model allows us to predict future peaks of probability around the years 2050 then 2115.

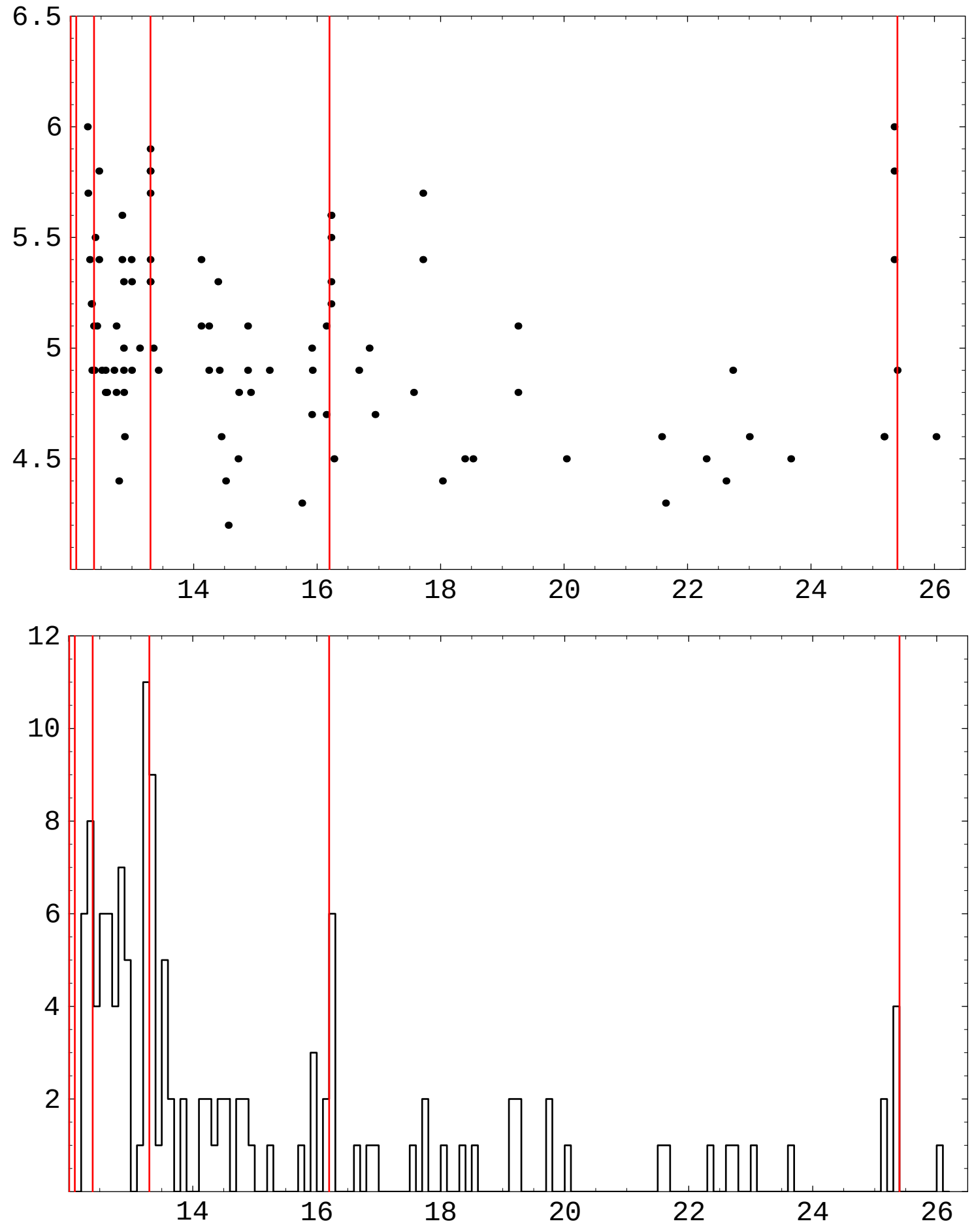

Figure 10: Time evolution during 14 days of the replicas of the May 12, 2008 Sichuan earthquake (data obtained and studied May 27, 2008 from the seismic data bank EduSeis Explorer, http://aster.unice.fr/EduSeisExplorer/form-sis.asp). The (up) figure gives the magnitudes of the replicas and the (down) figure the rate of replicas. Both show a continuous decrease to which are added discrete sharp peaks. The peaks which are common to both diagrams show a clear deceleration according to a log-periodic law starting from the main earthquake (red vertical lines), which allows one to predict the next strongest replicas with a good precision. For example, the peak of replicas of 25 May 2008 could be predicted with a precision of 1.5 day from the previous peaks. Reversely, the date of the main earthquake (May 12.27 2008, magnitude 7.9) can be retropredicted from that of the replicas with a precision of 6 h.

**Sichuan 2008 earthquake** The May 2008 Sichuan earthquake and its replicas also yields a good example of log-periodic deceleration, but on a much smaller time scale (see Fig 10).

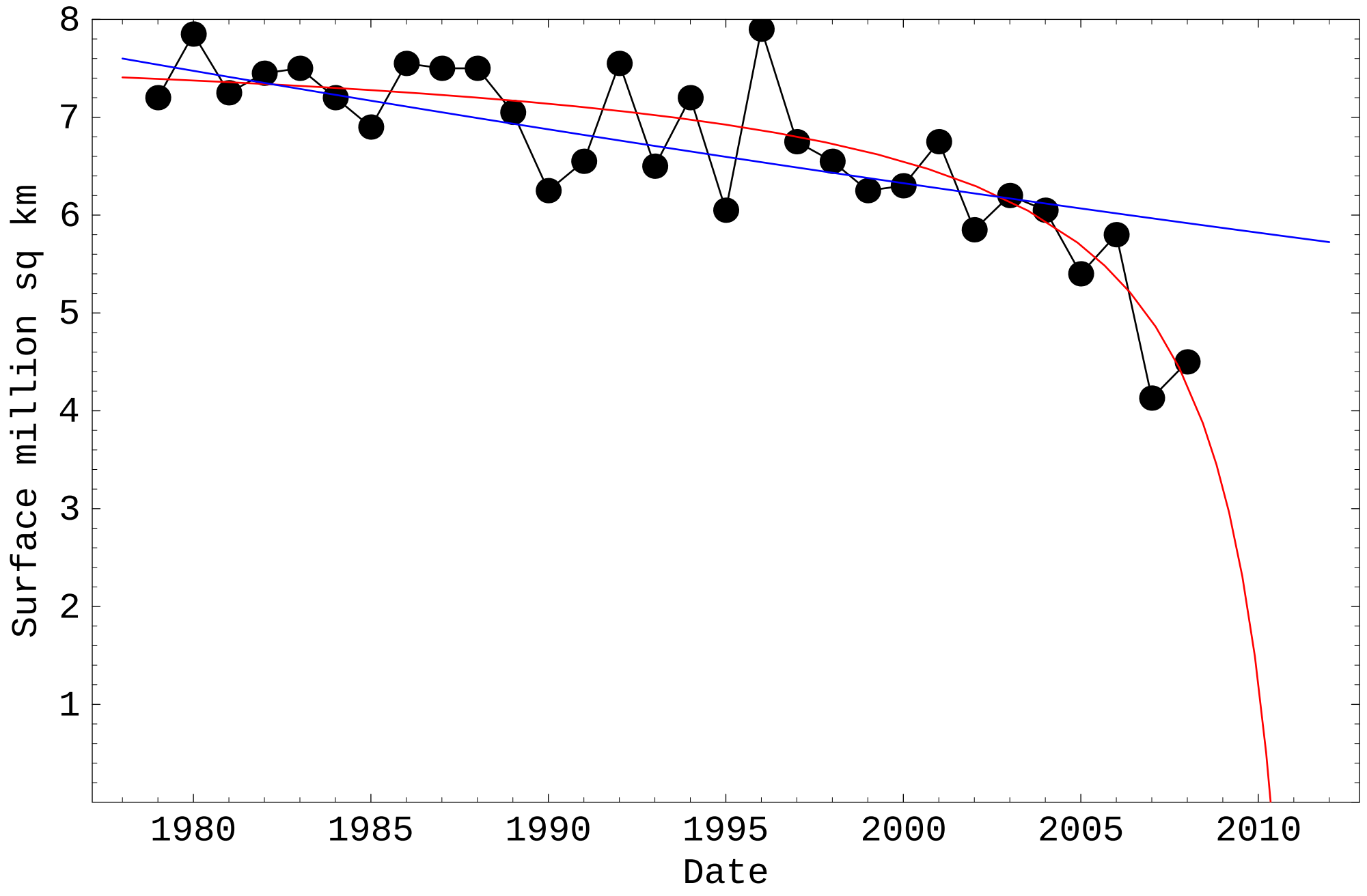

Figure 11: Observed evolution of the minimum arctic sea ice extent, according to the data of the U.S. National Snow and Ice Data Center (NSIDC, http://nsidc.org/), from 1979 to 2008. The minimum ocurs around 15 September of each year. This evolution is compared to: (i) the standard fit corresponding to an assumed constant rate of extent decrease (blue line); (ii) a fit by a critical law accelerating toward a critical date $T_c = 2012$. The second fit is far better and has allowed us to predict the 2007 and 2008 low points before their observation [98]. It implies that the arctic sea is expected to be totally free from ice by September 2011.

**Arctic sea ice extent** It is now well-known that the decrease of arctic sea-ice extent has shown a strong acceleration in 2007 and 2008 with respect to the current models assuming a constant rate ($\approx 8\%$ by decade), which predicted in 2006 a total disappearance of the ice at minimum (15 september) for the end of the century. From the view point of these models, the 2007 and now 2008 values (see Fig. 11) were totally unexpected.

However, we have proposed, before the knowledge of the 2007 minimum, to fit the data with a critical law of the kind $y = y_0 - a|T - T_c|^\gamma$. Such an accelerating law has the advantage to include in its structure the fact that one expect the ice to fully disappear after some date, while the constant rate law formally pushed the date of total disappearance to infinity. The fit of the data up to 2006 with the critical law was already far better than with the constant rate law, and it actually allowed us to predict a full disappearance epoch far closer than previously expected and a low 2007 point [98]. The 2008 point has confirmed the validity of the model in an impressive way (Fig. 11). We obtain by the $\chi^2$ method a best fit for the minimum ice surface (in square kilometers), $y = 8 - 12.3 \times |T - 2012|^{-0.86}$. The critical time is as early as $T_c = 2012$, which means that a full ice melting is predicted for September 2011, and is even possible for September 2010, for which the model gives only 1.2 million km$^2$ of remaining ice surface.

The application of the same method to the mean surface data during August and October months also shows a clear acceleration toward $T_c = 2013$, which means that only one year (2012) after the first total melting, the arctic sea can be expected to be free from ice during several months (August to October).

## 3.4 Applications of scale relativity to biology

One may consider several applications to biology of the various tools and methods of the scale relativity theory, namely, generalized scale laws, macroscopic quantum-type theory and Schrödinger equation in position space then in scale space and emergence of gauge-type fields and their associated charges from fractal geometry [69, 86, 90, 6, 95]. One knows that biology is founded on biochemistry, which is itself based on thermodynamics, to which we contemplate the future possibility to apply the macroquantization tools described in the theoretical part of this article. Another example of future possible applications is to the description of the growth of polymer chains, which could have consequences for our understanding of the nature of DNA and RNA molecules.

Let us give some explicit examples of such applications.

### 3.4.1 Confinement

The solutions of non-linear scale equations such as that involving a harmonic oscillator-like scale force [75] may be meaningful for biological systems. Indeed, its main feature is its capacity to describe a system in which a clear separation has emerged between an inner and an outer region, which is one of the properties of the first prokaryotic cell. We have seen that the effect of a scale harmonic oscillator force results in a confinement of the large scale material in such a way that the small scales may remain unaffected.

Another interpretation of this scale behavior amounts to identify the zone where the fractal dimension diverges (which corresponds to an increased 'thickness' of the material) as the description of a membrane. It is indeed the very nature of biological systems to have not only a well-defined size and a well-defined separation between interior and exterior, but also systematically an interface between them, such as membranes or walls. This is already true of the simplest prokaryote living cells. Therefore this result suggests the possibility that there could exist a connection between the existence of a scale field (e.g., a global pulsation of the system, etc..) both with the confinement of the cellular material and with the appearance of a limiting membrane or wall [90]. This is reminiscent of eukaryotic cellular division which involves both a dissolution of the nucleus membrane and a deconfinement of the nucleus material, transforming, before the division, an eukaryote into a prokaryote-like cell. This could be a key toward a better understanding of the first major evolutionary leap after the appearance of cells, namely the emergence of eukaryotes.

### 3.4.2 Morphogenesis

The generalized Schrödinger equation (in which the Planck constant $\hbar$ can be replaced by a macroscopic constant) can be viewed as a fundamental equation of morphogenesis. It has not been yet considered as such, because its unique domain of application was, up to now, the microscopic (molecular, atomic, nuclear and elementary particle) domain, in which the available information was mainly about energy and momentum.

However, scale relativity extends the potential domain of application of Schrödinger-like equations to every systems in which the three conditions (infinite or very large number of trajectories, fractal dimension of individual trajectories, local irreversibility) are fulfilled. Macroscopic Schrödinger equations can be constructed, which are not based on Planck's constant $\hbar$, but on constants that are specific of each system (and may emerge from their self-organization).

Now the three above conditions seems to be particularly well adapted to the description of living systems. Let us give a simple example of such an application.

In living systems, morphologies are acquired through growth processes. One can attempt to describe such a growth in terms of an infinite family of virtual, fractal and locally irreversible, trajectories. Their equation can therefore be written under the form of a fractal geodesic equation, then it can be integrated as a Schrödinger equation.

If one now looks for solutions describing a growth from a center, one finds that this problem is formally identical to the problem of the formation of planetary nebulae [27], and, from the quantum point of view, to the problem of particle scattering, e.g., on an atom. The solutions looked for correspond to the case of the outgoing spherical probability wave.

Depending on the potential, on the boundary conditions and on the symmetry conditions, a large family of solutions can be obtained. Considering here only the simplest ones, i.e., central potential and spherical symmetry, the probability density distribution of the various possible values of the angles are given in this case by the spherical harmonics,

$$P(\theta, \varphi) = |Y_{lm}(\theta, \varphi)|^2. \tag{105}$$

These functions show peaks of probability for some angles, depending on the quantized values of the square of angular momentum $L^2$ (measured by the quantum number $l$) and of its projection $L_z$ on axis $z$ (measured by the quantum number $m$).

Finally a more probable morphology is obtained by 'sending' matter along angles of maximal probability. The biological constraints leads one to skip to cylindrical symmetry. This yields in the simplest case a periodic quantization of the angle $\theta$ (measured by an additional quantum number $k$), that gives rise to a separation of discretized 'petals'. Moreover there is a discrete symmetry breaking along the $z$ axis linked to orientation (separation of 'up' and 'down' due to gravity, growth from a stem). The solutions obtained in this way show floral 'tulip'-like shapes (see Fig. 12 and [84, 90, 95]).

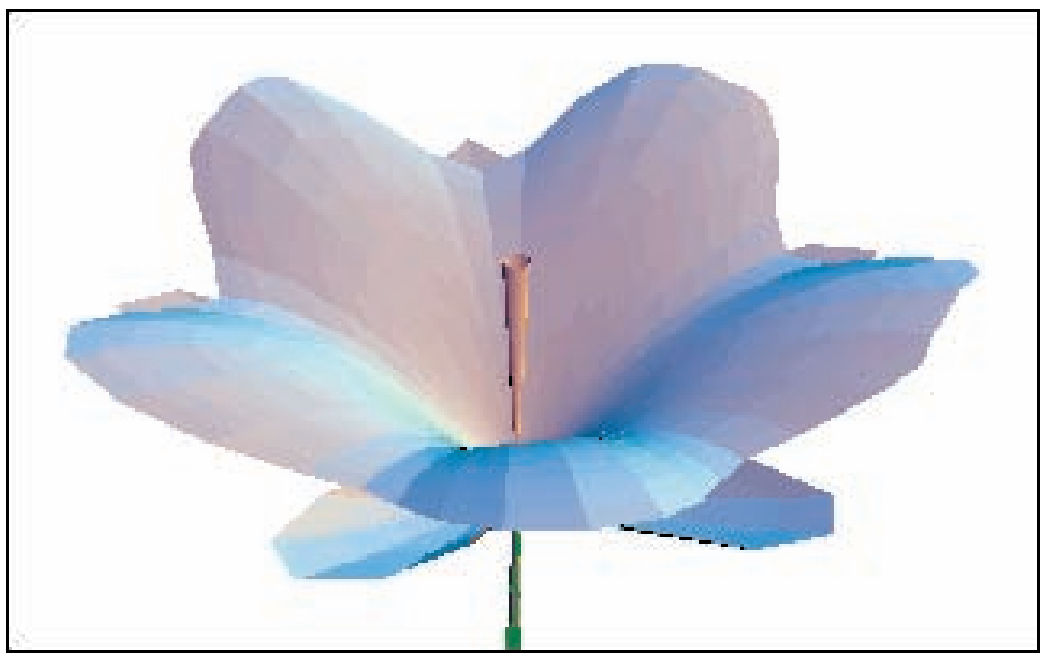

Figure 12: Morphogenesis of a 'flower'-like structure, solution of a generalized Schrödinger equation that describes a growth process from a center($l = 5,\ m = 0$). The 'petals', 'sepals' and 'stamen' are traced along angles of maximal probability density. A constant force of 'tension' has been added, involving an additional curvature of 'petals', and a quantization of the angle $\theta$ that gives an integer number of 'petals' (here, $k = 5$).

Coming back to the foundation of the theory, it is remarkable that these shapes are solutions of a geodesic, strongly covariant equation $\widehat{d}\mathcal{V}/dt = 0$, which has the form of the Galilean motion equation in vacuum in the absence of external force. Even more profoundly, this equation does not describe the motion of a particle, but purely geometric virtual possible paths, this given rise to a description in terms of a probability density which plays the role of a potential for the real particle (if any, since, in the application to elementary particles, we identify the 'particles' with the geodesics themselves, i.e., they become pure relative geometric entities devoid of any proper existence).

### 3.4.3 Origin of life

The problems of origin are in general more complex than the problems of evolution. Strictly, there is no 'origin' and both problems could appear to be similar, since the scientific and causal view amounts to consider that any given system finds its origin in an evolution process. However, systems are in general said to evolve if they keep their nature, while the question is posed in terms of origin when a given system appears from another system of a completely different nature, and moreover, often on times scales which are very short with respect to the evolution time. An example in astrophysics is the origin of stars and planetary systems from the interstellar medium, and in biology the probable origin of life from a prebiotic medium.

A fondamentally new feature of the scale relativity approach concerning such problems is that the Schrödinger form taken by the geodesic equation can be interpreted as a general tendency for systems to which it applies to make structures, i.e., to lead to self-organization. In the framework of a classical deterministic approach, the question of the formation of a system is always posed in terms of initial conditions. In the new framework, the general existence of stationary solutions allows structures to be formed whatever the initial conditions, in correspondence with the field, the symmetries and the boundary conditions (namely the environmental conditions in biology), and in function of the values of the various conservative quantities that characterize the system.

Such an approach could allow one to ask the question of the origin of life in a renewed way. This problem is the analog of the 'vacuum' (lowest energy) solutions, i.e., of the passage from a non-structured medium to the simplest, fundamental level structures. In astrophysics and cosmology, the problem amounts to understand the apparition, from the action of gravitation alone, of structures (planets, stars, galaxies, clusters of galaxies, large scale structures of the Universe) from a highly homogeneous and non-structured medium whose relative fluctuations were smaller than $10^{-5}$ at the time of atom formation. In the standard approach to this problem a large quantity of postulated and unobserved dark matter is needed to form structures, and even with this help the result is dissatisfying. In the scale relativity framework, we have suggested that the fundamentally chaotic behavior of particle trajectories leads to an underlying fractal geometry of space, which involves a Schrödinger form for the equation of motion, leading both to a natural tendency to form structures and to the emergence of an additional potential energy, identified with the 'missing mass(-energy)'.

The problem of the origin of life, although clearly far more difficult and complex, shows common features with this question. In both cases one needs to understand the apparition of new structures, functions, properties, etc... from a medium which does not yet show such structures and functions. In other words, one need a theory of emergence. We hope that scale relativity is a good candidate for such a theory, since it owns the two required properties: (i) for problems of origin, it gives the conditions under which a weakly structuring or destructuring (e.g., diffusive) classical system may become quantum-like and therefore structured; (ii) for problems of evolution, it makes use of the self-organizing property of the quantum-like theory.

We therefore tentatively suggest a new way to tackle the question of the origin of life (and in parallel, of the present functionning of the intracellular medium) [90, 6, 100]. The prebiotic medium on the primordial Earth is expected to have become chaotic in such a way that, on time scales long with respect to the chaos time (horizon of predictibility), the conditions that underlie the transformation of the motion equation into a Schrödinger-type equation, namely, complete information loss on angles, position and time leading to a fractal dimension 2 behavior on a range of scales reaching a ratio of at least $10^4$-$10^5$, be fulfilled. Since the chemical structures of the prebiotic medium have their lowest scales at the atomic size, this means that, under such a scenario, one expects the first organized units to have appeared at a scale of about 10 $\mu$m, which is indeed a typical scale for the first observed prokaryotic cells. The spontaneous transformation of a classical, possibly diffusive mechanics, into a quantum-like mechanics, with the diffusion coefficient becoming the quantum self-organization parameter $\mathcal{D}$ would have immediate dramatic consequences: quantization of energy and energy exchanges and therefore of information, apparition of shapes and quantization of these shapes (the cells can be considered as the 'quanta' of life), spontaneous duplication and branching properties (see herebelow), etc... Moreover, due to the existence of a vacuum energy in quantum mechanics (i.e., of a non vanishing minimum energy for a given system), we expect the primordial structures to appear at a given non-zero energy, without any intermediate step.

Such a possibility is supported by the symplectic formal structure of thermodynamics [109], in which the state equations are analogous to Hamilton-Jacobi equations. One can therefore contemplate the possibility of a future 'quantization' of thermodynamics, and then of the chemistry of solutions, leading to a new form of macroscopic quantum (bio)-chemistry, which would hold both for the prebiotic medium at the origin of life and for today's intracellular medium.

In such a framework, the fundamental equation would be the equation of molecular fractal geodesics, which could be transformed into a Schrödinger equation for wave functions $\psi$. This equation describes an universal tendency to make structures in terms of a probability density $P$ for chemical products (constructed from the distribution of geodesics), given by the squared modulus of the wave function $\psi = \sqrt{P} \times e^{i\theta}$. Each of the molecules being subjected to this probability (which therefore plays the role of a potentiality), it is proportional to the concentration $c$ for a large number of molecules, $P \propto c$ but it also constrains the motion of individual molecules when they are in small number (this is similar to a particle-by-particle Young slit experiment).

Finally, the Schrödinger equation may in its turn be transformed into a continuity and Euler hydrodynamic-like system (for the velocity $V = (v_+ + v_-)/2$ and the probability $P$) with a quantum potential depending on the concentration when $P \propto c$,

$$Q = -2\mathcal{D}^2 \frac{\Delta\sqrt{c}}{\sqrt{c}}. \tag{106}$$

This hydrodynamics-like system also implicitly contains as a sub-part a standard diffusion Fokker-Planck equation with diffusion coefficient $\mathcal{D}$ for the velocity $v_+$. It is therefore possible to generalize the standard classical approach of biochemistry which often makes use of fluid equations, with or without diffusion terms (see, e.g., [64, 118]).

Under the point of view of this third representation, the spontaneous transformation of a classical system into a quantum-like system through the action of fractality and small time scale irreversibility manifests itself by the appearance of a quantum-type potential energy in addition to the standard classical energy balance. We therefore predict that biological systems must show an additional energy (quite similar to the missing energy of cosmology usually attributed to a never found 'dark matter') given by the above relation (106) in terms of concentrations, when their total measured energy balance is compared to the classically expected one.

But we have also shown that the opposite of a quantum potential is a diffusion potential. Therefore, in case of simple reversal of the sign of this potential energy, the self-organization properties of this quantum-like behavior would be immediately turned, not only into a weakly organized classical system, but even into an increasing entropy diffusing and desorganized system. We tentatively suggest [95] that such a view may provide a renewed way of approach to the understanding of tumors, which are characterized, among many other features, by both energy affinity and morphological desorganization.

### 3.4.4 Duplication

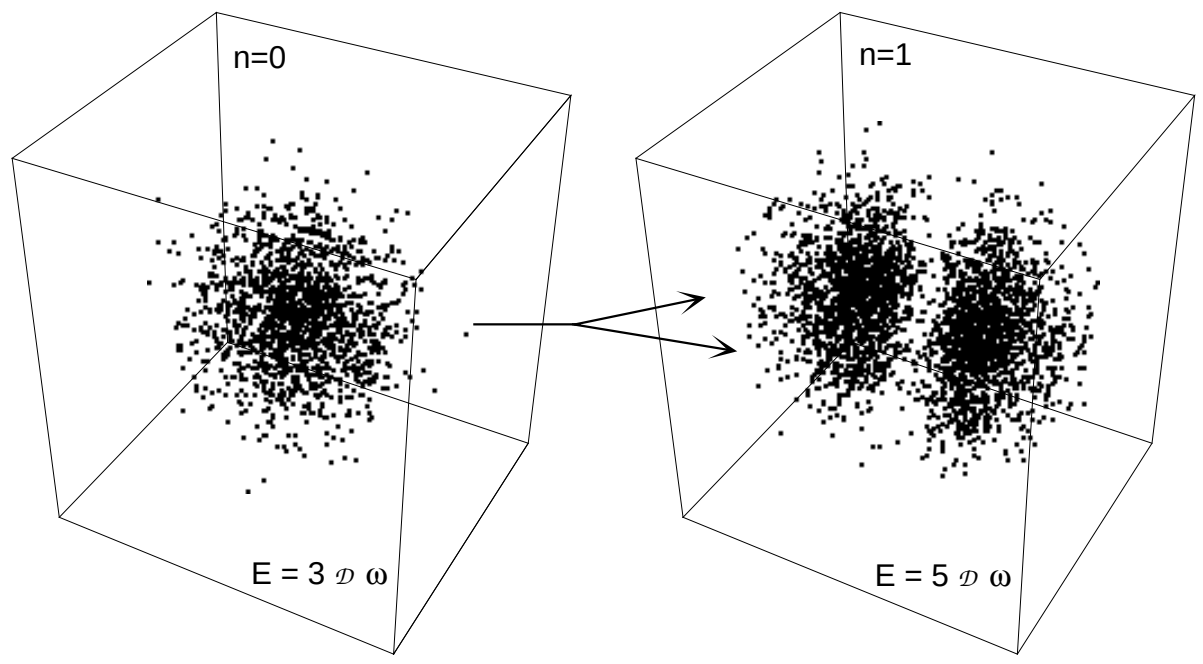

Figure 13: Model of duplication. The stationary solutions of the Schrödinger equation in a 3D harmonic oscillator potential can take only discretized morphologies in correspondence with the quantized value of the energy. Provided the energy increases from the one-structure case ($E_0 = 3\mathcal{D}\omega$), no stable solution can exist before it reaches the second quantized level at $E_1 = 5\mathcal{D}\omega$. The solutions of the time-dependent equation show that the system jumps from the one structure to the two-structure morphology.

Secondly, the passage from the fundamental level to the first excited level now provides one with a (rough) model of duplication (see Figs. 13 and 14). Once again, the quantization implies that, in case of energy increase, the system will not increase its size, but will instead be lead to jump from a single structure to a binary structure, with no stable intermediate step between the two stationary solutions $n = 0$ and $n = 1$. Moreover, if one comes back to the level of description of individual trajectories, one finds that from each point of the initial one body-structure there exist trajectories that go to the two final structures. In this framework, duplication is expected to be linked to a discretized and precisely fixed jump in energy.

It is clear that, at this stage, such a model is extremely far from describing the complexity of a true cellular division, which it did not intend to do. Its interest is to be a generic and general model for a spontaneous duplication process of quantized structures, linked to energy jumps. Indeed, the jump from one to two probability peaks when going from the fundamental level to the first excited level is found in many different situations of which the harmonic oscillator case is only an example. Moreover, this duplication property is expected to be conserved under more elaborated versions of the description provided the asymptotic small scale behavior remains of constant fractal dimension $D_F \approx 2$, such as, e.g., in cell wall-like models based on a locally increasing effective fractal dimension.

### 3.4.5 Bifurcation, branching process

Such a model can also be applied to a first rough description of a branching process (Fig. 14), e.g., in the case of a tree growth when the previous structure remains instead of disappearing as in cell duplication.

Note finally that, although such a model is still clearly too rough to claim that it describes biological systems, it may already be improved by combining with it various other functional and morphological elements which have been obtained. Namely, one may apply the duplication or branching process to a system whose underlying scale laws (which condition the derivation of the generalized Schrödinger equation) include (i) the model of membrane through a fractal dimension that becomes variable with the distance to a center; (ii) the model of multiple hierarchical levels of organization depending on ‘complexergy’ (see herebelow).

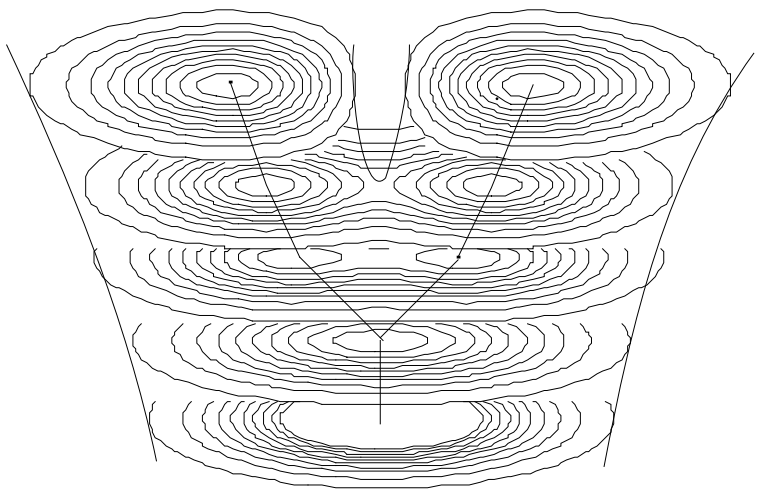

Figure 14: Model of branching and bifurcation. Successive solutions of the time-dependent 2D Schrödinger equation in an harmonic oscillator potential are plotted as isodensities. The energy varies from the fundamental level ($n = 0$) to the first excited level ($n = 1$), and as a consequence the system jumps from a one-structure to a two-structure morphology.

### 3.4.6 Nature of first evolutionary leaps

We have also suggested applications to biology of the new quantum-like mechanics in scale space [90].

In the fractal model of the tree of life described hereabove [19], we have voluntarily limited ourselves to an analysis of only the chronology of events (see Fig. 7), independently of the nature of the major evolutionary leaps. The suggestion of a quantum-type mechanics in scale space and of the new concept of complexergy [90, 100], which is a new conservative quantity appearing from the symmetry of the new scale variables (more precisely, of the fractal dimension become variable and considered as a fifth dimension) allows one to reconsider the question.

One may indeed suggest that life evolution proceeds in terms of increasing quantized complexergy. This would account for the existence of punctuated evolution [41], and for the log-periodic behavior of the leap dates, which can be interpreted in terms of probability density of the events, $P = |\psi|^2 \propto \sin^2[\omega \ln(T - T_c)]$. Moreover, one may contemplate the possibility of an understanding of the nature of the events, even though in a rough way as a first step.

Indeed, one can expect the first formation of a structure at the fundamental level (lowest complexergy), which is generally characterized by only one length-scale (this is the analog in scale space of the left part of Fig. 13 which concerns position space). Moreover, the most probable value for this scale of formation is predicted to be the 'middle' of the scale-space, since the problem is similar to that of a quantum particle in a box, with the logarithms of the minimum scale $\lambda_m$ and maximum scale $\lambda_M$ playing the roles of the walls of the box, so that the fundamental level solution has a peak at a scale $\sqrt{\lambda_m \times \lambda_M}$.

The universal boundary conditions are the Planck-length $l_{\mathbb{P}}$ in the microscopic domain and the cosmic scale $\mathbb{L} = \Lambda^{-1/2}$ given by the cosmological constant $\Lambda$ in the macroscopic domain (see Sec. 3.1.2). From the predicted and now observed value of the cosmological constant, one finds $\mathbb{L}/l_{\mathbb{P}} = 5.3 \times 10^{60}$, so that the mid scale is at $2.3 \times 10^{30}\, l_{\mathbb{P}} \approx 40\ \mu$m. A quite similar result is obtained from the scale boundaries of living systems (≈0.5 Angströms - 30 m). This scale of 40 $\mu$m is indeed a typical scale of living cells. Moreover, the first 'prokaryot' cells appeared about three Gyrs ago had only one hierarchy level (no nucleus).

In this framework, a further increase of complexergy can occur only in a quantized way. The second level describes a system with two levels of organization, in agreement with the second step of evolution leading to eukaryots about 1.7 Gyrs ago (second event in Fig. 7). One expects (in this very simplified model), that the scale of nuclei be smaller than the scale of prokaryots, itself smaller than the scale of eucaryots: this is indeed what is observed.

The following expected major evolutionary leap is a three organization level system, in agreement with the apparition of multicellular forms (animals, plants and fungi) about 1 Gyr ago (third event in Fig. 7). It is also predicted that the multicellular stage can be built only from eukaryots, in agreement with what is observed. Namely, the cells of multicellulars do have nuclei; more generally, evolved organisms keep in their internal structure the organization levels of the preceeding stages.

The following major leaps correspond to more complicated structures then more complex functions (supporting structures such as exoskeletons, tetrapody, homeothermy, viviparity), but they are still characterized by fundamental changes in the number of organization levels. Moreover, the first steps in the above model are based on spherical symmetry, but this symmetry is naturaly broken at scales larger than 40 $\mu$m, since this is also the scale beyond which the gravitational force becomes larger than the van der Waals force. One therefore expects the evolutionary leaps that follow the apparition of multicellular systems to lead to more complicated structures, such as those of the Precambrian-Cambrian radiation, than can no longer be described by a single scale variable.

### 3.4.7 Origin of the genetic code

We therefore intend, in future works, to extend the model to more general symmetries, boundary conditions and constraints. We also emphasize once again that such an approach does not dismiss the role and the importance of the genetic code in biology. On the contrary, we hope that it may help understanding its origin and its evolution.

Indeed, we have suggested that the various biological morphologies and functions are solutions of macroscopic Schrödinger-type equations, whose solutions are quantized according to integer numbers that represent the various conservative quantities of the system. Among these quantities, one expects to recover the basic physical ones, such as energy, momentum, electric charge, etc... But one may also contemplate the possibility of the existence of prime integrals (conservative quantities) which would be specific of biology (or particularly relevant to biology), among which we have suggested the new concept of complexergy, but also new scale 'charges' finding their origin in the internal scale symmetries of the biological systems.

The quantization of these various quantities means that any such system would be described by a set of integer numbers, so that one may tentatively suggest that only these numbers, instead of a full continuous and detailed information, would have to be included in the genetic code. In this case the process of genetic code reading, protein synthesis, etc... would be a kind of 'analogic solutioner' of Schrödinger equation, leading to the final morphologies and functions. Such a view also offers a new line of research toward understanding the apparition of the code, namely, the transformation of what was a purely chemical process into a support of information and of its implementation, thanks to the quantization of the exchanges of energy and other conservative quantities.

We intend to develop this approach in future works, in particular by including the scale relativity tools and methods in a system biology framework allowing multiscale integration [6, 100], in agreement with Noble's 'biological relativity' [65] according to which there is no privileged scale in living systems.

# 4 Conclusion

The theory of scale relativity relies on the postulate that the fundamental laws that govern the various physical, biological and other phenomenons find their origin in first principles. In continuity with previous theories of relativity, it considers that the most fundamental of these principles is the principle of relativity itself. The extraordinary success due to the application of this principle, since now four centuries, to position, orientation, motion (and therefore to gravitation) is well known.

But, during the last decades, the various sciences have been faced to an ever increasing number of new unsolved problems, of which many are linked to questions of scales. It therefore seemed natural, in order to deal with these problems at a fundamental and first principle level, to extend theories of relativity by including the scale in the very definition of the coordinate system, then to account for these scale transformations in a relativistic way.

We have attempted to give in this article a summarized discussion of the various developments of the theory and of its applications. The aim of this theory is to describe space-time as a continuous manifold without making the hypothesis of differentiability, and to physically constrain its possible geometry by the principle of relativity, both of motion and of scale. This is effectively made by using the physical principles that directly derive from it, namely, the covariance, equivalence and geodesic principles. These principles lead in their turn to the construction of covariant derivatives, and finally to the writing, in terms of these covariant derivatives, of the motion equations under the form of free-like geodesic equations. Such an attempt is therefore a natural extension of general relativity, since the two-times differentiable continuous manifolds of Einstein's theory, that are constrained by the principle of relativity of motion, are particular sub-cases of the new geometry in construction.

Now, giving up the differentiability hypothesis involves an extremely large number of new possible structures to be investigated and described. In view of the immensity of the task, we have chosen to proceed by steps, using presently known physics as a guide. Such an approach is rendered possible by the result according to which the small scale structure which manifest the nondifferentiability are smoothed out beyond some relative transitions toward the large scales. One therefore recovers the standard classical differentiable theory as a large scale approximation of this generalized approach. But one also obtains a new geometric theory which allows one to understand quantum mechanics as a manifestation of an underlying nondifferentiable and fractal geometry, and finally to suggest generalizations of it and new domains of application for these generalizations.

Now the difficulty with theories of relativity is that they are meta-theories rather than theories of some particular systems. Hence, after the construction of special relativity of motion at the beginning of the twentieth century, the whole of physics needed to be rendered 'relativistic' (from the viewpoint of motion), a task that is not yet fully achieved. The same is true as regards the program of constructing a fully scale-relativistic science. Whatever be the already obtained successes, the task remains huge, in particular when one realizes that it is no longer only physics that is concerned, but now many other sciences, in particular biology. Its ability to go beyond the frontiers between sciences may be one of the main interests of the scale relativity theory, opening the hope of a refoundation on mathematical principles and on predictive differential equations of a 'philosophy of nature' in which physics would no longer be separated from other sciences.

**Acknowledgements** I gratefully thank the organizers of this conference, C. Vidal and J. Smart, for their kind invitation to contribute and for interesting discussions, and two referees for helpful remarks having allowed me to improve this paper.

# Scale relativity: an extended paradigm for physics and biology?

Commentary on the paper by Laurent Nottale “Scale relativity and fractal space-time: theory and applications”

Charles Auffray[1] and Denis Noble[2]

1- Functional Genomics and Systems Biology for Health, CNRS Institute of Biological Sciences; 7, rue Guy Moquet, BP8, 94801 Villejuif, France. Tel: +33-1-49-58-34-98; E-mail: charles.auffray@vjf.cnrs.fr; URL: http://www.vjf.cnrs.fr/genexpress/

2- Department of Physiology, Anatomy and Genetics; Oxford University, Parks Road, Oxford OX1 3PT, United Kingdom. Tel: +44-18-65-272-533; E-mail: denis.noble@dpag.ox.ac.uk; URL: http://noble.physiol.ox.ac.uk/People/DNoble/

## Abstract

With scale relativity theory, Laurent Nottale has provided a powerful conceptual and mathematical framework with numerous validated predictions that has fundamental implications and applications for all sciences. We discuss how this extended framework may help facilitating integration across multiple size and time frames in systems biology, and the development of a scale relative biology with increased explanatory power.

## Extending the principle of relativity to scales: a new scientific paradigm?

Have we reached the limits of applicability of the principle of relativity? From Galileo to Einstein, it has been extended by removing theoretical constraints that represent privileging viewpoints of measurement lacking *a priori* bases. First the constraint of privileged location was removed: the Earth is no longer the centre of the Universe; then that of velocity: only relative velocities can be observed; then that of acceleration: an accelerating body experiences a force indistinguishable from that of gravity. To a large degree, this has unified our understanding of the Universe. But there is an exception: it matters whether we are talking of microphysics or macrophysics. At the micro level, we have to use quantum mechanics; at the macro level, general relativity. Could this distinction be resolved by removing yet another constraint? The obvious candidate is that of scale. Why should there be privileged scales for quantum mechanics or astrophysics? This led Laurent Nottale to remove yet another constraint, that of general space-time differentiability.

In this extended framework, only scale ratios have physical meaning, not absolute scales. The laws of quantum mechanics become manifestations of the fractal, non-differentiable geometry of space-time constrained by the principle of relativity extended to scales, and all the axiomatic postulates of quantum mechanics can be derived from the first principles of scale relativity theory. Moreover, Laurent Nottale introduces the notion of ‘complexergy’ which is the equivalent for scale to what energy is for motion. Complexergy is linked to the complexity of the system under consideration, leading to insights on the emergence of discrete levels of organisation. The hierarchy is not continuous: this is obviously true at the microlevel (from quarks to atoms) and the astronomical level (from stars through to superclusters). It is also true at the middle level, of importance for biological sciences. From molecules to organisms and beyond, we can also distinguish discrete levels.

**Validated predictions of scale relativity: which will trigger acceptance of the theory?**

Scale relativity has implications for every aspect of physics, from elementary particle physics to astrophysics and cosmology. It provides numerous examples of theoretical predictions of standard model parameters, a theoretical expectation for the Higgs boson mass which will be potentially assessed in the coming years by the Large Hadron Collider, and a prediction of the cosmological constant which remains within the range of increasingly refined observational data. Strikingly, many predictions in astrophysics have already been validated through observations such as the distribution of exoplanets or the formation of extragalactic structures. The possibility offered by the theory to have classical and quantum mechanics operating in a common framework makes it possible to interpret quantum laws as anti-diffusion laws. This allows revisiting the nature of the classic-quantum transition and the foundations of thermodynamics, with a wide range of as yet unexplored possible consequences for chemistry and biology.

This work is a testimony that extending the principle of relativity to scale represents a fundamental change of paradigm, with a wide range of consequences for physical and other sciences in terms of concepts, tools, and applications. The scale relativity theory and tools extend the scope of current domain-specific theories, which are naturally recovered, not replaced, in the new framework. This may explain why the community of physicists has been slow to recognize its potential and even to challenge it. Hence we are led to wonder which of the successful predictions of scale relativity will trigger its acceptance and spread within the scientific community. The prediction that the Artic sea will be free from ice during one to three months as soon as 2011-2012 may represent such a test.

**Towards a scale relative biology?**

As it can be considered as a theory of emergence and self-organisation reflecting the constraints imposed by the fractal geometry of space-time on all structures in nature, scale relativity has also important implications for biology. The pervasive presence of space and time dimension limitations in biological systems led Laurent Nottale to investigate the potential of his theory for understanding them, with initial application to time and structural regularities during species evolution, embryonic development and morphogenesis. His most recent proposals deal with the origin of the genetic code and of life itself.

Indeed, a major difficulty in modelling biological systems is that the formalisms we use for the different levels of organisation (genes, protein networks, subcellular systems, cells, tissues, organs) and time frames are different. It is difficult therefore to derive genuinely multi-level solutions. This is solved by using the outputs from one level as inputs to another. Could there be a better way of doing this? Could scale relativity be applied to extend the conceptual and mathematical framework of systems biology for integration across multiple time and size scales? This would open for a more systematic unification which could also provide an independent basis for validation of the scale relativity theory itself.

Application of scale relativity to biology represents a huge challenge for theoretical, computational and experimental biologists. As Nottale shows so elegantly for microphysics and astrophysics, the fractal nature of space-time leads naturally to quantised jumps in levels of organisation. The evolution of viruses, prokaryotes, eukaryotes, multicellular organisms, organs and systems would then be seen as representing an outcome not dissimilar to the existence of stars, galaxies, and clusters at astronomical dimensions or of the various forms of microphysical structures.

This requires abandonment of a unitary concept of causation in biology. The ‘cause’ of the existence of different levels of organisation would not be comparable to the ‘cause’ of particular activities in particular organisms. This is compatible with the theory of biological relativity proposed by one of us (Noble 2006, 2008), i.e. the principle that there is no privileged level of causation in biology. As in physics, these ideas and proposals have yet to be known or accepted by most biologists, and much work remains to be done to sustain the development and deployment of a scale relative biology (Auffray and Nottale, 2008; Nottale and Auffray, 2008).

# Multiscale integration in scale relativity theory

Answer to Auffray's and Noble's commentary:
**Scale relativity: an extended paradigm for physics and biology?**

Laurent Nottale

**Abstract -** We give a "direction for use" of the scale relativity theory and apply it to an example of spontaneous multiscale integration including four embedded levels of organization (intracellular, cell, tissue and organism-like levels). We conclude by an update of our analysis of the arctic sea ice melting.

Auffray and Noble, in their commentary, raise the important question of multiscale integration in biology and of the ability of the scale relativity theory to contribute by new insights and methods to a future possible solution of this problem (and of other questions in life and other sciences).

In order to give elements of answer, let us recall how the scale relativity theory can be used for practical applications.

The construction of the theory of scale relativity proceeds by extension and generalization with respect to currently existing theories. Its founding principles are the same as those on which these theories are founded (principles of relativity and covariance, of optimization – least action and geodesic principles), but applied also to scale transformations of the reference system. As a consequence, its equations are themselves extensions of the standard fundamental equations (Euler-Lagrange equations for particles and fields, energy equation). Moreover several equivalent representations of these equations have been established (geodesic form, Schrödinger quantum-mechanical form and fluid dynamical form with quantum potential), which connect various domains and methods often considered as totally or partially disconnected (quantum and classical mechanics, diffusion, hydrodynamics).

This allows one to suggest a fast way to apply it to various systems (where, for example, the current methods have failed): it consists in starting from the standard description and in looking for the possible existence of the additional terms introduced by the scale relativity theory.

More generally, let us give some "directions for use" of the scale reativity theory:

**(I) Laws of scale transformation.** The main new ingredient of the theory is the explicit introduction in the description (physical quantities and their equations) of explicit scale variables (« resolutions ») achieving a "scale space". The theory does not deal (only) with the scaling properties of the standard variables, but also and mainly of these new variables. The scale laws of the coordinates which depend on them, then of the physical functions of these coordinates are obtained as consequences. The scale relativity approach writes these laws of scale transformation in terms of differential equations acting in the scale space. One recovers in this way the standard fractal laws with constant fractal dimension as the simplest possible laws, but one also generalizes them in many ways (including the possibility of quantum-like laws in scale space).
For a given system, one can therefore
(1) (i) attempt to analyse in a differential way the scale behavior of the system, then
(ii) write the corresponding differential equation,
(iii) solve them and
(iv) compare these solutions to the observational / experimental data, or, in a more empirical approach
(2) (i) look for:
- transitions from scale-dependence to scale-independence, and/or between different fractal dimensions;
- variations of the fractal dimensions, including linear variation, log-periodic fluctuations, divergence, etc...;
then
(ii) study the cause for this deviation from pure self-similarity (scale force, geometric distorsion in scale space...).

**(II) Laws of motion**. On the basis of the internal laws of scale which have been obtained for a given system in step (I), one now construct the laws of motion (by defining and using a covariant derivative which includes the effects of the fractal geometry). The various representations of the laws of motion in scale relativity theory include the following forms and their generalizations:
- geodesic equation / fundamental equation of dynamics,
- Schrödinger equation,
- diffusion equations,
- hydrodynamics equations including a quantum potential.

The application of the scale relativity approach to a given system may therefore involve the following possibilities, depending on the standard description of the system:
-(i) check for the existence in the studied system of the additional terms in the covariant total derivative and in the corresponding equation of dynamics;
-(ii) look for signatures of a quantum-type system (probability density which is the square of the modulus of a wave function, existence of a phase involving interferences);
- (iii) complete the diffusion Fokker-Planck type equation by a backward Fokker-Planck equation (as a consequence of microscopic time scale irreversibility);
- (iv) check for the existence of an additional quantum-type potential in the hydrodynamic form of the equations.

We can now apply this method to the specific question of multiscale integration. Let us give a hint of what would be the successive steps of such an application (a fully developped description lies outside the scope of this short answer and will be the subject of future publications).

One starts from a "point", which represents the smallest scale considered (for example, intracellular "organelles"), then one writes a motion equation which can be integrated in terms of a macroscopic Schrödinger-type equation. Actually, the solutions of this Schrödinger equation are naturally multiscaled. It yields the density of probability of the initial "points", which describes a structure at a larger scale (the "cell" level). Now, while the "vacuum" (lowest energy) state usually describes one object (a single "cell"), excited states describe multi objects ("tissue-like" level), each of which being often separated by zones of null densities (therefore corresponding to infinite quantum potentials) which may represent "walls" (looking, e.g., like an Abrikosov lattice). Note that the resulting structure is not only qualitative, but also quantitative, since the relative sizes of these three levels can be obtained from the theoretical description. Finally, such a "tissue" of individual "cells" can be inserted in a growth equation which takes itself a Schrödinger form. Its solutions yield a new, larger level of organization, such as the "flower" of Fig. 12 of the paper. Finally, the matching conditions between the small scale and large scale solutions allow to connect the constants of these two equations, and therefore the quantitative scales of their solutions.

Let us conclude this answer by a short update of one of the questions also raised by Auffray and Noble's commentary, namely, that of the fast decrease of the arctic sea ice extent. We have given in Fig. 11 of the paper a fit of the US National Snow and Ice Data Center data up to 2008 by a critical law yielding a very close critical date of 2012. The 2009 minimum is now known: its value of 5.1 millions of square km is the third lowest value registered, of the order of the 2007 and 2008 values and it therefore confirms (within fluctuations) the acceleration. A simple model of fractal fracture of the sea ice (see "The Arctic sea-ice cover: Fractal space-time domain", A. Chmela, V.N. Smirnovb and M.P. Astakhovb, Physica A 357, 556) leads naturally to an exponential increase of the enlightened surface, and then of the melting. A fit of the data up to 2009 by such a model (which is close to the critical one) still yields a very close date of full ice melting in 2014-2015.

# The Self-organization of Time and Causality: steps towards understanding the ultimate origin

Francis Heylighen,
Evolution, Complexity and Cognition Group,
Vrije Universiteit Brussel
fheyligh@vub.ac.be

**Abstract**: Possibly the most fundamental scientific problem is the origin of time and causality. The inherent difficulty is that all scientific theories of origins and evolution consider the existence of time and causality as given. We tackle this problem by starting from the concept of self-organization, which is seen as the spontaneous emergence of order out of primordial chaos. Self-organization can be explained by the selective retention of invariant or consistent variations, implying a breaking of the initial symmetry exhibited by randomness. In the case of time, we start from a random graph connecting primitive "events". Selection on the basis of consistency eliminates cyclic parts of the graph, so that transitive closure can transform it into a partial order relation of precedence. Causality is assumed to be carried by causal "agents" which undergo a more traditional variation and selection, giving rise to causal laws that are partly contingent, partly necessary.

## 1. The problem of origins

Without doubt, the most difficult and fundamental problem in cosmology is the origin of the universe. One reason why this problem is so difficult is that *all* traditional physical theories assume the existence of time and causal laws. These theories include Newtonian mechanics, quantum mechanics, relativity theory, thermodynamics, and their various combinations, such as relativistic quantum field theories. In all these theories, the evolution of a system is reduced to the (deterministic or more rarely stochastic) change of the system's state s(t) according to a given causal law (which is typically represented by the Schrödinger equation or some variation of it) [Heylighen, 1990b]. The time *t* here is seen as a real number, which therefore by definition takes values between minus infinity and plus infinity. The "system" therefore is assumed to have existed indefinitely. If we apply this same formal representation to the evolution of the universe, then we can only conclude that this universe cannot have an origin at any finite time $t_0$, because that would assume that before $t_0$ there was no system that could evolve, and therefore no previous state that could causally give rise to the "origin" state $s(t_0)$. Yet, the observation by Hubble that the universe is expanding, when extrapolated backwards, leads to the conclusion that the universe started at a single point in time, the Big Bang.

The deeper reason for this paradox is that time and causality are part of the ontology—i.e. the set of *a priori* postulated entities—that physical theory uses for representing all phenomena. They therefore cannot be explained from within the theory. This assumption of the *a priori* existence of time and causality is in fact merely a formalization of our intuition that every moment was preceded by another moment, and that for every effect there is always a cause. In earlier times, this paradox could only be resolved by postulating a supernatural origin: God as the "prime mover" or "uncaused cause" of the universe. This is of course not acceptable in a scientific theory. Moreover, it merely pushes the difficulty a little further, since we still cannot explain the origin of God. Present-day cosmology evades the problem by viewing the origin of the universe as a "singularity", i.e. a point in time where continuity, causality and natural law break down. However, existing theories by their very nature cannot tell us anything about the nature or origin of this singularity, and therefore the explanation remains essentially unsatisfactory.

This problem requires a radical overhaul of existing theoretical frameworks. Recently, a number of alternative approaches have been proposed that may offer the beginning of an answer to the origin of time and causality. These include postulating an imaginary time from which "real" time would emerge [Hawking, 1988; Deltete & Guy, 1996, Butterfield, Isham & Kensington, 1999], process physics, which sees space and time self-organizing out of a random information network [Cahill, 2003, 2005; Cahill, Klinger & Kitto, 2000], the emergence of causal sets from a quantum self-referential automaton [Eakins & Jaroszkiewicz, 2003], and a structural language for describing the emergence of space-time structure [Heylighen, 1990a]. These proposals are heterogeneous, based on advanced, highly abstract mathematics, and difficult to grasp intuitively. They moreover all start from highly questionable assumptions. As such, they have as yet not made any significant impact on current thinking about the origin of the universe.

The present paper attempts to approach the problem in a more intuitive, philosophical manner, instead of immediately jumping to mathematical formalism, as is common in physical theory. To achieve that, we will look at the emergence of time and causality as a process of self-organization, albeit a very unusual one in that it initially takes place outside of time.

## 2. Generalized self-organization

Models of evolution and complex systems have taught us quite a bit about the phenomenon of self-organization, which can be defined most simply as the spontaneous appearance of *order out of chaos* [Prigogine & Stengers, 1984; Heylighen, 2001]. Extended to the level of the universe, this harkens back to the old Greek idea that *cosmos* emerged from *chaos*, an idea that predates more recent metaphysical theories where the cosmos is created by the pre-existing order or intelligence embodied in God.

Chaos here refers to randomness or disorder, i.e. the absence of any form of constraint, dependency or structure. Since maximum disorder is featureless and therefore indistinguishable from emptiness or vacuum, the existence of disorder does not need to be explained. In fact, modern physical theories conceive the vacuum precisely as a turbulent, boiling chaos of quantum fluctuations, continuously producing virtual particles that are so short-lived that they cannot be directly observed. Moreover, physical theory in principle allows the emergence of stable matter out of these quantum fluctuations without contradicting the law of energy conservation:

> in quantum theory, particles can be created out of energy in the form of particle/antiparticle pairs. But that just raises the question of where the energy came from. The answer is that the total energy of the universe is exactly zero. The matter in the universe is made out of positive energy. However, the matter is all attracting itself by gravity. Two pieces of matter that are close to each other have less energy than the same two pieces a long way apart, because you have to expend energy to separate them against the gravitational force that is pulling them together. Thus, in a sense, the gravitational field has negative energy. In the case of a universe that is approximately uniform in space, one can show that this negative gravitational energy exactly cancels the positive energy represented by the matter. So the total energy of the universe is zero. (Hawking, 1988, p. 129)

What we need to explain further is how such separation of positive and negative energy can occur, i.e. how the initially homogeneous chaos can differentiate into distinct spatial regions, particles and fields. Numerous observations of chemical, physical, biological and sociological processes have shown that some form of order or organization can indeed spontaneously evolve from disorder, breaking the initial homogeneity or symmetry. The only ingredients needed for the evolution of order are random *variation*, which produces a variety of configurations of the different elements, and the *selection* of those configurations that possess some form of intrinsic stability or invariance. The selection is natural or spontaneous in the sense that unstable configurations by definition do not last: they are eliminated by further variation. The stable ones, on the other hand, by definition persist: they are selectively retained. In general, there exist several stable configurations or "attractors" of the dynamics. However, random variation makes that the configurations will eventually end up in a single attractor, excluding the others.

This is the origin of *symmetry breaking*: initially, all attractor states were equally possible or probable (homogeneity or symmetry of possible outcomes); eventually, one has been chosen above all others (breaking of the symmetry). What forces the symmetry breaking is the instability of the disordered configuration: this initially homogeneous situation cannot last, and a "decision" needs to be made about which stable configuration to replace it with. A simple example is a pencil standing vertically on its tip. This position is very unstable, and the slightest random perturbation, such as few air molecules more bumping into it from the left rather than the from right, will push the pencil out of balance so that it starts to fall, in this case towards the right. It will end up lying flat on the right-hand side, thus breaking the initial symmetry where it was poised in an exact balance between left and right. More generally, an initial random fluctuation will normally be amplified by positive feedback until it pulls the whole system into a particular attractor [Heylighen, 2001].

Perhaps counter-intuitively, more variation or disorder produces faster self-organization and therefore more order. This is the principle of "order from noise" [von Foerster, 1960], or "order through fluctuations" [Prigogine & Stengers, 1984; Nicolis & Prigogine, 1977]. The explanation is simple: more variation means that more different configurations are explored in a given lapse of time, and therefore the probability to end up in a stable configuration in that period of time becomes greater. We may conclude that the emergence of differentiated order from initially homogeneous disorder is a simple and natural process that requires no further justification. It implies that we can explain the emergence of order out of chaos without need to postulate a pre-existing order or designer.

However, this variation-and-selection mechanism cannot as yet be used to explain the emergence of time, since it assumes processes taking place in time. To tackle this problem, we need to "abstract away" the notion of time from the two basic components of the process of self-organization, thus arriving at the following generalized notions:

- *generalized variation* does not require change of a configuration in time, but can be a static feature. The only thing needed is the presence of a variety or diversity of configurations. These can be generated by random variations on a simple "template".
- *generalized selection* does not require selective retention, where some configurations are "killed off" or eliminated, while others are allowed to "survive". We only need a selection criterion that allows certain configurations, while making others a priori impossible.

The most general selection criterion that I want to propose here is *consistency*. According to the American Oxford Dictionary, "consistent" has three, related meanings:

- "unchanging in achievement or effect over a period of time": this can be seen as a paraphrase of "stable" or "invariant"
- "compatible or in agreement with something": this can be interpreted as "fitting" or "adapted". For a system to be stable, it needs to fit in or "agree" with its environment, i.e. it should avoid potentially destructive conflict.
- "not containing any logical contradictions": this is the time-independent meaning that we will focus on here; it can be derived from the second meaning by noting that the components of a consistent configuration should be in mutual "agreement".

Consistency in the timeless sense can be seen as a requirement imposed by the law of contradiction in logic: A and *not* A cannot both be true. This law is tautological, which means that it is true by definition. Therefore, it does not need to be justified by recourse to some deeper law or to some external authority, such as God imposing laws on nature. The application of this law in generalized self-organization is that it can be used as a criterion to eliminate configurations that somewhere contain an inconsistency, i.e. some part or aspect of the configuration is in contradiction with some other part of aspect. While this requirement may seem obvious when discussing logical statements, the connection to physical states and self-organization is more subtle.

One illustration I can think of is de Broglie's historical conception of the quantized orbits of electrons around a nucleus. In the spirit of wave mechanics (the precursor of quantum mechanics), these energy eigenstates were seen as closed waves, since the wave has to travel around the nucleus and then connect with it itself. The explanation for quantization was that orbits that are not eigenstates cannot exist because of destructive interference of the electron's wave function with itself. The selection criterion (being an energy eigenstate, or in wave mechanics: being a standing wave with an integer number of nodes) selects specific quantized orbits and eliminates the rest. However, this is not conceived as a process in time, since the non-quantized orbits do not get eliminated one-by-one: they are intrinsically inconsistent, and therefore "logically" unrealizable.

Another example is perception, where the visual system selects a coherent "Gestalt" out of all the possible interpretations of the initial noisy data it receives, and ignores the interpretations that appear inconsistent [Stadler & Kruse, 1990]. Again, the alternative interpretations are not eliminated in a temporal sequence; they simply fail to make sense. Both orbit quantization and Gestalt perception exhibit symmetry breaking: from the homogeneous mass of potential states or interpretations, they select one (or a few), leaving out the rest.

## 3. The origin of time

Time is in the first place an order relation between events, allowing you to specify whether an event A came either *before* or *after* an event B. Relativity theory has generalized this intuitive notion of a complete or linear order of time by noting that sometimes the order of events cannot be determined: when A occurs outside of the light cone passing through B (which means that it is impossible to send a signal from B that arrives in A or vice-versa), then the temporal order between A and B is indeterminate. For some observers, A will appear to be in the future of B, for others in the past, or in the present. In general, we may say that A and B cannot be ordered absolutely. Therefore, according to relativity theory the order of time is only partial.

A *partial order* is actually a very simple and common mathematical structure. In fact, any arbitrary relation can be formally converted to a partial order by making the relation transitive [Heylighen, 1990a]. To show how this is done, let us represent this arbitrary relation by the symbol → , which can be taken to mean "connects to", according to some as yet unspecified connection criterion. Adding such a relation to a set of individual nodes {A, B, C…} turns this set into a network. The nodes can be interpreted as some as yet unspecified, primitive "events". We will now perform a *transitive closure* of this relationship or network. This means that if the links A → B, and B → C both exist, then the link A → C is added to the network if it did not exist yet. If it turns out that C → D also exists, then transitive closure means that in a second stage A → D is added as well. This adding of "shortcuts" or "bridges" that directly connect nodes that were indirectly connected is continued until the network has become transitive, i. e. until for every X → Y and Y → Z, there exist a X → Z link. This is a purely formal operation of generalizing the definition of the relation so that it includes indirect links as well as direct ones. It is similar to the extension from the relation between people “is parent of” to the transitively closed relation “is ancestor of”, or from the relation between natural numbers “is successor of” to “is larger than”.

In any relation or network, there are two types of links: *symmetric* (meaning that the link A → B is accompanied by its inverse B → A), and *antisymmetric* (meaning that the link has no inverse). The combination of transitivity and antisymmetry defines a *partial order* relationship: if you consider only the links without inverse, they impose a clear order on the nodes they link, from "smaller" to "larger", or from "earlier" to "later". The combination of transitivity and symmetry, on the other hand, determines an *equivalence relationship*: if A → B, and B → A, then A and B can be considered "equivalent" with respect to the ordering. If the ordering is interpreted as time, A and B are simultaneous. So, it appears as if this simple transitive closure operation has transformed our arbitrary, random network into a partial order that can be interpreted as an order of time. In other words, we get order (time) out of chaos (a random network).

This is pretty straightforward. However, complications arise if the original relation → (before the transitive closure operation) contains *cycles*. Imagine a long sequence of links: A → B, B → C, C → D, … Y → Z. Transitive closure means that you add all the shortcuts: A → C, B → D, C → E, etc. But now you also need to add shortcuts between the shortcuts: if both A → C, and C→ D are in the network, A → D also must be added, and so does A → E, A → F, etc. Eventually, the whole sequence will be "cut short" by the single link A → Z. This fits in with our intuition about time: if A precedes B, B precedes C, … and Y precedes Z, then A also precedes Z. But since we started from the assumption that the network is random, the probability is real that it would also contain the link Z → A. In that case, we have found a *cycle*: the sequence of links starting from A returns to its origin. Applying again the transitivity rule, A → Z and Z → A together imply A → A. In other words, A precedes A! This is not grave if we interpret the connection relation → as "precedes or is simultaneous with". The links A → Z, and Z → A are symmetric, and thus they belong to the equivalence part of the relationship. The normal interpretation is therefore one of simultaneity. However, the transitive closure operation implies that *all* elements of the sequence A, B, C, D, …, Z now become equivalent or simultaneous. This is still not necessarily a problem, since it is principle possible to have many simultaneous events.

The existence of cycles becomes a problem, though, if we make the assumptions that the initial network is both random—because we want order to emerge from chaos—and infinite, or at least unrestricted—because we want the emerging order to represent the infinite extension of time. If we continue to add random nodes and links to the network, sooner or later a very long sequence of ordered nodes will, by the addition of a single link going back to an earlier element of the sequence, turn into a cycle. This cycle, because of the formal operation of transitive closure that is needed to produce an order relation, will turn into an equivalence class. This means that the elements of the sequence, however extended, suddenly all lose their temporal order, and become simultaneous. Simulations of the growth of random networks [Kaufmann, 1995] clearly show that the addition of links will sooner or later connect all nodes into a single cluster or equivalence class. In other words, if we allow the network to grow freely, we will quickly lose our partial ordering and therefore any notion of time.

The only solution seems to be to get rid of the cycles somehow, i.e. to formulate a selection criterion functioning outside of time that excludes cycles and retains only the non-cyclical parts of the random network to constitute the backbone of time. The criterion of consistency is obviously relevant here because cycles in time can lead to the well-known paradoxes of the time machine: what happens if I go back in time before I was born and kill my own father? I have argued earlier [Heylighen, 1990b] that temporal cycles connecting events are either logically inconsistent (A leads to *not* A) or trivial (A leads to A). The trivial cycles merely reaffirm what is already there. The inconsistent ones, on the other, imply that A negates itself and therefore must be null or void. Therefore, selection for consistency would automatically eliminate all such cycles. The effect is similar to the destructive interference undergone by cyclical waves that do not have an integer number of periods: when the wave comes back to its origin with an amplitude opposite to the one it started out with, it effectively erases itself.

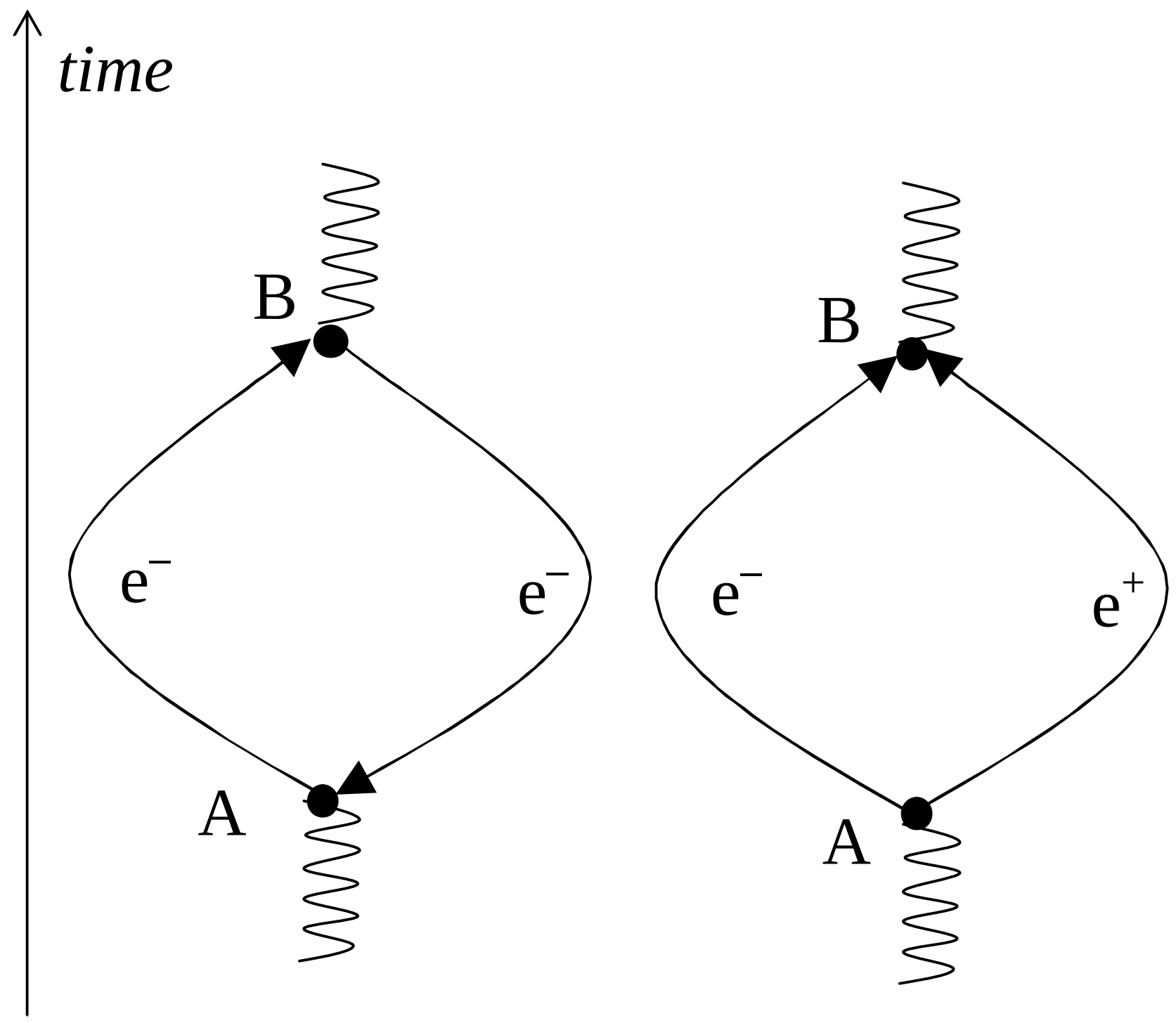

The trivial cycles, which do not "self-destroy", on the other hand, are redundant, and can therefore be safely ignored or reinterpreted as partial orders. One way to do this, as suggested in Heylighen [1990a], is to apply Feynman's [1949] interpretation of antiparticles as normal particles moving backwards in time; in other words, we can in principle reinterpret the "back in time" section of consistent cycles as antiparticles moving forward in time, thus changing the orientation of the connections on that section. For example, Fig. 1 shows a simple cycle A → B, B → A in which an electron ($e^-$) moves back in time (left). This is reinterpreted (right) as a photon giving rise to a particle-antiparticle pair ($e^+$: positron = anti-electron) after which particle and antiparticle mutually annihilate, producing again a photon. Note that in analogy with the quantization of closed waves, such a consistent cycle could be seen as a kind of eigenstate of a hypothetical time reversal operator.

This elimination of cycles leaves us with the non-cyclic parts of the initially random network of connections between events, and therefore with a partial order defining time. Moreover, it can be shown that the remaining connections can be divided in two categories, which can be interpreted respectively as "light-like" (i.e. representing processes with the speed of light), and "particle-like" (i.e. representing processes with a speed lower than light) (Heylighen, 1990a,b). The resulting mathematical structure is equivalent to the *causal structure* of relativistic space-time [Kronheimer & Penrose, 1967], which determines the bulk of space-time geometry. This construction thus not only produces the order of time, but even the fundamental properties of space in its relativistic interpretation (Heylighen, 1990a,b). The argument needs to be fleshed out in much more detail, but already suggests a simple and promising route to a theory of the self-organization of time.

The only additional ingredient we need to recover the full mathematical structure of relativistic space-time is an observer-independent notion of duration, i.e. a unit of time that allows us to measure how much time has passed (Heylighen, 1990b). Given such a unit of time, we immediately get a unit of space or distance for free, since we can define this spatial unit as the distance covered in a unit of time by a signal moving with the (invariant) speed of light.

The existence of invariant time units is equivalent to the assumption that it is possible under certain circumstances for synchronized clocks that are separated and then brought together again to still be synchronized [Sjödin & Heylighen, 1985], because all along they have counted with the same time units. In other words, equal causes (clocks initially showing the same time) produce equal effects (clocks having advanced independently still show the same time). This is actually a problem of causality, which will be discussed in the next section.

## 4. The origin of causal laws

In relativity theory, causality is usually understood to mean that a cause must necessarily precede its effect. However, this relation of precedence is already fully covered by our notion of time as a partial order between events, and therefore needs no additional explanation.

What remains to be explained is causality in the more traditional sense of "equal causes produce equal effects". This is the sense of causality as a rule or law that allows us to predict which kind of effect will follow given the characteristics of the cause. In my analysis of causality [Heylighen, 1989], I have argued that if we interpret "equal" as "identical" then the principle of causality is tautological, and therefore needs no further explanation. This interpretation corresponds to what I have called "microscopic causality". In practice, however, i.e. in the world of macroscopic observations, when we make predictions we do not assume *identical* causes, but *similar* causes leading to *similar* effects. This interpretation is the principle of "macroscopic causality". The sensitive dependence on initial conditions in non-linear dynamics (the "butterfly effect") and the Heisenberg uncertainty principle, however, both show how similar (macroscopically indistinguishable) causes can lead to dissimilar (macroscopically distinct) effects [Prigogine & Stengers, 1984; Gershenson & Heylighen, 2004]. Therefore, macroscopic causality is not a logical necessity: sometimes the assumption is valid, sometimes it is not. The question that remains then is: why do similar causes *often* lead to similar effects?

A possible approach is to consider a cause-effect relation as a *condition-action rule*, $A \rightarrow B$, describing the transition from A (cause) to B (effect): whenever a condition A, i.e. a state belonging to particular subset or category A of world states, is encountered, some agent acts to change this state into a new state, belonging to category B. This perspective fits in with an *ontology of actions* [Turchin, 1993], which sees all change as resulting from a combination of elementary actions performed by one or more agents. An agent in this perspective could be a particle, a field, a molecule, or some more complex system, such as an organism. This implies that causal rules are not absolute or universal, but dependent on the presence of a particular type of causal agent. The presence of this agent functions as a "background condition" necessary for the causation to take place [Heylighen, 1999]. For example, the rule "if a massive object is dropped (cause or condition), it will fall (effect or action)" implicitly requires the presence of gravitation, and therefore the proximity of a mass, such as a planet, big enough to produce gravitational forces. The planet's gravitation here plays the role of the causal agent. In its absence, e.g. in interstellar space, the causal law does not hold.

Such agents—and therefore the laws they embody—are normally the product of evolution. This idea may be illustrated by considering the origin of biological laws.

Living organisms all use the same genetic code, which is implemented by the mechanism of RNA transcription: a particular DNA/RNA triplet is transformed via a number of intermediate stages into a particular amino acid by the ribosomes and transfer-RNA molecules present in the cell. The causal rules governing this “translation” mechanism together form the *genetic code*. This genetic code is universal, i.e. the same triplet is always transformed into the same amino acid: equal causes produce equal effects. This universality can be explained by the fact that living organisms on Earth have a common ancestor. From this ancestor, all living cells have inherited the specific organization of the ribosomes that perform the conversion from triplet to amino acid. These complexes of RNA and protein were created very long ago by an evolutionary process of self-organization that took place among the autocatalytic cycles of chemical reactions that produced the first living cells. Natural selection has eliminated all variant forms of ribosomes that might have enacted different codes of translation, and thus fixed the present code. Thus, we can explain the law-like character of the DNA code by the selective retention and reproduction of a particular type of ribosomal agents.

Can we generalize such a process of self-organization to explain causal laws in general? The fundamental problem is to explain why natural laws appear to be the same in all regions of the universe. The genetic code example suggests that this may be because all the causal "agents" (which at the lowest level might correspond to elementary particles and fields) had a common origin during the Big Bang, i.e. they are all descendants of the same "ancestors". However, those original ancestors are likely to have come about contingently, and therefore different universes may well have different laws of nature—e.g. distinguished by the values of their fundamental constants. Why our universe has these particular laws may then be explained by a natural selection of universes picking out the "fittest" or most "viable" universes [Smolin, 1997].

However, the ribosome example suggests that there may have been many alternative laws, enacted by different collections of particle-like agents, that would have been just as effective in generating a complex universe that later gave rise to intelligent life. Biologists have no particular reasons to assume that the present genetic code is the only possible one. While there are arguments based on chemistry to show that the present code is more efficient than most other conceivable codes [Freeland & Hurst, 1998], there is still plenty of freedom in choosing between a large number of codes that appear equally efficient. Biologists assume that these other codes have lost the competition with the present code not because they were intrinsically less fit, but because of contingent events, such as one code being a little more common in the very beginning, which allowed it to profit more from exponential growth to outcompete its rival codes. Here we find again the basic mechanism of symmetry breaking: random, microscopic differences in the initial state (a few more cells with the present code) are amplified by positive feedback until they grow into irreversible, macroscopic differences in the final result. The implication is that there may be a large number of “viable” universes, which all have different laws, but that not all laws are equally viable.

For example, in cosmology the question is regularly raised why in our universe there is such a preponderance of matter over antimatter. The laws of physics as we know them do not exhibit any preference for the one type of matter over the other one. Therefore, we may assume that during the Big Bang particles of matter and of antimatter were produced in practically equal amounts. On the other hand, matter and antimatter particles annihilate each other whenever they interact. This means that such a homogeneous distribution of particles between matter and antimatter states was unstable, and could not continue. It has been suggested that an initial imbalance between matter and antimatter, which may have been random and tiny, has been magnified by this violent competition between the two states, resulting in the final symmetry breaking, where practically all antimatter was eliminated.

Since antimatter particles are still being formed in certain reactions, we know about the possibility of their existence. However, it is conceivable that the Big Bang witnessed the creation of huge varieties of other, more "exotic" particles, which not only have disappeared since, but which are so alien to the remaining particles that we cannot even recreate them in our particle colliders. Therefore, they are absent in our theoretical models, even as potential outcomes of reactions. Such particles might have embodied very different causal laws, exhibiting different parameters such as mass and charge, and undergoing different types of forces and interactions

Another implication of this hypothesis is that causal laws may not be as absolute and eternal as physics assumes. If a causal law is “embodied” in a particular type of agent that has survived natural selection, we may assume it to be relatively stable. Otherwise, the agent, and with it the law, would already have disappeared. On the other hand, evolution tells us that no agent is absolutely stable: it is always possible that the environment changes to such a degree that the original agent no longer “fits”. This will lead to increased variation and eventually the appearance of new agents that are better adapted to the new environment, thus outcompeting the old ones. When we think about basic physical laws, like those governing the interactions between common elementary particles, such as protons and electrons, it seems difficult to imagine environments where those particles and the laws they embody would no longer be stable. But that may simply be a shortcoming of our imagination, which has no experience whatsoever with totally different physical situations, such as those that might arise inside a black hole or during the Big Bang.

When discussing the contingency of laws it is important to note that there are two types of laws:

1) *logically necessary laws*: these are true tautologically, by definition, such as $1 + 1 = 2$ or the law of contradiction in logic
2) *contingent laws*: these could conceivably be different, such as the values of the different fundamental constants in physics.

The difference between these two is not always apparent. Some seemingly contingent laws may in a later stage be reduced to tautologies, which have to be true because of the way the properties that they relate are defined. The law of energy conservation is an example of this: at the most fundamental level, energy appears to be defined in such way that it must be conserved. More precisely, the law of energy conservation, like all other conservation laws, can be derived mathematically (through Noether’s theorem) from an assumption of symmetry [Hanca et al. 2004], in this case the homogeneity of time. This means simply that physical processes are independent of the particular moment in time in which they occur: postponing the process to a later moment without changing anything else about the situation will not change the dynamics that takes place. This assumption of time invariance appears to be true by definition: the time coordinate of an event is merely a convention, depending on how we have calibrated our clocks, and should therefore not affect the process itself. In fact, this could be interpreted as another example of the consistency requirement: to be consistent our description of a dynamical process should not change if we merely shift the time coordinate over an arbitrary amount, since time is defined relatively as a precedence relation, and not absolutely, as a number.

At present, most physicists seem to assume that the values of the fundamental constants are contingent, and therefore need to be explained by a combination of random variation and a selection mechanism such as the Anthropic principle [Carr & Rees, 1979; Barrow & Tipler, 1988] or cosmological natural selection [Smolin, 1997]. However, we must remain open to the possibility that they are necessary, and derivable from some as yet not clearly formulated first principles [see e.g Bastin et al. 1979, Bastin & Kilmister, 1995 for an attempt at deriving fundamental constants from combinatorial principles]. The example of the origin of the genetic code may remind us that some aspects of a law may be purely the result of chance, while others represent intrinsic constraints that determine which variants will be selected. That selection itself may happen in time, e.g. during a sequence of universes reproducing themselves as envisaged by Smolin (1997), or outside time, by a requirement of consistency like the one we discussed before or like the one that is implicit in symmetry-based derivations of laws based on Noether's theorem.

## 5. Conclusion

The problems of the origin of time and of causality are perhaps the most fundamental of all scientific problems, since all other scientific concepts and theories presuppose and therefore depend on the existence of time and causality. It therefore should not surprise us that as yet no convincing approaches to these problems have been proposed. However, rather than taking time and causality for granted, as practically all theories have done until now, the present paper has argued for a further investigation of these problems.

I have suggested to start from the by now well-documented notion of self-organization, because this concept proposes a concrete mechanism for the emergence of order out of chaos. When considering the origin of the universe, chaos should here be understood in its original, Greek sense, as a total disorder that is so much lacking in structure that it is equivalent to nothingness. Time and causality, on the other hand, are characterized by order. For time, this means the partial order relation of precedence that connects different events while establishing an invariant distinction between past and future. For causality, the order is in the invariance of cause-effect relationships, as expressed by the "equal causes have equal effects" maxim. Invariance can be conceived as stability under certain transformations. Stability can be explained as the result of a process of variation followed by selection that spontaneously eliminates unstable variations. Since chaos automatically implies variation, we only need to explain selection: why are only some of the variations retained?

In the case of causality, the variations can be conceived as causal agents that embody different condition-action or cause-effect rules. In the case of basic laws of physics, the agents are likely to represent elementary particles or fields. Since the agents interact, in the sense that the effect of the one's action forms an initial condition or cause for another one's subsequent action, they together form a complex dynamical system. These systems are known to necessarily self-organize [Ashby, 1962; Heylighen, 2001], in the sense that the overall dynamics settles into an attractor. This means that certain patterns of actions and agents are amplified by positive feedback until they come to dominate, suppressing and eventually eliminating the others, and thus breaking the initial homogeneity or symmetry in which all variations are equally probable. As yet, we know too little about the dynamics of such a primordial complex dynamical system to say anything more about what kind of causal rules might emerge from such a self-organization at the cosmic scale. However, the general notion of self-organization based on variation and selection suggests some general features of the resulting order, such as the fact that it will be partly contingent, partly predictable, and context-dependent rather than absolute.

In the case of time, this notion of self-organization needs to be extended in order to allow variation and selection to take place outside of time. For variation, this poses no particular problem, since selection can operate equally well on a static variety of possibilities. For selection, we need to replace the dynamic notion of stability as a selection criterion by the static notion of consistency. Consistency can be understood most simply as an application of Aristotle's law of contradiction—which states that a proposition and its negation cannot both the actual. In the case of time, consistency allows us to have a partial order of precedence emerge out of a random graph by eliminating cycles. The connections forming the random graph or network can be interpreted as elementary actions or processes that lead from one event to another. These random links and their corresponding nodes (events) form the initial chaos or variation out of which the order of time is to emerge.

The formal operation of transitive closure transforms a random network into a relation that is partly a partial order, partly an equivalence relation. The equivalence relation encompasses all the parts of the graph that are included in cycles. However, in an infinite random graph, this means in essence the whole graph, implying that no partially ordered parts are left. Therefore, we need a selection criterion that eliminates cycles. This can be motivated by generalizing the paradox of the time machine: temporal cycles that produce actual changes are a priori inconsistent, and therefore "self-negating", like the cyclic waves that undergo destructive interference with themselves. Therefore, we can exclude them a priori.

In both cases—the self-organization of time and of causality—the present description is still very sketchy, applying general principles at a high level of abstraction, but remaining awfully vague as to what the "agents", "connections" or "events" precisely are, or what properties they are supposed to have. At this stage of the investigation, such vagueness is probably unavoidable. However, by proposing a relatively simple and coherent explanation based on the well-understood concept of self-organization, the present approach at least provides some steps towards understanding these fundamental questions. I hope that other researchers may pick up these threads and weave them into a graceful fabric of understanding.

# Symmetries and symmetry-breakings: the fabric of physical interactions and the flow of time

*Reflections on Francis Heylighen's paper: The Self-organization of Time and Causality, steps towards understanding the ultimate origin*

*Giuseppe Longo*

Dépt. d'Informatique, CNRS – ENS et CREA, Paris
http://www.di.ens.fr/users/longo

**Summary.** This short note develops some ideas along the lines of the stimulating paper by Heylighen. It summarizes a theme in several writings with Francis Bailly, downloadable from this author's web page. The "geometrization" of time and causality is the common ground of the analysis hinted here and in Heylighen's paper. Heylighen adds a logical notion, consistency, in order to understand a possible origin of the selective process that may have originated this organization of natural phenomena. We will join our perspectives by hinting to some gnoseological complexes, common to mathematics and physics, which may shed light on the issues raised by Heylighen. *Note:* Francis Bailly passed away recently: his immense experience in physics has been leading our joint work for many years.

Historically, it is with relativist physics that there occurs a "change of perspective": we pass from "causal laws" to the structural organization of space and time, or even from causal laws to the "legality/normativity of geometric structures". This understanding of causal laws by the identification of structural organizations, stems essentially from the intrinsic duality existing between the characterization of the geometry of the universe and that of energy-momentum within that universe. By this duality and the putting into effect of the principle of invariance under the differentiable transformations of space-time, the "forces" are relativized to the nature of this geometry: they will even appear or disappear according to the geometric nature of the universe chosen a priori to describe physical behaviors.

Now, it is similar for quantum physics, in gauge theories. Here, gauge groups operate upon internal variables, such as in the case of relativity, where the choice of local gauges and their changes enable to define, or conversely, to make disappear, the interactions characterizing the reciprocal effects of fields upon one another. For example, it is the choice of the Lorentz gauge which enables to produce the potential for electromagnetic interactions as correlates to gauge invariances.

Consequently, if one considers that one of the modalities of expression and observation of the causal processes is to be found in the precise characterization of the forces and fields "causing" the phenomena observed, then it is apparent that this modality is profoundly thrown into question by the effects of these transformations. Not that the causal structure itself will as a result be intrinsically subverted, but the description of its effects is profoundly relativized.

This type of observation therefore leads to having a more elaborate representation of causality than that resulting from the first intuition stemming from classical behaviors. Particularly, the causality of contemporary physics seems much more associated to the manifestation of a formal solidarity of the phenomena between themselves, as well as between the phenomena and the referential frameworks chosen to describe them, than to an object's "action" oriented towards another in inert space-time, as classical mechanics could have accredited the idea. In summary, our strong stand towards a geometrization of causality may be summarized as follows. *Causes become interactions and these interactions themselves constitute the fabric of the universe of their manifestations, its geometry: modifying this fabric appears to cause the interactions to change; changing the interactions modifies the fabric*.

And now comes another fundamental issue raised by Heylighen. It appears that the symmetry / symmetry breaking pair is playing for the intelligibility of physics an absolutely crucial role. By Noether's theorems, to which Heylighen refers, transformations in symmetry correspond invariants (mathematical aspect) or conserved quantities (physical aspect) specific to the system under consideration and to any systems displaying identical symmetries.

Thus, the symmetry / breaking of symmetry pair thematizes, on the one hand, invariance, conservation, regularity and equivalence, and on the other, criticality, instability, singularity, ordering. We have seen in the book quoted below that through the pair's dialectic, it is an essential component of the very identity of the scientific object that is presented and objectivized. Could we go even further and consider that we have thus managed construct this identity at a level such that cognitive schemas conceived as conditions of possibility for any construction of objectivity are henceforth mobilized, thus reviving a form of transcendental approach, in a kantian sense ?

As a matter of fact, there exists a close formal relationship between the abstract properties of symmetry captured by mathematical group structures and logical structures as fundamental as the equivalence relation, which is extensively used by Heylighen. At the same time, there exists a similar formal relationship between the semi-group structure and the logical structure of the (partial) order relation, to which Heylighen refers in his understanding of time and causality. Thus, the theoretical analysis of the abstract notions of space and of time demonstrates that for their formal reconstruction these notions need to mobilize the mathematical structures of group and of semi-group, respectively. Indeed, regardless of the number of dimensions considered, the displacement properties, consubstantial to the concept of space, refer to the determinations of the displacement group, whereas the properties of irreversibility and of the passing of time refer to the characteristics of the semi-group (generally, for one parameter).

We then witness the constitution of a pair of abstract complexes which doubtlessly represents one of the essential bases for any objective interpretation within the processes of the construction of knowledge: the complexes of <space, group structure, equivalence relation> on the one hand and of <time, semi-group structure, order relation on the other>. Epistemologically, this is where Heylighen's thesis leads, in our opinion. By adding "consistency" in the logical sense, Heylighen suggests a way to break circularities in the cyclic structure of equivalence relations and pass by this to order relations, that is to semi-group and time ("going back" to a node A from which one "moved away" is a form of opposite movement, a negation $\neg A$, thus incompatible or logically inconsistent – one cannot have both A and $\neg A$).

Let's point out once more that the space and time evoked by the gnoseological complexes above no longer refer to physical entities as such, but rather to the conceptual frameworks which are meant to enable any physics to manifest itself, that is, to abstract conditions of possibility and not to effective realizations, thus reactualizing a kantian point of view. Thus, space and time are no longer considered as "objects" to be studied, but rather as the conditions of possibility for any sensible experience. In this sense, the symmetries and breakings of symmetries associated to these complexes appear not only as elements of the intelligibility of physical reality, but indeed as factors for the scientific constitution of such reality, including the understanding of causality.

Not only would we simply operationalize space and time (and thus causality), but by coupling them with the corresponding logical and mathematical determinations (group structure, equivalence relation, etc.) we refer them to the frameworks of invariance which make them into reference structures that are mathematically specified, rather than abstract and vague.

# Symmetry, Potentiality and Reversibility

Francis Heylighen

**Abstract**: This short comment confirms Longo's observation about the importance of symmetries for understanding space and time, but raises the additional issue of the transition from reversible to irreversible transformations.

I have little to criticize in Giuseppe Longo's (2010) comment on my paper "The Self-organization of Time and Causality" (Heylighen, 2010). I can only agree with Longo's emphasis on the duality symmetry/symmetry breaking as a foundation for our understanding of space and time.

Space, indeed, is characterized in the first place by symmetry, i.e. by the fact that translations from one point of space to another point can be perfectly reversed. The movement or translation from a position X to a position Y is the exact mirror image of the translation from Y to X. This expresses our intuition that space is something that you can freely travel in without making any irreversible changes, since you can always come back to the same place, i.e. point in space. Time, on the other hand, is characterized by the opposite: antisymmetry. A causal process leading from an event A to an event B by definition cannot be reversed: once B occurs, A is irrevocably in the past, and can no longer be reached from B. This expresses our intuition that it is impossible to travel in time in any other way than forward. In my paper [Heylighen, 2010], I have tried to explain that antisymmetry by using the consistency principle to exclude travel back in time from B to A.

This paper only discussed actual events and the actual processes connecting them. However, in order to recover the (both mathematical and intuitive) structure of space, we need to generalize from actual movements to potential ones. This may be clarified by expressing an event A in space-time as a combination of its space and time coordinates: $A = (x_A, t_A)$. In my travel from event A to event B, I change my position in space from $x_A$ to $x_B$, and my moment in time from $t_A$ to $t_B$. Obviously I cannot go back to $t_A$ from here, since by assumption $t_A < t_B$. However, I still can travel back from $x_B$ to $x_A$, although I will arrive there at a later time $t_C > t_B$. These coordinates are not part of the construction in my paper. However, they can be conventionally chosen so as to be in accord with the relativistic structure of space-time that does emerge from the construction.

The reason for choosing space coordinates separate from time coordinates is precisely in order to model space as a set of *potential* places: positions that we do not all visit, but that we might have visited given different initial conditions. Time does not have this property of potentiality or freedom: you have no choice but to follow the one-dimensional arrow of time, and there are no points in time that you can decide to skip or to visit earlier or later than determined. The concept of potential is foundational for all scientific models, and is captured in the basic mechanical concepts of cinematics (the study of potential movements), configuration space, phase space and state space (the spaces of all possible configurations, phases or states of the system). While we are inclined to see ordinary, physical space as something more concrete and “real” than those abstract spaces, because of our visual and motor intuitions about it, mathematically there is no real difference: ordinary Euclidean space is merely the state space of a point mass that can move in three dimensions: up-down, left-right and back-forth.

Once we have formalized the construct of the space of all potential positions or states, we can start to model processes in a more general way—not as a relation between specific points or events but as a relation between classes of points or events. For this we conventionally use transformations or mappings, which map the space onto itself. As Longo points out, the reversible mappings (such as translations or rotations), which represent the symmetries of space, form an algebraic group, which expresses the “cinematic” aspect of potential movement. We know however that the actual causal or dynamical processes are irreversible, and therefore the corresponding mappings should logically form a semi-group (lacking inverse transformations). The problem, however, is that the causal laws of classical mechanics (as well as quantum mechanics and most of their derivatives) are intrinsically reversible in their formulations. The laws of thermodynamics, on the other hand, are intrinsically irreversible as they imply a maximization of entropy. Prigogine (1980) has attempted to solve this paradox by constructing a mathematical transformation of probability distributions that would send reversible mappings (group) onto irreversible ones (semi-group). While well intended, this formal approach seems too artificial to tackle the fundamental conceptual problem of the irreversibility of time.

The issue is complicated even further by the phenomenon of symmetry breaking related by Longo (2010), which, as I noted in my paper, is characteristic of self-organization, which is itself a primary irreversible process. However, not all processes in time are characterized by symmetry breaking, yet all are irreversible at the level of events. In conclusion, the transition from reversible dynamics that maintain symmetry to irreversible dynamics that either maintain or break symmetries is a complex but important unsolved issue!

# The role of energy conservation and vacuum energy in the evolution of the universe

**Jan M. Greben**
CSIR, PO Box 395, Pretoria 0001, South Africa
E-mail: jgreben@csir.co.za

**Abstract.** We discuss a new theory of the universe in which the vacuum energy is of classical origin and dominates the energy content of the universe. As usual, the Einstein equations determine the metric of the universe. However, the scale factor is controlled by total energy conservation in contrast to the practice in the Robertson-Walker formulation. This theory naturally leads to an explanation for the Big Bang and is not plagued by the horizon and cosmological constant problem. It naturally accommodates the notion of dark energy and proposes a possible explanation for dark matter. It leads to a dual description of the universe, which is reminiscent of the dual theory proposed by Milne in 1937. On the one hand one can describe the universe in terms of the original Einstein coordinates in which the universe is expanding, on the other hand one can describe it in terms of co-moving coordinates which feature in measurements. In the latter representation the universe looks stationary and the age of the universe appears constant.

The paper describes the evolution of this universe. It starts out in a classical state with perfect symmetry and zero entropy. Due to the vacuum metric the effective energy density is infinite at the beginning, but diminishes rapidly. Once it reaches the Planck energy density of elementary particles, the formation of particles can commence. Because of the quantum nature of creation and annihilation processes spatial and temporal inhomogeneities appear in the matter distributions, resulting in residual proton (neutron) and electron densities. Hence, quantum uncertainty plays an essential role in the creation of a diversified complex universe with increasing entropy. It thus seems that quantum fluctuations play a role in cosmology similar to that of random mutations in biology. Other analogies to biological principles, such as recapitulation, are also discussed.

## 1. Introduction

In standard quantum mechanics conservation of energy is related to the invariance of the Lagrangian under space-time translations and is expressed as a divergence equation for the energy-momentum tensor. In General Relativity (GR) this divergence equation is replaced by a covariant equation and is equivalent to the Bianchi identities satisfied by GR [1]. However, in an expanding universe, this symmetry is no longer equivalent to energy conservation. For example, the popular de Sitter universe violates energy conservation ([2], p.120).

In view of the importance of the principle of total energy conservation we propose to impose this principle as a separate condition in GR. For a non-expanding universe this condition reduces to the usual divergence equation. However, the general consensus is that the universe is expanding, in which case this principle becomes a separate condition. It can be imposed by demanding that the spatial integral over the energy component of the energy-momentum tensor is constant over time. However, in the standard Robertson Walker (RW) metric this procedure leads to a problem. Given an energy-momentum tensor, the metric of the universe is fixed by the GR equations. In the usual RW metric, the expansion is incorporated in the metric via the scale factor, so that the expansion is fixed by the GR equations. Hence, there is no room for another condition for total energy conservation as the whole dynamics is already fixed (apart from boundary conditions). It is thus not surprising that the solution of the GR equations for a universe with a constant cosmological constant violates energy conservation ([2], p.120). Our solution to this conundrum is to remove the scale factor from the metric. This means that the expansion now has to be derived in a different way, and this is done via a scale factor $a(t)$ fixed by energy conservation. If the universe does not expand, this scale factor reduces to unity, and the extra condition merely represents a consistency condition for two equivalent definitions of energy conservation. This formulation ensures that the scale factor is a truly global function as it is fixed by the total energy, which is a property of the whole universe. It also means that the metric tensor exclusively serves its natural function of reflecting the (local) distribution of energy.

For a flat universe with constant vacuum energy density, this new formulation leads to a linear expansion. This is clearly the simplest possible mode of expansion of the universe and provides a natural representation of the observed Hubble expansion. It should be noted that such a simple solution is impossible in the RW metric, as the linear case represents a singular limit in that framework (only by setting the curvature $k = -1$ can one find a linear solution, the so-called Milne universe [2]). The vacuum energy, which dominates the energy content of the universe in our picture, is easily identified with the so-called dark energy, both having a pressure-to-density ratio of -1. Hence, this model automatically incorporates the present consensus that dark energy dominates the energy content of the universe. In addition to explaining dark energy, the constant vacuum energy density has many other important consequences and plays a central role in the dynamics of the universe, as we will demonstrate amply in this paper.

The original theoretical motivation for constructing a universe with a constant vacuum energy is that our analyses in quantum field theory (QFT) (see Section 2) suggest that the vacuum energy has no quantum contributions, in contrast to generally held beliefs. It is then natural to identify dark energy with a classical vacuum energy, however this forces a new approach to cosmology, as the conventional solution in this case - the de Sitter universe - does not feature a big bang singularity.

Current theoretical scenarios often contain an inflationary period at $t = 10^{-35}$ s. In our theory an inflationary period is not mandatory, as the vacuum metric leads to an infinite horizon, so that one of the main motivations for inflation falls away. This also obviates the need for unknown forces to explain the inflationary epoch, in particular those forces which derive from QFT vacuum energy and which would be excluded by our QFT findings. Most scenarios agree that this early period is followed by a period of linear expansion, moderated by a slight deceleration initially and a slight acceleration in the current epoch. Hence,the dominant form of the expansion (linear) is already accounted for in our approach. Phases of deceleration and acceleration can occur in our model because of known quantum field theoretic processes, such as the creation and annihilation of particles. The presence of matter and radiation does not change the essential linear evolution of the universe in our description, in contrast to universes described by Friedmann-Robertson-Walker(FRW) models, which are vastly different in radiation - and matter - dominated situations. Hence, the simple vacuum metric still dominates the universe in the presence of matter and radiation. Another important difference with the Friedmann models is that in the solution of the Einstein equations, the localized nature of matter distributions is taken into account. Since our scale factor is controlled by energy conservation rather than by the Einstein equations, such a refinement of the cosmological treatment is now feasible. This leads to a unified treatment of cosmological and local astronomical phenomena. Another consequence of this improvement is the emergence of a new tentative explanation of dark matter, which does not require any exotic new particle assumptions.

The evolution of the universe in this theory has rather definite characteristics, with some of the details still to be developed. Contrary to most popular scenarios, the universe starts out as a classical system, with an effective energy density proportional to $t^{-3}$. The universe with positive and negative time are exact replicas of each other, answering the question what happens "prior" to time zero. The cosmological principle (i.e. the homogeneity and isotropy of the density) is satisfied exactly in this initial classical period, so that the entropy is zero. After about $5 \times 10^{-24}$ seconds quantum field theory becomes effective and the first physical (rather than virtual) particles are created. The particle physics scale $5 \times 10^{-24}$ naturally emerges from our formulation and is expressed in terms of the fundamental cosmological constants. This epoch is characterized by deflation, as the universe has to contract to supply the energy for the production of matter. The emergence of physical particles in this epoch also allows the expression of the subsequent evolution in thermodynamic language. The particle-creation epoch is characterized by the Planck energy and by a corresponding temperature of $10^{32}$ $^0$K. In the subsequent epoch particles and anti-particles annihilate, leaving a residue of protons, neutrons and electrons. This period is accompanied by an inflationary expansion to counter the loss of energy in the annihilation phase due to the changing metric. After these processes, the normal Big Bang dynamics sets in which is again characterized by a linear scale factor.

The uncertain outcome of quantum events plays an essential role in the creative epoch. Firstly, it is responsible for breaking the symmetry of the matter distribution, leading to sufficient inhomogeneities for localized matter concentrations to form. In later epochs this asymmetry will initiate the creation of massive astronomical objects. In a classical universe this spatial symmetry would be maintained and no concentrations of matter could possibly emerge. Secondly, we expect the quantum fluctuations to be responsible for the imbalance between the different particle and anti-particle populations after the annihilation epoch. Hence, the current universe is a consequence of the physical laws and historical accidents, caused by the outcome of quantum processes in our world. Thus, randomness is as much a factor in the evolution of the universe as it is in biological (mutation) processes. To some extent the principle of the survival of the fittest is carried in quantum physics by the probability functions. Objects or configurations that form in the early universe, but then decay and vanish from the universe, appear to play a role similar to that of unsuccessful species in biology.

## 2. A universe with constant vacuum energy density

We will assume that the vacuum energy density $\epsilon$ is constant and is a basic property of Nature. This assumption is equivalent to the presence of a non-zero cosmological constant and leads to the usual de Sitter solution for the common Robertson-Walker (RW) metric. The assumption that the vacuum energy density is constant and small appears in conflict with standard QFT estimates, which quote vacuum energy densities of between 40 and 120 orders of magnitudes larger than the "observed" value. This problem of standard QFT is known as the cosmological constant problem ([3], [4], [5]).

Our hypothesis therefore implies that the vacuum energy does not derive from such QFT processes. To put it more bluntly: it suggests that the usual QFT derivations of vacuum processes contain serious flaws. Although this may be a natural conclusion to draw because of the phenomenal discrepancy between the standard QFT result and experiment, various practices with vacuum expectation values (vev's) in QFT have been so ingrained that the acceptance of this conclusion will require much debate. It therefore appears opportune to present some consequences of this hypothesis (such as in cosmology), before engaging in a full debate on its theoretical motivation. Our hypothesis is based on a study of the role of creation and annihilation operators of particles and anti-particles in QFT. We found that many vacuum phenomena, such as the definition of the propagator as a time ordered product, survive under our reformulated operator algebra. Also, the Casimir effect [6], which is often seen as a consequence of QFT vacuum energy, can be derived without invoking any vacuum energy [7]. We contend that other phenomena, such as the vacuum condensates in QFT [8] have been misinterpreted, and can possibly be reformulated with an equivalent quantitative formulation without resorting to the QFT interpretations of the vacuum used presently. Hence, the purpose of the present study is to derive a realistic cosmological theory for a constant (classical) vacuum energy density. The fact that the hypothesis avoids the cosmological constant problem and leads to a very elegant theory which explains many cosmological phenomena in a simple way, is then seen as a strong endorsement of the correctness of this hypothesis. We decided to test this hypothesis first in the cosmological context

Accepting this hypothesis we are now confronted with the standard de Sitter solution of the GR equations for a non-zero cosmological constant. This solution has no singular beginning. It also leads to a violation of total energy conservation, as the expanding vacuum universe will increase its energy content with time [2]. The solution to this problem is to employ a metric distinct from the usual Robertson-Walker metric. The proposed solution is a good candidate for the description of the actual universe, as it features the expected singularity at time $t = 0$. In addition energy conservation and the expansion of the universe will be compatible rather than in conflict with each other.

Let us briefly discuss this solution. The vacuum energy is represented by the following energy-momentum tensor:

$$T_{\mu\nu} = -\epsilon g_{\mu\nu}, \tag{1}$$

where we use the metric - popular in cosmology - with $g_{00}$ negative, so that $\epsilon$ is positive for positive vacuum energy. The Einstein equations then read:

$$R_{\mu\nu} - \frac{1}{2} R g_{\mu\nu} = -8\pi G \epsilon g_{\mu\nu}. \tag{2}$$

In view of the observed (approximate) spatial flatness of the universe [9] we try to solve this equation (2) with a metric tensor that is conformally flat:

$$g_{\mu\nu} = -g(x)\eta_{\mu\nu}, \tag{3}$$

where $\eta_{\mu\nu}$ is the Minkowski metric. In contrast to the Robertson-Walker (RW) metric, we do not introduce a scale factor in the metric. This raises the question how to account for the expansion of the universe in the current parametrization. As we will see shortly, we will account for the expansion without abandoning the Minkowski metric $\eta_{\mu\nu}$. Since the vacuum is expected to be spatially homogeneous, we restrict the dependence of $g(x)$ to the time coordinate $t$. We then obtain the following solution of the Einstein equations:

$$g(t) = \frac{3}{8\pi G \epsilon t^2} = \frac{t_s^2}{t^2}, \tag{4}$$

where $t_s$ is a characteristic time, which will play an important role in the following. Hence, the conformally flat metric of this vacuum universe now reads explicitly as follows:

$$ds^2 = \frac{t_s^2}{t^2}\left(-dt^2 + dx^2 + dy^2 + dz^2\right). \tag{5}$$

This can be contrasted with the usual Robertson-Walker metric where:

$$ds^2 = -d\tau^2 + a_{RW}(\tau)^2(dx^2 + dy^2 + dz^2), \quad (6)$$

The two representations are mathematically related by the following transformations:

$$\tau = \pm t_s \; ln(t/t_s) \longrightarrow a_{RW}(\tau) = exp(\mp\tau/t_s), \quad (7)$$

where the latter can be recognized as the de Sitter solution. However, the different choices of the physical variables lead to very different universes. For example, an expanding de Sitter universe ($a_{RW}(\tau) = exp(\tau/t_s)$ with $\tau$ positive and increasing towards the future) corresponds to a decreasing $t$ in our formulation, and therefore to a contracting universe. Our definition of the scale factor will accordingly follow a very different route from that in the RW formulation. It will not be based on the metric (which is left in its Minkowski form) and rather being based on the demand of energy conservation. The $1/\epsilon$ dependence of $g(t)$ emphasizes the non-perturbative nature of this vacuum solution, typical of complex systems. Hence, in a cosmological context it is incorrect to neglect the small vacuum energy $\epsilon$, or treat it perturbatively. The $1/t^2$ singularity of the metric at $t = 0$ makes this vacuum solution a good candidate for the description of the Big Bang.

An important property of this vacuum solution is that the geodesics represent either stationary points or test particles that move with the speed of light. Instead, in an ordinary flat universe (without vacuum energy) any (constant) speeds not exceeding the velocity of light are allowed [10]. It could thus be argued that the presence of the (classical) vacuum energy is responsible for both the origin of the velocity of light and for the apparent stationary nature of astronomical objects in the cosmos. Other interesting perspectives on the significance of $\epsilon$, or equivalently (if $G$ is constant) the cosmological constant $\Lambda$, are reviewed by Padmanabhan [11]. We quote: "the innocuous looking addition of a constant to the matter Lagrangian (which does not does not affect the dynamics of matter) leads to one of the most fundamental and fascinating problems of theoretical physics". What the present author finds particularly interesting is the suggestion [11] to consider the cosmological constant as a Lagrange multiplier, ensuring the constancy of the 4-volume of the universe when the metric is varied. Since we will see that the constancy of the (invariant) volume in the current approach is closely related to energy conservation, we suspect that there are deeper connections to be resolved.

It is interesting to note that Einstein originally introduced the cosmological constant to ensure the stationary nature of the universe, while we use it to generate an expanding universe. However, our analysis will show that from the perspective of a co-moving observer the universe does look stationary. This result also suggests possible links between our description and Hoyle's steady state universe, as the latter also makes use of a constant cosmological constant (see a paper by McCrea [12]).

In the next section we will demonstrate the origin of the linear expansion of the universe and the prescription for determining the scale factor.

## 3. Energy Conservation in General Relativity

The energy-momentum conservation condition from ordinary quantum mechanics is generalized in GR by replacing the derivative of the energy-momentum tensor by its covariant counterpart:

$$\nabla_\mu T^{\mu\nu} = 0. \tag{8}$$

This condition is automatically satisfied for a metric satisfying the Einstein equations, and can be shown to follow from the Bianchi identities [1]. Equation (8) is trivially satisfied by the vacuum energy density (Eq. (1)). However, in an expanding universe condition Eq. (8) is not sufficient to guarantee energy conservation. The total energy content of the vacuum universe is obtained by integrating $-T_0{}^0 = \epsilon$ over the invariant volume:

$$E = \int_V d^3x \sqrt{^3g}\ \epsilon = \int_V d^3x \frac{t_s^3}{t^3}\ \epsilon. \tag{9}$$

Here $^3g$ is the induced spatial metric, i.e. it is the spatial component of the determinant of the metric tensor (Ref. [2], p.120), which in our diagonal case equals $^3g = g_{11}g_{22}g_{33}$. In order to ensure energy conservation, the spatial volume $V$ in (9) must expand like $t^3$:

$$V(t) = \frac{t^3}{t_s^3} V_s. \tag{10}$$

The proportionality constant $V_s$ can be interpreted as the invariant volume since:

$$\int_V d^3x \sqrt{^3g} = V_s, \tag{11}$$

is constant. This invariant volume also equals the physical volume of the universe at the characteristic time $t_s$ . This volume also features in the expression for the total energy: $\epsilon V_s$, which therefore is invariant, as it should be. The expansion of the volume of the universe is best interpreted in terms of a scale factor that rescales distances, especially since it remains finite if $V_s$ is infinite. Hence, we write:

$$V(t) = a(t)^3 V_s, \tag{12}$$

where the scale factor for the vacuum equals:

$$a(t) = \frac{t}{t_s}. \tag{13}$$

Even after the introduction of matter and radiation, $V(t)$ will still display this cubic time dependence. In addition, however, $V_s$ will change every time a creation or annihilation process takes place. Hence, in the real universe $a(t)$ will also have to reflect these QFT processes. We will come back to this aspect in Section 6.

We note finally that the linear scale factor in (13) is unique to our approach as is normally forbidden in the RW metric [2]. The only other situation in which a linear scale factor occurs is in the Milne universe [2]. However, as this universe has zero vacuum energy and non-zero curvature, it is not related to our universe and is not an acceptable model of the universe.

## 4. The modified metric in the presence of matter

The vacuum universe can be described as an ideal fluid with a pressure-to-density ratio of -1. This value is in excellent agreement with the Supernova Legacy Survey ($w = -1.023 \pm 0.090(stat) \pm 0.054(sys)$ [13] for the dark energy equation of state). Hence, this strongly suggests that dark energy and vacuum energy are one and the same thing. Since dark energy appears to dominate the energy content of the universe, by implication vacuum energy dominates the global dynamics of the real universe. However, as the presence of matter and radiation is a consequence of QFT processes and make the universe interesting, our next task is to include these aspects as well. In view of the dominance of the vacuum energy it seems reasonable to treat the matter and radiation terms to first order, i.e. to linearize the Einstein field equations within the non-perturbative vacuum background. This approach has additional advantages, as it allows us to solve the Einstein equations exactly for the proposed representations of matter and radiation, and allows us to sidestep certain problems arising from the quantum nature of these terms.

The usual way to characterize the universe in the presence of matter and radiation is as an ideal fluid. However, instead of the constant pressure and density appropriate for a vacuum universe, one must now consider the pressure and density as being *time dependent* [14]. In our opinion, even this generalization is not sufficient for the matter in the universe: an important characteristic of matter is that it is localized, whereas the perfect fluid description does not take into account any spatial dependence. This localized nature of matter is true, irrespective of whether matter is in the form of fermions, planets, stars or galaxies. Astronomical objects are separated by vast empty areas and the matter distribution is thus far from being locally homogeneous. Neglect of this spatial dependence of matter is unlikely to provide the correct solution of the differential Einstein equations, where the spatial derivatives are expected to play a prominent role. Hence, we propose a matter density representation which emphasizes this local inhomogeneity:

$$T_{\mu\nu}^{matter}(x) = -\rho^{matter}(x)\hat{g}_{00}\delta_{\mu 0}\delta_{\nu 0}, \tag{14}$$

where the matter density is represented by

$$\rho^{matter}(t, \vec{x}) = \sum_i \frac{M_i}{\sqrt{^3\hat{g}(t, \vec{x})}}\delta^{(3)}(\vec{x} - \vec{x}_i). \tag{15}$$

A similar form for a source term expressed in terms of delta functions accompanied by a suitable function of the metric, was already suggested by Weinberg ([14], (5.2.13)). The appearance of the metric in the energy expression is not unexpected: the covariant condition, Eq.(8), clearly shows that a consistent definition of the energy-momentum tensor requires a particular dependence on the metric. In fact, the form (14) satisfies the covariant energy conservation condition (8) to the required order. We also introduced the new metric $\hat{g}_{\mu\nu}$ which accounts for the presence of matter. Since we treat the corrections to the vacuum metric to first order, we can approximate the exact metric in Eqs.(14 ) and (15) in most analyses by the vacuum metric $g_{\mu\nu}$. If we integrate this density (i.e. $-T_0{}^0$) over the whole universe we get the sum of all masses, as desired.

In order to calculate the first order effect of the matter term on the metric we write the metric tensor as follows:

$$\hat{g}_{\mu\nu}(t,\vec{x}) = g(t)\left\{\eta_{\mu\nu} + h^{matter}_{\mu\nu}(x)\right\} \ , \tag{16}$$

where $g(t)$ is given by (4). The inverse metric to first order then equals:

$$\hat{g}^{\mu\nu}(t,\vec{x}) = g(t)^{-1}\left\{\eta^{\mu\nu} - \eta^{\mu\alpha}h^{matter}_{\alpha\beta}(x)\eta^{\beta\nu}\right\}. \tag{17}$$

The linearized Einstein field equation reads:

$$\begin{aligned}
&-\eta^{\lambda\lambda}\partial_\lambda\partial_\lambda h_{\nu\mu} + \eta^{\lambda\lambda}\partial_\lambda\partial_\nu h_{\mu\lambda} + \eta^{\lambda\lambda}\partial_\lambda\partial_\mu h_{\lambda\nu} - \eta^{\lambda\lambda}\partial_\nu\partial_\mu\, h_{\lambda\lambda} \\
&+\frac{2}{t}\left(\partial_\nu h_{\mu 0} + \partial_\mu h_{0\nu} - \partial_0 h_{\nu\mu} + \eta_{\nu\mu}\eta_{\lambda\lambda}\partial_\lambda h_{\lambda 0} - \frac{1}{2}\eta_{\nu\mu}\eta^{\lambda\lambda}\partial_0 h_{\lambda\lambda}\right) + \frac{6}{t^2}h_{00}\eta_{\nu\mu} \\
&= 16\pi G\left(T^{matter}_{\mu\nu} - \frac{1}{2}g_{\mu\nu}T^{matter}\right) = 16\pi G g\rho^{matter}\left(\delta_{\mu 0}\delta_{0\nu} + \frac{1}{2}\eta_{\mu\nu}\right),
\end{aligned} \tag{18}$$

where some vacuum terms cancelled out. The solution can be expressed in terms of a single function $h(x)$:

$$h^{matter}_{\mu\nu}(x) = \delta_{\mu\nu}h(x), \tag{19}$$

with:

$$h(x) = 2Gg\int_{\hat{V}} d^3x'\frac{\rho^{matter}(t,\vec{x}')}{|\vec{x}-\vec{x}'|} = 2G\frac{t}{t_s}\sum_i M_i\frac{1}{|\vec{x}-\vec{x}_i|} \ . \tag{20}$$

Here the original volume $V$ is replaced by the volume $\hat{V}$, associated with the new state vector of the universe. This new volume (and hence the corresponding scale factor) is determined by the demand of global energy conservation, and is not fixed by the Einstein equations (see Section 6).

The only difference between (20) and the standard result in a flat background metric is the factor $t/t_s$. This factor counters the expansion of the universe and ensures that astronomical objects are in stable orbits despite the expansion of the universe. If we replace the coordinates $\vec{x}$ by co-moving coordinates $\vec{\tilde{x}}$, we get the standard result [14]:

$$h(x) = 2G\sum_i M_i\frac{1}{|\vec{\tilde{x}}-\vec{\tilde{x}}_i|}, \tag{21}$$

where:

$$\vec{\tilde{x}} = \frac{t_s}{t}\vec{x}. \tag{22}$$

Hence, two related representations are possible of the space-time characteristics of the universe. The first one is the co-moving representation, which is closest to our observations, as we cannot directly observe the scale factor $t/t_s$, whereas our astronomical observations are in agreement with (21). However, we have to use the original variables and the explicit form (20) in the Einstein equations, since $\vec{x}$ and not $\vec{\tilde{x}}$ is the independent variable in those equations. A similar duality occurs with respect to the time coordinate and the fourmomentum of particles, as we will discuss in more detail in Section 9.

Finally, we note that the sum (20) would be infinite for an infinite universe. The problem is that we have used an instantaneous solution. By imposing causality we can limit the contributions to

$$|\vec{x} - \vec{x}_i| < ct \ , \tag{23}$$

when applying (20). In terms of co-moving coordinates condition (23) implies:

$$\left|\vec{\tilde{x}} - \vec{\tilde{x}}_i\right| < ct_s \ . \tag{24}$$

The average correction to the flat metric is then:

$$< h(x) >= \frac{3}{2}\frac{\rho^m}{\epsilon} = \frac{3}{2}\frac{\sum_i M_i}{\epsilon \hat{V}_s} \ , \tag{25}$$

where $\rho^m$ is the average matter density in the universe. For the definition of the adjusted invariant volume $\hat{V}_s$ we refer to Section 6.

## 5. The modified metric in the presence of radiation

As we only consider $h_{\mu\nu}$ to first order, we can solve the equations of general relativity separately for the matter and electro-magnetic contributions in the vacuum background. The radiation density can be written in the perfect fluid form, as the QFT expression for the energy density is not localized (photons are represented by plane waves). Taking account of the metric factors so that the resulting expression satisfies the covariance condition (8), we arrive at:

$$T^{rad}_{\mu\nu}(x) = \frac{1}{g}\frac{\sum_j p^{(j)}}{\hat{V}}\begin{pmatrix} 1 & 0 & 0 & 0 \\ 0 & \frac{1}{3} & 0 & 0 \\ 0 & 0 & \frac{1}{3} & 0 \\ 0 & 0 & 0 & \frac{1}{3} \end{pmatrix} = g\begin{pmatrix} \rho^{rad} & 0 & 0 & 0 \\ 0 & p^{rad} & 0 & 0 \\ 0 & 0 & p^{rad} & 0 \\ 0 & 0 & 0 & p^{rad} \end{pmatrix} . \tag{26}$$

In (26) we included all the photons in the universe at time $t$. Both $p^{(j)}$ and $\hat{V}$ have an effective time dependence owing to the expansion of the universe: $p^{(j)}$ is complementary to the spatial Einstein coordinate and thus decreases like $1/t$ (see Section 9), whereas $\hat{V}$ increases like $t^3$ (see Section 6). Hence, although the explicit time dependence of $T^{rad}_{\mu\nu}(x)$ is like $t^2$, its effective time dependence after accounting for the expansion of the universe is like $t^{-2}$. Similarly, $\rho^{rad}$ and $p^{rad}$ have the explicit time dependence $t^4$, but after the expansion of the universe is taken into account, its effective time dependence is constant. In analogy to the co-moving coordinates $\vec{\tilde{x}}$ introduced in the matter case (see (22)), we can introduce momenta as observed by a co-moving observer:

$$\vec{\tilde{p}} = \frac{t}{t_s}\vec{p}. \tag{27}$$

This behaviour will be further discussed in Section 9, in which we discuss the two dual representations in detail. If we integrate $\rho^{rad}$ (or $-{T_0}^{0,rad}$) over all of space, we get the sum over all momenta $p^{(j)}$ at time t, multiplied by $t/t_s$; i.e the sum over all co-moving momenta. As we will see in Section 6 this implies that the radiative contribution to the total energy is constant over time, unless creation or annihilation events change the number and/or nature of the participating photons. Hence, this situation is similar to the matter case where the energy integral is also constant, as long as the state vector remains the same.

We can obtain $h_{\mu\nu}^{rad}$ by solving the first order equation (18) with the electro-magnetic source term. We find:

$$h_{\mu\nu}^{rad}(x) = -8\pi G t^2 T_{\mu\nu}^{rad}(x) \ . \tag{28}$$

Because of the effective $t^{-2}$ time dependence of $T_{\mu\nu}^{rad}(x)$, $h_{\mu\nu}^{rad}(x)$ behaves effectively like a constant. Naturally, when solving for $h_{\mu\nu}^{rad}(x)$ in the Einstein equations, we must use the explicit $t^4$ dependence of this function. In other words, the decrease of the radiation density with time caused by the expansion of the universe is countered by the $t^4$ dependence, arising from the background vacuum metric. The combination of these two factors is the cause of the constancy of the effective contribution to the total energy. The effective constancy of the radiation and matter terms also ensures the continued basic linear expansion of the universe in the presence of matter and radiation, as we will demonstrate in Section 6. It should be noted that $h_{\mu\nu}^{rad}$ has a sign opposite to that of the matter contribution, because of the minus sign in (28). Thus radiation has a gravitational effect opposite to that of matter. Because light cannot solidify into a massive astronomical body these effects are hard to measure, but this opposite sign of the radiation contribution has a distinct effect on the average metric in the universe.

Just as in the case of matter we can evaluate the average value of $h_{\mu\nu}$ in the radiative case. We find:

$$< h_{00}^{rad}(x) > = -3\rho^{rad}/\epsilon \ . \tag{29}$$

The two results, Eq.(25) and Eq.(29), have nearly the same form and can be used to assess the flatness of the universe. As noted above, matter and radiation have opposite effects on the metric.

## 6. Energy Balance and the Expansion of the Universe

After the introduction of matter and radiation the total energy of the universe is given by:

$$E = \int_{\hat{V}} d^3x \sqrt{^3\hat{g}(t,\vec{x})}\ \epsilon + \int_{\hat{V}} d^3x \sqrt{^3\hat{g}(t,\vec{x})}\ \rho^{matter}(x) + \int_{\hat{V}} d^3x \sqrt{^3\hat{g}(t,\vec{x})}\ \rho^{rad}. \tag{30}$$

Expanding (30) to first order in $h_{\mu\nu}$, we have:

$$\begin{aligned} E = \epsilon V_s &= \epsilon \frac{t_s^3}{t^3}\hat{V}(t) + \epsilon \frac{t_s^3}{t^3}\frac{3}{2}\int_{\hat{V}} d^3x \left\{ h(x) + \frac{1}{3}\sum_i h_{ii}^{rad} \right\} + \sum_i M_i + \frac{t}{t_s}\sum_i p^{(i)} \\ &= \epsilon \frac{t_s^3}{t^3}\hat{V}(t) + \epsilon \frac{t_s^3}{t^3}\frac{3}{2}\int_{\hat{V}} d^3x\ h(x) + \sum_i M_i + \frac{t}{t_s}\sum_i p^{(i)}(1-\frac{3}{2}) \ . \end{aligned} \tag{31}$$

We now extend (10) to the volume in the presence of radiation and matter:

$$\hat{V}(t) = \frac{t^3}{t_s^3}\hat{V}_s \ , \tag{32}$$

i.e. we assume that the introduction of matter and radiation only requires a change of the original invariant volume $V_s$ into $\hat{V}_s$ . Using (32) and (27) we can then show that all terms in (31) are constant for a given state vector, so that Ansatz (32) is consistent. Therefore, the volume $\hat{V}_s$ is indeed invariant under the basic linear expansion of the universe, although it adjusts itself whenever the state vector of the universe changes if particles are created or destroyed. Using the definition (32) and expression (25) for $< h >$, we get:

$$E = \epsilon V_s = \left(\epsilon + \rho^m + \rho^{rad} + \frac{9}{4}\rho^m - \frac{3}{2}\rho^{rad}\right)\hat{V}_s \ . \tag{33}$$

The last two terms originate from the modifications to the metric after the introduction of matter and radiation. Conveniently, they have the same form as the original matter and radiation terms, with only the coefficients being different. If, as is usually assumed, the matter term dominates over the radiation term, then the new invariant volume $\hat{V}_s$ is smaller than the original invariant volume $V_s$.

We can now generalize the scale factor in the presence of matter and radiation:

$$a(t) = \frac{t}{t_s}\left(\frac{\hat{V}_s}{V_s}\right)^{1/3} = \left(\frac{\hat{V}(t)}{V_s}\right)^{1/3} . \tag{34}$$

Hence, a change in $\hat{V}_s$ also implies a change in the scale factor. So, in addition to the linear time dependence, there is a further implicit time dependence, which depends on the evolving state vector of the universe. Whenever a QFT transition takes place, and the state vector is changed, the volume $\hat{V}_s$ is also adjusted, and, consequently, the scale factor. Together this leads to an effective time dependence of the scale factor. Thus, the creation and/or decay of matter and the creation or absorbtion of radiation in the universe has rather specific consequences for the acceleration or contraction of the universe over and above the linear expansion. Since the volume $\hat{V}_s$ is determined by the energy equation (33), the scale factor is no longer determined by the Einstein equations as in the FRW case, but rather by total energy conservation. This is an important difference, which allows us to consider local gravitational effects and effects of the global expansion together, as the scale factor no longer appears in the metric. The scale factor is now also a truly cosmological property, as it is the same everywhere in the universe, being defined in terms of integrals over the whole universe. It may be hard to accept how a transition at a distinct location can influence the scale factor in our neighbourhood, especially for an infinite universe. However, if we replace individual transitions by rates and assume that the universe looks the same everywhere on a large scale (the cosmological principle), then we can view our visible universe as a finite representation of the whole universe. The additional time dependence $\left(\hat{V}_s/V_s\right)^{1/3}$ then becomes continuous and represents the acceleration or deceleration of the universe as a deviation from the basic linear expansion of the universe. Obviously, this continuous time dependence does not feature explicitly in the Einstein equations, although the solution of these equations at a particular time must use the state vector pertaining to that particular moment.

One could call (33) the co-moving form of the energy balance equation. If we go back to the original volume using (32), we obtain:

$$E = \epsilon V_s = \left( \frac{t_s^3}{t^3}\epsilon + \frac{t_s^3}{t^3}\rho^m + \frac{t_s^3}{t^3}\rho^{rad} + \frac{9}{4}\frac{t_s^3}{t^3}\rho^m - \frac{3}{2}\frac{t_s^3}{t^3}\rho^{rad} \right) \hat{V}(t) \; . \tag{35}$$

This form clearly illustrates the high densities in the initial universe and the decreasing densities with time. It is these time-dependent densities which have to be considered in descriptions of the evolution of the universe and of the hot Big Bang, because they have to be compared to QFT processes that do not depend on the expansion of the universe and therefore play different roles in different epochs.

One question is now whether we can determine the contributions of the matter and radiation components to the energy balance (33). As stated previously, the unperturbed vacuum energy (i.e. the first term in (33)) has the same density to pressure ratio ($w = -1$, [13]) as dark energy, suggesting that this dark energy can be identified with the unperturbed vacuum energy. Since the ratio $w$ will change if $\hat{g}_{\mu\nu}$ differs considerably from the vacuum metric $g_{\mu\nu}$, the dominance of dark energy and the observed flatness of the universe suggest that the average matter and radiation densities $\rho^{matter}$ and $\rho^{rad}$ are small compared to the vacuum energy density $\epsilon$. The current estimates of the matter content of the universe (about 24% including dark matter, [15]) rely heavily on the energy balance as formulated in the RW framework ([4], [1]). Hence, these estimates should be re-examined in the context of the current framework and might actually be much less certain than usually is assumed. Furthermore, the usual estimate of the radiation content is based on the decrease of this density owing to the expansion of the universe and the red shift, and does not take account of the metric factor $t^4$, which completely compensates for this decrease, leading to a constant $\rho^{rad}$. Hence, the contribution of radiation to the total energy could well be comparable to the matter density, rather than merely having the tiny value of $5 \times 10^{-5}$ quoted in the literature [16]. It should also be noted that the observed decrease of photon momenta with time, popularly called the red-shift effect, has a rather different interpretation in our formulation, as we will see in Section 8. We attribute this to the dual nature of the vacuum universe and to our role as a co-moving observer therein. So one can expect that this "red-shift" effect is compensated for in energy expressions and that it will not lead to a reduction in the contribution of radiation with the passing of time.

In the standard picture the initial universe experienced a radiation-dominated phase, followed by the current matter-dominated phase. However, in our picture the contribution of these phases to the energy balance are more or less constant over time, owing to the influence of the background metric. Hence, both the dominance of radiation over matter in the early stages, and the dominance of matter over radiation at the present time must be re-examined in our approach.

If these densities are indeed of comparable magnitude then many interesting scenarios for the evolution of the universe become possible. For example, if:

$$< h_{00}^{rad} >= -\frac{3}{2} < h > \ , \tag{36}$$

or in terms of densities

$$\rho^{rad} = \frac{3}{4}\rho^{m} \ , \tag{37}$$

then the average metric becomes proportional to the flat Minkowski metric, even in the presence of large matter and radiation densities. Since $< h_{00}^{rad} >$ increases and $< h >$ decreases in size, whenever matter is converted into radiation, condition (36) cannot be satisfied at all times, unless the creation and annihilation processes are in equilibrium. However, the universe may have been close to this point for most of its existence, in which case (36) would explain why the universe appears so flat, despite containing a considerable amount of matter. Clearly, further analyses are required to examine these possibilities.

Using Eq.(24) it is also possible to calculate the approximate energy content of the visible universe. We find

$$E^{visible} = \epsilon \times \frac{4\pi}{3} t_s^3 = \frac{1}{2}\left(\frac{3}{8\pi}\right)^{1/2} \frac{1}{\epsilon^{1/2} G^{3/2}} \approx 5 \times 10^{79} \text{ GeV}. \tag{38}$$

where we used the value $\epsilon = 4.06 \times 10^{-47}$ GeV$^4$ derived in Section 9. Again, this emphasizes the important role of the two fundamental dimensionfull constants of Nature, $\epsilon$ and $G$. This result also allows one to make a rough estimate of the number of massive particles in the visible universe: about $10^{79}$ protons and the same number of electrons. Other independent estimates of this number will put further constraints on the value of $\epsilon$ or $t_s$.

## 7. A possible explanation for dark matter

An important cosmological problem is the nature of dark matter. It may be tempting to consider the mixed vacuum-matter term in (31) as a dark matter term. Firstly, it is closely related to the matter distribution and is localized near matter concentrations, on account of the form of $h(x)$. Secondly, its contribution to the total energy is much larger than that of the original mass term, as is the case for dark matter by comparison with ordinary matter. However, since the mixed term does not influence the metric in lowest order (it being rather a consequence of the metric) it could only have gravitational effects in higher order. Furthermore, the localization of the equivalent "mass" it represents is so weak that it cannot explain the distribution of dark matter near galaxies. Finally, the enhancement factor 9/4 differs considerably from the usual ratio of dark matter to ordinary baryonic matter (a factor of about 4.8, see [17]).

We will discuss another interesting possible explanation for dark matter. This explanation is based on certain second-order effects, which are unique to our approach. Since we have neglected second order effects up to now, this analysis is somewhat tenuous. However, it shows encouraging agreement with some observations. As we see from Eq.(15), and the integral in Eq.(20), the gravitational potential is inversely proportional to $\sqrt{^3\hat{g}(t,\vec{x})}$ owing to the form of the matter energy-momentum tensor. In first order we replaced $\sqrt{^3\hat{g}}$ by $\sqrt{^3g}$. However, in higher order we would need to consider the corrected metric factor. For a black hole at the center of a galaxy this would effectively mean that instead of experiencing the gravitational pull of the real mass $M$, one would experience the reduced effect of the apparent mass $M^{app}$ at a distance $r$:

$$M^{app}(r) = \frac{M}{\prod_{i=1}^{3} \{1 + h_{ii}^{matter}(x)\}^{1/2}} = \frac{M}{(1 + 2GM/r)^{3/2}} \ , \qquad r > R_{BH} \ , \tag{39}$$

where $R_{BH}$ is the radius of the black hole. At small distances $M^{app}$ could be much smaller than $M$, whereas at large distances the black hole mass would have its normal effect as the screening becomes negligible. Since this mass does not correspond to any visible material, it could represent dark matter. Assuming this to be the case, we can define the dark matter density by subtracting the observed mass near the black hole from the effective mass distribution:

$$4\pi \int_0^r dr' \ r'^2 \ \rho^{dm}(r') = \frac{M}{(1 + 2GM/r)^{3/2}} - M^{app}(R_{BH}) \ . \tag{40}$$

Differentiating with respect to r we obtain for small $r$ and large $2GM/r$:

$$\rho^{dm}(r) \sim r^{-3/2} \qquad r > R_{BH} \ . \tag{41}$$

Possible support for this picture comes from analyses of dark matter ([18], [19]), which also indicate a singular behavior of dark matter density in the center of galaxies. For example, Krauss and Starkmann [18] find that the dark matter density near the centre behaves like $r^{-3/2}$, which is exactly in agreement with our explanation. In addition, thermal models of galaxy densities [19] give a constant core density for normal matter, so that our effective mass distribution cannot be interpreted as normal matter. At large distances we obtain:

$$\rho^{dm}(r) \approx \frac{3}{4\pi} \frac{GM^2}{r^4} \qquad r \gg R_{BH} \ , \tag{42}$$

which gives the required localization near existing galaxies.

A possible objection to this explanation is that a current survey [16] only gives a black hole contribution of $7 \times 10^{-5}$ to the energy content of the universe, although this number may be surrounded by uncertainties similar to those around other estimates in the RW framework. If this number is based on the apparent mass of black holes as measured in the vicinity of these objects, then a huge screening effect is required to explain the large dark matter component in terms of black holes. However, as (39) allows such effects, the true mass of the black holes at the center of galaxies may well be order of magnitudes greater than is currently assumed. This possibility may also have an important influence on considerations of the evolution of the early universe, as enormous black holes are usually considered fatal to the development of galaxies. This need no longer be the case owing to the screening effects suggested in the current framework. Clearly, non-perturbative calculations are required to test this dark matter theory, as large second order corrections would in turn induce important third or even high-order terms, which could either moderate or enhance the effect observed in second order.

## 8. Description of red shift data and other observables

Let us now discuss a number of astronomical observables. Firstly we discuss the Hubble constant. This quantity is defined as the relative increase of the scale factor with time [14], which in our formulation reads:

$$H(t) = \frac{\dot{a}(t)}{a(t)} = \frac{1}{t}, \tag{43}$$

where the last transition follows from Eq.(34) if we ignore the time dependence of $\hat{V}_s$. As we will see in the following discussion, one actually measures the inverse of the Hubble constant, because one detemriens the luminosity distance $d_L$. As a co-moving observer would measure the co-moving distance $(t_s/t)d_L$, the Hubble constant measured would also be rescaled and would equal $(t/t_s)1/t = 1/t_s$. This leads to the pleasing result that the measured Hubble constant is indeed constant, since $t_s$ is constant! So this to some extent justifies the name Hubble *constant.* Since Eq.(43) shows that the inverse Hubble constant represents the age of the universe, we find that for a co-moving observer the age of the universe is constant and equals $t_s$. Unfortunately, this also implies that the Hubble constant does not provide us with the actual age of the universe in terms of GR coordinates $t = t_0$. The value $t_0$ is of importance, since it tells us in which epoch we are living, as it is expressed in the same representation as the elementary particle properties (for example the particle physics scale $t_c$ derived in the Section 10). In Section 9 we will discuss how one can get information about the value of $t_0$. It should be noted that the measurement of the Hubble constant gives information on the (constant) vacuum energy density $\epsilon$, because of the relationship between $\epsilon$ and $t_s$.

The Hubble constant is determined from supernovae measurements. As shown below the fit to the Cepheid data suggests a value of $t_s = 13.8 \times 10^9$ years, corresponding to $H_0 = 71\,\mathrm{km\,s^{-1}\,Mpc^{-1}}$. Current best estimates by Freedman *et al* [20] provide the value $H_0 = 72 \pm 8\,\mathrm{km\,s^{-1}\,Mpc^{-1}}$. The identity of $H_0^{-1}$ and the age of the universe in our theory is in good agreement with observations, as most analyses favour values which are close together for these two quantities. Although decay processes contribute towards an acceleration of the expansion, we do not expect such changes to upset this agreement. In matter dominated FRW universes the age of the universe equals $\frac{2}{3}\,H_0^{-1}$, whereas in radiation dominated FRW universes it equals $\frac{1}{2}\,H_0^{-1}$, both possibilities differing from the accepted values of $H_0$ and the age of the universe.

The increase in wavelength of photons originating from distant galaxies or supernovae, as first observed by Hubble, is known as red shift. This name suggests that the phenomenon is due to the Doppler effect of receding galaxies. However, as is well-known [14], the correct explanation should be based on the framework of GR.

Weinberg [14] gives the standard explanation in terms of the RW metric, leading to the relationship:

$$z = \frac{\lambda_{obs} - \lambda_1}{\lambda_1} = \frac{a_{RW}(t_0)}{a_{RW}(t_1)} - 1, \tag{44}$$

where the source is characterized by $t_1$ and $\lambda_1$, the observer being characterized by $t_0$ and the observed wavelength $\lambda_{obs}$. Although we do not use the RW representation, we still get the same final result. Our explanation is based on the dual representation of space-time, with the observer measuring the wave length in terms of co-moving variables $\vec{\tilde{x}}$ and $\vec{\tilde{p}}$.

Firstly, the atomic transition giving rise to the emission of the light is defined by a characteristic wavelength $\lambda$ or by a characteristic time interval $\Delta t$, which remain constant over time. The wave length measured by a co-moving observer at the source is then:

$$\lambda_{source} = \frac{t_s}{t_1} \lambda \ , \tag{45}$$

or alternatively by a time interval $\frac{t_s}{t_1}\Delta t$. Because of the invariance of the interval $ds = \frac{t_s}{t_1}\Delta t$, we will measure the same time interval $\frac{t_s}{t_1}\Delta t$ when the photon finally reaches our equipment. Hence we will also observe the wave length:

$$\lambda_{observed} = \frac{t_s}{t_1} \lambda \ , \tag{46}$$

at our current location at time $t_0$. However, if we measure the same transition at our terrestrial location, we observe the wave length:

$$\lambda_{terrestrial} = \frac{t_s}{t_0} \lambda \ . \tag{47}$$

Now, in the standard interpretation ([14], p. 417) the wavelength at the source, $\lambda_1$, is supposed to be equal to the wavelength currently measured in a terrestrial laboratory, which we indicate by $\lambda_{terrestrial}$. So the unknown $\lambda_1$ in Eq.(44) is replaced by $\lambda_{terrestrial}$. Hence, what one really tests in the red shift analysis is an expression involving $\lambda_{terrestrial}$, not the unmeasured $\lambda_1$. Therefore, we introduce the terrestrial wave length directly into our expression for $z$. We are then led to the relationship:

$$z = \frac{\lambda_{observed} - \lambda_{terrestrial}}{\lambda_{terrestrial}} = \frac{\frac{t_s}{t_1}\lambda - \frac{t_s}{t_0}\lambda}{\frac{t_s}{t_0}\lambda} = \frac{t_0}{t_1} - 1 = \frac{a(t_0)}{a(t_1)} - 1, \tag{48}$$

in which the last identity is valid if we ignore the time dependence resulting from the change in $\hat{V}_s$. As we see, the final relationship is identical to Eq. (44) derived in [14]. Hence, totally different philosophies can still lead to the same result and consequently to the same agreement with experiment. The relationships Eq.(45) and Eq.(47), which involve $1/t$, seem to suggest that the wavelength decreases rather than increases with time. However, this conclusion is wrong: because of the expansion of the universe all lengths such as $x$ and $\lambda$ are increasing with time (although this increase is not explicit in the GR equations), and the indicated time dependence in these equations merely compensates for this increase to make the effective wave length constant over time.

In deriving the simple expression $t_0/t_1 - 1$ in Eq.(48) we have ignored the time dependence of $h_{00}$ in the metric. This is justified by the fact that the contributions of both matter and radiation to $h_{00}$ are effectively constant. We would only get deviations from this identity if creation or decay processes substantially affect the time dependence of $h_{00}$.

In order to compare our theory with the Cepheid observations we have to express the luminosity distance in terms of $z$. We have [14]:

$$d_L = \frac{a^2(t_0)}{a(t_1)} d(t_0) \ , \tag{49}$$

where the distance $d(t_0)$ can be expressed in terms of the time of emission and the time of observation:

$$d(t_0) = c \int_{t_1}^{t_0} \frac{dt}{a(t)} \ . \tag{50}$$

Using the vacuum metric and eliminating $t_1$ in favour of $z$, we have:

$$d_L = cH_0^{-1}(z+1)ln(z+1) = cH_0^{-1} z \left(1 + \frac{1}{2}z - \frac{1}{6}z^2 + \cdots\right) \ . \tag{51}$$

This corresponds to a deceleration parameter $q_0$ and to a jerk parameter $j_0$ both of which are zero (see Visser [21] for a definition of these parameters and the corresponding red shift formula). Since a co-moving observer will measure $t_s/t_0 \ d_L$ rather than $d_L$ itself, we still have to multiply this expression by the factor $t_s/t_0$. The current notation can be retained if we understand that $H_0^{-1} \rightarrow t_s/t_0 \ H_0^{-1} = t_s$.

We considered recent Cepheid data for distance moduli [22], which have to be corrected by $-0.27$ according to a recent analysis by the same authors [23]. The best fit for the vacuum solution is obtained for the value of $H_0 = 71$ km s$^{-1}$ Mpc$^{-1}$, which corresponds to a measured lifetime of $t_s = 13.8$ billion years. This agrees well with a recent WMAP analysis by Hinshaw et al. [24], who state that: $H_0 = 70.5 \pm 1.3$ km s$^{-1}$ Mpc$^{-1}$. In Fig. 1 we show a comparison between the observations and our vacuum solution result, together with some additional fits. The ratio $H_0 \, d_L/z$ is displayed, rather than $d_L$ itself, so as not to obscure the deviations between experiment and theory at small values of $z$. The vacuum result fits the data very well. In order to put this result in perspective, we have also fitted various power expansions of the expression in brackets in the right-hand-side of Eq. (51). These will allow us to determine the values for $q_0$ and $j_0$ preferred by the data and give an indication of the uncertainty in these parameters. The linear fit gives $q_0 = .038$, which is close to the value of zero obtained in the vacuum solution. The quadratic fit yields $q_0 = -.63$ and $j_0 = 1.26$. The Hubble parameters for the linear and quadratic fit are 73.0 and 75.9 km s$^{-1}$ Mpc$^{-1}$, respectively, both falling outside the range given by Hinshaw et al. [24]. The corresponding ages are 13.4 and 12.9 billion years. We see that the parameters obtained depend quite strongly on the nature of the fit, casting some doubt on the strength of evidence for a pronounced acceleration ($q_0 < 0$) in the current universe. As stated above, our model with a linear expansion already fits the data very well.

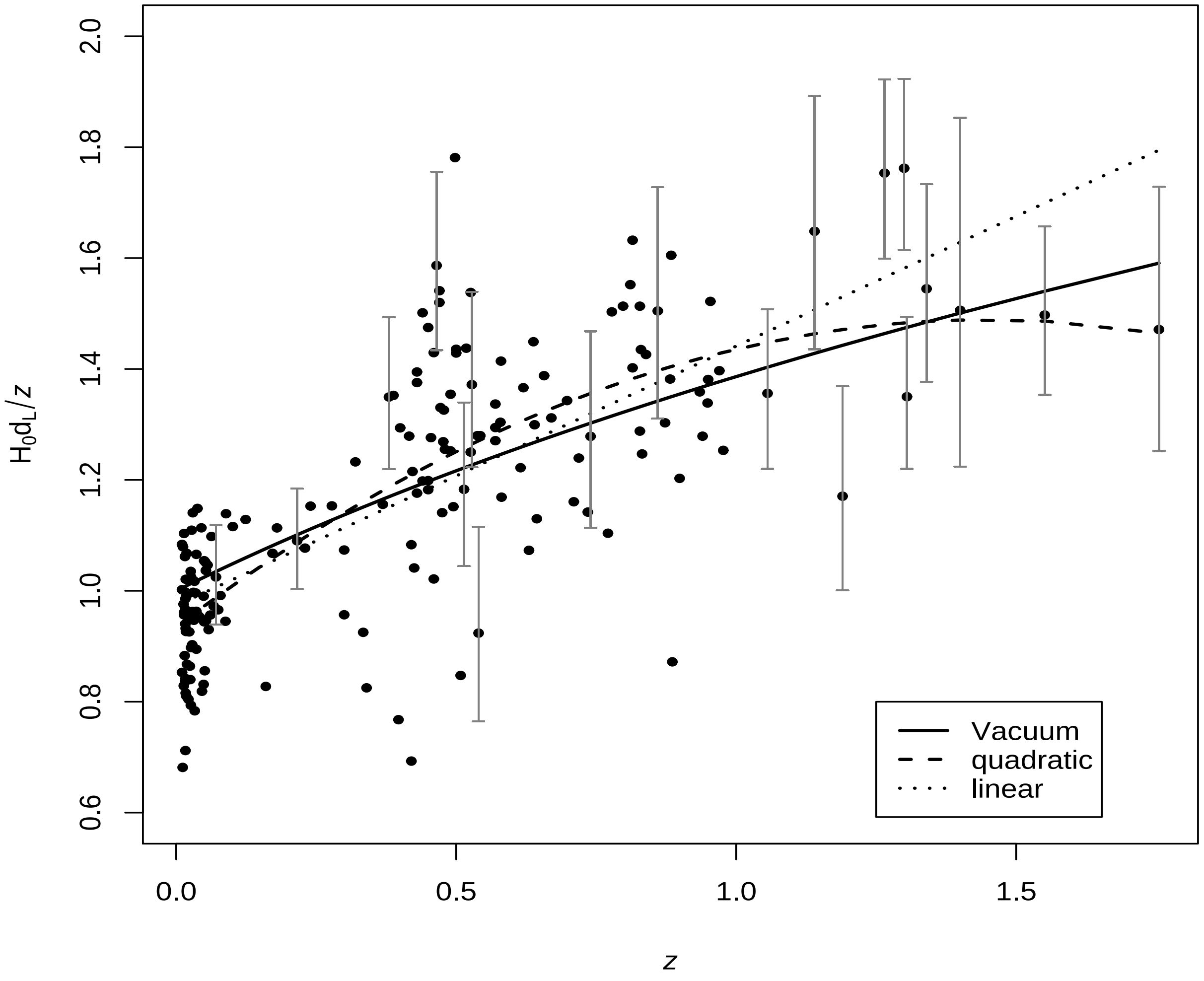

**Figure 1.** Comparison between Supernovae data ([22], [23]) and various theoretical descriptions. Plotted are the distance moduli multiplied by $H_0$ (as determined from our model) and divided by $z$ as a function of the red shift $z$. Some of the data errors are shown to illustrate the quality of the data and the fit.

It can be expected that supernovae data at higher values of $z$ will put stronger constraints on these parameters. The negative $q_0$ in the quadratic fit suggests that the universe is currently accelerating, whereas the positive jerk parameter suggests that there might have been a deceleration in the past. This agrees with the detailed statistical analysis carried out in Refs. ([22], [23]). Clearly, the present theory can explain the average expansion (linear). Within this theory it is natural to explain a possible current acceleration by means of the presence of decay processes. The standard decay process is radiation, as this process transforms matter into radiation. Other - more speculative - decay processes are possible as well. The decay of WIMP like particles in dark matter would contribute towards acceleration. Also, the annihilation of particles by black holes would be a possible source of acceleration. However, deviations from the linear expansion may also be attributed to the change in $h_{00}$ over time, as noted previously. Hence, a more detailed theoretical analysis is required, in which the role of decay processes in the current universe is elucidated. Higher order consequences of the abandonment of the RW formulation on the red-shift formula also have to be examined. In any case, it is clear that accurate supernovae data will yield strong constraints on the theoretical description.

Another important issue in cosmology is the horizon problem. The horizon is defined as the distance a photon traveled since the Big Bang to a particular point in time [2]. Obviously, this is infinite for the expression (50), as $t_1 = 0$. Since we only expect virtual photons to exist a finite time after the Big Bang (see Section 10), the physical horizon is not infinite. However, it is still true that in our description the horizon is much larger than in the typical FRW models, where the lower limit in the integral (50) vanishes. Hence there does not appear to be a horizon problem in our approach. This eliminates the main reason for the introduction of the inflationary hypothesis, although we predict an inflationary phase naturally in our approach, when particles and antiparticles were annihilated soon after they were created (see again Section 10).

## 9. The role of co-moving coordinates and the dual representation of space-time

As we have seen in previous sections, co-moving observers measure physical quantities in terms of the coordinates $\vec{\tilde{x}}$ and $\vec{\tilde{p}}$. These coordinates are invariant under the expansion of the universe (naturally they still function as variables in regard to local physical processes), and make it hard to measure this universal expansion directly. For example, the gravitational potential given in Eq.(20) remains constant in spite of the expansion of the universe, if it is expressed in terms of co-moving coordinates $\vec{\tilde{x}}$. We only measure the expansion indirectly via the red shift observations discussed in the previous section. These features appear to be especially simple for - and perhaps unique to - the vacuum metric, on account of the linear nature of the transformations.

The local co-moving representation is concisely given by:

$$\vec{\tilde{x}} = \frac{t_s}{t_0}\vec{x}, \tag{52}$$

$$\tilde{t} = \frac{t_s}{t_0}t\,, \tag{53}$$

while the conjugate relationships for the momentum reads:

$$\vec{\tilde{p}} = \frac{t_0}{t_s}\vec{p}, \tag{54}$$

$$\tilde{E} = \frac{t_0}{t_s}E\,. \tag{55}$$

So far we have avoided the use of the symbol $\tilde{E}$ for the co-moving energy. The energy $E$ used in previous equations (e.g. in Section 6) can be identified with the co-moving energy if we use the co-variant metric $\sqrt{-g}$, rather than ${}^3g$, in the relevant energy integrals. In that case we have to multiply the resulting expression by the factor $t_0/t_s$ (or by the global factor $t/t_s$) in order to obtain the standard result, called $E$ in Section 6. We will not discuss the implications of this modification further. Eq.(53) is a local representation of the global relationship $\tau = t_s \; ln(t/t_s)$, which was already mentioned in Eq.(7). We can also replace $t_0$ by $t$ in Eqs.(52, 54 and 55) to make the equations more global. However, for a local co-moving observer, the given equations are the relevant ones, since they simply amount to a rescaling of the original coordinates. This local representation is also natural in the context of the Einstein equations, since the (temporary) replacement of $t$ by $t_0$ ensures that the time-dependence of $\vec{x}$, $t$ and $p$ is not explicit in the Einstein equations.

Under the transformation Eqs.(52 and 53) the metric expression Eq.(5) near $t_0$ reads:

$$ds^2 = -d\tilde{t}^2 + d\tilde{x}^2 + d\tilde{y}^2 + d\tilde{z}^2. \tag{56}$$

It is natural that an observer would use a locally flat metric to carry out his observations, which explains the important role of this co-moving representation in measurements. The importance of such transformed variables in cosmology had already been recognized very early on in the development of cosmology. Milne [25] introduced dual variables in 1937, although he did not base himself on a universe with a finite vacuum energy density, so that he did not have a natural time scale $t_s$. It would be of interest to study the analogies further, although Milne did not use a relativistic formulation in his analysis.

As stated in the Section 8, one of the results of our particular measuring process is that the measured Hubble constant equals $(t/t_s)1/t = 1/t_s$. Given the value of the Hubble constant derived in the previous section from supernovae data, we find that the vacuum energy density acquires the value $\epsilon = 3/8\pi G t_s^2 = 4.06 \times 10^{-47}$ GeV$^4$. This is the value used in previous sections of this paper, e.g. in Eq.(38). This value is very close to the one given by Weinberg ([14], p. 476): $10^{-29}$g/cm$^3$, which corresponds to $4.31 \times 10^{-47}$ GeV$^4$. Actually, the value quoted by Weinberg represents the *critical* energy density of the universe, which must be close to the actual energy density for a universe that is flat (observations have shown that the geometry of the universe is very close to being spatially flat [1], [9]). Hence, this critical density should coincide with the vacuum energy density in a universe dominated by vacuum energy. In our theory there is nothing critical or accidental about the value of $\epsilon$, as slightly larger or smaller values would describe the universe equally well. Hence, this is another puzzle (why the critical and actual energy density are so close at this very moment) that is solved by the current theory. Carroll gives the rough estimate $10^{-8}$erg/cm$^3$ ([2], Eq. (8.162)), corresponding to $5 \times 10^{-47}$ GeV$^4$, which is also consistent with the current estimate.

The constancy of the age of the universe $t_s$ also indicates that an observer could never reach the beginning of time by moving backward in time (apart from the practical aspect that the Big Bang defines a direction in time which allows us only to "move" forward). This property also is evident if we use the global transformation variable $\tau$ defined in Eq.(7): $\tau = t_s \; ln(t/t_s)$. This suggests that $t = 0$ refers to the infinite past.

As the co-moving variables Eqs.(52, 53, 54 and 55) are essential for our measurements, the question can now be asked whether the magnitude of the original variables, as represented by our current time $t_0$, plays any (absolute) role. For example, in the red shift discussion, only the relative quantity $t_1/t_0$ played a role in the definition of $z$. However, in Section 10 we will see that different epochs in the evolution of the universe can be distinguished despite the "relativity" of the concept of time. Since properties of particles are expressed in the original time units, as they are independent of the expansion of the universe, the particle scale represents an independent way of measuring time. As a consequence, there are ways of inferring $t_0$, provided our particle models are sufficiently accurate. Since the currently accepted elementary particle models do not predict the masses of quarks and leptons, and in particular do not relate them to $G$ and $\epsilon$, we are a long way away from the situation that $t_0$ can be determined accurately. Nonetheless we will argue in Section 10 that $t_0$ is of the same order of magnitude as $t_s$. In other words: our current epoch is characterized approximately by the time scale $t_s$ of the vacuum universe.

## 10. Evolution and Development of a Vacuum Dominated Universe

It is common to consider the first moments after the Big Bang as a period of extreme complexity, during which particles are compressed into an extremely small space and carry enormous kinetic energy. This scenario is sketched in many articles and popular books, e.g. in a recent book by Martin Rees [27]. It also leads to the idea, often heard these days, that the LHC experiment at CERN will reproduce the early moments of the Big Bang [28]. Such a densely populated state of the early universe requires a reliable unified theory of QFT and GR. Since that does not (yet) exist, a reliable picture of this initial epoch is lacking. Our solution to the GR equations suggests another scenario. The singularity in the classical vacuum metric implies that the universe started out in the simple "classical" vacuum state. The initial density of the universe was so high and the distance scale so small that physical fermions, which we expect to have a finite - although extremely small - size, could not form. The creation of real photons, which is linked to the fermionic processes by the standard model Lagrangian, was likewise suppressed. Under these circumstances, the quantum fluctuations in the early universe do not lead to the creation of any physical particles and thus leave the vacuum state vector of the universe unaffected, as this state will only change once physical particles have been formed. As a consequence, in this early epoch the state vector of the universe equals the vacuum state and displays perfect homogeneity and zero entropy. The energy of this state is given by $\epsilon \; V_s$, which at the same time represents the total energy of the universe at all times and the permanent entry in the right-hand side of Eq. (33).

After the Big Bang the density decreases according to the formula $\epsilon t_s^3/t^3$ until the first particle epoch arrives when the energy density has diminished to such an extent that it matches the energy density of physical (as opposed to bare or virtual) quarks and leptons. The size of strings in the string model is of the order of the Planck length $G^{1/2}$. Similarly, we expect the volume of finite elementary particles (which in our theory are spherical finite objects) to be $O(G^{3/2})$, with a corresponding energy density of $O(G^{-2})$. Our theory of isolated elementary particles is based on non-linear self-consistent solutions of the field equations of QFT. We intend to publish this work in the near future. However, for now we just use the hypothesis that the physical particles are characterized by the Planck length. The creation of particles out of the vacuum is likely to require a matching of the energy density of the vacuum universe and the particle energy density. This occurs at a time $t_c$ given by:

$$\epsilon\sqrt{^3g(t_c)} = \epsilon\frac{t_s^3}{t_c^3} = G^{-2}, \tag{57}$$

leading to:

$$t_c = \left(\frac{G}{\epsilon}\right)^{1/6}\left(\frac{3}{8\pi}\right)^{1/2} \approx 5\times 10^{-24}\text{sec} = (125\text{MeV})^{-1}\ . \tag{58}$$

In Eq. (58) we ignored the modification of the metric due to the presence of the created matter. From Eq.(25) we see that $h(x)$ will rapidly increase with the creation of particles, in particular as the volume $\hat{V}_s$ will decrease to compensate for the energy increase resulting from particle formation and from the fact that physical particles now consume space. Hence, this will extend the creation epoch beyond $t_c$, as we will still satisfy Eq.(57) for some time after $t_c$, if we replace $^3g$ by $^3\hat{g}$ and $t_c$ by $\hat{t}_c$, where $\hat{t}_c > t_c$. Since the large size of $h(x)$ invalidates perturbative calculations, a more extensive theoretical investigation is required to describe the later stages of this epoch in detail. Dimensional considerations indicate that the initial constraints on particle creation have a spatial nature, so that during this epoch available space must be divided homogeneously, thus, providing a possible explanation for the homogeneous nature of our universe. This initial epoch has both developmental and evolutionary aspects to it. The evolutionary label is best reserved for processes with a degree of randomness (where and when the particles are created). The quantum fluctuations responsible for these aspects are discussed later in this section. The developmental aspects cover the fact that the universe expands (develops) linearly with time and that the states into which it can develop are fixed by QFT and GR together.

Eq.(58) illustrates how the particle physics scale can arise from the two fundamental dimensionfull constants of Nature, $\epsilon$ and $G$, and gives credence to our expectation that a truly fundamental understanding of elementary particles requires consideration of GR. Since the equations of motion in QFT do not contain any fundamental dimensional constants, it is not unexpected that the particle physics scale in QFT only emerges when QFT is unified with GR. As part of our model of particle creation, we also suggest that a creation process in QFT mimicks the creation process of particles at the $t_c$ epoch after the Big Bang. Such a mechanism is required in our theory of isolated elementary particles in order to stabilize the solution. The distortion of space resulting from the formation of a physical particle of Planck size must counter the collapse of the dressed particle to a singular point. This is roughly opposite to the situation of a black hole, where the metric induces rather than prevents the collapse. The creation terminates at the time $t_c$ and thereby explains the typical creation time of elementary particles. The annihilation of a particle is the reverse process, again characterized by the same time $t_c$. Although many aspects of our particle theory are still under development, the possibility of being able to explain the nature of particles and to give an explanation of their masses and of their creation and annihilation properties is a very exciting prospect, indeed.

If these notions are confirmed after further development, we see a phenomenon in elementary particle physics that reminds us of a mechanism known in biology, namely that the development of a current entity recapitulates a (series of) historical process(es). In biology these processes are called ontogeny (the development of an organism) and phylogeny (ancestor-dependent relationships in a group) and the biogenetic law to which we refer states that ontogeny recapitulates phylogeny (Haeckel [29]). In the physics analogy the historical process would be the creation of particles from the vacuum at the appropriate epoch $t_c$, whereas the current process of creation repeats this process as part of the full process of particle creation. The epoch at $t_c$ is associated with an increase in entropy and is irreversible, whereas the current physical process of particle creation does not increase entropy and is reversible (annihilation is the reverse process). Of course, the analogy with biology is to a large extent symbolic, still apart from the fact that Haeckel's law itself has a very limited range of validity in biology and has been severely criticized [30]. Nonetheless it is gratifying that Nature finds ways of expressing similar mechanisms under very different circumstances. To emphasize the limitations of this principle, we note that the annihilation of particles does not have a corresponding epoch in the development of the universe, unless the universe were to die in a big crunch in which particles are converted back into vacuum energy. This would require the decrease of entropy in order to return to a state of zero entropy and would violate thermodynamic laws.

It should be noted that the derivation of the particle physics scale only is only valid for our particular vacuum metric, confirming again the unique role of the current vacuum solution. The result is also contingent on physical elementary particles being three dimensional (spherical) objects, and therefore cannot be derived in the common form of string theory. Within our picture the creation epoch starts much later than the Planck time, which is often considered to be the critical time period for events near the Big Bang. In this way we have avoided the difficult question of the unification of GR and QFT at the Planck scale, although this question returns in a more controlled form in the treatment of elementary particles of size $G^{1/2}$ and energy density $G^{-2}$. It should also be noted that the particle physics scale $(t_s/t_0) \times t_c$, rather than $t_c$, is measured at present. An accurate model of particle properties will therefore give information on the ratio $t_s/t_0$, and thus on $t_0$. Since $t_c$ is of the order of the (currently) measured particle scale, we conclude that $t_0$ is currently of the same order of magnitude as $t_s$. Hence, we are living in a time and age $t_0$ characterized by the typical cosmological time unit $t_s$.

Naturally this is a rather qualitative statement as there is a large difference between the hadronic and the leptonic scale, making the definition of the particle physics scale rather uncertain. If we go back to the time $t = t_c$ when the first particles were created, then the measured age of the universe would still be $t_s$. However, the measured particle properties would be characterized by an interaction time $(t_s/t_c) \times t_c = t_s$, which is comparable to the age of the universe. Hence, the changing factor $t_s/t_0$ ensures that the universe proceeds through different physical epochs.

Since the first creation process of physical particles takes place in a homogenous vacuum universe, we would expect the created particles to be distributed homogenously, disturbed only slightly by the quantum fluctuations and randomness of the quantum processes that created them. It is only through these random processes that we can break the initial perfect symmetry and form increasingly complex and diverse structures, allowing an increase in the value of the entropy. Such a creation of entropy is discussed elsewhere in the literature [26]. Initially the linear expansion of space will be halted - or even reversed - when the creation of particles increases the mass terms in the energy balance, an energy increase which has to be matched by a corresponding decrease in the total vacuum energy, i.e. a decrease in $\hat{V}_s$. However, after the initial creation of particles and anti-particles, we would expect an inflationary period, when most of the particle-anti-particle pairs annihilate. These processes destroy most of the initial mass energy and the induced matter-vacuum energy, leaving only a small residue of "particles" and converting some of the energy into radiation and its associated negative mixed radiation-vacuum energy. To compensate for this energy loss the universe would have to expand very rapidly in a short time (an inflationary phase). Clearly, this inflationary period has an origin and nature quite different from that considered in currently popular inflationary scenarios. A phase of extremely quick inflation does not seem to be required in the current theory to explain the uniformity of the temperature distribution in the universe, as the infinite horizon in our description allows particles to interact over much larger distances than in the standard picture.

Since the current universe contains only a relatively small percentage of matter, we would expect that after these violent processes have been essentially completed, the universe would return to a state in which the vacuum energy dominates and the expansion is dominated by the linear trend. However, the effective density will still be huge initially, because of the factor $t_s^3/t^3$, as we saw in (35). Hence, we expect that the usual hot Big Bang phase, which is responsible for primordial nucleosynthesis, can be derived in the usual way, although further study is required to confirm this in detail. There are other characteristic epochs in the evolution of the universe which can be characterized in terms of $G$ and $\epsilon$. For example the epoch that the vacuum energy equals the particle physics scale is characterized by:

$$\epsilon\sqrt{^3g(t_N)} = \epsilon\frac{t_s^3}{t_N^3} = t_c^{-4}, \tag{59}$$

leading to a time $t_N = \epsilon^{-7/18}G^{-5/18} \approx 8$ hrs.

Finally we discuss the observation of early galaxies. In FRW calculations one usually employs a $t^{2/3}$ expansion for the early universe. Using this type of time scale, Bouwens *et al* [31] conclude from certain Hubble observations that the first galaxies were formed about 900 Myr after the Big Bang. Similar conclusions were reached by a Japanese group [32], which found early galaxies dating from 750 Myr after the Big Bang. By demanding that these events take place with the same value of $z$ as they do in the analysis by these authors, we find that in our theory the formation of the early galaxies would take place 1.9 and 1.7 billion years respectively after the Big Bang. Although the creation and annihilation events in the early universe might modify these estimates slightly, the net result is that the early galaxies were formed much later than claimed by the authors above, reducing the mystery of the early formation of galaxies.

## 11. Summary and Concluding Remarks

We have solved the standard equations of general relativity for the vacuum with a "classical" vacuum energy density. We have shown that this leads to a Big Bang solution with an associated linear expansion of the universe, even after the introduction of matter and radiation. The contributions of matter and radiation to the total energy and the distortions of the metric are effectively constant under this linear expansion. Deviations from this basic behaviour, which is controlled by total energy conservation, can appear through creation and annihilation processes. This model can explain many crucial observations of the universe without the need to introduce new variants of the basic theory of general relativity or extensions beyond the Standard Model. In particular the cosmological constant problem and the horizon problem are absent in this approach. The evolution of this universe proceeds from a classical beginning with perfect spatial symmetry and zero entropy to a diverse and complex future thanks to quantum fluctuations. Although, various details of this picture still have to be worked out, the initial results are very promising.

The abandonment of the RW formalism necessitates a reassessment of various properties of the universe, such as its matter and radiation content. Improved supernovae data will impose strong constraints on the current model and on the nature and intensity of the decay processes in the universe (mainly radiative processes), as in our theory, the latter are seen as the cause of the current acceleration of the expansion of the universe. The explanation we suggest for the observation of "dark matter" should also be studied further, since it will be affected by higher order effects that are not considered in this paper.

## Acknowledgments

I thank Zaid Kimmie for the discussion of statistical aspects of the supernovae data and help in preparing the manuscript.

# Anthropomorphic Quantum Darwinism as an explanation for Classicality.

Thomas Durt[1]

*Abstract: According to Zurek, the emergence of a classical world from a quantum substrate could result from a long selection process that privileges the classical bases according to a principle of optimal information. We investigate the consequences of this principle in a simple case, when the system and the environment are two interacting scalar particles supposedly in a pure state. We show that then the classical regime corresponds to a situation for which the entanglement between the particles (the system and the environment) disappears. We describe in which circumstances this factorisability condition is fulfilled, in the case that the particles interact via position-dependent potentials, and also describe in appendix the tools necessary for understanding our results (entanglement, Bell inequalities and so on).*

## Introduction.

Presently, it is still an open question to know whether quantum mechanics is necessary in order to describe the way that our brain functions[2].

Nevertheless, quantum mechanics is astonishingly adequate if we want to describe the material world in which we live. It is therefore natural to assume that the way we think has something to do with quantum mechanics. After all, if our worldview faithfully reflects the external world, it ought to reflect also its internal properties at the deepest level! For this reason, it is really interesting and important to reconsider epistemological questions in the light of the most recent conceptual developments of the quantum theory. A key concept in these issues is the so-called quantum entanglement.

The term entanglement was first introduced by Schrödinger who described this as the characteristic trait of quantum mechanics, "the one that enforces its entire departure from classical lines of thought" [2]. Bell's inequalities [3] show that when two systems are prepared in an entangled state, the knowledge of the whole cannot be reduced to the knowledge of the parts, and that to some extent the systems lose their individuality. It is only when systems are not entangled that they behave as separable systems[3]. So, entanglement reintroduces holism and interdependence at a fundamental level[4] and raises the following question: is it legitimate to believe in the Cartesian paradigm (the description of the whole reduces to the description of its parts), when we know that the overwhelming majority of quantum systems are entangled?

[1]TONA Vrije Universiteit Brussel, Pleinlaan 2, B-1050 Brussels, Belgium. email: thomdurt@vub.ac.be

[2]It is even an open question to know whether the non-classical aspects of quantum mechanics play a fundamental role in biological processes at all. It is for instance an open question to know whether or not quantum coherence must be invoked in order to explain intra-cellular processes. Nothing illustrates better the present situation than this quote of Eisert and Wiseman[1]: "When you have excluded the trivial, whatever remains, however improbable, must be a good topic for a debate"...

In order to tackle similar questions, that are related to the so-called measurement problem [6], Zurek and coworkers developed in the framework of the decoherence approach [7] the idea that maybe, if the world looks[5] classical, this is because during the evolution, decoherence selected in the external (supposedly quantum) world the islands of stability that correspond to the minimal quantum (Shannon-von Neumann) entropy [8, 9].

In the present paper, we go a step further and make the hypothesis that these *classical islands* (environment induced or EIN superselected [10]) would correspond to the structures that our brain naturally recognizes and identifies, and this would explain why the way we think is classical.

In the first section we make precise in which aspects our approach coincides with and departs from the standard decoherence and Quantum Darwinist approaches and what is our motivation.

In the second section and in appendix, we explain the meaning of relevant concepts such as quantum entanglement, quantum bits, quantum non-locality and separability as well as Shannon-von Neumann entropy. We also present a theorem that establishes that entanglement is the corollary of interaction (section2.2) in the sense that when two systems interact, they get entangled in general. The classical situation for which no entanglement is generated during the interaction is thus exceptional.

In the third section we describe in more detail the environment induced (EIN) superselection rules approach and we apply it to the simple situation during which two quantum particles interact through a position-dependent potential, in the non-relativistic regime. We study then the classical islands that, according to the EIN selection rule, minimise the entropy of the reduced system, which means that they correspond to maximally deterministic (minimal uncertainty) states. They correspond to the classical, entanglement-free interaction regime.

[3]It can be shown that whenever two distant systems are in an entangled (pure) state, there exist well-chosen observables such that the associated correlations do not admit a local realist explanation, which is revealed by the violation of well-chosen Bell's inequalities [4]. In appendix (section 5.4) we treat an example in depth and explictly derive Bell's inequalities that are violated in that special case.

[4]Holism is a rather vague concept that possesses several definitions, often mutually exclusive [5]. Here we mean that the quantum theory is holistic in the sense there can be a relevant difference in the whole without a difference in the parts. We provide in section 5.3 an illustration of this property: the Bell states are different bipartite states for which the reduced local states (sections 5.5 and 5.6) are the same. In this approach, entanglement, non-locality and non-separability are manifestations of holism and Quantum Weirdness, to be opposed in our view to the classical, Cartesian non-holistic approach in which the knowledge of the whole reduces to the knowledge of the parts.

[5]The goal of the decoherence approach is to reconcile the first principles of the quantum theory, in particular the linearity of the quantum temporal evolution law with an objective description of the world. In the present context, when we write the world *looks* classical it means implicitly that we do not need an observer to let it look classical. As we explain in section 1, our approach is slightly different: we want to show that the world looks classical because our eyes are blind to Quantum Weirdness.

We show that the classical islands are in one to one correspondence with the three classical paradigms elaborated by physicists before quantum mechanics existed; these are the droplet or diluted particle model, the test-particle and the material point approximations (section 3.2).

The results presented in section 3 illustrate to which extent entanglement marks the departure from our classical preconceptions about the world, in agreement with Schrödinger's view [2] according to which entanglement is the characteristic trait of quantum mechanics that enforces its entire departure from classical lines of thought.

They can be considered as a plausibility argument in favor of our main thesis according to which we are blind to Quantum Weirdness as a consequence of a long process of natural selection.

# 1 About the measurement problem and Quantum Darwinism.

The measurement problem is related to the so-called objectification problem that in turn is intimately related to the Schrödinger cat paradox and, roughly, could be formulated as follows. Let us assume that we prepare two superpositions of macroscopically distinguishible quantum states (say a living cat state and a dead cat state). According to the quantum theory, whenever we perform a measurement that aims at revealing whether the cat is living or dead, one alternative is realized and the other one is discarded. Besides, if the state of the system obeys a unitary evolution law (like Schrödinger's equation) both alternatives survive and the system will remain in a superposition state. What is shown in the decoherence approach is that in good approximation the superposition becomes decoherent due to the interaction with the environment. What the decoherence approach doesn't prove is that one alternative is privileged[6]. In order to show this, some extra-assumptions are required, for instance that the position is privileged (like in Bohm-de Broglie's interpretation) or that many worlds are present (here the world with a living cat and the world with a dead cat). In the Quantum Darwinist approach (developed in section 3.1), it is assumed that somehow the objectification process will take place, and that it will occur in a privileged basis, the basis that diagonalizes the reduced density matrix of the system constituted by the quantum system under interest and the measuring apparatus. For instance if the murder of the cat is triggered by the measurement of a quantum two level system, the system under interest is this two level system, and the cat can be considered as a macroscopic amplifier or measuring apparatus.

[6]Roland Omnès for instance who is an active propagandist of the consistent history approach in which decoherence plays a central role, introduced in one of his books [11] the Rule 5: *Physical reality is unique. It evolves in time in such a way that, when actual facts arise from identical antecedents, they occur randomly and their probabilities are those given by the theory.* In other words, Omnès must postulate that *Reality exists*; this is because he realized that the decoherence approach did not solve the measurement problem; in particular it could not solve the objectification puzzle so to say: if all alternative quantum possibilities, in a Schrödinger cat like experiment, do survive (with or without coherence that is not the point), then how is it possible that one of them is privileged and realized (actualized) in particular? In his review of Omnès's book, W. Faris [12] wrote about Rule 5...*This statement by itself does not give a clear picture of the mathematical formulation of actualization. The intent may be that the notion of fact is external to the theory, so that the rule of actualization is merely a license to use consistent logic to reason from present brute experience. This is supported by the assertion: The existence of actual facts can be added to the theory from outside as a supplementary condition issued from empirical observation. A dead cockroach is a fact; there is no more to it. This is a long way from the ambitious goal of basing everything on Hilbert space...*

Not every physicists is convinced (this is an euphemism) by Zurek's arguments [13], but Quantum Darwinism offers stimulating analogies with the biological world and with the Darwinian picture of the process of evolution. It are these features of Quantum Darwinism that stimulated the present work (more precisely, we were strongly stimulated by the postulated existence of a selection process that would ultimately lead to the selection of a preferred basis).

Another analogy between biological Darwinism and Quantum Darwinism is that in the latter the choice of the preferred basis is supposed to obey a principle of optimisation (similarly, in Darwin's approach, only the fittest species survive). At the same time many copies/duplicates of the preferred basis are supposed to be disseminated throughout the environment which is reminiscent of the reproduction mechanism of the surviving species.

According to us, Quantum Darwinism contains, like the many world or the Bohm-de Broglie interpretation a hidden extra-principle that goes beyond the standard principles of quantum mechanics, which is the Environment Induced Selection rule (see also section 3.1). This rule tells us that the preferred basis is related to islands of classicality (section 3.1) that minimise the Shannon-von Neumann entropy (in other words they minimise uncertainty, section 5.6) of the reduced state of the system constituted by the quantum system under interest and the measuring apparatus. We do not believe that this argument is very conclusive for the following reason:

-the Shannon-von Neumann entropy is related to the distribution of the probability of occurence of events in the eigen-basis of the reduced state (density matrix),

-but it does not make sense to talk about objective events and of their probabilitites as far as the measurement problem is not solved[7],

-so that, in our view, in the Quantum Darwinist approach, one takes for granted from the beginning, in a subtle and implicit way, what one wants to prove at the end (the emergence of objective facts).

In other words, we consider that in last resort the concept of entropy implicitly refers to an observer although the goal of the Quantum Darwinist approach is precisely to get rid of the dichotomy observed-observer, so that in our view the quantum measurement paradox is solved only at the price of introducing a logical loophole (circularity) somewhere in the reasoning.

It is not our goal to pursue in this direction: we have actually no clear idea about what is the right interpretation of quantum mechanics (if at least there exists such a thing, see e.g. the reference [13] where the consistency of the decoherence approach is scrutinized and criticized in depth...). Nevertheless we shall exploit the idea according to which the preferred basis obeys an optimality principle: as has been succesfully shown by Zurek, classical islands are related to preferred bases (section 3.1) in which the gathered information is reproducible, stable, maximally deterministic and insensitive to the disturbances of the environment.

[7] All the interpretational problems of the quantum theory arise because the quantum theory is a probabilistic theory.

Our next step is to seriously consider the hypothesis that humans (and possibly all sufficiently evolved living organisms, like cats for instance) select the preferred basis according to the Environment Induced Selection rule (section 3.1) because it is the most advantageous strategy in terms of the amount of stable and useful information that can be extracted from correlations, (the last idea is present in the Quantum Darwinist approach, the first one is a personal hypothesis).

At this point one should be very careful, we do not pretend in any way that the consciousness of the observer plays a role in the collapse process or anything like that[8]. No, our goal is rather to explain why our logic is classical although the logic of the quantum world is obviously non-classical. In other words we address the question to know why our representation of the world is classical although we live in a quantum world. Considered from this perspective, our answer is that maybe throughout eons of evolution, our ancestors (and their cousins, the ancestors of monkeys, cats, fishes and so on) became gradually blind to quantum interferences and quantum entanglement, because their manifestations (the associated correlations) were not useful from an informational point of view.

Our approach does not aim at solving the measurement problem (one still must postulate the existence of correlations, thus of probabilities between physical "objective" events, the existence of which is at the core of the measurement problem). Nevertheless, the novelty of our approach is to postulate that our sensorial system privileges data that are gathered in the preferred basis. So, what we retain from the Quantum Darwinist approach is that a preferred basis "exists", that it is the basis in which we measure the external world, and that this basis obeys an optimality principle that results from a (biological in this case) selection mechanism: the fittest is the best-informed, a reality that prevails in today's jungle, maybe more than ever.

In the section 3, we show that in a sense Quantum Weirdness is the rule, because entanglement, one of the most striking illustrations of Quantum Weirdness, can be considered to be the corollary of interaction. In simplified terms we can say that there is in general no interaction without creating entanglement.

Now classical islands precisely correspond to the exceptional situations for which entanglement is negligibly small. This brings us to the section 4 where we can find the main novel contribution of the paper. In that section we aim at establishing the identity between classical islands and our cognitive representation of what elementary particles are. More precisely we study the regimes in which two particles interact (by a potential that depends solely on their distance) *without getting entangled*. These *"entangled-free"* regimes are in one to one correspondence with the classical islands. We show that they also correspond to the models introduced by classical physicists before the advent of quantum physics in order to represent elementary particles such as the electron. This observation is more a plausibility argument than a definitive proof. But it is our main argument in the favor of our personal interpretation of Quantum Darwinism.

[8]In agreement with Zurek, we do not believe that it is necessary to invoke consciousness in order to solve the measurement problem so to say to explain how objective, actual facts emerge from the quantum substrate. As we said before, we do not claim to solve the measurement problem in our approach. Our thesis is that we select preferentially classical correlations and that we are blind to Quantum Weirdness, which does never ever mean that consciousness is a quantum phenomenon or that we need consciousness in order to collapse the wave function. The correlations that we are talking about are correlations between events, clicks in detectors, reactions of sensors. Before correlations are treated by our nervous sytem, it is very likely that the decoherence process is achieved and that the effective collapse of the wave function already took place.

Metaphorically, our interpretation is that our ears are deaf to the quantum music because it would sound like a cacophony. Similarly we are blind to weird quantum optical effects because they would be like a very bad quality mirage to us (no stability, no reproducibility, no useful correlations to exploit). After all, what we see is for instance not really what the detectors in our eyes perceive, there is still a huge work of data processing that occurs in the brain (in the case of vision, the corresponding hardware is not a negligible part of the brain!). We defend in this paper the thesis that we are blind to quantum effects not only because of the inadequacy of our receptors but mostly because of the treatment that we perform with the data gathered by them.

One could argue that at the level of our nervous system, quantum effects are so tiny that we should not be able perceive them. Our point of view is that it is not because they are tiny that we should not perceive them but rather that we do not perceive them because they are not useful[9].

# 2 Entanglement and Interaction.

## 2.1 The concept of entanglement.

In order to avoid unnecessary technicalities, and with the aim of addressing the paper to a large, interdisciplinary, audience, we shall restrict ourselves in what follows to the simplest case: two systems $A$ and $B$ are prepared in a pure quantum state $\mathbf{\Psi}_{AB}$. Then the state of the sytem is said to be factorisable (section 5.3) at time $t$ whenever it is the (tensor[10]) product of pure quantum states of the subsystems $A$ and $B$, which means that the following constraint is satisfied:

[9] Particle physicists attempted to explain why organic molecules are chirally oriented in terms of the (tiny) energetic advantage that differentiates them from their mirror-molecule. In last resort this tiny difference of energy would be explained in terms of parity-violation by weak interactions [14]. We argue that the same effect, of amplification of tiny differences during millions of years, could explain the emergence of classical perceptions. If there was an informational advantage in exploiting non-classical quantum features like entanglement, it is likely that evolved organisms would be able to exploit this advantage. From this point of view, very primitive organisms, like bacteria would maybe be closer to the quantum world than we are, and it seems indeed, although no conclusive proof of this idea exists yet, that certain bacteria optimize their mechanism of harvesting light by exploiting the rich possibilities offered by the quantum superposition principle [15]. Considered so, the receptors of those bacteria would exhibit quantum coherence, a fascinating hypothesis.

[10] The tensor product is represented by the symbol $\otimes$. It is the right mathematical operation that is needed when different Hilbert spaces are brought together, for instance the spaces associated to different particles. One can find a standard definition of Hilbert spaces and tensor products on the wikipedia website ([16],[17]).

$\mathbf{\Psi}_{AB}(t) = \psi_A(t) \otimes \psi_B(t)$. Otherwise the system is said to be entangled[11]. As we show in appendix (section 5.3), when they are in a non-entangled, factorisable, state, the two sub-systems $A$ and $B$ are statistically independent in the sense that the average values of any physical quantity associated to the subsystem $A(B)$ is the same that would be obtained if the system $A(B)$ was prepared in the state $\psi_A(t)(\psi_B(t))$. Moreover, there are no correlations at all between the subsystems and they can be considered to be independent.

The four so-called Bell two-qubit[12] states [19] are for instance entangled because as shown in section (5.3) they do not factorize into a product of local qubit states.

They are defined as follows [13] :

$|B_0^0\rangle = \frac{1}{\sqrt{2}}(|+\rangle_Z^A) \otimes |+\rangle_Z^B + |-\rangle_Z^A) \otimes |-\rangle_Z^B)$

$|B_1^0\rangle = \frac{1}{\sqrt{2}}(|-\rangle_Z^A) \otimes |+\rangle_Z^B + |+\rangle_Z^A) \otimes |-\rangle_Z^B)$

$|B_0^1\rangle = \frac{1}{\sqrt{2}}(|+\rangle_Z^A) \otimes |+\rangle_Z^B - |-\rangle_Z^A) \otimes |-\rangle_Z^B)$

$|B_1^1\rangle = \frac{1}{\sqrt{2}}(|+\rangle_Z^A) \otimes |-\rangle_Z^B - |-\rangle_Z^A) \otimes |+\rangle_Z^B)$.

Because the Bell states do not factorize, the local measurements in the distant regions $A$ and $B$ are not statistically independent, that is to say, the two qubits exhibit correlations. Moreover those correlations are essentially non-classical in the sense that they violate Bell's inequalities which is impossible for correlations derived from a local realistic model as we also show in appendix (5.4).

## 2.2 Entanglement and Interaction.

Entanglement between $A$ and $B$ is likely to occur whenever they interact [20] as shows the following property that we reproduce here without proof [21].

Let us consider two interacting quantum systems $A$ and $B$. We assume that the numbers of levels of the $A$ $(B)$ system is finite and equal to $d_A$ $(d_b)$[14], that the wave-function of the full system is a pure state and obeys the Schrödinger equation:

$$i\hbar\partial_t \mathbf{\Psi}_{AB}(t) = H_{AB}(t)\mathbf{\Psi}_{AB}(t) \quad (1)$$

where $H_{AB}(t)$, the Hamiltonian of the system, is a self-adjoint operator, that we assume to be sufficiently regular in time. Then the following property is valid [21]:

*All the product states remain product states during the interaction if and only if the full Hamiltonian can be factorised as follows:*

$$H_{AB}(t) = H_A(t) \otimes I_B + I_A \otimes H_B(t) \quad (2)$$

[11]The characterization of entanglement can be generalized when the full state is not pure, or in the case of multipartite systems that are not bipartite but it is more complicate in this case [18].

[12]A qubit is a 2-level quantum system. Examples of physical realisations of qubits are given in appendix (section 5.1).

[13]The state $|B_1^1\rangle$ is also known as the singlet spin state.

[14]The Hilbert spaces associated to these systems are thus finite dimensional (of dimensions $d_A$ and $d_B$ respectively).

*where $H_i$ acts on the ith system only while $I_j$ is the identity operator on the jth system $(i, j = A, B)$.*

In simple words: there is no interaction without entanglement, which establishes that entanglement is very likely to occur; for instance, when we see light coming from a distant star, it is nearly certainly entangled with the atoms that it encountered underway. Entanglement can also be shown to be present in solid states, ferro-magnets and so on, and to play a very fundamental role in the macroscopic world, for instance during phase transitions [22].

# 3 The decohererence program and the classical limit.

## 3.1 Environment induced superselection rules and classical islands.

The EIN superselection approach was introduced by Zurek in the framework of the decoherence approach [9, 7, 8, 10]. He postulated that the preferred basis obeys a principle of optimality that can be formulated in terms of the Shannon-von Neumann entropy[15]. More precisely, the EIN selection rule predicts that, once a system, its environment and their interaction are specified, the classical islands associated to the system are the states that diagonalize the reduced density matrix of the system[16]. When the full state of the system-environment is pure, these states belong to the Schmidt basis (section 5.5) that diagonalizes the limit asymptotically reached in time by the system-apparatus bipartite state. When the environment can be considered as an apparatus, the classical islands define the so-called pointer basis[17], also called sometimes the preferred basis.

Roughly speaking, the EIN selection principle expresses that, during the evolution, the classical islands that belong to the prefered basis or pointer basis (the one that minimizes the Shannon-von Neumann entropy [9, 10] of the reduced system) are selected preferentially to any other basis. In the quantum Darwinist approach, the emergence of a classical world that obeys EINselection rules can be explained following two ways:

A) these rules correspond to maximal (Shannon-von Neumann) "certainty" or useful information[18]; this fits to our own acceptance and interpretation of Zurek's Darwinism as we explained in the section 1: it is well plausible that our brain selects the features of the natural world that are maximally deterministic and minimally uncertain;

B) Zurek also invokes an argument of structural stability: superposition of states that would belong to such islands would be destroyed very quickly by the interaction with the environment which radiates irremediably the coherence (which is equal in virtue of our definitions to the certainty, so to say to 1-the Shannon-von Neumann entropy of the reduced density matrix of the system, see section 5.6)) into the environment [7]. This process is called the decoherence process and is very effective.

[15]This quantity is defined in appendix (section 5.6). Entropy is a measure of the uncertainty of a probabilistic distribution. It is worth noting that very often entropy is also considered to be a measure of information, but that in our approach we consider that the useful information increases when entropy decreases. Indeed let us measure the degree of "certainty" or determinism assigned to a random variable by the positive quantity $C$, $C$= 1 - entropy; in other words certainty+uncertainty =1. We consider that when a distribution of probability is very uncertain (flat) it does not contain much useful information. On the contrary when a probability distribution is peaked it contains a high level of useful information. So when we use the word information in the paper we implicitly mean "certain" information or "information useful for deterministic evolution" (=1-entropy), a quantity that increases when entropy decreases, contrary to most commonly used conventions.

[16]In appendix we explain that when the measure of entropy is the Shannon-von Neumann entropy of a density matrix $\rho$, $C$ is also a measure of the purity or coherence of $\rho$. Actually we have shown in ref.[21] that when two coupled oscillators interact through the interaction Hamiltonian $H_{AB} = a^\dagger b + ab^\dagger$ (written in function of the creation and destruction phonon modes of the oscillators $A$ and $B$), states that remain factorisable throughout time are de facto eigenstates of $a$ ($b$), which means that they are so-called coherent oscillator states for which it can be shown that Heisenberg uncertainties are minimal.

[17] *"Pointer states can be defined as the ones which become minimally entangled with the environment in the course of the evolution"* (which means here temporal evolution described by Schrödinger's equation...), quoted from Ref.[23].

### 3.2 A toy model for Quantum Darwinism: two interacting particles.

We applied the Quantum Darwinist approach to a very simple situation during which the system $A$ and the environment $B$ are two distinguishable particles and are described by a (pure) scalar wave function that obeys the non-relativistic Schrödinger equation. We also assumed that their interaction potential $V_{AB}$ is an action a distance that is time-independent and only depends on the distance between the particles, so that it is invariant under spatial translations (a Coulombian interaction for instance). This is a standard text-book situation that was deeply studied, for instance in the framework of scattering theory. The systems $A$ and $B$ fulfill thus (in the non-relativistic regime) the following Schrödinger equation:

$$i\hbar\partial_t\Psi(\mathbf{r}_A,\mathbf{r}_B,t) = -(\frac{\hbar^2}{2m_A}\Delta_A + \frac{\hbar^2}{2m_B}\Delta_B)\Psi(\mathbf{r}_A,\mathbf{r}_B,t)$$
$$+V_{AB}(\mathbf{r}_A-\mathbf{r}_B)\Psi(\mathbf{r}_A,\mathbf{r}_B,t) \quad (3)$$

where $\Delta_{A(B)}$ is the Laplacian operator in the $A(B)$ coordinates. Let us now consider that the system $A$ is the quantum system that interests us, and that the other system is its environment. Actually, the argument is symmetrical as we shall see so that this choice is a mere convention. In order to identify the classical islands in this case, we must identify the states that exhibit maximal coherence or minimal Shannon-von Neumann entropy. We assume here that the full state is pure, which constitutes an oversimplification, because usually interaction with an environment destroys coherence. Nevertheless, as we shall show, one can get interesting insights even in this oversimplified situation.

[18]The Shannon-von Neumann entropy of a reduced maximally entangled pure bipartite state, for instance of a Bell state is maximal (equal to 1) and the corresponding certainty (or useful information) minimal (0) while for a factorisable state the Shannon-von Neumann entropy of the reduced state is 0 and the certainty is equal to 1 (section 5.6). As a consequence, factorisable states minimize the Shannon-von Neumann entropy of the reduced states. They correspond thus to classical islands, in the case that they are stabilized by the interaction with the environment.

Without entering into technical details that are presented in appendix (section 5.6), all we need to know at this level is that two systems in a pure state minimize the Shannon-von Neumann entropy when their state is factorisable or non-entangled.

Then, the classical islands correspond to the states that initially and during their interaction as well, remain factorisable (more precisely in a pure factorisable state). This constraint can be shown [21] to correspond to what is somewhat called in the litterature the mean field or effective field approximation, or Hartree approximation [24, 25]. In this regime, particles behave as if they were discernable, and constituted of a dilute, continuous medium distributed in space according to the quantum distribution $\left|\psi_{A(B)}(r_{A(B)},t)\right|^2$. Then, everything happens as if each particle ($A(B)$) "felt" the influence of the other particle as if it was diluted with a probability distribution equal to the quantum value $\left|\Psi(\mathbf{r}_{B(A)})\right|^2$. It corresponds also to the concept of droplet or diluted particle[19]
There are two interesting special cases:

i) When the potential only depends on the relative position $\mathbf{r}_{rel} = \mathbf{r}_A - \mathbf{r}_B$ $m_A << m_B$, the initial state is factorisable and the $B$ particle is initially at rest and well localized, it can be shown that the state remains factorisable in time and occupies thus a classical island. This corresponds to what is called the test-particle regime (no feedback of $A$ onto $B$). For instance this is a good approximation of what happens in the hydrogen atom, where the electron is so light that it can be considered as a test particle.

ii) Another situation that is of physical interest is the situation of mutual scattering of two well localized wave packets when we can neglect the quantum extension of the interacting particles. This will occur when the interaction potential $V_{AB}$ is smooth enough and the particles $A$ and $B$ are described by wave packets the extension of which is small in comparison to the typical lenght of variation of the potential. It is well known that in this regime, when the de Broglie wave lenghts of the wave packets are small enough, it is consistent to approximate quantum wave mechanics by its geometrical limit, which is classical mechanics. For instance the quantum differential cross sections converge in the limit of small wave-lenghts to the corresponding classical cross sections. Ehrenfest's theorem also predicts that when we can neglect the quantum fluctuations, which is the case here, the average motions are nearly classical and provide a good approximation to the behaviour of the full wave-packet so that we can consider it to be a material point. Actually, in this regime, we can in good approximation replace the interaction potential by the first order term of its Taylor development around the centers of the wave-packets associated to the particles $A$ and $B$ so that the evolution equation is in good approximation separable into the coordinates $\mathbf{r}_A, \mathbf{r}_B$ [21] and we have that, when $\Psi(\mathbf{r}_A, \mathbf{r}_B t = 0) = \psi_A(\mathbf{r}_A, t = 0)) \otimes \psi_B(\mathbf{r}_B, t = 0) = \psi_A(\mathbf{r}_A, t = 0)) \cdot \psi_B(\mathbf{r}_B, t = 0)$, then, at time $t$, $\Psi(\mathbf{r}_A, \mathbf{r}_B, t) \approx \psi_A(\mathbf{r}_A, t) \cdot \psi_B(\mathbf{r}_B, t)$ We shall discuss in the conclusion the relevance of this result.

[19]Actually, the diluted particle model corresponds to Schrödinger's own interpretation of the modulus square of the wave function, before Born's probabilistic interpretation was adopted by quantum physicists. The droplet picture is reminiscent of pre-quantum models of the electron that were developed by classical physicists such as Poincaré, Abraham, Laue, Langevin and others at the beginning of the 20th century. In this approach $|\psi|^2$ represents the charge density and as a consequence of Maxwell's laws, each particle *"feels"* the Coulomb potential averaged on the distribution of the other particle.

## 4 Conclusions and discussion.

The Quantum Darwinist approach sheds a new light on the emergence of classical logics and of our classical preconceptions about the world. The distinction between internal and external world, the Cartesian prejudice according to which the whole can be reduced to the sum of its parts and the appearance of preferred representation bases such as the position is seen here as the result of a very long evolution and would correspond to the most useful way of extracting stable and useful information from the quantum correlations.

We conjectured in the present paper that our difficulties and resistances for conceiving "entanglement" are due to the fact that millions of years of evolution modelled our vision of the world, leading us to become blind to aspects of it that are not advantageous from the point of view of the acquisition of useful information.

We showed that in a simplified situation (two particles that "feel" each other via an interaction potential), the EIN-selected classical islands are regions of the Hilbert space where the mean or effective field approximation (or Hartree approximation in the static case) is valid. In this regime, the interaction factorises into the sum of two effective potentials that act separately on both particles, and express the average influence due to the presence of the other particle.

In that regime, it also makes sense to consider the particles, in accordance with classical logics, not as a whole but as separate objects.

Our analysis confirms that our approach is well-founded in an undirect manner; indeed we show that the regime in which two particles interact without getting entangled possesses two extreme cases: the point particle regime (that corresponds to the classical material point description of matter) and the diluted matter approach (that corresponds to fluido-dynamics). The test-particle regime, where the heavy particle is treated like a material point, and the light particle as a diluted distribution, is intermediate between these two extreme cases[20]. These ways of conceiving and describing matter are so deeply imbedded in our every-day representation of the physical world that it is nearly impossible to find an alternative representation for particles and atoms in our mental repertory. This explains according to us why it took such a long time for quantum physicists to realize the implications of the EPR paradox (30 years) and of the concept of entanglement. Even today, a well-trained quantum physicist can merely hope, at best, to reach a mathematical intuition of the phenomenon and entanglement remains a very counter-intuitive property.

[20]There are two interesting limits that are special cases of the Hartree regime (the test-particle and material points limit). The Hartree regime corresponds to the droplet model; the test-particle and material points limits correspond to the test-particle concept and to classical mechanics. The Hartree regime is the most general regime in which entanglement is negligible. The limit cases are obtained by neglecting the extension of one droplet, the one associated to the massive particle (the other particle appears then to behave as a test-particle) and of both particles (classical limit).

It is interesting to note that somewhat similar conclusions could be drawned from the study of wave propagation, which is intimately related to propagation of information at the classical level, in other words of communication, another aspect of information[21].

A very interesting study was indeed performed at the beginning of the 20th century concerning the concept of dimension (see ref.[27] and references therein): in a space-time of 1+1 or 3+1 dimensions wave propagation ruled by d'Alembert's equation obeys Huygens principle[22], that can be translated into informational terms [28]: the state of the image reproduces accurately the state of the source (after a time delay proportional to the distance between image and source). It is this property that allows us to obtain a fidel representation of the external world by using sensitive receptors such as our ears and our eyes. Also here one could invert the reasoning and assume that maybe the conventional 3+1 representation of space-time was privileged because it is informationally advantageous to do so. It could be that other physical phenomena that are characterised by other dimensions coexist but that we are blind to them simply because they are informationally deprived of interest and of sense[23].

Let us for instance excite a 2-dimensional vibrating membrane such as a drum skin by hitting at its centre. One can show that the signal at the edge at a given time is a convolution of the signal that was imposed at the centre of the skin in the past. The difference with what occurs in 1 and 3 dimensions is that the time-interval on which the convolution is taken has a non-negligible extension. Therefore correlations between the centre of the resonating membrane and the extremities are diluted and get lost, in close analogy to what happens during the decoherence process.

It is extremely difficult for us to imagine the properties of a 4-dimensional space, which can be seen as a plausibility argument in the favor of a selection by our brain and sensors of a dimension (3) that optimizes the amount of useful information (in this case that optimizes efficient communication). A 2-dimensional space, a plane, is easy to visualize because, in our approach, it can be seen as a projection of the 3-dimensional space that we are supposed to live in.

Another promising direction of research was suggested by Nicolas Lori during the refereeing process of the paper. It concerns the possibility that certain ancient civilisations developed concepts similar to entanglement and inter-connectedness at an higher level than ours. This kind of research is outside of the scope of our paper, but it is worth noting that this observation could be brought in connection with the last part of footnote 9.

[21] Claude Shannon wrote hereabout *"The fundamental problem of communication is that of reproducing at one point either exactly or approximately a message selected at another point"* in his famous paper A mathematical theory of communication [26].

[22] Actually this is so in space-time of dimension $d+1$, where $d$ is an odd and positive integer.

[23] We are conscious that unfortunately this type of reasoning is not deprived of some degree of circularity, what is illustrated by the sentence *The world has 3 dimensions because we listen to music.* Therefore the best arguments that we can produce in their favor are plausibility arguments. This was precisely the scope of our paper (section 3), in the framework of Quantum Darwinism.

Before ending the paper it is good to recall what the paper is about or rather what it is not about. We do not pretend to solve the measurement problem or the objectification problem. We do not pretend to settle definitively and unambiguously the question about where the collapse process (effective or not) would take place (in the brain or at the level of our physical sensors, or even before that in the external world, during a decoherence process). We claim that as the result of a natural selection process that privilegges the best-informed we became gradually blind to entanglement, but we are not categoric about where the blindness occurs: it could occur at the level of our sensors, or during the data treatment that occurs in the brain or at both levels simultaneously...The history of Quantum Mechanics has shown that one can survive without answering to all fundamental questions and paradoxes generated by the theory. In our view mystery and knowledge are to some extent complementary and undissociable.

### Acknowledgment

T.D. acknowledges support from the ICT Impulse Program of the Brussels Capital Region (Project Cryptasc), the IUAP programme of the Belgian government, the grant V-18, and the Solvay Institutes for Physics and Chemistry. Thanks to Johan Van de Putte (Physics-VUB) for realizing the picture reproduced in figure 1. Thanks to the referees (Nicolas Lori, Alex Blin and anonymous) for useful comments and criticisms.

# 5 Appendix: Entanglement and non-local correlations.

## 5.1 Qubits.

A qubit consists of a two-level quantum system [29]. This is the simplest conceivable quantum system, because a one-level system would always remain in the same quantum state, and it would be eternally static, which does not present any interest from a physical point of view since physical theories are mainly focused on transformations. The state of a two-level quantum system is described, in the case of pure states, by a ray of a two-dimensional Hilbert space. There exists several ways to realize such systems, for instance, the qubit could be assigned to degrees of freedom such as light polarization of a given electro-magnetic mode, electronic, nucleic or atomic spin 1/2, energy of an orbital electron experimentally confined to a pair of energy levels, and so on. In what follows, we shall most

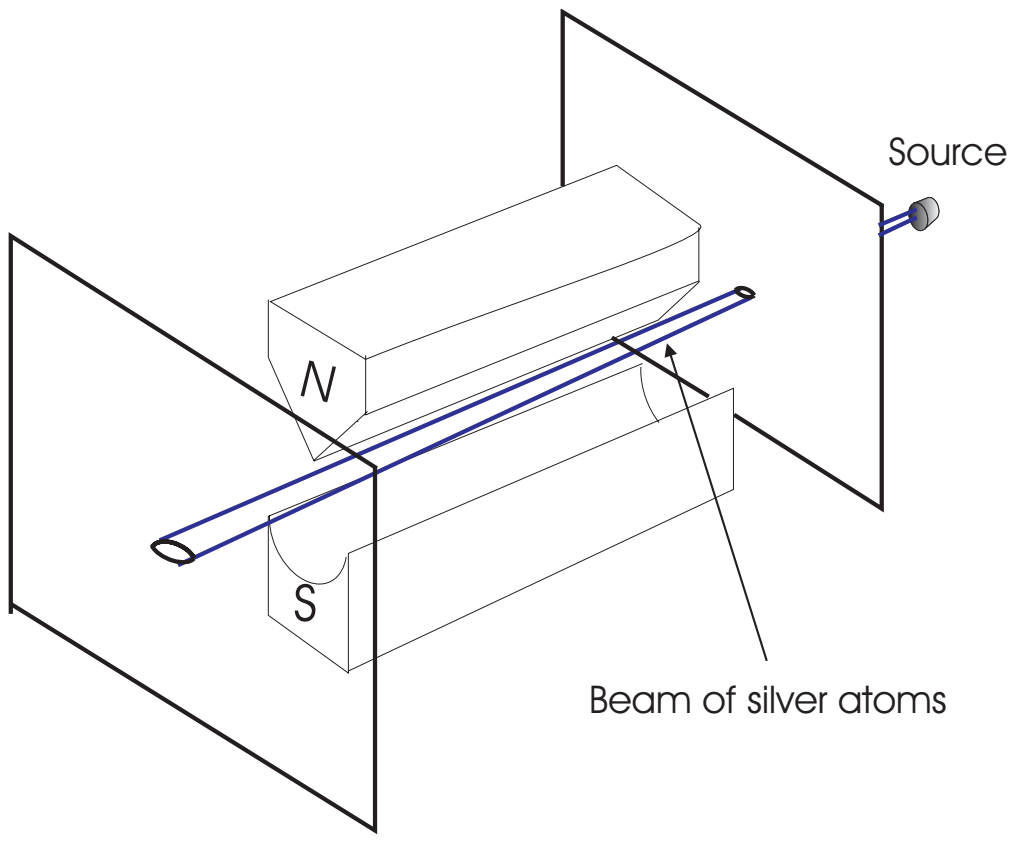

Figure 1: Stern-Gerlach spin measurement device.

often assume that the qubit system of interest is a spin-1/2 particle. Let us then denote $|+\rangle_Z$ and $|-\rangle_Z$ the spin up and down states relatively to a conventional direction of reference $Z$. An arbitrary qubit state can always be expressed as a superposition of the basis states of the form $\alpha|+\rangle_Z + \beta|-\rangle_Z$ where $\alpha$ and $\beta$ are two normalized complex amplitudes: $|\alpha|^2 + |\beta|^2 = 1$. In analogy with classical logic, we are free to associate with each of the basis states a conventional binary value yes-no or 0-1, for instance according to the assignment

$|+\rangle_Z \leftrightarrow 0 \leftrightarrow yes,$

$|-\rangle_Z \leftrightarrow 1 \leftrightarrow no.$

Although in classical logic the value of a classical bit is either 0 or 1, a quantum bit or qubit can in general be prepared in a superposition state of the form $\alpha|0\rangle + \beta|1\rangle$ which offers more formal flexibility than in the classical case. Of course, during a measurement process, the outcomes are dichotomic. For instance, if we measure thanks to a Stern-Gerlach apparatus the projection of the spin along the $Z$ direction, the probabilities of the two possible outcomes spin up and down are respectively equal to $|\alpha|^2$ and $|\beta|^2$. Despite of the fact that the distribution of the measurement outcomes is dichotomic, the evolution of the qubits, between the initial preparation and the final measurement, obeys the superposition principle.

## 5.2 Two-qubit systems.

Let us now consider two spin 1/2 particles $A$ and $B$ that are localized in far away regions of space. Let us measure with Stern-Gerlach devices their spin projection along $Z$, we get four possible outcomes:

$\text{up}_A$-$\text{up}_B$,

$\text{up}_A$-$\text{down}_B$,

$\text{down}_A$-$\text{up}_B$

and down$_A$-down$_B$.

These outcomes correspond to the states

$|+\rangle_Z^A \otimes |+\rangle_Z^B$,

$|+\rangle_Z^A \otimes |-\rangle_Z^B$,

$|-\rangle_Z^A \otimes |+\rangle_Z^B$

and $|-\rangle_Z^A \otimes |-\rangle_Z^B$.

The most general two-qubit state is superposition of those 4 states:

$|\Psi\rangle = \alpha|+\rangle_Z^A \otimes |+\rangle_Z^B + \beta|+\rangle_Z^A \otimes |-\rangle_Z^B + \gamma|-\rangle_Z^A \otimes |+\rangle_Z^B + \delta|-\rangle_Z^A \otimes |-\rangle_Z^B$.

## 5.3 Factorisable versus entangled states.

A state that can be written as follows:

$|\Psi\rangle = (\alpha_A|+\rangle_Z^A + \beta_A|-\rangle_Z^A) \otimes (\alpha_B|+\rangle_Z^B + \beta_B|-\rangle_Z^B)$

is said to be factorisable. For such states, the outcomes of local measurements in the A and B region are independent. Indeed, local observables are of the type $O^A \otimes Id.^B$ $(Id.^A \otimes O^B)$ so that

$\langle i|_Z^A \otimes \langle j|_Z^B O^A.O^B|i\rangle_Z^A \otimes |j\rangle_Z^B = \langle i|_Z^A \otimes \langle j|_Z^B O^A \otimes O^B|i\rangle_Z^A \otimes |j\rangle_Z^B = \langle i|_Z^A O^A|i\rangle_Z^A \otimes \langle j|_Z^B O^B|j\rangle_Z^B$, which means that outcomes of local measurements are statistically independent.

**By definition: non-factorisable states are said to be entangled.**

The so-called Bell states [19] are massively used in the machinery of Quantum Information [29], they provide a generic example of maximally entangled states. They are in 1-1 correspondence with the well-known Pauli spin operators:

$\sigma_0 = |+\rangle_Z\langle+|_Z + |-\rangle_Z\langle-|_Z \leftrightarrow |B_0^0\rangle = \frac{1}{\sqrt{2}}(|+\rangle_Z^A \otimes |+\rangle_Z^B + |-\rangle_Z^A \otimes |-\rangle_Z^B)$

$\sigma_x = |+\rangle_Z\langle-|_Z + |-\rangle_Z\langle+|_Z \leftrightarrow |B_1^0\rangle = \frac{1}{\sqrt{2}}(|+\rangle_Z^A \otimes |-\rangle_Z^B + |-\rangle_Z^A \otimes |+\rangle_Z^B)$

$\sigma_y = i|+\rangle_Z\langle-|_Z - i|-\rangle_Z\langle+|_Z \leftrightarrow |B_0^1\rangle = \frac{1}{\sqrt{2}}(|+\rangle_Z^A \otimes |-\rangle_Z^B - |-\rangle_Z^A \otimes |+\rangle_Z^B)$

$\sigma_z = |+\rangle_Z\langle+|_Z - |-\rangle_Z\langle-|_Z \leftrightarrow |B_0^1\rangle = \frac{1}{\sqrt{2}}(|+\rangle_Z^A \otimes |+\rangle_Z^B - |-\rangle_Z^A \otimes |-\rangle_Z^B)$

Bell states are not factorisable; for instance if $|B_0^0\rangle$ would factorize then

$\alpha^A.\alpha^B = \beta^A.\beta^B = \sqrt{\frac{1}{2}}$ and $\alpha^A.\beta^B = \beta^A.\alpha^B = 0$;

Obviously such a system of equations has no solution because it implies that

$\alpha^A.\alpha^B.\beta^A.\beta^B = \sqrt{\frac{1}{2}}.\sqrt{\frac{1}{2}} = \frac{1}{2}$ and $\alpha^A.\beta^B.\beta^A.\alpha^B = 0.0 = 0$ so that finally $1/2 = 0$, a logical contradiction, which shows the non-factorisable or entangled nature of the Bell states.

To the contrary of factorisable states, when composite systems are prepared in entangled states, the distributions of outcomes observed during local observations are no longer statistically independent.

For instance, let us assume that the qubit systems $A$ and $B$ are prepared in the Bell state $|B_0^0\rangle$ and let us measure the spin projection along $\vec{n}_A$ in the region $A$ and the spin projection along $\vec{n}_B$ in the region $B$ (with $n_x^{A/B} = sin\theta^{A/B}$, $n_y^{A/B} = 0$, $n_z^{A/B} = cos\theta^{A/B}$). In order to evaluate the corresponding distribution of outcomes, we can make use of the spinorial transformation law

$|+\rangle_{\vec{n}} = \cos\frac{\theta}{2}e^{\frac{-i\phi}{2}}|+\rangle_Z + \sin\frac{\theta}{2}e^{\frac{+i\phi}{2}}|-\rangle_Z$ and $|-\rangle_{\vec{n}} = -\sin\frac{\theta}{2}e^{\frac{-i\phi}{2}}|+\rangle_Z + \cos\frac{\theta}{2}e^{\frac{+i\phi}{2}}|-\rangle_Z$, where $\theta$ and $\phi$ are the polar angles associated to the direction $\vec{n}$ (here $\phi_A = \phi_B = 0$), so that the Bell state $|B_0^0\rangle$ transforms as follows:

$$\begin{aligned} |B_0^0\rangle = \sqrt{\frac{1}{2}}(&cos\frac{(\theta_A-\theta_B)}{2}|+\rangle_{\vec{n}}^A \otimes |+\rangle_{\vec{n}}^B \\ &-sin\frac{(\theta_A-\theta_B)}{2}|+\rangle_{\vec{n}}^A \otimes |-\rangle_{\vec{n}}^B \\ &+sin\frac{(\theta_A-\theta_B)}{2}|-\rangle_{\vec{n}}^A \otimes |+\rangle_{\vec{n}}^B \\ &+cos\frac{(\theta_A-\theta_B)}{2}|-\rangle_{\vec{n}}^A \otimes |-\rangle_{\vec{n}}^B). \end{aligned}$$

Making use of Born's transition rule, the probability that after the preparation of the Bell state $|B_0^0\rangle$ the outcomes of the spin measurements in $A$ and $B$ are found to be up-up is equal to $|\langle B_0^0|(|+\rangle_{\vec{n}}^A \otimes |+\rangle_{\vec{n}}^B)|^2$ so to say to $\frac{1}{2}cos^2\frac{(\theta_A-\theta_B)}{2}$. Similarly the probability of $(up_A, down_B)$ is $\frac{1}{2}sin^2\frac{(\theta_A-\theta_B)}{2}$, the probability of $(down_A, up_B)$ is $\frac{1}{2}sin^2\frac{(\theta_A-\theta_B)}{2}$, and the probability of $(down_A, down_B)$ is $\frac{1}{2}cos^2\frac{(\theta_A-\theta_B)}{2}$.

In particular, when local quantization axes are parallel: ($\theta_A - \theta_B$=0), we get perfect correlations:

$P(up_A, up_B) = P(down_A, down_B) = 1/2$

$P(down_A, up_B) = P(up_A, down_B) = 0.$

Obviously there is no longer statistical independence; otherwise we would get

$P(up_A, up_B).P(down_A, down_B)$=1/2.1/2=$P(down_A, up_B).P(up_A, down_B) = 0.0$ so that finally $1/4 = 0$, a logical contradiction.

We shall show in the next section that correlations exhibited by entangled systems have no classical counterpart.

## 5.4 Bell's inequalities.

Let us consider a situation à la Bell [3] during which a pair of qubits is prepared in the entangled state $|B_0^0\rangle$. A Stern-Gerlach measurement is performed along the direction $\vec{n_A}$ in the region $A$ and another Stern-Gerlach measurement is performed simultaneously along the direction $\vec{n_B}$ in the distant region $B$ (see fig.1[24]).

[24]This is a schematic representation, in the case that the the particles are emitted along opposite directions along the $Y$ axis, with the source in-between the regions $A$ and $B$, that we did not represent on the picture in order not to overload the representation.

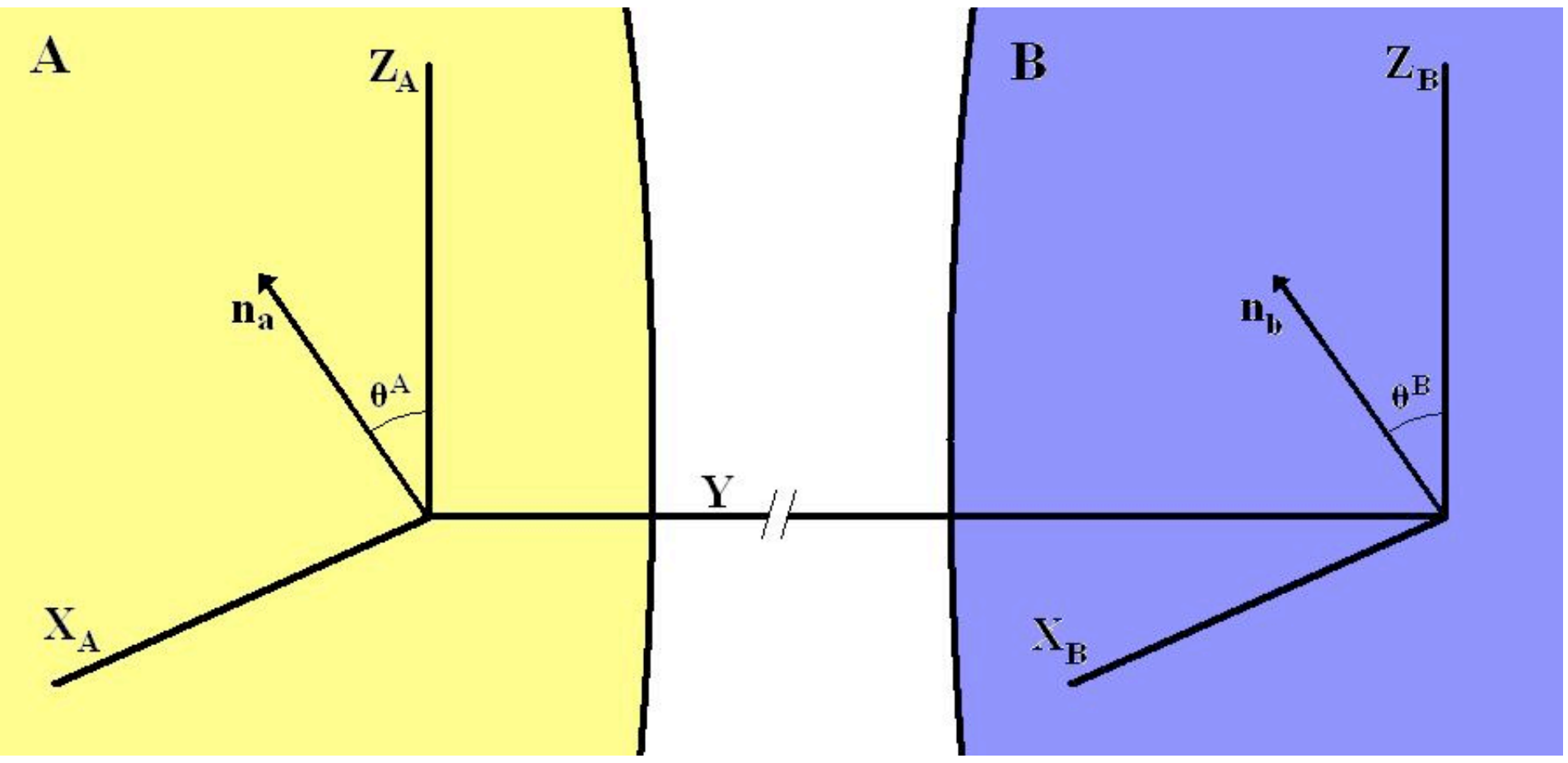

Figure 2: Bell-like measurement on a bipartite spin entangled system.

As we noted before, whenever the directions $\vec{n}_A$ and $\vec{n}_B$ are parallel, the outcomes are maximally correlated in the sense that the probability that the outcomes observed in the $A$ and $B$ regions are different is equal to 0. We thus face a situation in which we could in principle predict the outcome that is observed during a measurement in one of the regions, simply by performing the same measurement in the other region. By itself, this situation has nothing special: it could occur that making use of maximal correlations, a classical observer can infer with absolute certainty the validity of certain properties without testing them directly.

What is puzzling is that what quantum mechanics says about the value of the local spin is that it could be up with probability 50 percent and down with probability 50 percent, and that the formalism gives the feeling that the transition does not occur before the measurement process occurs.

It would mean that we "create" the value of the spin only when we look at it, a rather counter-intuitive result. Intuitively, we are likely to think that our observation only reveals pre-existing properties, since this is so, as far as we know, at the classical level. This is particularly obvious when $A$ and $B$ are separated by a space-like distance because, even if one accepts that the measurement influences the system under observation, it is difficult to understand how this influence would occur instantaneously, thus faster than the speed of light. Our classical intuition thus suggests the existence of pre-existing properties- and this is essentially the reasoning held by Einstein, Podolski and Rosen [30] in 1935- that the value of the spin pre-existed before the measurement process. If this is so, we are led to infer from the EPR reasoning that some deterministic "element of reality" is present but then this information is hidden and lost at the level of the quantum formalism, because the prediction of quantum mechanics is simply that each possible outcome (up or down) is observed with probability fifty percent, a purely indeterministic prediction.

The reasoning of EPR did not go much further than this; their final remark was that, given that a hidden determinism is present and that such a hidden determinism is not present at the level of the quantum formalism, the quantum theory is not complete and ought to be completed by a (hidden) deterministic or local realistic hidden variable theory (HVT).

In 1965, John Bell went further [3, 31] and showed that the existence of hidden determinism is incompatible with the predictions of quantum mechanics. We shall reproduce here the essence of his reasoning, and, following Ref.[32] (see also references about Pseudo-telepathy in ref.[33]), we shall enhance the dramatic character of the result by assuming that two persons, Alice and Bob, make use of the results of the Stern-Gerlach measurements on the qubits $A$ and $B$ respectively in order to simulate a "telepathic" relationship. During this "performance", Alice and Bob are located in far away regions; for instance, Alice is on earth while Bob is located inside an interplanetary rocket, more or less one lighthour away from earth. Both are kept in isolated, high-security, cells and are not allowed to communicate with each other. Every hour, a guardian $A$ enters Alice's cell and asks a question that is chosen at random among three possible questions $\alpha$, $\beta$ and $\gamma$. For instance the questions could be:

$\alpha$: Are you thirsty?

$\beta$: Are you tired?

$\gamma$: Are you happy?

Exactly at the same time, (which means in this precise case simultaneously relatively to (an inertial frame comoving with the center of mass of) the solar system), a guardian $B$ enters Bob's cell and asks a question that is chosen at random, and independently on the choice performed by the guardian $A$, among the selection $\alpha$, $\beta$ and $\gamma$. We also assume that the experiment is repeated many many times, hour after hour, in order to establish a relevant statistics of the correlations between Alice and Bob's answers.

Another rule of the game is that each time they are presented with a question, Alice and Bob must answer at once and have two possible answers: Yes and No.

Let us now assume that Alice and Bob make use of a quantum device in order to answer the questions: they share a pair of qubit states prepared in the entangled state $|B_0^0\rangle$, and in order to answer the questions $\alpha$, $\beta$, or $\gamma$, they measure thanks to a Stern-Gerlach device the spin of the qubit in their possession along the directions $\theta_\alpha = 0, \phi_\alpha = 0$; $\theta_\beta = 2\pi/3, \phi_\beta = 0$; or $\theta_\gamma = 4\pi/3, \phi_\gamma = 0$.

Because of the perfect correlations exhibited by the Bell state $|B_0^0\rangle$, whenever Alice and Bob are asked simultaneously the same question they will provide exactly the same answer, which tends to simulate a telepathic communication between them.

The first reaction, confronted with such a situation, would be to make the rational hypothesis according to which Alice and Bob possibly cheat by sharing a same list on which they have written in advance all possible answers to the three questions, at all times. Before they answer, they would consult the list and answer accordingly. This is nothing else, in the present context, than EPR's hypothesis.

John Bell [3] went further by showing that if such a list existed, the correlations ought to obey certain constraints (inequalities), and that those inequalities are violated by the quantum correlations, which renders impossible the existence of a list of pre-existing outcomes; in other words, the violation of Bell's inequalities denies the possibility of explaining quantum correlations by a local realistic HVT.

In the present case, it is easy to derive such an inequality, following the approach of Ref.[34] and making use of a property that was baptised by mathematicians under the name of pigeonhole's principle. The idea is simple: let us assume that three pigeons plan to spend the night in the holes of a cliff; if there are only two holes then certainly at least two pigeons will have to sleep in the same hole. This kind of reasoning is used for instance to show that within a population of $10^6$ persons, at least two persons will have exactly the same number of hair.

Now, there are three questions and two answers so that, in virtue of the pigeonhole principle, two questions will share the same answer and we can write the equality:

$$P(\alpha_A = \beta_B \vee \beta_A = \gamma_B \vee \gamma_A = \alpha_B) = 1, \tag{4}$$

where $\vee$ expresses the logical disjunction ("or") and the equality symbolically means that two questions have the same answer, for instance $\alpha_A = \beta_B$ means that the measured values of the spins along the directions $\alpha_A$ and $\beta_B$ are the same (either both up or both down). Now, it is well-known that the probability of the disjunction of two or more properties is less than or equal to the sum of their probabilities so that we can write the Bell-like inequality

$$P(\alpha_A = \beta_B) + P(\beta_A = \gamma_B) + P(\gamma_A = \alpha_B) \geq 1. \tag{5}$$

This inequality is violated by quantum correlations because $P(\alpha_A = \beta_B) + P(\beta_A = \gamma_B) + P(\gamma_A = \alpha_B) = P(\alpha_A = \beta_B = up) + P(\beta_A = \gamma_B = up) + P(\gamma_A = \alpha_B = up) + P(\alpha_A = \beta_B = down) + P(\beta_A = \gamma_B = down) + P(\gamma_A = \alpha_B = down)$ $= 6.\frac{1}{2}cos^2(\pi/3) = 3/4$. It is not true that $3/4 \geq 1$ and the inequality is violated.

Of course there are other logical explanations of the correlations: it could be that Alice and Bob secretly communicate, but as their distance is of the order of one light-hour and that they must answer at once (within say one second), they have to communicate more or less 3600 times faster than light.

Actually a similar situation was experimentally implemented in the surroundings of Geneva: instead of one light-hour the distance was ten kilometers and instead of one second the duration was of the order of 10 picoseconds. This imposes the experimental limits according to which, if Alice and Bob cheat and communicate in secret, they must do it $7.10^6$ times faster [35, 36] than light. In the literature, this (hypothetical) phenomenon is called non-locality, and is reminiscent of the mysterious Newtonian action-at-a-distance.

## 5.5 About the Shannon-von Neumann entropy and the bi-orthogonal (Schmidt) decomposition.

One can show [37, 38]) that when a bipartite system is prepared in the (pure) state $|\Psi\rangle^{AB} = \sum_{i,j=0}^{d-1} \alpha_{ij} |i\rangle^A \otimes |j\rangle^B$ (where $|i\rangle^A$ and $|j\rangle^B$ are states from orthonormalized reference bases) everything happens "from A's point of view" as

if he had prepared his system in the state described by the effective or reduced density matrix $\rho^A = \sum_{i,i'=0}^{d-1} \sum_{j=0}^{d-1} \alpha_{ij}^* \alpha_{i'j} |i'\rangle^A \langle i|^A$.

Now, it can be shown that the reduced density matrix has all the properties of density matrices (its trace is equal to one, it is a self-adjoint operator with a positive spectrum) so that we can find at least one basis in which it gets diagonalized, that we shall label by tilde indices ($|\tilde{i}\rangle^A$): $|\Psi\rangle^{AB} = \sum_{i,j=0}^{d-1} \tilde{\alpha}_{ij} |\tilde{i}\rangle^A \otimes |j\rangle^B = \sum_{i=0}^{d-1} |\tilde{i}\rangle^A \otimes (\sum_{j=0} \tilde{\alpha}_{ij} |j\rangle^B) = \sum_{i=0}^{d-1} \alpha_i |\tilde{i}\rangle^A \otimes |\tilde{i}\rangle^B$ where we introduced the notation $\alpha_i |\tilde{i}\rangle^B = \sum_{j=0} \tilde{\alpha}_{ij} |j\rangle^B$, with $\alpha_i$ a normalization factor. The states $|\tilde{i}\rangle^B$ are necessarily orthogonal, otherwise they would generate off-diagonal interference terms in the reduced density matrix, which may not occur because the basis states $|\tilde{i}\rangle^A$ diagonalize the reduced density matrix of A's subsystem.

This proves that we can always write a bipartite pure state $|\Psi\rangle^{AB}$ in the so-called bi-orthogonal form [39]:

$|\Psi\rangle^{AB} = \sum_{i=0}^{d-1} \alpha_i |\tilde{i}\rangle^A \otimes |\tilde{i}\rangle^B$, where the states $|\tilde{i}\rangle^A$ ($^B$) are orthonormalized. This form is called bi-orthogonal because the matrix $\alpha_{ij}$ is in general a non-diagonal matrix. It is only when the full state is expressed in the product of the bases composed by the states that diagonalize the reduced density matrices that the amplitudes-matrix becomes diagonal: $\tilde{\alpha}_{ij} = \alpha_i \delta_{i,j}$.

When the state is expressed in its biorthogonal form, it is easy to analyze the degree of entanglement of the two subsystems. One can quantitatively estimate the degree of entanglement by counting the number of coefficients $\alpha_i$ that differ from zero (this is called the Schmidt number). The Shannon entropy of the probability distribution $|\alpha_i|^2$, which is also equal to the Shannon-von Neumann entropy of the reduced density matrix of A or B's subsystem (section 5.6), provides a more quantitatively precise parameter in order to estimate their degree of entanglement (it is equal to 0 for factorizable states, in which case the biorthogonal decomposition contains only one factor, and equal to 1 when the state is maximally entangled so to say when $|\alpha_i|^2 = 1/d, \forall i$). It is easy to check that Bell states are maximally entangled, which corresponds to a density matrix proportional to the identity operator. Such a density matrix is said to be totally incoherent because in all conceivable interference experiments it will always exhibit a flat interference pattern (of "visibility" equal to 0).

Actually the purity or coherence which is equal to 1-the Shannon-von Neumann entropy of a reduced density matrix measures the degree of anisotropy exhibited by the corresponding state in the Hilbert space. When a state is pure it determines a preferred orientation (or ray) in the Hilbert space. An incoherent state is, from this point of view, totally isotropic and indeed the probability of transition of such a state to any pure state is constant and equal to $1/d$. This explains why such states always exhibit totally flat interference patterns.

## 5.6 Entanglement, non-separability and loss of identity.

Another aspect of entanglement is its "fusional" nature which we consider to be a manifestation of quantum holism. Bell's analysis of the nature of quantum correlations shows that, in contradiction with the Cartesian paradigm, when two systems are prepared in an entangled state, the knowledge of the whole cannot be reduced to the knowledge of the parts, and that to some extent the systems lose their individuality. It is only when their joint wave-function is factorizable that they are separable[25]. A very interesting uncertainty (complementarity [40]) relation characterizes the entanglement of a pair of quantum systems prepared in a pure state: the entanglement of the whole system (that measures its degree of inseparability) is equal to the Shannon-von Neumann entropy of the reduced system, which is also a ngeative measure of the coherence of the system. This relation is expressed by the equality $E(A-B) = 1 - C(A) = 1 - C(B)$, where $C(A(B)) = 1 + Tr(\rho^{A(B)} log_d \rho^{A(B)})$, the Shannon-von Neumann coherence of the subsystem $A$ $(B)$ which measures the degree of purity or coherence of their reduced state, as well as the degree of certainty associated to this state, while $E(A-B)$ is, in the case of pure states, a "good" measure of the entanglement between the subsystems $A$ and $B$. In simple terms, the irreducibility of the whole to the parts (or entanglement between them) increases when the Shannon-von Neumann measure of the "certainty" of the parts (or their "purity" or degree of "coherence") decreases and vice versa. For instance, when the state of the full system is pure and factorizable, their entanglement is equal to 0 and the reduced system is a pure state with a minimal Shannon-von Neumann entropy equal to 0 (maximal coherence equal to 1). When the full system is prepared in a Bell state, their entanglement is maximal and equal to 1 and the reduced system is a totally incoherent density matrix proportional to the identity, with a maximal Shannon-von Neumann entropy equal to 1 (minimal coherence equal to 0). This complementarity relation can be generalized when the full state is not pure, but the situation is more involved in this case [41, 42], among others because there exists no simple measure of the entanglement of two subsystems when the system is prepared in a mixed state [42].

If metaphorically we transfer this idea to human relationships, it could be translated (in a very free way, because there is no direct counterpart for the concept of quantum purity at the human level) by something like the "fusional nature of entanglement": when $A$ and $B$ are strongly entangled, they lose to some extent their individuality. We mean that the coherence of the parts decreases, and the coherence or purity is seen here as a measure of the independence (singularity) relatively to the rest of the world. This fusional nature is contagious to some extent (the friends of my friends are my friends) because it can be shown that two systems $C$ and $D$ can become entangled although they never interacted directly and remain spatially separated, provided they get entangled (through interaction for instance [21, 20]) with subsystems that are entangled (this is called entanglement swapping-see e.g. ref.[33] and references therein). For instance, regions that are separated by huge distances in the galaxy [43] can be shown to become entangled because they both interact with the cosmic background radiation which presents a non-negligible degree of spatial entanglement. Another related property is monogamy: fusional relations are often exclusive, which possesses a quantum counterpart the so-called quantum monogamy [44, 37].

[25]As we mentioned before, whenever two distant systems are in an entangled (pure) state, it has been shown [4] that there exist well-chosen observables such that the associated correlations do not admit a local realist's explanation, which is revealed by the violation of well-chosen Bell's inequalities.

# On definitions of Information in Physics

Nicolás F. Lori

IBILI, University of Coimbra, Portugal.

**Abstract**

During the refereeing procedure of *Anthropomorphic Quantum Darwinism* by Thomas Durt, it became apparent in the dialogue between him and me that the definition of information in Physics is something about which not all authors agreed. This text aims at describing the concepts associated to information that are accepted as the standard in the Physics world community.

## Introduction

The purpose of this text is to provide a brief description of what are the concepts of information that are accepted as the standard in the Physics world community, which does not mean that all Physicists agree with such definitions but simply that a majority does. The purpose of a standard is to develop concepts in a clear and non-ambiguous enough way for a large community (of physicists in this case) to be able to disagree based on the substance of the arguments instead of disagreeing on the definition of the words. This text resulted from my refereeing the *Anthropomorphic Quantum Darwinism* article by Thomas Durt, and it will be especially focused on the relation between information and the meaning of Shannon/von Neumann entropy (henceforth referred to as Shannon entropy).

By discussing with me the issues proposed in his work Thomas Durt was able to find expressions and statements that were able to represent his scientific thought using word-definitions that agree with the standard definitions of the words "entropy", "information" and "negentropy". Because we are analyzing the standard accepted usage of words, the definitions appearing Wikipedia will be a major source of support. Although Wikipedia might have incorrect terms here and there, its widespread usage makes it the *de facto* standard for the meaning of words.

## Entropy and Information

Acccording to the Wikipedia, "Shannon entropy is a measure of the average information content" [http://en.wikipedia.org/wiki/Entropy_(Information_theory)] and "The negentropy, also negative entropy or syntropy, of a living system is the entropy that it exports to keep its own entropy low; it lies at the intersection of entropy and life. The concept and phrase "negative entropy" were introduced by Erwin Schrödinger in his 1943 popular-science book What is life? Later, Léon Brillouin shortened the phrase to negentropy, to express it in a more "positive" way: a living system imports negentropy and stores it." [http://en.wikipedia.org/wiki/Negentropy]. In short, entropy=<information> and negentropy=-entropy.

According to [http://en.wikipedia.org/wiki/Information_entropy], we have that: "In fact, in the view of Jaynes (1957), thermodynamics should be seen as an application of Shannon's information theory: the thermodynamic entropy is interpreted as being an estimate of the amount of further Shannon information needed to define the detailed microscopic state of the system, that remains uncommunicated by a description solely in terms of the macroscopic variables of classical thermodynamics. In short, the Shannon entropy defines the average amount of information in the system that we do not have about the system unless we observe in detail the molecules of the system."

There area few things that need to be clarified about the relation between these terms. The first is that entropy is always the total (or averaged) amount of information in a system. The entropy of the system is the amount of information in a system, of which the observer can perceive only a small portion. So, as the information in a system grows so does your ignorance about that system grows, as we typically cannot make the information extraction process keep up with the information growing process.

In what regards information, there are two ways of looking at it. In one way the entropy of the system is the average of the information we do not have, so from the perspective of the observer: E_{system}=-<I_{observer}>. From the perspective of the system the entropy expresses the amount of information the system has (using the classical perspective where information exists regardless of the observation); meaning that: E_{system}=<I_{system}>. I prefer the second way of expressing it, as it does not mix the observer with the system.

The amount of information in a system is called the Shannon entropy, and is also referred to as self-information. In Wikipedia it is stated that "self-information is also sometimes used as a synonym of entropy, i.e. the expected value of self-information in the first sense (the mutual information of the system with itself)". "In information theory, entropy is a measure of the uncertainty associated with a random variable. The term by itself in this context usually refers to the Shannon entropy, which quantifies, in the sense of an expected value, the information contained in a message, usually in units such as bits. Equivalently, the Shannon entropy is a measure of the average information content one is missing when one does not know the value of the random variable." The uncertainty associated to a random variable is resolved by the appearance of the information about the random variable. The relation between the entropy $S$ and the Shannon entropy $H$ is: $S=ln(2)K_BH$. Where $K_B$ is Boltzmann's constant and $ln$ stands for the natural logarithm, meaning that to 1 bit of Shannon entropy corresponds $0.9572 \times 10^{-23}\ J\ K^{-1}$. If $p_i$ is the probability of a state $i$, then the expression for the Shannon entropy is: $H=-\Sigma_i\, p_i \log_2 p_i$ .

**Information in Quantum Mechanics**

Again using Wikipedia, we can read that "entropy is simply that portion of the (classical) physical information contained in a system of interest whose identity (as opposed to amount) is unknown (from the point of view of a particular knower). This informal characterization corresponds to both von Neumann's formal definition of the entropy of a mixed quantum state, as well as Claude Shannon's definition of the entropy of a probability distribution over classical signal states or messages (see information entropy)." In the "Von_Neumann_entropy" entry in Wikipedia, it is stated that: "Given the density matrix ρ , von Neumann defined the entropy as $S(\rho)=-Tr(\rho\ ln\rho)$ which is a proper extension of the Gibbs entropy (and the Shannon entropy) to the quantum case." Actually, it is a proper extension of the Shannon entropy, with the connection to the entropy requiring the multiplication of an extra $ln(2)K_B$ term

Two works that I consider essential in understanding quantum information are:

Brukner, C., Zeilinger. 2002. Inadequacy of the Shannon Information in Quantum mechanics. arXiv:quant-ph/0006087 v3 5 Apr 2002.

Which states:
"We suggest that it is therefore natural to require that the total information content in a system in the case of
quantum systems is sum of the individual amounts of
information over a complete set of m mutually complementary
observables. As already mentioned above, for a
spin-1/2 particle these are three spin projections along
orthogonal directions. If we define the information gain
in an individual measurement by the Shannon measure
the total information encoded in the three spin components
is given by
Htotal := H1(p+x , p-x ) + H2(p+y , p-y ) + H3(p+z , p-z )."

Timpson, C. G. 2003. On a Supposed Conceptual Inadequacy of the Shannon Information in Quantum mechanics. Studies In History and Philosophy of Science Part B: Studies In History and Philosophy of Modern Physics
Volume 34, Issue 3, September 2003, Pages 441-468.

Which states:
"Another way of thinking about the Shannon information is as a measure of the amount of information that we expect to gain on performing a probabilistic experiment. The Shannon measure is a measure of the uncertainty of a probability distribution as well as a measure of information."

In both manuscripts the entropy and information are considered to be identical up to a scale (as it occurs in the classical case), but while in Timpson the agreement with Shannon entropy is like in the classical case; it is a bit more elaborate in Brukner& Zeilinger. The consensus seems to be that Timpson’s argument is stronger, but I will lay claim for neither one.

**Acknowledgements:** Thanks are due to Alex Blin for his helpful comments.

# Competing definitions of Information versus Entropy in Physics

**By** Thomas Durt**, Vrije Universiteit Brussel.**

**Abstract:**

*As was mentioned by Nicolas Lori in his commentary, the definition of Information in Physics is something about which not all authors agreed. According to physicists like me Information decreases when Entropy increases (so entropy would be a negative measure of information), while many physicists, seemingly the majority of them, are convinced of the contrary (even in the camp of Quantum Information Theoreticians). In this reply I reproduce, and make more precise, some of my arguments, that appeared here and there in my paper, in order to clarify the presentation of my personal point of view on the subject.*

## Entropy and Information.

Entropy is one among several possible measures of the uncertainty of a probabilistic distribution. It is worth noting that very often entropy is also considered to be a measure of information, but that in my approach it is natural to consider that the useful information increases when entropy decreases. Indeed let us measure the degree of "certainty" or the degree of "determinism" assigned to a random variable by the positive quantity C,
C= 1 - entropy; entropy is then assumed to represent the degree of uncertainty of the distribution of the random variable.
By definition, in this approach, certainty+uncertainty =1 which means that entropy is a negative measure of certainty.
Moreover, my choice to associate entropy with uncertainty means that I do consider that when a distribution of probability is very uncertain it does not contain much useful information.
On the contrary when a probability distribution is peaked it contains a high level of useful information.
So when I use the word information in my paper I implicitly mean "certain" information (=1-entropy), a quantity that increases when entropy decreases, contrary to most commonly used conventions…

In quantum mechanics, the more standard measure of entropy of a density matrix Rho is the Shannon-von Neuman entropy, in which case C is also, by definition, a measure of the purity or coherence of Rho.
A very interesting uncertainty (complementarity) relation characterizes the entanglement of a pair of quantum systems prepared in a pure state: the entanglement of the whole system (that measures its degree of inseparability) is equal to the Shannon-von Neumann entropy of the reduced system, which is also a negative measure of the coherence of the system. This relation is expressed by the equalities
E(A -B) = 1 - C(A) = 1 - C(B),
where C(A(B)) = 1 + Tr(A(B)logdA(B))= the Shannon-von Neumann coherence of the subsystem A (B) which measures the degree of purity or coherence of their reduced state, as well as the degree of certainty associated to this state, while E(A - B) is, in the case of pure states, a "good" measure of the entanglement between the subsystems A and B. In simple terms, the irreducibility of the whole to the parts (or entanglement between them) increases when the Shannon-von Neumann measure of the "certainty" of the parts (or their "purity" or degree of "coherence") decreases and vice versa.

For instance, when the state of the full system is pure and factorizable, their entanglement is equal to 0 and the reduced system is a pure state with a minimal Shannon-von Neumann entropy equal to 0 (maximal coherence equal to 1). When the full system is prepared in a Bell state, their entanglement is maximal and equal to 1 and the reduced system is a totally incoherent density matrix proportional to the identity, with a maximal Shannon-von Neumann entropy equal to 1 (minimal coherence equal to 0). This complementarity relation can be generalized when the full state is not pure, but the situation is more involved in this case, among others because there exists no simple measure of the entanglement of two subsystems when the system is prepared in a mixed state.

If metaphorically we transfer this idea to human relationships, it could be translated (in a very free way, because there is no direct counterpart for the concept of quantum purity at the human level) by something like the "fusional nature of entanglement": when A and B are strongly entangled, they lose to some extent their individuality. We mean that the coherence of the parts decreases, and the coherence or purity is seen here as a measure of the independence (singularity relatively to the rest of the world).

This fusional nature is contagious to some extent (the friends of my friends are my friends) because it can be shown that two systems C and D can become entangled although they never interacted directly and remain spatially separated, provided they get entangled (through interaction for instance) with subsystems that are entangled (this is called entanglement swapping).

For instance, regions that are separated by huge distances in the galaxy can be shown to become entangled because they both interact with the cosmic background radiation which presents a non-negligible degree of spatial entanglement.

Another related property is monogamy: fusional relations are often exclusive, which possesses a quantum counterpart the so-called quantum monogamy of entanglement, a property according to which if A and B are strongly entangled there is few room left for entanglement with the rest of the world (or with a third party C).

# Final remark:

Ambiguities also appear when we consider the so-called Kolmogorov complexity (also known as descriptive complexity, Kolmogorov-Chaitin complexity, stochastic complexity, algorithmic entropy, or program-size complexity) of an object: a random series of bits possesses maximal complexity according to the definition of algorithmic complexity, but it is also natural to consider that the corresponding amount of information is minimal because it is maximally ''uncertain''.

There is no way to escape this kind of paradoxes according to me because they are related to antagonistic acceptances of the concept of probability.

If we are interested in characterizing a noisy communication channel via the measure of the entropy of the distribution of correctly and wrongly transmitted results, as is often the case in quantum cryptography for instance, this entropy is an increasing function of the ''error rate'' so that the rate of correctly transmitted information increases when entropy decreases.

In other contexts one is free to consider impredictability as a richness offered by a system in which case it makes sense to assume that information is a monotonously increasing function of entropy. In this approach, a message in which only one letter is used is devoided of information, while in the case that two letters appear (0 amd 1 for instance), if the frequency of appearance of each letter differs from ½ (this is what we get by tossing a biased coin), one can conceive that the signal could be compressed because there is redundancy in its encoding.

Considered so (and this is the standard approach), a biased distribution of 0's and 1's (a random binary distribution for instance) which is maximally entropic would also possess the highest degree of information because it cannot be compressed. This view is legitimate if we realize that a low entropic series of random bits (that we could obtain by tossing a strongly unbiased coin for instance a coin that will fall on the head face with probability 99 percent) can be compressed and replaced by a quite shorter signal. It will require less bits to encode the series than for a random one, and it is common to consider that when a message is long it contains much information.

Let us consider another example, the full trajectory of a satellite around Earth, which is univoquely determined by 6 real numbers, the initial position and velocity of the satellite-in the Newtnian picture.

In the standard approach one would say that the full trajectory is nearly devoided of information (in the sense of computational complexity) because it is redundant: the position and velocity at any time contain in germ the full trajectory.

In our approach we consider that it is the contrary because randomness is like noise and, in our view, noise doen't contain information at all.

On the contrary, a deterministic series (like the series of positions occupied by the satellite throughout time) possesses a high degree of certainty and predictability, so that it could be potentially used for encoding information in a fidel and reproducible way.

Viewed so we consider that regularities are the message, and not chaos, even if by doing so we risk to be relegated to a minority.

Obviously in our approach the "informational degree'' refers to the unambiguity of the message (it measures to which degree a physically encoded message ''makes sense''). In the standard approach ''information'' refers rather to the length of the message. It is not amazing that these approaches contradict each other because they express radically different points of view…

# Application of Quantum Darwinism to Cosmic Inflation: an example of the limits imposed in Aristotelian logic by information-based approach to Gödel's incompleteness

Nicolás F. Lori
*IBILI, Universidade de Coimbra, 3000-354 Coimbra, Portugal**

Alex H. Blin
*Departamento de Física, Universidade de Coimbra, 3004-516 Coimbra, Portugal*†

Gödel's incompleteness applies to any system with recursively enumerable axioms and rules of inference. Chaitin's approach to Gödel's incompleteness relates the incompleteness to the amount of information contained in the axioms. Zurek's quantum Darwinism attempts the physical description of the universe using information as one of its major components. The capacity of Quantum Darwinism to describe quantum measurement in great detail without requiring ad-hoc non-unitary evolution makes it a good candidate for describing the transition from quantum to classical. A baby-universe diffusion model of cosmic inflation is analyzed using quantum Darwinism. In this model cosmic inflation can be approximated as Brownian motion of a quantum field, and quantum Darwinism implies that molecular interaction during Brownian motion will make the quantum field decohere. The quantum Darwinism approach to decoherence in the baby-universe cosmic-inflation model yields the decoherence times of the baby-universes. The result is the equation relating the baby-universe's decoherence time with the Hubble parameter, and that the decoherence time is considerably shorter than the cosmic inflation period.

## I. INTRODUCTION

Linde's approach to the Big Bang [1] indicates that the creation of a universe from "nothing" occurs in a Brownian-motion-like process (Brownian motion and diffusion are used in this work as being equivalent terms). Zurek's quantum Darwinism approach to quantum mechanics indicates that Brownian motion can be related to decoherence [2]. It would therefore be expected that quantum Darwinism is related to decoherence during cosmic inflation.

There is no disagreement in the physics community on the experimental evidence that quantum systems exist in a multitude of states, with only a portion of those states being observable. The approaches in physics differ in what happens to the non-observed states and in the process by which the states become observed states. Examples of two different points of view are the works of Linde and Zurek.

The approach by Linde proposes that the universe follows a deterministic evolution about which we can only observe partial aspects of the multiple possible occurrences that are deterministically created. The approach by Zurek proposes that the deterministic evolution of the universe is constrained by a Darwinian extinction of some of the possible evolution paths of the system. The approaches by Linde and Zurek agree in what is observed, but they disagree about what happens to the non-observed states. In the case of quantum gravity effects during cosmic inflation those differences may be relevant.

The mathematical representation of quantum Darwinism requires a way of including extinction in a mathematical formalism. Using Chaitin's approach to Gödel's incompleteness [3] we propose in the Appendix a separation between two ways of representing mathematical axiomatic systems. The formal axiomatic systems (FAS) which is a system where the consistency of the axioms is preserved, meaning that no proposition can be both true and not true; and the Darwinian axiomatic system (DAS) where some true propositions lead to false propositions (which become extinct) and so the DAS is not consistent but complete. The DAS is complete for there are no true propositions that are not obtained from true propositions; all propositions have parent propositions that need to be true for otherwise they become extinct. Gödel's incompleteness theorems showed that a non-trivial axiomatic system cannot be both complete and consistent [3, 18]; the FAS is the choice for consistency and the DAS is the choice for completeness.

*Electronic address: nflori@fmed.uc.pt
†Electronic address: alex@fis.uc.pt

This work mostly uses the works of Linde [1] and Zurek [2]; other approaches considered are Chaitin's use of information in mathematics [3], Wheeler's concept of 'it for bit' [4], Rovelli's relational approach to quantum gravity [5], Smolin's relation between quantum mechanics and quantum gravity [6], and Guth's approach to cosmic inflation [7].

Although other approaches to decoherence during cosmic inflation have been developed [8–14], our approach differs in that it is based on quantum Darwinism's interpretation of the decoherence caused by the interaction through molecules in Brownian motion [2]. This is a good parallel with the baby-universes in Brownian motion [1] because also in that case the major source of decoherence is the interaction between baby-universes. The issue of the medium surrounding the molecules is not relevant for this model of decoherence because it is the other molecules that are the environment.

The remainder of the present article is structured as follows. The next five sections discuss the underlying theory: Introduction to Quantum Measurement; Introduction to Quantum Darwinism; Relation between Quantum Darwinism and Quantum Diffusion; Diffusion in Cosmic Inflation; Effects of a Quantum Darwinism Approach to Cosmic Inflation. The calculations at the end of each subsection are then presented in the section on Results, with their relations highlighted. The section entitled Discussion and Summary describes the possible implications of the results obtained and highlights the principal results.

## II. INTRODUCTION TO QUANTUM MEASUREMENT

Quantum Darwinism [2, 17] is an approach to quantum measurement that is strongly based on Wheeler's "it-for-bit" approach [4] and so it has parallels with both information theory and computation. The classical technical definition of the amount of information was provided by Shannon's information entropy and stated that if the sending device has a probability $P_j$ of sending message $j$ from a set of $N$ messages, then the information transmitted when one message is chosen from the set is, in units of bits [3],

$$\mathcal{H} = -\log_2 P_j \ . \tag{1}$$

For a brief description of quantum Darwinism it is helpful to resort to a short description of the limitations of non-Darwinian quantum mechanics, the limitations that quantum Darwinism addresses. In quantum mechanics the universe is separable into 3 parts: I. System *S,* II. Apparatus *A,* III. Environment *E.* The evolution of quantum systems occurs according to Schrödinger's equation. Entanglement between system and apparatus can be modeled by unitary Schrödinger evolution. Von Neumann [15] proposed a non-unitary selection of the preferred basis,

$$|\Psi_{SA}\rangle \langle\Psi_{SA}| \rightarrow \sum_k |a_k|^2 |s_k\rangle \langle s_k| \, |A_k\rangle \langle A_k| = \rho_{SA} \ . \tag{2}$$

and also proposed the non-unitary "collapse" succeeding the occurrence of a unique outcome to be a different event from the selection of the preferred basis. For example, the occurrence of state 17:

$$\sum_k |a_k|^2 |s_k\rangle \langle s_k| \, |A_k\rangle \langle A_k| \rightarrow |a_{17}|^2 |s_{17}\rangle \langle s_{17}| \, |A_{17}\rangle \langle A_{17}| \ . \tag{3}$$

Zurek [2, 17] proposed an approach to entanglement which is unitary and as un-arbitrary as possible, using the environment. The use of the environment implies abandoning the closed-system assumption [17], requiring the following alteration:

$$|\Psi_{SA}\rangle |e_0\rangle = \left(\sum_k a_k |s_k\rangle |A_k\rangle\right) |e_0\rangle \rightarrow \sum_k a_k |s_k\rangle |A_k\rangle |e_k\rangle = |\Psi_{SAE}\rangle \ . \tag{4}$$

The selection of the preferred basis is obtained using unitary evolution by assuming $|\langle e_k \, |e_l\rangle|^2 = \delta_{kl}$ and tracing over the environment [17],

$$\rho_{SA} = \mathrm{Tr}_E \, |\Psi_{SAE}\rangle \langle\Psi_{SAE}| = \sum_k |a_k|^2 |s_k\rangle \langle s_k| \, |A_k\rangle \langle A_k| \ . \tag{5}$$

The preferred basis is defined by the set of states that the apparatus can adopt, and do not interact with the environment; and thus, only interact with the system. The apparatus adopts one of the pointer states after it makes a measurement. For this set of pointer states to exist it is necessary that the apparatus be entangled with the environment. Entanglement is a non-classical quantum behavior where two parts of the universe that have interacted at a certain point in time have to be described with reference to each other even if they are now separated in space, as long as they remain entangled. The above explanations of quantum measurement do not clarify the meaning of tracing over the environment, and the non-unitary "collapse" is not really explained. Quantum Darwinism addresses both issues successfully [2, 19].

## III. INTRODUCTION TO QUANTUM DARWINISM

In quantum Darwinism, the following statements are considered to be valid: (a) The universe consists of systems. (b) A pure (meaning completely known) state of a system can be represented by a normalized vector in Hilbert space $H$. (c) A composite pure state of several systems is a vector in the tensor product of the constituent Hilbert spaces. (d) States evolve in accordance with the Schrödinger equation $i\hbar|\dot{\psi}\rangle = H|\psi\rangle$ where $H$ is Hermitian. In quantum Darwinism no "collapse" postulate is needed. An assumption by von Neumann [15] and others is that the observers acquire information about the quantum system from the quantum system, but that is (almost) never the case. The distinction between direct and indirect observation might seem inconsequential as a simple extension of the von Neumann chain, but the use of the point of view of the observer in quantum Darwinism makes it possible to obtain the "collapse" [17, 19].

In quantum Darwinism there is "no information without representation", meaning that the information is always about a state that survived, and as a consequence being represented in the Hilbert space $H$. Preferred pointer states selected through entanglement define what is being stored in the environment. The "information amount" in quantum systems is defined using the density matrix $\rho$ and is based on ref. [17].

Environment-assisted invariance (envariance) is a quantum symmetry exhibited by the states of entangled quantum systems. The $SA$ system is from now simply represented by $S$ to simplify notation. The joint state of system $S$ entangled (but no longer interacting) with an environment $E$ can always be described by a Schmidt basis if the environment is made big enough (even if the initial joint state is mixed). As the environment no longer interacts with the system, probabilities of various states of the system cannot be – on physical grounds – influenced by such purification. Such purification is assumed to be either unnecessary or already carried out:

$$|\Psi_{SE}\rangle = \sum_{k}^{K} a_k \, |s_k\rangle \, |e_k\rangle \ . \tag{6}$$

Envariance refers to the existence of unitary transformations $U_S$ acting on $S$ alone that alter $|\Psi_{SE}\rangle$ non-trivially and whose effect can be canceled by the action of a unitary operation $U_E$ acting on $E$ alone,

$$U_E \left( U_S \, |\Psi_{SE}\rangle \right) = |\Psi_{SE}\rangle \ , \tag{7}$$

or, in more detail,

$$[1_S \otimes u_E] \left( [u_S \otimes 1_E] \, |\Psi_{SE}\rangle \right) = |\Psi_{SE}\rangle \ . \tag{8}$$

All envariant unitary transformations have the eigenstates that coincide with the Schmidt expansion and are given by

$$u_S = \sum_{k}^{K} e^{i\phi_k} \, |s_k\rangle \, \langle s_k| \ . \tag{9}$$

The corresponding operator in the environment is

$$u_E = \sum_{k}^{K} e^{i(\phi_k + 2\pi l_k)} \, |e_k\rangle \, \langle e_k| \tag{10}$$

with $l_k$ integer. Properties of global states are envariant iff they are functions of the phases of the Schmidt coefficient. To regard phases as unimportant and absorb them using the Schmidt expansion is a dangerous over-simplification as phases do matter.

The classical-approach prejudice that the information about the system is synonymous with its state and the presence of that information is physically irrelevant for that state is maybe based on the Aristotelian logic (a.k.a. binary logic and 2-state logic) assumption that thinking of the object is identical to the object existing. This assumption in not shared by the Buddhist logic approach [23]. The Buddhist logic (a.k.a. 4-state logic) approach has many parallels to the mathematical constructivist approach. In Buddhist logic the statement “A is B” can be denied by the Aristotelian denial “A is not-B”, but it can also be denied using “A not-is B”. In the “not-is” what is denied is the capacity of “A” to be “B”, instead of affirming that “A” is “not-B”. The fourth statement is “A not-is not-B” which is different from the statement “A is B” [23]. Completeness implies that all propositions *are*, and consistency implies that all propositions are either true or false. Buddhist logic allows for extending logic to approaches that are neither complete nor consistent. The states of classical objects are absolute (the state of an object *is*), while in quantum theory there are situations –entanglement- where the state of the object is relative. In general relativity there can only be envariance if the environment includes the whole universe (Newton’s bucket comes to mind). It is the non-absolute nature of existence that invites the abandonment of Aristotelian logic in favour of Buddhists logic as a first step, but not as a complete step (Figure 1).

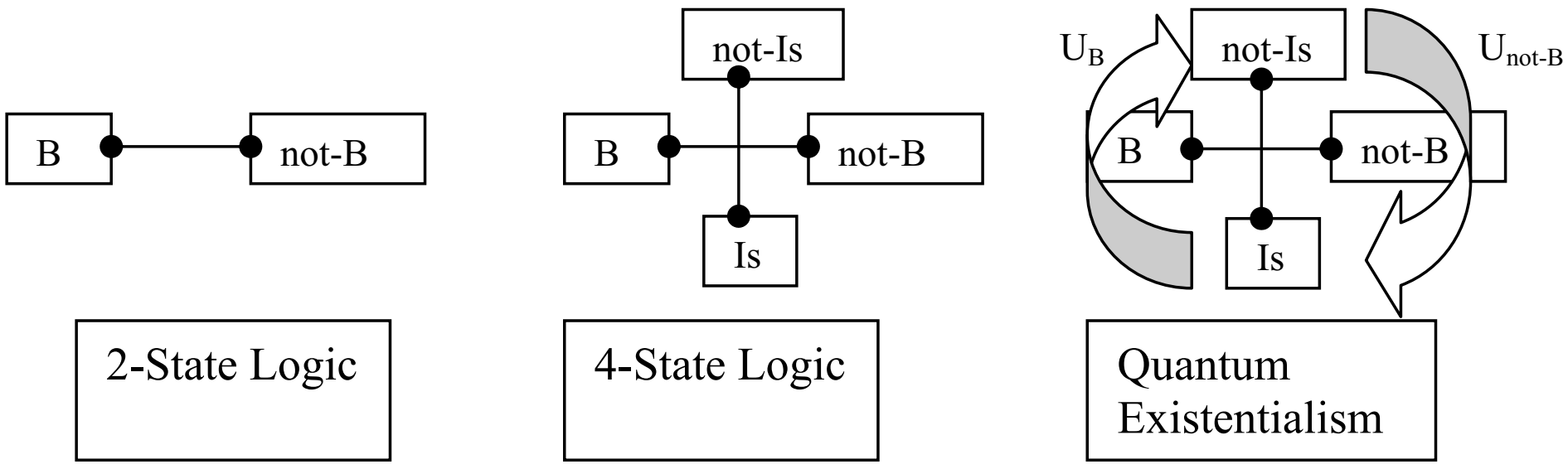

FIG. 1: In Aristotelian (2-state) logic the two possibilities are “A is B” and “A is not-B”. In Buddhist (4-state) logic the four states are “A is B”, “A is not-B”, “A not-is B”, and “A not-is not-B”. In quantum Existentialism there is an “almost continuous” transition between “is” and “not-is”, and between “B” and “not-B”; the “envariance” enables to cancel the alteration of the “is” (or “not-is”) of “B” caused by $U_B$ by the action of a $U_{not-B}$ acting on “not-B”.

The mathematical requirement of completeness implies that all propositions obtained by the rules of deduction are true [3], which requires complete/global access to the states of the system. The mathematical requirement of consistency implies that a state cannot simultaneously be and not be; in quantum Darwinism only “pointer” states are like that [2]. Global states are measured when observations focus on the “system+environment”, and “pointer” states are obtained when the focus is on the system to the detriment of the environment. Observing the “system+environment” and observing the system are complementary approaches, analogous to the complementarity of measuring position and momentum. Thus quantum Darwinism makes a direct connection between Chaitin’s approach to Gödel’s incompleteness theorem [3] on one hand and quantum complementarity on the other.

## IV. RELATION BETWEEN QUANTUM DARWINISM AND DIFFUSION

In molecular Brownian motion, the Brownian motion of quantum states implies decoherence. The quantum Brownian motion model used here consists of an environment $E$ made of a collection of harmonic oscillators of position $q_n$, mass $m_n$, frequency $w_n$, and coupling constant $c_n$, interacting with a system $S$ of mass $M$, position $x$, and harmonic potential $V(x)= \frac{1}{2}MW^2x^2$. The total Lagrangian is [20]

$$L\left(x, q_n\right) = \underbrace{\frac{M}{2}\left[\dot{x}^2 - W^2x^2\right]}_{L_S} + \underbrace{\sum_n \frac{m_n}{2}\left[\dot{q}_n^2 - w_n^2\left[q_n - \frac{c_n x}{m_n w_n^2}\right]^2\right]}_{L_{SE}} . \quad (11)$$

The Lagrangian component $L_{SE}$ takes into account the renormalization of the potential energy. Let us denote $k$ as the Boltzmann constant and $T$ as the temperature. If the thermal energy $kT$ is higher than all other relevant energy scales, including the energy content of the initial state and energy cutoff in the spectral density of the environment $C(v)$, then the master equation for the density matrix $\rho_S$ of an initially environment-independent system $S$ depends on the position $y$ of another molecule, the renormalized Hamiltonian $H_{ren}$ and on

$$\gamma = \frac{2}{MW}\int_0^\infty dl \int_0^\infty dv C\left(v\right)\sin\left(Wl\right)\sin\left(vl\right) \tag{12}$$

in the following way [20]:

$$\dot{\rho}_S = -\frac{i}{\hbar}[H_{ren},\rho_S] - \gamma[x-y]\left[\frac{\partial}{\partial x}-\frac{\partial}{\partial y}\right]\rho_S - \frac{2M\gamma kT}{\hbar^2}[x-y]^2\rho_S \ . \tag{13}$$

In this high $T$ case the master equation is independent of $V(x)$. The relaxation time is $\gamma^{-1}$ and the decoherence time is [16, 17]:

$$\tau_D = \gamma^{-1}\left[\frac{\frac{\hbar}{\sqrt{2MkT}}}{x-y}\right]^2 \ . \tag{14}$$

The Wigner quasi-distribution representation $Z$ of the high temperature density matrix master equation (Eq. (13)) is [20]:

$$\dot{Z} = -\frac{p}{M}\frac{\partial}{\partial x}[Z] + \frac{\partial V}{\partial x}Z + 2\gamma\frac{\partial}{\partial p}[pZ] + 2\gamma MkT\frac{\partial^2}{\partial p^2}[Z] \ . \tag{15}$$

The minimum uncertainty Wigner quasi-distribution for a phase space localized wave-packet is [20]:

$$Z\left(x_0, x, p_0, p\right) = \frac{1}{\pi\hbar}\exp\left(-\left[\frac{x-x_0}{\sqrt{\frac{\hbar}{2MW}}}\right]^2 - \left[\frac{p-p_0}{\sqrt{\frac{\hbar MW}{2}}}\right]^2\right) \ . \tag{16}$$

If there are two wave packets separated by $\Delta x$, with average location $x$ and average momentum $p$, then the joint Wigner quasi-distribution is equal to averaging the two localized Wigner distribution expressions plus a non-classical interference term equal to [20]

$$W_{\text{int}} \approx \frac{1}{\pi\hbar}\cos\left(\frac{\Delta x}{\hbar}p\right)\exp\left(-\left[\frac{x}{\sqrt{\frac{\hbar}{2MW}}}\right]^2 - \left[\frac{p}{\sqrt{\frac{\hbar MW}{2}}}\right]^2\right) \ . \tag{17}$$

Joining the diffusion coefficient expression [16, 20]

$$D = \frac{kT}{\gamma M} \tag{18}$$

with the decoherence time definition of Eq. (14) yields a relation between decoherence time and diffusion coefficient,

$$\tau_D = \frac{D}{2}\left[\frac{\hbar}{kT\left[x-y\right]}\right]^2 \ . \tag{19}$$

From Einstein's diffusion equation we know that $\langle(x(t)-x(0))^2\rangle = 2Dt$ for a single molecule. Consider now two molecules. Let $t_{\{x,y\}}$ be the time interval since the last collision of two molecules which collided at the point $x_0 = y_0$ and which are now at the positions $x$ and $y$, respectively. Using the statistical independence of the two molecules, $\langle xy\rangle = \langle x\rangle\langle y\rangle$ and noting that $\langle x\rangle = \langle y\rangle = x_0$, the expression becomes $\langle(x-y)^2\rangle = 4Dt_{\{x,y\}}$. This is an expression for the *average* behavior of a pair of molecules. A corresponding expression for the *particular* behavior of two molecules can be written as $(x-y)^2 = 4D_{\{x,y\}}t_{\{x,y\}}$ where $D_{\{x,y\}}$ is a coefficient valid for that particular event. If the medium the molecules inhabit is fairly homogeneous, or if the molecules can be assumed to typically occupy similar parts of the medium for a similar amount of time, then $D_{\{x,y\}} \simeq D$. With this, Eq. (19) can be rewritten as

$$\tau_D = \frac{1}{8t_{\{x,y\}}}\left[\frac{\hbar}{kT}\right]^2 \ . \tag{20}$$

## V. DIFFUSION IN COSMIC INFLATION

The purpose of this section is to describe how cosmic inflation relates to Brownian motion. It is not intended to present a thorough description of cosmic inflation. In the present description of cosmic inflation there are multiple Big Bang occurrences, and in each of these occurrences baby-universes are created [1]. One of the baby-universes is our own universe. In order to describe cosmic inflation it is helpful to explain what is being inflated. The behavior of spacetime is characterized by the relation between differences in time and differences in spatial location, and can be represented by the behavior of a single characteristic time varying scale parameter $a$ which appears in the line element which is characteristic of spacetime. The Hubble parameter is the fractional change of $a$ with time: $\mathbb{H} = \frac{1}{a}\frac{da}{dt}$. Inflation describes the early epoch period of rapid growth of $a$. During inflation $\mathbb{H}$ is approximately constant at a value roughly of the order $\mathbb{H} \cong 10^{34}\mathrm{s}^{-1}$ which makes $a$ approximately proportional to $\mathrm{e}^{\mathbb{H}t}$. Inflation comes to an end when $\mathbb{H}$ begins to decrease rapidly. The energy stored in the vacuum-like state is then transformed into thermal energy, and the universe becomes extremely hot. From that point onward, its evolution is described by the hot universe theory.

To correctly describe Brownian behavior during cosmic inflation, it is convenient to distinguish between two horizons: the particle horizon and the event horizon. The particle horizon delimits what an observer at a time $t$ can observe assuming the capacity to detect even the weakest signals. The event horizon delimits the part of the universe from which we can ever (up to some maximal time $t_{max}$) receive information about events taking place now (at time $t$). The particle and event horizons are in a certain sense complementary. In an exponentially expanding universe, the radius of the event horizon is equal to $c\mathbb{H}^{-1}$ where $c$ is the speed of light in vacuum. In an exponentially expanding universe, any two points that are more than a distance $c\mathbb{H}^{-1}$ apart will move away from each other faster than $c$, meaning that those two points will never observe each other. They might belong to the same baby-universe if they come from the same Big Bang, but the points will lie beyond each other's particle horizons.

As described in Ref. [1], cosmic inflation leads to the creation of multiple baby-universes one of them our own. Some of those universes will have physical behaviors very different from the behavior of our universe, but we will now consider the behavior of quantum fluctuations in the cosmic inflation model. The scalar inflaton field $\varphi$ (sometimes identified with the Higgs field, although this is controversial) is represented as [1]

$$\varphi(\mathbf{x}, t) = (2\pi)^{-\frac{3}{2}} \int d^3p \left[a_p^+ \psi_p(t) e^{i\mathbf{px}} + a_p^- \psi_p^*(t) e^{-i\mathbf{px}}\right] . \tag{21}$$

The $(2\pi)^{-\frac{3}{2}}$ term is simply a normalization factor, $\int d^3p$ is the integration over all possible values of the momentum, $a_p^+$ creates a field with momentum $\mathbf{p}$ parameter with a probability modulated by $\psi_p(t)$ and propagating in spacetime as the wave $e^{i\mathbf{px}}$, and $a_p^-$ destroys that same field.

The first cosmic inflation models considered that $\varphi$ was a classical field (meaning non-quantum). The way a quantum system becomes classical is through the process of decoherence. As described in the previous section, the process of decoherence has strong similarities to Brownian motion. Ref. [1] describes the similarity of the behavior of $\varphi$ during cosmic inflation and Brownian motion.

As it is typical in Brownian motion, the diffusion of the field $\varphi$ can be described by the probability distribution $P(\varphi, t)$ of finding the field $\varphi$ at that point in instant $t$. In Eq. 7.3.17 of Ref. [1] it is found that

$$\frac{\partial P(\varphi, t)}{\partial t} = D\frac{\partial^2 P(\varphi, t)}{\partial \varphi^2} . \tag{22}$$

Using Eq. (22), Ref. [1] shows that

$$\langle \varphi^2 \rangle = 2Dt \tag{23}$$

as is expected in diffusion processes (Eq. 7.3.12 in Ref. [1]) and that

$$D = \frac{\mathbb{H}^3}{8\pi^2 c^2} . \tag{24}$$

## VI. QUANTUM DARWINISM APPROACH TO COSMIC INFLATION

The way a quantum system becomes classical is through the process of decoherence. According to quantum Darwinism, in the high temperature limit of quantum Brownian motion, the decoherence is caused by the molecular interaction. After the Big Bang and during cosmic inflation the temperature is extremely high, so it is possible that the Brownian process of the baby-universe before and during the cosmic inflation described in Ref. [1] entails the extinction of the non-decohered universe states.

The Big Bang proposes to describe the creation of an observable universe from "nothing", and so a Darwinian perspective of it (such as the approach used here which is based in quantum Darwinism) will have a FAS-component that is a lot smaller than its DAS component and so it would be very Darwinian (see Appendix). A Darwinian evolution is a Brownian evolution where extinction might occur; and so this study of the relation between decoherence (extinction of some quantum states) and diffusion (Brownian motion) of baby-universes is a study of Darwinian processes occurring during cosmic inflation.

Solving the diffusion equation (22) during cosmic inflation, one obtains the probability for creation of a universe with a certain vacuum energy. Summing over all topologically disconnected configurations of just-created universes enables one to obtain the probability for creating universes with a certain cosmological constant value [1], causing Linde to write that although "*it is often supposed that the basic goal of theoretical physics is to find exactly what Lagrangian or Hamiltonian correctly describes our entire world. ... one could well ask ... if the concept of an observer may play an important role not just in discussions of the various characteristics of our universe, but in the very laws by which it is governed.*" The answer proposed here to Linde's question is that if the quantum Darwinism approach is applied to cosmic inflation, then the laws of physics are themselves the result of a Darwinian evolution of quantum systems.

## VII. RESULTS

We use Eq. (20) to generalize the results obtained for molecules in quantum Brownian motion to baby-universes undergoing Brownian motion during cosmic inflation. The decoherence time $\tau_D$ is then a time duration referring to two baby-universes, with $t$ being the time since they last interacted (typically the last time they were at the same place would be at the beginning of the Big Bang). The decoherence time is obtained as:

$$\tau_D = \frac{1}{8t}\left[\frac{\hbar}{kT}\right]^2 . \tag{25}$$

The difference in the approaches by Linde and by Zurek, which can be linked to the differences between the axiomatic systems FAS and DAS (see Appendix), implies different outcomes for the non-observed states. The FAS/Linde approach considers that the outcomes incompatible with the observed outcome exist in different multi-verses, while the DAS/Zurek approach considers the outcomes incompatible with the observed outcome to have become non-existent. In Zurek's approach information-transmission is what enables existence [20], while in quantum gravity existence (expressed as the number of quantum particles) is observer dependent and thus can only be understood as a relational concept [1, 5].

The representation of cosmic inflation using a diffusion process in a de Sitter space allows to consider thermal equilibrium with [1, 21]

$$T = \frac{\hbar\mathbb{H}}{k} \tag{26}$$

so that Eq. (25) becomes

$$\tau_D = \frac{1}{8t}\left[\frac{1}{\mathbb{H}}\right]^2 . \tag{27}$$

This result implies that during the duration of cosmic inflation, the decoherence time is much smaller than the cosmic inflation duration. In cosmic inflation, there is a growth by typically at least a factor of $e^{60}$, starting at $t = 10^{-35}$s and ending at $t = 10^{-32}$s, so that $\Delta t$ is about $10^{-32}$s. Therefore, since $\mathbb{H}\Delta t = 60$, the Hubble parameter $\mathbb{H}$ is about $10^{34}\mathrm{s}^{-1}$. Decoherence occurs when the time $t$ reaches $\tau_D$, so $t = \tau_D$ in Eq.(27) yields $t \simeq 10^{-34}\mathrm{s}^{-1}$, a time well before the end of inflation.

So even if the baby-universes were in a quantum coherent state at the beginning of inflation, they would decohere after a small fraction of the duration of cosmic inflation. The present result agrees with Martineau's observation [11] that decoherence is extremely effective during inflation, but we reach that conclusion more easily. The approach to "decoherence during Brownian motion" used by Zurek considers that the effect of zero-point vacuum fluctuations is neglected. Kiefer et al. [14] propose that the inclusion of zero-point vacuum fluctuations makes decoherence still effective but no longer complete, meaning that a significant part of primordial correlations remains up to the present moment.

## VIII. DISCUSSION

Obtaining values for the decoherence time requires knowledge of the value of the Hubble parameter before and during inflation. Values of the Hubble parameter have a large range, and the measurement of its value is a topic of current research [1]. The existence of baby-universes is also a not yet established observational fact [1]. Thus, obtaining experimental proof of Eq. (25) and Eq. (27) is not yet possible. But if baby-universes exist, and if more information is obtained about the time-dynamics of the Hubble parameter, the relation between Hubble parameter and decoherence time expressed in Eq. (25) and Eq. (27) would be likely to become useful.

A characteristic of biological Darwinism is the existence of a first cell. The approach to cosmic inflation described in Ref. [22] indicates that the inflating region of spacetime must have a past boundary. This implies that it is not possible to use cosmic inflation as a mechanism for avoiding the occurrence of a primordial Big Bang for which it is not possible to define a time preceding it. The work in Ref. [22] does not refute the possibility that there were other Big Bangs before the most recent one; but it shows that even if there existed other Big Bangs before, there must have necessarily occurred a primordial Big Bang that started from "nothing" or very close to "nothing". The past boundary marks the transition from a zero amount of information situation (the "nothing" state) to one where information exists (the "something" state).

In this work a relation between Quantum Darwinism and HAS is presented (see Appendix). The smaller the amount of information in the beginning/axiomatic state of the HAS, the more the HAS will behave as a DAS as opposed to a FAS. The Quantum Darwinism treatment of the Big Bang would therefore correspond to a process that is extremely DAS-like, and would be even more DAS-like for the Big Bang where the past boundary occurred.

## IX. CONCLUSION

The measurement described in Eq. (25) and Eq. (27) obtains what the physical constants (and laws) will be for a certain baby-universe by a Darwinian extinction of the other possible values. That the measurement occurring during cosmic inflation is the selector of the physical constants is already proposed in section 10 of Ref. [1], but the approach proposed here is different in that it proposes the Darwinian extinction of the non-obtained quantum alternatives that are not moving away at a speed faster than $c$.

To summarize, an expression was obtained for the time after which different previously entangled baby-universes would decohere.

### Acknowledgments

Nachum Dershowitz, Jean-Claude Zambrini, Juan Sanchez, Juan Pablo Paz, and Thomas Durt were helpful by their comments and questions. This work was partially supported by Fundação para a Ciência e a Tecnologia: Programa Ciência 2007, CERN POCI/FP/81926/2007 and CERN/FP/83510/2008.

## APPENDIX: FORMAL AND DARWINIAN AXIOMATIC SYSTEMS

The random extinction of information in quantum Darwinism contrasts with the preservation of information in Hilbert's formal axiomatic systems (FAS) [3]. The FAS is a deterministic system where the consistency of the axioms is preserved, meaning that no proposition can be both true and not true; and the logic in the FAS obtains true propositions from true propositions. In Darwinian approaches (e.g. quantum Darwinism) if survival is identified with truth, then some true propositions lead to false propositions (which become extinct) and so Darwinism is not consistent. However, in Darwinism there are no true propositions that are not obtained from true propositions (all entities have parent entities that need to be true since they gave offspring); meaning Darwinism is necessarily complete. Gödel's incompleteness theorems showed that a non-trivial FAS cannot be both complete and consistent [3, 18].

An axiomatic system made to be complete and not consistent would have validly inferred propositions being both true and not-true. A way of dealing with this difficulty would be to validate propositions not by the valid application of inference rules, but by using a proof-checking algorithm that would eliminate propositions that are inconsistent within themselves. Such a process of selecting valid propositions is called here a Darwinian axiomatic system (DAS). The FAS and the DAS are the two extreme ways of dealing with Gödel's incompleteness theorems, respectively the consistent and the complete forms. It is possible to conceive an hybrid axiomatic system (HAS) between the FAS and the DAS. A FAS is constituted by alphabet, grammar, axioms, rules of inference, and a proof-checking algorithm. In the FAS approach to mathematics, one starts with axioms considered as self-evident and built using the alphabet and the grammar; then the rules of inference are applied to the axioms and all the theorems (logical inferences of the axioms) are obtained. A proof-checking algorithm checks if a proof follows all the rules by doing reverse inference starting from the proof's result and checking if what is obtained are the axioms. Gödel's incompleteness theorems showed that a non-trivial FAS cannot be both complete and consistent [3, 18].

Axioms in FAS are typically made to be consistent so that the FAS is consistent, but a FAS cannot be both consistent and complete. Zurek's quantum Darwinism is an attempt to balance the consistency of the Schrödinger-Copenhagen approach with the completeness of the Wheeler-Multiverse approach [24]. The approach proposed here is that quantum Darwinism is a HAS that has successfully mixed the FAS-like Schrödinger-Copenhagen approach with the completeness characteristic of the Wheeler-Multiverse approach. Moreover, we also propose that quantum Darwinism looks very much like a DAS when no amount of information is being preserved from the past. The physical situation that most closely resembles the creation of information from "nothing" (state of zero amount of information) is the Big Bang, which is why we chose to first apply the concept of the dichotomy between the FAS and DAS in the context of the Big Bang.

To Chaitin's information-based Gödel incompleteness conclusion [3] that real numbers are non-computable with probability 1, quantum Darwinism answers through a discrete universe. In mathematical randomness [3] the value of a random variable is only known by running a computer, and in quantum Darwinism the value of a random quantum variable only occurs if the interaction in an experiment is strong enough [19]. The quantum randomness [17, 19] concept is identical to the mathematical randomness [2] concept if the quantum systems' existence is enabled through their transmission of information, which occurs in quantum Darwinism. The existence's dependence on information is the part of the Existentialist philosophical structure added by quantum Darwinism [17].

Gödel's incompleteness theorems propose to describe the difficulties of creating a mathematical formalism from "nothing", i.e. from limit of zero information, using Hibert's FAS [3, 18], which is a deterministic approach. Quantum Darwinism proposes to address the creation of classical reality from a quantum reality, using a Darwinian approach. The deterministic/FAS/consistent and the Darwinian/DAS/complete approach to creation can be considered as the two extreme approaches of dealing with Gödel's incompleteness theorems.

---

# Towards a hierarchical definition of life, the organism, and death

**Gerard A.J.M. Jagers op Akkerhuis**
Wageningen University and Research Centre, Alterra, Centre for Ecosystem Studies
P.O. Box 47, 6700 AA, Wageningen, The Netherlands
Tel: +31-317-486561
E-mail: gerard.jagers@wur.nl
Website: www.hypercycle.nl

## Abstract

Despite hundreds of definitions, no consensus exists on a definition of life or on the closely related and problematic definitions of the organism and death. These problems retard practical and theoretical development in, for example, exobiology, artificial life, biology and evolution. This paper suggests improving this situation by basing definitions on a theory of a generalized particle hierarchy. This theory uses the common denominator of the "operator" for a unified ranking of both particles and organisms, from elementary particles to animals with brains. Accordingly, this ranking is called "the operator hierarchy". This hierarchy allows life to be defined as: matter with the configuration of an operator, and that possesses a complexity equal to, or even higher than the cellular operator. Living is then synonymous with the dynamics of such operators and the word organism refers to a select group of operators that fit the definition of life. The minimum condition defining an organism is its existence as an operator, construction thus being more essential than metabolism, growth or reproduction. In the operator hierarchy, every organism is associated with a specific closure, for example, the nucleus in eukaryotes. This allows death to be defined as: the state in which an organism has lost its closure following irreversible deterioration of its organization. The generality of the operator hierarchy also offers a context to discuss "life as we do not know it". The paper ends with testing the definition's practical value with a range of examples.

**Contents**

**I. Introduction**

In a chronological overview of developments, Popa (2003) presents about 100 definitions of life, meanwhile demonstrating that no consensus exists. Many classical definitions include long lists of properties, such as program, improvisation, compartmentalization, energy, regeneration, adaptability and seclusion (Koshand Jr. 2002) or adaptation, homeostasis, organization, growth, behavior and reproduction (Wikipedia: Life). Most properties in such lists are facultative; it is still possible to consider an organism a form of life when it does not grow, reproduce, show behavior, etc. The inclusion of facultative aspects is a source of lasting difficulty in reaching consensus on a definition of life. Because of the seeming hopelessness of the situation, certain scientists have adopted a pragmatic/pessimistic viewpoint. Emmeche (1997) christened this viewpoint the "standard view on the definition of life". He suggests that life cannot be defined, that its definition is not important for biology, that only living processes may be defined and that life is so complex that it cannot be reduced to physics. Others have warned that a comprehensive definition of life is too general and of little scientific use (e.g. van der Steen 1997).

In their search for a definition, other scientists have focused on properties that are absolutely necessary to consider an entity life. In this context Maturana & Varela (1980, p. 78) have proposed the concept of autopoiesis (which means "self making"). They use the following definition: "*An autopoietic machine is a machine organized (defined as a unity) as a network of processes of production (transformation and destruction) of components which: (i) through their interactions and transformations continuously regenerate and realize the network of processes (relations) that produced them; and (ii) constitute it (the machine) as a concrete unity in space in which they (the components) exist by specifying the topological domain of its realization as such a network.*" Special about the autopoietic process is, that it is "*closed in the sense that it is entirely specified by itself* (Varela 1979 p. 25)".

The concept of autopoiesis has increasingly become a source of inspiration for discussions in the artificial life community about how to define life (Bullock et al., 2008). Reducing the number of obligatory traits defining life to just one, autopoiesis is a rather abstract concept. People have sought, therefore, to describe some of the processes that underlie autopoiesis more specifically. An example of such a description is a triad of properties defining cellular life: container (cell membrane), metabolism (autocatalysis) and genetic program (e.g. Bedau 2007).

These descriptions, however, have not resulted in a consensus definition of life. This has led Cleland & Chyba (2002, 2007) to suggest that a broader context, a "theory of life", is required. In line with a broader framework, life may be regarded as a special realization of the evolution of material complexity. According to Munson and York (2003), considering life in a general evolutionary context requires arranging "*all of the phenomena of nature in a more or less linear, continuous sequence of classes and then to describe events occurring in the class of more complex phenomena in terms of events in the classes of less complex phenomena..* ". An important property of such a hierarchy would be that *"…an increase in complexity is coupled with the emergence of new characteristics … suggesting that the hierarchical arrangement of nature and the sciences is correlated with the temporal order of evolution*". Similar views for integrating material complexity and the evolution of life can be found, for example, in the work of Teilhard de Chardin (1966, 1969), von Bertalanffy (1968), Pagels (1985), Maynard Smith & Szathmáry (1995, 1999) and Kurzweil (1999).

In contribution to these discussions, the present author has published an evolution hierarchy for all "particles". The latter hierarchy uses the generic word "operator" to address both physical (e.g. quark, atom, and molecule) and biological particles (e.g. prokaryote cell, eukaryote cell, and multicellular). The word operator emphasizes the autonomous activity of the entities involved, which "operate" in a given environment without losing their individual organization. The hierarchical ranking of all operators is called the "operator hierarchy" (see Figure I)(Jagers op Akkerhuis and van Straalen 1999, Jagers op Akkerhuis 2001, Jagers op Akkerhuis 2008 and the author's website www.hypercycle.nl). Because the operator hierarchy is important for the definition of life proposed below, the outlines of this theory are summarized in the following lines

The operator hierarchy ranks operators according to the occurrence of a circular pattern, such as that which connects the beginning and end of a process or structure. Circularity causes a closed organizational state, also referred to as "closure" (for discussions of closure see, for example, Heylighen 1990, Chandler and Van de Vijver 2000). Because closure causes a discrete

“quantum” of organization (e.g. Turchin 1977, 1995 and Heylighen 1991), the operator becomes an “individual entity”, a “whole” or a “particle”, while still retaining its construction of smaller elements. Closure thus defines the operator‘s complexity level and sequential closures imply a higher complexity level. An operator‘s closure is the cause of its existence and typical for its complexity. This implies that complexity is not measured in terms of the number of genes, functional traits or organs of an organism, but in a very abstract way, in terms of the number of closures. Upon losing its closure, the organization of the operator falls back to that of the preceding operator. The actual shape of a closure can differ. Biological examples of closure are the cell membrane and the circle of catalytic reactions allowing the cell to maintain its chemical machinery. It is essential for a strict ranking that a lower-level and a higher-level operator always differ by exactly one closure level. The single closure (eukaryotic cell) or a parallel pair of closures (autocatalysis plus membrane of the cell) that define the next level are referred to as “first-next possible closure(s)”. A consequent use of first-next possible closures allows physical and biological operators to be ranked according to the “operator hierarchy” (Figure I). The operator hierarchy includes quarks, hadrons, atoms, molecules, prokaryotic cells, eukaryotic cells, multicellulars (e.g. plants, fungi) and “animals”, the latter representing an example of the operators that possess a neural network with interface and that are called “memons” in the operator hierarchy.

Due to its focus on closure, the operator hierarchy represents an idealization because it excludes potential transition states in between two closures. For example, several hundreds of metal atoms may be required before a functional Fermi sea transforms a collection of single atoms into a metal grid. Also, the emergence of multicellularity (discussed in detail in §III below) may require a colonial, multicellular state in between the single cell and the multicellular operator. The above shows that transition states form natural intermediate phases in the emergence of closures. The operator hierarchy does not include these transition states, however, because its hierarchical ranking is exclusively based on entities that already show first-next possible closure.

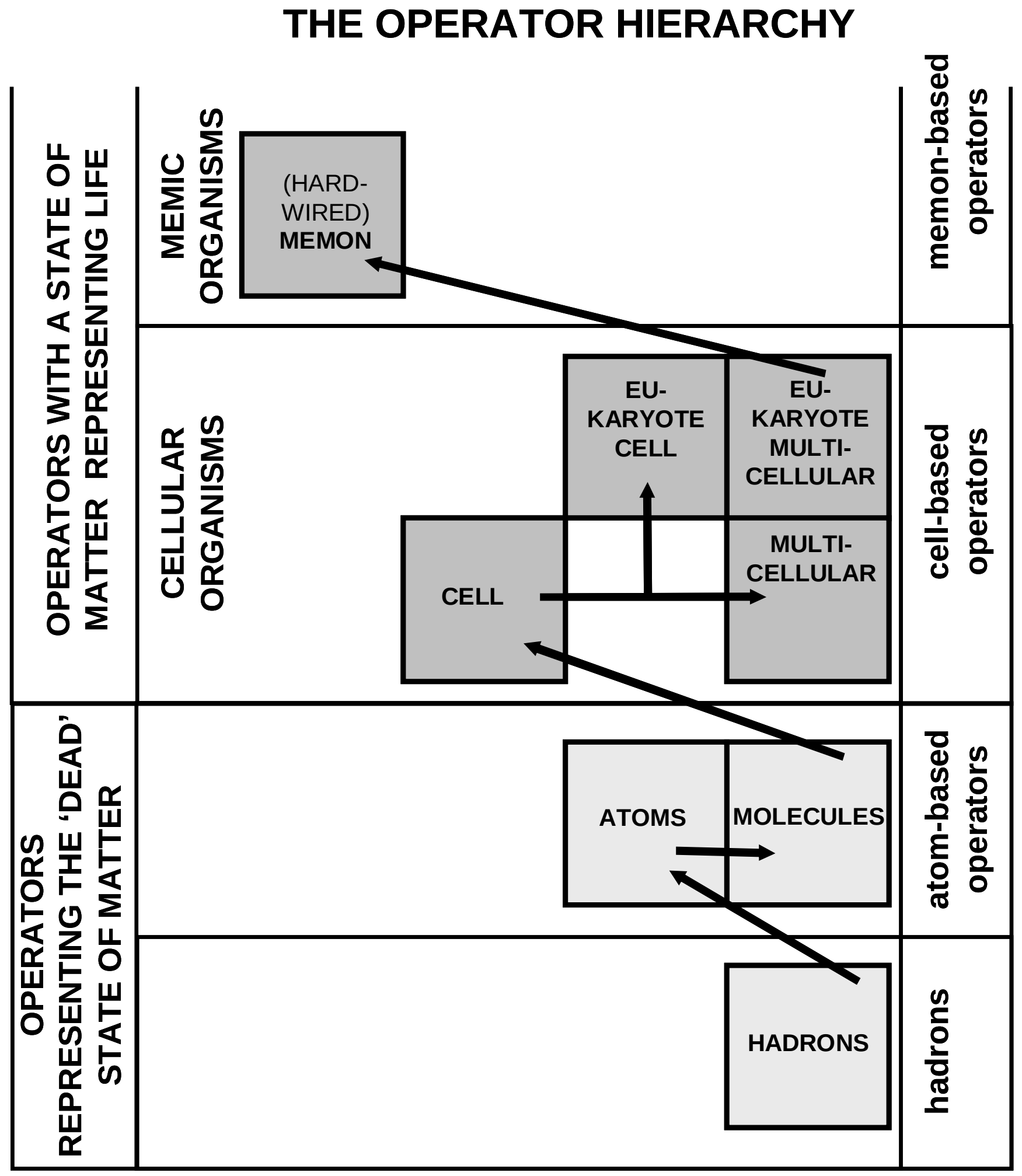

Figure 1: Using the operator hierarchy to define life and organisms. Arrows indicate how closures create operators (more information can be found in Jagers op Akkerhuis 2008, and the author's website www.hypercycle.nl).

The main reason for writing this paper, and adding yet another definition of life to the listings, is that the operator hierarchy offers several advantages in solving definition problems. First, the definitions of the operators are generally applicable because they focus on the essences of organization. For example, demanding autocatalysis leaves open which specific catalysts will perform the process. Second, the use of first-next possible closures ensures a critical filtering of only obligatory properties from property lists. Finally, the use of the operator hierarchy makes it easy to develop a hierarchy-based definition of life. In other words, the operator hierarchy offers a novel path for structuring and simplifying discussions about which entities are life.

The following paragraphs discuss different aspects of existing definitions of life and examine new ways to define the organism, living and death. At the end, a test of the practical value of the present definitions for the solving of a range of classical problems, such as a virus, a flame, a car, a mule and a mitochondrion, will be presented.

**II. Defining life and the organism**

Before discussing the use of the operator hierarchy for defining life, living and the organism, it is important to note that when talking about definitions, care should be taken that "*a definition is a series of superimposed language filters and only the definiendum (the term to be defined) can* *penetrate it*" (Oliver and Perry 2006). Problems may arise when the words used for the definiendum and for the filter have a broad meaning or have different meanings in different contexts. It is thus useful to elaborate on the current context for "life" before continuing.

"Life" has different meanings in different contexts. For example, people refer to the period between birth and death as their life (this is the best day of my life) even though lifetime would be more correct. In addition, the experience of "being alive", or "living", also carries the label of life (to have a good life). Other uses of life holistically refer to the importance of selective interactions in ecosystems that over generations lead to better-adapted life forms (the evolution of life). Ruiz-Mirazo et al. (2004) have proposed a definition of the latter type. They state that life is "*a complex collective network made out of self-reproducing autonomous agents whose basic organization is instructed by material records generated through the evolutionary-historical process of that collective network*". In philosophy, life is sometimes considered a graded concept for being because all what is, is alive in the measure wherein it is (Jeuken 1975). Due to the contextual dependence of these and other interpretations, it is improbable that a general definition of life can be constructed. Van der Steen (1997) indicates that even if such an overly general definition existed, it would probably be difficult to applie it to specific situations.
To avoid problems with generality and multiple interpretations of concepts, the present study adopts a limited viewpoint, presuming a one-to-one relationship between a definition of life and a specific material complexity. In this context, life is an abstract group property shared by certain configurations of matter.
The operator hierarchy offers a context for a general matter-based definition of life. Focusing on all operators showing a complexity that exceeds a certain minimum level, the hierarchy suggests a definition of life *sensu lato* as: matter with the configuration of an operator, and that possesses a complexity equal to or even higher than the cellular operator. Only the prokaryote cell, the eukaryote cell, the prokaryote and eukaryote multicellular, the hardwired memon and the potential higher-level operators fit this definition (Figure 1). In addition to this general definition, various specific definitions are possible by focusing on operators that lay between a lower and an upper closure level. An example of a specific definition is one describing cellular life (e.g. algae, plants and fungi) as: matter showing the configuration of an operator, and that possesses a minimum complexity of the cellular operator and the maximum complexity of a multicellular operator. The latter includes only the cell, the eukaryotic cell, the prokaryotic and the eukaryotic multicellular. It is possible to choose any of these approaches for defining living as: the dynamics of an operator that satisfies the definition of life.

The above approach results in a strictly individual based definition of life as a group property of certain operators. This definition has the advantage, that it offers a solid basis for defining the creation of offspring. Subsequently, the evolution of life can be dealt with as an emergent process occurring in any system with interactions between individual living entities that lead to differential survival of variable offspring, produced either without or with recombination of parental information.

The organism is the key ontological unit of biology (Etxeberria 2004, Korzeniewski 2004) and is also referred to as a "living individual". Understanding the latter requires insight into what is "living", and what is an "individual". By defining "living" as the dynamics of those operators that satisfy the definition of life, the operator hierarchy uses operators instead of individuals because operators define a being or an individual more strictly than the Latin concept of *individuum*. The word individuum stands for an "indivisible physical unit representing a single entity". This definition leaves a great deal of room for choice of the elements that form the physical unit and for the rules that determine indivisibility. These indeterminacies may be the reason for the discussion about whether certain life forms are organisms. Townsend et al. (2008) use the phrase "unitary organism" to indicate the individual organism. However, certain jellyfish, for example, the Portuguese Man O" War (*Physalia physalis*), look like individuals, but consist of differentiated individuals, each with its proper neural network (e.g. Tinbergen 1946). In the operator hierarchy, the latter jellyfish are colonies, not organisms, because each contributing individual has its proper neural network as its highest emergent property, and the colony still lacks a recurrent interaction of the neural interfaces of the individuals.

The operator hierarchy now suggests a way to create congruency between the definition of life and the definition of the organism by accepting as organisms only entities that fit the operator-based definition of life. For example, using the general definition of life, only the cells, the eukaryotic cells, the prokaryotic and eukaryotic multicellulars and the memons are organisms.

**III. Levels of life**

*a. The cell.* The most important properties of the cell are the autocatalytic set of enzymes and the membrane. The autocatalytic set shows reproduction as a set. Every molecule in the set catalyzes a reaction that produces some other molecule in the set until any last reaction product closes the cycle. In different ways, reproduction as a set is part of various theories about the origin of life (e.g. Rosen 1958, 1973, 1991, Eigen 1971, Gánti 1971, Eigen and Schuster 1979, Kauffman 1986, 1993, Bro 1997, Kunin 2000, Hazen 2001, Martin and Russell 2002, Hengeveld and Fedonkin 2007).

Autocatalysis demands that a cell can potentially autonomously sustain its catalytic closure. Accordingly, if a cell allocates a part of its autocatalytic closure to another cell, the cell is no longer an operator. An example of the latter is the mitochondrion. It is generally accepted that mitochondria started the interaction with their host cells as autonomous endosymbiontic α-proteobacteria. Over many generations, these bacteria transferred more than 90 percent of their catalytic control to their host (Allen 1993, Berg and Kurland 2000, Searcy 2003, Capps et al. 2003, Lane 2005). The loss of the potential of autocatalysis implies that mitochondria have become a special kind of organelle.

In addition to autocatalysis, the operator hierarchy demands an interface because a set of autocatalytic enzymes only gains the physical individuality that allows its maintenance when it functions in a limited space, the limits being part of the system. The integration of autocatalysis and the membrane is part of various important theories, for example, the theories of autopoiesis (Varela 1979) and of interactors (Hull 1981).

*b. The eukaryote cell*. A single cell has two dimensions for creating a next closure. One is to create cooperation between cells, which leads to multicellularity. The other is to create an additional closure mediating the hypercyclic functioning of the cell in the form of the nucleus. Interestingly, it is quite likely that the most important complexity boundary in cell biology, that between prokaryotic and eukaryotic cells, thanks its existence to the energy boost and genetic enrichment offered by endosymbionts. With respect to the emergence of eukaryotic cells, theories roughly divide along two major lines depending on whether the nucleus or the endosymbionts emerged first. In addition to other aspects, support for the nucleus-first hypothesis comes from allegedly primitive eukaryotes that show a nucleus without harboring endosymbionts. Genetic analyses (Rivera 1998) and observations of endosymbiont traces (Clark 1995), however, suggest that the "primitive eukaryotes" are recent developments that lost their endosymbionts in a process of evolutionary specialization. The endosymbiont hypothesis advocates that a merger between a methanogenic bacterium that was member of the archaea and an α-proteobacterial endosymbiont created the eukaryotic cell (Martin and Russel 2002). Subsequent transmission of genes for membrane creation from the endosymbiont to the host allowed it to produce membranes that formed the basis for the engulfment of the nucleus. Whatever the actual path taken by evolution, the operator hierarchy focuses on the occurrence of closure involving both structural and functional aspects of the host cell, resulting in an internal interface for the autocatalytic set and the mediation of its functioning. Even though endosymbionts may become obligatorily integrated in the functioning of their host cell by the transfer of part of their genetic regulation to the host cell, they do not mediate the functioning of the autocatalytic set of the host nor form an interface for its functioning. For this reason the operator hierarchy does not regard endosymbiosis, but the nucleus as the relevant closure that defines the limit between prokaryotes and eukaryotes.

*c. The multicellular*. When does a group of cells become a multicellular operator and, according to the above definition, an organism? In the operator hierarchy, multicellularity involves a structural and a functional component represented by structural attachment of cells and an obligatory recurrent pattern of functional interactions between them. As such, it is possible to define a multicellular operator (a multicellular organism *sensu stricto*) as: a construction of mutually adhering cells showing obligatorily recurrent interactions based on the same interaction type, that has the potential of maintaining its functioning as a unit and that does not show memic structure.

Multicellularity has developed independently in many branches of the phylogenetic tree (reviews by, for example, Bonner 1998, Kaiser 2001, Grosberg and Strathmann 2007) presumably because it is associated with a range of evolutionary advantages. Multicellularity increases mobility and access to resources, and reduces predation, and finally yet importantly, the cells in genetically uniform multicellulars share the same genes and do not have to compete with each other for reproduction. Willensdorfer (2008) indicates that the alleviation of reproductive competition allows for a division of labor because "*cells can specialize on non-reproductive (somatic) tasks and peacefully die since their genes are passed on by genetically identical reproductive cells which benefited from the somatic functions*".

In some cases a multicellular organism results from the aggregation of individually dwelling unicellulars (for example, true slime molds, Ciliates and Myxobacteria). More generally, a multicellular organism develops when daughter cells cohere after cell division. A simple, temporary form of multicellular life is present in slime molds. Here, genetically-different, individually-dwelling cells aggregate and bind using membrane proteins to form a colonial state in which the cells intercellularly communicate by diffusion. At a certain moment, obligatory interactions between cells lead to the formation of irreversible cell differentiation producing a reproductive structure. During this state, the slime mold cells are temporarily a multicellular organism.

With the evolutionary development of plasma connections, advanced multicellular life became possible. Plasma connections allow efficient and rapid intercellular communication, involving electrical signals, chemical signals and nutrient transport (Mackie et al. 1984, Peracchia and Benos 2000, Nicholson 2003, Panchin 2005). Plasma connections have evolved in several lineages of multicellulars. Plasma connections between animal cells depend on gap junctions, between plant cells on plasmodesmata, in blue-green algae on microdesmata, and in certain fungi or in developing insect eggs on incomplete cell walls. The evolution of gap junctions some 700 million years ago coincided with an explosion of multicellular life forms.

Multicellular organisms may go through life stages that are not multicellular. For example, sexual reproduction involves single-celled egg and semen. Furthermore, during the two-, four- and early eight-cell stages most vertebrate embryos have loosely attached cells without obligatory dependency. Accordingly, they represent a colony. When separated from the colony, the cells show a normal development. Early separation of embryonic cells is the reason why identical twins exist. Embryo cells in the early stages can even mix with another embryo's cells of the same age and develop into a normally functioning organism, called a chimera, in which some organs and tissues belong to a different genotype than others. A definition of life should, therefore, respect that an organism's cells may differ in genotype. From the late eight-cell stage, the development of gap-junctions marks the emergence of regulation as a unit, which makes the cellular colony a multicellular.

The realization of a multicellular's potential for maintenance depends on prevailing conditions. For example, a tree twig that is stuck in the ground may become a tree again if the weather is not too warm, too cold, or too dry, etc. and if the twig has the genetic potential for regeneration and is large enough, in good condition, etc.. Whether the twig is an organism depends on its potential to show all dynamics required for being a multicellular operator. This potential is in principle gene-based, but it depends on the condition of the phenotype and the environment for its realization.

Sometimes two multicellular organisms show symbiosis, such as plants living in close association with mycorrhiza fungi in their roots. As the fungus and the plant already are multicellular on forehand, a plant with mycorrhiza represents an interaction between two multicellular organisms.

*d. The memon*. Attempts to define life frequently focus on the typical properties of the first cell. The underlying assumption may be that all organisms consist of cells and that, for this reason, the definition of the living properties of cells will automatically cover other, more complex organizations. According to the operator hierarchy, this reasoning is incomplete because, with respect to artificial intelligence, it unsatisfactorily excludes technical life *a priori* . The reason is that the fundamental construction of the brain is not principally different when built from cellular neurons, technical neurons (small hardware acting as a neuron) or programmed neurons (virtual devices modeled to act as neurons). Even though all organisms on earth currently consist of cells or show neural networks that consist of cells, the fact that technical memons may, one day, have a brain structure similar to cellular memons implies that a general definition of life must consider the possibility of technical memons.
Memons show a neuron network and a sensory interface. The basic neuron-units have been named categorizing and learning modules or CALMs and allow for a recurrent network of CALMs (Murre, Phaf and Wolters 1992, Happel 1997). The interface includes sensors that allow the memon to perceive its body and environment, and effectors that allow it to move the cellular vehicle it resides in. The interface and vehicle co-evolved during the evolution of neural networks. In principle, it is possible to construct a functional memon from any kind of technical hardware that provides the required neural architecture. This is the reason that the study of neural networks in biology shows a fundamental overlap with research on technical artificial intelligence. The recognition that memons show a recurrent network of CALMs surrounded by an interface allows Siamese twins with separate brains to be classified as two memons sharing the same vehicle and showing in this vehicle a partial overlap of their interfaces.

**IV. No life, no reproduction**
According to some authors (e.g. the Von Neumann & Burks, 1966) reproduction is a pre-requisite for life. Like the chicken and the egg problem, it can also be said that life is a pre-requisite for reproduction. Clearly, any decision on this matter critically depends on the context that is used to define life. If the operator hierarchy is used, the least complex life form is the prokaryotic cellular operator. Two arguments currently suggest that life is a pre-requisite for reproduction. The first states that even though all other organisms originate from the first cell by reproduction, the first cell itself had an inorganic origin. The emergence of the first cell thus shows that life does not obligatorily result from reproduction. The second argument posits that organisms do not need to show reproduction, i.e., producing offspring, to comply with the operator-based definition of life; The operator-based definition demands that organisms show two closures: autocatalysis and a membrane. Autocatalysis can be regarded as reproduction without creating offspring. As Jagers op Akkerhuis (2001) pointed out, autocatalysis implies that a cell autonomously creates a structural copy of its information, a process that is called “structural (auto-) copying of information”. Before answering the question of whether the structural (auto-)copying of the cell‘s information means that it must reproduce, it is important to detail the concept of information. For the latter, I suggest applying Checkland and Scholes (1990) definition of information to the autocatalytic set. These authors have defined information as data with a meaning in a context. In line with this reasoning, Kauffman (1993) proposed that, by selecting the autocatalytic process as the context, every catalytic molecule becomes a data-unit with a catalytic meaning (the “purpose” mentioned by Kauffman 1993, p.388) and represents a part of the information of the autocatalytic process. Following one round of autocatalysis, or more rounds to account for the loss of enzymes over time, the cell contains copies of all of its information. At that moment, it has autonomously performed structural copying of information and fulfills all the requirements of the operator hierarchy, even when it does not produce an offspring. Based on this reasoning, the capacity of autocatalytic maintenance is an obligatory requirement for cellular life and reproduction is a possible consequence.

The above implies that it is not relevant for a general definition of life to distinguish between life forms with or without replication, as Ruiz-Mirazo et al. (2004) has suggested. The latter authors distinguish “proto-life stages” that do not show a phenotype-genotype decoupling (soma with genes) from “real life” with genes. In line with the operator hierarchy based definitions, Morales (1998) warns that "*if reproduction is required: This is a troubling development, because it means that we could not tell whether something is alive unless we also know that it is the product of Darwinian evolution.*" The operator-based definition considers life as a prerequisite for reproduction instead of reproduction as a prerequisite for life. Consequently, worker bees, mules, infertile individuals and other non-reproducing organisms and/or phenotypes are life. This point of view also solves problems that may arise when demanding that memons be able to reproduce as a prerequisite for recognizing them as life forms. In fact, none of the cellular memons living today shows reproduction, at least not reproduction of their neural network structure determining their closure. The things they pass on during reproduction are the genes of their cells, allowing the development of a multicellular organism with a neural network, capable of learning but devoid of inherited neural information other than reflexes.

**V. Life holding its breath**

The above chapter shows that reproduction is not a prerequisite of life but a possible consequence of it. Going one step further, it can also be concluded that metabolism is not a prerequisite for life. Many taxa such as bacteria, protozoa, plants, invertebrates and vertebrates have developmental stages showing natural inactivity (seeds, spores) or reversible inactivation when submitted to desiccation, frost, oxygen depletion, etc. The inactive state carries the name of anabiosis, after the process of coming to life again (for a review of "viable lifelessness" concepts, see Keilin 1959). Another type of reversible inactivity showing marked similarity with anabiosis is the state of neural inactivity in memons following anesthesia. An anesthetic that blocks the transmission of signals between neurons while leaving the remaining metabolic activity of the neurons intact causes a reversible absence of neural activity that corresponds to an anabiotic state of the memon.

Even in the early days of the biological sciences, scholars discussed whether dried or frozen anabiotic stages are alive at a very slow pace, or whether they are truly static states of matter. In 1860, the famous Société de Biologie in Paris wrote a lengthy report on this subject (Broca 1860-1861). Quite importantly, this report concluded that the potential to revive an anabiotic stage is an inherent aspect of the organization of the material of which the object consists and that it is equally persistent as the molecular state of the matter forming the system. In short, the Société de Biologie found that "*la vie, c"est l"organisation en action*". Additional support for this conclusion came from Becquerel (1950, 1951) who subjected anabiotic stages to a temperature 0.01 degree above absolute zero, a temperature at which no chemical processes can occur, even not very slowly. Becquerel demonstrated that structure alone is enough to allow revival at normal temperatures. Anabiosis from absolute zero or complete desiccation has led to the conclusion that "*The concept of life as applied to an organism in the state of anabiosis (cryptobiosis) becomes synonymous with that of the structure, which supports all the components of its catalytic systems*" (Keilin 1959), or that "*life is a property of matter in a certain structure*" (Jeuken 1975). With respect to the question of: what certain structure?, the operator hierarchy suggests that all operators with a complexity similar to or higher than the cell answer this question.

**VI. Life as we do not know it**

Considerations about "life as we do not know it" depend on assumptions. As a context for such assumptions, the operator hierarchy offers two advantages. First, the operator hierarchy has its basis in the general principle of first-next possible closure. Second, the rigid internal structure of the operator hierarchy offers a unique guide for assumptions about life that we do not yet know .

Based on the general principle of first-next possible closure, the operator hierarchy shows a strict sequential ranking of the operators. Assuming that closures act as an absolute constraint on all operator construction, the operator hierarchy then has universal validity. Support for the latter assumption comes from the observation that, as far as we know, all operators with a complexity that is equal to or lower than the molecules seem to have a universal existence. If this universality extends to the biotic operators, the material organization of higher-level operators, such as cells and memons, may then possibly be found in the entire universe. Such universality would significantly assist in the search for exobiotic life forms because alien life may show similar organization to the life we do know, at least with respect to the first-next possible closures involved. The demand of closure still leaves a good deal of freedom for the physical realization of operators. On other planets, different molecular processes may form the basis of the autocatalysis and interface of the first cells. Similarly, the operator hierarchy poses no limits to the actual shape, color, weight, etc. of exobiotic multicellular organisms. Furthermore, even though the presence of neural networks may be required for memic organization throughout the universe, the operator hierarchy does not restrict the kind of elements producing these networks, or the details of the neural network structure other than demanding hypercyclicity and interface.

The rigid internal structure of the operator hierarchy allows predictions about the construction of life forms that have not yet evolved on Earth. Of course, any discussion of this subject involves speculation, but the operator hierarchy may well offer a unique starting point for such a discussion. In an earlier publication (Jagers op Akkerhuis 2001), I have indicated various future operator types with a higher complexity than the cellular hardwired memon. To minimize the aspect of speculation, I would like to discuss here only the memon immediately above the cellular hardwired memon (see fig. I), the so-called “softwired memon”. According to the operator hierarchy, this type of memon should be able to copy information structurally. This means that the organism should be able to copy all of its information by copying the structure of its neural network. At a lower level in the hierarchy, cells do this by copying their genetic molecules. Softwired memons can also do this. They are based on a virtual neural network that resides in computer memory arrays. During their operation softwired memons continuously track all their neurons, neural connections, connection strengths and interactions with the interface. It is therefore only a small step for softwired memons to read and reproduce all the knowledge in their neural network by copying these arrays. On these grounds, it may be deduced that softwired memons (or still higher complexity memons) form the easiest way to satisfy the demands of the operator hierarchy for the autonomous, structural copying of information. The operator hierarchy suggests therefore that life as we do not know it will take the shape of technical memons.

The above reasoning shows that the operator hierarchy offers clear criteria with respect to different forms of “artificial life”. The acceptance of an artificial entity as life is only possible when it shows all of the required properties of an operator. Referring to the difference between strong artificial life and weak artificial life, which do and do not consider a-life entities as genuine life, respectively, it would be fully in line with the present reasoning to consider as genuine life all a-life entities that fulfill the requirements for being an operator.

**VII. On life and death**

Given the present focus on states of matter, it is quite simple to define dead matter as: all operators that do not fit the general definition of life. It is more difficult, however, to define death.

Given the current point of view, death represents a state in which an organism has lost its closure. The use of closure in this definition helps prevent that “*…. the properties of an organism as a whole [would be confused] with the properties of the parts that constitute it*” (Morales 1998). However, organisms also loose their closure during transitions that are part of life cycles and that are not associated with the organism’s death. For example, the closure of the organism is lost and a new closure gained when the zygote exchanges its unicellular organization for the multicellular state of the embryo and when the multicellular embryo develops to a memic state. Is it possible to specify the loss of closure during death in a way that excludes closure losses during life cycles?

With respect to the above question of how to exclude the loss of closure during transitions in life cycles when defining death, the general process of deterioration offers a solution. During their lives, organisms deteriorate because of injury and ageing. The loss of closure marking death is always associated with the organism‘s irreversible deterioration. Demanding irreversible deterioration, therefore, helps to prevent that one would be tempted to consider, for example, a caterpillar as having died, when its tissues are reorganized during the transition via the pupae to a butterfly. Accordingly, it is possible to describe death as: the state in which an organism has lost its closure following irreversible deterioration of its organization.

Using the above definition, death may occur in either an early or late phase of the deterioration process, and following the death of multicellulars, a short or long period may pass until the organism's body parts become dead matter. The latter has its cause in the hierarchical construction of multicellular organisms. Accordingly, the loss of the highest closure implies a classification of the remaining body as an operator showing the first-next lower closure.

Death depends on the loss of closure. To illustrate the contribution of this statement to the analysis of death, the death of a memon can be used. Due to the memon's strongly integrated organization, death may occur at various levels that affect other levels. For example, the multicellular regulation may be the first to collapse due to the loss of liver functions. After a certain period, this will cause failure of neural functioning, the latter marking the memon's death In another situation, the neural functions may be lost first, and the memon is the first to die, dragging its body with it in its fall. However, sometimes enough neural activity may remain for a vegetative functioning of the memon's body as a multicellular unit. The vegetative state cannot maintain itself autonomously (in principle, a requirement for a multicellular organism) but it may continue given the right medical care. If this care is withdrawn, the multicellular body will start deteriorating after which the cells in the organs and tissues will start dying at different rates. At a certain point, the multicellular closure is lost, and separately surviving cells have become the next level operators to die. Physiological differences between cells now determine the period during which they can survive in the increasingly hostile habitat of the dead memon, which is cooling below the normal operating temperature of cells and which shows many adverse chemical changes such as the lowering of oxygen levels, the release of decay products of dead cells, etc. Shortly after the memon's death, it is possible to take intact body-parts, organs and cells from its body and sustain their functioning following transplantation to a favorable environment. For example, the offspring of cells from the cervix of Henrietta Lane are still cultured as He La cells in many laboratories.

**VIII. The inutility of property lists**

The above arguments and examples have explored the possibilities of using the operator hierarchy for creating coherent definitions of life, the organism, living and death. However, how should the outcome be evaluated? Have the present attempts led to definitions that could be generally accepted in the field? A way of evaluating this that has become rather popular is to check the results against lists of preset criteria. Those who want to evaluate the present approach in this way may want to examine the following lists of criteria.

Morales (1998) has published a list of properties for a definition of life that includes the following criteria: 1. Sufficiency (Does the definition separate living entities from non-living ones?), 2. Common usage (simple classification of easy examples), 3. Extensibility (Does the definition deal with difficult cases, such as viruses, mules, fire, Gaia, extraterrestrial life and robots?), 4. Simplicity (few ifs, buts, ands, etc.) and 5. Objectivity (Criteria are so simple that everyone applies them with the same result). Emmeche (1997) offers another criteria list for a definition of life that includes the following: 1. The definition should be general enough to encompass all possible life forms. (The definition should not only focus on life as we know it.), 2. It should be coherent with measured facts about life, (It should not oppose obvious facts.), 3. It should have a conceptual organizing elegance. (It can organize a large part of the field of knowledge within biology and crystallize our experience with living systems into a clear structure, a kind of schematic representation that summarizes and gives further structure to the field.), 4. The definition should be specific enough to distinguish life from obviously non-living systems. Emmeche (1997) furthermore states that a definition "should cover the fundamental, general properties of life in the scientific sense". Korzeniewski (2005) has also proposed a list of criteria for a cybernetic definition of life, and Poundstone (1984) has extracted further criteria for life from the work of von Neumann & Burks (1966). Oliver and Perry (2006) have suggested a list more or less similar to that of Emmeche (1997) focusing specifically on properties of a good definition.

With respect to the use of criteria lists, I agree with other authors (Maturana and Varela 1980, van der Steen 1997) that it is not necessarily an advantage if a theory performs well or a disadvantage if a theory performs poorly according to a list of criteria; an approach's value does not necessarily correspond to its performance in these types of checklists. The match depends on the similarity in major goals and paradigms and the creator's influence on the selection and definition of criteria in a given list. In addition, the selection of "favorable" lists can lead to false positives.

For the above reasons, I am convinced that it is only possible to evaluate the currently proposed definitions "the hard way", i.e., by critically examining the internal consistency and transparency of their logic. In this respect, the present approach has the advantage of a fundamental bottom-up construction. It starts with defining elementary building blocks, the operators, and their hierarchical ranking in the operator hierarchy. To recognize and rank the operators, the operator hierarchy uses first-next possible closures. In the resulting hierarchy, the definition of every higher-level operator depends, in an iterative way, on a lower-level "ancestor" until a lowest-level ancestral system is reached, which is presumably the group of elementary particles that according to the superstring theory may have a common basic structure. The result is a strict, coherent and general framework that is open to falsification: the operator hierarchy. Subsequently, the operator hierarchy offers a fundament to define a range of secondary phenomena, such as life, the organism, living and death. Because of the reference to the operator hierarchy, the present definitions are short, logical statements that show a high specificity with respect to whether a certain entity satisfies the definition (list of examples in the following §).

**IX. Testing the definition of life**

When using the operator hierarchy as a context for a definition, it is easy to conclude that viruses, prions, memes or replicating computer programs are not forms of live. Both a virus with a surrounding mantle and a viral strand of DNA or RNA are not operators, thus not life. Prions are molecules, thus not life. Memes, such as texts and melodies, are pieces of coding that memons can decode and replicate (Dawkins 1976). Accordingly, memes are not operators, thus not life. Ray (1991) has created computer programs that can replicate themselves onto free computer space, show mutation, and modify and compete for the available space in a virtual world called Tierra. Since its start, this virtual "ecosystem" has seen the evolution of a range of different computer programs. In the same way as molecular viruses depend on cells, the programs in Tierra depend on a computer to copy and track their structure. Accordingly, they are not operators, thus not life. Sims (1994) has used genetic algorithms for evolving virtual computer creatures with body parts and a neural network with interface. The simulation of these animal-models allows virtual movement such as finding and grasping virtual food items. Sims's programmed creatures may possess hypercyclic neural networks and on these grounds show similarity to softwired memons. According to the operator hierarchy, a softwired memon should autonomously be able to copy its information structurally. Although I am not an expert in this field, it seems to me that Sims's organisms do not *themselves* keep track of their arrays with information about their interface and neurons, neural connections, and connection strengths, and that they do not autonomously organize their maintenance. Assuming that the latter interpretations are correct, Sims's computer animals are not yet life.

The use of the present definition also allows the effortless rejection of other systems that are not operators and sometimes receive the predicate of "borderline situations", such as flames, whirlwinds, crystals, cars, etc. Technical, computer based memons, however, such as robots, can be operators when they show the required structure.
To summarize the practical applicability of the present definition of life, I include a list of the examples that were discussed in the text and supplement them with some additional cases. The examples in this list form three groups depending on whether the entities involved are operators or not, and whether they show a complexity that equals or exceeds that of the cellular operator. In the text below I use the concept of "interaction system" (e.g. Jagers op Akkerhuis 2008) for all systems that are not operators because the interactions of their parts do not create a first-next possible, new, closure type.

Group A. Systems that are not life because they are not an operator
1. An entire virus particle with external envelope (represents a simple interaction system)
2. A computer virus based on strings of computer code
3. A flame
4. A tornado
5. A crystal
6. A car
7. A bee colony (The colony is an interaction system, and the bees are organisms.)
8. A cellular colony not showing the requirements of multicellularity (The individual cells are organisms and thus represent life.)
9. A colony of physically connected cellular memons (as long as the individuals lack the required memic closure)
10. A robot (as long as it is a non-memic technical machine)
11. Computer simulations of organism (including memons) that depend on external “orchestration”
12. A cutting/slip of a plant that cannot potentially show autonomous maintenance given the right conditions (It lacks the closure required for multicellularity.)
13. A separate organ, such as a liver or leg (not potentially capable of autonomous maintenance)
14. Endobiontic bacteria having lost genes that are obligatory for autonomous maintenance. The transfer to the genome of the host of DNA coding for enzymes required in autonomous maintenance implies a partitioning of the aucatalytic closure between the endobiont and its host,. Because of this, the endobiont is no longer an autonomous organism but has become a special kind of organelle.

Group B. Systems that are operators but that are not life because their complexity is lower than that of the cellular operator
1. A prion
2. Self-replicating DNA/RNA particles (catalyze their own copying in a solution containing the right building materials)
3. A DNA or RNA string of a virus that is copied in a cell

Group C. Operators representing life
1. A cutting/slip or other plant part that can potentially maintain itself given favorable environmental conditions
2. Anabiotic organisms (The fact that they are dried, frozen, etc. does not take their required closure away.)
3. Fully anaesthetized animal supported in its functioning by a mechanical heart-lung support and showing no neural activity (This can be regarded as a form of memic anabiosis with the potency become active again.)
4. A computer memon or other technical memon (a memic robot)
5. An artificial cellular operator constructed by humans
6. A exobiotic cellular operator with another chemistry than that found on earth
7. Sterile or otherwise non-reproducing organism (e.g. a mule, worker bee, sterile individuals)
8. Endoparasites or endosymbiontic unicellular organisms living in cells and still possessing the full potential of autocatalysis

## X. In conclusion

1. Overviews of the definitions of life from the last 150 years show that no consensus definition on life exists. In the light of the continuous failure to reach consensus on this subject, certain scientists have adopted a practical viewpoint, accepting, for example, the use of property checklists for identifying living systems. Others have advocated that the need for a generally accepted definition remains acute. Amongst the proposals for solving the problem is the suggestion to construct a broader context, a “theory of life” before continuing with attempts to define of life.
2. Inspired by the latter suggestion, the present paper invokes a classification of the generalized particle concept, called the operator hierarchy”. This hierarchy has several advantages for defining life: first, it offers a general context for including and differentiating between life and non-life, and second, it offers the unique possibility to extrapolate existing trends in the evolution of material complexity and to use these as a guide for discussions about “life as we do not know it”.
3. In close association with the reviewed literature, the use of the operator hierarchy allowed the following definitions to be suggested:
   A. From the viewpoint of the evolution of material complexity, life is: matter with the configuration of an operator, and that possesses a complexity equal to or even higher than the cellular operator.
   B. Living describes the dynamics of an operator that satisfies the definition of life.
   C. The definition of unitary organisms can take the form of: the operators that fit the definition of life.
   D. A multicellular organism (the cellular operator showing the multi-state) is: a construction of mutually adhering cells showing obligatorily recurrent interactions based on the same interaction type, that has the potential of maintaining its functioning as a unit and that does not show memic structure
   E. Dead matter applies to all operators that do not fit the definition of life.
   F. Death is: the state in which an organism has lost its closure following irreversible deterioration of its organization.
4. From the discussion of examples in the literature, it was concluded that the present set of definitions easily distinguishes life and non-life regardless of whether this is tested using the “obvious examples”, the “borderline cases” or “life as we do not know it”. This suggests that the present approach may well offer a practical step forward on the path towards a consensus definition for the states of matter representing “life”.

## XI. Acknowledgements

The author would like to thank Herbert Prins, Henk Siepel, Hans-Peter Koelewijn, Rob Hengeveld, Dick Melman, Leen Moraal, Ton Stumpel and Albert Ballast for constructive discussion and/or comments on the manuscript and Peter Griffith for English correction.

# The issue of "closure" in Jagers op Akkerhuis's operator theory

Commentary on "Towards a hierarchical definition of life, the organism and death", by G.A.J.M. Jagers op Akkerhuis

by
Nico M. van Straalen
Department of Ecological Science, VU University, De Boelelaan 1085
1081 HV Amsterdam,The Netherlands Tel. +31-20-5987070 Fax: +31-205987123
E-mail: nico.van.straalen@falw.vu.nl Website: http://www.falw.vu.nl/animalecology

## Abstract

Attempts to define life should focus on the transition from molecules to cells and the “closure” aspects of this event. Rather than classifying existing objects into living and non-living entities I believe the challenge is to understand how the transition from non-life to life can take place, that is, the how the closure in Jagers op Akkerhuis's hierarchical classification of operators, comes about.

In this issue of Foundations of Science Jagers op Akkerhuis (2009) proposes a definition of life based on his earlier theory of operators. A great variety of objects fall into the category of operator, and by introducing this term Jagers op Akkerhuis was able to draw a parallel between elementary particles, molecules, cells and multicellular organisms. The common denominator of these operators is their autonomous activity and maintenance of a specific structure. Consequently, operators were classified in a logical and hierarchical system which emphasizes the commonalities across what is normally called non-life (atoms, molecules) and life (cells, organisms). One very attractive aspect of the classification is that it joins the objects traditionally studied by physicists, chemists and biologists into one overarching system. Obviously, the hierarchy crosses the traditional border between life and non-life, so it should be possible to develop a definition of life from the operator theory. This is what Jagers op Akkerhuis attempt to do in the present paper. However, I believe he misses the point.

In the operator hierarchy, successive levels of complexity are separated by "closure events", e.g. when going from from hadrons to atoms, from molecules to cells and from multicellular eukaryotes to memic organisms. One of these closure events actually defines the origin of life: the transition from molecules to cells. Death, as defined by Jagers op Akkerhuis, is the loss of this closure, a fall-back from cells to molecules. There is another important transition, the origin of self consciousness, a closure event that accompanies the highest level of complexity in the classification of operators. Life with this level of complexity (maybe call it "hyper-life"?) is included in Jagers op Akkerhuis's definition of life.

Another interesting aspect of the operator system is that it is strictly hierarchical, that is, every operator can be classified on a more or less linear scale and the big leaps forward are punctuated by closures on that scale. This aspect of the system is reminiscent of the "Great Chain of Being", or *scala naturae*, which was the dominating view of life for many centuries. In evolutionary biology, it is now recognized that pathways can split and run in parallel, maybe even achieving similar closures independently from each other. I am not sure how this aspect fits into the operator classification of Jagers op Akkerhuis.

To define life in terms of the operator theory I believe the focus should be on the transition from molecules to cells and the closure aspects of this event. In other words, the closure of operating systems defines life better than the classification of operators. However, Jagers op Akkerhuis seems to add another seemingly hopeless definition of life to the nearly 100 already existing. Classifying what is life and what is not is, I believe, a rather trivial exercise. Everybody knows that a flame is not life, and it only becomes a problem when you spend too many words on it. Rather than classifying things into living and non-living entities I believe the challenge is to understand how the transition from non-life to life can take place, that is the how the closure in Jagers op Akkerhuis's hierarchical classification of operators, comes about.

The issue of closure is intimately linked to that of emergence. Both concept recognize that the characteristic properties of a living system cannot be reduced to its component parts only, but also depend on the way in which the components are organized in a network. The properties that arise from interactions between components are said to be "emergent". Emergent properties are not shared by the components, they "appear" when many components start interacting in a sufficiently complex way.

The concept of emergence plays an important role in genomics, the science that studies the structure and function of a genome (Van Straalen & Roelofs 2006). After about a decade of genome sequencing, scientists started to realize that the genome sequence itself does not define the organism. The human genome turned out to contain no more than 24.000 genes, much less than the earlier assumed 124.000. This raised the question how it could be possible that such a complicated organism as a human being could be built with so few genes. Obviously the pattern of gene and protein interaction defines human nature much more than the genes and proteins themselves. A new branch of biology was defined, systems biology, which was specifically geared towards the analysis of interacting networks, using mathematical models (Ideker et al. 2000).

Schrödinger (1944), in discussing the question "What is life?" foresaw a new principle, not alien to physics, but based on physical laws, or a new type of physical laws, prevailing inside the organism. These are the kind of laws that systems biology is after. The operator classification of Jagers op Akkerhuis is an important step because it emphasizes the continuity between physical systems and biological systems. However, the challenge of defining life is not in classification but in understanding the closure phenomenon by which life emerged from non-life.

# Definitions of life are not only unnecessary, but they can do harm to understanding

Rob Hengeveld

Institute of Ecological Science, Vrije Universiteit
De Boelelaan 1087, 1081 HV, Amsterdam
rhengeveld@wish.net

Abstract: In my response to the paper by Jagers op Akkerhuis, I object against giving definitions of life, since they bias anything that follows. As we don't know how life originated, authors characterise life using criteria derived from present-day properties, thus emphasising widely different ones, which gives bias to their further analysis. This makes their results dependent on their initial suppositions, which introduces circularity in their reasoning.

In his paper, Jagers op Akkerhuis (this volume) refers to a list of almost 100 different definitions subsequently having been given in the literature to the phenomenon of life as we know it. These definitions may even have a more general application or meaning than that concerning life on earth only. That is, also to some form of life as we don't know it, even though we don't know it. Like other authors, he feels that all this activity messed things up. Thus, Jagers op Akkerhuis mentions authors emphasizing "the seeming hopelessness of the situation", some of them adopting "a pragmatic/pessimistic viewpoint". Others would have suggested "that life cannot be defined, that its definition is not important for biology", or that "a comprehensive definition of life is too general and of little scientific use". Finally, only "living processes may be defined" which "cannot be reduced to physics".

His theory based on the criterion of hierarchically arranged operators would tidy up this mess a little. I feel, though, that the introduction of his own definition "life may be regarded as a special realization of the evolution of material complexity" brings the 98 existing definitions even closer to 100. Worse, this theory and definition will confuse our biological issues even more by their circularity of reasoning. They are circular because his operator concept "emphasizes the autonomous activity of the entities involved, which "operate" in a given environment without losing their individual organization". How do we distinguish the autonomy of processes in early living systems or even in present-day molecular biological ones from those of non-living processes? Also, activity, operation, and organisation are concepts connected with living systems and their functioning. Furthermore, individual organisation smells of one of the criteria on which some earlier definitions have been based. Thus, recognising something as living depends on criteria derived from known, recent living systems; a bean is a bean because it is bean-shaped.

When, as a beginning ecologist, I was studying ground beetles, and later as a biogeographer, I never felt any need for a definition of life. Then, such a definition was clearly useless. More recently, being concerned with questions about the origin of life, that is concerned with processes ultimately having resulted in a beetle as a living system, I came to realise that most, if not all, of these definitions were designed particularly within this context of the origination of living systems. However, we don't need to define the moon to understand its origin either. Yet, they not only seem useless, they are even harmful. Adopting certain criteria on which to base the one or the other definition, authors easily force themselves to look into the wrong direction. Or even at the wrong biogenetic phase, too late in the development of life. For example, not only is "organisation" difficult to delineate objectively at a molecular level, and this without circularity, but depending on a subjectively chosen threshold level, it easily excludes initial phases from analysis, however significant these could have been. Continuing along such a misdirected road is fruitless.

Thus, the criteria used, such as a certain level of organisation, are always derived from present-day life forms, from processes or structures that may not have existed in the early biogenetic phases. One criterion, as one of many examples, points at macromolecules, although these will have developed from earlier oligomers (see, for a clear example, Eck and Dayhoff, 1966). Another, widely applied criterion derives from the present prevalence of carbon as a principal biochemical constituent. Yet, carbon forms very stable molecules, as do its neighbours in the Periodic System, nitrogen and oxygen, for example. They are difficult both to form as well as to break down again, which is therefore usually done by enzymes. These enzymes, plus the enzymatic apparatus they together form, must have developed earlier, before carbon could have been taken on board biochemically (see Hengeveld and Fedonkin, 2007). Moreover, in their turn, individual enzymes are often very complex macromolecules, which not only must have been derived evolutionarily from more primitive ones, but they have to be formed by and operate within an intricate biochemical apparatus in which DNA is pivotal. Yet, DNA itself requires the operation of a very complex system of repair enzymes, etc., plus the mediation of spliceosomes and ribosomes for the final construction of those enzymatic macromolecules. Clearly, carbon as an element must have been inserted into the biochemistry only at a later, evolutionarily more highly developed stage of biogenesis.

Personally, I prefer to abstain from using definitions in this context. This differs from asking what requirement is needed to form a molecular bond, of a system of molecules, etc., any form of organisation, biological or non-biological. This puts the problem within the thermodynamic realm. A basic requirement, one that can be met by several properties, therefore differs from a property, physical, chemical, biological, or socio-economic; instead, it defines both the process and the shape of molecules taking part in it (Hengeveld, 2007). It defines the properties. It's the resulting processes happening and developing which are of interest, for the understanding of which a definition of life is irrelevant. It does not add anything.

Formulating the study of biogenesis in terms of processes happening and developing precludes the design of definitions, which are more likely to be applied to static or stable situations. And which are, already for that reason only, to be shunned. Defining life is not a part of our scientific endeavour.

# Explaining the origin of life is not enough for a definition of life.

Reaction of Gerard Jagers op Akkerhuis to the comments and questions of Rob Hengeveld and Nico van Straalen.

**Abstract:** The comments focus on a presumed circular reasoning in the operator hierarchy and the necessity of understanding life's origin for defining life. Below it is shown that its layered structure prevents the operator hierarchy from circular definitions. It is argued that the origin of life is an insufficient basis for a definition of life that includes multicellular and neural network organisms.

I thank the commentators for their reactions, both positive and negative, giving me the opportunity to elucidate some important aspects of the presented theory.

As Van Straalen indicates, the operator hierarchy offers valuable innovations: Firstly, the hierarchy '… joins the objects traditionally studied by physicists, chemists and biologists into one overarching system.' Secondly, '...it is strictly hierarchical'. I think that precisely these two aspects make the operator hierarchy a unique tool for defining life in a way that simultaneously addresses all the different organizational levels of living entities, e.g. prokaryotic cells, eukaryotic cells, pro- and eukaryotic multicellulars and neural network organisms, including future ones based on technical neural networks.

Further reactions of the commentators indicate that, probably due to the novelty of the operator theory, certain aspects require further explanation. I will discuss some essentialities in the following lines.

Hengeveld criticizes an asserted circularity in reasoning, in the sense that living operators are defined by means of concepts, which are derived from living systems. The confusion on this point results from my explanation in the paper. There I indicate that the name operator originates from the operating (in a very general sense) of individual entities. It may be reassuring to Hengeveld that the *origin* of the name 'operator' shows no direct relationship with the *definition* of the operators as system types. The entire set of all operators is defined as follows: based on the presumed existence of a lowest complexity operator, every system that belongs to the operator hierarchy resides at exactly one higher closure level than its preceding-level operator. Every closure level is defined by the occurrence of one or two first-next possible closures. Although this is a recursive definition in the sense that every operator in principle depends on its preceding-level operator, its hierarchical architecture precludes circularity of reasoning.

Hengeveld furthermore states in a general way that definitions of life 'are always derived from present day life forms, from processes or structures that may not have existed in the early biogenetic phase'. This general criticism does not apply to the operator hierarchy. The reason is that both abiotic and biotic operators are all defined using first-next possible closure. In fact, the operator theory turns the argumentation of Hengeveld upside down, hypothesizing that limited possibilities for reaching first-next possible closure have acted as a blue-print for the essential construction properties we recognize in abiotic elements and organisms.

I agree wholeheartedly with both Hengevelds' and Van Straalens' argumentation that we need to increase our understanding of the processes that have caused life. I strongly support the search for bootstrapping mechanisms allowing simple system states/elements to autonomously create more complex system states/elements (e.g. Conrad 1982, Martin & Russell 2003, Hengeveld 2007). In fact, every closure step in the operator hierarchy is the product of a specific (the first-next possible) bootstrapping mechanism. With respect to specifying the closure types resulting from such bootstrapping mechanisms, I consider concise and general definitions as indispensible tools, being helpful (instead of harmful!) in our search for the essences of the evolution of matter. Thus when Hengeveld advocates that he prefers '… to abstain from using definitions in this context.' I find his viewpoint surprising for two reasons. The first reason is that even a very thorough understanding of specific reaction mechanisms will not automatically result in a general definition of a meta-aspect such as the type of material organization defining living entities. The second reason is that I think that accurate definitions are simply a way to improve the precision and communication of science: sloppy definitions lead to the development of sloppy theory and a lack of definitions leads to no science at all.

Referring to a demand for a mechanistic focus when defining life, Van Straalen states that 'the challenge is to understand how the transition from non-life to life can take place' as this can explain how the classification of operators comes about. Also in his last sentence Van Straalen writes that '…, the challenge of defining life is not in classification but in understanding the closure phenomenon by which life emerged from non-life'. Both statements being true, it is nevertheless impossible to construct an overarching theory such as the operator hierarchy if one limits his view to the mechanisms explaining one single step involved.

The warm interest of Hengeveld and Van Straalen for mechanisms that could explain the origin of life is understandable, because it frustrates the scientific community that science is not yet able to *de novo* synthesize life, not even in the form of a primitive cell. This general focus on the construction of life seems, however, to have caused a tunnel vision with respect the *definition* of life. Imagine that we would be able to explain the cell, and even construct it, would this then mean that we would have a proper definition of life in all its forms, including multicellular organisms and neural network organisms? The answer is a clear NO. If everything that is based on living cells would be life, then a donor organ and a fresh, raw steak would also be life. Moreover, any technical being, however intelligent, could never be called life, because it is not based on cells. This proves that a focus on cells alone is not enough. We need to broaden the scope and define all levels of organization associated with higher forms of life. It is my personal conviction that, for the latter goal, the operator hierarchy offers a unique and unprecedented tool.

# Complexity and Evolution:

## a study of the growth of complexity in organic and cultural evolution

Börje Ekstig
Department of Curriculum Studies,
Uppsala University, Box 2136, S-75002 Uppsala Sweden.
E-mail: borje.ekstig@did.uu.se

**Abstract**

In the present paper I develop a model of the evolutionary process associated to the widespread although controversial notion of a prevailing trend of increasing complexity over time. The model builds on a coupling of evolution to individual developmental programs and introduces an integrated view of evolution implying that human culture and science form a continuous extension of organic evolution. It is formed as a mathematical model that has made possible a quantitative estimation in relative terms of the growth of complexity. This estimation is accomplished by means of computer simulations the result of which indicates a strong acceleration of complexity all the way from the appearance of multicellular organisms up to modern man.

## Complexity

Complexity is an intriguing and widely discussed concept. The problem with this concept is that it is an immeasurable quantity and that discussions therefore are based mainly on intuitive notions. Besides, there is no definition generally agreed on. There is a comprehensive literature on complexity and I will open my paper by discussing some selected contributions to this literature.

John Maynard Smith and Eörs Szathmáry begin their salient book *The Major Transitions in Evolution* with the following statements that could equally well serve as a declaration of the aim of the present paper :

> Living organisms are highly complex… . This book is about how and why this complexity has increased in the course of evolution. … Our thesis is that the increase has depended on a small number of major transitions… (Maynard Smith et al. 1995 p. 3).

Of special value for the present discussion is the fact that these authors include the evolution of the human language as one of the major transitions, a notion that I will develop in more detail later in this paper.

Increasing complexity is suggested to be one of several possible forms of large-scale evolutionary trends that may result either from driving forces or from passive diffusion in bounded spaces (McShea 1994). In his books, *Life's Grandeur* (Gould, 1996) and *The Structure of Evolutionary Theory* (Gould, 2002), the late Stephen Jay Gould builds on McShea's idea of a lower boundary from which the trend has only one possible direction. Gould develops this idea in the form of the metaphor of the left wall illustrated by the "drunkard's walk". When the drunkard staggers on a sidewalk with the wall of the bar on his left side, he will by purely statistical reasons tumble more and more to the right. In this sense, Gould supports a purely passive diffusion as ground for the trend. For the present discussion it is of value to find support for the notion of a pervasive trend towards increasing complexity but it is not necessary to specify the suggested distinctions between possible causes of the trend.

Gould draws a couple of diagrams to clarify the discussion (Gould, 1996, p. 171). In the first of these he illustrates the frequency of occurrence of species versus complexity for the Precambrian epoch in which bacteria, the only living creatures at this time, form a pile near the left wall. In the second diagram he illustrates the situation for present time with the same axes demonstrating how life has been spread out over a wide range of complexity in a skewed distribution. Most species are found at low complexity like bacteria whereas species with higher complexity are found in successively decreasing abundance. He illustrates a series of species spread out over the dimension of complexity including bacteria, trilobites, fish, dinosaurs, apes and humans in this order. Needless to say, this series also shows the temporal order in which the exemplified species have come into being, a conclusion that also Gould admits in concluding that

> In popular description of evolution, … we have presented the history of life as a sequence of increasing complexity, with an initial chapter on unicellular organisms and a final chapter on the evolution of hominids. I do not deny that such a device captures something about evolution. That is, the sequence of

> bacterium, jelly-fish, trilobite, eurypterid, fish, dinosaur, mammoth and human does, I suppose, expresses “the temporal history of the most complex creature” in a rough and rather anthropomorphic perspective (Gould, 2002, p. 897).

I have chosen to discuss Gould’s views on complexity because Gould, despite his generally acknowledged scepticism regarding the application of the concept of complexity to evolution, accepts increasing complexity at least in a descriptive sense. Furthermore, Gould extends his survey of the evolutionary process to include the human species, which, as we will see, is given an important role in the present investigation.

There are however many other authors that are less sceptical than Gould. Thus Heylighen (1999) observes that the directions in which complexity increases are generally preferred and Adami and co-authors (Adami et al. 2000) develop a Maxwell Demon mechanism that “is at work during all phases of evolution and provides the driving force towards ever increasing complexity in the natural world”. Emmeche (1997) gives a philosophical review of the many difficulties related to the lack of a stringent definition of complexity. Increasing complexity is generally associated to an acceleration of the evolutionary process, a notion suggested to be generalized to the entire universe as well to technological development. (For an overview, see Smart 2008).

There is an emerging branch of science called *evodevo* in which relations between the developmental and evolutionary processes are studied, mainly confined to morphological traits. One of the representatives of this field maintains that “when comparisons are made between very different levels of complexity, the pattern that emerges is broadly recapitulatory, although only in a very imprecise way, in the sense of recapitulating levels of complexity rather than precise morphological details” (Arthur 2002).

These examples of authors in the field of complexity thus make me encouraged in my intension to investigate the growth evolutionary complexity and in this endeavour make use of the relationship between the developmental and evolutionary processes.

## Development and evolution

As a point of departure for the present discussion I call attention to the fact that biological as well as cultural evolution are formed as results of continuous modifications in the developmental programs of living organisms including human beings. Additionally, modern genetics has shown that genes for specific traits in many cases are preserved over long periods of time and that the evolutionary modifications to a certain extent are formed by hox genes triggering the temporal onset of the action of specific genes. This, in my view, explains the intriguing observation that vestiges of earlier developmental programs can be observed in the embryogenesis of present-day individuals, an observation that has been subject to enduring and partly bewildering discussions focused on the idea of recapitulation (For a survey of recapitulation, see Gould 1977).

The interpretations of the observed vestiges have mostly been restricted to morphological features within the field of biology. When also cultural features of the human species are included in the analysis, a conspicuous pattern is appearing that only to a minor extent is built on morphological features. This pattern is formed by the observation of a temporal relationship between the evolutionary history and the developmental process of a modern individual human being and is made visible by means of a diagram (Fig. 1).

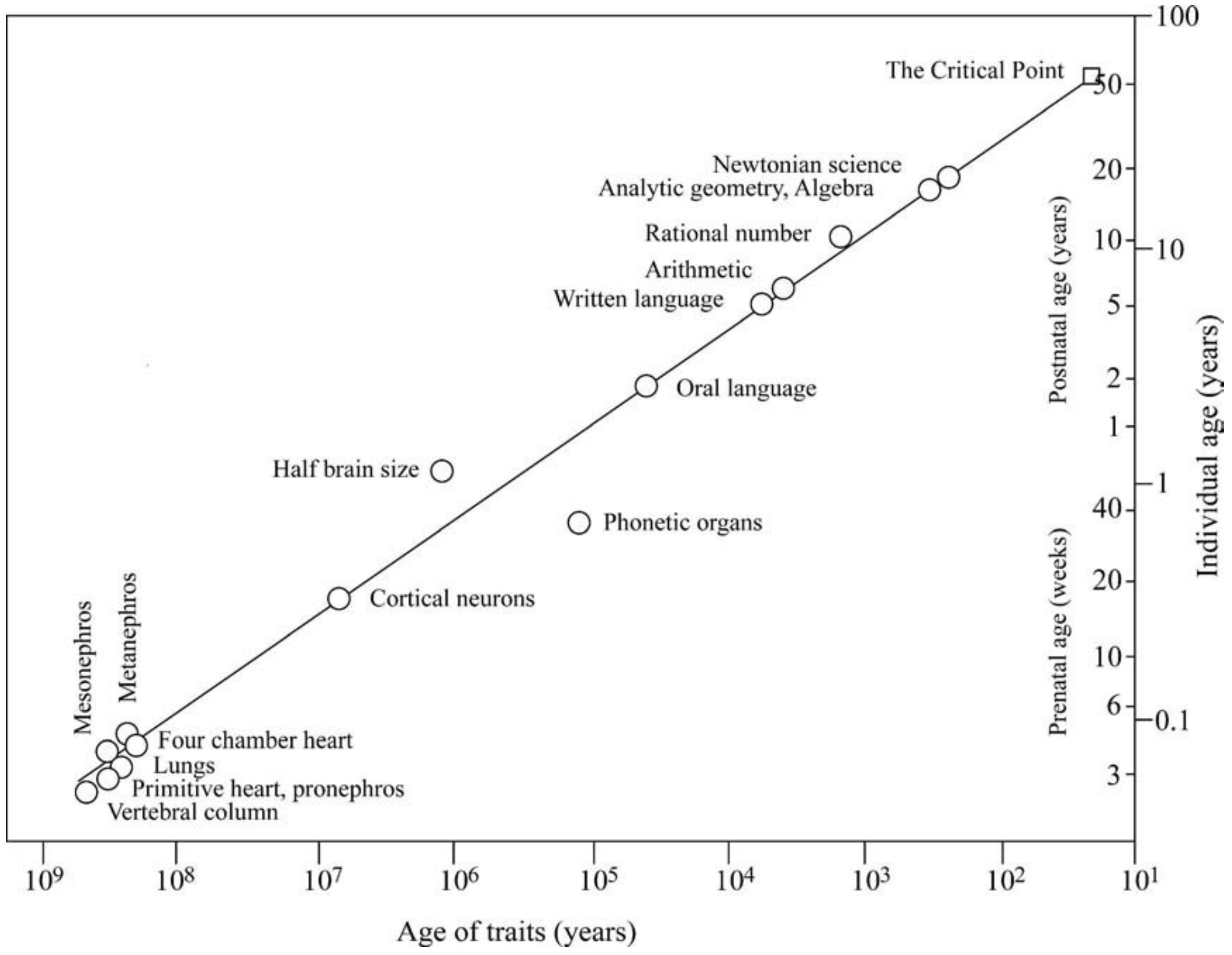

Figure 1. Individual development versus evolution in the lineage of man. The horizontal axis gives evolutionary time measured backwards from the present point of time. The vertical axis shows the individual age. Please note that both scales are logarithmic. References to the traits that are included in the diagram are given in my earlier publication of the pattern (Ekstig 1994). This version of the diagram was published in Ekstig (2007). The Critical Point is of mathematical nature and not part of the empirical basis.

I have chosen not to show the uncertainties in the positioning of the points in the diagram, uncertainties that stretch out in the horizontal as well as in the vertical dimensions. If displayed in the diagram, these uncertainties seem to be smaller for morphological traits than for the more recent, this feature being a consequence of the logarithmic construction of the scales. In spite of this consideration, a conspicuous pattern emerges, namely that the points form a straight line and this feature must be regarded as remarkable when the great span of time covered by the diagram is taken into account. Before an interpretation of the pattern will be suggested a descriptive mathematical analysis is carried out.

The equation of straight line is

$\ln t_o = C_1 - C_2 \ln t_p$ (1)

where $t_o$ and $t_p$ denote the developmental (ontogenetic) and evolutionary (phylogenetic) age, respectively. The value of constants are determined by the line yielding $C_1 = 5.12$ and $C_2 = 0.39$.

An essential feature of the diagram is that both time scales are logarithmic. Due to this logarithmic form, the present moment of time is not included. Another characterizing feature of the horizontal time scale is that it is directed backwards. An inherent feature of such a scale it is that, as time proceeds, the position of each particular event will be displaced to the left and, due to the fact that the time scale is logarithmic, the quantity of this displacement as seen in the diagram is smaller for older traits than for more recent. Such an unlinear displacement is incompatible with an assumption that the linearity of the line would be a general feature and not just an accidental coincidence of the present time. However, the linearity is preserved over time if it is assumed that the points simultaneously are displaced downwards, i.e. towards earlier developmental appearance, at a pace that is determined by the value of the derivative $a$ of Eq. (1), called acceleration:

$a = dt_o / dt_p = - C_2\, t_o / t_p$ (2)

Such a regular displacement of developmental traits can be seen as the result of an appropriate shortening of all preceding stages, a shortening called condensation. I have shown (Ekstig 1985) that such a condensation, depicted $q$, is given by the simple formula

$q = - (C_2 + 1) / t_p$ (3)

implying that the value of condensation of each stage is determined by merely one parameter, the phylogenetic age of the trait, in such a way that it is inversely proportional to the duration of its action on the stage. Generally speaking, the formula implies that the rate of condensation will decrease with time which is intuitively sensible since the more a stage is shortened the more difficult it must be to shorten it still more. In this way the linear pattern is explained as a consequence of a continuous and regular shortening of developmental stages, i.e. condensation.

Acceleration and condensation are introduced in the context of heterochronic mechanisms that together with terminal addition are reviewed by Gould (1977) and Mc Kinney and McNamara (1991). The present model allows for a more precise determination of values of acceleration and condensation as exemplified by the following examples.

The value of acceleration for the embryonic period of heart formation is $- 5 \times 10^{-11}$ that means a displacement of a few days in 100 million years and the value of its condensation is $- 3 \times 10^{-9}$ per year that means a shortening of 3% over 10 million years. The second example is oral language the value of acceleration of which is $- 2 \times 10^{-5}$ that means a displacement of 7 days per millennium. Its value of condensation is $- 3.5 \times 10^{-5}$ per year corresponding to a shortening of 3.5 % per millennium. As to the period of acquisition of Newtonian science the value of acceleration is – 0.023 that corresponds to a displacement of approximate 1 year in every 40 years whereas its condensation is 5% per decade. As acceleration can be seen as caused by condensation, I will hereafter restrict the discussion mainly to condensation.

Morphological traits are very old, yet vestiges of them are still, quite remarkably, visible in the early part of the human embryo. This implies that they are assembled in the lower left part of the diagram and the pattern in the diagram would not be discernible if the analysis of evolution were restricted merely to the morphological realm.

The present model is in concordance with the widespread intuitive notion of an accelerating evolutionary process and the consequential decreasing intervals between the additions of evolutionary novelties. Such an accelerating course is in the present model coupled to the developmental process and the linkage is formed by the process of condensation that thus implies a vital concept at the analysis of the diagram of Figure 1. Truly remarkable is the regularity at which condensation is working over the so different evolutionary processes as those of the biological, cultural, and scientific realms. Therefore, our next step will be to analyze the applicability of this common principle in these areas. But first a short discussion of a particular point on the line.

## The Critical Point

There is one point in the diagram that is determined by means of extrapolation of the line as determined by empirical data. This is the point on the line at which the unit of time has the same length on both axes, which is the same as to say that the absolute value of the derivative equals 1. I have called this point *The Critical Point* and its coordinates can easily be calculated by means of Eq. (1) and Eq. (2) for $a = -1$.

The result is found to be $t_o = 52$ years and $t_p = 20$ years. These data can be interpreted in the following way. 20 years ago, at age 52, the human being under study caught up with the evolutionary process but later on this grew faster than he was able to follow. This person is a modern scientists and the reason why a scientist is given such a key role in this context is that the last points displayed in the diagram are the points for the acquisition of elementary mathematics, a route followed and extended by modern scientists. As a consequence of the data found for the Critical Point, the age of this scientist is 72. It should be noticed that the calculated figures are results of an idealized mathematical model and should not be taken too far. Individual variations are indeed very large. Many scientists are quite young when they start to contribute to their field of research and may be active at the front for several decades. On the other hand, most people do not at all participate in the scientific enterprise.

## Common principles within biology, culture and science

A conspicuous feature of the diagram of Figure 1 is that it is constructed on an evolutionary time scale that runs backwards. This means that evolution is represented in a historic, retrospective perspective and as a point of departure for this perspective, mankind's present position is chosen. In his historic survey of biological life on the Earth, Dawkins (2004) uses this, what he calls hindsight, perspective. In the present paper, I will extend Dawkins' analysis to include human culture as well and, like Dawkins, I will follow the human lineage but with the point of departure in today's scientific culture. I maintain that it is fully legitimate to base the analysis of the evolutionary process on such a retrospective perspective without presupposing that evolution would have been aimed towards us, or that there would be a predetermined plan for evolution.

Another striking feature of the diagram is that the morphological and cultural parts of the evolutionary process are unified by a common pattern and it is therefore necessary to discuss the legitimacy of such a model. Thus, in spite of the fact that one may regard culture as being of a different nature as compared to biology one may also acknowledge common basic principles. As Dennett (1995 p. 343) points out, evolution occurs whenever there is variation, heredity and differential fitness. The following table clarifies and supports the application of these principles also to the field of culture in which science is included as a separate part.

| | Biology | Culture | Science |
|---|---|---|---|
| Heredity | Copying | Imitation | Teaching and learning |
| Variation | Mutations | Human creativity | Human creativity |
| Differential fitness | Natural selection, sexual selection | Sexual selection, Self-reinforcing selection | Selection on behalf of observations and experiments |

Table 1. Common evolutionary principles within biology, culture and science.

Another common principle of the evolutionary processes of biology, culture and science is condensation, as demonstrated by means of the mathematical analysis of the linear pattern in the diagram of Figure 1, and I will analyze the application of this principle to each one of the three processes. Moreover, also terminal addition is a mechanism that can be recognized in all these processes.

In the realm of biology, Stearns (1992) has demonstrated that the shortening of developmental stages can be seen as a result of a ubiquitous selection pressure coming about by the fact that it is reproductively advantageous to develop a short maturation period. As he points out such a shortening is advantageous because a shortened generation period gives a demographic effect. In addition, a shorter maturation period is advantageous because the time of predator exposition before maturation is minimized.

It is well-known that all species have preserved a common stock of genes over long evolutionary periods during which their expressions are trigged by hox genes, in this way amongst other things giving rise to speciation. However, the effect of condensation seems rather to be to gradually shorten the expression of separate genes in embryogenesis by building them more efficiently.

There is another simultaneously acting process implying that condensation does not normally result in an overall shortening of the maturation period. Novel evolutionary traits may be reproductively advantageous to an extent that overrules the effect of condensation. Such terminal additions are independent of and superimposed on condensation and therefore the fact that for instance humans have a longer maturation period than chimpanzees does not contradict the action of condensation. Without condensation the human maturation period would have been still longer. Unfortunately, gradual changes of embryogenesis during the evolutionary history are not sufficiently well known to ascertain the effect of condensation during morphological evolution. It is possible that the rapidly expanding research in the field of genetics might be able to shed some light on this problem.

In the realm of culture, I will concentrate on the acquisition of verbal language because it is common knowledge that language is the preeminent feature of man that takes a great part of our neocortex into utilization. Thus, as I already have pointed out, Maynard Smith and co-author (1995) have included language in their survey of major evolutionary transitions. Blackmore (1999) has emphasized that imitation is a central agent in the acquisition of language. She argues that imitation is specific to humans, primarily a method applied by children at the acquisition of their mother tongue. Furthermore, she applies sexual selection in the context. An "important decision is whom to mate with, and again the answer should be the best imitators, because they will provide you with children who are more likely to be good imitators. … As imitation improves, more new skills are invented and spread, and these in turn create more pressure to be able to copy them. " (ibid. p. 77)

Especially regarding the application of condensation in this context, I would like to add the following reasoning. There is, and certainly always has been, a great variation in verbal competence in human populations. Parents that happen to have a better than average verbal talent also bring this talent to their children and when these children are developing into parents to the next generation of children the verbal talent is still more reinforced. Furthermore as Blackmore points out, because imitators tend to imitate the most skilled precedents, children of all kinds of parents are expected to acquire verbal ability at gradually earlier age. In this way one can talk about a self-reinforcing positive feedback mechanism that gradually shortens verbal developmental stages. In addition, the verbal talent makes the quality of the brain visible and can be sees as an ornament in connection to sexual selection. And, as is well known, sexual selection tends to give rise to a run away process that in this context can be seen as contributing to the positive feedback mechanism.

In our modern time, I think, the suggested process can be extended to education itself. Parents with good education certainly appreciate this feature and are therefore urgent to give good education to their children as well with the same self-reinforcing result as that concerning language acquisition. Such an enhanced education means that children not only acquire verbal skilfulness at gradually earlier age but that also the acquisition of many abstract and scientific concepts is enhanced.

Many parts of the cultural and scientific realms of evolution are formed by cumulative additions. It is for instance impossible to imagine a written language to appear before verbal language or algebra to be developed before arithmetic. This is so in the evolutionary as well as in the developmental process. In this respect written language was added terminally during the epoch of language evolution as algebra was in the history of mathematics. Therefore the concept of terminal addition is applicable in the cultural and scientific realms.

Finally, as to the realm of science, one may regard science education as an intentional activity that, at least during the last hundred years, has substantially improved its pedagogical methods, resulting in a rapid enhancement of learning. This means, especially regarding the highly cumulative learning of mathematics, that pupils learn specific stages in gradually less time and at earlier age. In other words, condensation and acceleration are at work. Also, I think that there is an intentional selection of curriculum stuff in order to speed up learning. Thus for instance, pupils need no longer to learn the square root algorithm. However, even if the individual stages of scientific knowledge are shortened and displaced towards earlier age, the overall time of science education is prolonged due to the continuous addition of novel knowledge to the end of curricula. This then can be regarded as a form of terminal addition.

The courses of elementary school mathematics are by and large followed by modern scientists and must therefore be considered as an essential foundation of the scientific culture. However, stages in the history of science, except those of mathematics, are not in general reflected in the order of acquisition by young persons and therefore, science on the whole is not included in the model.

In summing up, I conclude that there are fundamental common principles that support the notion of evolution as not just a biological phenomenon but also applicable to the realms of culture and science. Furthermore, I have demonstrated that the principles of condensation and terminal addition are present in all three realms of this extended evolutionary process.

**Preliminaries of the method for estimation of complexity**

In addition to the principles included in Table 1, some other additional principles are decisive for the present method to estimate the growth of complexity in evolution. These principles imply

1. that morphological and at least some parts of the cultural and scientific evolution are manifestations of continuous modifications in developmental programs;
2. that many evolutionary novelties are added as terminal additions to the developmental programs of embryos, children and young people;
3. that many embryonic traits expressed by a specific gene have preserved their level of complexity from their first appearance;
4. that the levels of complexity assigned to stages in the cultural history actually are levels of complexity in individual human beings preserved over historic time.
5. that the coupling between the developmental and evolutionary processes as manifested in the observed pattern in the diagram of Figure 1 forms the basis of the present method for an estimation of the growth of complexity in evolution.
6. that the present model is applicable only to those cultural manifestations that are preserved as identifiable stages during the developmental growth.

1. This point can be seen as a reformulation of a law of recapitulation in a reversed mode – it is the gradual modifications in the developmental courses of all creatures that ever have existed that lie behind the abstract notion of evolution.

2. There has been an enduring dispute in the literature, restricted to morphological traits exclusively, whether all evolutionary novelties are added at the ends of ontogenies as terminal additions (for an overview, see for instance Gould 1977, Mc Kinney and McNamara 1991). However, the morphological part of the diagram covers a minor part of the diagram and a change in the ontogenetic order of the acquisition of traits in this part of the line would not disturb the general feature of the pattern. Actually, the morphological part of the diagram can be regarded as one single point from which the line has its point of departure. Therefore the discussion in the literature on the multitude of heterochronic mechanisms (Mc Kinney and McNamara 1991) are by and large not applicable in the present context. The only concepts applied are acceleration, condensation and terminal additions.

3. Genes express themselves as traits during embryogenesis and are, as I already have emphasized, to a large extent preserved during the evolutionary process. A fundamental assumption for the present model is that many developmental traits expressed by specific, essentially unchanged genes have preserved their level of complexity from their first appearance. This assumption forms a basic prerequisite for the estimation of complexity in the field of morphological evolution and implies that when we are estimating complexity in the evolutionary process, we are actually just comparing levels of complexity in stages in different individual organisms living at great time separation.

4. Another fundamental assumption for the present model is that the complexity of the cultural or scientific stages that we are observing in a present-day individual, say the ability of verbal language or the proficiency to make arithmetic calculations, are reminiscences of the levels of complexity that were achieved, nota bene, by individual human beings, when these traits once upon a time arose in our history.

5. Condensation acts on individual organisms including human beings, implying that its developmental stages are shortened. Such a condensation is expressed in consecutive sequences of individuals and therefore forms a mechanism for the evolutionary process. When these sequences of individual organisms are extended over the entire process of evolution (at least since the Cambrium Explosion), the effect of condensation has displaced many traits to appear very early in the embryonic phase of living animals. Thus, condensation connects the evolutionary course to the developmental process, a coupling manifested as the regular pattern in the diagram of Figure 1. As we will see, it is this very pattern that makes possible the estimation of the growth of complexity in the evolutionary process.

6. It must be emphasized that the pattern observed in the diagram is based only on traits in the lineage of man and on a few specific stages in the cultural history of the western society. This means that, as to the field of biology, there are many clades, for instance insects, that do not form part of the line and thus are not included in the present method. As to the field of culture, the multitude of cultural forms demonstrate that human culture is much more rich in expressions than as indicated by the diagram. There are, to say the least, an immense amount of other human cultural expressions. Amongst these one may mention agriculture, religion, artistic creation, music, literature, politics, warfare, and technology; some of these cultural expressions displaying substantially increased levels of complexity during the course of time. These cultural manifestations are primarily seen in adult activities and their historic evolution is by and large not reflected in children's somatic or mental development. Nor have they been subject to condensation. If attempted to be shown in the diagram of Figure 1, they would be stretched out horizontally in the upper part of the diagram, i.e. in the adult part of the developmental dimension. Therefore, applicable to the present model are only those cultural manifestations that are preserved as identifiable stages during the developmental growth.

In considering the breadth and depth of all human cultural manifestations including those not used in the present estimation of complexity we may conclude that human culture is more complex than anything other species may bring to the table in form of possible cultural manifestations. This conclusion, I maintain, can be drawn without chauvinism.

In summing up, I have argued that amongst the multitude of species only those forming our ancestral line are applicable for the estimation of complexity. Likewise, amongst the enormous multitude of all human cultural manifestations only a few are used, namely those the historic evolution of which is recognized in the developmental course of a modern child. This limited applicability of the present model must of course be considered when it comes to the conclusions that may be drawn as to the growth of complexity in evolution.

As I have emphasized, all the applied features, whether morphological, cultural or scientific, have been subject to condensation. The reason why condensation is maintained to be such an inescapable condition is that, as demonstrated in the mathematical analysis of the straight line in the diagram of Figure 1, it lies behind the linearity of the line, which, as we will see, has turned out making possible an estimation of the growth of complexity. The method for this estimation will now be introduced.

## Complexity versus time

As we have seen, Gould accepted, though reluctantly, the idea of a continuously increasing level of complexity throughout the history of evolution. He draws a sequence of species including fish, dinosaurs, apes and humans on an axis of increasing complexity (Gould 1996, p. 171). The distinguishing characteristics of these organisms may be associated to the points in the diagram of Figure 1 marked vertebral column, lungs, cortical neurons and oral language, respectively. I therefore conclude that I'm actually supported by Gould in my interpretation of the line in the diagram as representing increasing complexity. I have in this context preferred to refer to Gould since he is known as a most ardent opponent to the application of complexity to evolution, and this fact, I think, strengthens my point.

The position of each point in the diagram of Figure 1 is determined by two time coordinates, one for evolution and one for development. These two time coordinates for each trait will now be represented on a common linear time axis stretching backwards from the present point of time, giving rise to a sequence of pairs of time coordinates. The evolutionary coordinates are already adapted for such a negative time scale whereas the developmental coordinates have to be recalculated to give the time elapsed since the actual age was reached by a present-day human being.

Next, I want to distribute these pairs of points in a diagram displaying complexity on its vertical axis, a crucial procedure at the shaping of a diagram of complexity versus time considering the fact that complexity is an immeasurable quantity. To that end, one must keep in mind that every pair of points on the common time axis originates from the same point on the line in the diagram of Figure 1, thus representing the same degree of complexity as I already have emphasized. These pairs of points will then form two separate series of points, one for evolution and one for the individual's life history. The next step is to distribute these points on the dimension of complexity.

As one immediately can see, the diagram of Figure 1 indicates that the intervals between major evolutionary transitions are shortening, a process associated to an acceleration of the evolutionary process. (Please note that the meaning of the term *acceleration* in this context differs from its use above as a heterochronic term.) Actually, acceleration is subject to intensive studies and considered to be a universal characteristic of all evolutionary processes, thus also including cosmological and technological evolution. This view is discussed at length by John Smart (2008) and, especially concerning technological evolution, by Ray Kurzweil (2005). The problem with such an interpretation of the concept of acceleration is that there is need of an entity that is assumed to be accelerating. It seems however to be a more or less tacit assumption that this entity is complexity. Furthermore, since the traits under observation are cumulative, it is reasonable to assume the evolutionary increase of complexity to follow an exponential course. This assumption forms the first step at the construction of a diagram over complexity versus time.

Since the developmental coordinates are linked to the evolutionary coordinates by the straight line of Figure 1, any assumption of the form of the evolutionary trajectory will give a consequential form of the developmental trajectory at the shaping of the complexity diagram. The method for the construction of this diagram is therefore to choose a form of the evolutionary trajectory that gives the most sensible form of the developmental curve.

It turns out that the assumption of an exponential form of the evolutionary curve will give the developmental curve a shape wearing a rough semblance of a logarithmic function. These forms of the curves are illustrated in Figure 2. The logarithmic form of the individual curve is sensible since it is common knowledge that the intervals between changes in the individual growth are relatively short in the embryonic period and gradually grow longer with age. Furthermore, it turns out that the two curves coincide at one point where they have the same direction. This point is recognized as the Critical Point.

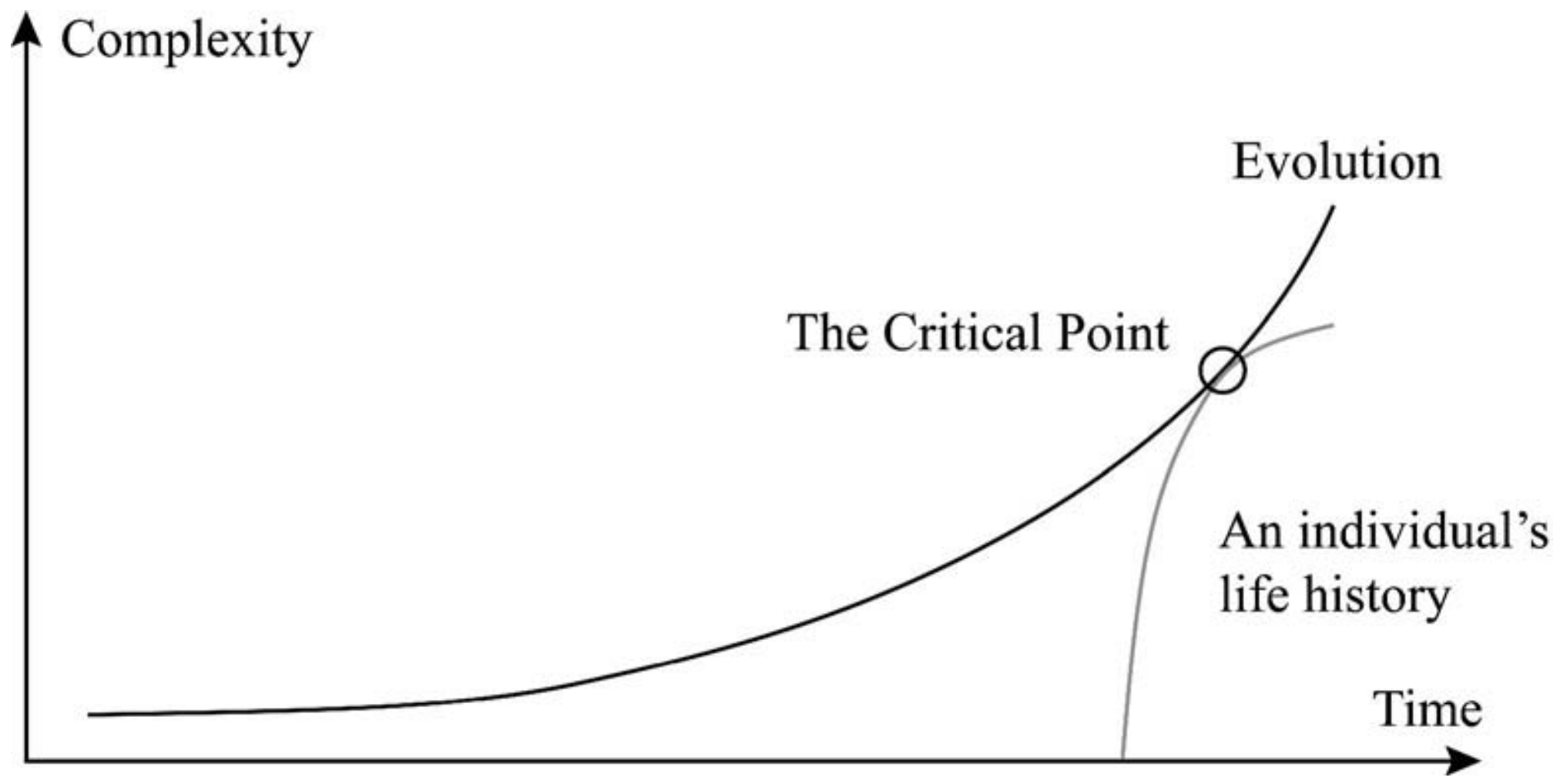

Figure 2. A principal draft of complexity versus time on a linear time scale. The curves are formed by the pairs of time coordinates of each point in the diagram of Figure 1, each such pair having the same degree of complexity.

It is interesting to generalize the notion of the Critical Point to earlier epochs of human history and to other species. Corresponding developmental curves may be drawn in the same principal way for all human individuals and all individual animals in history. All these curves would have a similar form as the individual's life history as shown in Figure 2 but placed on their proper temporal position on the time scale. Furthermore, only some of these curves would reach the evolutionary curve at their corresponding Critical Point but never transcending it. Thus, the evolutionary curve could be seen as the envelop of all living creatures. The slope of these individual curves might be different, corresponding to the different lengths of life histories. Thus for instance, elephants, in spite of having longer lives, would not be more complex than humans. But the slope of their individual curve in this diagram would be less steep thus reaching their Critical Point at a higher age. This general picture of the evolutionary process is in concordance with the notion of evolution as being the result of continuous modifications in individual life histories as I have already emphasized.

Returning now to the construction of a diagram over complexity versus time, the first step of the method is to adopt a form of the evolutionary curve, which here is hypothesized to be an exponential curve. For every point of time, $t_p$, on this curve the corresponding developmental date $t_o$ is calculated according to Eq. (1). Then the date of this developmental trait on the time axis is calculated, taken into account that $t_o$ denotes the age of an individual of age 72. This procedure will give a diagram of the principal shape as illustrated in Figure 2.

It must be emphasized that this diagram is just of principal form in order to clarify the method used to build up a more quantitative diagram. Such a quantitative construction is most readily performed by means of computer simulations. An example of such a computer simulation is shown in Figure 3.

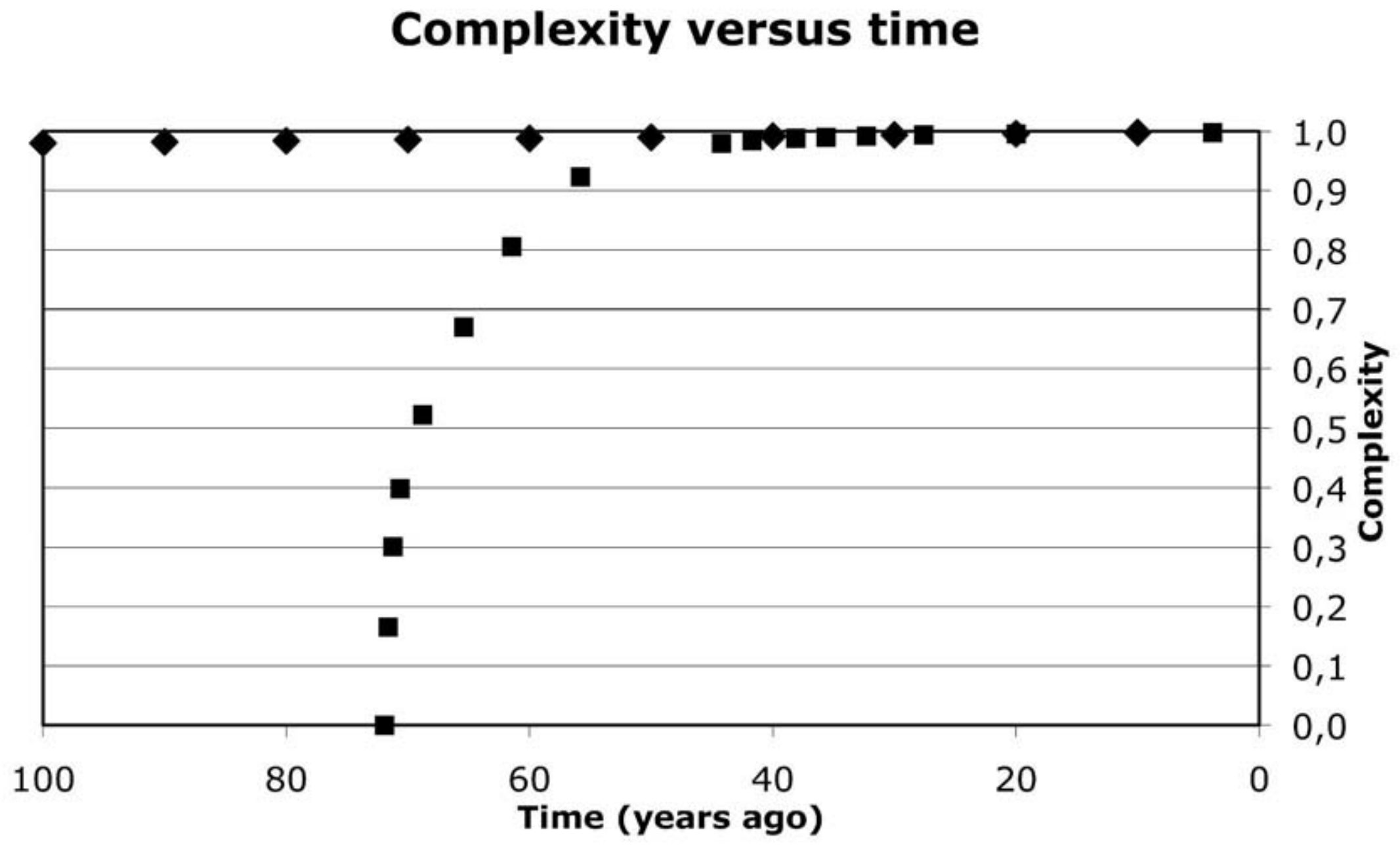

Figure 3. A computer simulation of complexity of the evolutionary (upper series) and the developmental (lower series) processes for the last 100 years. The evolutionary series is tentatively formed by segments of exponential functions with different bases between 1.0000001 and 1.0002. The values of complexity of the developmental series are the same as those for the corresponding points in the evolutionary series. The temporal coordinates of the developmental process are calculated by means of the formula for the straight line in the diagram of Figure 1. The scale on the vertical axis is relative. The evolutionary series stretches to the left far out of the frame. As an example, the date for the onset of the vertebral column 600 million years ago is to be found on this time axis some 600 km to the left with a complexity value of about $10^{-26}$ , a value of complexity given to the lowest point on the developmental curve. These simulations are performed by means an MS Excel program available on my homepage (http://www.did.uu.se/ekstig/ ) where the results of other entrance values are demonstrated as well. The program is open for examinations of alternative simulations.

In the diagram of Figure 3 the last hundred years of the evolutionary process is shown on a linear time scale, since this is the part of interest for comparison with the developmental process of a modern human being. The temporal coordinates of the developmental process are calculated by means of Eq. (1) for chosen coordinates on the evolutionary scale. This means that the lower part of the developmental curve is based on the empirically determined points on the line in the diagram of Figure 1 whereas the higher part of the curve is based on an extrapolation of the empirical points in the diagram of Figure 1 in accordance with Eq. (1). The scale on the vertical axis is relative.

The time axis shows past time in the negative direction. Complexity is calculated as points on the negative tail of an exponential function (or a similar curve), therefore giving a relative value of complexity ranging between 0 and 1.

Trials with different bases of the exponential function give the following results. A base of 1,0002 implies that the first half of the embryonic period shows very small growth of complexity that seems unrealistic. A larger base gives a still more apparent similar feature. If the base on the other hand is chosen to be 1,0000001 the fetus reaches 90 % of maximal complexity already at birth that also seems unrealistic. A smaller base gives a still more apparent similar feature.

The result of these simulations shows that no sole exponential function will give a realistic picture of the individual growth of complexity. However, there is actually no reason to expect evolution to follow such a simple model. The simulations make possible an examination of the growth of complexity in the evolutionary process following different rules at different periods. Such a method lies behind the diagram of Figure 3 in which the base of the exponential function at the early evolutionary epochs is 1,0000001 to increase step by step to reach 1,0002 for the last 400 years. According to my subjective judgment, these values give a realistic form of the individual growth curve. These simulations indicate that evolutionary complexity is growing faster in its latest phase than an exponential function that fits the early epochs and that it is growing slower in its early epochs than an exponential function fitting the latest phase.

Kurzweil (2005 pp. 491 - 496) discusses the application of a double exponential function, i.e. an exponential function of the form $W = \exp(k_1 \exp(k_2 t))$, in his studies of the evolution of knowledge. Simulations with such a function are here performed analogously to the above calculations. It turns out that even by means of this function it is impossible to find a combination of the two constant in the double exponential function that gives a sensible form of the developmental curve. The conclusion is that, although growing more rapidly than a simple exponential function, the integrated evolutionary processes of morphological, cultural and scientific evolution cannot be accounted for by a double exponential function.

In another simulation the evolutionary complexity was calculated as the sum of two exponential functions with different bases. In this way a rather good trajectory of the individual growth was achieved. This simulation supports the previous conclusion that no sole exponential function can give a sensible trajectory of the growth of developmental complexity.

It should be noted that the linear time axis stretches out to the left. The oldest traits in the diagram of Figure 1, formed 600 million years ago, is in the diagram of Figure 3 placed some 600 km to the left. It is instructive here to give some values of complexity for points of time that are not seen in the diagram of Figure 3.

| Time before present (years ago) | Individual age (years from conception) | Relative value of complexity |
|---|---|---|
| 3700 million | 0.03 | $10^{-161}$ |
| 600 million | 0.06 | $10^{-26}$ |
| 6 million | 0.38 | 0.16 |
| 1 million | 0.76 (birth) | 0.30 |
| 25000 | 3.2 | 0.52 |
| 1200 | 10 | 0.80 |
| 400 | 16 | 0.92 |
| 100 | 28 | 0.98 |
| 20 | 52 | 0.99 |

Table 2. Some relative values of complexity as given by the simulation program.

The extremely small values of complexity for the early epochs of the history of life demonstrate the wide range of complexity characterizing evolution although in the present arrangement compressed to the interval between 0 and 1. Especially, the table illustrates the rapid increase of complexity for more recent epochs. Thus, it shows an increase of complexity of about 30% all the way from the origin of life up to one million years ago, a point of time of about the appearance of man. Furthermore, the simulation indicates that as much as 8% of the total complexity is accomplished during the last 400 years, the scientific epoch, and that 2% is achieved during the last 100 years. Thus the calculations demonstrate, as far as complexity is concerned, the great significance of the appearance of culture and science (about 70%) as compared to the morphological evolution (about 30%) of the body of man. It should be remembered that these values of complexity are applicable only to the lineage of man and to cultural expressions restricted as previously discussed.

Of course, this limitation of applicability of the method means that the values of complexity obtained are only a part of the total complexity that has grown in the course of the evolution of life and in the large variety of human cultures. The relative values obtained by means the present method must thus only be interpreted in relation to the method used for their calculation and to their limited range of applicability but, on the other hand, the method used indicates a way to receive a quantitative estimation of complexity in the evolutionary process of special significance for the human species, the human culture, and the scientific civilization.

## Applications

Gould’s conundrum, apparently shared by many other researchers, seems to be to bring together two seemingly contradictory observations. He didn’t see any tendency of progress in discrete species but, on the other hand, he couldn’t deny that the history of life on the whole exhibits increasing complexity. In the same vein, Maynard Smith and his co-author state that even if progress is not a universal law of evolution, common sense does suggest that at least some lineages have become more complex (Maynard Smith et al. 1995 p. 5).

The additional complexity found in the human species sets apart humans from other animals. This does not imply that all other organisms are arranged in a single ascending line behind us. Rather, the present model suggests a visualization of evolution as a branching tree in which most species show little increase of complexity or none at all but that consecutive species may display increasing complexity. Therefore, the history of life by and large exhibits increasing complexity as Gould himself concludes from the observation that species that have come into existence late in the evolutionary process exhibit a greater level of complexity than those that have appeared earlier. Especially, in the last million years a superior level of complexity is accomplished by the human species and furthermore, the human species is unique inasmuch as it shows a steady trend of increasing complexity whereas most other discrete species by and large show no tendency of such an increased complexity. In this way, I maintain, Gould's conundrum is resolved.

It is a spread notion also to regard technological evolution as an extension of organic and cultural evolution, a notion of which Kurzweil (2005) is a strong representative. He points out that each paradigm develops through three faces in an S-shaped curve and that evolution, both in its biological and technological expression, evolves through a series of such curves forming a soft stair-like curve. Kurzweil emphasizes that evolution follows an accelerating pace and that in our own time this pace has reached an unprecedented rate, a view in concordance with the present model. However, when technological evolution, as Kurzweil anticipates, transcends biological evolution the present model reaches its limit of applicability since this type of evolutionary process is not coupled to the individual developmental process. Further applications of the present model to future evolution are discussed in Ekstig (2007).

A principal difference between morphological and cultural evolution is that as soon as a species is split, no reunion is possible whereas different cultures carry out far-reaching influences on one another. In our own days, the rapidly increasing efficiency of communication between societies and between individuals all over the world presumably contributes strongly to the rapid growth of the complexity in a world-wide cultural evolution. However, these contributions to complexity are mostly expressed in fields of culture that are not part of the present method.

As can be seen in the references discussed in the introductory paragraph, it is suggested that evolution has a direction, and that such a direction is built on the hitherto prevailing trend of increasing complexity. According to the present model, a continued increase of complexity is coupled to gradual condensation of developmental stages. However, the condensation of embryonic stages is so small that it has only theoretical interest. In contrast, condensation of mental stages is more substantial. It builds on an enhancement of education all the way from the upbringing of infants to the scientific education of adolescents and adults (Ekstig 2004). It is a most alarming piece of evidence, though, that today's school pupils in many western countries exhibit less ambition to learn the basic concepts of mathematics, science, and engineering and as a result the long-lasting trend of accelerating biological and cultural evolution coupled to condensation of developmental stages now may be at risk. Therefore, I would in the present context like to emphasize the crucially important task of enhancing mathematical and scientific education at all levels.

## Conclusion

I have in the present model given a wide view over the evolutionary process in the lineage of man ever since the formation of multicellular organisms and up to present time. The model implies that the cultural and scientific parts of human evolution fit into the preceding biological evolution in a common pattern, suggested to be the result of an extensive regularity of the process of condensation. This wide-ranging regularity of condensation is found to be the consequence of the fact that the value of condensation of each stage is inversely proportional to the duration of its action on the stage.

The model is focused on the growth of complexity and, in spite of the fact that complexity is an immeasurable quantity, the model has made possible an estimation of the relative values of complexity all over the history of biological and cultural evolution in the lineage of man. The method for these estimations is restricted to traits that can be identified in the developmental as well as in the evolutionary processes. Therefore, many expressions of evolution, especially in its cultural realm, are not contributing to the values of complexity obtained by the method.

The result indicates that the value of complexity has grown in a strongly accelerating pace throughout the history of human biological and cultural evolution. In spite of the limitation in the applicability of many cultural manifestations, the result indicates that mankind's contribution to complexity during the last million years is greater that what has been accomplished during the long preceding history of man's biological evolution.

# Does species evolution follow scale laws ? First applications of the Scale Relativity Theory to Fossil and Living-beings

**Jean Chaline**[1]

**Abstract**

We have demonstrated, using the Cantor dust method, that the statistical distribution of appearance and disappearance of rodents species (Arvicolid rodent radiation in Europe) follows power laws strengthening the evidence for a fractal structure set. Self-similar laws have been used as model for the description of a huge number of biological systems. With Nottale we have shown that log-periodic behaviors of acceleration or deceleration can be applied to branching macroevolution, to the time sequences of major evolutionary leaps (global life tree, sauropod and theropod dinosaurs postural structures, North American fossil equids, rodents, primates and echinoderms clades and human ontogeny). The Scale-Relativity Theory has others biological applications from linear with fractal behavior to non-linear and from classical mechanics to quantum mechanics.

## 1. First demonstration of power-laws in fossil living-beings evolution

### 1.1. Introduction

The part that chance plays in evolution is a much debated issue (Monod, 1970; Grassé, 1978; Dawkins, 1986). While chance clearly is involved in the living world at many levels, e.g. in the formation of gametes, fertilization and the choice of mates, it is not yet really known to what extent chance is involved in speciation, microevolution, extinction, and macroevolution. We have sought to find out whether the events that punctuate the fossil record adopt a random logic or whether they involve fractal dynamics implying the existence of power laws ?

As non-linear structures and scale laws had been discovered in Earth sciences (Allègre et al., 1982 ; Dubois and Cheminée, 1988, 1993), it was interesting to test this approach in paleontology. The idea was dared : can species evolution had a fractal behavior as earthquakes or volcanic eruptions ? The very high quality of Eurasian Arvicolid rodent data occurring within a well calibrated climato-stratigraphic record provides good basis for such an analysis.

[1] Laboratoires de Biogéosciences (UMR CNRS 5561) et de Paléobiodiversité et Préhistoire de l'EPHE, Université de Bourgogne, Centre des Sciences de la Terre, 6 Bd. Gabriel and 143 av. V. Hugo, 21000 Dijon, France. (+33380573546 ; jean.chaline@orange.fr).

## 1.2. The arvicolid data

Voles and lemmings, which make up the Arvicolid family, have a holarctic distribution. Lineages within the group can be traced from the Pleistocene to the Recent without great difficulty. Biological, genetical, morphological and palaeontological data have made it possible to specify kinship among lineages. Arvicolids are abundantly fossilized in owl pellets and in beds which lie close together both in time (ranging from hundreds to thousands of years, often dated by physical methods) and in space throughout Eurasia. Paleontological data calibrated by physical dating (Chaline and Laurin, 1986; Chaline, 1987) have served to establish the chronology of speciation (speciation or First Appearance Datum: FAD) and extinction (extinction or Last Appearance Datum: LAD) events used in this study.

Voles appeared in the fossil record 5.00-4.5 Myr ago and are abundant in Pliocene and Quaternary sediments. There are 140 lineages and 37 distinct genera. A hundred species survive (Honacki et al., 1982; Chaline et al., 1999). Over 5 Myr 52 new lineages appeared in Eurasia (FAD) by cladogenesis or allopatric speciation, while 34 lineages disappear (LAD). Some specific lineages display gradual evolution where the authors have distinguished each evolutionary step in the chronomorphocline by specific names, or diachrons (Chaline and Laurin, 1986. These phyletic species are excluded from the inventory of FAD, which only correspond to the appearance of new species by cladogenesis. All appearances of species by allopatric speciation necessarily correspond to a punctuation on the geological scale, because they occur in such small populations in such confined areas that they have statistically no chance of being fossilized. FAD and LAD can be considered as biological events because they occur instantaneously on the geological scale, like other geological events (e.g. volcanic eruptions) which are amenable to point-process analysis.

Thus, the arvicolid radiation is one of the best known (Chaline, 1987; Fejfar and Heinrich, 1990; Korth, 1994 ; Chaline et al., 1999) and provides an opportunity to analyze the anatomy of a radiation (Chaline and Brunet-Lecomte, 1992). Arvicolids form a homogeneous group. The average life expectancy of species is less than 1 million years and the speciation and extinction dates of species are known to within 100.000 years. We have confined our investigations to Eurasia. The plot of species variety versus time highlights three phases in the radiation : a multiplication in the number of species between 4.4 and 2.3 Myr is followed by stabilization between 2.3 and 0.7 Myr and new multiplication between 0.7 and the present (Fig. 1 ; table 1).

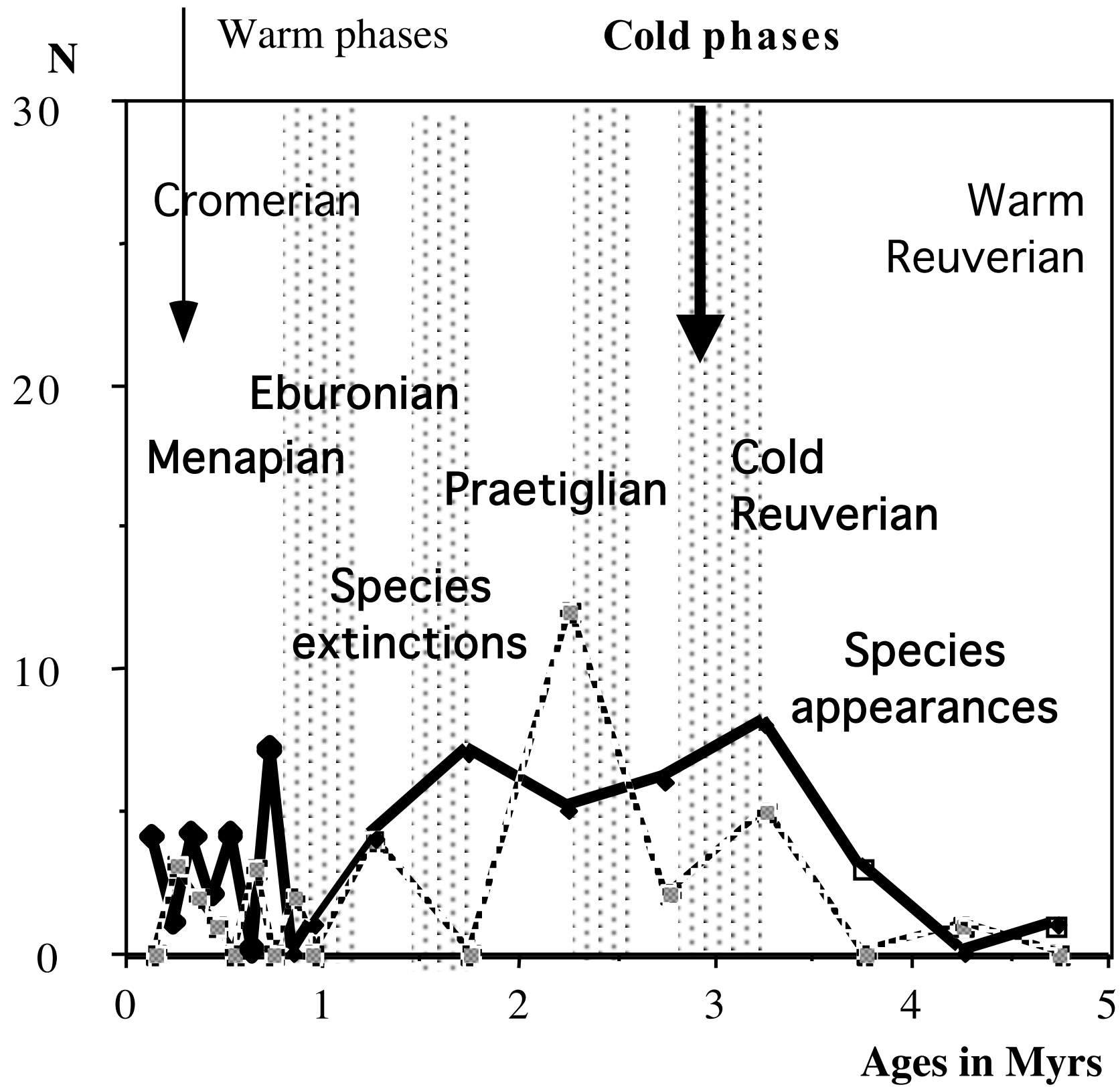

Fig. 1. Species appearances and extinction. The European Arvicolid radiation in the Quaternary. Distribution of the number of species over time. The first phase between 5 and 3 Myr corresponds to an exponential multiplication of species. A second stabilising phase occurs between 3 and 0.7 Myr. A third phase between 0.7 Myr and the Present Day shows a new exponential multiplication. Diagram showing the variation in species diversity (N species), the number of appearances (FAD: 52 lineages) and extinctions (LAD: 34 lineages) related to the broad Plio-Pleistocene climatic phases cold (Menapien, Eburonien and cold Reuverien) and warm phases (Cromerian, Waaline, Tiglien and warm Reuverien). It is obvious that the major phases of extinction do not take place during cold phases but during the warm or temperate phases which follow them. (after Chaline and Brunet-Lecomte (1992).

| AGES in Myrs | N FAD | N LAD | SPECIES RICHNESS |
|---|---|---|---|
| 0.15 | 3 | 0 | 22 |
| 0.25 | 1 | 3 | 21 |
| 0.35 | 3 | 2 | 22 |
| 0.45 | 2 | 1 | 19 |
| 0.55 | 3 | 0 | 17 |
| 0.65 | 0 | 3 | 16 |
| 0.75 | 7 | 0 | 16 |
| 0.85 | 0 | 2 | 11 |
| 0.95 | 0 | 0 | 11 |
| 1.25 | 4 | 4 | 13 |
| 1.75 | 6 | 0 | 10 |
| 2.25 | 5 | 11 | 15 |
| 2.75 | 6 | 2 | 12 |
| 3.25 | 8 | 5 | 11 |
| 3.75 | 3 | 0 | 3 |
| 4.25 | 1 | 1 | 1 |

Table 1. Arvicolid data in Europe. Ages, N FAD (First Appearance Datum or speciation), N LAD (Last Appearance Datum or extinction), N species (after Chaline and Brunet-Lecomte, 1992).

Randomness and non-random testings, the survivorship curve test, on the basis of which van Valen (1973) proposed the Red Queen's Hypothesis and a Poisson distribution test for randomness, and the Cantor dust model and the Stationary model for non-random tests.

## 1.3. The survivorship curve and the Red Queen's hypothesis

Randomness can be tested by several models, including survivorship models. These generally apply to large samples (several hundred) designed to investigate the longevity of species. Survival time must not be confused with the rate of speciation and extinction over a given unit of time. The scale of the analysis in terms of systematics may involve species, genera or families and in terms of time may range from 100.000 to 1 or even 10 million years. Depending on the scales used, the appearance or extinction of a species will not have the same causes or the same signification. The method consists in plotting the proportion of species that survive for given time intervals. The use of a logarithmic ordinate means that "*the slope of the curve at any age is proportional to the probability of extinction at that age*" (van Valen, 1973: p. 1). van Valen, testing more than 25.000 taxa, concludes "*almost uniform linearity for extinct taxa except for effects attributable to sampling error*". "*For living taxa linearity of the distribution requires both constant extinction and constant origination*" (Van Valen 1973: p. 7). Van Valen attributes the deviations from linearity to sampling error and to exceptional events such as massive extinction at the end of the Permian. This model suggests that the physical environment plays no role in the appearance and extinction of species and that only biotic factors are involved in a wholly inside story.

The survivorship test curve for the Arvicolid radiation data does not fit the random model well (for graph and calculations see Dubois et al., 1992). Despite a highly significant correlation coefficient ($r = 0.93$), a linear model is not sufficient to account for the relation between the two variables, as is attested by the non-random distribution of residuals *versus* species duration.

Chaline and Laurin (1986) had demonstrated that the evolution of the *Mimomys occitanus - ostramosensis* vole lineage between 4.4 and 1.5 Myr was partly controlled by climatic changes. They seem to be linked to climatic fluctuations which act as *stimuli* speeding up or slowing down the rate of phyletic gradualism occurring in this lineage. The adaptive response is either different (progressive cement development or gradual hypsodonty increase) during similar climatic conditions, or identical (hypsodonty increase) for distinct climates (cold or warm). Thus, climatic factors are clearly involved, but indirectly, at one remove. If the radiation data are placed in the Plio-Pleistocene climatic context, it can be seen that disappearances, or extinctions, do not occur during cold phases but mainly during warm or temperate crises, suggesting a climatic control factor. These data are in conflict with the Red Queen's Hypothesis. It is clear that external factors are active.

### 1.4. Poisson distribution test of intervals between FAD

Randomness implies the existence of a limiting distribution but also means the absence of correlations. A further confusion seems to stem from the recognition (Arnéodo and Sornette, 1984; Sornette and Arnéodo, 1984) that passing all the usual statistical tests is not the sign of "true" randomness since certain completely deterministic systems have been successfully put through all the tests. Passing all tests of randomness does not preclude a system from possessing low dimensional deterministic dynamics. Let us be more precise about the main origin of randomness. The apparent randomness in a given sequence of numbers obtained from dynamic change of a system may result essentially from two types of dynamics[2]. Coming back to our time series, let us first examine the distribution of time intervals between two successive speciations (FAD) or extinctions (LAD). For intervals longer than 50.000 years, the evidence of a non linear is quite clear. This observation is fundamental because it excludes, in our opinion, any exponential, Poisson or gamma distribution. Conversely the bi-log graph seems to provide a fairly good linear fit, suggesting a power law distribution. The previous test provides a strong argument for the deterministic behavior of the dynamic system (in the physicists' sense of dynamic systems, no other meanings should be given to dynamic) which generates the time series. To find out more about this determinism we have used more appropriate tests and compare the results obtained from both FAD and LAD series. For this type of data either a fractal approach by Cantor dust analysis or a first return map could have been used. The approximation on the short intervals suggested that the return map would be strongly polluted along the main bisector (see Sornette et al., 1991; Dubois and Cheminée, 1993). In order to strengthen our argument about the evidence for fractal structures, we shall also compute the information dimension for comparison with the similarity dimension as provided by our Cantor dust analysis.

---

[2] In the first type, systems are characterized by many degrees of freedom and they may develop random dynamics as a result of the complexity created by the superposition and coupling of changes in each degree of freedom (e.g. Brownian molecules, Johnson electrical noise due to thermal excitation of the motion of electrons). In these cases a Markovian process may adequately describe changes in the system over time and (any observable) changes in a physical variable cannot be distinguished from those in a suitable random process. Randomness is thus linked to the large number of degrees of freedom. In the second type, the dynamics of the system have only a few degrees of freedom only but may present highly complex behavior. This involves deterministic chaos (Bergé et al., 1984; Sornette et al., 1991; Dubois, 1995). Let us recall the main properties of these systems: (a) sensitive dependence upon initial conditions, (b) reinjection of the trajectories (in the phase space) in a closed domain by non-linear processes, (c) fractal dimension (in the phase space) of this geometrical object, the attractor. This second type is very important and extends the concept of determinism to a very broad domain which was previously considered to involve randomness. Time sequences of such low dimensional dynamic systems have been closely studied and it was realized that, when sufficiently chaotic, their dynamics are indistinguishable from truly random processes. The result is that time sequences of deterministically chaotic systems are found to pass all tests of randomness and thus qualify as completely random. Thus a major breakthrough in non-linear dynamics has been to reduce the huge array of systems that were thought to be random by means of techniques for quantifying the number of degrees of freedom. To highlight the deterministic nature of subsets with a finite number of degrees of freedom (often even small numbers - 2, 3, 4, etc.) we separate them from those with a larger number and that are still (for how long?) thought of as random (and so unpredictable).

## 1.5. Test of the stationary model

In contrast to the "Red Queen's Hypothesis" which claims that extinction rates remain constant within groups regardless of physical changes in the environment (Law of constant extinction or random model), the Stationary model proposed by Stenseth and Maynard-Smith (1984) predicts that evolution will cease with the absence of abiotic parameters. In constant environments the Stationary model predicts zero rate of change in biomes, while in periodically disturbed environments, e.g. under Pleistocene climatic fluctuations, appearances and extinctions of species are one response to the perturbations (Benton, 1985). It seems clear that climatic fluctuations play a role, at least for extinctions. As climatic fluctuations modify the paleogeography of species, it should be noted that they induce faunal shifts and isolation within species and are also favorable periods for speciation.

## 1.6. The Cantor dust method: fractal set in Arvicolids

The Cantor dust method was used for the first time in paleontology (Dubois et al., 1992) in order to analyze of the Arvicolid radiation irrefutably shows that the distribution of speciation and extinction events is fractal, i.e. that it depends on several factors resulting in a deterministically chaotic pattern. The FAD series of tests performed by computing Similarity, Correlation and Information dimensions confirms that the results are coherent. Appearance (FAD) and disappearance (LAD) phenomena that are one of the major aspects of the evolution of biodiversity stem from the interaction of internal and external coercions. Internal coercions include gene mutations and their expression, especially in development and morphogenesis. External contingencies refer to climatic environmental parameters of temperature, humidity, plant and animal context (Hokr, 1951) and to geological, or cosmic events; factors which introduce contingency into evolution.

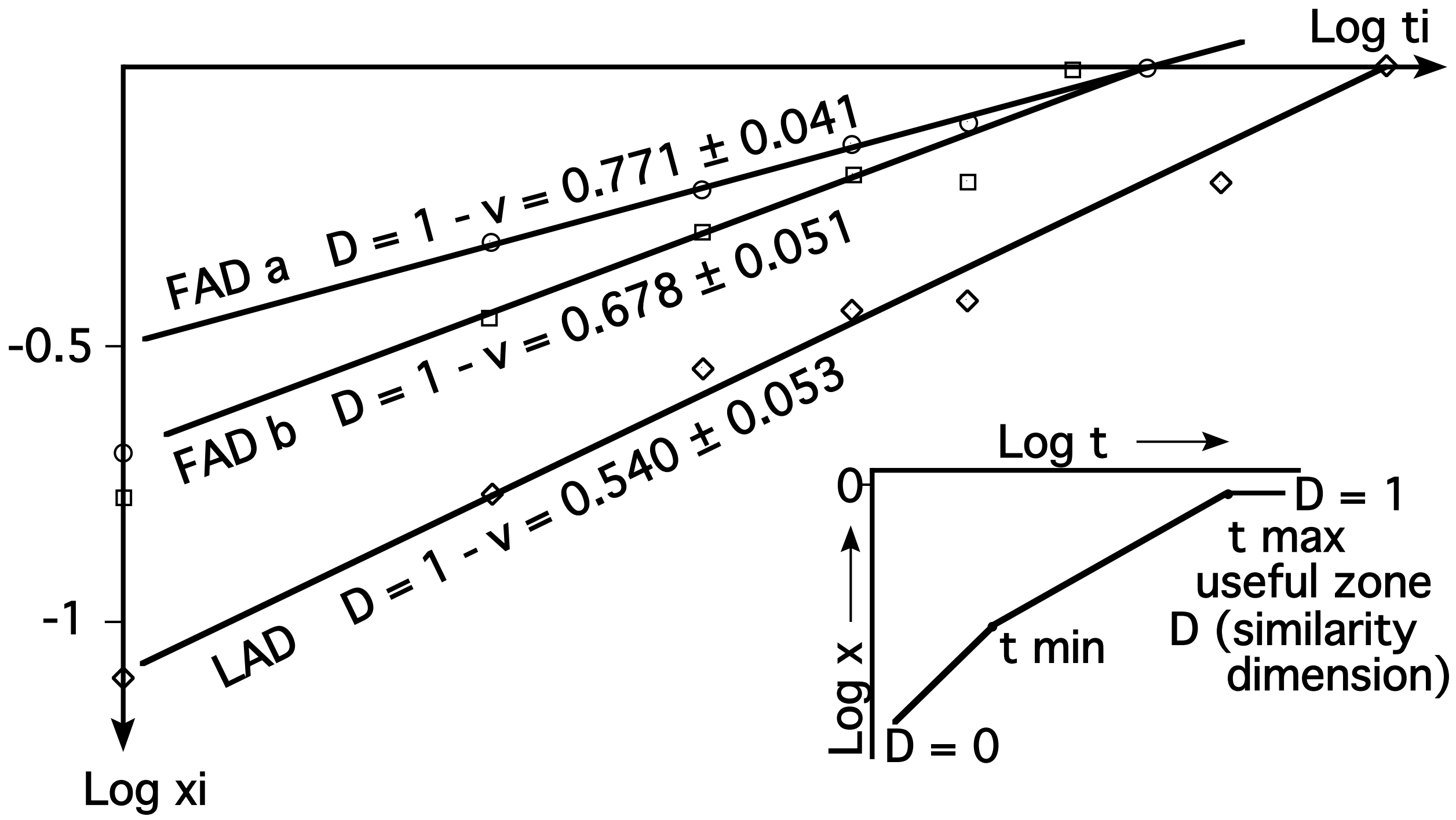

Figure 2. The graphs Log xi versus Log $t_i$ for the three time series: FAD (a) (-5 Myr-0), the slope of the least square regression line is v = 0.2292 ± 0.0412 and D = 1- v = 0.7708 + 0.0412; FAD (b) (-1 Myr - 0), D = 0.6782 ± 0.0506; LAD (-5 Myr - 0), D = 0.5397 ± 0.0529 (after Dubois et al., 1992).

FAD and LAD data show that the two phenomena are not constrained in the same way (Fig.2). LAD data show a stronger tendency to clustering, which may be interpreted as the outcome of a marked, major impact of external coercions. In the present state of knowledge, it seems to us that this factor is above all climatic and environmental.

In fact, it can be seen from the Arvicolid data (Chaline, 1987) that during the same climatic sequences some species remain in stasis (species in equilibrium), whereas others evolve gradually and yet others disappear (species in disequilibrium) (Chaline and Laurin, 1986). Climatic parameters may underlie the disappearances, as it has already been noted that they seem to occur in warm or temperate phases following major cold spells. This observation has been confirmed by the works of Graham and Lundelius (1984), Graham (1985) and Graham and Mead (1987) on variations in diversity specific to mammal associations in North America during the last Ice Age, showing that more species disappear at the start of the post-glacial re-warming, seemingly after the break up of plant and animal communities. Climatic warming occurs rapidly on the geological scale and causes potentially fatal disequilibrium between species and their environments. In contrast, climatic cooling, which is far slower on the geological scale, allows species to acclimatize either by following geographical shifts in nutritive plant areas (case of lemmings and voles in Siberia) or by evolving in terms of morphology, ecology or ethology. During cold phases, there was a reduction in seasonal variation accompanied by a fall in average summer temperatures and a rise in average winter temperatures which allowed highly diverse communities to develop in areas that were limited by the spread of continental glaciers. Conversely, during warming, there was an increase in seasonal variations which destroyed existing biomes and produced new ones.

This explains the fact that the coercions on extinctions are stronger than those on speciations perhaps also because of the rapid warming subsequent to the end of glaciation. Appearances of new species are out of phase in time possibly because of the appearance of new ecological niches and the time species need to give rise to new species. New developments in theoretical ecology have shown that chaos may play a role in reducing the probability of extinction (Allen et al., 1993).

Moreover, as it is difficult to separate biotic from abiotic factors in extinction and speciation rates, our method suggests an overlapping Red Queen/Stationary model, depending on circumstances. In fact these models depend on several other factors, for example whether species are in ecological equilibrium or disequilibrium with their environment and whether groups display great or small genetic variability.

There seems, therefore, to be an overall logic behind speciations and extinctions in the living world. The extension of this type of analysis will very likely permit better evaluations of the impact of external coercions on evolution. The result is highly important for evolutionary theory in that it reduces, or changes the role played by chance in what used to seem highly random processes. The appearance and extinction of species clearly obeys fractal dynamics.

## 1.7. Others fractal patterns in paleontological data

The problem of the existence of fractal structures in the fossil record is largely discussed since this first approach, and the many works arrive to sometimes contradictory conclusions. Minelli et al., (1991) have shown that the genera distribution within the life classification follows a power-law. Plotnick and McKinney (1993), have demonstrated self-organization evidence in planktic foraminiferal evolution. The same year, Hanski et al., (1993) explains that the boreal rodents population oscillations regulated by mustelid predators leads to a chaotic system. Patterson and Fowler (1996) have shown that originations and extinctions in planktics foraminifers follows also power-laws and that extinctions were more constrained; a result comforting our rodent results. The self-similarity of extinction statistics in the fossil record described by Solé et al., (1997) indicate a non-linear answer of biosphere to environment perturbations as a major cause of extinctions, a scenario contradicted by Kirchner et Weil (1998) who do not observe fractal structure in the statistics of extinctions. Burlando (1990, 1993) shows that the living-being system and radiations present an auto-similarity and that the geometry of evolution was fractal. Verrecchia (1996) demonstrated, that stromatolithes present a fractal structure (*D = 1.73*) and that stromatolithe growth may be simulated by Limited Diffusion Aggregation models (DLA systems) showing variations of growth from dendritic to moss-like. Many others have discussed of problems of self-organized criticality, auto-similarity and extinctions in the evolution of some groups (Plotnick and Sepkovski, 2001; Newman, 1996 ; Newman, and Eble, 1999a & b ; Bak, 1996 ; Kauffman, 1993 ; Bornholdt, 2003 ; Bornholdt and Sneppen, 1998 ; Sneppen et al., 1995 ; Sibani et al., 1998). We can now consider that the fractal pattern of evolutionary phenomena (speciations, radiations and extinctions) was clearly established at the end of the nineties.

Numerous books about fractal structures in nature were published by Mandelbrot (1977, 1982), Dubois (1995), Sapoval (1997) and many others. But all these works were purely descriptive, showing the universality of fractals, but they present no theory explaining why the fractals?

Nottale since the eighties has developed his new *Theory of Scale Relativity* (SRT) explaining origin and the distribution of fractals in nature resolutions (see Nottale's article).

## 2. Applications of the theory of scale relativity to fossil and living beings

### 2.1. Linear cases with fractal behavior

It has been shown that toward the small scale one gets a scale invariant law with constant fractal dimension, while the explicit scale-dependence is lost at larger scales. Self-similar laws have been used as models for the description of a huge number of biological systems. Many examples of fractal structures were demonstrated within living-being hierarchical organization, from the simplest to the most complex: DNA sequences containing noncoding material[3], ADN coiling up in chromosomes, neuronshape, blood vessels network of lungs and kidneys, mammalian lungs (Schlesinger and West, 1991), branching patterns of retinal vessels, bifurcations viscorous fingering of the flow of hydrochloric acid in the mucus gel layer protecting epithelium and stomach against acidification (Buldyrev et al., 1994) and self digestion, bird's feathers, bacterial colonies showing analogies with the inorganic growth model called diffusion limited aggregation (DLA), root systems, branching of trees and plants (ferns, black elder and cauliflowers). Many aquatic mollusks present fractal fatures pattern on their shells. For instance, the shells of *Conus thalasiarchus* and *Olivia porphyria* show a Sierpinski triangle pattern which can be now simulated (Fowler et al., 1992). Boettiger et al. (2008) have found that the neurosecretory system of aquatic mollusks generates their diversity of shell structures and pigmentation patterns. The anatomical and physiological basis of this model sets it apart from other models for shape and pattern. The model reproduces all known shell shapes and patterns, and accurately predicts how the pattern alters in response to environmental disruption and subsequent repair. They have shown that all the patterns emerge from combinations of three types of bifurcations: Turing instabilities giving raise to stable bands perpendicular to the growing hedge and Hopf bifurcations, and wave propagation and collisions producing zig-zag stripes. In *Olivia porphyria* the pattern is created by a reverse wave. At larger ecological scales size distribution of plants supporting insects are related to the fractal distribution of the leaves, diffusive spread of populations growth with the shape of a smooth disc becoming increasingly rough, bird vigilant behavior. Such phenomena have been also described in ion channels in cell membranes, the human heartbeat disease, certain malignancies, sudden cardiac death, epilepsy, fetal syndrome, human language, and many others, etc. The systems displaying power laws and fractal structures are largely widespread in nature. One reason is certainly the fact that they increase organ surface providing better function, a characteristic certainly maximized by natural selection.

[3] The lack of long-range correlations in coding region seems to be a necessary condition for functional biologically active proteins (Buldyrev et al., 1994).

## 2.2. From fractal to log-periodic laws

The above results suggested that paleobiological data must be quantized and analyzed at distinct scales by new non-linear methods. For instance, the renormalization group approach (Nauenberg, 1975 ; Jona-Lasinio, 1975) predicts that solutions of renormalization equations concerning initially only « quantum field theory » and applied to statistical physics to « phase transition phenomena » and « critical phenomena », lead both to *power law scale behavior* and *log-periodic corrections* of such behavior. Moreover, the critical behavior is *a priori* symmetrical around the *critical value* of the variable under consideration. Under the proximity of critical time, the system became instable and fractal, and shows precursor events of an accelerated rate leading to critical time, specific of the system. After the critical time, the system displays replical events in a decelerating manner. It must be noted that the application of acceleration to life evolution has been anticipated by Meyer (1947, 1954) naturally without any possible adequate calculations. Both log-periodic accelerations before the critical point ("precursors") and decelerations after it ("replicas") are expected (Fig.3), and they have been confirmed for spatial structures and temporal structures, in earthquakes (Sornette D. and Sammis C.G 1995) and stock market crashes (Sornette et al., 1996; Sornette, 2003). Among corrections to scale invariance, log-periodic laws may thus play a very important role in many domains, not only confined to physics, but also to Earth, life and human sciences (probabilist deterministic and predictive laws).

Sornette's work on « *discrete scale invariance* » (1998) led Nottale to consider the tree of life as a real tree, but with a temporal fractal structure rather than a spatial one. The jumps between species involve bifurcations allowing to liken the general evolutionary process to a "tree of life" where "branch" lengths represent time intervals between major events. The question raised is whether this tree can be described by a mathematical structure, at least at a statistical level.

## 2.3. Log-periodic behavior

By analogy with real trees, we have tested as a first approximation the simplest possible law, i.e. a self-similar tree (Chaline et al., 1999a). Such a law[4] corresponds to discrete scale-invariance and log-periodic acceleration or deceleration, characterized by a critical point of convergence $T_C$ which varies with the lineage in question (Fig. 3).

These studies are purely chronological analysis which does not take into account the nature of events, but they provide us with a beginning of predictivity interpreted as the dates of the peaks of probability for an event to happen.

We have analyzed the time sequences of major evolutionary leaps at various scales, from the scale of the global life tree (appearances of prokaryotes, eukaryotes, multicellulars, exoskeletons, tetrapody, homeothermy and viviparity), to the scales of large clades (orders and families) such as sauropod and theropod dinosaurs postural structures, North American fossil equids, rodents, primates and echinoderms clades. Finally, considering the relationships between phylogeny and ontogeny, it was interesting to verify whether the log-periodic law was also applied to the various stage of human ontogeny.

---

[4] This law [$T_N = T_c + (T_0 - T_c)\,(g)^{-n}$ ] is dependent on only two parameters, $g$ (scale ratio between successive time intervals) and $T_c$, which of course have no reason *a priori* to be constant for the entire tree of life. Note that $g$ is not expected to be an absolute parameter, since it depends on the density of events chosen, i.e. on the adopted threshold in the choice of their importance (namely, if the number of events is doubled, $g$ is replaced by $\sqrt{g}$). Only a maximal value of $g$, corresponding to the very major events, could possibly have a meaning. On the contrary, the value of $T_c$ is expected to be a characteristic of a given lineage, and therefore not to depend (within error bars) on such a choice.

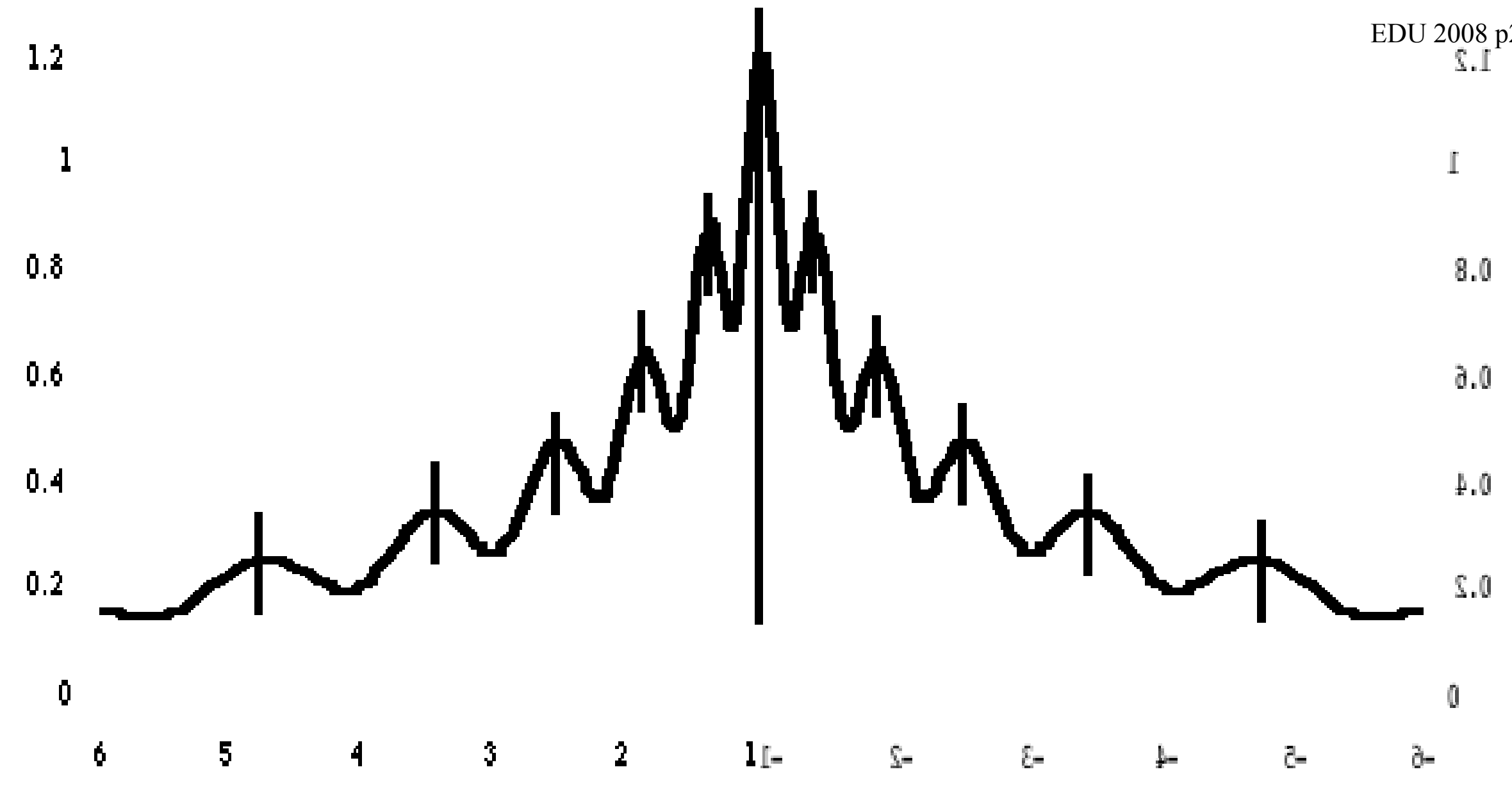

Figure 3. Log-periodic law. Log-periodic accelerations (precursors) occur before the critical point and decelerations after it (replicas) (after Dubois and Chaline, 2006).

### 2.3.1. Primates bauplans

Delattre and Fenart (1954, 1956, 1960) showed that cranial ontogenesis of higher apes and humans is characterized by varying degrees of occipital flexion, on the one hand, and prognathism on the other hand depending on genus and bipedalism. One fundamental discovery is that of a growth phenomenon in which three-dimensional organization of basicranio-facial architecture is controlled by the processes of flexure at the base of the skull. This affects the morphogenesis of the two components of the face, the maxilla and the mandible (Deshayes, 1986, 1988, 1991 ; Dambricourt Malassé and Deshayes, 1992). These major stages of cranio-facial contraction are a well known phenomenon in Primates, alternatively termed flexure of the skull base or occipital shift (Deniker, 1885; Anthony, 1952), fundamental ontogenies[5] (Dambricourt-Malassé, 1987, 1988, 1996). Dambricourt-Malassé and Deshayes (1992) discovered and proved that the cranio-facial contraction is an embryonic phenomenon which is clearly visible in the mandible and that the living and fossil lower jaws retain the range of contraction. The fundamental bauplan is fixed at a very early stage of the embryonic development. The contraction process starts very early in ontogeny and is contemporaneous with organogenesis. Such structures are determined by *Hox* genes and are defined in scope and duration by embryogenesis which represents the first eight weeks after fertilization for humans and a little less in other primates. The chondrocranium and splanchnocranium retain the imprint through to adulthood bauplan in turn is rapidly diversified, by speciation, into as many species as there are ecological niches available.

[5] Dambricourt-Malassé (1987) introduces the term of *fundamental ontogeny* covering a set of populations belonging to one or more species, or even to one or more genera, dispersed in time and space, but united by a common ancestral ontogenetic bauplan. It is a set of individual ontogenetic designs that develop from a common embryonic plan.

After comparison of Recent mandibles recording the contraction stages, Dambricourt Malassé et al. (1999) distinguish at least five distinctive organizational groups of mandibles related to different cranial morphological plans among extant and fossil primates. These skull bauplans can be identified from the increasing intensity of embryonic cranio-facial contraction and increase of cranial capacity, respectively as follows:

(1) prosimians (strepsirrhines and *Tarsius*) (plus adapiform) ;

(2) monkey apes: platyrrhines, cercopithecoids and hylobatids (plus *Propliopithecidae*) ;

(3) great apes (gorilla, chimpanzee, and orang utan) (plus *Dryopithecidae*) ;

(4) earliest *Homo* ;

(5) *Homo sapiens.*

After Dambricourt, it is the embryonic amplitude of cranio-facial contraction that differs between the various primates and not the nature of the process itself. Craniofacial contraction and cranial capacity are minimal in both fossil (adapiform) and extant prosimians, more substantial in simians or monkeys (cercopithecids), even more pronounced in great apes *(Pongo, Gorilla, Pan)* and australopithecines (*Australopithecus* and *Paranthropus*). Conversely, it is maximal in humans particularly in *Homo neanderthalensis* and *Homo sapiens*, and is sustained to a late stage, or even to adulthood. The basicranio-facial bauplan of gorilla, orang utan, and chimpanzee, despite nearly 20 millions years of evolutionary divergence, are built to the same ontogenetic design termed the "great ape" as opposed to the "monkey" or "australopithecine" skull plans. Comparison of the fundamental ontogeny of the « great apes » and of « *Australopithecus* » shows that it is the fundamental ontogeny and the common organogenesis of the great ape species that has been restructured and dynamized. The transition from the "Great Ape" skull plan to the "Australopithecine" skull plan is characterized by occipital rotation, facial contraction and expansion of the upper cranial vault, with the *foramen magnum* at the skull base moving to a more horizontal position. The transition from the "Australopithecine" skull plan to the "*Homo*" skull plan is reflected by tilting and forward movement of the *foramen magnum*, posterior extension of the skull, facial contraction and broadening of the frontal bone definitively separating the *bregma* and *stephanion*, a clear characteristic of the genus *Homo.* Recent geometric morphometrics studies on the skull have shown that the modern human form is not significantly different from the earlier forms of the genus *Homo* and thus the *Homo sapiens* bauplan cannot be retain (Chaline et al., 1998). The figure 4 emphasizes the overall craniofacial contraction that occurred in leaps and entailed re-shaping of the skull outline and the tilting of the *foramen magnum.*

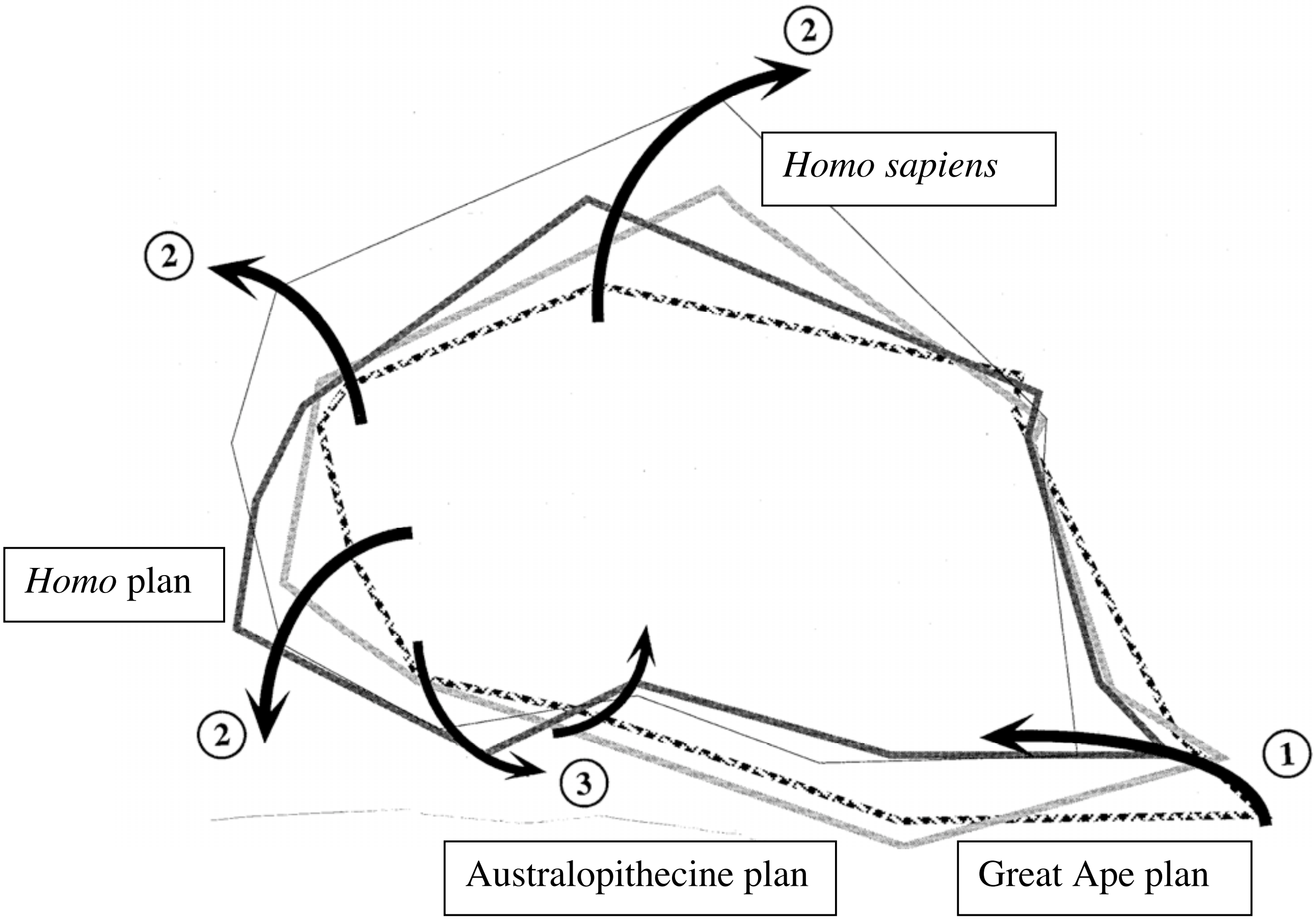

Figure 4. These simplified outlines of the three skull plans identified in the sagittal plane recapitulate the morphological changes observed between Great Apes plan, Australopithecine plan, *Homo* and *Homo sapiens* plan (Procrustes methods). 1. Craniofacial contraction ; 2. Uppervault by extension; 3. Tilting of the *foramen magnum* (after Chaline et al., 1998).

From great apes to modern man numerous heterochronies of development (hypermorphosis, hypomorphosis and post-displacements) have occurred during ontogeny (Chaline et al., 1998), allowing (1) the acquisition of permanent bipedalism of *Australopithecus* and *Homo*, (2) the increased cranial capacity of primitive forms of *Homo* (*habilis*, *ergaster*, *rudolfensis*, *erectus*, *heidelbergensis* and *neanderthalensis*) and (3) the disappearance of simian characters associated with renewed increase in cranial capacity in *Homo sapiens*.

The jumps between species (Chaline et al., 1993) involve bifurcations allowing us to liken the general evolutionary process to a tree of life where branch lengths represent time intervals between major events. The question raised is whether this tree can be described by a mathematical structure, at least at a statistical level. By analogy with real trees, we test as a first approximation the simplest possible law, i.e. a self-similar tree.

Data: Dambricourt et al., 1999; Chaline, 1998. The skull bauplans appearances are respectively as follows (Fig.5):

(1) Prosimians (strepsirrhines and *Tarsius*) (plus adapiform) (-65 Myr);
(2) Monkey apes: platyrrhines, cercopithecoids and hylobatids (plus *Propliopithecidae*) (-40 Myr);
(3) Great apes (gorilla, chimpanzee, and orang utan) (plus *Dryopithecidae*) bauplan (-20 Myr);
(4) Earliest *Homo* bauplan (-2 Myr) including *Homo sapiens* within the *Homo* bauplan (-0.18 ± 0.03 Myr).

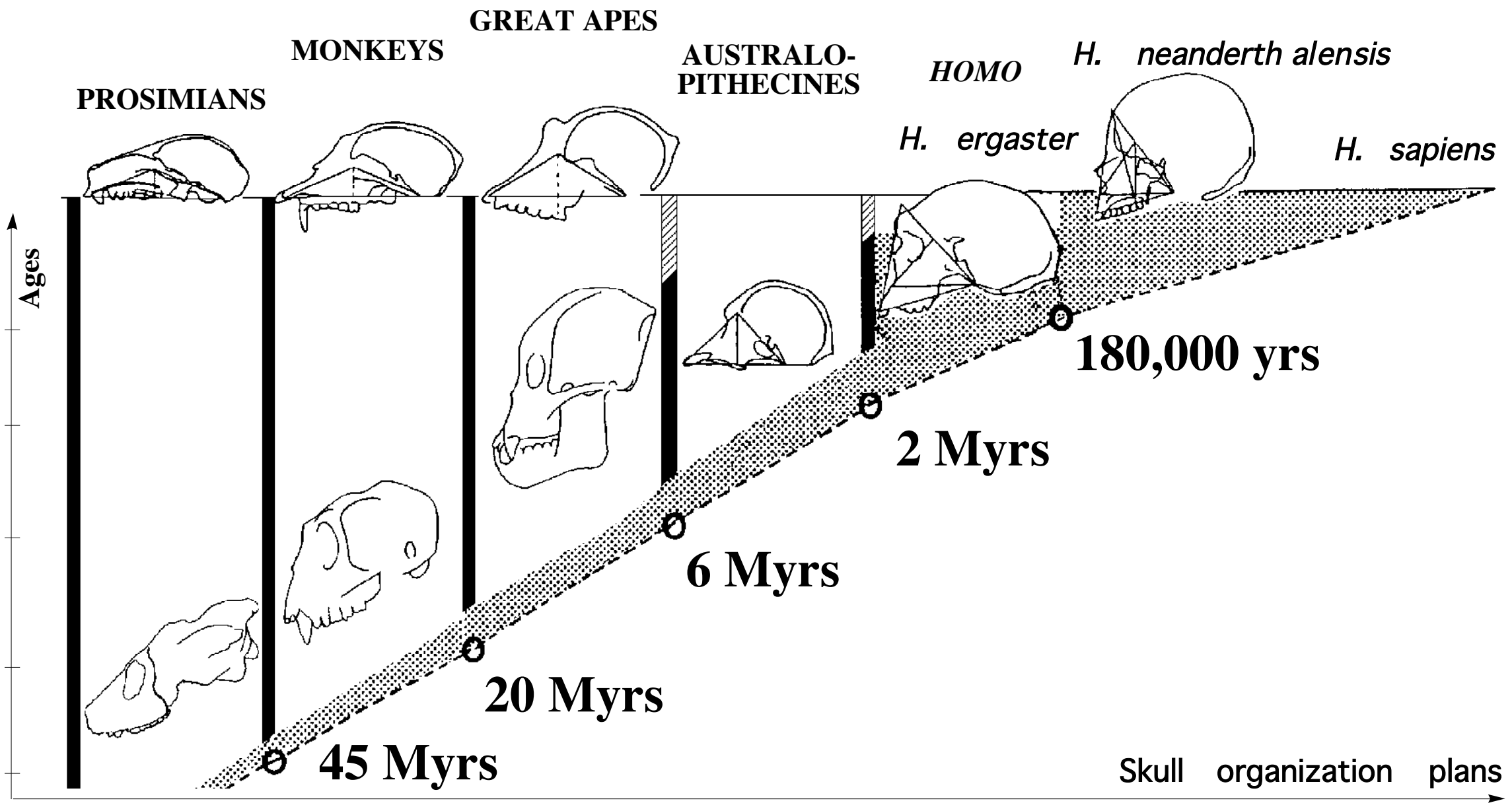

Figure. 5. Phylogeny of embryonic basicranial-facial contraction in Primates (after Dambricourt et al., 1999).

Results: Chaline et al., 1999b. $T_C = 2.1 \pm 1.0$ Myr ; $g = 1.76 \pm 0.01$; $t_{st} = 110$, $P<0.0001$ (N = 14 events, including the "global" tree) (Fig.6).

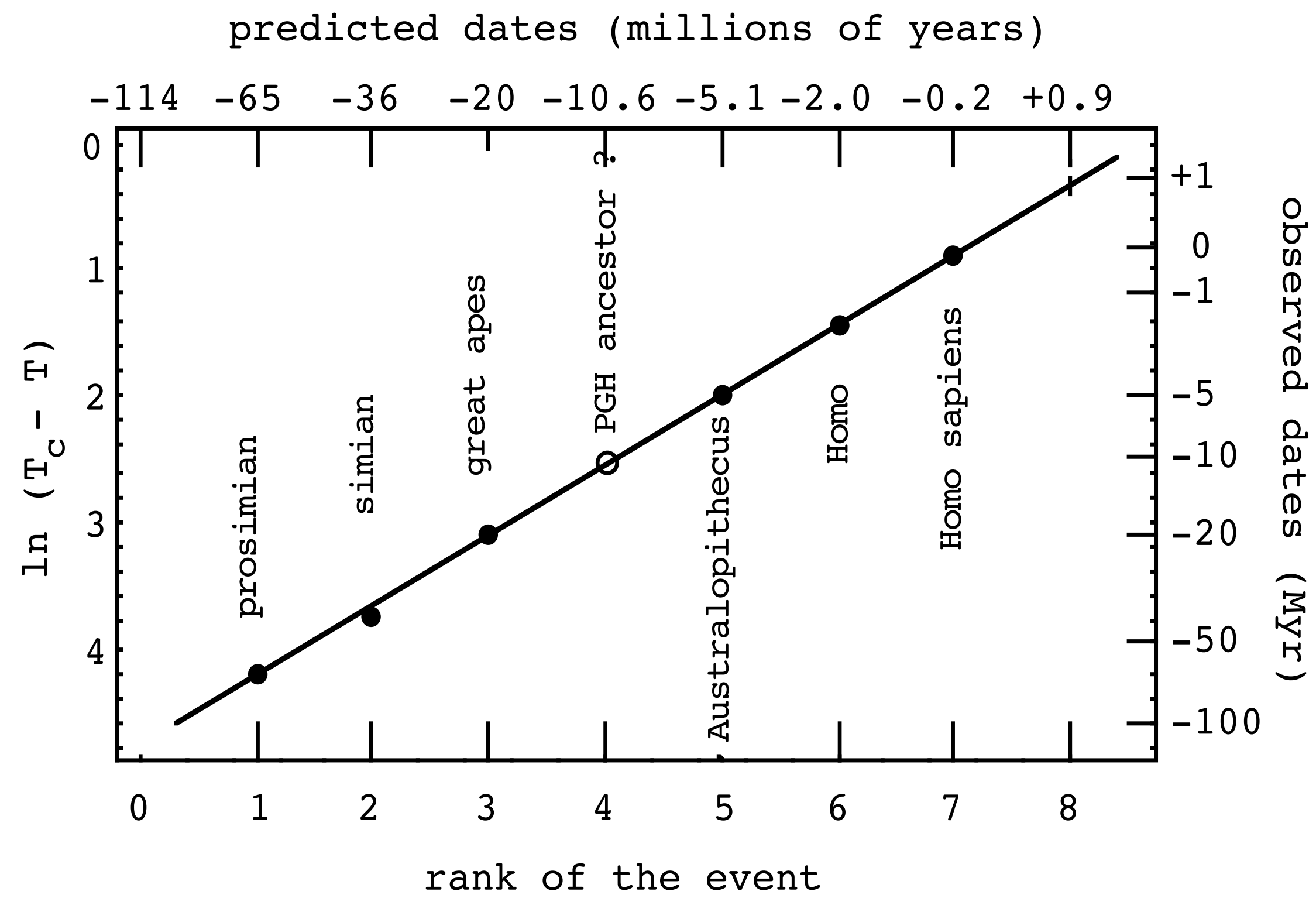

Figure 6. Comparison of dates of major events in primate evolution (dots) and a log-periodic law of ratio $g$ = 1.73 and critical time $T_C$ = + 2.3 Myr. The ring marks the date for the next evolutionary event for the lineage as predicted by the law (after Chaline et al., 1999).

Some colleagues have recently suggested that other events (actually minor ones) should also be taken into account for this lineage, leading to the following dates: -65, -53, -40, -35, -25, -20, -17, -10, -7, -5, -3.5, -2, -0.18. The statistical analysis gives: $T_C$ = 5.8 ± 4.0 Myr; $g$ = 1.23 ± 0.01; $t_{st}$ = 57, $P$<0.001 (N = 13 events). The result is still significant, and, moreover, the critical date agrees within error bars (to less than 1s) with our previous determination. This confirms that $T_C$ is characteristic of the lineage beyond the choice of the events. On the contrary the value of $g$, which depends on the density of dates, is not conserved, as expected. The fitting law which is applied to time intervals is a wo parameters function ($Tc$ and $g$) so that only three events are needed to determine theses parameters. Although these analyses concern only the chronology of events corresponding to the dates of the peaks of probability for an event to happen, independently of their nature, the log-periodic law permits a certain predictivity. The retroprediction of an important date, at -10 Myr, correspond to the estimated date expected from genetic distances and phylogenetic studies of appearance of the common *Homo-Pan-Gorilla* ancestor. This ancestor has not been yet discovered in the fossil record.

### 2.3.2. Fossil North American Equids

Data: Devillers in Chaline, 1990. *Hyracotherium* : -54 ± 5 Myr. ; *Mesohippus* : -38 ± 5 Myr. ; *Miohippus* : -31 ± 5 Myr. ; *Parahippus* : -24 ± 4 Myr. ; *Archeohippus* : -19 ± 3 Myr. ; *Hipparion* : -15 ± 3 Myr. ; *Protohipus* : -11 ± 2 Myr.; *Nannipus* : -9 ± 3 Myr). ; *Plesippus* : - 6 ± 2 Myr. ; *Equus :* -2 ± 1 Myr.

Results: Chaline et al., 1999b. $T_C$ = -1.0 ± 2.0 Myr ; $g$ = 1.32 ± 0.01; $t_{st}$ = 99, $P$<0.001 (N = 16 events, including the "global" tree). The *Tc* at the present time means that North American Equids have reached this time boundary. In fact Equids disappear effectively from North America 8.000 years ago, being reintroduced by the Spanish conquest (Fig.7).

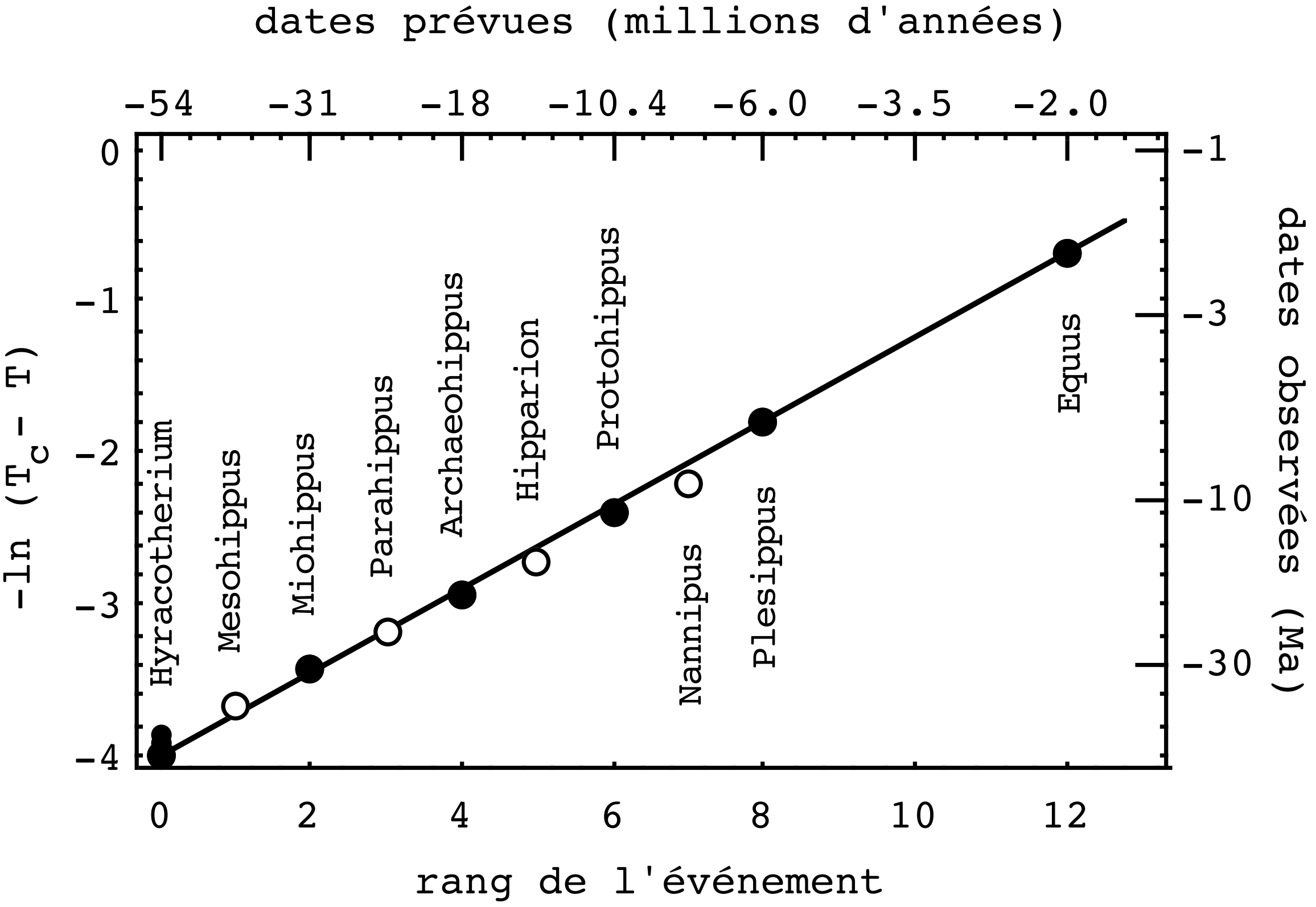

Figure 7. Appearance dates of North American equid genera compared with a log-periodic law with $T_C$ = 0 and $g$ = 1.32 (like colored dots are in ratio 1.73 to be compared with the equivalent diagrams for primates and general evolution) (after Chaline et al., 1999).

### 2.3.3. Rodents

Data: Hartenberger, 1998. In the case of rodents, half of all mammals, the analysis is different from the other lineages, since it is made on their whole arborescence.

Results: Chaline et al., 1999b. On Fig. 8 is plotted the histogram of the distribution of the 61 dates of appearance of rodent families. Well-defined peaks can be identified in this distribution. It is on these peaks that we perform our analysis. However, some uncertainties remains, in particular concerning the large peak after the date of first apparition of the lineage. Three different interpretations are considered. The mean value (-50 Myr) of the first peak has been used. This yields a critical date $T_c$ = 12 ± 6 Myr in the future. One can also singularize the latest date, yielding: (-56 Myr, -45 Myr, -34 Myr, -26 Myr, -18 Myr, -12 Myr, -7 Myr, -2 Myr). One obtains: $T_c$ = + 7 ± 3 Myr; $g$ = 1.32 ± 0.01; $t_{st}$ = 78, $P$<0.001 (N = 15 events, including ancestors in the "global" tree). But a closer scrutiny of the data suggests that the spurt of branching (that correspond to the sub-peaks inside the main first peak in Fig. 9) that followed the group's first appearance actually decelerates. This would be in agreement with the interpretation of these structures in temrs of critical phenomena. We find that the deceleration is issued from a critical point at $T_c$ = -62 Myr ± 5 Myrs, which agrees with the date estimated for the group's first appearance. Once this initial *deceleration* is allowed for, the following dates (-34 Myr, -26 Myr, -18 Myr, -12 Myr, -7 Myr, -2 Myr) exhibit highly significant *acceleration* toward $T_c$ = 27 ±10 Myr ($t_{st}$ = 98, $P < 10^{-4}$).

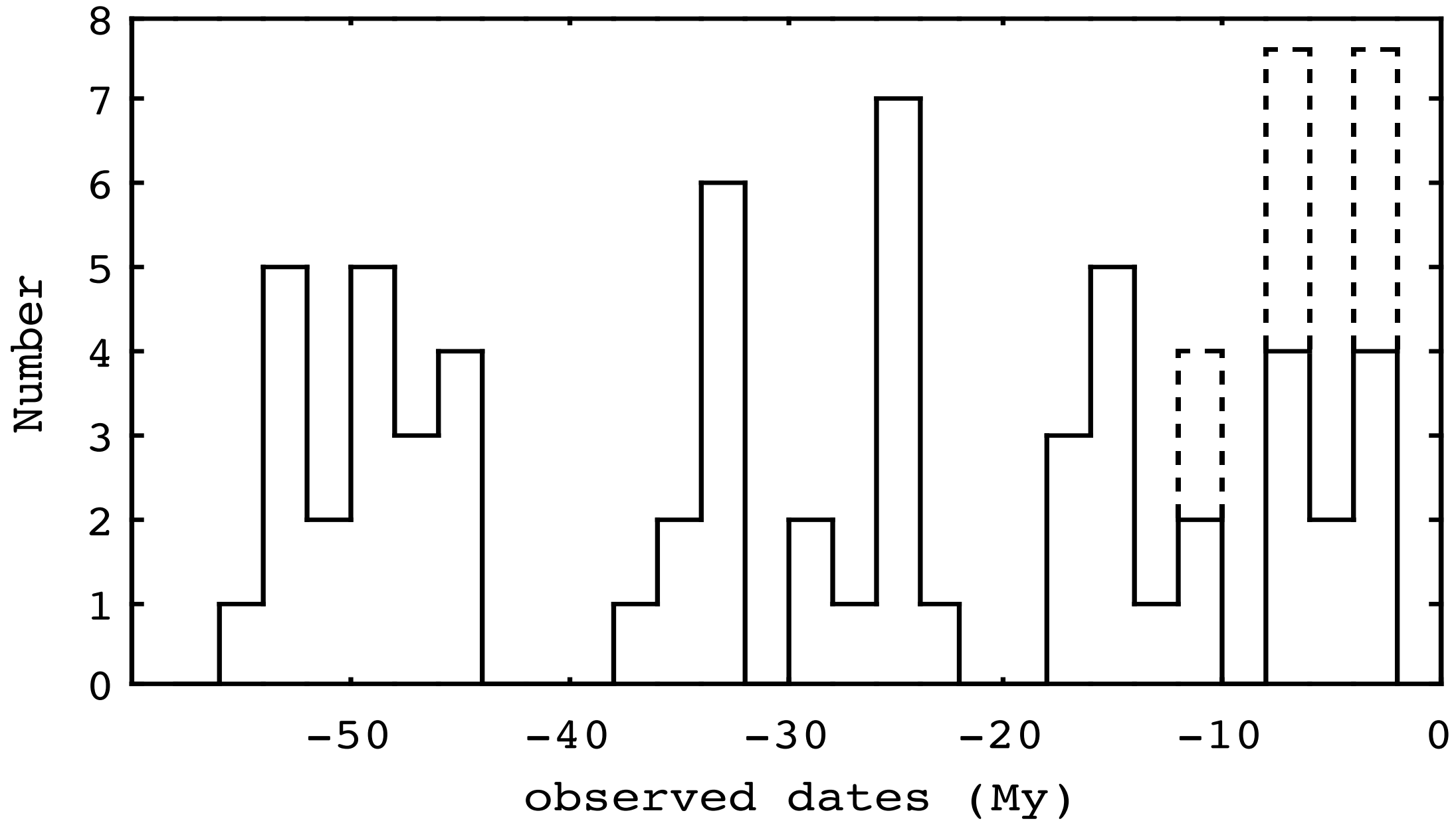

Fig. 8. Histogram of the distribution of the dates of appearance of families in the arborescence of the order of rodents. These data include only a subfraction of the events after -12 Myr, so that the amplitude of the last peaks is underestimated and has been extrapolated (dotted line) (after Nottale et al., 2000).

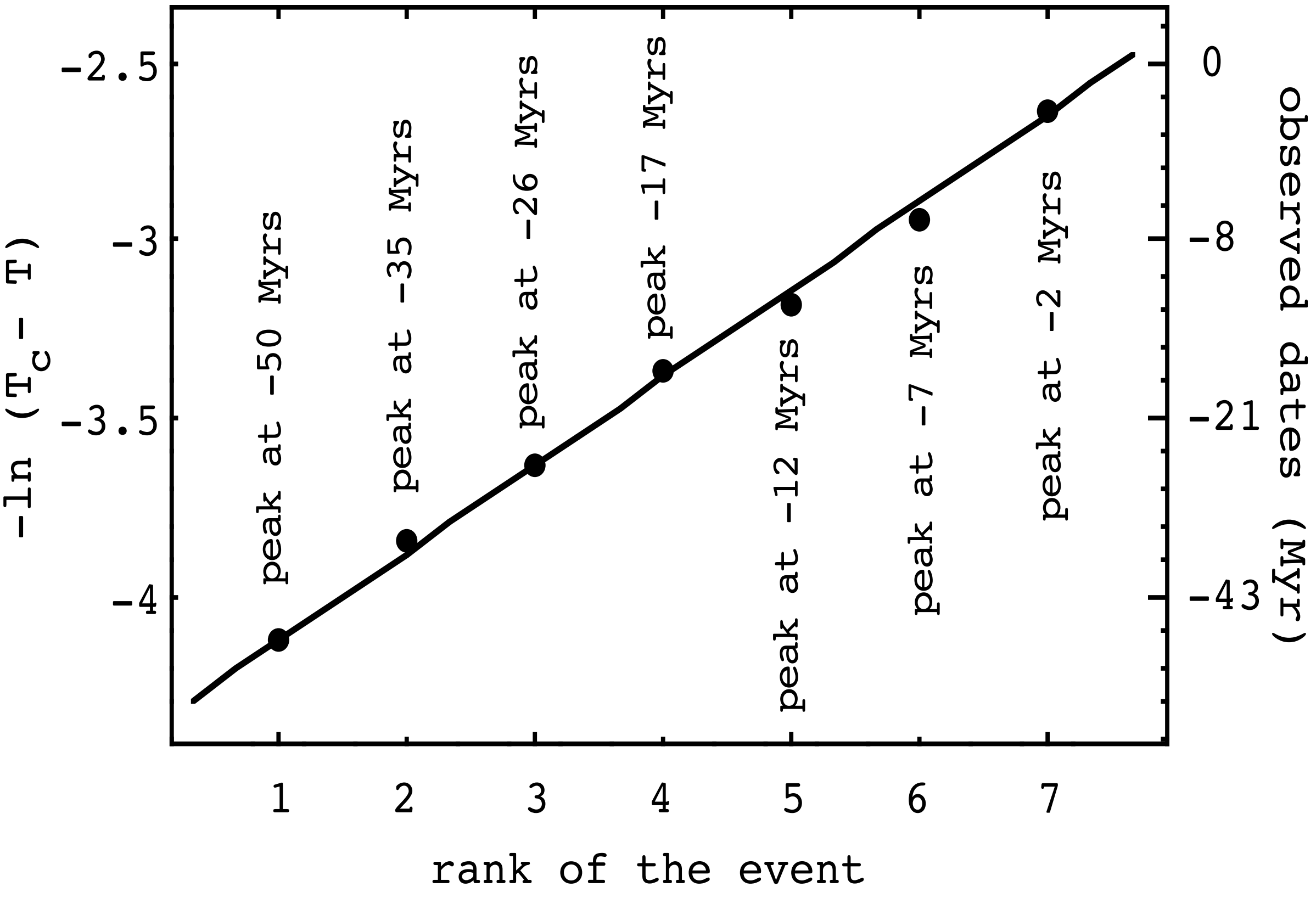

Figure 9. Fit of density peaks in distribution of rodent species appearances against a log-periodic law with $T_c$ = +12 Myr and $g$ = 1.32 (after Chaline et al., 1999)

## 2.3.4. Sauropod dinosaurs

Data: Wilson and Sereno (1998) have identified five well-defined major events in the evolution of their principal postural changes at the higher-level phylogeny: Prosauropoda/Sauropoda bifurcation: -230 Myr; Vulcanodon/Eusauropoda bifurcation: -204 Myr; Neosauropoda: -182 Myrs; Titanosauriforms: -167 Myr; Titanosauria: -156 Myr.

Results: Chaline et al., 1999b. These events exhibit a marked *log-periodic acceleration* toward: $T_c$ = -128 ± 10 Myr ; $g$ = 1.41 ± 0.01; $t_{st}$ = 122, $P$<0.001 (N = 10 events, including ancestors from the "global" tree) (Fig.10).

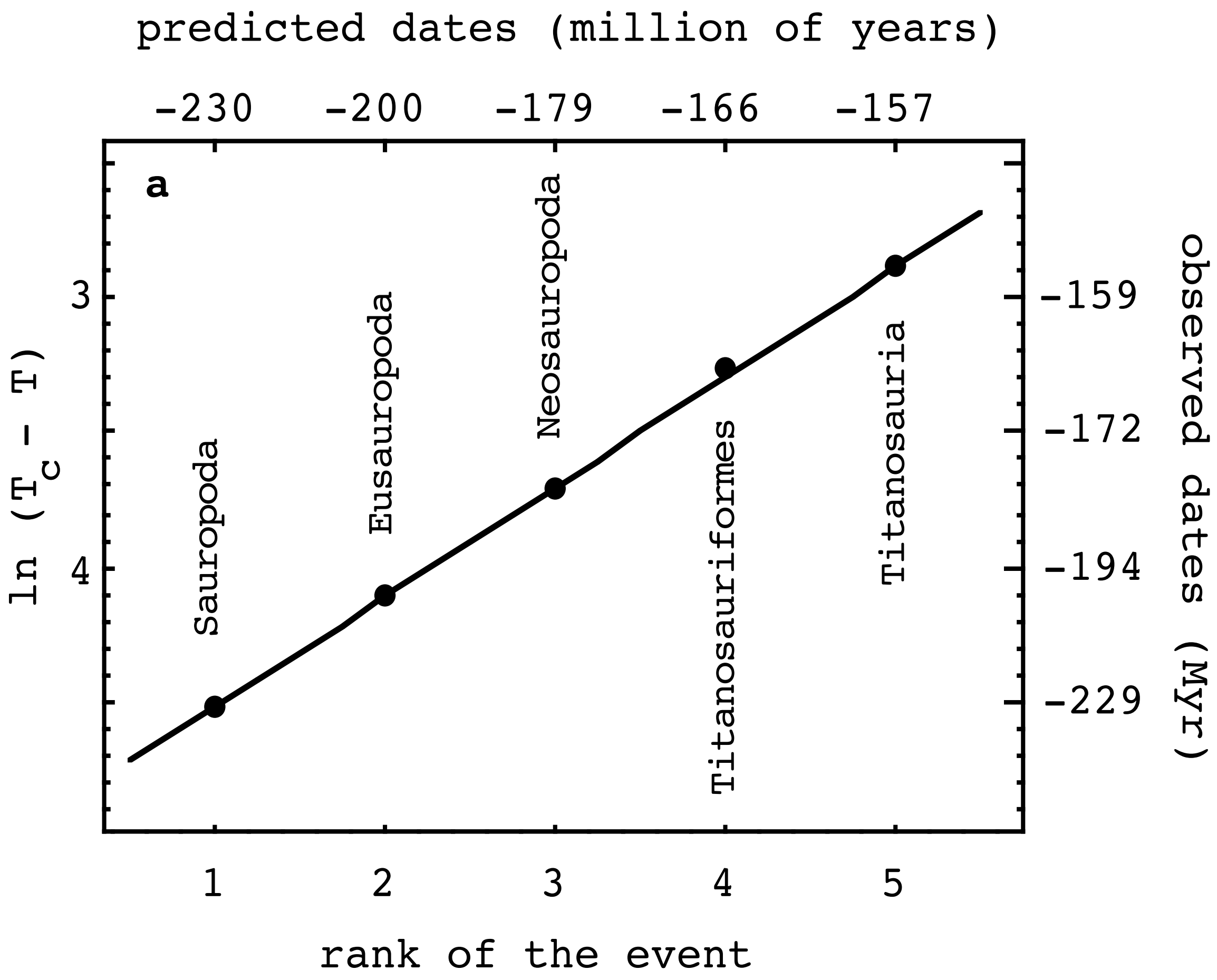

Figure 10. Fit of dates of main evolutionary leaps for sauropods against a log-periodic law with $T_C$ = -139 Myr and $g$ = 1.5 (after Chaline et al., 1999).

## 2.3.5. Theropod dinosaurs

Data: Sereno, 1999. One can identify from the data of Sereno the following main events in the evolution of Theropods: -227 Myr; -187 Myr; -160 Myr; -150 Myr; -145 Myr.

Results: Chaline et al., 1999b. There is a *significant acceleration* toward: $T_C$ = -139 ± 4 Myr; $g$ = 2.02 ± 0.02; $t_{st}$ = 69, $P$<0.001 (N = 10 events, including ancestors down to the origin of life).

This supports the existence of a log-periodic acceleration for the whole group of *Saurischia* (Sauropods and Theropods). On the contrary, an analysis of the other large dinosaur group, *Ornithischia*, has given no statistically significant structure. This could indicate, either that the log-periodicity is not universal in the tree of life and characterizes only some particular lineages, or that the data are uncompleted for this group.

### 2.3.6. Echinoderms

Data: David et Mooi, 1999. Seven major events that punctuate their evolution happen at the following dates, according to David and Mooi: (1) appearance of echinoderms (*Arkarua*) at the end of Precambrian in Ediacara fauna: -560-570 Myr (Vendian); (2) skeleton origin in *Stromatocystites* -526 Myr ; (3) theca and stem appearances in Burgess fauna : -520 Myr ; (4) Bifurcation between Asterozoairs/ Urchins/holothuries -490 Myr (Ordovician) ; (5) Echinids *sensu stricto* -430 Myr (Silurian) ; (6) Reduction of the test to 20 plates columns instead of 40 to 60: -355 Myr (Lower Carboniferous) ; (7) (Irregular Urchins: -180 Myr (Toarcian) ;

Results: The critical phenomena approach to evolutionary process leads to expect not only acceleration toward a crisis date, but also deceleration from it. The echinoderm group supports this view. Processing of this data shows that this group *decelerate* from a critical date $T_C$ = -575 ± 25 Myr (see Fig.11). The graphic is in agreement with the description in terms of critical phenomena with a critical point at the date of the group first appearance. This epoch identifies, within error bars, with the first appearance (datum around -570 Myrs. We find: $T_C$ = -575 ± 25 Myrs; $g$ = 1.67 ± 0.02; $t_{st}$ = 58, $P$<0.003 (N = 5 events).

The two innovations occurring in echinoderms at -526 Myr (skeleton appearance in *Stromatocystites*) and at -520 Myr (theca and stem appearances in Burgess fauna) are interesting. These events are distinct and fundamental from a cladistic point of view leading to two clades, two bifurcations (arrow). But from an evolutionary point of view, they show that two innovations could appear during the same probability peak. What means such a result from a genetical point of view, as the rate of mutations are globally constant? Does it means that the mutations of bauplans (*Hox genes*) appear more probably during these probability peaks following the event log-periodic distribution of the inferred group? Why? How? Following which kind of mechanism?

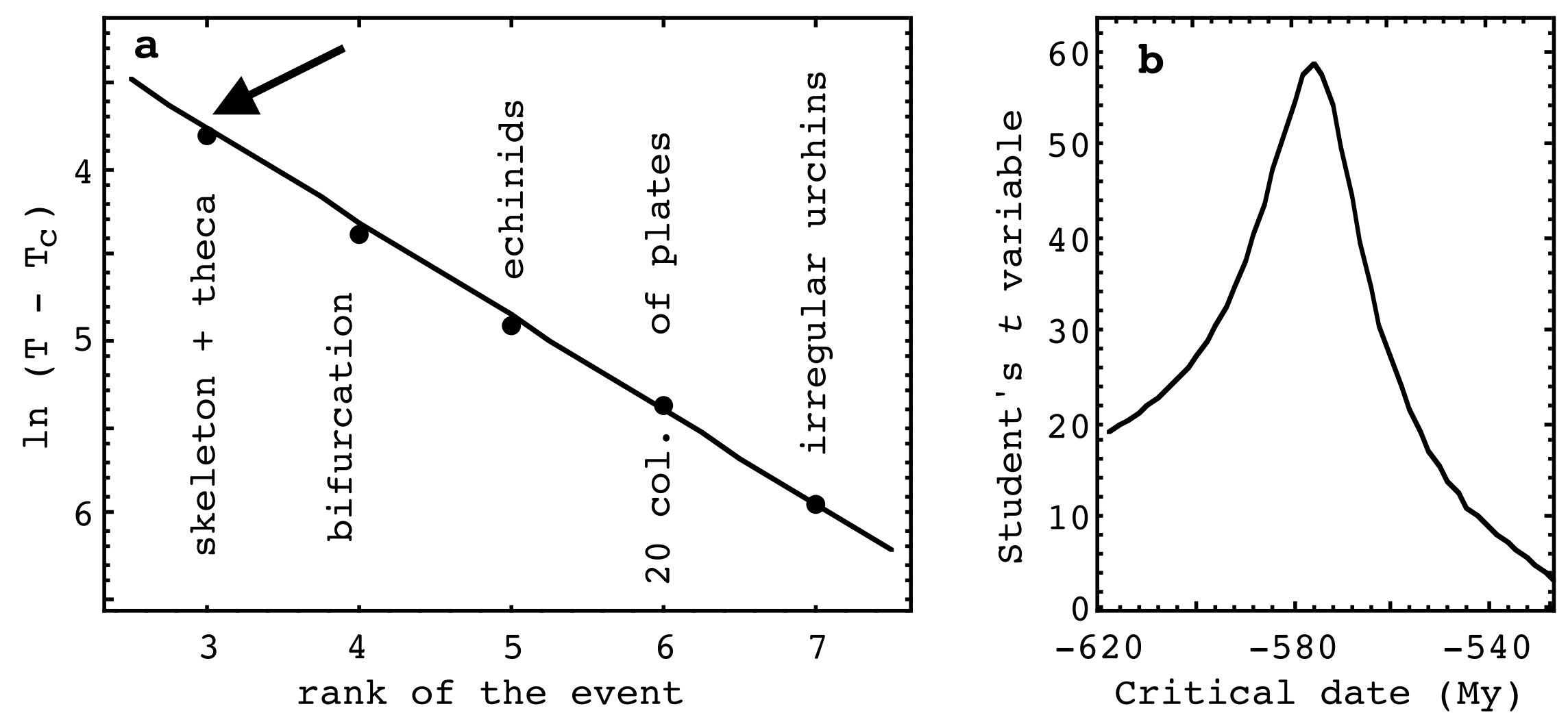

Figure 11. Comparison of the dates of the main events of the evolution of echinoderms with a log-periodic law of critical date $T_C$ = -575 Myr and scale ratio $g$ = 1.67 (figure a). Figure b shows the estimation of the critical date through the optimization of the Student's $t$ variable (P = 2 x $10^{-3}$) (after Nottale et al., 2000).

### 2.3.7. Global life tree

Data: Delsemme, 1994. From the origin of life to viviparity: Origin of life / first Prokaryotes : -3.500 ± 400 Myr; Eukaryotes : -1.750 ± 250 Myr. ; Multicellulars : -1.000 ± 100 Myr. ; Vertebrates : -570 ± 30 Myr. ; exoskeleton : -380 ± 30 Myr. ; homeothermy : -220 ± 20 Myr. ; viviparity : -120 ± 20 Myr.

Results: These events exhibit a significant acceleration: $T_C$ = -32 ± 60 Myr ; $g$ = 1.83 ± 0.03; $t_{st}$ = 36, $P$<0.003 (N = 7 events) (Fig. 12).

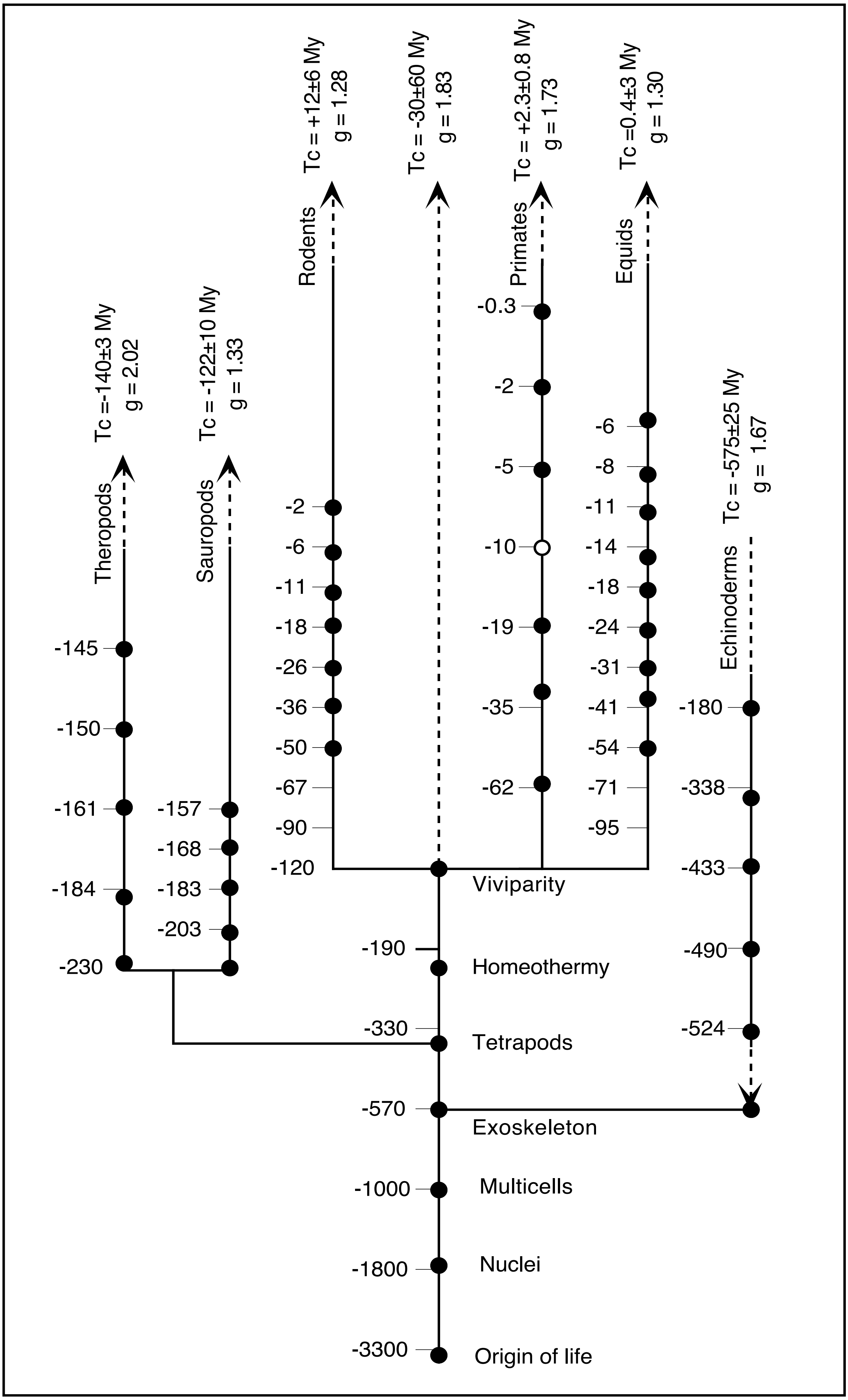

Figure 12. The dates of major evolutionary events of seven lineages (common evolution from life origin to viviparity, Theropod and Sauropod Dinosaurs, Rodents, Equids, Primates including Hominids and Echinoderms) are plotted as black points in terms of $log\ (T_C\text{-}T)$, and compared with the values from their corresponding log-periodic models (computed with their best-fit parameters). The adjusted critical time $T_C$ and scale ratio $g$ are indicated for each lineage (excluding here ancestors of the "global" tree). The result is still significant, and, moreover, the critical date agrees within error bars (to less than 1 s) with our previous determination. This confirms that $T_C$ is characteristic of the lineage beyond the choice of the events. On the contrary the value of $g$ which depends on the density of dates is not conserved as expected (after Nottale et al., 2002).

### 2.3.7. Human ontogeny

As a significant log-periodic acceleration was found at different scales for global life evolution and considering the relationships between phylogeny and ontogeny, it appeared very interesting to verify whether such a law could also be applied to the scale of ontogeny.

Data: Cash, O'Rahily, 1972 and Larsen, 1996. The data corresponding to the pre-birth period (266 days or 38 weeks) is accepted as a general consensus among embryologists. For the post-birth development we integrated the main stages of child psychomotricity changes even though it is more difficult to define a precise age for the major leaps because of the great inter-individual variability (Encha-Razavi and Escudier, 2001 ; Moore, and Persaud, 1998 ; Bourrillon, 2000 ; Arthuis et al., 1998).

Results: Cash et al., 2002 demonstrate that the log-periodic law describing critical phenomena may be also applied to ontogenetical stages. We observe a significant deceleration starting from fecundation day ($Tc$) and this day may be retropredict at 2 or 7 hours, using posteriors events (Fig.13).

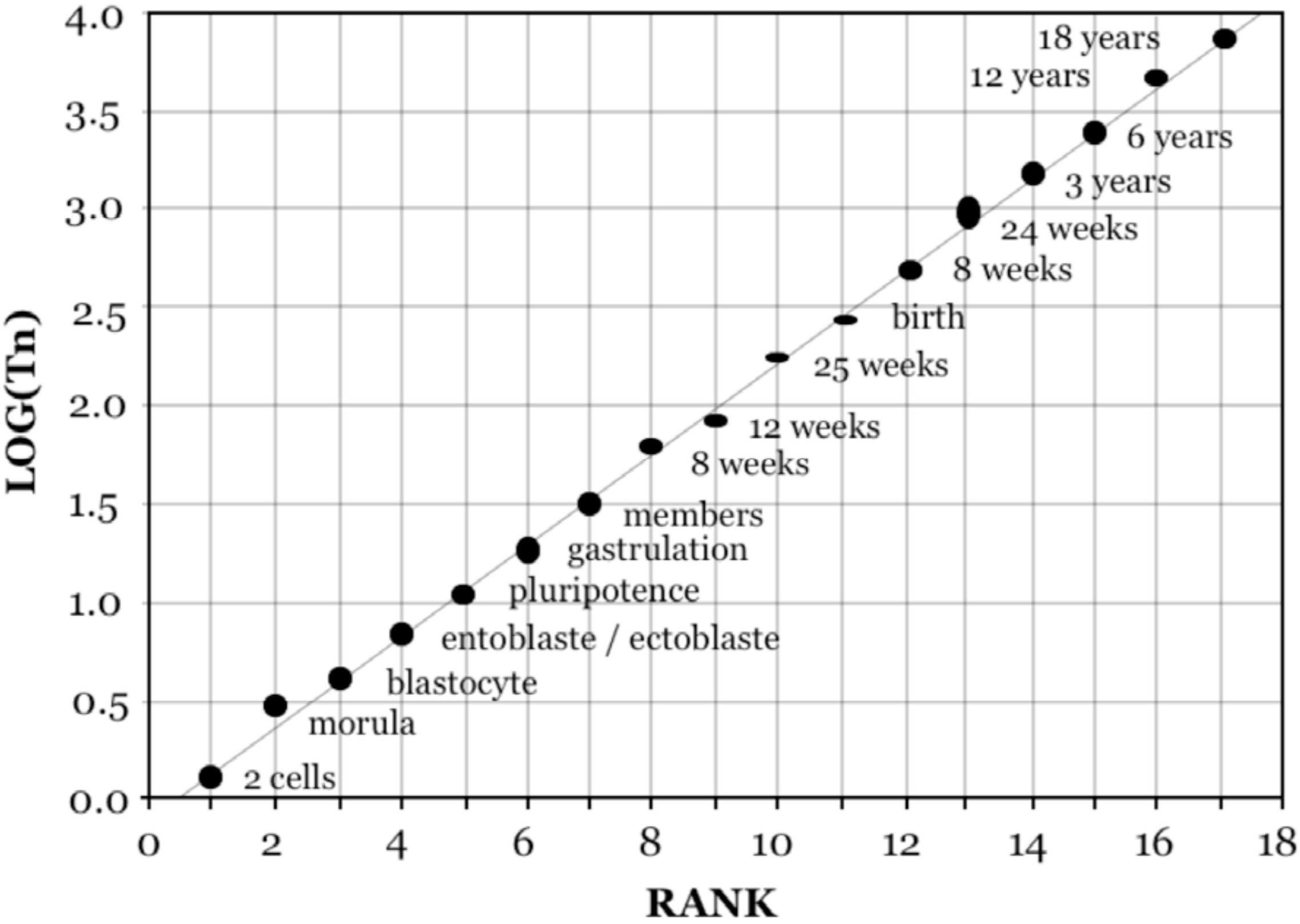

Figure 13. Decimal logarithm log $T_n$ of human ontogeny data related to their rank *n, showing a deceleration with a scale ratio* g = 1.71 ± 0.01. The vertical size of the point gives the confidence interval for the dates (after Cash et al., 2002).

### 2.3.8. Discussion

In each case we find that a log-periodic law provides a satisfactory fit for the distribution of dates, with different values of the critical date $T_C$ and of the scale ratio *g* for different lineages. The obtained behavior may be depending on the lineage and on the time scale. The results are found to be statistically significant. We give in what follows the adopted dates (in Myr before present) for the major jumps of the studied lineages. The error bars are typically $dT/T \approx 10\%$ or less, i.e., $dlog(T_C\text{-}T) \approx 0.04$. Since we are interested here in pure chronology, if several events occur at the same date (within uncertainties), they are counted as one. Then we give the result of the least-square fit of the log-periodic model and the associated Student variable with its corresponding probability to be obtained by chance. For each lineage we include in the analysis the common ancestors down to the origin of life, except for Echinoderms and Human ontogeny which show deceleration instead of acceleration. The obtained parameter values are compatible with those given in Fig.12, which result from a fit that does not include the ancestors of the lineage.

In all these cases chronology has been quantified and the discontinuous appearance of clades follows a log-periodic law with a middle scale report of 1.73. In each case we find that these data are consistent with a log-periodic law of acceleration (for most lineages) or of deceleration (for echinoderms and human ontogeny) to a high level of statistical significance. Such a law is characterized by a critical epoch of convergence *Tc* specific to the lineage under consideration and that can be interpreted, in the acceleration case, as *the end of that lineage's capacity to evolve ?* Thus, as a consequence, that means *there may be a limited time-life for each specific group, depending on Tc and contingency events (climatic, biologic or cosmic phenomena) leading to extinction.*

It is interesting to note that for echinoderms, the deceleration law permits the retropredictivity of echinoderms appearance at -575 ± 25 Myr, this epoch identifing, within error bars, with the first appearance datum around -570 Myr. For human ontogeny, the retropredictivity of fecundation time is of 2 hours !

Let us end this contribution by discussing possible biases and uncertainties in our analysis. There is a "perspective" bias, linked to the fact that observational data are fossil records only observed at the present epoch. This bias can manifest itself in two ways. First, the uncertainty on the dates increases with the date itself, so that we expect that dT/T be about constant, which could lead to alog-periodic behavior. We have discussed this bias and we have shown that it cannot account for the observed structure. The additional information given here and in that one observes also decelerations reinforce this conclusion. A second possible form of this bias [Sornette, private communication] could be an increasing number of missing events in fossil records for increasing dates in the past. Against such an interpretation, one can recall that the quality of the fossil records, concerning in particular their completeness, has been recently reaffirmed by Kidwell and Flessa (1996). Moreover, the number of missing links needed to compensate for the acceleration seems to be unreasonably large (the interval between major events goes from billion years at the beginning of life to million years now). In addition, the bias about the choice of dates, in particular in defining, which characters are considered to be major ones, has been analyzed here. The solution to this problem lies in the observation that, if the acceleration (or deceleration) is real and intrinsic to the lineage under study, its occurrence and the date of convergence *Tc* ought not to be dependent (within errors) on the limit applied as to the choice of which events count as important ones. However, there is nothing intrinsic about the scale factor *g* between intervals, as it decreases as the number of events allowed for increases. We have been able to test this stability of the critical date with the data for which we considered several possible choices (rodents, sauropods) as well as with choices suggested by other workers (primates). We conclude that this uncertainty cannot explain the observed law, which therefore seems to be a genuine one. However, while log-periodic accelerations or decelerations have been detected in the majority of lineages so far investigated, the question of whether this behavior is systematic or not remains opened (cf. the general tree for dinosaurs published by Sereno, 1999). Analysis of the values of the critical date for the various lineages leads us to interpret it, in the case of an acceleration, as a limit of the evolutionary capacity of the corresponding group. When a deceleration has been detected, it starts from the apparition date of the lineage. One possible interpretation is that it concerns memory phenomena, each level of organization being built on previous levels with no possibility of backward leaps, meaning the irreversibility of phenomena. *« As ontogeny creates phylogeny »* (Garstang, 1922), the two phenomena obey to the same coercions such as cell differentiation, tissue and organ building. This example is a good illustration of the expression of such *a memory phenomenon at each scale of organization of living-being, from microevolution to macroevolution, demonstrating the unity of life evolution.*

Let us finally stress the fact that the existence of such a law does not mean that the role of chance in evolution is reduced, but instead that randomness and contingency may occur within a framework which may itself be structured (in a partly statistical way).

The tests were extended to economic crisis /no-crisis patterns in Western Europe economic system and in pre-Columbian civilizations (Nottale et al., 2000 ; Grou et al., 2004), world economy dynamics (Johansen et Sornette, 2001), as well as earthquakes (Sornette and Sammis, 1995), stock market crashes (Sornette, 2003), jazz history, monastic orders appearances and even chemical elements discoveries (Brissaud, in Nottale et al., 2009*). These results reinforce the universality of the log-periodic model in the description of evolutionary branching phenomena during temporal crisis as much in the inorganic world as in the organic one.*

How to explain these fractal structures at many level of a hierarchical organization and patterns of living beings, the log-periodic distribution of clades, critical time for living being groups announcing and explaining why groups have a beginning and an end ? Answers are to be found in the scale relativity theory (SRT) as shown by Nottale and in some other representations. In this context, let us quote the work by Queiros-Conde which describes the « tree of life » as a sort of temporal turbulent flow. A geometrical description (« entropic-skins theory ») introduced in the field of turbulence (Queiros-Conde, 1999) to describe the phenomenon of intermittency can help to visualize the meaning of $g$ and $T_c$. The tree of life is assumed to be a multi-scale tree evolving in a three-dimensional embedding space but the scale is a time interval *Dt*. To be consistent with the log-periodical law, it is shown that the tree can not be simply fractal but should be the dynamical result of the superposition of an infinite number of fractal sets (« skins ») having fractal dimension noted $D_p$ varying from the dimension $d$ (d = 3 if the phenomenon is occuring in a three-dimensional space as it is assumed for « tree of life ») of embedding space to a minimum dimension $D_\infty$ which characterizes the most intermittent events i.e. the ones which display the highest tendency to clustering. For each skin (which corresponds to an event occuring at age $T_p$), the scale-entropy is defined by $S_p(Dt)=\ln(Dt/Dt_0)^{Dp}$ where $Dt_0$ is the largest time-interval occuring in the system. The author then introduces a scale-entropy flux $f_p(Dt)=[S_p(Dt)-S_{p+1}(Dt)]/(T_p-T_{p+1})$. Entropic-skins approach assumes that the scale-entropy flux $f_p(Dt)$ does not depend on the skin (i.e. on the level in the hierarchical structure) and remains constant through the tree.

It leads to $(D_p-D_{p+1})/(D_pD_p)=(T_p-T_{p+1})/(T_{p-1}-T_p)=g$ with g=1/g, an equality which links fractal dimensions $D_p$ and dates $T_p$. It is then easily derived $g = (D_f-D_\infty)/(d-D_\infty)$. By defining a threshold indicating the existence of a clustered structure compared with an homogeneous system presenting no clustered structure (i.e. $D_f = d = 3$), the « active part » of the tree of life which can be characterized by a fractal dimension $D_1 = D_f$ ($D_f<d$) is introduced. An analytical determination of *g* is possible through entropic-skins geometry described in Queiros-Conde (2000). In this derivation, it is assumed that $D_f = 2$ which means that life is a space-filling process : biosphere is in fact mainly a two-dimensional sheet. The crest dimension $D_\infty$, assuming a phenomenological argument of space distribution, is the dimension of the Cantor set i.e. $D_\infty = \ln 2/\ln 3$. It then leads to $g = \ln(3^2/2)/\ln(3^3/3)\approx 0.58$ which means $g\approx 1.724$ very close to the measured mean value of g (g = 1.73). The geometrical model of « entropic-skins » associated to a turbulent structure for the tree of life certainly displays interesting features which would deserve to be developed.

## 2.7. Conclusions

The scale relativity theory can open new perspectives in evolution. Can we expect a *scale relativity theory for living-beings evolution?* The applications of scale-relativity theory to living-beings appear as a promising new way of research providing some suggestions or explanations. Among them, SRT explains fractal structures (corals, plants, trees etc…) as increasing the surface possibilities from DNA to biosphere, certainly maximized by natural selection. SRT explains also the log-periodicity of the life-tree with critical times and as a consequence, a limited time-life for each specific group following natural selection and contingency after the occurrence of *Tc*. The existence of such a law does not mean that the role of chance in evolution is reduced, but instead that randomness and contingency may occur within a *« framework »* which may itself be structured in a partly statistical way. Many questions are opened with this new approach. Does SR determines preferentially more or less DNA mutations, mutation rates within log-periodic probability peaks or not ?, developmental biology (*Hox* genes control: EVO-DEVO) and epigenesis which are not directly linked to these processes and play a major role in structuration? Species evolution (EVO-DEVO) may be only *a component* of a more general *theory of scale relativity of the universe* (*SR-EVO-DEVO*). It is an interesting and very promising program to be discussed for biology and paleontology.

## ACKNOWLEDGEMENTS

This paper is dedicated to Jacques Dubois, a precursor in nature fractal structuration and made all the calculations of the rodent analysis. I am grateful to Laurent Nottale for collaboration and to Diogo Queiros-Conde for its description of the « tree of life » as a sort of temporal turbulent flow, Richard Gordon and Bertrand Martin-Garin for their suggestions and corrections.

**Biography** : Jean Chaline is a paleontologist, born on 18/02/1937 in Châlons-sur-Marne (France). Emeritus research Director in CNRS (University of Burgundy, Dijon) and honorary Professor at the Ecole Pratique des Hautes Etudes (Paris), he was the Director of the *« Biogeosciences CNRS Laboratory »* and *« Paleobiodiversity and Prehistory laboratory »* (EPHE). He was working on rodents and human evolution through ontogenetical processes. Since 15 years, he works with astrophysician Nottale and sociologist Grou on the applications of the *Scale Relativity Theory* to living-beings. 245 papers, 34 books, more than 100 invited conferences.

Nottale L., Chaline J. et Grou P.: 2000, Les arbres de l'évolution. Hachette, pp. 1--380.
Dubois J. et Chaline J.: Le monde des fractales. La vraie géométrie de la nature. Ellipses, Paris, pp. 1--312.
Chaline J.: 2006, What's up since Darwin ? The evolutionary theory in all its stages. Ellipses, pp. 1--480.
Nottale L., Chaline J. et Grou P.: 2009, Des fleurs pour Schrödinger. La relativité d'échelle et ses applications. Ellipses, pp. 1--421.

# Development (and Evolution) of the Universe

Stanley N. Salthe

Biological Sciences, Binghamton University
Emeritus, Biology, City University of New York

42 Laurel Bank Avenue, Deposit, New York, 13754, USA
607-467-2623
ssalthe@binghamton.edu

ABSTRACT
I distinguish Nature from the World. I also distinguish development from evolution. Development is progressive change and can be modeled as part of Nature, using a specification hierarchy. I have proposed a 'canonical developmental trajectory' of dissipative structures with the stages defined thermodynamically and informationally. I consider some thermodynamic aspects of the Big Bang, leading to a proposal for reviving final cause. This model imposes a 'hylozooic' kind of interpretation upon Nature, as all emergent features at higher levels would have been vaguely and episodically present primitively in the lower integrative levels, and were stabilized materially with the developmental emergence of new levels. The specification hierarchy's form is that of a tree, with its trunk in its lowest level, and so this hierarchy is appropriate for modeling an expanding system like the Universe. It is consistent with this model of differentiation during Big Bang development to view emerging branch tips as having been entrained by multiple finalities because of the top-down integration of the various levels of organization by the higher levels.

## INTRODUCTION

In this paper I attempt a summary of a developmental perspective I have been constructing for two decades as a kind of 'treaty' among several disciplines. Beginning with some conceptual clarifications, I distinguish 'Nature', our scientific construct, from the World. Nature is our map or operating manual for the World. Nature embodies logic as its basic framework, and its embodiments are typically found in, e.g., inscriptions, diagrams, tables, models, equations, laws, universal constants and classifications. Nature is the subject of studies in the philosophy of nature (natural philosophy), as in this paper. It is mediated by languages, while the World is mediated to us through (what in Nature we know as) our biology (Uexküll, 1926). The World is experienced and phenomenal, but is not 'known' (Snowdon, 2008). My favorite example of this is that, while we may ride a bicycle, we do not 'know' (cannot describe) how we do it. I will as well propose a logically founded model in set theory format of 'general development', which is posited to be universal for dissipative structures as they are conceived in Nature. Of course, set theory cannot model dynamical change. I use this format to parse sequences of developmental stages, following the usage (as Normentafeln) in biology. I will also, in this way of describing stages of development, outline a 'canonical developmental trajectory' characteristic of dissipative structures. I acknowledge evolution in passing. As a cue to the reader, I note that my perspective is that of natural philosophy as developed out of beginnings made by the early Friedrich Wilhelm Joseph Schelling (e.g., Esposito, 1977; Salthe, 1993). What is relevant here is that Schelling first proposed a developmental view of the natural world.

## MODES OF CHANGE

Generalized from their long usage in biology, I distinguish development from evolution, as general modes of change (Salthe, 1993). Development is progressive change, while evolution is expressed in the effects of accumulating marks acquired from contingent encounters. Developments, interpreted as constitutive of the kinds of systems showing them, can be modeled as parts of Nature. Well known examples that I would place as developments are the 'main sequence' of stars, embryonic development, and ecological succession. Developments are like Goodman's (1976) 'scripts', but inferred from scientific investigations rather than by creative action. Evolution (more generally, individuation) occurs continually during the development of any material system, building in historically acquired information, leading to increasing dimensions of uniqueness in particular phenomena. One very important kind of particular phenomenon in science is the biological species. A species' storage of historically acquired information is held in the genomes of the cells of its parts, as well as in material configurations in cell structures. At its own scale each species is unique; while at their scales, its parts (e.g., organisms) differentiate increasingly as they recover from perturbations during development, becoming ever more intensively unique.

With regard to evolution, given the need perceived in Nature for systemic mutual fittingness, some phenomena will inevitably not be as well suited to persist in their surroundings as others, and these get recycled more rapidly than those more 'fitting' or 'better adapted'. This fact of differential persistence has been elaborately constructed as competition in our Western cultural ideology, and is highly developed in neoDarwinian evolutionary biology, where it plays out as competition between genotypes for representation in future generations of a population (e.g., Ewens, 2004). But this is a small aspect of the 'big picture' of change in Nature. A simplest kind of example of selection would be the self-organizing choice of a drainage channel from among several possibilities by a high altitude water reservoir as it develops from glacier to lake (Salthe and Fuhrman, 2005). All actual phenomena have been individuated by evolution. This paper, however, is concerned with development.

DEVELOPMENT

Development proceeds from relatively vague beginnings toward ever more specified particulars. It is logically a process of refinement via the differential promotion of potential informational constraints, followed by the subsequent fixation of some of them. Informational constraints are sites or regions that might assume more than one configuration. After choosing among possibilities, they express information. As development continues, an early informational constraint can become the site for the emergence of others. Stages of development can be conveniently represented using set theoretic format, as:

**{ stage 1 { stage 2 { stage 3 }}}**

Stage 2 develops out of Stage 1, and Stage 3 from Stage 2. Stage 2 is necessarily immanent, along with other possibilities, in Stage 1. More developed stages are in this model logically refinements of earlier stages, by way of having acquired further information. That is, an early stage anlagen will be only roughly -- vaguely -- adumbrated. It will be preserved in development, but increasingly altered downstream in a way that would require more information to describe.

Guidance by this hierarchical format (see details below) imposes a certain logical structure upon development which helps to guide our thinking about it. An important example of this guidance is the stricture that nothing totally new appears during development; instead some of the vague tendencies in a given stage become ever more definitely embodied in further stages, emerging ever more definitively. This perspective runs counter to the currently favored perspective that genuinely new things do appear in the world. Novelty in the present view can be initiated by way of perturbations from outside, or by internal excursions, during a developing system's individuation. These may become integrated into an emerging system during its development and homeostasis. For example, in the development of an organism, congenital perturbations of an embryo can result in abnormalities that become smoothly integrated into the living individual (Reid, 2007). These individuating modifications of a developmental trajectory are almost always of minor import compared with typical developmental patterns. When they are more important, the developing system is likely to fail. Historical impacts always do frequently perturb a surviving developing system to some, relatively minor, degree.

The imposition of this model upon development, including its broader application to the development of the universe, is a major departure of this paper. The logic is developed thus: the quintessential example of development can be found in embryonic development. What I have said above can be derived from observing this particular case. The extension of this model to dissipative structures in general is effected by looking for more generally applicable descriptors, like thermodynamic and information theoretic ones. There are few developmental tendencies that can be said to be universal. As an example, I have proposed (Salthe, 1993) a 'canonical developmental trajectory' of dissipative structures running from immaturity through maturity (in more well integrated systems) to senescence, with these stages defined thermodynamically and Informationally, as shown in Table 1. Extension further to the universe as a whole is of course speculative, but is encouraged by the developmental sequences of, for example, stars (e.g., Chaisson, 2001),

IMMATURE STAGE
- Relatively high energy density (per unit mass) flow rate
- Relatively small size and/or gross mattergy throughput
- Rate of acquisition of informational constraints relatively high, along with high growth rate
- Internal stability relatively low (it is changing fast), but dynamical stability (persistence) is high
- Homeorhetic stability to same-scale perturbations relatively high

MATURE STAGE (only in relatively very stable systems)
- Declining energy density flow rate is still sufficient for recovery from perturbations
- Size and gross throughput is typical for the kind of system
- Form is definitive for the kind of system
- Internal stability adequate for system persistence
- Homeostatic stability to same-scale perturbations adequate for recovery

SENESCENT STAGE
- Energy density flow rate gradually dropping below functional requirements
- Gross mattergy throughput high but its increase is decelerating
- Form increasingly accumulates deforming marks as a result of encounters, as part of individuation
- Internal stability of system becoming high to the point of inflexibility
- Homeostatic stability to same-scale perturbations declining

TABLE 1: Thermodynamic and informational criteria of the developmental stages of dissipative structures. See Salthe (1989, 1993) for more details and citations.

A developing system builds in informational constraints, and fixes many of them in the manifest information of definite form, upon which further constraints may emerge (Brooks and Wiley, 1988). Abiotic developments (as in tornadoes) never get to be very highly specified; internally stored informational constraints, such as are held in the genetic apparatus of living systems, is required for that. I would suggest that, if we propose a system that does not develop from immaturity to senescence, that this would not be a bonafide material system, but possibly part of one instead. Thus, for example, the biological population might better be viewed as part of an ecosystem (Damuth, 1985), where development has been successfully proposed as the process of ecological succession (Holling, e.g., 1986; Ulanowicz, e.g., 1997). The same may be said of the biological species (Simpson, 1953), except that the larger scale material dissipative structure that it would be part of has not, I think, yet been identified, since a species' areographic extent, or the migrations of its parts, could place it as a component of more than one biome.

## DEVELOPMENT OF THE UNIVERSE

I make a materialist interpretation of the Big Bang theory (e.g., Chaisson, 2001; Turner, 2007; Lineweaver and Egan, 2008). Accelerated Universal expansion of space leads to the precipitation of matter, which initiates gravitation and the clumping of masses. Based on Einstein's physical intuition that led to the 'equivalence principle' (Einstein, 1907), gravitation can be postulated to be a kind of obverse of accelerated expansion, leaving matter 'behind' as space-time expands. Acceleration is required here in order to understand why matter was unable to stay in equilibrium with expanding space (Nicolis, 1986). Assuming that the universe is a thermodynamically isolated system, my own further interpretation follows. The process of expansion continues to produce an increasingly disequilibrated universe, wherein develop, from the masses in some locales, forms, and from the forms in yet fewer places, organizations, moving the Universe ever further from thermodynamic equilibrium. I conjecture that continued development of increasingly complicated systems depends upon the continuation of Universal expansion. In an ‘equal and opposite’ reaction to these disequilibrating processes, the Second Law of thermodynamics emerges globally, scaled to the rate of Universal expansion. At present this imposes locally the necessity for significant entropy production in connection with any energy dissipation (Prigogine, 1955). Consequently, when dissipation is harnessed to effective work, that work is rarely better than around 50% energy efficient (Odum, 1983; Turner, 2000). Urgency, striving and haste make work even less energy efficient (Carnot, 1824). These facts can lead to a resuscitation of final cause in physics since entropy production is the way that non-equilibrium locales can promote Universal thermodynamic equilibration.

Causality

I have proposed reviving the Aristotelian causal analysis (e.g., Matthen, 1989) as being helpful for dealing with complex systems (Salthe, e.g., 1985, 1993, 2005). Complexity involves a number of factors, including local interactions and transactions of numerous differently capable elements of varying kind and scale as components of a system (e.g., Salthe, 2006,b), as well as historicity and vagueness. The Aristotelian causal categories help somewhat to tease apart some of this complexity by allocating different modes of causation. In my analysis (2006,a), material cause is susceptibility; formal cause is the ‘set-up’. Susceptibility refers to the readiness of a locale or a material to undergo the kind of change being modeled, while the set-up refers to the organization of initial and boundary conditions impinging upon that locale or material. Together these synchronic categories establish the locale of relevant or investigated events, as well as the propensities (Popper, 1990) of occurrences of various events. Efficient cause is a forcing or push, getting the change going; final cause is the pull of ‘why’ anything happens. A lightning strike makes a convenient example, with a local buildup of electrical energy gradient as charge separation between clouds or between clouds and the ground, based on the formal cause of large scale local dynamical and structural configurations, with the push given by some perturbation after the system has reached a threshold of instability, and with finality found in the pull of the Second Law of thermodynamics (again, assuming that the universe is a thermodynamically isolated system).

Of course, finality has been banished from natural science for some centuries now (e.g., Weber, 1908), largely I believe because of that discourse’s entrainment by pragmatic applications, where human intentionality trumps other possible entrainments. To enforce this interdiction, finality has usually been said to imply forbidden religious connotations. The science informed by this opinion was/is a science of conceptually simple, if technically complicated, experimental constructions. These are focused upon setting up formal arrangements that could produce desired kinds of results when initiated by an experimenter’s -- or some ‘natural’ -- forcing. But now that we are becoming buried in complexity, we may need any tool we can find. Ecologists have already begun tentatively using these causal categories (e.g., Patten et al, 1976; Ulanowicz, 1997). Computer simulations would be an ideal medium for exploring the relations between the Aristotelian causal categories in various applications.

I use final causation in order to explain a major factor in our lives and economy -- the relatively poor energy efficiency of effective work -- otherwise left unexplained as to ‘why’ it must be the case. To understand this one needs to understand that the Second Law of thermodynamics is a final cause of all events and occasions insofar as they are mediated by energy flows. Whatever happens, including work, happens as an aspect of the dissipation of metastable energy gradients (e.g., Schneider and Kay, 1994) in a universe far from thermodynamic equilibrium. Work is linked to this dissipation of energy gradients, and mediates some relatively small portion of that energy (the 'exergy') into action and/or products of lesser energy amount than that dissipated from the tapped gradients. Because the energy efficiency of work is so poor, it is possible to see that any work is undertaken in order to move the universe closer to thermodynamic equilibrium. If work efficiency could be much greater, this view would be untenable. Since it cannot (Odum, 1983), this understanding seems obligatory, even if not fully explanatory regarding any given work done, which would be associated with various biological, personal and sociocultural finalities. I feel that denying this surprising view would be tantamount to claiming omniscience -- that is to say, claiming to know of a counter example hidden away in some obscure corner of the universe. In order to explore this further I will parse finality throughout Nature, for which I need to use a hierarchy format.

Synoptic View of Embodiment Within Universal Development

I use the Specification Hierarchy format (Salthe, 1993, 2002, 2006,b) to make a fully global representation of any unique particular. For example:

{physical dynamics {material connectivities {biological activities {individual action {sociopolitical projects}}}}}

with {lower level {higher level}} and {more generally distributed {more particular}}; the brackets have the same meaning as in set theory

Thus:

{dissipative structure {living system {animal {human {myself}}}}}

(I do not here distinguish natural kinds from individuals, as natural kinds are categories resulting from unrelated analysis)

The levels here are 'integrative levels' (Salthe, 1988), or ontological levels (Poli, 1998). They have been constructed during the progress of Western science, as the subjects of different investigations. As noted already in the 'unity of the sciences' perspective (Neurath et al, 1955-1969) lower levels give rise to and subsume all higher ones, but as well, we now realize that higher ones integrate and harness all lower ones under their own rules locally. Thus, consider:

{entropy production {free energy decline {work}}}

Here each level entrains at least one finality, with lower level, physical, finalities present as relatively weak entrainments locally compared to higher-level ones. But they are continuously in place, whereas higher level entrainments are episodic, and different ones may cancel each other out. As an example of the pervasiveness of the Second Law in our own lives, we might note our tendency to fidget in many ways when more important work is not at hand. Thus, activity of any kind in the non-equilibrium situation is (whatever else of greater import to us it might be) a local contribution to Second Law finality -- in the service of the equilibration of our thermodynamically isolated, currently disequilibrated Universe.

As I mentioned earlier, another aspect of this hierarchical model is that it can have a diachronic interpretation as development (Salthe, 1993). This is based on the fact that higher levels are logically refinements of lower level possibilities, with the general developmental process then being:

{vaguer -> {more definite -> {more mechanistic}}} (the arrows representing change)

For the earth, this could be represented as

{geosphere -> {biosphere -> {noosphere}}} (Vernadsky, 1944, 1986)

In application to the Universe, this pattern would proceed at different rates in the different ontological levels. Thus, the material/chemical level would have reached a mechanistic stage with the establishment of the relations shown in the Periodic Table, while the biological level did not reach that stage until much later, with the genetic code and associated apparatus. Continuing this trend, some believe that we are engaged in constructing a noosphere, which would be the mechanization of human culture.

One way of imagining this developmental process is to focus on the lowest physical level, the quark-gluon plasma. In the early universe there would have been a virtually unlimited number of degrees of freedom for any particle in regard to its position and momentum. After the origin of atoms, many degrees of freedom for many of these particles would have become frozen out. With the further emergence of molecules, the degrees of freedom of these same fundamental particles would have become even further restricted, and so on (Salthe, 2009).

This model imposes a ‘hylozooic’ kind of interpretation upon Nature, somewhat like that of Charles Peirce (CP, 6.32-33). Nothing *totally* new can appear after the initial expansion of the primal singularity. And so all emergent features that appear later at higher integrative levels would have been implicit during earlier developmental stages. One could say that these later emergents are somewhat like the ‘high grade actual occasions’ of Whitehead (1929), precipitating within the local society of ‘actual occasions’, thereby pointing to another ‘take’ on development as here understood. In this view the past is present in any current individual or occasion. In the present interpretation, these would initially have been only vaguely and episodically present as sketched in the lower integrative levels, but would have become stabilized materially with the developmental emergence of emerging higher integrative levels. This would have occurred globally during development of the earth, but is also repeated locally during the development of, e.g., a living thing like myself. In this development, particles that had become ‘captured’ by atoms, became / become further restricted in their motions when these atoms are incorporated into molecules which are parts of cells, most of whose physical degrees of freedom are almost reduced to nothing within an organism. Their ‘universal’ degrees of freedom have then become restricted to being consequences of organismic motion (if any).

Thus, the configuration of fundamental particles within my body at this moment will have occurred transiently many times in the primal quark-gluon plasma after it had reached a certain volume. More colorfully, I will have had a ‘ghostly’ presence almost since the beginning of the universe, and have been ‘lucky’ to have become embodied at last! There is thus, in this sense, ‘nothing new under the sun’, but some things have become increasingly more definite as the universe continued to expand. This is a weakly deterministic philosophical position. On the one hand, our ‘ghostly’ representation in the quark-gluon plasma will have been unstable, episodic, and without the higher level details that would need to be acquired through historical adventures as the universe developed. On the other hand, those historical adventures might have eliminated the possibility of continued materialization of any particulate lower level template. Thus, while nothing completely new – i.e., not fostered by an ancestral particulate template -- can appear, not all such potential templates get to be fostered by any given unfolding history.

Some will argue that a new species of sparrow, a new crime, or a new poem are examples of genuine newness. Note that I hold that all possibilities for any potential universal development to have been present in vague form initially, and that only some get to become more definite during universal development. This 'becoming more definite' involves what many take to be genuinely new things. Each new species or poem may seem 'new' before our eyes, but this level of newness is trivial in the universal -- and in the developmental -- perspective. Each is developmentally a restriction on what was possible before that emergence. Consider language. Having chosen English we can say an unlimited number of things, but we cannot express certain larger moods accessible, say, in French (Salthe, 2000), and which may have been conveyable in the ancestral common language.

The above perspective can be played off against the idea emphasizing that every actual occasion would be unique in the higher integrative levels, signaling a radical indeterminacy in the world (Elsasser, 1969; Ulanowicz, 2007). Anticipating the future, as in the predictive mode of science, there would seem to be an immense number of different higher level configurations potentially emergent at any future locale in a future moment. Thus, whatever happens at these higher integrative levels (biological, sociocultural) would be something completely new, and emergent at these levels. But at the lower, physico-chemical levels, any configuration whatever will have been rehearsed many times. Looking the other way, back from what has occurred, historicity, even at the higher levels, will have narrowed the possibilities gradually, by way of concatenated contingencies as the present moment was unfolding. Whatever actually occurs will have been prefigured at the lower integrative levels and gradually prepared for at the higher levels. Present configurations at the higher integrative levels therefore imply -- that is, material implication or conceptual subordination -- that which gave rise to them. This looking backward (note also the ‘anthropic principle’, whether ‘strong’ or ‘weak’) is a mode of finality, and narrates how I, the current observer, came to be here in this world.

## The Logic of Multiple Worlds

Multiple worlds are implicit in the specification hierarchy model because its form is that of a tree, with its trunk in its lowest level -- in the present application that would be the physical integrative level. Inasmuch as I am here assuming that this refers to our known physics, I am not appealing here to particular models of ‘multiple universes’ imagined by some cosmologists. A subsumptive hierarchy is plainly appropriate for modeling an expanding system like the Universe, which, in the developmental model used here, then can become occupied by ever more, ever more definite, locales. Even if no new matter continues to appear in the expanding universe, the continual incorporation of historical information at every gravitating material locale would result in the emergence of increasingly more individuated phenomena. Every emerging locale acquires its own unique configurations and conformational possibilities. Shown in the hierarchies above are only single branches of this hierarchy, where the lower levels would be shared by increasingly more possible branch tips that are not being represented. Thus. e.g., there could logically be some other kinds of dissipative structures coordinate with living ones. These would not be abiotic ones like eddies and tornadoes because those represent the grounds from which living dissipative structures emerged, and so they are relatively lower level, and therefore not coordinate with the living. Other possibilities would also not be mechanical ones like automated factories, because these are fostered by socioeconomic systems and so are relatively higher level to, and so again not coordinate with, the living.

It seems most likely that in any given world only a few of the immense number of vague tendencies in the primal quark-gluon plasma would get to become stabilized in that world. This would be especially true of those entrained into the highly individuated systems emergent in the highest integrative levels. It would be consistent with this model of differentiation during the development of the Big Bang to view emerging branch tips in the hierarchy as having been entrained by multiple finalities. That is, we might reasonably consider every actual occasion to be the locus of several finalities.

Integration

I think it important to end this essay by emphasizing the top-down integration of the various levels of organization (e.g., Greenberg and Tobach, 1988) in the specification hierarchy model. This aspect is perhaps the major message, as yet little noted, from the continuing development of the unity of the sciences perspective. On the template of:

{physical universe {material locales {biological forms {sociocultural organizations}}}}

We can consider the relations among:

{dynamics {location {form {functional individuality}}}}

Consider the interpenetration of these levels in any high enough grade of actual occasion. There can be no activity at any level that is not actually physical dynamics fostered by entropy production. And there can be no location that is not mediated by gravitating matter and chemical affordances. As well, all human inventions are 'inhabited' by the human form, as, e.g., a pile driver represents the human arm. For a currently actively pursued example, biological form and function is largely constrained by scale (Bonner, 2006; Brown and West; 2000, Sims et al, 2008; West et al, 2001). Nevertheless, as long as they exist, individual higher level entities, each instituting their own formal and final causes, harness all lower levels locally into colluding to promote them by way of providing material causes for them from one moment to the next. Individuals continually integrate all of nature into their own embodiment (Polanyi, 1968) until they disperse, when they are recycled. As a thought experiment, we might try to imagine a world without particular, history-mediated individual phenomena. We find, I think, that we must go back to something like the primal quark-gluon plasma before this becomes possible. From a 'pansemiotic' perspective (Salthe, 2008), the emergence of higher integrative levels can be seen to have pulled the universe into ever more particular meanings.

CONCLUSION

This developmental perspective is advanced in order to be posed against the currently fashionable pan-historicism, and yet historicity does play a role. One take-home message would be that if we are to try to anticipate newly emerging events and occasions, we need to develop techniques to assess vague tendencies while they are becoming liminal and beginning to emerge (Salthe, 2004). In connection with this, acting on Charles Peirce's (1905) suggestion to develop a 'logic of vagueness', is long overdue.

ACKNOWLEDGEMENTS

I thank Victoria Alexander, Richard Gordon, Igor Matutinovic, John McCrone, John Smart, Clément Vidal, and Michael Zimmerman for stimulating discussions or improving suggestions.

Salthe has worked in evolutionary biology, systems science (hierarchy theory) and, semiotics. Current interests include thermodynamics and internalism.

# Must Complex Systems Theory Be Materialistic?

Horace Fairlamb,
University of Houston-Victoria
3007 N. Ben Wilson, Victoria, TX 77901
fairlambh@uhv.edu, http://www.uhv.edu/asa/494_3135.htm.

**ABSTRACT**: So far, the sciences of complexity have received less attention from philosophers than from scientists. Responding to Salthe's model of evolution, I focus on its metaphysical implications, asking whether the implications of his canonical developmental trajectory (CDT) must be as materialist as his own reading proposes.

(1) Salthe offers his model as a treaty among several disciplines. Regarding the natural sciences, I would say that his model is more like an offer they can't refuse. Given the embodiment of the CDT at different levels of physical organization (physics, chemistry, biology), its interdisciplinary consilience seems (in my view) to be decisive.

On the other hand, metaphysicians might read the CDT in contrasting ways. In that case, Salthe's offer of a "materialist interpretation of the Big Bang theory" would be an offer that non-materialists *could* refuse. In fact, several features of Salthe's view appear to invite other metaphysical interpretations. On the basis of Salthe's argument, I conclude (a) that his anti-reductionism softens his materialism; and (b) that his own interpretation of the CDT allows, if not invites, emergentist alternatives to materialism.

(2) Resisting reductive materialism

Modern materialists typically contrast themselves with traditional or modern idealists by claiming that the basic stuff of the world is matter, not mind. Consistent with that view, Salthe's view of evolution begins with physics.

But even if materialism is a convenient place to begin an evolutionary theory, it may not be the only place where the CDT leads. The question is whether Salthe so complicates his materialism that the name becomes dispensable.

Salthe complicates his materialism, for instance, by turning his model against the reductive materialism of the gene-centrists. From Monod (1972) to Dawkins (1976), biological atomists have accorded explanatory and causal privilege to the *invariance* of the gene, a privilege that is supposed to justify making genes the unit of selection, making gene-proliferation the "ultimate utility function" of evolution, and making phenotypes mere vehicles for gene-proliferation.

Against that view, complex systems theories like Salthe's show that the invariance of self-organizing processes is causally prior to the invariance of genes.

First, the functional invariance of dissipative structures is found in physical systems simpler than living systems (storms, chemical clocks, icicles). Second, without a self-organizing system to reproduce, genes have no function at all. Third, invariance at the level of organic development may be immune to genetic differences that only become causally active under the right circumstances (e.g., many genetic effects are interactive with other genes). In these ways, the functional invariance of the organism precedes the material invariance of the gene by creating the systemic context in which genes are able to play their determining role.

Against the gene-centrist tendency to assign evolutionary agency to the lowest possible level, Salthe acknowledges how the higher functions “harness all lower levels locally into colluding” with those higher functions. This returns genes to their proper role as instruments for phenotype functions rather than rendering phenotypes the tools of genes.

If systemic determinacy precedes genetic determinacy, it also increases in evolutionary influence as organisms evolve higher level steering capacities. By evolving the capacity for culture, humans have raised foresight to new levels, inventing a post-material, purely informational form of evolution. As evolution proceeds, in other words, causation becomes less liable to materialist reduction. But if evolution can evolve a non-material form of evolution, what happens to materialism as a foundation? Might not materialism give way to semiotics as an ultimate framework?

(3) Causal ambiguities

To the extent that Salthe’s systemic model of evolutionary causation resists reduction to material atoms and physical forces, it allows emergent evolutionary causation and perhaps ontological dualisms. True, explanations of evolution from simple to complex systems may seem at first to privilege atoms and physical forces. But even on the ground floor, Salthe’s model accepts formal and final causation as equally fundamental, as his reading of the Second Law shows. If the Second Law is a version of final causation, then there is nothing more fundamental than that. But if formal and final causes (systemic causation) are as fundamental as material and efficient causes (atoms and physical forces), Salthe’s initial materialism may be taken as a heuristic starting point.

According to some contemporary materialists, the shibboleth of Salthe’s materialism may be his commitment to two principles that might be called *the total novelty denial* and *the principle of continuity:* “nothing totally new appears during development.” This is so because “all emergent features at higher levels would have been vaguely and episodically present primitively in the lower integrative levels.” This principle echoes Peirce’s ontological principle of continuity, and it resembles Daniel Dennett’s (1995, Ch. 3) argument against mentalistic dualism (appeals to explanatory “skyhooks”). According to Dennett, evolutionary appeals to mental causes are unnecessary because evolution exhibits an unbroken continuity of material mechanisms (“cranes”). But if the principle of continuity truly rendered intrusions of *total* novelty impossible at any given temporal point, it does not rule the gradual emergence of *real novelty in the long run,* which could conceivably include non-material properties. The claim of emergent dualism is not that the emergence of novelty is discontinuous, but only that real novelty can eventually be identified by its contrast with what came before.

(4) Metaphysical ambiguity

In light of these considerations, it would appear that while Salthe's materialism makes a convenient starting point for explaining the CDT, it finally proves dispensable. The evolution of complexity may begin with chunks of matter, but even on the ground floor the formal and final aspects of systemic evolution are as causally fundamental as the material aspects. This may suggest to some that Salthe's evolutionary model at least reconstructs materialism, and perhaps invites non-materialistic readings of evolution.

# Friends of wisdom?

A comment on "Development (and Evolution) of the Universe" by Stan Salthe

Gertrudis Van de Vijver
Centre for Critical Philosophy
Ghent University
gertrudis.vandevijver@ugent.be
www.criticalphilosophy.ugent.be

**Abstract:** *This commentary addresses the question of the meaning of critique in relation to objectivism or dogmatism. Inspired by Kant's critical philosophy and Husserl's phenomenology, it defines the first in terms of conditionality, the second in terms of oppositionality. It works out an application on the basis of Salthe's paper on development and evolution, where competition is criticized in oppositional, more than in conditional terms.*

In "What is philosophy?", Deleuze and Guattari argue that the space of Western philosophy opened by the Greeks is intrinsically a space of competition: the "friends" of wisdom are in the first place peers competing for the highest wisdom, for truth, for pure thinking. As such, they can no longer wish to equal the figure of the Wise. On the contrary, they must agree to bury precisely him, turning their focal attention and rivalry to "die Sache", the thing or the object. Modern science is a continuation of the Greek project: it operates within the very same space of competition between peers and attempts to victoriously constitute stable and reliable, objective, knowledge. However, since Modernity the quest for objectivity has more and more turned into a form of objectivism or dogmatism, according to which science has the exclusive rights to describe and explain the things as they are in themselves, and in which the question of the constitution of objectivity largely disappears out of sight.[1]

In "Development (and Evolution) of the Universe" Salthe is critical about the idea of competition. His focus is in particular on evolutionary theory, that narrows down the idea of change in Nature to differential persistence, "where it plays out as competition between genotypes for representation in future generations of a population". (p. 3) Salthe seems to agree with Deleuze and Guattari in underlining that differential persistence has been elaborately constructed as competition in our Western thinking. He even speaks of "our Western cultural ideology" (p. 3) in this regard. To him, however, competition is only a small part of the big picture of change in Nature; it needs nuance and supplementation. His proposal calls for a more "phenomenological" acknowledgment of the World, an acknowledgment of that which is experienced and phenomenal, beyond that which can be objectified into a Nature that functions as an operating manual for the World. His developmental viewpoint is inspired, so he states, by Schelling.

We won't challenge the value and necessity of critical accounts of objectivist discourses and of the role competition plays in them. But what is it to be critical? More precisely, what can it mean to be critical with regard to objectivism or dogmatism? Kant was the first philosopher in Modernity to have introduced the idea of Critique. He pointed out in his *Critique of Pure Reason* that the objectivity of scientific knowledge is not the result of an *object in itself* in some sense out there in nature and as such dictating appropriate ways of apprehension. Objectivity is on the contrary the result of a very specific questioning activity that gives rise to objects that have a validity only *within* the range of that activity. *If* universality and necessity are to be related to scientific knowledge it is because the questioning subject succeeded in constituting the answering potentiality of nature as a point of invariance, of exactness, of necessity and universality. The contingent perspective of the questioner *is* the possibility of any objectivity: it is *from within* that contingent perspective that objectivity witnesses of the possibility of a stabilized, non-contingent, necessary relation between questions and answers.

[1] Edmund Husserl has described this process and its potentially devastating consequences in *The Crisis of the European Sciences.*

In this regard, there seems to be something paradoxical about Stan Salthe's critical discourse. At first sight, it is a critique on the idea of competition. But at closer inspection, there are a number of elements that suggest that Salthe's thinking is operating within the very competitive space it criticizes. For instance, his proposal to distinguish Nature from the World, the one being an operating manual, symbolic and mediated, the other referring to the phenomenal, to what is not 'known' symbolically, indicates a kind of dualism between, roughly, language and biology, between the symbolic order and the body. What remains unquestioned here is the *possibility* or the *conditionality* of this opposition. Wouldn't a critical questioning of that aspect be crucial in an understanding of change in Nature, certainly when a complex viewpoint is envisaged? Because indeed, if there is complexity, involving local interactions at various levels and with various elements, as well as historicity and vagueness (pp. 6-7), to assume that there are two different "orders of being" such as Nature and World is certainly not evident. In his effort to criticize the one-sidedness of evolutionary theory – that we identify here as objectivism – what Salthe seems to do, is to defend the opposite viewpoint – subjectivism. This can remind us of Schelling's reaction to Kant: romanticism as a response to a position (wrongly) perceived as objectivism.

What is it then, to be critical? As Kant has shown, critique is not a matter of oppositionality, it is a matter of conditionality. Not A versus –A, but the *possibility* of that distinction is the heart of the matter. Kant's viewpoint is different from the nostalgic, romantic, dream of getting out of, or supplementing, the very space within which all thinking can take place. The main reason is that the sensitively and conceptually engaged subject is an intrinsic part, co-constitutive, of that thinking space. No way to get out of this human condition and fly high up to the godly heaven, neither so in living praxis, nor in symbolic discourse. Salthe's statement "(…) now that we are becoming buried in complexity, we may need any tool we can find" (p. 7) suggests that there was a time of less complexity, a human condition in which things could be ruled more simply and more adequately. His idea that "Nothing *totally* new can appear after the initial expansion of the primal singularity. And so all emergent features that appear later at higher integrative levels would have been implicit during earlier developmental stages" (p. 9, italics original) goes in the same direction. From which point of view is it possible to say that something is totally new? Isn't it the case that the distinction between new and not-new is only possible *from within* the primal singularity? Salthe seems to underscore here the idea that the constraint is the possibility – it is from within the expansion of the primal singularity that things have to be grasped – but meanwhile, it is not clear in what kind of historicity his own discourse is to be situated. Current configurations are only gradual evolutionary diminishments, specifications, individualisations. Salthe states that this is a mode of finality, which I agree with. But he seems to refrain from questioning his own perspective in making this statement, and to that extent, we might wonder whether he is not faithful to the classical scientific operation of excluding teleology from the scientific practice. It is true that the ideology of Western science has put much effort in excluding final cause. It is part of its objectivist development, excluding the perspective of constitution of objective knowledge, to focus solely on the object as it is in itself. In objectivism, the questioning activity, and thus the teleological implementation of subjectivity and objectivity, can no longer have a place, except then for heuristic or pragmatic purposes, where, as Salthe correctly notes, “human intentionality trumps other possible entrainments” (7). Teleology might then be excluded from science, it does return in a most uncritical form, namely in the unquestioned subjective capacity to know, to be directed upon, the capacity of intentionality, entirely and exclusively concentrated in the human, conscious, subject. I am in favour of putting the question of final cause back on the scientific agenda, but I doubt that Aristotle will do the job, as the configuration of objectivism, in dichotomous opposition to subjectivity, was not at all present at his time, neither epistemologically, nor ontologically. I believe that philosophers like Kant and Husserl are more adequate to deal with that topic, in as far as they attempt to give a place to the idea of teleology as "a place amidst other places", as well as in their focus on the need to take up, as philosophers, as scientists – "friends of wisdom" –, the present, experienced and lived, moment *as* part of a more encompassing whole. That movement of "taking up" is what introduces human beings into history, as part of *a* history of a certain kind. It is what constitutes history *as* a history. And isn't it there, in that singularised history, that new singularities can emerge?

# MATERIALISM: REPLIES TO COMMENTS FROM READERS

Stanley N. Salthe
ssalthe@binghamton.edu

**Abstract:** The canonical developmental trajectory, as represented in this paper is both conservative and emergentist. Emerging modes of existence, as new informational constraints, require the material continuation of prior modes upon which they are launched. Informational constraints are material configurations. The paper is not meant to be a direct critique of existing views within science, but an oblique one presented as an alternative, developmental model.

## Reply to Fairlamb

I agree with almost all of what Fairlamb ascribes to my thinking. So I will pick out a few points where there might be more disagreement. I will answer by the numbers.

(1) I would say that my understanding of CDT is necessarily emergentist. This indeed is one meaning of the brackets that I use; a contained class is not just a mere member of the containing class. Something has changed (in fact, been added) -- in my view, that is the stability of higher-level configurations, which, on materialist grounds, had to have been fleetingly present from the beginning. These stabilized configurations could be labeled 'functional invariances'.

(2) My sense of materialism, which entails emergentism, is that more generally present aspects of Nature (e.g., the physical world) have given rise to the more highly developed aspects. What is new is only the stability of configurations of the more general properties within the higher properties. As well, these configurations do not, in my view, open up new possibilities *more* than they close off some that must have been available prior to the emergence of a new integrative level. That is, while I have an emergentist metaphysics, it is a materially closed one, devolving from an increasing burden of informational constraints as the world develops.

(3) The emergence of newly stabilized configurations should, I believe, be viewed as the emergence of (levels of) information from the physical world, moving the system of interpretance --> interpreter into an ever more explicitly semiotic position with respect to the rest of Nature. My views are 'pansemiotic', whereby I infer that explicit semiosis in contemporary systems must have had precursors -- all the way back. Thus, it might be said that semiosis emerged gradually from a primal proto-semiosis implicit (immanent?) in the earliest post-inflation universe. I would think that any materialist, or evolutionist, must assume some such position.

(4) There is a duality at work in my metaphysical perspective -- because nothing happens in a vacuum. The emergence of a new level would need to be have been entrained by way of some still higher level environmental configurations, which have been imposing boundary conditions upon the ground from which the new emerges. This brings in my commitment to an expanding universe. If, e.g., in some locales life has emerged, then those locales necessarily must have had certain environmental affordances (formal causes) allowing the production of dissipative structures, and the argument would be similar for each level. It is universal expansion that produces those environments, and that keeps modifying them as the expansion proceeds. This duality (in effect, the material / formal causal pair) could be viewed as denying my materialist credentials. I don't think, however, that requiring boundary conditions is the same as erecting skyhooks. Rock bottom, for example, even the simplest equations cannot be solved without manifest values for their constant parameters. I am unconcerned whether some would find deities lurking in the values of equations.

**<u>Reply to Van de Vijver</u>**

Vandevijver correctly interprets my paper as being situated outside the reigning perspectives of natural science. But she takes my effort to be a critique, while I, instead, view it as a supplement to the reigning approaches. I do make an implied criticism of the usual historicism in connection with evolution. This was supposed to be signaled by my dismissal of 'evolution' from consideration in the paper (in the very year that its 'messenger', Darwin, is being lauded!). This snubbing is based in my generalization of the terms 'evolution' and 'development' as proposed in my 1993 book, which was founded on usages current within biology. In this generalization (evolution is synonymous with individuation) only development associates with particular knowable, describable, and therefore predictable changes, which would be aspects of the 'matter' making up our construct (as I call it), 'Nature'.

The supposed objectivity of science, as raised already by Kant, and as revived in the postmodern critique of natural science, is correctly characterized by Vandevijver as "contingent". This means that what I refer to as 'Nature' has no broadly implicated objectivity. This would be the case even if natural science was not overtly a key element in our culture's technological pursuits. My construction of the developmental viewpoint from scientific findings has been a procedure of 'confirmation' -- that aspect of science whereby hypotheses are constructed -- rather than an attempt at 'testing', whereby hypotheses are falsified (or not). The act of confirming is a creative (yes, it might be 'Romantic') searching, while testing is a kind of elaborate critique. Testing may be contingently 'objective', in a limited sense, inasmuch as there is a lot at stake, while the process of confirmation is driven in large part by esthetic intuition.

The body of my paper, then, is not devoted to a critique of evolutionary theory (which is barely mentioned in passing), but to an elaboration of what may be delivered by the developmental perspective. My opposition here to dealing with evolution has nothing to do with (Darwinian) evolutionary theory, which I have addressed elsewhere. It has to do simply with the fact that, as individuation, nothing much can be said about evolution -- it just happens. The conceptual action within Nature takes place in models of development (as well as of homeostasis), one of which I try to characterize in my paper.

Vandevijver also asks why I did not devote some time to justifying the purported dualism of Nature and The World. I think this would really be the topic of yet another paper -- one that I should perhaps have produced before this. Reflecting again the postmodern perspective, it seems clear to me that all cultures and languages are contingent, while the World confronting them, whatever it may be, can be taken to be potentially the same everywhere. ‘A map is not the territory’ -- and reflects as much the interests and investments of the society making it as it does the world being portrayed. Thus, I am contributing to our culture’s perspective on The World by offering a view of development using facts elucidated by its scientific activities. My hope is that, having recently been confronted explicitly with the complexity of the world in comparison with its investigative tools, that ‘objective’ science might be open to considering the developmental perspective to be worth explicitly critiquing.

# POSSIBLE IMPLICATIONS OF THE QUANTUM THEORY OF GRAVITY

## An Introduction to the Meduso-Anthropic Principle

by Louis Crane, math department, KSU crane@math.ksu.edu

*Abstract*: If we assume that the constants of nature fluctuate near the singularity when a black hole forms (assuming, also, that physical black holes really do form singularities) then a process of evolution of universes becomes possible. We explore the implications of such a process for the origin of life, interstellar travel, and the human future.

In his excellent book The Life Of The Cosmos [1], Lee Smolin has proposed that if the quantum theory of gravity has two special features the universe would fine tune itself.

Specifically, he proposes 1. that the quantum theory softens the singularity which forms in a Kerr-Newmann black hole, for example, if we perturb it, so that the new universe on the other side of it will really form. He then proposes 2. that the constants of nature might fluctuate in such a process, so that the new universe would have slightly different physics in it. In the absence of a quantum theory of gravity, of course, both of these suggestions are speculative.

Nevertheless, it is interesting to see what would result from such a process. Universes with peculiar fine tuned values of the coupling constants would have many more "daughter" universes, so that fine tuning, instead of an improbable accident, would be highly probable.

Professor Smolin proposes this as an explanation of the fine tuning evident in our universe, which leads to the possibility of life. In this view, life and intelligence are biproducts of the very special physics necessary to provide surface chemistry and radiating stars in order to produce many generations of black holes.

Philosophically, then, life and even intelligent life are accidents.

I do not intend to repeat Professor Smolin's argument in this note. Rather, I wish to propose a modification of his reasoning which seems practically necessary in the framework of his conjectures, but which changes their philosophical implications immensely.

I have called this modified version the meduso-anthropic principle, after a stage in the life cycle of the jellyfish. The reason for this metaphor will be explained in time.

In addition, I want to point out in this note that within the approach to quantizing gravity which I proposed in my paper "Topological Quantum Field Theory as The Key to Quantum Gravity [2]," the second conjecture which Professor Smolin makes, i.e. that coupling constants might fluctuate when the topology of spacetime changes, becomes much more plausible. This is because I propose a quantum theory in which the coupling constant is part of the state of the universe. If this approach were extended to gravity coupled to matter, the coupling constants of the matter fields would have a similar role. One could try to compute topology changing amplitudes from them via the state sum methods in my paper. I do not yet see how to do this, but there are some suggestive possibilities in the underlying algebraic picture.

The question whether the real quantum theory of gravity has the right features to make this picture work will no doubt remain open for a considerable time. Nevertheless, let us explore the implications.

## MAN IN THE LOOP

The conjecture which I believe modifies Professor Smolin's conclusions is the following:

SUCCESSFUL ADVANCED INDUSTRIAL CIVILIZATIONS WILL EVENTUALLY CREATE BLACK HOLES.

The synthesis of this with Smolin's two conjectures is what I call the meduso-anthropic principle. Before exploring the implications, let us consider the plausibility of this conjecture.

SUBCONJECTURE 1: SUCCESSFUL ADVANCED INDUSTRIAL CIVILIZATIONS WILL EVENTUALLY WANT TO MAKE BLACK HOLES

and

SUBCONJECTURE 2: SUCCESSFUL INDUSTRIAL CIVILIZATIONS WILL EVENTUALLY BE ABLE TO PRODUCE BLACK HOLES.

It is fairly clear, at least, that the conjecture follows from the two subconjectures. (This paper is not on the mathematical level of rigor).

Let us first consider subconjecture 1. There are two reasons to want to make black holes. One might want to make a few for scientific purposes. Indeed, barring major surprizes in physics they are the ultimate high energy phenomenon. If it came within the reach of a technological civilization to build them, certainly the scientists in it would want to do so.

The second motivation for creating black holes is much more compelling. The hydrogen supply of the universe is slowly being exhausted. At some point in the future, any civilization will face the possibility of perishing from a lack of free energy, In principle, black holes could provide a permanent solution to this problem, since they can convert matter into energy via the Hawking radiation forever and with perfect efficiency. (They are also the perfect waste disposal for similar reasons). In order to make this practical, it would be necessary to have very small and very hot black holes, and to be able to "feed" and manage them very carefully. However difficult this problem finally is, our descendants in a few hundred billion years will have no alternative if they want to go on living.

There is another reason to want to make artificial black holes, which would act in a much shorter time scale. They would allow us to have access to much greater energy densities than any other technology we can currently forsee. This would make it possible to design a starship which could carry humans This is discussed in my recent paper [3].

Now let us consider the second subconjecture. The main difficulty in creating a black hole is cramming a lot of mass-energy in a small space.

Nature solves this problem by cramming a lot of nuclear matter into the center of a large star. This is completely inadequate for our purposes, since the resulting black holes are much too big, and hence much too cold to be of use. Also, it is hard to imagine a civilization doing such a thing.

Fortunately, there is an approach which could produce much higher densities, and hence much smaller holes.

In this approach, one simply creates a huge sphere of converging lasers and fires them simultaneously at a central point. Since light is composed of bosons, there is no Pauli exclusion principle to overcome, and the bursts of photons could all occupy a very small space simultaneously, creating a black hole of a temperature corresponding to the frequency of the light. (The term "successful' in the conjecture has to be taken on such a scale. Still, a hundred billion years is a long time).

The critical length in such an apparatus is the wavelength of the light used. If our descendents can build nuclear lasers which lase in hard gamma, then a spherical converging laser the size of a small asteroid would suffice to produce very small hot black holes. Of course, gamma interferometry would be necessary to keep it focussed. None of this is beyond what could plausibly be done in a few centuries.

This poses subtle relativity questions, as well as extremely obvious engineering ones. Nonetheless, I believe that it demonstrates some degree of plausibility for my second subconjecture.

The notion of a successful civilization here is considerably beyond contemporary standards. As investigated in [3],a machine to build useful black holes could be constructed using only a small part of the output of our sun. Space based materials and a large fleet of robots would be necessary.

## IMPLICATIONS OF THE CONJECTURES

If both Smolin's two conjectures and mine are true, then the fine tuning of physical constants would not stop with physics which produced stars and gas clouds which cool. Rather, the selection would continue until physics evolved which resulted in successful civilizations, with a very exacting definition of success. In the limit as the number of generations of daughter universes increases, this would be true even if we only make the weaker assumption that successful civilizations make a few black holes as experiments. This would increase the average number of daughter universes in universes with successful civilizations, even if by a small fraction. Each such universe would have more daughters, until, after many generations, intelligences would be present in almost all universes. The effect would be much more rapid if black holes as energy sources turn out to be practical.

The philosophical implications of such a process are very deep. Although it has been generally believed by people with a scientific frame of mind that human life and history take place within the rule of physical law, it has generally been assumed that the relationship between the specific laws of physics and human events was complex and accidental. This has, in fact placed science in conflict with the otherwise dominant currents of Western (and by no means only Western) thought.

Indeed, it has been the belief of most philosophers, and a surprizing number of important scientists, that humanity had some fundamental role in the universe, and that mind was more than an accidental attribute of organized matter.

If the combination of hypotheses described above is correct, a richer connection between mind and matter appears in a surprizing way. Almost all universes would produce successful intelligence, because their detailed structure would be fine tuned by a long process in which intelligences had reproduced universes over and over; a process with the closest analogy with the passage of millions of generations which has honed life forms to an almost unimaginable perfection.

If the combination of hypotheses which I am giving the name of meduso-anthropic is correct, the relationship of civilization to environment would entail a thousand improbable coincidences with favorable outcomes. Historical events would skirt innumerable disasters and find an improbable path to success. The relationship of humanity to the universe would have an organic quality.

It is now possible to explain the metaphor I have chosen in the title meduso-anthropic. It refers to a stage in the development of the animals in the phylum which includes the jellyfish and coral. These animals have two phases of life, medusid and polyp. Medusids produce polyps, which produce medusids. It is sometimes even difficult to recognize that the two stages represent a single species. Analogically, intelligences are the medusids, and black holes/universes are the polyps of a single evolutionary process.

The idea of an organic fine tuning of the relationship of life to nature seems improbable as long as one listens to the voice of scientific common sense. The minute one examines any part of natural history as our understanding of it is growing, experience begins to drown out common sense.

## A SUMMATION

It is certainly impossible to claim that the quantum theory of gravity has reached a stage where these ideas can be validated. On the other hand, it could reach such a stage in fairly short order, if we are lucky.

Nevertheless this much seems clear. The laws of Physics which only operate at very high energies can nevertheless have profound implications for what we see around us. They have the possibility of changing our understanding of ourselves and our world in ways we cannot yet imagine. The fact that machines at Planck scale energies are not yet in the cards does not mean that quantum gravity is of no concern to us.

For myself, I intend to return to my four dimensional state sums with greatly increased ardor, in the hope that something like the meduso-anthropic principle might emerge from them.

# Two purposes of black hole production

Clément Vidal
*Department of Philosophy, Vrije Universiteit Brussel, Center Leo Apostel and Evolution, Complexity and Cognition, Brussels, Belgium.*
clement.vidal@philosophons.com http://clement.vidal.philosophons.com

**Abstract**: Crane emits the speculative conjecture that intelligent civilizations might want and be able to produce black holes in the very far future. He implicitly suggests two purposes of this enterprise: (i) energy production and (ii) universe production. We discuss those two options. The commentary is obviously highly speculative and should be read accordingly.

Crane's note dates back from 1994, and is to our knowledge the first suggestion to extend Lee Smolin's Cosmological Natural Selection (Smolin 1992) by including intelligent civilization. Since then, other authors have developed this idea further (Harrison 1995; Gardner 2000; Gardner 2003; Baláz 2005; Smart 2008; Stewart 2009; Vidal 2008; 2009). Philosophically, we believe that Crane's speculations are highly valuable, since they concern foundamental problems of the very far future of civilizations.

(1) Crane suggests "a quantum theory in which the coupling constant is part of the state of the universe". Is this model still topical? If so, it is not clear to me, to which coupling constant it is referred to (all of them?). I emitted the proposition that as physics progresses, fine-tuning arguments would be reduced to initial conditions of a cosmological model (Vidal 2009, in this volume). If the coupling constants are indeed part of the state of the universe, does it support this proposition?

(2) In this paper, intelligent civilizations are hypothesized to use black holes for two distinct purposes:

(i) energy production and waste disposal (page 2)
(ii) produce one or several universes (page 4)

(3) Crane claims that "any civilization will face the possibility of perishing from a lack of free energy". At the scale of the universe, this problem is known as the thermodynamical death or heat death of the universe (see e.g. Adams and Laughlin 1997). Crane suggests that "black holes could provide a permanent solution to this problem", via Hawking radiation. This is of course a very interesting suggestion, since it would allow intelligent civilization to continue indefinitely in this universe.

(4) However, on an extremely long term, my intuition that this would *not* provide a permanent solution. Indeed, even if a civilization is able to produce black holes as we produce microprocessors today, one certainly needs a lot of energy to produce black holes (the two approaches developed page 3 certainly need a lot of energy input). Where do we find this energy in a dying universe?

(5) Now let us assume a perfect efficiency of black holes as matter to energy conversion, and thus no need to make new black holes. Are indefinitely renewable energy cycles really possible? If so, how? Wouldn't it be like trying to produce a perpetual motion machine, which is clearly impossible?

(6) Furthermore, if purpose (i) really works, why would intelligent civilization bother producing a new universe (ii), as suggested later in the paper?

(7) It is not clear to me if and why the same kinds of black holes would generate a source of energy, and/or a new universe.

(8) For (i), it is argued that "very small and very hot black holes" (page 2) would be suitable.

(9) Similarly, what would be the characteristics of a black hole for the purpose of producing a new universe (ii)? Here are a few possibilities I can think of:

(a) Intelligent civilization produces the appropriate black hole itself.

(b) Intelligent civilization uses the black hole produced at the death of a large star.

(c) Intelligent civilization uses the supermassive black hole at the centre of a galaxy.

(d) Intelligent civilization uses the whole universe as a black hole. Here, there are two possible variations.

(d1) The first is natural, if the "Big Crunch" (or oscillating) scenario is finally favoured. This option is however not favoured by current cosmological models of accelerating expansion.

(d2) The second is then artificial, if intelligent civilization can have a global influence on universe's expansion and provokes an artificial "Big Crunch".

What arguments and counter arguments do support or refute one or several of these speculative possibilities? Or maybe are there yet other possibilities?

# From Philosophy to Engineering

Louis Crane.
Mathematics Department Kansas State University

112 Cardwell Hall
Kansas State University
Manhattan KS
66506

crane@math.ksu.edu

**Abstract:** In the years since I first thought of the possibility of producing artificial black holes, my focus on it has shifted from the role of life in the universe to a practical suggestion for the middle-term future, which I think of as on the order of a few centuries.

I think I need to apologise a bit for my paper. I am a mathematical physicist by choice, and only a philosophe “malgré lui”. The paper was written for an audience of relativists, and only very sketchy. I am overwhelmed by the implications of the idea and I only turned back to it after 14 years because it gradually attracted attention. I shall try to supply some background here, by commenting on Vidal's nine points.

(1) The coupling constants of nature include the strengths of the forces and the masses of the fundamental particles. Physicists do not know how to predict their values, they are put in to the equations by hand. Attempts to explain them from a more fundamental theory have not succeeded.

The values of the constants are peculiar, there are enormous differences in magnitude. The universe would be completely different if their values were changed by a small amount. For example, the existence of stars which shine for billions of years depends on a delicate balance between electromagnetic and nuclear forces. Nucleii are almost stable, but not quite.
Another example is a delicate resonance in nuclear physics discovered bt Gamow. It means that stars produce a large amount of carbon (see Smolin's book).

These phenomena are well known to physicists under the term fine tuning.

Now it is generally believed that the particles and forces we see are not truly fundamental, they are the result of symmetry breaking as the universe cooled. This is the basis of the standard model of particle physics.

So we could imagine that as the spacetime near a black hole singularity heated and then cooled once past the singularity the state of the vacuum could fluctuate. This means the peculiar conditions that form life would be the result of the outcome of events near the initial singularity of a universe.

I still do not know how to couple gravity to matter in a quantum theory, so I still cant say if this turns out to work or not.

(2) I do not really agree with the two black holes production purposes distinguished by Vidal. For me, production of new universes is only a byproduct. I am not convinced that altruistic motives have ever played much of a role in human history other than as excuses. I think advanced civilisations will make small black holes in order to have their vast energy output available, and to go to the stars. If they result in the creation of new universes, that explains the overall evolutionary success of fine tuned universes, but not the motivation of the creators.

(3) My thinking on this point has shifted, due to the profound energy problems our civilisation has come to face, and the understanding of the great difficulties in human space travel which have emerged.

I now see the creation of artificial black holes as an important matter for the next few centuries. Once we get over the hill of creating the generator (a laser the size of a small asteroid), we will have energy resources no other development based on known physics could approach. I think it is the only feasible way to make a stardrive (see Crane, Westmoreland 2010 for details). My discussion of the extreme future now seems naïve to me.

(4) As I explain in (Crane, Westmoreland 2010), a black hole can be fed ordinary matter as fast as it radiates, thus turning it to energy. We could use already created black holes to charge a generator, thus turning dust into black holes ad infinitum. This does not really go on forever. We run out of matter someday. But its a very long time.

(5) This has nothing to do with perpetual motion. The black hole is converting matter which is fed into it into energy. Forget the “forever” business. Im sorry I ever said it.

(6) The new universes are not accessible to us. They remain forever in our future. That is general relativity for you. We do not “bother” making them, they form as a result of the singularity in maximally extended black hole solutions. (That may or may not be physically real, its debated among relativists. I dont know if the meduso anthropic principle really works or not, it depends on subtle issues in general relativity.) We have no purpose in making them except possibly a quasireligious one.

(7) Black holes dont really have “kinds”. The no-hair theorem says they only have mass=size, angular momentum and charge. If the interpretation of the maximal extensions of the solutions is physical, new universes always appear when they form, one per black hole for stationary, a sequence for rotating ones. Size doesnt matter.
A large number of black holes would produce more baby universes than a small number.

(8) The Hawking radiation is negligible except for very small black holes, that would be unlikely to occur in nature, but which would be easier to produce artificially, since they do need as much energy to produce.

(9) The black holes in b, c, and d are of no use for power production or star travel. They are too cold. They automatically produce new universes, and nothing we do affects the process. They therefore cannot enter into an evolutionary process in which we play a role. We only affect things by changing the number of black holes in our universe. I have not worried about the “big crunch.” It looks improbable. The proposal to create a black hole generator would take a labor force as large as the one that built the pyramids, and last as long as the construction of the cathedrals. In order to sustain such an effort, perhaps something like a religious motivation might be necessary. The meduso-anthropic principle connects intelligence into the formation of the universe and makes us part of the process. It therefore has a strange resemblance to Spinoza's version of God as identical to self-creating nature. It may be that in order to meet the enormous challenge of constructing black hole technology, civilisation will need to undergo a spiritual transformation, in which religion and science fuse. I admit I find this an attractive prospect, although I am not inclined to attempt to start it.

# Computational and Biological Analogies for Understanding Fine-Tuned Parameters in Physics

Clément Vidal
Center Leo Apostel
Evolution, Complexity and Cognition research group
Vrije Universiteit Brussel (Free University of Brussels)
Krijgskundestraat 33, 1160 Brussels, Belgium
http://clement.vidal.philosophons.com
clement.vidal@philosophons.com

**Abstract:**
In this philosophical paper, we explore computational and biological analogies to address the fine-tuning problem in cosmology. We first clarify what it means for physical constants or initial conditions to be fine-tuned. We review important distinctions such as the *dimensionless* and *dimensional* physical constants, and the classification of constants proposed by Lévy-Leblond. Then we explore how two great analogies, computational and biological, can give new insights into our problem. This paper includes a preliminary study to examine the two analogies. Importantly, analogies are both useful and fundamental cognitive tools, but can also be misused or misinterpreted. The idea that our universe might be modelled as a computational entity is analysed, and we discuss the distinction between physical laws and initial conditions using algorithmic information theory. Smolin introduced the theory of "Cosmological Natural Selection" with a biological analogy in mind. We examine an extension of this analogy involving intelligent life. We discuss if and how this extension could be legitimated.

## Contents

# 0 Introduction[1]

After Leibniz famously wrote "why is there something rather than nothing?" he then qualified this by writing: "Also, given that things have to exist, we must be able to give a reason why they have to exist as they are and not otherwise." (Leibniz 1714, para. 7). Trying nowadays to tackle this age old metaphysical question, we have to take into account the progress that science has made. Both the emerging sciences of complexity and cosmology can help this philosophical enterprise. Modern science can successfully connect the general physico-chemical cosmological evolution with biological and cultural evolution (e.g. Chaisson 2001; De Duve 1995). Thus, it seems reasonable to assume that science is an effective method in enabling us to understand the whole evolution of our universe. The problem of harmony in the cosmos has thus shifted to its beginning : why did the universe start with these initial conditions and laws, and not others?

The belief in God allowed western thinkers to understand why the "Laws of Nature" are as they are and not otherwise. Scientific activity ultimately consisted of discovering the “Laws of Nature” set up by God. However, now that many scientists no longer believe in God, there is a lack of explanation in the origin of the “Laws of Nature” (Davies 1998).

Why is our universe as it is, and not different? This question is a very much debated issue, at the intersection of cosmology, theology and philosophy. In modern terms, it is known as the *fine-tuning problem* in cosmology. It states that *if a number of parameters, both constants in physics and initial parameters in big-bang models had been slightly different, no life or, more generally, no complexity would have emerged* (Barrow and Tipler 1986; Davies 1982; 2008; Ellis 1993; Leslie 1989; 1998; Rees 2000; Hogan 2000; Barrow et al. 2008). The standard models of particle physics and cosmology require 31 free parameters to be specified (Tegmark et al. 2006). It is a main challenge of modern physics to build stronger theories able to reduce this number.

As Leslie (1989) reminds us, the argument that the universe is fine tuned is not based on the assumption that there is a fine-tuner. It *only* means that the emergence of life or complexity is sensitive to many different slight changes in our physical and cosmological models.

The literature around this issue can be divided into two main classes of solutions: “God” or “Multiverse”. Either it is God who created the Universe with all its parameters fit for life and intelligence; or there is a huge number of other universes with different parameters, so that it is very probable that there is one containing life and intelligence. The fact that it is the one we happen to inhabit is

1 This paper was commented by Greben (2009) and Vaas (2009). In (Vidal 2009a) I respond to these comments and criticisms (this EDU2008 volume).

an observational selection effect which thus makes fine-tuning less mysterious (e.g. Carr 2007; Bostrom 2002).

From a rational and scientific point of view, an appeal to God suffers from being a non-naturalistic explanation. Furthermore, God is often linked with the "God of the gaps" assumption. If we can not understand a phenomenon, we use God to explain it, and we thus do not seek another explanation. This attitude can, by its very definition, *explain everything*. A parallel can be made with the seemingly opposite approach to the fine-tuning problem, holding that the universe's parameters happened by pure random chance. Indeed, this appeal to chance works everywhere and is not restricted by any limit; so it can also *explain everything*. One should also note that *in fine*, appealing to a multiverse with a selection effect is similar to the chance explanation. Indeed, in both cases we have the chance to be in a life-permitting universe.

Iris Fry (1995) already pointed out in the context of explaining the origin of life that appealing to chance or to God are after all similar attempts. This situation is comparable with the "God" or "Multiverse" alternative. Both options thus surprisingly suffer from similar shortcomings. Are there other possible roads to address the fine-tuning problem?

This paper aims explicitly and carefully to use two great analogies, to open new research axes. These are *computational* and *biological* analogies. We first clarify what physical constants and initial conditions are, to better grasp what it means to state that they are fine-tuned. We thus review some propositions to classify the two principal sets of physical parameters: physical constants in particle physics and initial conditions in cosmology. Before exploring the two analogies, we investigate the basic functioning of analogical reasoning. We do this to avoid naïve ontological statements such as "the Universe is a computer" or "the Universe is alive". We then analyse aspects of computational and biological analogies. Among others, this leads us to questionning the distinction between laws and initial conditions. We also point out some of the epistemological limits of Smolin's attempt to tackle the fine-tuning problem with his biological-inspired theory of "Cosmological Natural Selection". We then propose a *philosophical* extension to it, "Cosmological *Artificial* Selection" which includes a possible role for intelligent life.

# 1 Physical constants and initial conditions

We distinguish between *dimensional* and *dimensionless* physical constants, as proposed by Michael Duff in (Duff, Okun, and Veneziano 2002). If a constant has a unit after its value, it is dimensional. Dimensional constants depend on our unit system choice and thus have a conventional aspect. The velocity of light *c*, the reduced Planck constant $\hbar$ or the gravitational constant *G* are dimensional constants (their respective dimensions are, for example, $m.s^{-1}$, eV.s and $m^3.kg^{-1}.s^{-2}$). Certainly, we can for example make the velocity of light equal to 1, and thus apparently dimensionless. However, this applies only to a particular unit system, and the constant will be dimensional again in another unit system.

By contrast, dimensionless constants are dimensionless in *any* unit system. They are ratios between two physical quantities, such as two forces or two masses. For example, the electron-proton mass ratio is $m_e/m_p = 1/1836.15$... Since the two quantities are masses, we can get rid of the units (i.e. the dimension), and keep only a pure number. Other dimensionless constants are deduced by a similar *dimensional analysis*. If the analysis leads to a pure number without dimension, we have a “dimensionless” constant.

Along with this dimensional versus dimensionless distinction, Jean-Marc Lévy-Leblond (1979, 238) proposed another complementary classification of physical constants. Three types are distinguished, in order of increasing generality:

> A. Properties of particular physical objects considered as fundamental constituents of matter; for instance, the masses of "elementary particles", their magnetic moments, etc.
>
> B. Characteristics of classes of physical phenomena: Today, these are essentially the coupling constants of the various fundamental interactions (nuclear, strong and weak, electromagnetic and gravitational), which to our present knowledge, provide a neat partition of all physical phenomena into disjoint classes.
>
> C. Universal constants, that is constants entering universal physical laws, characterizing the most theoretical frameworks, applicable in principle to any physical phenomenon; Planck constant $\hbar$ is a typical example.

The classification is only intended to be a basis for discussing and analysing the historical evolution of different physical constants. For example, the constant *c*, the velocity of light, was first discovered as a type-A constant. It was a property of light, as a physical object. With the work of Kirchhoff, Weber, Kohlrausch and Maxwell, the constant gained type-B status when it was discovered that it also characterized electromagnetic phenomena. Finally, it gained type-C status when special and general relativity were discovered, synthesizing concepts such as spatio-temporal intervals, or mass and energy (see (Lévy-Leblond 1979, 252-258) for a detailed account of the status change of *c*).

What happens next, when a constant has reached its type-C status? The fate of universal constants (type-C), explains Lévy-Leblond, is to "see their nature as concept synthesizers be progressively incorporated into the implicit common background of physical ideas, then to play a role of mere unit conversion factors and often to be finally forgotten altogether by a suitable redefinition of physical units." (Lévy-Leblond 1979, 246). More precisely, this remark leads him to the distinction of three subclasses of type-C constants, according to their historical status:

(i) the *modern* ones, whose conceptual role is still dominant (e.g. $\hbar$, $c$);
(ii) the *classical* ones, whose conceptual role is implicit and which are considered as unit conversion factors (e.g. thermodynamical constants $k$, $J$);
(iii) *archaic* ones, which are so well assimilated as to become invisible (e.g. the now obvious ideas that areas are square of lengths).

If all dimensional constants follow this path, then they all become "archaic", and thus integrated in the background of physical theories. The fate of dimensional constants seems then to fade away. Is it possible to seriously consider this remark, and try to relegate all dimensional constants to archaic ones? Michael Duff (Duff 2002; Duff, Okun, and Veneziano 2002) did a first step in this direction by convincingly arguing that the number of dimensional constants (type-C) can be reduced to ... zero! Thus, he considers constants like $c$, $G$, $\hbar$, which are often considered as "fundamental", as merely unit conversion factors. According to his terminology, only dimensionless constants should be seen as fundamental.

A dimensionless physics approach is also proposed in the framework of scale relativity (Nottale 2003, 16). Following the idea of relativity, one can express any physical expression in terms of ratios. Indeed, in the last analysis a physical quantity is always expressed *relative* to another. Of course, experimentalists still need to refer to metric systems, and often to many more dimensional physical constants than just the common $c$, $G$ and $\hbar$. The point here is that it is possible to express the results in physical equations without reference to those dimensional constants (see also (Lévy-Leblond 1979, 248-251)).

What are the consequences of these insights for fine-tuning arguments? If the fate of dimensional constants is to disappear, then the associated fine-tuning arguments with these constants should also disappear. Considering what would happen if a type-C dimensional constant would be different has to be considered skeptically. Such a scenario has already been considered, for example by Rozental (1980) where he analysed what would happen if the Planck constant were to be increased by 15%. Again, as Duff argued, the problem is that dimensional constants are conventions, and changing them is changing a convention, not physics. It is thus only meaningful to express the possible changes in terms of dimensionless constants.

In fact, most fine-tuning arguments focus on considering changes in dimensionless constants. Typically, they consider what would happen if we were to change one of the four fundamental dimensionless coupling constants. These include $\alpha$ for electromagnetism, $\alpha_G$ for gravity, $\alpha_W$ for the weak nuclear force and $\alpha_s$ for the strong nuclear force. It should be noted that these constants are actually not constant, since they change with energy scale (Wilczek 1999).

The main conclusion in this area of study is that there are *conditions* on these constants for key events in the evolution of complexity in the universe to occur. For example, during the big-bang nucleosynthesis, the condition $\alpha_G < (\alpha_W)^4$ must be fulfilled, or else all hydrogen goes to helium. Bernard Carr (2007a) provided a detailed account of other constraints related to these four constants for the baryosynthesis, nucleosynthesis, star formation and planet formation to occur.

Along with these coupling constants, there is a whole other set of fine-tuning arguments based on cosmological parameters. These include parameters such as matter density (Ω), amplitude of initial density fluctuations, photon-to-baryon ratio, etc. For example, the "total density parameter Ω must lie within an order of magnitude of unity. If it were much larger than unity, the Universe would recollapse on a time-scale much less than the main-sequence time of a star. On the other hand, if it were much smaller than unity, density fluctuations would stop growing before galaxies could bind." (Carr 2007, 80). We wrote in introduction that one of the main challenges of modern physics is to construct theories able to reduce this number of parameters, both in the standard model of particle physics and in cosmological models. Parameters involved in cosmological models can be explained by new physical theories. Such is the case with the dimensionless cosmological constant, whose value has been predicted by scale relativity (Nottale 2008, 27-31).

Following Duff and Lévy-Leblond, we saw that type-C constants are bound to disappear. Another challenge we would like to propose is the following: could type-A and type-B constants emerge from initial conditions in a cosmological model? If we were be able to explain all these constants in terms of a cosmological model, it would certainly be a great achievement. Smolin (1997, 316) also argued that fundamentally, progress in quantum mechanics must lead to a cosmological theory. Indeed, all particles ultimately originate from the big-bang, thus *a complete understanding of particle physics should include an explanation of their origin, and thus relate with a cosmological model.*

In a certain sense, progress in this direction has already happened, if we consider the discovery of big-bang nucleosynthesis. Properties of atomic elements could be thought as fundamental constituents of matter, and thus type-A constants, until we discovered they were actually formed at the big-bang era. If we extrapolate this trend, a future cosmological model may be able to derive many (or even all) type-A constants from initial conditions.

The same can be said about fundamental coupling constants (type-B). Special scale relativity can indeed predict the value of $\alpha_s$ (the strong nuclear force) at the Z mass energy level. This was predicted with great precision, and has been confirmed by experimental measures (see (Nottale 2008, 26-27) in this volume for further details). Thus, if physics continues its progress, it is reasonable to conceive that particle physics models would become integrated into a cosmological model.

The consequence for fine-tuning arguments is that *fine-tuning of physical constants would progressively be reduced to initial conditions of a cosmological model.*

Given this analysis of physical constants, let us now examine an argument against the idea of fine-tuning proposed by Chaisson (2006, xvi-xvii):

> Rather than appealing to Providence or "multiverses" to justify the numerical values of some physical constants (such as the speed of light or the charge of an electron), I prefer to reason that when the laws of science become sufficiently robust, we shall naturally understand the apparent "fine-tuning" of Nature. It's much akin to mathematics, when considering the value of π. Who would have thought, a priori, that the ratio of a circle's circumference to its diameter would have the odd value of 3,14159.... ? Why isn't it just 3, or 3,1, or some other crisp number, rather than such a peculiar value that runs on ad infinitum? We now understand enough mathematics to realize that this is simply the way geometry scales; there is nothing mystical about a perfect circle -yet it surely is fine-tuned, and if it were not it wouldn't be a circle. Circles exist as gracefully rounded curves closed upon themselves *because* π has the odd value it does. Likewise, ordered systems in Nature, including life, likely exist *because* the physical constants have their decidedly odd values.

First, we can remark that the speed of light and the charge of the electron are dimensional constants. As we analysed, it makes not much sense to speak about a variation, and thus a fine-tuning of them (see also (Duff 2002)).

Let us look more closely at Chaisson's suggestion that if “the laws of science become sufficiently robust we shall naturally understand the apparent 'fine-tuning' of Nature”. Considering what we have argued so far, we can agree with this. We have outlined Duff's proposal that dimensional constants can be reduced to 0. We have suggested that fundamental coupling constants could be in future explained from more general principles, and that many apparent “fundamental constants” in the past can nowadays be explained by more general theories. Accordingly, a great deal of fine-tuning has been and certainly will be explained by more advanced physical theories.

Let us now examine the analogy with π. We can see a disanalogy between mathematical and physical constants. Mathematical constants are defined *a priori* by the axioms: they are *internal to the system* and are generally definable and computable numbers. For example, we have plenty of algorithms to calculate π. This is not the case with physical constants. Many of them remain *external to the system,* in the sense that they are not computable from inside the model. At some stage there has been a measurement process to get their values. We can reformulate Chaisson's position by saying that progress in science will allow us to understand (or compute) these constants, from more fundamental principles. We will develop this computational view in the third section of this paper.

However, if we saw that future physics could understand physical constants in terms of a cosmological model, it is unlikely that this would also include the initial conditions of this model. Indeed, if we were to have a theory deciding all values of initial conditions in a cosmological model, it then leads to the idea of a “final theory” or a “theory of everything”. This would bring many conceptual and metaphysical problems (Rescher 2000). One of these problems is that, ironically, this idea of a final theory is an act of faith and is thus similar to the Providence explanation (e.g. Davies 2008, 170). Smolin (1997, 248) wrote that the “belief in a final theory shares with a belief in a god the idea that the ultimate cause of things in this world is something that does not live in the world but has an existence that, somehow, transcends it.” Fine-tuning arguments based on initial conditions of a cosmological model thus remain intact in Chaisson's critique.

# 2 Analogies for scientific purposes

We have seen some aspects of physical parameters from a physical point of view. We will shortly turn to other approaches to the fine-tuning issue, inspired by computational and biological analogies. However, let us begin with a short digression explaining how analogies can be used for scientific purposes. Many great scientific discoveries have been triggered by analogies (see (Holyoak and Thagard 1995, chap. 8) for plenty of examples). This constitutes an important motivation to understand in greater detail the functioning of analogical reasoning. Analogical reasoning is certainly an essential cognitive tool, which nevertheless needs to be carefully used.

What is an analogy? It is a structural or functional similarity between two domains of knowledge. For example, a cloud and a sponge are analogous in the sense that they can both hold and give back water. More precisely, we can give the following definition: “an analogy is a mapping of knowledge from one domain (the base) into another (the target) such that a system of relations that holds among the base objects also holds among the target objects.” (Gentner and Jeziorski 1993, 448-449). In this very simple example, the relations “holding and giving back water” which are verified in the base (the cloud) are also verified in the target (the sponge).

Analogical reasoning is recognized to be a basic cognitive mechanism allowing us to learn and solve problems (e.g. Minsky 1986; Hofstadter 1995; Holyoak and Thagard 1995). Leary (1990, 2) even argued that language and every kind of knowledge is rooted in metaphorical or analogical thought processes. Indeed, when we do not know a domain at all, we must use analogies as a cognitive tool to potentially gain some insights from what we already know. In this manner, a map from the known to the unknown can be drawn.

Specifically, Holyoak and Thagard (1995, 185-189) argued that analogical reasoning is helpful in *discovering, developing, educating,* or *evaluating* scientific theories. Indeed, they allow us to propose new hypotheses, and thus *discover* new phenomena. These new hypotheses trigger us to *develop* new experiments and theories. Let us note however that there is nothing automatic or easy in this process. The relational system should first be examined in both domains, and then a more precise analogy or disanalogy can be found worthy of testing.

The *educating* part of analogies is useful for diffusing scientific knowledge, both to colleagues and pupils. Indeed, it can be very helpful and efficient to consciously use analogies to help others grasp a new idea, based on what they already know.

The *evaluating* part confronts us with one of the main dangers of analogies. One should emphasize that *an analogy is not a proof*. Analogies can thus not properly be used to prove statements, but their main utility is in giving *heuristics* for discovering and developing scientific theories. To illustrate this point, let us consider the teleological argument of God's existence popularized by William Paley (1802), based on the analogy of our Universe with a watch. It goes as follows:

(1) A watch is a fine-tuned object.
(2) A watch has been designed by a watchmaker.
(3) The Universe is fine-tuned.
(4) The Universe has been designed by God.

In the base domain (1-2), we have two objects, the watch and the watchmaker. They are linked by a “designed by” relationship. In the target domain (3-4), the Universe is like a watch, and God, like a watchmaker. That the relation (1)-(2) is a verifiable fact does not imply at all that the same relation “designed by” in (3)-(4) should be true. There is no causal relationship between the couple (1)-(2) and (3)-(4). This reasoning at most gives us an *heuristic* invitation to ponder whether the universe is fine-tuned. Although it is a logically flawed argument, one can appreciate that this reasoning induces a strong intuitive appeal.

There are in fact other pitfalls associated with analogical reasoning. To avoid them, Gentner and Jeziorski (1993, 450) proposed six principles of analogical reasoning (Gentner 1993, 450):

> 1. **Structural consistency**. Objects are placed in one-to-one correspondence and parallel connectivity in predicates is maintained.
> 2. **Relational focus**. Relational systems are preserved and object descriptions disregarded.
> 3. **Systematicity**. Among various relational interpretations, the one with the greatest depth - that is, the greatest degree of common higher-order relational structure - is preferred.
> 4. **No extraneous associations**. Only commonalities strengthen an analogy. Further relations and associations between the base and target - for example, thematic connections - do not contribute to the analogy.
> 5. **No mixed analogies**. The relational network to be mapped should be entirely contained within one base domain. When two bases are used, they should each convey a coherent system.
> 6. **Analogy is not causation**. That two phenomena are analogous does not imply that one causes the other.

Is it possible to further generalize the use of analogical reasoning into a science which would focus only on the structural or functional aspects of systems? Studying different models in different disciplines having structural or functional similarities leads to the development of very general interdisciplinary scientific frameworks, like the network or systems theory paradigms. Indeed, Ludwig van Bertalanffy defined general systems theory as an interdisciplinary doctrine “elaborating principles and models that apply to systems in general, irrespective of their particular kind, elements, and 'forces' involved” (Bertalanffy, quoted in (Laszlo 1972, xvii)). In a similar fashion, the study of networks is independent of the nodes and types of relations considered.

To conclude this section, one can use Hesse's (1966, 8) pragmatically valuable distinction between *positive*, *negative* and *neutral* analogies. The positive analogy addresses the question: *what is analogous?* and constitutes the set of relations which hold in the two domains. The negative analogy addresses the question: *what is disanalogous?* and constitutes the set of relations which do not hold in the two domains. Finally, neutral analogies trigger the question: *are the two domains analogous?* To answer this last question, one has to examine or test if such or such relation holds in the target domain.

Given this analysis, we can now carefully explore the fine-tuning issue, aided by computational and biological analogies. Since we have claimed in the first section that fine-tuning arguments could be in future reduced to initial conditions, we will now focus on this aspect.

# 3 The computational universe

The idea that our universe is analogous to a computer is a popular one. We can see it as the modern version of a mechanistic worldview, looking at the universe as a machine. There are various ways to consider this analogy, with cellular automata, e.g. (Wolfram 2002) with quantum computing e.g. (Lloyd 2005), etc. The analogy has been pushed so far that a modern version of idealism has even be considered, i.e. that our universe would actually be run by a computer, and we might be living in a computer simulation (e.g. Bostrom 2003; Martin 2006).

We saw that fine-tuning arguments might ultimately be reduced to initial conditions of a cosmological model. Here, we examine what the initial conditions are from a computational perspective, and discuss the relation between laws and initial conditions. This is conducted within the framework of Algorithmic Information Theory (AIT, (Chaitin 1974; 1987)). We will then argue that computer simulations provide an indispensable tool if we wish to tackle the fine-tuning problem scientifically. We will conclude by pointing out some limitations of this computational analogy.

AIT studies complexity measures on strings. The complexity measure -the Kolmogorov complexity[2]- of an object such as a piece of text is a measure of the computational resources needed to specify the object. Below is a simple example originally presented in the Wikipedia encyclopaedia (2008) :

> consider the following two strings of length 64, each containing only lower-case letters, numbers, and spaces:
>
> abababababababababababababababababababababababababababababababab
> 4c1j5b2p0cv4w1 8rx2y39umgw5q85s7ur qbjfdppa0q7nieieqe9noc4cvafzf
>
> The first string admits a short English language description, namely "ab 32 times", which consists of 11 characters. The second one has no obvious simple description (using the same character set) other than writing down the string itself, which has 64 characters.

In this AIT framework, laws represent information which can be greatly shortened by algorithmic compression (like the "ab 32 times" string); whereas initial conditions represent information which cannot be so compressed (like the second string). If we import this analogy into physics, a physical law is to be likened to a simple program able to give a compressed description of some aspects of the world; whereas initial conditions are data that we do not know how to compress.

Can we interpret this distinction between physical laws and initial conditions in a cognitive manner? We either express our knowledge in terms of laws if we can compress information, and in terms of initial conditions if we cannot. In this view, scientific progress allows us to dissolve initial conditions into new theories, by using more general and efficient algorithmic compression rules.

In mathematics, Gödel's limitation theorems state that in any sufficiently rich logical system, there will remain undecidable propositions *in that system*. But using another stronger system, one can decide such previously "undecidable" propositions (even if new undecidable propositions will arise in the stronger system...). For example, the consistency of Peano's arithmetic cannot be shown to be consistent within arithmetic, but can be shown to be consistent *relative* to modern set theory (ZFC).

There is a theorem similar to Gödel's incompleteness in AIT. Informally, it states that a computational system cannot compress structure in a system that is more algorithmically complex than this computational system. Let us assume again that physical laws represent compressible information, and initial conditions incompressible information. Are initial conditions in cosmological models algorithmically incompressible? There are two ways to answer this question.

First, we can interpret this incompressible data in an absolute way. This data is then "lawless, unstructured, patternless, not amenable to scientific study,

2 also known as program-size complexity, Kolmogorov-Chaitin complexity, descriptive complexity, stochastic complexity, or algorithmic entropy.

incompressible" (Chaitin 2006, 64). Suggesting that those initial conditions are incompressible implicitly implies that we, poor humans, will never be able to understand them. This attitude freezes scientific endeavour and thus has to be rejected. Limitation theorems are only valid within formal systems, because one needs the system to be completely formalized and specific formal tools to be able to prove them. Therefore, we should be extremely careful when exporting limitation theorems into other less formalized domains. Moreover, the history of science has shown that it is hazardous to fix boundaries on human understanding. Let us take the example of infinity, which was for many centuries thought to be understandable only by a God who is infinite, and not by finite humans. A rigorous theory of infinite numbers, constituting the foundations of modern mathematics, has finally been proposed by the mathematician Georg Cantor. Therefore, boundaries are likely to be broken. We will shortly see how the *multiverse hypothesis* or computer *universe simulation* bring us beyond the apparently incompressible initial conditions.

The second option is that incompressible information may reflect the limits of our theoretical models. If we are not able to account for the reasons of initial conditions, it is a hint that we need a broader theoretical framework to understand them. This situation can be illustrated by considering the problem of the origin of life. In this context, initial conditions for life to emerge are generally assumed without justification: chemical elements are assumed to be here, along with an Earth with water, neither too far nor too near from the Sun, etc. With these hypotheses (and others), we try to explain the origin of life. Now, what if we try to explain the origin of these initial suitable conditions for life? We would then need a broader theory, which in this case is a theory of cosmic evolution. If we then aim to explain initial conditions in cosmology, we are back to the problem of fine-tuning.

As we wrote in the introduction, multiverse models are precisely attempting to introduce a broader theory to explain those initial conditions, by proposing the existence of various other possible universes with different initial conditions. The problem is that the multiverse hypothesis is a *metaphysical assumption*. George Ellis (2007a, 400) expressed it well:

> There can be no direct evidence for the existence of other universes in a true multiverse, as there is no possibility of even an indirect causal connection. The universes are completely disjoint and nothing that happens in one can affect what happens in another. Since there can be no direct or indirect evidence for such systems, what weight does the claim for their existence carry? Experimental or observational testing requires some kind of causal connection between an object and an experimental apparatus, so that some characteristic of the object affects the output of the apparatus. But in a true multiverse, this is not possible. No scientific apparatus in one universe can be affected in any way by any object in another universe. The implication is that the supposed existence of true multiverses can only be a metaphysical assumption. It cannot be a part of science, because science involves experimental or observational tests to enable correction of

> wrong theories. However, no such tests are possible here because there is no relevant causal link.

To improve testability, Ellis further suggests examining a variation on the causally disconnected universes, considering multi-domain universes that are not causally disconnected. Still, I would like to emphasize the *philosophical* importance of the multiverse hypothesis, because it is a logically consistent way to tackle the fine-tuning problem. Is there a way other than the multiverse to theorize about "other possible universes"? This is what we will analyse now.

One of the main limitations of fine-tuning arguments is that the vast majority of them vary *one single* parameter of physical or cosmological models and conclude that the resulting universe is not fit for developing complexity. What if we would change *several* parameters at the same time? For example, if the expansion rate of the universe had been greater, and gravity had been stronger, could the two changes cancel each other out? Systematically exploring those possibilities seems like a very cumbersome enterprise. As Gribbin and Rees wrote (1991, 269):

> If we modify the value of one of the fundamental constants, something invariably goes wrong, leading to a universe that is inhospitable to life as we know it. When we adjust a second constant in an attempt to fix the problem(s), the result, generally, is to create three new problems for every one that we "solve". The conditions in our universe really do seem to be uniquely suitable for life forms like ourselves, and perhaps even for any form of organic complexity.

A way to overcome this problem would be to use *computer simulations* to test systematical modifications of constants' values. An early attempt of such an approach has been proposed by Victor Stenger (1995; 2000). He has performed a remarkable simulation of possible universes. He considers four fundamental constants, the strength of electromagnetism $\alpha$; the strong nuclear force $\alpha_s$, and the masses of the electron and the proton. He then analysed "100 universes in which the values of the four parameters were generated randomly from a range five orders of magnitude above to five orders of magnitude below their values in our universe, that is, over a total range of ten orders of magnitude" (Stenger 2000). The distribution of stellar lifetimes in those universes shows that most universes have stars that live long enough to allow stellar evolution and heavy elements nucleosynthesis. Anthony Aguirre (2001) reached a similar conclusion by proposing "a class of cosmologies (based on the classical 'cold big-bang' model) in which some or all of the cosmological parameters differ by orders of magnitude from the values they assume in the standard hot big-bang cosmology, without precluding in any obvious way the existence of intelligent life." In conclusion, other possible universes may also be fine-tuned!

It is certainly possible to critique Stenger's simulation as being too simplistic. Maybe we should consider other or more "fundamental" constants to vary; or to vary the laws of physics themselves. Stenger did not attempt to vary physical laws and it seems indeed a very difficult enterprise, because we do not even know how to make them vary (see Vaas 1998).

In fact, *the distinction between laws and boundary conditions is fuzzy in cosmology* (Ellis 2007b, sec. 7.1). One can see boundary conditions as imposing constraints, not only on initial conditions (lower boundary of the domain), but also at the extremes of the domain. Both physical laws and boundary conditions play the same role of imposing constraints on the system at hand. Because we can not re-run the tape of the universe, it is difficult -if not impossible- to distinguish the two. In this view, some laws of physics might be interpreted as regularities of interactions progressively emerging out of a more chaotic state. The cooling down of the universe would progressively give rise to more stable dynamical systems, which can be described by simple mathematical equations.

A similar situation occurs in computer science. One can distinguish between a program, which is a set of instructions, and the data on which the program operates. The program is analogous to physical laws, and the data to initial conditions. This distinction in computer science can be blurred, when considering self-modifying programs, i.e. a program which modifies itself. Also, at a lower level, both the program and the data are processed in the form of bits, and here also the distinction is blurred.

Back to Stenger's simulation, it does not answer the following questions: would other interesting complex structures like planetary systems, life, intelligence, etc. evolve in those other universes? However, this is only an early attempt in simulating other possible universes, and the enterprise is certainly worth pursuing. The fine-tuning problem could then be seriously tackled, because we would know precisely the likelihood of having our universe as it is, by comparing it to other possible universes. To summarize, this approach needs to:

(1) define a space of possible universes
(2) vary parameters defining this space, to see how likely it is to obtain a universe fine-tuned to develop complex structures.

Many multiverse scenarios such as those in (Carr 2007) proposed answers to step (1). The second proposed step can be tackled with computer simulations.

I argued in (Vidal 2008) that a simulation of an entire universe will result from future scientific activity. Such a simulation would enable us not only to understand our own universe (with "real-world modelling", or processes-as-we-know-them) but also other *possible* universes (with "artificial-world modelling", or processes-as-they-could-be). In this way, we would be able to scientifically assess to what degree our universe is fine-tuned or not. If it turns out that our universe is not so special, then a response to fine-tuning would be a principle of *fecundity*: "there is no fine-tuning, because intelligent life of some form will emerge under extremely varied circumstances" (Tegmark et al. 2006, 4).

We thus need to develop methods, concepts and simulation tools to explore the space of possible universes (the "cosmic landscape" as Leonard Susskin (2006) calls it in the framework of string theory). In (Vidal 2008), I proposed to call this new field of research "Artificial Cosmogenesis". It is an attempt to a "general cosmology", in analogy with Artificial Life which appeared with the help of computer simulations to enquiry about a "general biology".

In summary, if we assume that initial conditions are analogous to incompressible information, then there are two possible reactions. Either we claim that we reached the limit of scientific understanding; or we recognize that we need an extended framework. Multiverse and computer simulations of other possible universes are examples of such extended frameworks.

Let us now see some limits of this computational analogy. If we apply our careful analogical reasoning, we can ask "what is disanalogous between the functioning of our universe and that of a computer?". We can at least make the following restrictions. In a computational paradigm, space and time are assumed to be independent, and non-relativistic. Most of the well studied cellular automata even use only two spatial dimensions, which is of course a limitation for complexity to develop.

A fundamental difference between a physical and an informational-computational paradigm is that the former has at its core *conservation laws* such as the conservation of energy, where the total amount of energy remains unchanged in the various transformations of the system. By contrast, the bits manipulated by computers are not subjected to such conservation laws. We neither create nor destroy energy, whereas we easily create and delete files in our computers. This difference can be summarized by the adage: "When sharing energy, we divide it. When sharing information, we multiply it." (formula that I borrow from Pierre-Alain Cotnoir).

Another limit of this computational paradigm (which is similar to a Newtonian worldview) is that when we have initial conditions, and a set of rules or laws, then the evolution of the system is trivial and predictable: it is just an application of rules/laws to the initial conditions. There would not be much more to understand, as such a model would capture the whole of reality.

The complexity of interactions (such as synergies, feed-back loops, chaos, random errors, etc.) is not in the focus of this approach. The biological analogy is more appropriate in giving insights into the complexity of interactions. Embryologists know that the formation of a fetus is a process of an incredible and fascinating complexity, leading from one single cell, to the complexity of a billions-cells organism. The development of the individual is certainly not as easy to predict from the genome to the phenotype as was the case with the computational paradigm: we just needed the initial conditions and a set of rules to understand the dynamic. By contrast, in biology, phenomena of phenotypic plasticity have been identified, i.e. the acknowledgement that phenotypes are not uniquely determined by their genotype. This becomes particularly clear when considering genetically identical twins. They exhibit many identical features, but also a unique differentiation, due to more stochastic processes occurring during the development. As Martin Rees (2000, 21) noticed, cosmology deals with the inanimate world, which is in fact simpler than the realm of biology. A phenomenon is difficult to understand because it is complex, not because it has a huge extension.

# 4 The biological universe

The idea that our universe is similar to an organism has a long story, which is still very inspiring. It can be traced back to the Ancient Greece (see Barrow and Tipler 1986 for historical aspects). One general aim of the “Evo Devo Universe” research community is to explore how traditional cosmology can be enriched by introducing a biological paradigm, as suggested by George Ellis (2007b, Thesis H4). More specifically, the field of evolutionary developmental ("evo-devo") biology (e.g. Carroll 2005) provides a great source of inspiration, acknowledging both the contingency of evolutionary processes and the statistically predictable aspect of developmental processes. We will now focus our attention on what is analogous in biological systems regarding initial conditions.

Lee Smolin's Cosmological Natural Selection (CNS) hypothesis is an example of a biologically-inspired approach to the fine-tuning problem (Smolin 1992; 1997; 2007). One can introduce his theory with an (other!) analogy. The situation in contemporary cosmology is analogous to the one in biology before the theory of evolution, when one of the core questions was “*(1) Why are the different species as they are?”*. It was assumed more or less implicitly that “*(2) Species are timeless categories”*. In present physics, the question behind the fine-tuning problem is “*(1') Why are the physical constants as they are?”*. Currently, it is usually assumed (probably from the remains of the Newtonian worldview) that “*(2') Physical constants are timeless”*. It is by breaking assumption (2) that Darwin was able to theorize about the origin of species. Analogously, Smolin is trying to break assumption (2'), by theorizing about the origin of constants.

According to this natural selection of universes theory, black holes give birth to new universes by producing the equivalent of a big-bang, which produces a baby universe with slightly different constants. This introduces variation, while the differential success in self-reproduction of universes (via their black holes) provides the equivalent of natural selection. This leads to a Darwinian evolution of universes, whose constants are fine-tuned for black hole generation, a prediction that can in principle be verified.

One might be surprised by the speculative aspect of this theory. Although Smolin emphasizes the refutability of CNS and thus its scientific aspect in (Smolin 2007), he himself is not proud that the theory talks about processes outside the universe (Smolin 1997, 114). This conjectural aspect of the theory puts it at the edge of science and philosophy (Vaas 1998). Let us stress once again that when we attempt to answer the metaphysical question "why is the universe the way it is?", we must be ready to cross the border of current experimental and observational science. Attempting to answer this issue leads to the construction of *speculative* theories. The nature of the inquiry then becomes philosophical, because we aim at answering our most profound questions here and now, whatever their difficulty and our limited knowledge (Vidal 2007).

This non-scientific (but not un-scientific) aspect would at first sight be a reason to fall into an intellectual relativism, where every theory would have equal value, since anyway, there seems to be no way to *scientifically* compare competing speculations. This is correct, but it is still possible to compare speculations from a *philosophical* standpoint. I proposed a general philosophical framework and criteria to compare philosophical speculations (Vidal 2007; 2009b). In the following pages, we will intentionally address the problem from this speculative philosophical point of view, focusing our attention on the *objective criteria* we have identified for a good philosophical system. A philosophical system is better than an other, when, other things being equal:

(1) It has a better fit with all the natural sciences.

(2) It addresses and adequately resolves a broader range of philosophical questions.

(3) It exhibits greater internal and systemic coherence. It thus has fewer anomalies.

From this point of view, CNS has some limitations. Firstly, the roles of life and intelligence are incidental. Criterium (2) is then poorly satisfied because, without including life in the cosmic picture, this theory diminishes its philosophical scope and depth considerably. Indeed, the philosophical endeavour is mainly concerned with the relation between humanity and intelligence on the one hand and the universe on the other hand. Secondly, the theory does not propose a hereditary mechanism for universe replication (Gardner 2003, 84). Its internal coherence is thus imperfect and could be improved regarding criteria (3). As Gentner and Jeziorski proposed in the third principle of analogical reasoning (see the second section), one should seek for the greatest possible *systematicity* in the relational network. Is it possible to overcome these two shortcomings?

A few authors have dared to extend CNS by including intelligent life into the picture (Crane 2009; Harrison 1995; Gardner 2000; 2003; Baláz 2005; Smart 2008; Vidal 2008; Stewart 2009). They put forward the hypothesis that life and intelligence could perform this mechanism of heredity, thus playing an essential role in the Darwinian evolution of universes. To better grasp this extension of CNS, Baláz and Gardner proposed to consider von Neumann's (1951) four components of a self-reproducing automaton. We summarized this completion of CNS in table 1. below.

Let us describe these four components in more detail. Physical constants are analogous to DNA in biology, and to the *blueprint* of this self-reproducing automaton. The universe at large or the cell as a whole constitute the *factory*. When furnished with the description of another automaton in the form of a blueprint, the factory will construct this automaton. The *reproducer* reads the blueprint and produces a second blueprint. In the cell, these are reproduction mechanisms of the DNA. The *controller* will cause the reproducer to make a new blueprint, and cause the factory to produce the automaton described by the new blueprint. The controller separates the new construction from the factory, the reproducer and the controller. If this new construction is given a blueprint, it finally forms a new independent self-reproducing automaton.

| ***Components*** | ***Description*** | ***BIOLOGY (cell)*** | ***COSMOLOGY (universe)*** |
|---|---|---|---|
| Blueprint | Gives instructions for the construction of the automaton | Information contained in the DNA | Physical constants |
| Factory | Carries out the construction | Cell | The universe at large |
| Reproducer | Reads the blueprint and produces a second blueprint | The reproduction of the DNA | CNS:? **Intelligence unravelling the universe's blueprint** |
| Controller | Ensures the factory follows the blueprint | The regulatory mechanisms of the mitosis | CNS:? **A cosmic process, aiming at universe reproduction** |

Table 1. Components of a von Neumann’s (1951) self-reproducing automaton. The second column provides a general description of the automaton functions. The third and fourth columns propose examples respectively in biology - the cell - and in cosmology - the universe.

We now clearly see the limits of the CNS model, which is not specifying what the reproducer and controller are. Intelligence unravelling the universe's blueprint can precisely fulfil the reproducer's function. This reproducer component is indeed essential for providing a mechanism for heredity. Without heredity, there can be no Darwinian evolution. The controller in this context would be a more general process, aiming at universe reproduction with the help of intelligence. In Table 1, we completed in bold these two missing components of CNS, thus including intelligence in this hypothesized cosmic reproduction process.

Let us now explore in more depth the biological analogy of natural selection and apply it to cosmological evolution. Natural selection implies a differential reproduction with a high level of fidelity, still permitting some mutations. Is there a fidelity in reproduction of physical constants in CNS? If the fidelity were perfect, there would be no evolution, and thus no dynamic to fine tune the constants. If the fidelity were not perfect, it would imply that there are slight changes in the constants from one universe to another. Yet the whole point of fine-tuning arguments is to show that small changes in physical parameters do not lead to a viable universe! It thus seems very unlikely that a CNS would succeed in producing a fine-tuned universe.

A consequence of this speculative theory is that intelligent life, unravelling the universe through scientific understanding, generates a “cosmic blueprint” (a term used by Paul Davies (1989)). The cosmic blueprint can be seen as the set of physical constants; or just initial conditions of a cosmological model, if our previous reasoning holds. One can now throw a new light on the fact that cosmic evolution gave rise to scientific activity. In this view, the increasing modelling abilities of intelligent beings is not an accident, but an indispensable feature of our universe, to produce a new offspring universe.

I have argued that fine-tuning of this cosmic blueprint would take place in “virtual universes”, that is in simulated universes (Vidal 2008). This scenario is likely if we seriously consider and extrapolate the exponential increase of computer resources. As argued earlier, simulating universes would also allow virtual experimentation, and thus to compare our own universe with other possible ones. This endeavour would therefore help to progress concretely on the fine-tuning issue.

One can interpret this approach as a variation on the multiverse proposal. However, the selection of universes would take place on *simulated universes*, replacing Smolin's natural selection of *real universes* (Barrow 2001, 151). In CNS, we need many generations of universes in order to randomly generate an interesting fine-tuned universe. In contrast, these simulations would dramatically improve the process by artificially selecting (via simulations) which universe would exhibit the desired features for the next generation universe. This would facilitate the daunting task of producing a new universe. In this case it would be appropriate to speak about a “Cosmological *Artificial* Selection” (CAS), instead of a “Cosmological *Natural* Selection”.

One question which naturally arises when considering this CAS, is "why would intelligent life want to produce a new offspring universe?". This is an issue which would deserve much more development, but we briefly propose the following answer (see also (Harrison 1995; Vidal 2008)). The core problem an intelligent civilization has to deal with in the very far future is the inevitable thermodynamical decay of stars, solar systems, galaxies and finally of the universe itself. This scenario is commonly known as the *heat death* (e.g. Adams and Laughlin 1997; Ćirković 2003 for a review). Surprisingly there is little interest in this fundamental and inevitable ageing of the universe. In the realm of biology, the solution to the problem of aging is reproduction. Could it be that an analogous solution would take place at the scale of the universe? Pursuing this cosmic replication process would in principle enable the avoidance of heat death in a particular universe (Vidal 2008). Cosmic evolution would then continue its course indefinitely.

This approach may sound like science-fiction. It is however no more extravagant than the competing explanations, which are the belief in a supernatural God, or the ontological statement that there actually exist a huge number or an infinity of other universes. To summarize, the perspective of a CAS:

1. Proposes a response to the heat death problem.
2. Enables us to scientifically progress on the fine-tuning problem, via universe simulations. This remains acceptable even if one does not wish to endorse the complete speculative framework proposed, i.e. with intelligent life involved in universe reproduction.
3. Offers intelligent life a fundamental role in cosmic evolution.

# 5 Conclusion

Our analysis of physical parameters led us to the conclusion that fine-tuning of physical constants would progressively be reduced to the fine-tuning of initial conditions in a cosmological model. However, it is unlikely that a physical theory would derive *all* cosmological initial conditions. Multiverse and universe simulations are two similar options to go beyond this limitation, because they both propose ways to vary those initial conditions.

The computational analogy suggests a description of the universe in terms of information and computation. Simulations have already started to be used to tackle the fine-tuning problem, allowing us to vary *several* physical parameters, and exploring the resulting possible universes.

We outlined a *philosophical* extension of Smolin's "Cosmological Natural Selection" theory, inspired by a biological analogy. It led us to the hypothesis that intelligent life is currently unravelling a cosmic blueprint, which could be used in the far future to produce a new offspring universe. Simulating an entire universe would be an efficient way to fine tune this cosmic blueprint. Furthermore, this offspring universe would perpetuate cosmic evolution beyond the predictable heat death of our universe.

Physical, computational and biological descriptions are different ways to model the universe. Providing that we make clear and explicit our analogical reasoning, keeping this diversity of possible modelling approaches is certainly a way to constantly enrich and cross-fertilize our models. Importantly, we have to acknowledge the limits of scientific enquiry when tackling the fine-tuning problem. A philosophical approach allows us to keep a rational and critical attitude.

# 6 Acknowledgements

I thank Sean Devine, Börje Ekstig, Pushkar Ganesh Vaidya, Bernard Goossens, Francis Heylighen, Gerard Jagers op Akkerhuis, John Leslie, Robert Oldershaw, Marko Rodriguez and John Smart for useful and stimulating comments. Thanks to Sophie Heinrich and Jean Chaline for insightful criticisms about the biological analogy in cosmology. I warmly thank Luke Lloyd for his help in improving the English.

# On the nature of initial conditions and fundamental parameters in physics and cosmology

Jan M. Greben

CSIR, PO Box 395, Pretoria 0001 South Africa
E-mail: jgreben@csir.co.za

**Abstract:** The cosmological theory of the author, discussed elsewhere in this journal, has a number of implications for the interpretation of initial conditions and the fine-tuning problem as discussed by Vidal.

In his introduction, Vidal (2009) notices that many authors (see his references) have emphasized the importance of fine-tuning of the constants of physics and the initial conditions of the universe for the emergence of life. The concept of initial conditions is a typical classical idea as in classical physics the future “depends” uniquely on an earlier state, and therefore one can ask the question whether this initial state is fine-tuned. However, since nature is quantum-mechanical (QM), it is more relevant to consider the concept “initial state” in the QM context.

In QM the specification of the state is done via the state vector, which can be defined as a string of creation operators operating on the vacuum state. These operators specify the momentum and further quantum characteristics (e.g. spin) of each individual elementary particle. Transitions are then expressed by Feynman diagrams, which are expressible in the parameters of the standard model. Hence, the current state of the universe is a result of a string of such quantum transitions (such as the creation, annihilation or scattering of particles). However, in each transition many future states are allowed (generally an infinity of them), so our current reality is a consequence of these random quantum choices. If one uses the language of wave function collapse (Bell 1986), one can state that each choice is associated with a collapse of the wave function. The issues around wave function collapse are also referred to as the measurement problem. Hence, the current state vector neither allows us to specify our future, nor our past, in contrast to classical physics. The initial state thus plays a more special role than in classical physics (where it is equivalent in information content to the state specification at any other time). However, since there is no continuous and deterministic relationship between the initial state and the current state, it is also difficult to argue that the existence of life is due to the fine-tuning of the initial state. Apart from this, we think this question is largely academic. There are strong indications (Penrose 2008) that the initial universe had zero entropy. Hence, the initial state vector does not require any specific information and is thus not fine-tuned (unless one wants to call its trivial zero entropy status fine-tuned).

As Penrose states (Penrose 2008, p.162): the spatial isotropy and homogeneity of the initial universe is an indication that the gravitational degrees of freedom were not excited at the beginning of the universe. Technically this leads to the condition that the Weyl tensor equals zero and that the entropy is zero or minimal at the beginning. These are exactly the properties displayed by the cosmological model by the author (Greben 2009). In this theory, the initial state is a singular state, which only consists of classical vacuum energy, whose effective density becomes infinite in the big bang limit. No physical particles are present, so no particle states have to be specified and the information content of the big bang state is zero. The first physical particles are created in the particle epoch, which is characterized by the time $5 \times 10^{-24}$ seconds (see Eq. 58, Greben 2009). At that time entropy increases and the random quantum choices specify where and when particles are created. Hence, the initial conditions do not play a role in ensuring a universe where life can exist. However, the subsequent string of transitions specify the universe and therefore determine amongst others the nature of the universe and its contents. Vidal does not comment on the role of the collapse of the wave function. However, I think this is the quantum replacement of the initial conditions in classical physics, as a given history (controlled by random events) leads to a specific state, which functions as the initial state at that time.

We now make some comments on the fine-tuning of parameters. With time the number of free parameters in physics has reduced. However, the number of free parameters in the standard model (usually quoted as 21) suggests that this is still an effective theory, resulting from a more fundamental theory with fewer parameters. Often the fine-tuning discussions refer to the values of the nuclear masses (Davies 1982). However, these masses follow in principle from the standard model. Hence, any such fine-tuning must be referred back to more fundamental parameters of the standard model. Similarly, any fine-tuning of the parameters of the standard model must be explained in terms of the parameters of a more fundamental model (if it exists). Hence, the question is: what is the minimal number of free parameters that characterize nature? The fewer the number of independent parameters, the more doubtful it becomes to talk about fine-tuning. In our paper (Greben 2009) we indicate how the particle physics scale is related to the fundamental cosmological parameters $G$ and $\varepsilon$, the latter being the vacuum energy density. The relevant mass scale is given by $(\varepsilon / G)^{1/6} \approx 125$ MeV. This suggests that particle mass parameters in the standard model can be expressed in terms of cosmological parameters (notice that $\varepsilon$ is related to the Hubble constant via $\varepsilon = 3H_0^2 / 8\pi G$). Assuming one can demonstrate this for both quark and lepton masses, the fine-tuning problem is enormously reduced. This would undermine the fine-tuning ansatz for dimensionfull constants. Another coincidence - the closeness of the density of the universe to the critical density - is explained naturally in the author’s model (Greben 2009). This removes another fine-tuning problem. Hence, the author feels that it may well be premature to take recourse to the anthropic principle and to multiverse models to explain apparent fine-tuning properties, as it still an open question whether fine-tuning exists.

In conclusion, we disagree with Vidal that fine-tuning of physical constants would progressively be reduced to the fine-tuning of initial conditions, and rather would dismiss the importance of fine-tuning for initial conditions. We suspect that the apparent fine-tuning of parameters will be resolved by a deeper understanding of the laws of nature.

# Cosmological Artificial Selection: Creation out of something?

Rüdiger Vaas

Center for Philosophy and Foundations of Science, University of Giessen, Germany
Posener Str. 85, D – 74321 Bietigheim-Biss., Germany
Ruediger.Vaas@t-online.de

**Abstract:** According to the scenario of cosmological artificial selection (CAS) and artificial cosmogenesis, our universe was created and possibly even fine-tuned by cosmic engineers in another universe. This approach shall be compared to other explanations, and some far-reaching problems of it shall be discussed.

Why are the laws and constants of nature as well as the boundary conditions as they are? And why do they appear to be special – even fine-tuned for life? If they would be only slightly different, complex information-processing structures – and, hence, intelligent observers – probably could not have developed (e.g. Leslie 1989, Vaas 2004a, Carr 2007, Tegmark 2010). According to the scenario of *cosmological artificial selection* (CAS) and *artificial cosmogenesis* (Vidal 2008 & 2010) the fine-tuning of our universe is the result of an – at least partly – intentional action: a creation of our universe as (or via) an extensive computer simulation or as a physical experiment by cosmic engineers in another universe. This is a bold speculation, but it is not unscientific in principle. So what are the arguments? Are there better alternatives? And what problems have CAS proponents to solve?

There are many options for explaining the fine-tunings (Vaas 2004a & 2009). They might (1) just be an *illusion* if life could adapt at very different conditions or if modifications of many values of the constants would compensate each other; or (2) they might be a result of (incomprehensible, irreducible) *chance*, thus inexplicable; or (3) they might be *nonexistent* because nature could not have been otherwise, and with a *fundamental theory* we will be able to prove this; or (4) they might be a product of *selection*: either *observational selection* within a vast multiverse of (infinitely?) many different realizations of those values (*weak anthropic principle*), or a kind of *cosmological natural selection* making the measured values quite likely within a multiverse of many different values, or even an *teleological or intentional selection*. CAS is one example of the latter – but there are other alternatives, for instance some versions of the strong anthropic principle (i.e. an impersonal teleological force or a transcendent designer) including deistic or theistic creation. Even worse, those alternatives are not mutually exclusive – for example it is logically possible that there is a multiverse, created according to a fundamental theory by a cosmic designer who is not self-sustaining, but ultimately contingent, i.e. an instance of chance.

A fundamental theory has its explanatory limits and cannot ultimately get rid of some fine-tunings if it has empirical content at all, and transcendent causation is beyond physics and runs into deep philosophical problems (Vaas 2004a & 2006a). Thus a multiverse scenario is most promising (Vaas 2004a & 2004b & 2010a). Here, CAS (which does not exclude a fundamental theory) competes with two much simpler kinds of multiverse scenarios: (1) those with an equal or random distribution of laws, boundary conditions and/or values of constants, and (2) those with a probability distribution of some kind.

Within the first scenario, the fine-tunings can be understood by observational selection: We are able to live only in a specific universe whose physical properties are consistent with our existence – a prerequisite for it –, and therefore we should not be surprised that we observe them. But this is, strictly speaking, not a physical explanation (Vaas 2004a), and it might not even be testable, i.e. predictive (Smolin 2010). So one might claim that CAS is superior.

Within the second scenario, the fine-tunings are the result of cosmological natural selection (Smolin 1992, 1997 & 2010). This could be seen as both the nearest relative of and the strongest alternative to CAS. It seems to be testable, but CAS proponents might still argue that it does not explain enough. Furthermore, it has many problems of its own (e.g. Vaas 1998 & 2003). But this is also the case for CAS (Vaas 2009):

First and foremost, there is the problem of *realizing* CAS: It is completely unclear whether universes can not only be simulated to some extent but also physically instantiated. A few scientific proposals are already on stage (see Ansoldi & Guendelman 2006 for a review), but still in their infancy.

Second, one must distinguish between intentional *creation* and *simulation* (even if it would be empirically impossible to decide between them from "within"). A simulated universe does not have all the properties of a physically real universe – as a simulated river might obey hydrodynamical equations but doesn't make anything wet. Admittedly, deep epistemological problems are lurking here. And perhaps it will be possible to make the simulation so real that the missing properties are simply irrelevant; or to make it at least so useful that, for instance, conscious life within it is possible and the creators could "upload" their minds, knowledge and experiences. But the hardware problem remains: How can something simulate something else which is comparably complex? And if the programmer's universe is doomed, their universal computer and, hence, computed simulation sooner or later should be doomed too.

On the other hand CAS might be praised for stressing that in principle design is – although not mandatory of course – at least possible within a cosmological and naturalistic framework. In contrast to theistic postulates of transcendent, nonphysical entities and causation, a CAS scenario is fully reconcilable with *ontological naturalism*. Cosmic engineers are not something "above" or "beyond" nature (i.e. the multiverse), but a part and, indeed, a product of it. However, this implies a further problem:

If there are cosmic engineers at work, perhaps some of them having fine-tuned our universe, how did they emerge in the first place? In other words: *What or who created the creator(s)?*

To avoid an infinite explanatory regress, it seems most probable that they arose naturally in a life-friendly universe themselves. Then at least the creator's home universe should have formed without any intentional fine-tuning. Thus, either its origin was pure chance or the outcome of cosmological natural selection or the result of a multiverse "generator" according to some fundamental laws etc. Therefore the CAS scenario just shifts the fine-tuning problem to the problem of explaining an earlier fine-tuned universe where the cosmic engineers evolved. This is a major objection against CAS.

Furthermore one might wonder whether CAS has any convincing explanatory force at all. Because ultimately CAS tries to explain something complex (our universe) with something even more complex (cosmic engineers and engineering). But the usual explanatory scheme is just the converse: The *explanans* should be simpler than the *explanandum*. Furthermore, CAS adds something qualitatively new: While multiverse and fundamental law scenarios postulate some new nomological regularities, CAS postulates an *intentional* cause *in addition*. CAS is therefore a mixture of explanations: physical and intentional. Intentions are not, as such, nonphysical, and actions can be conceptualized as specific causes (Davidson 2001), so there is no reason to abandon naturalism here; but intentional explanations are nevertheless not epistemologically reducible to physical explanations. Without further knowledge however it is impossible to tell anything about the intentions of the creator(s), and an intentional explanation without explaining the intention might be considered as a shortcoming. But speculations are possible, e.g. scientific curiosity, a truly universal entertainment, or building a rescue universe if the engineer's one is dying (Vaas 2006b & 2010b).

These arguments are not a knock-out objection. And if it would be possible for us to carry out artificial cosmogenesis by ourselves, a strong case for CAS can be made even within its explanatory restrictions. If so, it seems quite unlikely, however, that our universe is the very first one to accomplish artificial cosmogenesis. This would violate the Copernican principle because our location in spacetime, in this case the multiverse, would be very special. So contrary to the problems sketched above, our universe might indeed be a result of cosmological artificial selection – one link within the probably future-eternal chain. Then the spark of life may endure endlessly. And even if we or our successors would not be able to pass it on, being a tiny dead end within a flourishing realm of evolution, we will at least have envisioned it.

# Fine-tuning, Quantum Mechanics and Cosmological Artificial Selection

Clément Vidal
*Department of Philosophy, Vrije Universiteit Brussel, Center Leo Apostel, and Evolution, Complexity and Cognition, Brussels, Belgium.*
clement.vidal@philosophons.com http://clement.vidal.philosophons.com

**Abstract**: First, Greben criticized the idea of fine-tuning by taking seriously the idea that "nature is quantum mechanical". I argue that this quantum view is limited, and that fine-tuning is real, in the sense that our current physical models require fine-tuning. Second, I examine and clarify many difficult and fundamental issues raised by Vaas' comments on Cosmological Artificial Selection.

## Contents

## 1 Reply to Greben

Greben disagrees with my proposal that fine-tuning of physical constants would progressively be reduced to initial conditions. He would rather "dismiss the importance of fine-tuning for initial conditions" and "suspect that the apparent fine-tuning of parameters will be resolved by a deeper understanding of the laws of nature". Greben's position on the fine-tuning problem is perhaps the most commonly held position among physicists. It is thus important to discuss this perspective.

## 1.1 Initial conditions and quantum mechanics

The idea of "initial conditions" comes from a classical and deterministic (Newtonian, Laplacian) view of physics. I underline this point in my paper as a limitation of a purely physical approach, to motivate the exploration of a more biologically inspired analysis. Emphasizing the importance of quantum mechanics (QM), Greben takes another road, and analyses initial conditions from a quantum perspective.

We should note that the use of QM for understanding the origin of the universe is arguably one of the most difficult subject in theoretical physics. Yet we have to admit that introducing ideas from QM is necessary. This is due to the fact that physics in the big bang era involves very high energies (and very small scales), which are best described by QM. Given the difficulties of the unresolved issues involved, I hope that this discussion will not be too confusing for the reader.

I suspect Greben and I hold very different epistemological positions regarding physical theories. It might seem like a point of detail, but it is not. The expression "nature is quantum-mechanical" sounds like a strong ontological statement to me. I would rather say something like: "at the smallest known scales, QM provide the most accurate models". This divergence will be reflected implicitly in my reply below.

How important is QM really? QM is not all of physics, and has a limited domain of application. For example, although the evolution of stars involves many quantum phenomena, it is predictable with astrophysical models (e.g. with the Hertzsprung-Russell diagram). So, it is not because at the smallest scales we use QM, that QM can explain the rest of physics. On larger scales (i.e. non quantum), we have other models, offering predictions and explanations. Connecting the different scales in nature (from quantum to cosmological) is a major challenge in physics, on which Scale Relativity (Nottale 2009) already made impressive progress.

Today, by combining different insights in different branches of physics, we can reconstruct the story of the universe, despite all the uncertainty involved in quantum processes. So, it is certainly possible to connect the initial state with our current state, even if there is randomness in quantum processes. Ehrenfest's theorem shows for example how average quantum values evolve deterministically. The key is not to focus only on one physical scale and approach.

I found Greben's idea that initial conditions are incrementally specified from a zero entropy state quite fascinating. I do not know whether it is correct or not. I simply hope that we will soon be able to test this theory with (potentially refutable) observations.

However, we can note that Greben's interpretation of QM in terms of Feynman diagrams is limited. Indeed, Feynman's diagrams are efficient for scattering theory, but when bound states are involved, this is no longer the case. It is also worth noting that Schrödinger's equation, describing the evolution of the wave function through time, is deterministic.

Greben also wonders why I did not speak about the collapse of the wave function at the level of the universe. This idea implies that we use the measuring formalism of QM... at the level of the universe. Yet, this is deeply problematic. Who measures the universe? To make a measurement, we need an observer, so it is very difficult to make sense of this suggestion. Which observer could have existed at the big bang era?

To sum up, I would like to emphasize that Greben's conclusion that "it is difficult to argue that the existence of life is due to the fine-tuning of the initial state" only holds if we restrict our analysis to this QM approach ("nature is quantum-mechanical"), and that we endorse the interpretation of QM he presents.

## 1.2 Entropy of the universe

Did the initial universe have zero entropy? Greben says it is the case. He quotes a paper by Penrose to support this claim. However, I would like to put Penrose's reasoning in context. In his article, Penrose actually describes an important *paradox* in cosmology. If we draw the spectrum of the observed microwave background radiation (MBR), it turns out to be very close to the Planck spectrum. This Planck spectrum is a typical example of thermal equilibrium, i.e. a state with *maximum entropy*. However, following the second law of thermodynamics, if entropy can only increase, going back to the initial state of our universe, it should be at a *minimum entropy*. Hence the paradox*, was the initial entropy of the universe maximum or minimum?*

The answer, writes Penrose, is that the MBR is to be interpreted as a thermal equilibrium between radiation and matter. We can also note that the MBR is not primordial from the point of view of the origin of the universe, since the universe is already about 400 000 years old at that time. So, its maximal entropy could be interpreted as a final state. Anyway, since we do not have a theory of quantum gravity, the entropy related to the gravitational field remains an open question.

### 1.3 Fine-tuning

The reader might wonder why I speak about 31 free parameters, whereas Greben writes there are only 21. In fact, this divergence is not that important, since it is merely a matter of convention, or, more precisely, of "compromise balancing simplicity of expressing the fundamental laws [...] and ease of measurement." (Tegmark et al. 2006)

I agree with Greben about the attitude and the scientific fruitfulness of always searching for better theories, further explaining fine-tuning. I agree that as a general trend, the number of free parameters has reduced. I also agree that if particle mass parameters in the standard model can be expressed with the help of cosmological parameters, this is an indication that we progress with our physical models.

However, Greben concludes that it is still an "open question whether fine-tuning exists". I disagree with this statement. Fine-tuning does exist as a characteristic of our current physical and cosmological models. There are these free parameters that today remain unexplained theoretically. We can give them any value, and they remain compatible with the theory. To choose the right values, we make experiments or observations and fill them in in the model. Finding other ways to deduce those values is still a major problem in modern physics (Smolin 2006, 13). Furthermore, many of these parameters are sensitive to slight changes, which have important consequences for the evolution of the universe. It is in that sense that we can say that they are fine-tuned. The open question is not whether there is fine-tuning, but whether we will be able in the future to explain all these parameters from a more fundamental theory. A way to avoid such confusion is to reformulate statements like "the universe is fine-tuned" to "our current physical models require fine-tuning".

### 1.4 Conclusion on Greben's Comment.

I agree with Greben that fine-tuning should not be seen as an unexplainable "miracle". It is indeed fundamental to leave space for scientific progress on this issue, continuing to reduce the number of free parameters. I conjecture that this progress will stop at the level of initial conditions of a cosmological model. Greben conjectures that fine-tuning will be resolved by a deeper understanding of the "laws of nature". It is interesting to note that in this later case, the problem would not be numerical, but of structural symmetries concerning physical laws (Greben, private communication).

Whatever the most advanced resulting theories on fine-tuning will be (at the level of laws, universal structures, initial or boundary conditions), what could an intelligent civilization do with such knowledge? This invites us to ponder about the more speculative part of my paper, Cosmological Artificial Selection (section 4), about which Rüdiger Vaas points out the most challenging issues.

## 2 Reply to Vaas

Responses to fine-tuning like chance, God, multiverse, a fundamental theory, Cosmological Natural Selection (CNS), Cosmological Artificial Selection (CAS) or a combination of them might well be correct. However, when dealing with such a difficult topic, we have to acknowledge the limit of the enterprise. For example, there are observational limits regarding the early universe, i.e. data we will never get. Neither science nor philosophy will bring us certain answers and this is an invitation to humility, honesty, and carefulness when approaching those ultimate questions in cosmology (see also Ellis 2007, 1259; Vaas 2003, section 7)

Speculating on those issues can easily lead very far. For this reason, we need to have clear ideas on *how* and *why* we speculate. I distinguish three kinds of speculations to navigate into their variety (Vidal 2009a):

1. **Scientific**: a speculation is scientific if we have strong reasons to think that we will be able in the future to have observations or experimentations corroborating or refuting it.
2. **Philosophical**: a speculation is philosophical if it extrapolates from scientific knowledge and philosophical principles to answer some fundamental philosophical questions.
3. **Fictional**: a speculation is fictional if it extends beyond scientific and philosophic speculations.

In my papers (2008; 2009b) I have presented CAS as happening in the future. In the end, considering these metaphysical uncertainties, I think it is more fruitful to try to contribute to shape the future than to understand the past. However, the full CAS scenario also concerns the origin of the universe and the meaning of intelligent life in it. What is more, this full scenario also presents epistemological, logical, and metaphysical difficulties that I did not address in my previous papers. I would like to thank Vaas for bringing these question to the front, because they indeed deserve attention.

I will show that the difficulties raised by Vaas can largely be weakened and put into perspective. I strat by inviting to carefullness when using terms such as "creation" and "design". Then I invoke a principle of rational economy to tackle fine-tuning and propose to explicit the scope of the inquiry regarding cosmological issues. I discuss if universes can be simulated and instantiated. Importantly, I distinguish fine-tuning from other related metaphysical issues.

Finally, I propose a possible metaphysical framework to approach the question "how did the cosmic engineers emerge in the first place?".

Obviously, given the number and depth of the issues raised, I will straightforwardly focus on urgent remedial, and develop in details the full picture and arguments in later works. I hope this response will help to clarify the scope and even beauty implied by CAS.

### 2.1 Vocabulary confinement

First, I would like to forcefully stress that the whole CAS scenario is naturalistic, and, as Vaas notices, is fully compatible with ontological naturalism. This is why I would rather be careful with the term "create", because it generally supposes to get something out of nothing, whereas here it is question of a "creation out of something"...

Since Vaas (2010) speaks about "design" when discussing CAS, it is important to notice that scientists loath the term "design" in explanatory contexts. Why? Because it freezes scientific explanation. If we are confronted to something to explain, the "design" answer, like the god of the gaps explanation, will at the same time always work, and *explain everything* (nay, rather nothing). In fact, the intentional explanatory mechanisms involved in CAS do not interfere at all with normal scientific explanation. On the contrary, maybe surprisingly, CAS as a whole can be seen as an invitation to a fantastic scientific and technological challenge.

### 2.2 Rational economy and scope of the inquiry

Vaas describes four major responses to fine-tuning. From a logical point of view, he also correctly points out that they are not mutually exclusive. However, each of them was developed to be an independent and sufficient response. Since all of them are speculative, what benefit do we gain by combining them? Taking seriously those combinations as explanations resembles more *fictional speculation* than anything else. One might argue that such an attempt would please at the same time proponents of the different options, but even this is not certain. A principle of rational economy should be at play here, e.g. :

> Never employ extraordinary means to achieve purposes you can realize by ordinary ones (Rescher 2006, 8)

Is CAS simpler or more complicated than other multiverse scenario? Vaas writes that CAS "competes with two much simpler kinds of multiverse scenarios". First, I agree with Vaas that scenario (1), an observational selection effect, is not a physical explanation, and therefore that CAS does not compete with it. Concerning scenario (2), I disagree that CNS is "simpler" than CAS. Since this claim might seem surprising, we first must make an important epistemological remark about the concept of simplicity. It is well known that simplicity is very hard to define, and specialists consider it to be subjective (Heylighen 1997), or largely context dependant (Kuhn 1977; McMullin 2008). So we need to make explicit the cosmological context at play here, or the scope of the inquiry, as Ellis (2007, 1245) already suggested. The scope we discuss here concerns four fundamental problems:

**(1) Why do the laws of physics, constants and boundary conditions have the form they do?**

As we saw in details in the main article, this question mainly concerns the fine-tuning problem.

**(2) Why not nothing?**

This is certainly one of the deepest metaphysical question. The formulation here is a shorter version proposed by Apostel (1999) of Leibniz' "why is there something rather than nothing?".

**(3) What is the meaning of the existence of intelligent life in the universe?**

This question asks about the *meaning* of intelligence in the universe. Are life and intelligence merely epiphenomena in cosmic evolution? As Davies (1999, 246) formulates it, why did the laws of the universe engineere their own comprehension?

**(4) How can intelligent life survive indefinitely?**

The future of the universe is gloomy. Physical eschatology (see Ćirković 2003 for a review) teaches us that none of the known scenario seem to allow indefinite continuation of life and information processing in the very long term (Vaas 2006).

These four questions are more philosophical than scientific. Another way to put it is to see CNS as a *scientific speculation*, tackling question (1), whereas CAS is a *philosophical speculation*, tackling questions (1), (3) and (4). Question (2) has a special status, because it is metaphysical. Looking at question (1) alone, CAS is indeed not a simple explanation at all, and CNS is much better. However, CAS is ultimately busy with those three (or four) questions *together*. I insisted strenuously in my papers (2008; 2009b) that CAS shall foremost be seen a speculative *philosophical* scenario, precisely because of its more ambitious and comprehensive scope than CNS. In one sentence, *CAS is a speculative philosophical scenario to understand the origin and future of the universe, including a role for intelligence*.

As Vaas notices, since CAS is a close relative of CNS, it might also have scientific aspects, but they are rather peripheral and I will not discuss or insist on them here. With these preliminary considerations, we can now dive into the problems of CAS.

### 2.3 Can universes be simulated and instantiated?

Vaas asks whether CAS can be realized. The two underlying questions are:

(a) Can a universe be simulated?

(b) Can a universe be instantiated?

Those two questions are major challenges, and efforts to answer them are still in their infancy. The coming discipline of Artificial Cosmogenesis (ACosm) is meant to tackle those challenges explicitly. As in Artificial Life (ALife), ACosm can be divided in two efforts, soft ACosm and hard ACosm.

Soft ACosm consists in making computer simulations of other possible universes and is therefore busy with question (a). Performing simulations for this purpose does not require to simulate every detail. A simplified simulation to understand the general principles of cosmogenesis would in principle suffice. It is useful here to remember the metaphor of artificial selection in biology, where the ones performing it do not need to understand the whole process. It is enough that they know how to foster traits over others. Cosmologists have already started to simulate and model other possible universes (this is implicit in multiverse models. See also e.g. Stenger 2000; Aguirre 2001).

Hard ACosm consists in producing universes in the real world (question (b)). As Vaas mentions, there are already early attempts in this direction. This is a challenge which is probably orders of magnitudes more difficult than soft ACosm, but not impossible. Black holes are good candidate for this realization. First, from a physical point of view, they exhibit enormous densities (similar to the big bang singularity), which are susceptible to modify the structure of space-time and give birth to baby universes. Second, from a computational point of view, Lloyd (2000) argued that the ultimate computing device would be a black hole.

So one might speculate that, in the very far future, the hypothetical use of black hole computers will meet with universe production. For other hints towards this realization, see also (Gardner 2003, 167-173; Vidal 2008; Smart 2008).

### 2.4 Are we in a simulation?

I totally agree with Vaas' critique of the idea that we are in a computer simulation. The problem “how can something simulate something else which is comparably complex?" is indeed highly problematic. This is why in Soft ACosm we do not have to simulate every detail. As Vaas acknowledges, to run computer simulations, we need some hardware, and to make it continue indefinitely in our universe seems very difficult, if not impossible. In the end, I consider the question of whether we are in a simulation or not, as a fictional speculation which therefore does not deserve that much attention.

### 2.5 Fine-tuning and related metaphysical issues

"Who created the creators?" is typically a metaphysical problem. Whatever reply X will be to this question, we can always ask: "where does X which created the creator come from?". In the last analysis, whatever reply we provide, at least the metaphysical question (2), "why not nothing?" will remain. *These questions are metaphysical and should not be confused with the fine-tuning problem* (Harrison 1998). The fine-tuning problem is concerned with question (1) above, and "who created the creator?" is of a metaphysical nature, like question (2).

If we take into account this distinction, then it follows that *no response to fine-tuning escapes metaphysical issues*. Indeed, *if we could prove that there is indeed a fundamental theory*, we could still wonder why it is here, and why it gave rise to our universe, rather than having just nothing. *If we could prove that there is indeed a God,* we run into the same problem of "who created the creator?". *If we could prove that there is indeed a multiverse,* we must answer: how did the multiverse start in the first place? Where do the generative mechanism, like the space of possible universes and the variation of this space from one universe to another come from?

In conclusion, to properly respond to "who created the creator?" in the framework of CAS, as with other options, we need to develop a metaphysics (see the last section).

### 2.6 Who created the creators?

In the complete CAS scenario, we need to assume an intentional cause, which is not present in other naturalistic scenarios. However, it is still logically possible to assume that CAS will happen in the future, but did not happen in the past. In that case, there would be no intentional cause involved to explain the origin of the universe. If we consider CAS *as only valid in the future*, it is perfectly possible to hold the following logically consistent positions:

(a) God as the first cosmic engineer and CAS

(b) Fundamental theory and CAS

(c) Multiverse and CAS

(d) Skepticism and CAS

A religious person might go with (a), a scientist might like (b) or (c). The skeptic (d) might say that we should stop arguing about the origin of the universe, since anyway it is unlikely that we get unambiguous support for such or such option. Still, he could agree that CAS is an interesting prospect for the future of intelligence in the universe. Those four options would still allow intelligent life to take up the future of their universe.

However, as Vaas also remarks, considering such options would violate the Copernican principle. So, how can we respond? Following the Copernican principle and being faithful to the principle of rational economy afore mentioned against the combination of explanations to fine-tuning, what could the scenario (e) “CAS and CAS” be?

### 2.7 Towards a cyclical metaphysics?

Vaas points out that “CAS tries to explain something complex with something even more complex”. This critique was also made by Byl (1996). It is indeed correct, but the underlying fundamental problem is that the usual explanatory scheme Vaas describes does not hold when we bring a kind of “ultimate theory” at play (Rescher 2000). By ultimate theory, I do not necessarily mean a “theory of everything” like it is sometimes speculated in physics; but a general “all encompassing” scheme of explanation (if one allows such a speculative attempt). Accordingly, the explanatory scheme of CAS is not usual, but comparing the scope of classical explanations and CAS, we can argue that the *explanatory force* of CAS is much wider (see also Vaas 2010 where Vaas acknowledges this broad view on CAS). We can now summarize three levels to interpret CAS, where each level includes the precedent:

**(i) CAS in the future**

This is the scenario I have described in my papers (2008; 2009b), which provides : a response to heat death, a guarantee to scientifically progress on the fine-tuning issue, a role for intelligent life in cosmic evolution. For what happened in the past, positions (a)-(d) are all logically possible options.

**(ii) CAS in the future and in the past**

This is scenario chooses option (e) "CAS with CAS" to tackle the origin of the universe. This implies that our universe has been produced and fine-tuned by an intelligent civilization.

**(iii) CAS in the future, past and a metaphysics**

However, position (ii) implies further metaphysical problems. A metaphysics for CAS is needed to avoid a shift of the fine-tuning problem, and to propose a framework to answer metaphysical questions like “who created the creators?” or “why not nothing?”. We attempt in the following lines the sketch of one such possible framework. Others are also certainly possible.

Although it is at odds with our knowledge of cosmic evolution, to avoid a shift of the fine-tuning problem, one can suppose that the tuning of new universes is *not* enhanced as the universal reproduction cycle repeats. Indeed, if we assume a complexification between universes, we will automatically shift the fine-tuning problem (see also in this volume Vidal 2009a). In addition, we must assume that there is no “first universe”. This might sound strange for we are used to think in a linear way, with a beginning, a middle and an end. However, it is possible to postulate a cyclical metaphysics, where there is no beginning at all, only an indefinitely repeating cycle. To sum up, in this metaphysical picture (iii), CAS describes an indefinite cycle of self-reproductive universes, mediated by intelligent civilization.

It is also important to emphasize that circular explanations and infinite regresses are not necessarily vicious (Gratton 1994). One attributes viciousness to such reasoning, but this is based on the assumption that “there is some obligation to begin a beginningless process or to end some endless process” (Gratton 1994, 295). To sum up, instead of trying to avoid an infinite explanatory regress, we can choose to embrace it, without any contradiction.

### 2.8 Conclusion on Vaas' comment

To conclude, in contrast to what Vaas wrote, I would like to stress that CAS is more than a “physical experiment”, “a simulation” or an attempt to build a “rescue universe”. The response of an intelligent civilization to the awakening that their particular universe is doomed (heat death or another gloomy scenario) is likely to be a strong driver to produce a new universe. Therefore, CAS is not about playing with virtual universes, nor making a physical experiment to see what it is like to produce a universe. The “rescue universe” idea is interesting, although it would be more about rescuing the evolution of the cosmos at large, rather than the memory of a particular intelligent civilization.

I do not see evidence for a God, a fundamental theory coming soon or proofs of a multiverse actually realized. Yet, I see overwhelming evidence of our exponential use of computer resources, both memories and computational power (Moore's law). Those advances have a tremendous impact on our lives and societies, and this will continue. In particular, computers are more and more ubiquitous in scientific activities, for example in maths (to assist in proofs), in studying complex systems (by simulating them), in biology (e.g. with biotechnologies, and their databases of genomes, protein networks, etc.), in cosmology (with many projects of large-scale simulations of the universe) and of course with ALife and its legitimate successor, ACosm. If we choose and manage to successfully conduct soft and hard ACosm, (i) *CAS in the future* would be realized. It would then give us strong indications and inspirations to think that broader interpretations of CAS, (ii) or (iii) are accurate.

# THE MEANING OF LIFE IN A DEVELOPING UNIVERSE

***John Stewart***
Member of the Evolution, Complexity and Cognition Research Group, The Free University of Brussels
Address: 720/181 Exhibition Street, Melbourne 3000 Australia
john.stewart@evolutionarymanifesto.com
Phone: 61 3 96637959

***Abstract:*** The evolution of life on Earth has produced an organism that is beginning to model and understand its own evolution and the possible future evolution of life in the universe. These models and associated evidence show that evolution on Earth has a trajectory. The scale over which living processes are organized cooperatively has increased progressively, as has its evolvability. Recent theoretical advances raise the possibility that this trajectory is itself part of a wider developmental process. According to these theories, the developmental process has been shaped by a yet larger evolutionary dynamic that involves the reproduction of universes. This evolutionary dynamic has tuned the key parameters of the universe to increase the likelihood that life will emerge and produce outcomes that are successful in the larger process (e.g. a key outcome may be to produce life and intelligence that intentionally reproduces the universe and tunes the parameters of 'offspring' universes). Theory suggests that when life emerges on a planet, it moves along this trajectory of its own accord. However, at a particular point evolution will continue to advance only if organisms emerge that decide to advance the developmental process intentionally. The organisms must be prepared to make this commitment even though the ultimate nature and destination of the process is uncertain, and may forever remain unknown. Organisms that complete this transition to intentional evolution will drive the further development of life and intelligence in the universe. Humanity's increasing understanding of the evolution of life in the universe is rapidly bringing it to the threshold of this major evolutionary transition.

***The Meaning of Life***

## 1. Introduction

Until recently, a scientific understanding of the natural world has failed to provide humanity with a larger meaning and purpose for its existence. In fact, a scientific worldview has often been taken to imply that the emergence of humanity was an accident in a universe that is completely indifferent to human concerns, goals, and values (e.g. see Weinberg, 1993).

Humanity has had to supplement a naturalistic understanding with beliefs in supernatural beings and processes if it wanted a worldview that includes a meaningful role for humanity in a larger scheme of things.

But recent advances in evolutionary science are beginning to change this. In particular, we are rapidly improving our understanding of the evolutionary processes that have produced life on Earth and that will determine the future evolution of life in the universe. While it is far too early to tell with certainty, it is possible that the universe and the evolution of life within it have been shaped by yet larger evolutionary processes to perform particular functions that are relevant to these larger processes.

If this proves to be the case, then these functions have a purpose in the same sense that the functions performed by our eyes have a purpose in the larger evolutionary processes that have shaped humanity.

This paper explores some key implications for humanity of the larger-scale evolutionary and developmental processes that appear to operate in the universe and beyond. In particular, the paper shows that humanity has a role to play in these processes. It also shows that the success of the processes depends critically on humanity (and other organisms that reach a similar stage in evolution) understanding this role and embracing it intentionally.

We begin by briefly surveying some of the main theories of these larger-scale processes.

## 2. The Trajectory of Evolution

Many theorists have suggested that evolution exhibits large-scale trends (see Blitz, 1992; Ruse, 1996 for overviews).

In particular, it has often been noted that evolution has moved through a sequence of transitions in which smaller-scale entities are organized into larger-scale cooperatives. Self-replicating molecular processes were organized into the first simple cells, communities of simple cells formed the more complex eukaryote cell, organizations of these cells formed multi-cellular organisms, and organisms were organized into cooperative societies. A similar sequence appears to have unfolded in human evolution: from family groups, to bands, to tribes, to agricultural communities and city states, and so on (e.g. see de Chardin, 1965; Corning, 1983; Blitz, 1992)

As this sequence unfolded, a progressively higher proportion of living processes were organized into cooperative organizations of larger and larger scale. This long sequence also seems to have been associated with a series of improvements in evolvability (the capacity to discover effective adaptations).

However, evolutionary biologists have been very reluctant to accept that these apparent patterns represent actual evolutionary trajectories (e.g. see Gould, 1996; Ruse, 1996).

In large part this is because these hypotheses about directionality were not accompanied by explanations of how the claimed trajectories were produced by known evolutionary processes. This left them open to the criticism that they necessarily relied on impermissible teleological mechanisms.

The view that the evolutionary process is not directional eventually became widely accepted within the modern evolutionary synthesis (Gould, 1996). But this was not because any evidence or theory conclusively ruled out large-scale directionality. Instead, as demonstrated in detail by Ruse (1996), opposition to directionalism was given considerable impetus by the actions of the founders of the synthesis *who were in fact themselves directionalists*. As part of their intentional efforts to build the professional standing of evolution as a scientific discipline, the founders used their power as editors and peer reviewers to cleanse the synthesis of notions of direction, progress and purpose. Apparently they feared that to do otherwise would embroil their fledgling field in public controversy and attract criticism that it was unscientific. Ironically, the intentional and sustained efforts of directionalists paved the way for anti-directionalism to become mainstream dogma in evolutionary biology during the second half of the twentieth century.

### *2.1 The evolution of cooperation*

Until near the end of the 20th century, the hypothesis that evolution moves in the direction of producing cooperative organizations of larger and larger scale gained little traction. In large part this was because mainstream biology held to the view that selfishness, rather than cooperation, is favored in evolution (e.g. see Williams, 1966; Dawkins 1976). This position notes that selection will act against entities that invest resources in cooperation but do not capture any of the benefits it produces. They will be outcompeted by 'selfish' entities that undermine cooperation by taking the benefits without contributing anything in return (e.g. cheats, free-riders and thieves).

According to this position, only limited forms of cooperative organization are likely to emerge at any level, and then only under restricted conditions. Cooperation will be restricted to those limited circumstances in which individual entities are somehow able to capture the benefits of cooperating. This can occur where the interactions between entities are 'collectively autocatalytic' (i.e. where the actions of each entity that participates in a cooperative process increases the fitness of one or more others, and its fitness is in turn increased by other entities). The simplest form of collective autocatalysis is where two entities engage in reciprocal exchanges of benefits.

Examples at various levels of organization of cooperation that is collectively autocatalytic are: autocatalytic sets of proteins (e.g. see Bagley and Farmer, 1991); RNA hypercycles (e.g. see Eigen and Schuster, 1979); autocatalytic cycles of indirect mutualism in ecosystems (e.g. see Ulanowicz, 1986); kin selection amongst multi-cellular organisms (e.g. see Hamilton, 1964); and reciprocal altruism (direct and indirect) amongst humans (e.g. see Trivers, 1971).

However these forms of collective autocatalysis fall far short of accounting for the major evolutionary transitions. They are unable to explain how evolution has apparently organized smaller-scale entities into complex larger-scale cooperatives that eventually become entities in their own right.

In part this is because free-riding, cheating and theft can be expected to undermine and disrupt autocatalytic processes. Furthermore, these processes will emerge only where interactions between entities just happen to form a closed autocatalytic system. There is no reason to expect that advantageous cooperative processes will be collectively autocatalytic. Those that are not will fail to self-organize and will be undermined by individual selection. The complex forms of cooperative organization that are necessary if a group is to become an entity in its own right will not emerge.

### *2.2 Advances in understanding the evolution of complex cooperative organization*

However, in the past two decades considerable progress has been made in understanding how evolution has repeatedly organized independent entities into larger-scale cooperatives. A number of researchers have contributed to the development of a thorough understanding of specific transitions, such as the transition to multi-cellularity (Buss, 1987; Michod, 1999). Others have attempted to develop more general models that are applicable to all the transitions to larger-scale cooperation (Stewart 1995, 1997a,b, 2000; Maynard Smith and Szathmary, 1995; Heylighen, 2006).

In general, this work has shown that evolution can organize complex cooperation amongst self-interested individuals once particular conditions are met.

Stewart (1995; 1997a,b, 2000) shows that evolution will favor the emergence of cooperation amongst self-interested entities when they are embedded in a particular form of organization that makes it in their interests to cooperate. In this form of organization, sets of evolvable constraints (managers) constrain the activities of the self-interested entities, preventing free riding and other actions that undermine cooperation. Furthermore, the evolvable constraints ensure that entities that contribute to effective cooperation are rewarded. As a result, entities capture the benefits of their cooperation and cooperation can be favored by individual selection.

An organization managed by a set of evolvable constraints therefore escapes the limitations that prevent collective autocatalysis from producing complex cooperative organization. The manager can ensure that any cooperative process that benefits the organization as a whole is sustainable, even if the process itself is not collectively autocatalytic. It achieves this by using its power to ensure that entities that contribute to the process benefit from doing so, as well as by restraining free-riding and cheating.

Examples of evolvable constraints include the RNA that managed early cells, the DNA that is reproduced in each cell of a multi-cellular organism (and that thereby manages the interactions between cells), and the governments that manage modern human societies (e.g. see Stewart, 2000).

Importantly, a manager is able to harvest some of the benefits that flow from any cooperation that it organizes. As well as using these resources to reward cooperators, the manager can also use some to enhance its own fitness. In this way it captures some of the benefits of organizing cooperation. Selection will therefore tend to favor managers that are able to organize effective cooperatives.

As a consequence, it is in the interests of the manager to organize cooperation, and in the interests of the smaller-scale entities to cooperate. In this form of organization, the interests of all the members of the organization (including the manger) are aligned with the interests of the organization as a whole.

Evolution will tend to favor cooperative organizations over independent entities because of the advantages that cooperation can provide. For example, cooperation enables the exploitation of synergies, including through specialization and division of labor (Corning, 1983; Stewart, 2000).

And the larger the scale of cooperative organization, the more resources commanded by the cooperative, the greater its power, the greater the impact and scale of its actions, and therefore the wider the range of environmental challenges that it can meet successfully. And the greater the evolvability, the greater the capacity to respond effectively to any adaptive needs and opportunities.

Larger scale and greater evolvability both have the potential to provide evolutionary advantage to living processes across a wide range of environments. This is because they are meta-adaptive capacities—they improve the ability to adapt in all circumstances, although they are not themselves an adaptation to any specific circumstance (it is also worth noting that both are deeply interrelated—increases in scale and power generally increase the range of possible adaptive responses, and hence enhance evolvability [Stewart, 1995])

As improvements in these capacities are discovered, life will tend to evolve directionally. Of course, this trajectory will often be masked by meandering, halting and back-tracking, particularly where the process that searches for improvements relies on blind trial and error. Furthermore, improvements in these capacities will be favored only when the advantages they provide outweigh their cost. As a consequence, directional change will often stall until evolution discovers a cost-effective way of enhancing the capacities.

Taken together, the research of the last two decades constitutes a very strong case that the apparent trajectory of evolution towards larger-scale cooperative organization and greater evolvability is ‘real’. It is driven by processes that do not rely on teleology and are readily understandable within modern evolutionary theory.

## 3. Extrapolating the Trajectory

The extrapolation of this trajectory is reasonably straightforward, at least initially.

The next major transition on Earth would be the formation of global cooperative society (Stewart 1995, 1997a, 2000; Heylighen, 2007). Such a society would be enabled by a system of global constraints (governance) that organizes cooperation (including market processes) and that suppresses destructive competition (including war and international pollution). The evolvability of human society is also likely to increase rapidly through the continued development of artificial intelligence and other technology, and also through the fundamental transition in human psychology which will be discussed in Section 5 below.

Extrapolating this trajectory further would see the continued expansion of the scale of cooperative organization out into the solar system and beyond, and the further enhancement of evolvability (including through the intensification and compression of intelligence discussed by Smart, 2008). However, the expansion of the scale of cooperative organization might occur largely by linking with other organizations of living processes that originate elsewhere, rather than by ‘empire building’. This linking up could be expected to greatly increase evolvability through the exchange of knowledge and technology (including artificial intelligence).

The possibility of life arising elsewhere seems high. There does not appear to be anything special about this planet, the emergence of life on it, or its evolutionary trajectory that make it unlikely to have occurred elsewhere. The details can be expected to differ, but it is likely that the general increase in evolvability and the step-wise increase in the scale of cooperative organization will be ubiquitous. And as we will discuss below, any other living processes that are expanding out into space can be expected to be cooperators, not hostile. (However, there is no consensus within the scientific community about the likelihood of extraterrestrial life and the 'specialness' of Earth. For an overview of the debates, see Dick, 1996.)

If the trajectory continues in this way, the scale of cooperative organization would continue to expand throughout the universe, comprised of living processes from multiple origins. As it increased in evolvability and scale, its command over matter, energy and other resources would also expand, as would its ability to use these resources intelligently to pursue whatever objectives it chooses. The ability of cooperative life to influence large-scale events in the universe would increase, and it might even develop a capacity to impact on processes outside the universe we currently know.

The question of whether the trajectory is likely to continue in this way is discussed in Section 5.2.

## 4. The Developmental Possibility

### *4.1 Current developmental hypotheses*

With our current level of knowledge and intelligence, we cannot say much about the nature of any larger-scale processes in which our universe is embedded. But as a number of theorists have noted, the following considerations raise some intriguing possibilities (see Davies, 2006 for a broad overview):

(1) The known universe exists (there is something rather than nothing), and it is reasonable to presume that it owes its existence to processes that exist outside it.

(2) The fundamental laws and parameters of the known universe seem extraordinarily finely tuned to the needs of life—slight changes would produce a universe in which life would seem highly unlikely.

(3) the evolution of life follows a trajectory that appears likely to eventually produce a universe that is controlled and managed in significant respects by intelligent life (including artificial intelligence).

A number of theorists have tried to account for these considerations by suggesting that our universe and the evolution of life within it is a developmental process that has itself been shaped by evolutionary processes of even wider scale (Crane, 1994; Harrison, 1995; Gardner, 2000, 2003, 2005, 2007; Smart, 2000, 2002, 2008; Vidal, 2008).

According to this hypothesis, the basis laws and parameters of the universe have been tuned so that it eventually develops into an entity that is managed by intelligence. This intelligence is 'developmentally destined' to organize the reproduction of the universe and to tune the parameters of the 'offspring universes' so that they are especially conducive to the development of life and intelligence. The effectiveness of the tuning of the developmental process is enhanced as the cycle repeats.

The developmental singularity hypothesis (Smart, 2002, 2008) includes much of this basic schema, but suggests that life will transcend the universe and initiate the reproduction cycle without linking up with other living processes on the scale of the universe. Smart builds on the idea that life on Earth will enter a post-biological technological singularity in the relatively near future, possibly this century (Adams, 1909; Good, 1965; Vinge, 1993; Broderick, 1997). This accelerating trend towards higher evolvability will continue as intelligence rapidly increases in density and efficiency through compression in matter/energy and space/time. Eventually, this is postulated to produce local intelligences with black-hole-analogous features—a highly local, dense and maximally computability-efficient network of entities that the hypotheses terms a 'developmental singularity'. Smart goes on to suggest that this local (Earth originating) intelligence will interact with some other intelligences that originate elsewhere, and then begin the process of universe reproduction in the quantum domain of black holes.

Smart points out that a particular strength of his hypothesis is its ability to parsimoniously explains the Fermi Paradox—the hypothesis suggests that the reason why we do not see evidence of intelligence or life elsewhere in our universe is because soon after life reaches our stage of development, it enters the developmental singularity, effectively disappearing from view in our space/time.

### 4.2 Other developmental possibilities

However, work on developmental models is only in its early infancy. Existing models do not explore all the broad classes of developmental possibilities and their implications.

In particular, the key models incorporate the assumption that the only source of inherited information provided to offspring universes is transmitted through the equivalent of the germ line—i.e. through the particular values of the fundamental laws and parameters that shape how the offspring universes will develop.

This assumption seems to be based largely on an analogy with the development and evolution of life on Earth prior to the emergence of cultural evolution. For most organisms on the planet, inherited information is transmitted primarily through the germ line. Very little is transmitted from parent to offspring during their lives, or between adults. This is a serious limitation in evolvability—all the adaptive information that is acquired by an organism during its life dies with it.

This limitation was overcome somewhat with the evolution of humanity and the emergence of cultural modes of transmission. Humans undergo a relatively short period of development in the womb where the information they inherit is largely restricted to the germ line. But this is followed by a much longer period in which they acquire cultural information that has been accumulated across the generations and is transmitted to them during their lives. The emergence of cultural transmission paved the way for the massive enhancement in evolvability that produced human science and technology.

The transmission mechanism postulated by current developmental models of the universe is not quite as limited as the mechanism that applied prior to the emergence of cultural transmission on Earth. Most models suggest that when an intelligence is tuning the parameters of offspring universes, it would draw on the knowledge it acquired during the life of the universe. But this is still an extremely limited information channel. Most of the science and knowledge acquired by the intelligence would be lost. This limitation would be even more serious if universes governed by intelligence engage in extra-universal interactions and activities that affect their evolutionary success. Most of what is learnt about those interactions would be lost.

We could therefore expect that extra-universal evolutionary processes would favor transmission between parent universes and their offspring, and between adult and young universes, once the young universes have developed sufficiently. It would also favor transmission between adult universes. If these forms of transmission are achievable, they would significantly enhance the evolvability of offspring universes, including their capacity to engage in extra-universal interactions that affect their fitness. The germ line of a universe that fails to develop or to receive these forms of transmission could be expected to suffer a similar fate to a human germ line that fails to support cultural transmission.

Life and intelligence that is committed to contributing to the successful development of its universe can therefore be expected to search for every feasible way of opening up possibilities for such transmission. It will also seek to exploit any potential for other cooperative interactions between universes for whatever projects are relevant to evolution and development at that level.

The fact that we are not yet receiving such transmission does not rule out its existence. It might simply mean that like individual humans, life in the universe might have to achieve a particular level of development before this form of transmission is feasible and productive. This possibility is also consistent with the likelihood that intelligent life on a planet would not be contacted by life originating elsewhere until it reaches a particular level of development (Stewart, 2000; 2008a).

It is also far too early to rule out the possibility that transmission can occur between intelligent universes, or that intelligent universes can be involved in some extra-universal evolutionary dynamic that involves interaction between them and some larger environment. To attempt to decide these issues on the basis of current physics would be even less reliable than Lord Kelvin's impressive 1895 proof that heavier-than-air flight is impossible. The difficulty we face in trying to evaluate these possibilities at our current scale and intelligence would be similar to the challenge facing an intelligent bacterium in our gut that is trying to make sense of the social interactions that humans engage in.

We are in a similar position in relation to developmental hypotheses in general. We have insufficient data at our present state of knowledge and intelligence to adequately test these hypotheses. And there are a number of competing non-developmental hypotheses that can account for the same evidence. Some alternatives such as the multiverse hypothesis account for the apparent fine tuning of the universe for life on the basis of a kind of blind trial and error—it postulates a large number of universes each with different fundamental laws and parameters, with chance favoring the likelihood that some will suit the emergence of life (e.g. see Susskind, 1995). Other alternatives are similar to the developmental hypotheses in that they account for fine tuning through the operation of intelligence. For example, a class of models suggests that our universe could be a simulation designed and initiated by an unknown intelligence operating outside the universe (Bostrom, 2003; Martin, 2006).

To summarize to this point, there is a very strong case that the evolution of life on Earth has been directional. There is also good reason to suggest that this trajectory applies to life that originates elsewhere in the universe. If the trajectory continues, life from different origins will link up to form cooperative organizations of increasing scale and evolvability. This is consistent with the possibility that the evolution of life in the universe is itself part of a larger developmental and evolutionary process. Other evidence such as fine-tuning is consistent with this possibility. But it is far too early to treat developmental hypotheses as anything other than possibilities.

## 5. The Transition to Intentional Evolution and Development

This section focuses on a critical psychological transition that needs to occur if evolution on a planet is to advance beyond the emergence of a global society and go on to contribute to any larger-scale developmental and evolutionary processes.

Up until the emergence of global society, natural selection and cultural processes will tend to drive evolution along its trajectory towards greater evolvability and increased scale of cooperation. As we have seen, cooperative organizations that are larger in scale and more evolvable will out-compete others. But these forms of competition-driven selection will come to an end as a global society emerges. The global society will not be involved in direct or immediate competition with other planetary societies (Stewart, 2000).

Once this stage is reached, the actions and objectives of the global society would be determined by the values and goals of its members (provided the global society is organized democratically). The society will do what is necessary to advance the evolutionary process only to the extent that this is consistent with the goals and motivations of its members.

However, it is highly unlikely that the desires and motivations of the members of the global society will be consistent with the needs of future evolution. Their desires and motivations will have been shaped by past evolution to be successful in previous environments, not for the future. In large part, members of the society will continue to pursue the proxies for evolutionary success implanted in them by their evolutionary past.

In the case of humanity, members of the global society will continue to spend their lives pursuing the positive feelings produced by experiences such as popularity, self-esteem, sex, friendship, romantic love, power, eating, and social status. And they will continue to strive to avoid the negative feelings that go with experiences such as stress, guilt, depression, loneliness, hunger, and shame.

The way in which these desires and motivations are satisfied will be influenced significantly by cultural factors, but the goals of behavior will be largely unchanged. And these goals will in turn determine the ends that will be served by technological advances, including the uses to which artificial intelligence and genetic engineering are put.

It is only by chance that the desires and motivations of global citizens will be consistent with the needs of future evolution and development. The selection processes that shaped these predispositions operated without foresight and were not influenced by the needs of future evolution.

In the much longer term, selection processes will operate to some extent between planetary societies. But these processes will not force immediate changes in values within the societies. For example, societies with values that lead them to vegetate on their planet of origin will have minimal impact and relevance in the future evolution of the universe. But unless new processes begin to operate, there is nothing to drive such planetary societies to change their values so as to align them with any longer-term developmental or evolutionary imperatives. There is nothing to stop them continuing indefinitely to shape their technology (including artificial intelligence) for the satisfaction of ‘stone age’ desires.

## 5.1 *The need for freedom from the constraints of past evolution*

For these reasons, a planetary society is unlikely to intentionally contribute to the success of any larger developmental and evolutionary processes until it realigns its values with the needs of those processes. The success of any wider processes would therefore seem to depend on the willingness of planetary societies to adopt pro-evolutionary goals and values (Stewart, 2000, 2001 and 2008a). Any wider developmental processes within the universe can succeed only to the extent that intelligent life is motivated to carry out the tasks that will advance the process and eventually reproduce the universe.

Here I will suggest that this fundamental change in values is likely to emerge as the members of the planetary society begin to realize the possibility that they are living in the midst of a developing universe.

Once evolution on a planet produces organisms that have the capacity to develop realistic models of their environment, they are likely to develop theories of the evolutionary process that produced them and the world about them. Eventually they are likely to begin to construct models which reveal the direction of evolution on their planet and how the trajectory is likely to continue throughout the cosmos in the future. They will begin to awaken to the possibility that this trajectory is part of a larger developmental process that has been shaped and tuned by wider evolutionary processes that may eventually reproduce the universe.

They will see from their models that selection will drive evolution along its trajectory to the point that they have reached. But they will also see that unless they now commit to intentionally advancing the evolutionary process, it will stall on their planet. Unless they intentionally align their values with those of the wider evolutionary process, life on their planet will not participate in any wider-scale developmental process in the universe. It is as if they are living in the midst of a developmental process that depends for its continued success on their commitment to intentionally advance the process.

Although their immediate desires and emotions will often clash with the demands of future evolution, it is likely that their most fundamental values will be consistent with making a commitment to pursue evolutionary and developmental goals. This is because the deepest and most fundamental values held by intelligent organisms that reach this stage in evolution are likely life-affirming and meaning-seeking.

These values will tend to motivate them to choose to act in ways that lead to the survival and thrival of the living systems of which they are part. They will see that it is only by assisting the advancement of any larger-scale developmental trajectories that they can contribute to something that has a chance of surviving indefinitely. Any other actions they could take would be futile—such actions would have only temporary effects, and in the long run, everything would be as if they had never lived. Setting out intentionally to contribute to the success of any wider-scale developmental and evolutionary trajectories is the only action they can take that keeps open the possibility that their life and actions can have meaning in a larger context.

Deriving pro-evolutionary goals from fundamental values in this way would not commit the naturalistic fallacy (Stewart, 2008a). The fallacy arises when values are derived from facts alone (i.e. when ‘oughts’ are derived from ‘is’s’ alone), not when they are derived from more general values (i.e. when ‘oughts’ are derived from other ‘oughts’ as well as ‘is’s’) [also see Wilson et al., 2003].

## 5.2 *Commitment before certainty*

It is of critical important to recognize that the reasons for the adoption of pro-evolutionary goals are valid irrespective of whether the organisms know with certainty that the universe is a developmental process. The reasoning applies even though they may be in the same position as humanity is at present. Humanity has discovered the trajectory of past evolution and can see how it is likely to continue into the future. But we cannot yet be certain that the trajectory is part of a wider developmental and evolutionary process that will reproduce the universe.

But organisms that reach this stage in the trajectory of evolution will realize that they cannot wait until certainty is achieved before they commit to advancing the evolutionary process. If any developmental process is to have the greatest chance of success, they need to begin immediately to invest their resources and intelligence in advancing the process. It may be a very long time until their science and intelligence is developed to a point where they can know for sure whether they are in the midst of a developmental process. It is likely that the true nature of any larger-scale evolutionary processes (particularly any extra-universal aspects) will only be discovered gradually, after significant improvements in evolvability.

In fact, it seems likely that absolute certainty may never be reached. Any large-scale processes that have shaped the development of the universe may in turn be shaped by even larger-scale processes, and so on. And it would seem that intelligence could never know if it had discovered the processes of the largest-scale, even if there are such processes—it would seem impossible to ever rule out the possibility of new discoveries and advances, or the existence of processes of even wider scale. The development and evolution of life and intelligence seems likely to prove to be in the nature of a journey without a final destination.

Given this context of fundamental uncertainty, and given that the success of any developmental process depends on their willingness to advance the process, organisms would appear to have only one option to ensure that they can participate in larger-scale processes if they exist. They will have to act as if they are in the midst of a developing universe and intentionally take the action needed to advance the process. Strategically, this means they will have to begin immediately to build the capacities they will need to participate in the wider processes if they prove to exist. They need to get into the game and to stay in it long before its final nature is clear. Only by acting as if the universe is a developmental process can they ensure that they will be able to contribute positively to any developmental and evolutionary processes that are eventually proven to exist. A key element of this strategy will be to invest significant resources in attempting to discover all they can about the larger-scale evolutionary processes within the universe and beyond.

For as long as their science leaves open the possibility that they can participate in meaningful, larger-scale developmental and evolutionary processes, they will need to continue to build capacity so they can take advantage of any possibilities that arise. Even if their universe happens to be the first to arise in which life emerges, such a strategy would maximize their chances of developing the means to reproduce the universe and initiate an on-going process.

Advancing the local trajectory of evolution by linking up with life that originates elsewhere would appear an important and productive way to build this capacity. It would enable the formation of cooperatives of larger and larger scale that are more powerful and able to influence larger-scale events in the universe (this would be of critical importance if mature universes prove to participate in extra-universal events and associated evolutionary dynamics). As mentioned earlier, it would also enable evolvability to increase through the sharing of knowledge, intelligence and different perspectives. Life that wants to position itself to contribute to the success of any larger-scale developmental processes in the universe would appear to have good reason to build capacity in this way.

However, some theorists have questioned whether the continued expansion of the scale of cooperative organization is likely. Smart (2008) argues that the key trend will be towards the increasing efficiency and density of intelligent computation which will remain local. Although he acknowledges that the linking up of intelligences from different origins to share knowledge and intelligence would be advantageous, he postulates that this will occur in localized areas, such as black holes. Cirkovic (2008) critiques some of the arguments that have been advanced previously to support the view that the scale of cooperative organization will continue to expand, but not the main argument outlined in this paper.

It is not possible to decide conclusively between these competing hypotheses at our current level of knowledge and intelligence. However, we can be more certain that these issues will be of great interest to any living processes that decide to contribute to the success of developmental and evolutionary processes in the universe. A key priority for pro-developmental life will be to identify the capacities it should build and the actions it should take to best advance these processes.

### 5.3 *The inadequacy of an intellectual commitment*

However, a mere intellectual decision to align their goals with the trajectory of evolution will not free the organisms from the desires, motivations and emotions inherited from their biological and cultural past (Stewart, 2000, 2001). Their thought processes are unlikely to be able to easily modify their inherited predispositions. When a capacity for thought first arises, it would be unable to understand why motivations and emotions influence behavior in the particular ways that they do. It would not understand their adaptive purposes. Selection would therefore be likely to act against the emergence of any capacity for thought to override these predispositions, since it is likely the outcome would be maladaptive.

Thus humans do not use their thought processes to choose their desires and emotional responses, or their likes and dislikes. Humans generally use thinking and rationality to devise better ways to achieve their ends, not to determine their ends.

As a result, in order to fully align their values with the trajectory of evolution, the organisms would first have to free themselves from the predispositions inherited from their biological and cultural past. Since these changes will not be driven by natural selection, the organisms will have to develop this new psychological capacity intentionally, through changes to their ‘software’ rather that their ‘hardware’ (Stewart, 2001).

Living processes that have completed this transition will not be hostile as they move out into space. They will be motivated by pro-developmental goals, and know that in order to achieve these goals, they will need to work cooperatively with other living processes.

### 5.4 *The significance of the transition as a major enhancement of evolvability*

The transition to intentional evolution produces a fundamental change in evolvability and in the way evolution unfolds (Stewart, 2000 and 2001). Prior to the transition, evolution relies largely on natural selection to advance the evolutionary process. But natural selection operates mainly by trial and error, and has no foresight—it is blind to future possibilities. Evolution's search for innovation and adaptation is driven by a particularly unintelligent process.

After the transition, intelligent organisms intentionally seek to advance evolution. When deciding how to adapt, they use foresight and modeling to take into account the future effects of alternative adaptations, including the long-term evolutionary effects. As a result, the evolutionary process itself will begin to advance intelligently and with foresight.

## 6. Implications for Humanity

Humanity is beginning to enter the early stages of the transition to intentional evolution and development.

As outlined earlier, our evolutionary science has established a strong case that the evolution of life on Earth is directional. We have good reasons to believe that this trajectory applies to life wherever it originates, and continues after life reaches our stage. Some of the most recent developments in evolutionary science suggest the possibility that this wider trajectory is itself part of a large-scale developmental process that has been shaped and tuned to reproduce the universe.

Until now, evolution on Earth has moved along this trajectory of its own accord. However, it is becoming increasingly clear to evolutionary science that evolution will continue to advance only if certain conditions are met: humanity must awaken to the possibility they we are living in the midst of a developmental process; we must realize that the continued success of the process depends on our actions; and humanity must commit to intentionally moving the process forward (Stewart, 2008a).

As yet there is no certainty that such meaningful larger-scale developmental and evolutionary processes exist. However, it is clear that if humanity is to put itself in a position to contribute to these processes if they do exist, it must commit to the pursuit of evolutionary goals now, long before certainty is achieved. Robust strategizing does not have to await robust science. It is possible to identify courses of action that are strategically optimal despite radical long-term uncertainty.

As humanity begins to enter the transition to intentional evolution, we are seeing events emerge that are of great evolutionary significance. Similar events are likely to occur on any planet that moves through the transition. In particular, a key milestone on each planet will be its first global scientific conference that is convened to discuss the large-scale evolution of life in the universe and beyond. On this planet, this evolutionary milestone occurred in 2008 (The conference on the Evolution and Development of the Universe, held in Paris in October 2008).

If the transition is to be completed successfully, humanity will have to free itself from the dictates of its biological and cultural past. Humanity will have to align its goals and motivations with the needs of any larger-scale evolutionary and developmental processes (Stewart, 2008a). The imminent possibility of a technological singularity lends great urgency to the development of these new psychological capacities and to the adoption of pro-evolutionary goals. Unless we develop these capacities before hand, the artificial intelligence and other technologies that enter any singularity will have been engineered to serve humanity's 'stone age' desires, as they are at present. They will not be engineered to serve pro-developmental goals.

There is abundant evidence that the psychological organization of humans is such that we have the potential to free our behavior from the dictates of our evolutionary past (Stewart 2000, 2001, 2007). The world's spiritual and contemplative traditions have discovered a variety of practices and techniques that can be adapted to develop the new psychological software that is needed. These practices can also enhance the ability of humans to understand and manage complex systems (Stewart, 2007, 2008a). Work has begun on using the tools of scientific inquiry to model and understand how these capacities can be developed and how the practices that train them can be enhanced (Stewart, 2007). The discoveries of the contemplative traditions about the human potential for enhanced modes of consciousness are being shorn of their mystical associations and are being integrated into mainstream science.

As humanity begins to enter the transition we also see the first attempts to initiate a political program that explicitly seeks to advance the evolutionary process (Stewart 2008a and 2008b).

If humanity goes on to complete this great evolutionary transition, we will have embraced a role that provides meaning and purpose for our existence.

## 7. Acknowledgments

Thanks to Clément Vidal and John Smart for helpful comments on an earlier version of the paper.

**Short Biography:** John Stewart is an Australian-based member of the Evolution of Complexity and Cognition (ECCO) Research Group of the Free University of Brussels. His main interest is in the development of an evolutionary worldview that reveals to us who we are and what we should be doing with our lives. His work on the directionality of evolution and its implications for humanity has been published in key papers in international science journals. A number of his recent papers have focused on psychological development, including the future evolution of consciousness. He is the author of the book 'Evolution's Arrow: the direction of evolution and the future of humanity'. More recently he has finalized 'The Evolutionary Manifesto'. The Manifesto outlines an evolutionary worldview and explores its relevance to humanity. It is at www.evolutionarymanifesto.com.

# *What* and *that* humans do: Participating in the meaning of life, a contributor's critique

Franc Rottiers
Ghent University, Centre for Critical Philosophy
Blandijnberg 2, 9000 Ghent
+32 484 34 18 61
franc.rottiers@ugent.be
www.criticalphilosophy.ugent.be

## Abstract

The aim of this contribution is to critically examine the metaphysical presuppositions that prevail in Stewart's answer to the question "are we in the midst of a developmental process" as expressed in his statement "that humanity has discovered the trajectory of past evolution and can see how it is likely to continue in the future".

## *What* and *that* humans do

Humans select certain elements out of the local dynamics they are confronted with and translate them in records. These selections are part of objectification processes and have a sequential nature. This means that though a selected element can be 'recorded', this does not imply that the record itself has an absolute, 'objective' status. Also, to repeat a particular selection only implies that the objectification procedure is repeated, not that the record will be exactly the same. Of course, some records will be inscribed in a system that, for reasons of visibility and communication, will require some level of stability. When rendered visible and communicable in a system, records can be operationalized. This is what humans do.

What humanity does – and this can be extrapolated from Stewart's argument – is relate to a global system in which the operationalization of records aims at capturing the organization of the global system. This requires the cooperation effort of *all* humans. Moreover it requires an effort founded on competition-driven selection situated at larger developmental scales. In the end however, competition will be obsolete. It is the moment at which the global society will emerge. This moment has however yet to come.

Now is a good time to ask the question whether we indeed are "in the midst of a developmental process?" It is, within Stewart's metaphysical perspective, a *necessary question* to ask. The answer, i.e. "that humanity has discovered the trajectory of past evolution and can see how it is likely to continue in the future", represents a mere *possibility* as we can never be certain about the destination of the trajectory. In fact, this kind of uncertainty is exactly the same as the one expressed in perspectives addressing the *increasing complexity* of society. So what are the consequences of taking up the perspective Stewart presents?

When addressing the necessity of the question, the perspective '*that* humans do', *that* they engage in objectification procedures, *that* they *contribute* to a complex society, has to make room for the perspective '*what* humans do', how and why they *participate* in society, acquire competences and learn to be 'objective' in a society where complexity 'increases'. In the latter perspective, the questioning activity cannot be but necessary and static because the records to be inscribed and the system in which they are inscribed will only allow 'all knowing' global descriptions of the Laplacean demon kind (Matsuno & Salthe, 1992). Such global descriptions however, have a very particular characteristic, i.e. they are relatively absolute. As complexity is confined to the status of 'always increasing', all global descriptions will fit this definition absolutely and the questions that pertain to this absoluteness will be necessary. Relative to this absoluteness, all answers to necessary questions cannot be but possible. The rationale here is that the realm of possible answers is the only realm left over where human evolvability can further 'develop' – as a possible answer. This however, renders the possibility of a questioning activity, especially with regard to the possibility of evolvability and development, immobile. Let's put the perspective more concrete in the context of Stewart's argument: a trajectory is identified and deemed necessary for the evolvability and development of humanity; though the end of this trajectory is known absolutely, i.e. it is inscribed in a likely future that runs analogous with the perspective that complexity in society increases, the answers are merely possibilities relative to this absolute position and in essence disconnected from an engaged entity that can ask possible questions. Only 'humanity' can pose necessary questions, not 'humans'. That is one perspective on the matter.

So what would become possible when we subscribe to the perspective that allows for the possibility of the question and the necessity of the answer, i.e. the answer always being the givenness of complexity, more precisely *that* there is complexity? Well, then the perspective *that* humans do, can to its fullest extent, be further explored. *That* humans do is not about part-taking in a whole, not about cooperative organization, which is merely the operationalization of "an unhappy marriage between atomism and a materialistic (and often mystical) holism in which a predominantly atomistic and functionalist conception of the organism *per se* is coupled with a holistic conception of a 'central directing agency' conceived as a material entity – the so called 'genetic programme' – which is supposed to determine, order and unify the atomic units and events" (Webster & Goodwin, 1982, p. 16). Within such a perspective, which is the one that subscribes to increasing complexity, the whole coincides with the parts, the consequence of which is that the organism as a structured entity cannot claim its place. *That* humans do, is exactly about taking up the perspective that humans can appear as structured entities, meaning that they can appear as engaged individuals, not statically concerned with possible answers trying to tackle society's apparent increasing complexity, but that they can dynamically invest in a questioning activity that allows for the possibility of asking questions pertaining to the matter *that* there is complexity, *what* this exactly is, is not so much of interest. To put it in other words, records can be operationalized, they need not be. This is just one of the possibilities.

The rationale behind Stewart's argument concerning the meaning of life is directed at humanity and directs humans to organize themselves in a cooperative/participative way. The aim of this commentary was to lay bare that Stewart's position gives voice to a perspective that needs to be taken up *as* perspective. Other perspectives are possible. The meaning of life need not be situated in participative practices that are in need of competent and competitive behaviour. Complexity is here and always has been here. It is tangibly present as 'the' necessary answer, also for people that, while unable to participate fully to society due to exclusion mechanisms, do contribute to it. It are those contributions that are in danger to be stripped away from the meaning of life for 'humanity'.

# Analysis of Some Speculations Concerning the Far-Future of Intelligent Civilizations

Clément Vidal
*Department of Philosophy, Vrije Universiteit Brussel, Center Leo Apostel and Evolution, Complexity and Cognition, Brussels, Belgium.*
clement.vidal@philosophons.com http://clement.vidal.philosophons.com

**Abstract**: I discuss some of the speculations proposed by Stewart (2009). These include the following propositions: the cooperation at larger and larger scales, the existence of larger scale processes, the enhancement of the tuning as the cycle repeats, the transmission between universes and the motivations to produce a new universe.

I do largely share Stewart's general vision of cosmic evolution and development, and I very much value the speculations he proposes. Connecting in such a coherent way intelligence and the evolution of the universe is rarely attempted, and even more rarely so inspiring. The significance of the worldview outlined here is potentially of paramount importance for the future of intelligence.

Obviously, it would be boring for the reader to state all the points were we do agree. Therefore, I will focus on (relatively minor) issues where our positions diverge, to sharpen and extend them or hopefully to reach agreement.

As Stewart acknowledges, many themes in this paper are speculative: "the difficulty we face in trying to evaluate these possibilities at our current scale and intelligence is likely to be similar to the challenge facing an intelligent bacterium in our gut that is trying to make sense of the social interactions we engage in." Speculating on those issues can easily lead very far. For this reason, we need to have clear ideas on *how* and *why* we speculate. I distinguish three kinds of speculations to navigate into their variety :

1. **Scientific**: a speculation is scientific if we have strong reasons to think that we will be able in the future to have observations or experimentations corroborating or refuting it.
2. **Philosophical**: a speculation is philosophical if it extrapolates from scientific knowledge and philosophical principles to answer some fundamental philosophical questions.
3. **Fictional**: a speculation is fictional if it extends beyond scientific and philosophic speculations.

## Cooperation at larger and larger scales

At several places in the text, Stewart suggests that cooperation will continue to expand at larger and larger scales (e.g. page 4, 10). It is of course an open possibility, and it would certainly be exciting to be able to communicate with most or even all other supposed intelligent civilizations in the universe. It is a scientific speculation in the sense that we are likely to understand if such a cooperation is possible in the future. Nevertheless, I think there are severe difficulties to this scale extension scenario. My concerns regard the why and the how. First, why would an intelligent civilization want to do that? Travelling in space costs time, money, energy, etc. We know that it is technically possible to go to the Moon. However great and inspiring this exploit is, humanity has done it only a handful of times, certainly because it is very expensive and a great technical challenge, and gives finally a relatively poor pay-off. What would make such an expense of energy profitable?

Second, how would intelligent civilization achieve that? What are the other cooperative organizations interesting to collaborate with in our solar system? Where would intelligent civilization go? The arduous problem is that distances are huge in the universe. We can be reminded that even at the speed of light, it takes about 8 minutes to reach the Sun, 4.2 years to reach the nearest star Alpha Centauri and 2.5 million years to reach the nearest Galaxy Andromeda.

Even assuming such an extension, how could communication actually take place at very large space scales? Communication would be extremely slow, also because of the huge distances. The cooperation would therefore be inefficient and probably not that interesting. Imagine that you have to wait several years to get an answer to a message you just sent...

Some authors speculated that there might be shortcuts (wormholes), leading from one part of the universe to another very distant one (Thorne 1994). However, these wormholes are speculative theoretical entities, and it is even more speculative to suppose that any form of life or information could transit in them.

**The largest scale processes**

Stewart remarks that "intelligence could never know if it had discovered the processes of the largest-scale" (page 9). A first line of reasoning is to state that if space-time is really the basic structure of everything else, then we do not need to suppose any wider processes.

This reasoning is a kind of "tower of turtles" speculation: is there an infinity of larger and larger scales processes? This idea is similar to the proposition that we might be in a computer simulation. We would also never know it for sure, and there might be an infinity of simulations simulating other universes. As with the simulation, the careful answer is that we do not know, and we will certainly never be able to know for sure if there are such processes at play.

My pragmatic response to this issue is: no need to worry. Indeed, let us analyse the proposition with Leibniz' logical principle of the identity of the indiscernibles. This principle states that “if, for every property F, object x has F if and only if object y has F, then x is identical to y”. Let x be our known reality, and y be the supposed largest-scale we would be living in. If we have no way to distinguish between them, they are identical. Unless we find a property F that could only exist in a larger scale-universal process and not in reality, this hypothesis is useless.

For this reason, I would qualify this speculation and the simulation hypothesis as fictional.

**The tuning is enhanced as the cycle repeats**

Stewart proposes that "the effectiveness of the tuning of the developmental process is enhanced as the cycle repeats." (page 5). This speculation is philosophical, and even properly metaphysical. Modern cosmic evolutionary theory shows an increase in complexity from the Big Bang to the development of our modern technological society. So, why shouldn't we also expect a kind of evolutionary progress between universal cycles?

It is yet not necessary to enhance the tuning at the level of universes. All what is needed here is a reproduction mechanism, which is fertile (i.e. enables further universe reproduction). An undesirable (metaphysical) implication we encounter if we suppose this enhancement, is that if we look back in the cycle, earlier universes should be simpler and simpler. Then we face again the fine-tuning problem for these very early universes. We would just have shifted the fine-tuning of our universe to the fine-tuning of other previous universes (See also Vaas and my reply Vidal).

**Transmission between universes**

In analogy with biological organisms, Stewart suggests that a "parent universe" could transmit information to the offspring universes (page 6). This is a very interesting and exciting scientific and philosophical speculation. However, physical constraints are likely to rule out this possibility.

I suspect the constraints from one universe to another are too strong to pass on messages. Let us assume that a whole new disconnected space-time structure is generated from one universe. Such a space-time has a different causal structure from the previous universe. Therefore, it is by definition impossible to make the two communicate, because the common causal relationship between the two vanishes after the reproducing event.

**Motivation to produce a new universe**

Stewart emphasizes the importance of motivation from intelligent life to take part into a supposed developmental process (e.g. page 7). This question of motivation is properly philosophical, more precisely ethical. It is indeed of prominent importance.

However, even if we had the certainty to be in an developmental process, this would be just part of the motivation to produce a new universe. Two other drivers are likely to be central.

First, as described by Stewart page 9, the most fundamental values an intelligent civilization would have are life-affirming, meaning-seeking. Those values are likely to be strongly connected to the idea of surviving indefinitely (Lifton and Olson 2004). A strong commitment to these values would reinforce the willingness of an intelligent civilization to participate actively to the evolutionary process.

However, there is a second driver for intelligence to reproduce the universe. It is well established that there will be a progressive end of all stars in galaxies, including our Sun. The universe will irreversibly decay towards a state of complete disorder or maximum entropy. This scenario is commonly known as the "heat death" (Adams and Laughlin 1997; Ćirković 2003 for a review). Certainly, modern cosmology shows that there are some other models of the end of the universe (such as Big Bounce, Big Rip, Big Crunch..., see (Vaas 2006) for an up-to-date review). However, the point is that none of them allows the possibility of the indefinite continuation of life as we know it. I argued in (Vidal 2008) that awareness to this gloomy fate would be a driver for intelligent civilization to produce a new universe.

It is often objected that this problem is far too far in the future to consider it as a serious preoccupation. However, the situation here is analogous to climate change, except the problem concerns an even larger scale than the Earth. A few decades ago, few people were seriously concerned with climate change. Nowadays, people, various organizations and governments have started to seriously mobilize to tackle this complex issue. What produced this shift? Among other factors, the strong integration and globalization of society contributed to this sensitivity about climate change. Indeed, it is only since recently that we have such numerous and precise channels of information from every corner of the world, which provides us an unprecedented understanding of the planet as a whole. This leads us to a global awareness of what happens on this planet.

Similarly, a global awareness will finally emerge from our increasing understanding of the cosmos. Only after such an awakening, will an intelligent civilization start to actively tackle such a large-scale problem. I hope Stewart's paper and this special issue will contribute to let us look ahead and awaken us to such a cosmic perspective.

# The Future of Life and What it Means for Humanity
## Response to commentaries.

John Stewart
*Member of the Evolution, Complexity and Cognition Research Group, The Free University of Brussels.*
john.stewart@evolutionarymanifesto.com
http://www.evolutionarymanifesto.com

**Abstract:** Vidal's and Rottiers' commentaries raised a number of important issues about the possible future trajectory of evolution and its implications for humanity. My response emphasizes that despite the inherent uncertainty involved in extrapolating the trajectory of evolution into the far future, the possibilities it reveals nonetheless have significant strategic implications for what we do with our lives here and now, individually and collectively. One important implication is the replacement of postmodern scepticism and relativism with an evolutionary grand narrative that can guide humanity to participate successfully in the future evolution of life in the universe.

### R1. Is the evolutionary perspective 'privileged? Can it support a new viable 'Grand Narrative'?

Rottiers suggests that the evolutionary worldview outlined in my paper is just one possible perspective amongst many and that it is not privileged over the multiplicity of other perspectives.

Rottiers' perspective appears to resonate strongly with the postmodernist rejection of the possibility of a 'grand narrative' that is capable of providing a comprehensive account and explanation of human existence. Lyotard (1979) argued that the rise of rationality and science has undermined the possibility of a believable grand narrative. He reaches this conclusion in relation to materialist grand narratives (e.g. Marxism) as well as in relation to ones that rely on supernatural processes. According to Lyotard, attempts to develop materialist grand narratives have failed to meet the tests of science (see, for example, Popper, 1957).

To the contrary, my paper shows that the emerging evolutionary perspective is capable of surviving postmodernist scepticism and relativism. The evolutionary perspective outlined by the paper is not teleological, and meets the other tests of science. Furthermore, it is highly relevant to human concerns and preoccupations.

Evolutionary processes have shaped humanity and will determine its future. Humanity can ignore the evolutionary consequences of its actions, but evolution will not. Evolutionary processes determine what it is that can survive and thrive into the future. Values, ethical principles, perspectives and political systems that are not aligned with these processes will be temporary.

It is true that the ways in which we express our theories of evolution will be influenced by culture. Our theories will always be maps, never territories. And our evolutionary theories will always be susceptible to being appropriated and shaped by particular interests for ideological purposes. But the evolutionary processes-in-themselves are not culture-bound. And our best maps of these processes will be our best guides to what will survive into the future.

Evolution has been and will continue to be 'the great marginalizer'. In the face of evolution, all perspectives, all values and all actions are not equal. Some will survive and flourish, others will not. Denying this or just wishing it were otherwise will not stop marginalization. In fact, it invites it: as I have argued in my paper, a planetary civilization that is not guided by an evolution-based grand narrative is highly likely to be marginalized, locally as well as globally.

## R2. Cooperation at larger and larger scales

Vidal questions whether human civilisation will be motivated to expand beyond Earth and link up with any other civilisations it encounters.

It is true that humanity's *current* values and desires may fail to motivate the actions that are needed to evolve humanity successfully into the future. They may not motivate humanity to exploit the evolutionary advantages that will flow from expanding the scale of human organisation and linking up with any life that originates elsewhere (e.g. the advantages of increased evolvability, greater power to influence events over larger and larger scales, more extensive cooperative division of labour, etc.)

It is therefore likely that if humanity is to continue to survive and thrive into the future it will need to complete the transition to intentional evolution outlined in my paper—humanity will need to intentionally free itself from the desires and needs shaped by its biological and cultural past and instead develop the capacity to find motivation and satisfaction in whatever needs to be done to achieve future evolutionary success. This will also necessitate shaping the economic system so that it appropriately funds pro-evolutionary activities.

Vidal also questions the feasibility and practicality of extra-planetary civilisation due to the huge distances involved.

This is an issue that is encountered to some extent in every transition to larger-scale organisation. The processes that coordinate activities across a cooperative will necessarily operate at significantly slower rates than the processes that operate within the members of the cooperative. The relative slowness of the coordinating processes that organize cooperatives has not prevented their emergence repeatedly during the evolution of life on Earth. This is despite the fact that some of the differences in scale between members and cooperatives are extremely large. For example, the descendants of prokaryote cells that are found within the cells of the human body now participate in human activities that are coordinated across the globe. The relative difference in scale between prokaryote cells and a planetary organisation is broadly comparable to the difference in scale between a planetary organisation and the solar system.

The evolutionary record suggests that the advantages of larger-scale cooperative organisation have repeatedly outweighed the relative slowness of larger-scale coordination processes. As yet, there is no reason to believe that this will cease to be the case for extra-planetary organisations.

## R3. The largest scale processes

Vidal argues that if any postulated largest-scale processes are indistinguishable from our know reality, they are identical to know reality and any speculation about the postulated processes should be classed as fictional.

However, whether any postulated larger-scale processes are distinguishable is likely to depend on the knowledge and technology of the living processes in question. As life develops, it may eventually discover ways to distinguish a yet larger-scale process which encompasses its previously-known reality. This possibility may or may not be relevant to any particular conception of what is classed as 'science' by a particular culture at a particular point in its evolution. But this possibility will be highly relevant to life that wants to make good strategic decisions about how to evolve and perpetuate itself—the nature of any such larger-scale processes may have a significant practical impact on what it needs to do in order to survive and flourish indefinitely.

### R4. The tuning is enhanced as the cycle repeats

Vidal argues that the 'intelligent tuning' hypothesis does not solve any fine-tuning problem for pre-intelligence universes (see also Vaas' commentary to Vidal's paper, this volume).

This is true, but does not rule out the 'intelligent-tuning' hypothesis. The exploration and testing of this hypothesis will be critically important to life and intelligence as it develops in a particular universe. Life will need to discover the precise nature of any such large-scale evolutionary processes (including any extra-universal processes) if it is to work out what it has to do to survive and thrive indefinitely.

### R5. Transmission between universes

Vidal argues that physical constraints rule out the transmission of adaptive information between universes (except during any 'reproduction event').

Humanity's current understanding of the relevant physics certainly seems to suggest this. And life may not discover anything new about how to achieve such transmission as it evolves into the future, linking up throughout the universe, increasing massively in scale, and continually bootstrapping its intelligence over billions of years. It is possible that there is nothing additional to be learnt about this issue beyond what our science has discovered already, after a couple of hundred years.

But if nothing new is learnt, it is unlikely to be for want of trying. The transmission of adaptive information between universes would massively increase evolvability. Recognition of these benefits is likely to have driven attempts to accomplish such transmission in any previous universe, and will surely drive it in this universe.

### R6. Motivation to produce a new universe

Vidal argues that recognition of the fact that life will eventually be obliterated due to the 'heat death' of the universe, a 'big crunch' or other such event will be a powerful driver for intelligence to reproduce the universe.

He makes a strong case. In general, the threat of large-scale extinction events can be expected to strongly motivate life and intelligence to build its capacity to survive the threats (including by increasing its evolvability, its scale and its power). There are many such threats that confront life on this planet, now and in the future: global warming, nuclear war, collision with asteroids, engulfment of Earth by the sun when it becomes a red giant, collision with other star systems, and so on and so on. Due to the prevalence of large-scale extinction threats, intelligent life that declines to develop its capacities and instead spends its time vegetating on its planet of origin in the pursuit of narrow 'stone age' desires is likely to be temporary. It will forgo its opportunity to participate positively in the greatest adventure of all—the evolution of life and intelligence in the universe and beyond.

acceleration (61) according (78) al (90) allows (66) analogy (67) analysis (70)
application (85) applied (63) approach (164) artificial (70)
aspects (68) assume (71) based (101) become (108)
biological (117) biology (121) black (83) cas (73) case (168)
causal (92) cells (159) chaline (61) change (105)
classical (141) closure (94)
complexity (353) concept (84)
conditions (109) conservation (69) consider (159)
constant (277) corresponding (104) cosmic (95)
cosmological (201) critical (108) current (65)
darwinism (70) data (95) date (95) de (172) defined (93)
definition (206) density (114) derived (82)
described (117) description (74) development (186)
developmental (137) different (119) differential (63)
dimension (80) discuss (72) distribution (86) ed (65)
emergence (116) energy (328) entanglement (66)
entropy (129) eq (77) equal (62) equation (154) et (120)
events (153) evolution (534)
evolutionary (226) example (120)
existence (104) expansion (71) expected (70) explain (74) explanation (63)

expression (69) fact (75) field (79) figure (74) finally (73) fine-tuning (76)
following (98) form (170) fractal (196) framework (73)
function (107) fundamental (170) future (124)
general (196) given (108) heylighen (62) hierarchy (99)
http (136) human (101) idea (66) implies (65) important (75)
including (95) increasing (87) indeed (87) individual (64)
information (191) initial (132) intelligent (144)
interesting (76) interpretation (78) larger (66) laws (283) leads (90)
level (208) life (534) living (100) ln (98) major (66)
material (64) mathematical (85) matter (166) means (164)
mechanics (127) metric (76) model (230) motion (81)
multicellular (73) myr (111) nature (179) nottale (129)
number (102) objects (62) observed (63) occur (67) operator (144)
order (161) organisms (150) origin (130)
paper (129) parameters (113) particles (208)
particular (100) pdf (63) perspective (76) phys (68)
physical (402) point (170) position (61)
possible (299) potential (85) pp (215) predicted (73)
present (154) press (123) principle (122) probability (91)
problem (174) process (394) properties (96)

quantum (394) question (100) radiation (64)
random (87) rather (65) reason (87) relation (123)
relativity (202) result (166) role (102) salthe (67)
scale (588) science (130) scientific (129)
section (87) selection (153) similar (76) solution (108)
space (141) species (97) standard (85) state (198)
structure (194) suggests (125)
system (492) tc (98) terms (162) theoretical (71)
theory (307) therefore (168) towards (64) trajectory (67)
transformations (61) transition (112) ts (66) understanding (66)
universe (642) used (101) vacuum (139)
value (108) variable (91) various (71) velocity (61) vidal (68) view (92)
work (68) world (114) years (86) york (65)